# Precision Spectroscopy of the 2S-$n$P Transitions in Atomic Hydrogen

Lothar Maisenbacher

# Precision Spectroscopy of the 2S-$n$P Transitions in Atomic Hydrogen

**Lothar Maisenbacher**

Dissertation
an der Fakultät für Physik
der Ludwig–Maximilians–Universität
München

vorgelegt von
Lothar Maisenbacher

München, den 10.12.2020

**Note on this version**

This is a revised version of the thesis originally submitted in December 2020. Minor revisions include corrections identified during the preparation of subsequent publications. A list of changes is provided at the end of this document. The original thesis is available from the Ludwig-Maximilians-Universität München library at doi:10.5282/edoc.29054.

Revised: December 2025

# Zusammenfassung

Präzisionsspektroskopie an atomarem Wasserstoff ist eine wichtige Methode die Quantenelektrodynamik (QED) gebundener Systeme, einer der Bausteine des Standardmodells, zu testen. Im einfachsten Fall besteht ein solcher Test aus dem Vergleich einer gemessenen Übergangsfrequenz mit der Vorhersage der QED, welche für das Wasserstoffatom mit sehr hoher Präzision berechnet werden kann. Diese Berechnungen benötigen allerdings bestimmte physikalische Konstanten als Eingangsparameter, unter anderem die Rydberg-Konstante $R_\infty$ sowie den Protonenladungsradius $r_\mathrm{p}$, welche gegenwärtig beide zu einem großen Teil selbst durch Wasserstoffspektroskopie bestimmt werden. Für einen Test der QED ist es deshalb notwendig, die Übergangsfrequenzen von mindestens drei verschiedenen Übergängen zu bestimmen. Gleichermaßen ist ein Vergleich der aus Messungen verschiedener Übergänge bestimmten Werte für $R_\infty$ und $r_\mathrm{p}$ ein Test der QED.

Hierzu wurde in dieser Arbeit Laserspektroskopie der optischen 2S-$n$P-Übergänge durchgeführt. Da es sich bei diesen Übergängen um Ein-Photonen-Übergänge handelt, sind sie von einem anderen Satz an systematischen Effekten betroffen als Zwei-Photonen-Übergänge, auf denen die meisten anderen spektroskopischen Messungen an Wasserstoff basieren. Um zu einem Test der QED beitragen zu können, muss ihre Übergangsfrequenz mit einer relativen Unsicherheit in der Größenordnung von eins zu $10^{12}$ bestimmt werden, in absoluten Einheiten etwa auf 1 kHz. Dies ist etwa 10 000 Mal kleiner als die relativ große natürliche Linienbreite der 2S-$n$P-Übergänge, weshalb für eine erfolgreiche Messung sowohl ein sehr großes Signal-zu-Rausch-Verhältnis als auch ein detailliertes theoretisches Verständnis der Linienform der beobachteten Resonanz notwendig ist.

Die 2S-$n$P-Übergänge wurden an einem kryogenen Strahl aus Wasserstoffatomen, die optisch in den metastabilen 2S-Zustand angeregt wurden, untersucht. Der Atomstrahl wurde rechtwinklig mit zwei gegenläufigen Spektroskopie-Laserstrahlen gekreuzt, die die Atome weiter in den $n$P-Zustand anregten. Die Fluoreszenz des anschließenden, raschen spontanen Zerfalls diente als Messsignal. Die Anregung mit zwei gegenläufigen Strahlen führt zu zwei Dopplerverschiebungen der gleichen Größe, aber mit umgekehrten Vorzeichen, die sich damit aufheben. Eine nach der Geschwindigkeit der Atome aufgelöste Detektion erlaubte die Bestimmung eventuell verbliebener Dopplerverschiebungen, die im Rahmen der Messunsicherheit jedoch für beide unten vorgestellten Messungen ausgeschlossen werden konnten.

In einem ersten Experiment wurde der 2S-4P-Übergang untersucht. Quanteninterferenz zwischen benachbarten atomaren Resonanzen führte zu subtilen Verformungen der Linienform, die sich aufgrund der sehr hohen Auflösung bezogen auf die Linienbreite als signifikant herausstellten. Die durch diese Verformungen verursachten Linienverschiebungen konnten direkt beobachtet und mit einem auf Störungstheorie basierenden Linienformmodell entfernt werden. Somit konnte die Übergangsfrequenz mit einer relativen Messungenauigkeit von 4 zu $10^{12}$ bestimmt werden. In Kombination mit der sehr präzise gemessenen 1S-2S-Übergangsfrequenz erlaubte dies die zu dem Zeitpunkt präziseste Bestimmung von $R_\infty$ und $r_\mathrm{p}$ mittels Spektroskopie an atomarem Wasserstoff. Darüber hinaus wurde eine gute Übereinstimmung mit dem durch Spektroskopie an myonischem Wasserstoff bestimmten, sehr viel präziseren Wert für $r_\mathrm{p}$ festgestellt, welcher signifikant von den vorherigen Daten aus (elektronischem) Wasserstoff abweicht und damit zu Zweifeln an der Gültigkeit der QED geführt hatte. Der myonische Wert für $r_\mathrm{p}$ wurde seitdem von weiteren Experimenten bestätigt. Die 2S-4P-Messung wird im Anhang dieser Arbeit behandelt.

Trotz des hohen Signal-zu-Rausch-Verhältnisses war die Genauigkeit der 2S-4P-Messung durch die Zählrate des Messsignals limitiert. Um die Präzision weiter zu erhöhen, war ein Übergang mit kleinerer Linienbreite und ein verbessertes experimentelles Signal notwendig. Deshalb wurde mit der Untersuchung des 2S-6P-Übergangs, welcher eine dreimal kleinere natürliche Linienbreite bietet, begonnen. Der Atomstrahlapparat wurde modifiziert, wodurch eine entsprechende Reduktion der experimentell beobachteten Linienbreite und ein fast eine Größenordnung höherer Fluss an langsamen Atomen im Atomstrahl erreicht werden konnte. Zusammen mit einer Neukonstruktion des Detektors führte dies im Vergleich zur 2S-4P-Messung zu einem bis zu 16-fach höherem Signal und machte damit den Weg zu höherer Präzision frei. Um diese Präzision zu ermöglichen, war darüber hinaus eine Weiterentwicklung der Dopplerverschiebungsunterdrückung notwendig. Dazu wurde ein Faserkollimator entwickelt, der eine exzellente Strahlqualität der Spektroskopie-Laserstrahlen bei der neuen Übergangswellenlänge von 410 nm bietet.

Dies ermöglichte eine Messung der 2S-6P-Übergangsfrequenz mit einer statistischen Unsicherheit von 430 Hz, fünfmal niedriger als für die 2S-4P-Messung. Dies entspricht einer Unterdrückung der Dopplerverschiebung um sechs Größenordnungen. Bei dieser Präzision wird die Lichtkraftverschiebung durch die Beugung der Atome am optischen Gitter, welches durch die gegenläufigen Laserstrahlen erzeugt wird, signifikant. Diese Lichtkraftverschiebung wurde zum ersten Mal für die 2S-$n$P-Übergänge beobachtet und konnte durch ein hierfür entwickeltes Modell gut beschrieben werden. Die Größe aller anderen systematischen Effekte, mit Ausnahme der sehr genau bekannten Rückstoßverschiebung, wird mit jeweils kleiner als 500 Hz abgeschätzt. Die blinde Datenanalyse ist zum Zeitpunkt des Verfassens dieser Arbeit noch im Gange, weshalb noch keine Übergangsfrequenzen angegeben werden können. Die vorläufige Analyse lässt jedoch eine fünffache Verbesserung der Bestimmung von $R_\infty$ und $r_\mathrm{p}$ im Vergleich zur 2S-4P-Messung und eine zweifache Verbesserung im Vergleich zur momentan präzisesten Bestimmung an atomarem Wasserstoff erwarten. Damit liegt die Unsicherheit auf den bestimmten Wert für $r_\mathrm{p}$ innerhalb eines Faktors fünf der Unsicherheit des myonischen Wertes. Die 2S-6P-Messung ist zentraler Gegenstand dieser Arbeit.

# Abstract

Precision spectroscopy of atomic hydrogen is an important way to test bound-state quantum electrodynamics (QED), one of the building blocks of the Standard Model. In its simplest form, such a test consists of the comparison of a measured transition frequency with its QED prediction, which can be calculated with very high precision for the hydrogen atom. However, these calculations require some input in the form of physical constants, such as the Rydberg constant $R_\infty$ and the proton charge radius $r_\mathrm{p}$, both of which are currently determined to a large degree by hydrogen spectroscopy itself. Therefore, the frequency of at least three different transitions needs to be measured in order to test QED. Equivalently, a comparison of the values of $R_\infty$ and $r_\mathrm{p}$ determined from measurements of different transitions constitutes a test of QED.

To this end, laser spectroscopy of optical 2S-$n$P transitions has been performed in this work. As these transitions are one-photon transitions, they are affected by a different set of systematic effects than the two-photon transitions on which most other spectroscopic measurements of hydrogen are based. In order to contribute to the test of QED, their transition frequencies must be determined with a relative uncertainty on the order of one part in $10^{12}$, corresponding to approximately 1 kHz in absolute terms. This is in turn approximately a factor of 10 000 smaller than the relatively broad natural linewidth of the 2S-$n$P transitions, and a successful measurement requires both a very large experimental signal-to-noise ratio and a detailed theoretical understanding of the line shape of the observed resonance.

The 2S-$n$P transitions were probed on a cryogenic beam of hydrogen atoms, which were optically excited to the metastable 2S level. The atomic beam was crossed at right angles with counter-propagating spectroscopy laser beams, which further excited the atoms to the $n$P level. The fluorescence from the subsequent rapid spontaneous decay served as experimental signal. The excitation with two counter-propagating beams led to two Doppler shifts of equal magnitude, but opposite sign, which thus canceled each other out. A velocity-resolved detection was used to determine any residual Doppler shifts, which could be excluded within the measurement uncertainty for both of the measurements discussed below.

In a first experiment, the 2S-4P transition was probed. Quantum interference of neighboring atomic resonances produced subtle distortions of the line shape, which were found to be significant because of the very large resolution relative to the linewidth. The line shifts caused by the distortions were directly observed and could be removed by use of a line shape model based on perturbative calculations. With this, the transition frequency was determined with a relative uncertainty of 4 parts in $10^{12}$. In combination with the very precisely measured 1S-2S transition frequency, this allowed the, at the time, most precise determination of $R_\infty$ and $r_\mathrm{p}$ from atomic hydrogen. Moreover, good agreement was found with the much more precise value of $r_\mathrm{p}$ extracted from spectroscopy of muonic hydrogen, which had been in significant disagreement with previous data from (electronic) hydrogen, causing concern about the validity of QED. This result has since been confirmed by other experiments. The 2S-4P measurement is treated in the appendix of this thesis.

The 2S-4P measurement, despite its large signal-to-noise ratio, was limited by counting statistics. To improve precision, a transition with a narrower linewidth and an improved experimental signal was necessary. Hence, the study of the 2S-6P transition, which offers a three times smaller natural linewidth, was begun. The atomic beam apparatus was upgraded, resulting in a corresponding decrease of the experimentally observed linewidth, and a close to an order of magnitude larger flux of atoms in the low-velocity tail of the atomic beam. Together with a detector redesign, this led to an up to 16 times larger signal than for the 2S-4P measurement, opening the path to increased precision. The Doppler-shift suppression was also rebuilt to support such precision, including a fiber collimator developed for this purpose, which provides high-quality spectroscopy beams at the new transition wavelength of 410 nm.

This enabled a measurement of the 2S-6P transition frequency with a statistical uncertainty of 430 Hz, five times lower than for the 2S-4P measurement and corresponding to a suppression of the Doppler shift by six orders of magnitude. At this level of precision, the light force shift from the diffraction of atoms at the light grating formed by the counter-propagating spectroscopy beams becomes significant. This light force shift was directly observed for the first time for the 2S-$n$P transitions and found to be well-described by a model derived for this purpose. The size of all other systematic effects, except the very precisely known recoil shift, is estimated to be below 500 Hz each. The blind data analysis is ongoing at the time of writing and thus no transition frequencies can yet be given. However, a preliminary analysis suggests a five-fold improvement in the determination of $R_\infty$ and $r_\mathrm{p}$ as compared to the 2S-4P measurement, and a two-fold improvement over the currently most precise determination from atomic hydrogen. This places the uncertainty of the determined value of $r_\mathrm{p}$ within a factor of five of that of the muonic value. The 2S-6P measurement is treated in the main text of this thesis.

# Contents

# List of Figures

# List of Tables

# Chapter 1

# Introduction

## Hydrogen spectroscopy

Hydrogen spectroscopy [1] is fundamental to our understanding of physics, not least because of its instrumental role in the development of quantum mechanics [2] and its subsequent evolution into quantum electrodynamics (QED) [3]. The success of QED [4, 5], in turn, inspired quantum field theory (QFT), which forms the basis of the Standard Model of particle physics and includes QED as one of its building blocks. The Standard Model has been experimentally tested extensively, culminating in the discovery of the long predicted Higgs boson [6, 7]. Nevertheless, the Standard Model is known to be incomplete, as it only includes three out of the four fundamental interactions, missing a QFT of gravity. It also cannot describe, in its present form, experimental observations such as neutrino oscillations [8, 9], which imply a nonzero neutrino mass. Moreover, astrophysical observations such as galactic rotation curves [10] and the anisotropies of the cosmic microwave background [11] challenge our current understanding of matter and the evolution of the universe itself, which has lead to the hypotheses of dark matter and dark energy [12]. Traces of this physics beyond the Standard Model, or just new physics, are expected to be detectable in atomic and molecular physics experiments through interactions with various hypothesized fields and particles [13–18].

The role of hydrogen spectroscopy in this pursuit is to compare with ever-increasing experimental precision the frequency of various transitions, i.e., the energy difference between atomic levels, to the predictions of bound-state QED. This is because the QED predictions for the energy levels of the hydrogen atom, being the simplest of all stable atoms, can be calculated with an extreme precision, exceeding 12 digits for some transitions [4].

## The proton radius puzzle

However, some input from experiments in the form of physical constants is generally required to calculate those predictions, either because the needed constants cannot (yet) be calculated, such as the fine-structure constant $\alpha$, the electron-to-proton mass ratio $m_e/m_p$, and the proton charge radius $r_p$, or they serve as conversion factors from natural units to the system of units used in the laboratory, as is the case for the Rydberg constant $R_\infty$. For the current state of theory and experiments, $r_p$ and $R_\infty$ are at least partly determined from hydrogen spectroscopy[1], such that measurements of at least three distinct transitions are needed to compare the predictions with the experiments. Put another way, if QED is correct, the determinations of these constants from different measurements are expected to be consistent. It is thus no surprise that a novel, very accurate determination of the proton charge radius through spectroscopy of the 2S-2P transition in muonic hydrogen[2] [21, 22] was met with considerable interest. This is because it disagreed by four standard deviations ($\sigma$) with the previously established value (“2014 H world data”, adjustment 8 of [4]) determined from the spectroscopy of (electronic) atomic hydrogen (H). This situation is shown Fig. 1.1. Despite a flurry of theoretical activity, no widely accepted solution to this proton radius puzzle (PRP) has been put forward.

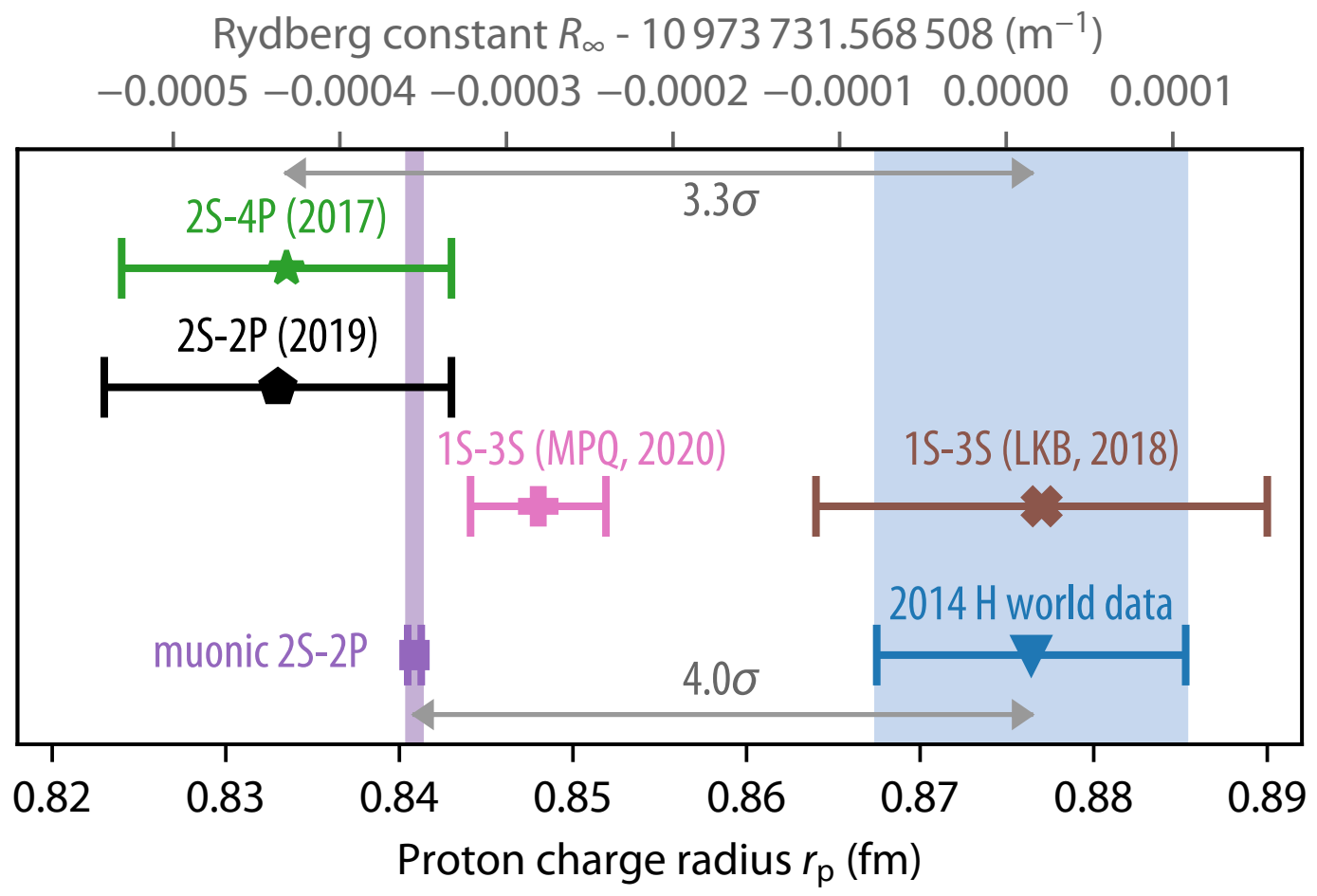

Figure 1.1: Determinations of the Rydberg constant $R_\infty$ (top axis) and proton charge radius $r_\mathrm{p}$ (bottom axis) from hydrogen spectroscopy. The 2014 H world data (adjustment 8 of [4], blue triangle and blue-shaded region) consists of the 16 most precise hydrogen spectroscopy measurements, including the 1S-2S measurement of [23], available as of 2014. The measurement of the 2S-2P transition in muonic hydrogen ([22], purple square and purple-shaded region) gives a much more precise value of $r_\mathrm{p}$, but disagrees by $4\sigma$ (gray arrow) with the 2014 H world data. Published in 2017 [24] and reproduced in Appendix A, the measurement of the 2S-4P transition frequency (green star) was the first of a new batch of measurements trying to resolve this discrepancy. The 2S-4P result agrees with the muonic result, but disagrees with the 2014 H world data by $3.3\sigma$. A 2019 microwave measurement of the 2S-2P transition ([25], black pentagon) gives a value of $r_\mathrm{p}$ that is very similar to the 2S-4P result. Two measurement of the 1S-3S transition, using different laser spectroscopy techniques, either favor the muonic result (MPQ, [26], pink plus) or the 2014 H world data (LKB, [27], brown cross). The results of the 2S-6P measurement discussed in this thesis are still blinded and thus not shown. The determined values of $r_\mathrm{p}$ and $R_\infty$ are highly correlated, allowing them to be shown in the same graph.

[1] $r_\mathrm{p}$ can also be determined from elastic electron–proton scattering [19, 20]. However, these determinations are currently less precise than determinations from hydrogen spectroscopy, and, as this work focuses on spectroscopic measurements, are not further discussed here.

[2] Muonic hydrogen is an atom formed by a proton and a negative muon. The muon is approximately 200 times heavier than the electron, and correspondingly the Bohr radius of muonic hydrogen is about 200 times smaller than in regular (electronic) hydrogen. The finite size of the proton therefore has a larger influence on the energy of the atomic S levels, enabling the much more precise measurement of $r_\mathrm{p}$ using muonic hydrogen.

## Measurement of the 2S-4P transition frequency

The proton radius puzzle also inspired precision measurement groups around the world to begin new, or improve existing, spectroscopic measurements of electronic hydrogen. This was also the case at the Laser Spectroscopy Division at the Max Planck Institute of Quantum Optics (MPQ), where the existing hydrogen spectroscopy apparatus was modified to allow for a measurement of the 2S-4P transition. This apparatus had previously been used to determine the 1S-2S transition frequency with a relative uncertainty of $4.2 \times 10^{-15}$ [23] through two-photon laser spectroscopy, during which the 2S-4P transition had been used to determine the speed distribution of the atoms. Conveniently, a corresponding laser system was thus already available.

The 1S-2S apparatus provides a cryogenic beam of hydrogen atoms in the metastable 2S level, which can then be further laser-excited to the higher-lying $n$P levels. This measurement comes with two primary challenges: first, the Doppler shift of the one-photon 2S-$n$P transition needs to be suppressed by many orders of magnitude, as the atoms are still moving at relatively high speeds. Second, in order to reach a competitive precision, the line center of the resonance needs to be determined to about one part in $N = 10\,000$ of its observed linewidth, which is comparatively large with $\Gamma_{\mathrm{F}} \approx 20\,\mathrm{MHz}$. $N$ is known as the line splitting, and is shown for various laser spectroscopy measurements of H in Fig. 1.2. Such a high line splitting requires both a large experimental signal-to-noise ratio (SNR) and a detailed theoretical understanding of the line shape. Those challenges were successfully met using sophisticated optics in the form of an active fiber-based retroreflector (AFR, [28]) to suppress the Doppler shift, a high-efficiency fluorescence detector with a high SNR, and a line shape model integrating distortions from quantum interference [29], respectively. With this, the 2S-4P transition frequency could be determined with an uncertainty of 2.3 kHz, significantly smaller than the PRP, which corresponds to 8.9 kHz in terms of the transition frequency. At the time, this also corresponded to a precision second only to the measurement of the much narrower 1S-2S transition [23]. Since the second-best measurement, which now had been improved, limits the determination of physical constants, combining the 2S-4P and the 1S-2S measurements lead to a much more accurate determination of $R_\infty$ and $r_{\mathrm{p}}$ (green star in Figs. 1.1 and 1.2) than from any other pair of H measurements. In fact, the resulting uncertainty is equivalent to that of the complete 2014 H world data. The determination disagrees by $3.3\sigma$ with the latter, but is in good agreement with the muonic value. In other words, the QED prediction of the 2S-4P transition frequency, using the 1S-2S and muonic 2S-2P measurement as input, agrees with the measured 2S-4P transition frequency within its 12 digits of accuracy. The result decreased the wiggle room of a theoretical explanation of the PRP based on hypothesized fundamental differences between the muon and electron, while increasing the likelihood of some underlying experimental problem in H spectroscopy.

The 2S-4P measurement was published in 2017 [24], after more than two years of data analysis, and is reproduced in Appendix A. The author, who joined the Laser Spectroscopy Division at MPQ in early 2014, took part in preparing and carrying out the experiment leading to the data set of [24], and did the final analysis of this data set as presented in [24]. A preliminary data analysis is given in the 2016 thesis of co-author Axel Beyer [30], who had been working on the project since 2010. A. B. and the author were jointly awarded the 2018 Helmholtz Prize for their work on the 2S-4P measurement.

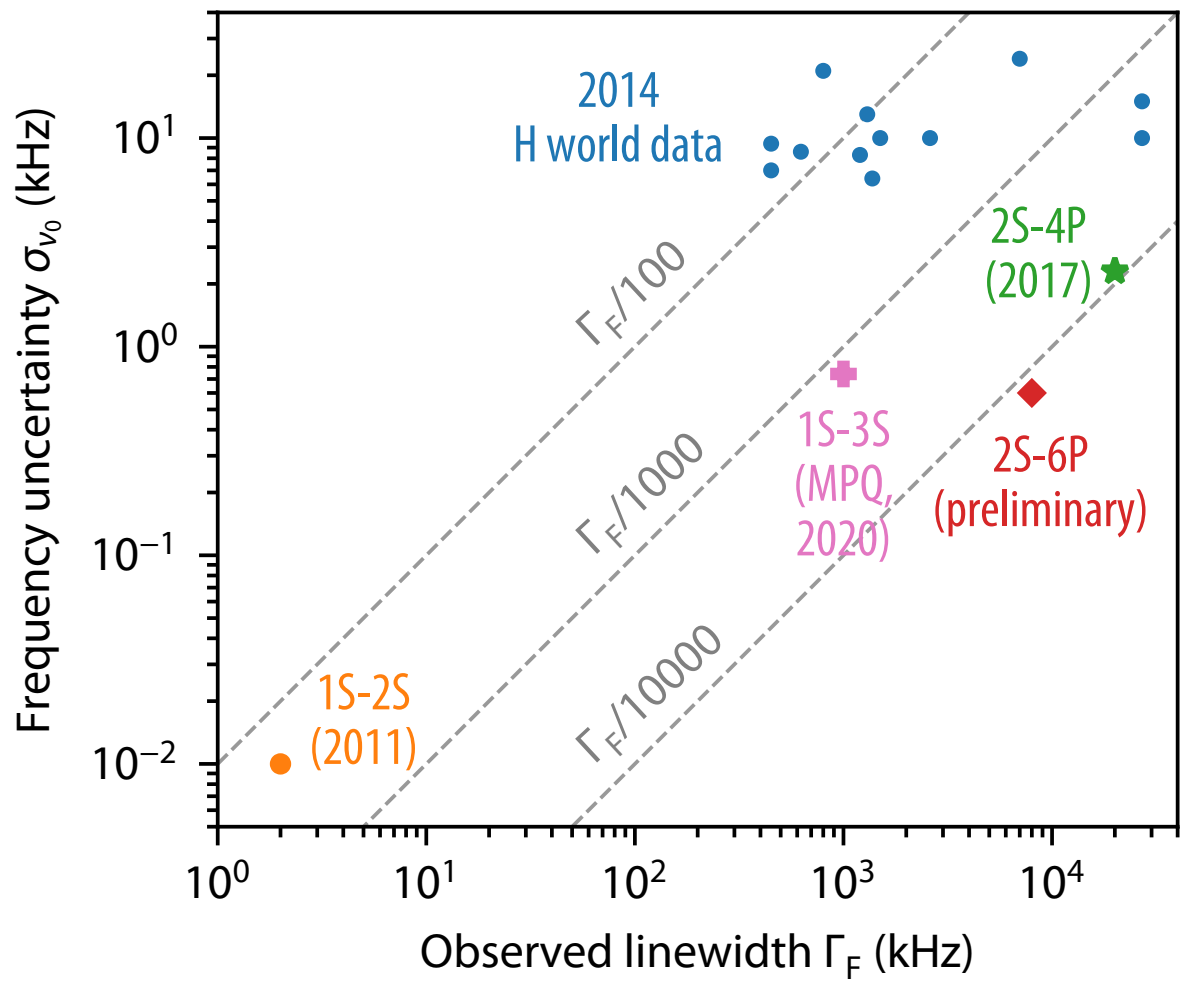

Figure 1.2: The frequency uncertainty $\sigma_{\nu_0}$ of selected laser-spectroscopy precision measurements of atomic hydrogen versus the observed linewidth $\Gamma_F$ (full width at half maximum). Measurements with a constant line splitting $N$, defined as $\sigma_{\nu_0} = \Gamma_F/N$, lie on diagonal lines, shown for $N = 100$, 1000, 10 000 (dashed lines). All laser spectroscopy results contained in the 2014 H world data ([31–34], small blue circles) are shown, including the 1S-2S measurement ([23], orange circle). The 1S-2S transition frequency has been determined with the by far smallest uncertainty, for which a moderate $N$ of 200 is sufficient, as the transition also has the by far smallest linewidth. The 2017 2S-4P measurement (Appendix A, green star) reaches a line splitting of almost $N = 10\,000$, but was recently surpassed in frequency uncertainty by the 2020 measurement of the narrower 1S-3S transition ([26], pink plus), for which an almost one order of magnitude lower line splitting was sufficient. The 2S-6P measurement (red diamond) discussed in this thesis is projected to reach a 50 % higher line splitting than the 2S-4P measurement, corresponding to a four-fold lower frequency uncertainty due to the narrower linewidth.

## Recent developments in hydrogen spectroscopy

Since the 2S-4P measurement, three more hydrogen spectroscopy results have been published: first, in 2018, an improved measurement [27] of the 1S-3S transition frequency from the Laboratoire Kastler Brossel (LKB, brown cross in Fig. 1.1). This result, again combined with the 1S-2S measurement, favors the 2014 H world data over the muonic and 2S-4P values. Second, an improved microwave measurement of the 2S-2P transition, i.e., the Lamb shift, became available in 2019 [25]. From this measurement, $r_p$ can be determined without the need for high-precision input from another H measurement, and gives a very similar result and uncertainty (black pentagon in Fig. 1.1) as the 2S-4P measurement. Third, very recently, a competing 1S-3S measurement [26] was completed across the hallway at MPQ (pink plus in Figs. 1.1 and 1.2), likewise favoring the muonic value over the 2014 H world data, but with a 2.4 times lower uncertainty than the 2S-4P result. This result is intriguing because it can be directly compared to the LKB result of the same transition without having to use any theory input, a comparison that is currently limited in statistical power by the LKB measurement.

All in all, the PRP remains not fully understood, but rather might fade away as higher accuracy measurements become available. Indeed, it may never be quite resolved in the sense that a common experimental factor shifting the measurements of the 2014 H world data is found, assuming that both theory and the muonic value are correct. In some sense, the determination of $r_p$ through H spectroscopy is a distraction from testing QED and searching for new physics, since there is one less free parameter when $r_p$ can be taken from the much more precise muonic measurement. Then, comparing the frequencies of any two different transitions constitutes a QED test, with the comparison of two measurement of the same transition serving as a test on accuracy of H spectroscopy itself.

### The road to increased precision using the 2S-6P transition

The work presented in this thesis follows the quest for ever increasing accuracy and thus an improved test of QED. Already during the work on the 2S-4P measurement it became clear that the large linewidth is a substantial limitation to its precision, as the measurement was limited by counting statistics despite its large signal-to-noise ratio. On the other hand, the techniques developed for this measurement, foremost the Doppler-shift suppression with the AFR, did not appear to be limiting factors just yet. A logical choice was then to move to a 2S-$n$P transition with a higher principal quantum number $n$, as the natural linewidth $\Gamma$ approximately decreases as $n^{-3}$ (see Section 2.2.1), while the measurement scheme itself can remain largely unchanged. Thus, using the 2S-6P transition instead of the 2S-4P transition results in a decrease in $\Gamma$ by a factor of 3.3, from 12.9 MHz to 3.90 MHz.

Moreover, the change in transition frequency from $\nu_{\text{2S-4P}} = 617\,\text{THz}$ to $\nu_{\text{2S-6P}} = 731\,\text{THz}$ leads to an increased sensitivity to $R_\infty$ and $r_p$ in combination with the 1S-2S transition frequency. This translates into an increased size of the PRP of 12.1 kHz in terms of the 2S-6P transition frequency. One way to see this is that the 1S-2S transition frequency contains a large correction to the 1S level energy from the finite size of the proton (see Eq. (2.1)), while the 2S-$n$P transition frequencies are relatively unaffected. Then, the measurement of the latter mainly determines $R_\infty$, which is proportional to the transition.

On paper, all that needed to be done to switch to the 2S-6P transition was thus a straightforward exchange of the laser system and some optics[1]. Unsurprisingly, things turned out not be quite that simple. First, a smaller natural linewidth $\Gamma$ is only of use if the observed linewidth $\Gamma_F$ is not limited by other effects, with the main culprit in this experiment being Doppler broadening through the divergence of the atomic beam. By adding a variable beam aperture, adapting the experimental geometry, and improving alignment diagnostics and procedures, an observed linewidth of down to ≈6 MHz was realized, a factor of three lower than the observed linewidth of the 2S-4P transition, as shown in Fig. 1.3.

The next challenge concerned the centerpiece of the AFR, a fiber collimator producing a Gaussian beam with very little aberrations [28], necessary to successfully suppress the Doppler shift. Unfortunately, because of the now shorter wavelength of 410 nm instead of 486 nm, the previously used collimator, assembled from off-the-shelf optics, could not be reused. At the same time, no off-the-shelf optics were available at the new wavelength and initial custom-made designs proved unsuccessful. This started a more than a year-long process of improving both optics design procedures and measurement procedures. Fortunately, it was at this point in early-2017 that Vitaly Wirthl joined the project, who took on a large share of this endeavor and at the time of writing is preparing a corresponding publication [35, 36]. The resulting collimator design, along with other improvements of the AFR, allowed for a complete suppression of the Doppler shift in the measurement of the 2S-6P transition, as deduced from the preliminary data analysis discussed below. The 2S-6P spectroscopy laser and the AFR are discussed in detail in Section 4.4.

[1] A single laser system at 820 nm can be used to drive the 2S-6P and the 1S-3S transition by using the second and fourth harmonic, respectively. Such a laser system already was set up by 2014, and used both as master laser for the 2S-6P measurement and as reference laser for the aforementioned 1S-3S measurement.

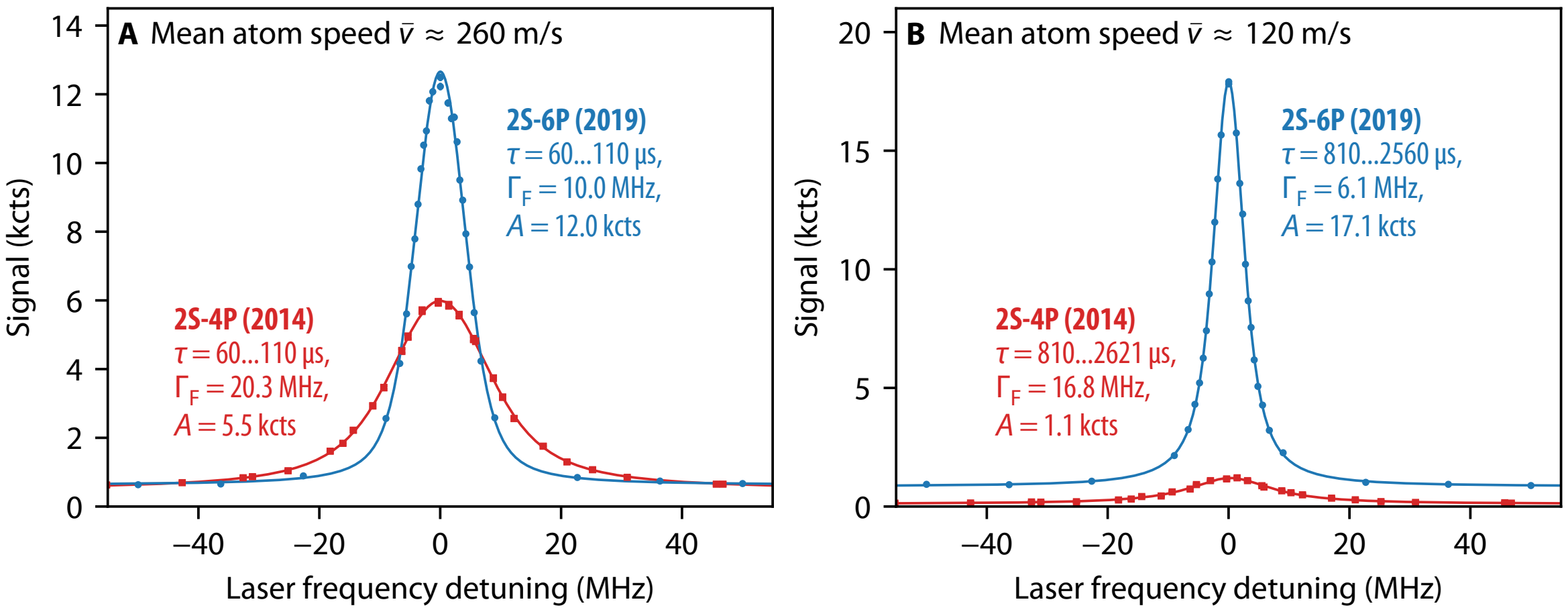

Figure 1.3: Comparison of the typical experimental fluorescence signals from a single line scan over the 2S-6P$_{1/2}$ resonance (blue points) and the 2S-4P$_{1/2}$ resonance (red squares). Two integration ranges of the delay time $\tau$ are shown (see Chapter 4), corresponding to the signal from atoms with a mean speed of (**A**) $\bar{v} \approx 260\,\mathrm{m/s}$ and (**B**) $\bar{v} \approx 120\,\mathrm{m/s}$, as determined for the 2S-6P transition. Voigt fits (colored lines) to the data reveal the observed linewidth $\Gamma_{\mathrm{F}}$ and the amplitude $A$. The 2S-6P data were recorded in 2019, using a spectroscopy laser intensity of $I = 3.9\,\mathrm{W/m^2}$ (peak intensity per direction, corresponding to $P_{\mathrm{2S\text{-}6P}} = 30\,\mu\mathrm{W}$), a flux of $3.1 \times 10^{17}$ atoms/s into the system, and a nozzle temperature of $T_{\mathrm{N}} = 4.8\,\mathrm{K}$. The 2S-4P data were recorded in 2014 and are part of the 2017 publication (see Appendix A), with the same line scan shown in Fig. 2 (B) therein. They were recorded with a laser intensity of $I = 2.8\,\mathrm{W/m^2}$, a more than five times higher flux of $1.7 \times 10^{18}$ atoms/s, and $T_{\mathrm{N}} = 5.8\,\mathrm{K}$. The probability for an atom to be resonantly excited to the $n$P level is here approximately proportional to $\mu^2 I/\Gamma$, where $\Gamma$ is the natural linewidth and $\mu$ is the dipole moment (see Tables 2.1 and B.1). For the parameters used here, this excitation probability is ≈20 % larger for the 2S-4P transition. The 2.2 (16) times larger signal and ≈2 (≈3) times lower observed linewidth of the 2S-6P resonance shown in (A) ((B)), despite lower atomic flux and excitation probability, is the result of improvements of the atomic beam formation and shaping and of the detector assembly that were implemented as part of this thesis.

While the linewidth of the 2S-$n$P transitions decreases with higher $n$, the sensitivity to stray electric fields and thus dc-Stark shifts of the transition frequency strongly increases. This is why for the 2S-6P measurement an in situ determination of the electric fields inside the apparatus is necessary, while for the 2S-4P measurement an estimation based on previous experiments was deemed sufficient. To this end, the detector assembly needed to be redesigned to reduce possible sources of stray fields and to accommodate electrodes. The latter allows the atoms themselves to be used as electric field sensors during the transition frequency measurement.

The redesign of the detector assembly also opened up the possibility to further improve the collection efficiency of fluorescence photons emitted by the decay of the 6P level, and thus improve the SNR of the measurement. If not only the efficiency, but also the solid angle of the detection is increased, line shape distortions from quantum interference (QI) can be reduced, as shown in the 2S-4P measurement (see Appendix A and [29]). In fact, the detection in the latter measurement was designed to maximize the QI effect while still providing a relatively large SNR. Having studied and understood the QI effect, it could now be minimized, while at the same time increasing the collection efficiency. This was achieved by using a modified detection geometry, and, most importantly, a photoemission material with a higher quantum efficiency. In fact, the signal was now so high that the detectors approached their maximum sustainable count rate, requiring more sophisticated diagnostics and electronics. However, as shown in Fig. 1.3, those efforts were rewarded with a substantially larger SNR than for the 2S-4P measurement, which allowed for a reduction of the number of atoms entering the system and of the spectroscopy laser intensity, which both decrease the influence of various systematic effects. The detector assembly is the topic of Section 4.6.

The suppression of the Doppler shift not only relies on the AFR, but also on the interrogation of atoms with different mean speeds $\bar{v}$. In this way, any remaining residual Doppler shift, which is proportional to $\bar{v}$, can be determined and removed by an extrapolation to zero speed. Experimentally, lower speeds are accessed through a delay time $\tau$ after the production of 2S atoms, which allows the faster atoms of the initially thermal speed distribution to escape, leaving only slower atoms to be interrogated. During initial 2S-6P measurements, still using the cryogenic hydrogen nozzle of the 2S-4P measurement and the same flux of atoms, it was however found that the number of slow atoms in the beam was much lower than expected for a thermal distribution. This observation holds true when taking the photoionization of the 2S atoms into account, to which such a depletion was previously attributed to in the 1S-2S measurement [37]. Through an investigation of various atomic fluxes, nozzle temperatures, and nozzle designs, atomic collisions removing slow atoms from the beam were identified as the main cause of the depletion. Remarkably, the depleted speed distribution is well-described by adding a single-parameter exponential suppression of slow atoms. Using the gained knowledge, a nozzle design and beam parameters were found for which considerably more slow atoms are present in the beam. This is why the signal shown in Fig. 1.3 (B), originating from atoms with a mean speed of $\bar{v} \approx 120\,\mathrm{m/s}$, is 16 times larger in the 2S-6P measurement than in the 2S-4P measurement, while the signal of faster atoms with $\bar{v} \approx 260\,\mathrm{m/s}$, shown in Fig. 1.3 (A), is 2.2 times larger. Section 4.5 describes the details of this investigation.

The ultraviolet 1S-2S preparation laser, used to excite the atoms to the metastable 2S level, was already employed during the 2S-4P measurement. The reliability of its in-vacuum enhancement cavity was a major limitation to the time of operation of the apparatus during this measurement, ultimately leading to a premature end of the final measurement run. Thus, the cavity underwent various design changes and improvements, as detailed in Section 4.3. The various upgrades of the experiment also necessitated a new data acquisition (DAQ), which during the 2S-4P measurement was still based on more than 20-year-old software. To this end, in 2015 the complete software and hardware of the DAQ was replaced as described in Section 4.7.

### The 2S-6P measurement

By March 2019, the experimental apparatus was finally ready to attempt a precision measurement of the 2S-6P transition. In the following five months, in total 3155 line scans of the 2S-6P resonance were recorded in three measurement runs (A–C, see Table 6.1), surpassing the ≈2400 line scans that make up the published data set of the 2S-4P measurement. Between those runs, the experimental apparatus was further optimized, and the recorded data were preliminarily analyzed. While many of the measurement procedures were automatized, the apparatus could not be operated unattended or remotely. Furthermore, on each measurement day, a series of alignment procedures were necessary, and thus spectroscopy data could gen-erally only be acquired after a few hours of preparation. As it tends to happen, this resulted in much of the precision data acquired in the evening and at night, as shown in Fig. 1.4.

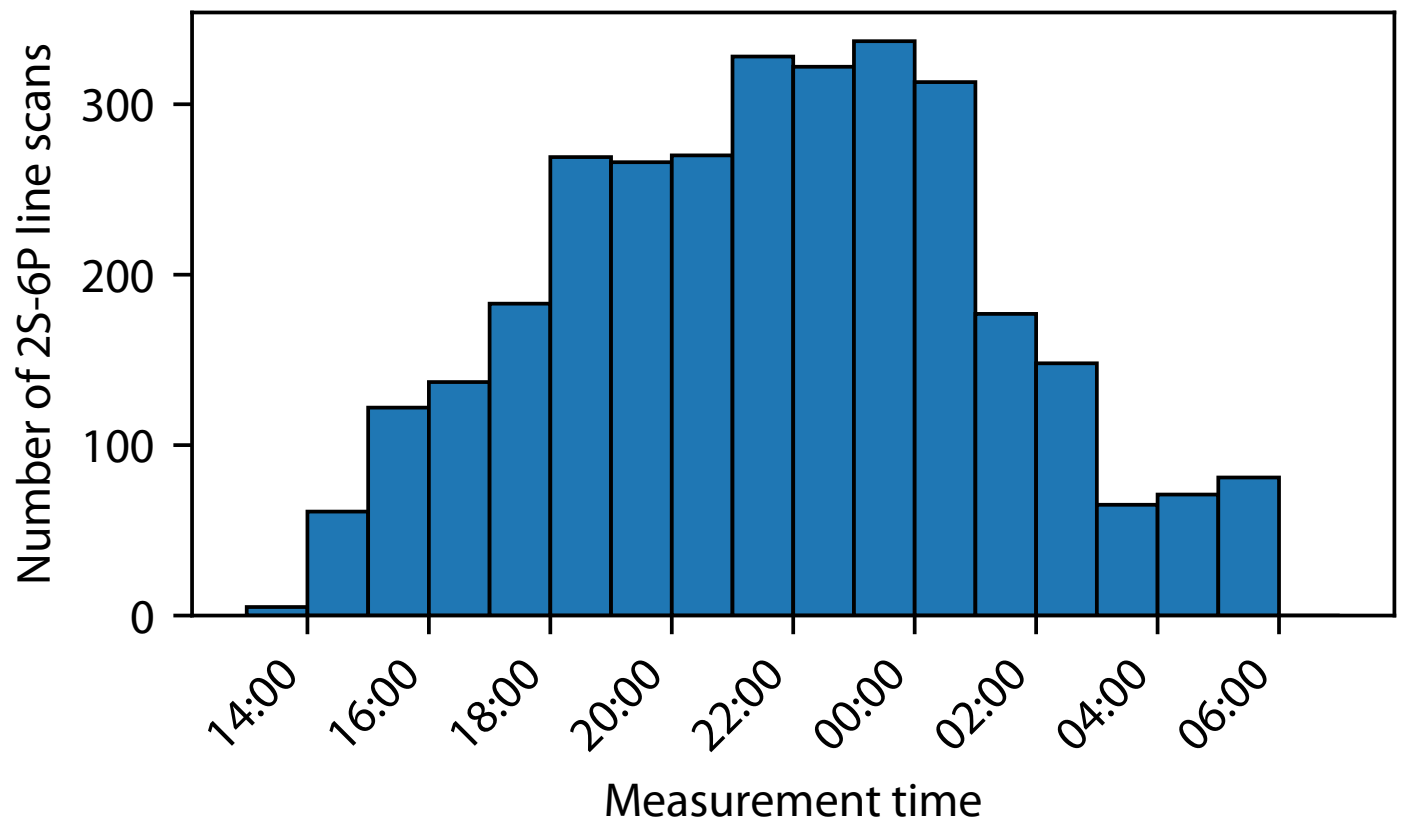

Figure 1.4: Measurement time, in the local timezone, of all 3155 line scans of the 2S-6P measurement, distributed over 25 measurement days.

For the initial run A, a spectroscopy laser power $P_{2\text{S-6P}}$ of 30 µW and 15 µW was used for the two fine-structure components of the 2S-6P transition, 2S-6P$_{1/2}$ and 2S-6P$_{3/2}$, respectively. This corresponds to a similar $n$P excitation fraction as for the powers used in the 2S-4P measurement. Because of the improvements outlined above, the 447 line scans of this run are sufficient to achieve a two times lower statistical frequency uncertainty than the total 2S-4P measurement. With this, a sub-1 kHz-uncertainty measurement of the transition frequency came into reach. However, as the statistical uncertainty decreases, the relative contribution of systematic uncertainties to the total uncertainty increases. The light force shift (LFS), which just happens to be on the order of 1 kHz and was relatively unimportant for the 2S-4P measurement, now took on the role of such a potentially limiting systematic effect. The LFS corresponds to the diffraction of the partially coherent matter waves of the atoms at the light grating created by the counter-propagating spectroscopy laser beams, which suppress the much larger first-order Doppler shift. Since the LFS scales approximately linearly with the spectroscopy laser power, it was decided[1] to reduce this power and thus trade statistics for a lower LFS.

Therefore, during run B, which consists of 2085 line scans, most scans were acquired at a three times lower laser power of 10 µW for the 2S-6P$_{1/2}$ transition and its equivalent for the 2S-6P$_{3/2}$ transition. For the remaining scans 20 µW and 30 µW, and their equivalents, were used, which can be used to constrain any remaining laser power dependency of the result. The statistics of the low-power scans are already sufficient to determine the transition frequency to ≈600 Hz, with the total necessary systematic corrections of approximately the same size. This corresponds to a line splitting of 13 000 relative the average observed linewidth of ≈8 MHz.

Unlike the line shifts from quantum interference, the LFS has not been directly observed for the 2S-$n$P transitions. Moreover, it depends on the only approximately known properties of the atomic beam, and the correct way to model the LFS has been under discussion since the 2S-4P measurement. The model of the LFS developed for this work, which is the topic of Chapter 3, predicts that the sign of the LFS changes when a small offset angle $\alpha_0$ is introduced between atomic beam and spectroscopy laser beams, which ordinarily are aligned to cross at a right angle. To test this model, such an offset angle was set for most line scans of run C. The frequency difference found between this data and data without such an offset angle is found to be in excellent agreement with the prediction of the LFS model.

[1]The author is grateful to Eric Hessels for very helpful discussions about the 2S-6P measurement during his visit to MPQ in May 2019.

At this point, the failure rate of the various devices, lasers, feedback loops, and so on started to exceed the repair rate, and the measurement was concluded at the end of August.

### Current status of the analysis of the 2S-6P measurement

Since then, the author has followed a two-pronged approach: first, a preliminary analysis of the 2S-6P measurement to quickly find inconsistencies or other problems that might require additional data to be taken or new systematic effects to be evaluated. Second, a description and careful analysis of the many changes made to the experimental apparatus and routine, and the various test measurements done, in rapid succession during 2017–2020. This is not merely useful for documentation purposes, but seems to be more and more a prerequisite for a complete analysis of 2S-6P measurement.

This is why the description of the experimental apparatus in Chapter 4 takes up the bulk of this thesis. It is mostly independent of the other chapters, but for the reader unfamiliar with hydrogen spectroscopy Chapter 2 introduces some general concepts and the 2S-6P transition in particular.

The analysis of the 2S-6P measurement is split over multiple chapters. Chapter 2 contains the derivation of the optical Bloch equations necessary to describe the dynamics and line shape of the 2S-6P transition. The motivation, derivation, and evaluation of the light force shift model fills Chapter 3. This model is especially dependent on an accurate description of the beam of metastable atoms, which is given in Chapter 5. This chapter also covers the analysis of the line scans.

Chapter 6 presents the preliminary results of the 2S-6P measurement. It is meant to give an overview over the data set, as the data analysis is still ongoing and the results are subject to change. To prevent experimenter bias, a blind analysis is performed, that is an unknown offset is added to the 2S-6P transition frequency. Only once the data analysis is thought to be complete will the offset be removed. Therefore, no value of the 2S-6P transition frequency can be given yet. However, the statistical significance, simulation corrections, and the size and uncertainty of systematic shifts are discussed in detail. The estimated frequency uncertainty of the 2S-6P measurement and the corresponding line splitting are shown in Fig. 1.2.

Finally, a conclusion and an outlook are given in Chapter 7.

# Chapter 2

# Theory

## 2.1 Energy levels of atomic hydrogen

The energy levels in atomic hydrogen (H) can be expressed as (see [4] and Appendix A)

$$E_{nlJ} = chR_\infty \left( -\frac{1}{n^2} + f_{nlJ}(\alpha, \frac{m_\mathrm{e}}{m_\mathrm{p}}, \dots) + \delta_{l0} \frac{C_\mathrm{NS}}{n^3}\ r_\mathrm{p}^2 \right), \tag{2.1}$$

where $n$, $l$ and $J$ are the principal, orbital, and total angular momentum quantum numbers, respectively. The first term describes the gross structure of H as a function of $n$. The Rydberg constant $R_\infty = m_\mathrm{e}\alpha^2 c/2h$ links the natural energy scale of atomic systems and the International System of Units (SI), which is used in the laboratory. $R_\infty$ thus connects the mass of the electron $m_\mathrm{e}$, the fine-structure constant $\alpha$, the Planck constant $h$, and the speed of light in vacuum $c$.

The second term in Eq. (2.1), $f_{nlJ}(\alpha, \frac{m_\mathrm{e}}{m_\mathrm{p}}, \dots) = X_{20}\alpha^2 + X_{30}\alpha^3 + X_{31}\alpha^3 \ln(\alpha) + X_{40}\alpha^4 + \dots$, accounts for relativistic corrections, contributions coming from the interactions of the bound-state system with the quantum electrodynamics (QED) vacuum fields, and other corrections calculated in the framework of QED. The electron-to-proton mass ratio $m_\mathrm{e}/m_\mathrm{p}$ enters the coefficients $X_{ik}$ through recoil corrections caused by the finite proton mass $m_\mathrm{p}$.

The last term in Eq. (2.1) with coefficient $C_\mathrm{NS}$ is the leading-order correction originating from the finite charge radius of the proton, $r_\mathrm{p}$, defined as the root mean square (RMS) of its charge distribution [4]. It only affects atomic S levels (with $l = 0$) for which the electron's wave function is nonzero at the origin. Higher-order nuclear charge distribution contributions are included in $f_{nlJ}(\alpha, \frac{m_\mathrm{e}}{m_\mathrm{p}}, \dots)$.

An instructive and comprehensive, if slightly outdated, overview of the derivation of Eq. (2.1), including effects of external fields and leading-order corrections from QED and nuclear effects, is given in [38]. The state-of-the-art QED calculations are summarized in [4]. From this, the uncertainty of the QED prediction for the 2S-6P transition frequency is found to be approximately 300 Hz, corresponding to a relative uncertainty of $4 \times 10^{-13}$.

Eq. (2.1) does not take into account the hyperfine structure (HFS), chiefly arising from the interaction of the nuclear magnetic dipole moment with the magnetic field generated by the electron. The HFS is associated with the additional quantum number $F$. While the HFS is not included in the description of [4], it is treated in detail in [39].

Throughout this work, numerical values of Eq. (2.1) in SI units are taken from the recent compilation of [40], with additional HFS corrections taken from [39] if necessary (see Section 6.2.4.6). In contrast to [4], [40] use the muonic value of the proton radius $r_\mathrm{p}$ from [22].

The constants entering $f_{nlJ}(\alpha, \frac{m_\mathrm{e}}{m_\mathrm{p}}, \dots)$ can be determined with sufficient accuracy from experiments other than hydrogen spectroscopy [4, 17, 18, 41, 42]. Thus, effectively two constants in Eq. (2.1) are left to be determined by hydrogen spectroscopy, $R_\infty$ and $r_\mathrm{p}$. This procedure has lead to a determination of $R_\infty$ with a relative uncertainty of $8.7 \times 10^{-12}$, using the measurement of the 2S-4P transition (see Appendix A) in combination with the 1S-2S transition frequency [23]. A recent measurement of the 1S-3S transition [26] has further reduced the relative uncertainty to $3.6 \times 10^{-12}$. Using the value of $r_\mathrm{p}$ determined by spectroscopy of muonic hydrogen [22] and likewise combining it with the 1S-2S transition frequency, $R_\infty$ can be determined with a relative uncertainty of $9 \times 10^{-13}$ [40]. This makes $R_\infty$ one of the most precisely determined constants of nature to date. As has been detailed in Chapter 1, comparing the values of $R_\infty$ and $r_\mathrm{p}$ determined from different combinations of spectroscopic measurements constitutes a powerful test of QED.

## 2.2 2S-$n$P transitions

### 2.2.1 General properties

The subject of this thesis is the study of the 2S-$n$P transitions ($n > 2$) in atomic hydrogen, specifically the 2S-6P transition. The 2S-$n$P transitions are transitions between electronic energy levels separated in energy by $h\nu_{\mathrm{A},0}$, where $\nu_{\mathrm{A},0}$ is an optical frequency on the order of hundreds of THz. The involved levels have orbital quantum numbers $l = 0$ and $l = 1$, respectively, and therefore are of different parity, which implies the existence of an electric dipole moment $\mu$. Thus, the transitions are dipole-allowed (E1) and can be excited by absorption of a single photon with energy close to $h\nu_{\mathrm{A},0}$. Fig. 2.1 shows the relevant level scheme for the 2S-6P transition.

The lower energy level is the metastable 2S level. This level is special in that its dominant[1] decay mechanism to the 1S ground level is not a dipole-allowed transition, but the simultaneous emission of two photons whose combined energies equal the energy difference between the 1S and 2S levels. This leads to a comparatively very long radiative lifetime of $\tau_{2\mathrm{S}} = 121.53(2)\,\mathrm{ms}$ [43], making the 1S-2S transition the narrowest optical transitions in atomic hydrogen with a natural linewidth of[2] $\Gamma_{2\mathrm{S}} = 1.31\,\mathrm{Hz}$.

In the experiment discussed in this thesis, the atoms typically traverse the entire experimental apparatus[3] within about 1 ms, leading to a decay of 0.9 % of the atoms in the 2S level. The interaction time during which the 2S-6P transition is probed is even shorter, with even the slowest atoms observed spending only approximately 60 µs traversing the laser beams[1], during which a fraction of $5 \times 10^{-4}$ of the 2S atoms decays. Therefore, in the discussions in this thesis related to the radiative decay of the $n$P level, both the 1S and 2S levels are treated as stable. That is, all decays and decay cascades result with the atom either in the 1S or 2S level, with the final decay thus a Lyman (Ly) or Balmer (Ba) decay, respectively. The 2S decay is however included in the simulations of the preparation of the atoms in 2S level by two-photon absorption from the 1S level, as described in Section 5.2.

[1] The 2S level can also decay through other, here negligible mechanisms [38]: the magnetic dipole (M1) transition to the 1S ground level with an associated lifetime of 2 days, and the electric dipole (E1) transition to the lower-lying $2\mathrm{P}_{1/2}$ level (which in turn rapidly decays to the 1S level), which has an associated lifetime of 20 years because of the small energy separation between the 2S and $2\mathrm{P}_{1/2}$ levels.

[2] Here the following unit conventions are used, inspired by [44]: lifetimes $\tau$ are given in units of s. Decay rates and linewidths $A, \gamma, \Gamma$ are either given in units of dcy/s (decays per second) or Hz (cycles per second), with the numerical value given by $1/\tau$ or $1/2\pi\tau$, respectively. Frequencies are given in rad/s (radians per second) or Hz, with the numerical value in rad/s given by $2\pi$ times the numerical value in Hz. Where possible, the symbol $\omega$ is used for the former case and the symbols $f, \nu$ are used for the latter case.

[3] The propagation length is $L + r_\mathrm{det} \approx 0.23\,\mathrm{m}$, the typical atom speed is $v_\mathrm{typ} = 200\,\mathrm{m/s}$ (see Chapter 4 for details).

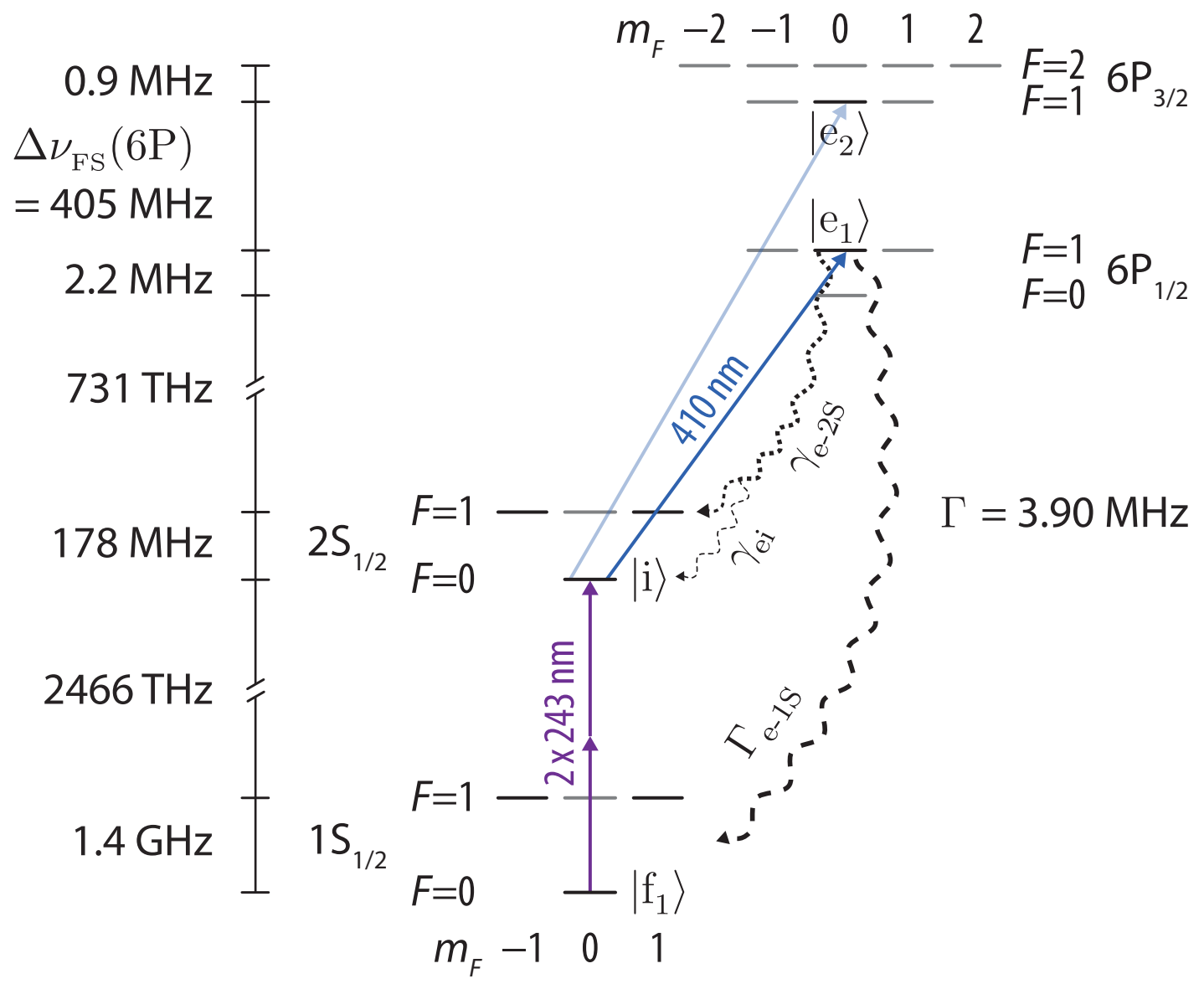

Figure 2.1: Level scheme (not to scale) for the spectroscopy of the 2S-6P transition, showing the relevant energy levels of atomic hydrogen and their couplings through laser light and radiative decay. The atoms are prepared in the metastable $2S_{1/2}^{F=0}$, $m_F = 0$ level ($|i\rangle$) by driving the two-photon transition from the $1S_{1/2}^{F=0}$, $m_F = 0$ ground level ($|f_1\rangle$) with the 1S-2S preparation laser ($\lambda_{1S\text{-}2S} = 243$ nm, purple) (see Section 2.2.6). The 2S-6P spectroscopy laser ($\lambda_{2S\text{-}6P} = 410$ nm, blue) probes either of the two transitions from $|i\rangle$ to the excited $6P_{1/2}^{F=1}$, $m_F = 0$ ($|e_1\rangle$) or $6P_{3/2}^{F=1}$, $m_F = 0$ ($|e_2\rangle$) level, which are separated in energy by the 6P fine-structure splitting $\Delta\nu_{FS}(6P) \approx 405$ MHz. Here, the resonant excitation to $|e_1\rangle$ is shown, with $|e_2\rangle$ only excited off-resonantly (light blue). The excited levels rapidly decay, directly or through cascades, to the 1S and 2S manifold with rate $\Gamma_{e\text{-}1S}$ and $\gamma_{e\text{-}2S}$, respectively, resulting in a natural linewidth of $\Gamma = 3.90$ MHz. Most decays (branching ratio $\Gamma_{e\text{-}1S}/\Gamma = 88.2\,\%$) are Lyman (Ly) decays leading to the 1S manifold, with direct decays from the 6P manifold (Ly-$\epsilon$ decays, branching ratio $\Gamma_{det}/\Gamma = 80.5\,\%$, energy $h\nu = 13.22$ eV) predominantly detected in the experiment. The remaining $\gamma_{e\text{-}2S}/\Gamma = 11.8\,\%$ of decays are Balmer (Ba) decays leading to the 2S manifold, with $\gamma_{ei}/\Gamma = 3.9\,\%$ and $\gamma_{ei}/\Gamma = 7.9\,\%$ leading back to the initial level $|i\rangle$ from $|e_1\rangle$ and $|e_2\rangle$, respectively. Energy levels shown in gray are only weakly coupled through decay cascades or optical pumping, with excitation of the $6P_{1/2}^{F=0}$ and $6P_{3/2}^{F=2}$ levels from $|i\rangle$ forbidden by angular momentum conservation.

The upper levels of the 2S-$n$P transitions are the two fine-structure manifolds $n\mathrm{P}_J$ with $J = 1/2$ and $J = 3/2$ and orbital quantum number $l = 1$. The fine-structure splitting $\Delta\nu_{FS}(n\mathrm{P})$ that separates these manifolds in energy approximately scales as $n^{-3}$. These levels can decay through dipole-allowed transitions to lower-lying S and D levels, directly or through a cascade reaching the 1S or 2S level. This includes direct decay to the 1S level, which because of the large energy difference bridged ($h\nu > (3/4)R_\infty$) is the dominant decay channel, with, e.g., 81 % (84 %) of decays for $n = 6$ ($n = 4$) being of this type. Consequently, the $n\mathrm{P}_J$ levels have a rather short radiative lifetime $\tau$ and thus broad natural linewidth $\Gamma$ compared to $nl$, $l \neq 1$ levels, where the decay to the 1S level is not dipole-allowed. For example, the linewidth of the 2S-6P transition is $\Gamma = 3.90\,\mathrm{MHz}$, while the 2S-6S and 2S-6D transitions have linewidths of 0.30 MHz and 1.34 MHz, respectively [33].

[1]The interaction time is approximately given by $T \approx 2W_0/v$, where $W_0$ is the $1/e^2$ beam radius of the 2S-6P spectroscopy laser beams through which the atoms fly and $v$ is the speed of the atoms. Here, $W_0 = 2.2$ mm, and atoms with speeds down to $v \approx 70$ m/s are observed (see Chapter 4).

While both the number of possible decays and the energy differences bridged increase for increasing $n$, the wave function overlap and thus the magnitude of the dipole moment $\mu$ for a given lower level decrease with increasing $n$. The Rabi frequency $\Omega$, for example, which is directly proportional to $\mu$ (see Section 2.3), is a factor of 2.4 lower for the 2S-6P transition than for the 2S-4P transition. In total, the smaller dipole moment dominates and the natural linewidths of the 2S-$n$P transitions decrease with $n$ approximately as $\Gamma \propto n^{-3}$. Thus, going from $n = 4$ to $n = 6$ leads to a decrease in $\Gamma$ by a factor of 3.3 from 12.9 MHz to 3.90 MHz. This decrease is the main reason why the 2S-6P transition was chosen over the previously studied 2S-4P transition (see Appendix A) for the improved measurement of the 2S-$n$P transition frequencies presented in the main text of this work.

Quantum interference (QI) between the two possible paths from an initial 2S level to a given final level through either fine-structure manifold gives rise to distortions of the observed resonance (see Section 2.3.2.3). If not properly taken into account, the distortions can cause systematic shifts of the observed transition frequency of up to $\Gamma^2/\Delta\nu_{\mathrm{FS}}(n\mathrm{P})$ [29, 45]. Because both $\Gamma$ and $\Delta\nu_{\mathrm{FS}}(n\mathrm{P})$ scale as $n^{-3}$, and the ratio $\Delta\nu_{\mathrm{FS}}(n\mathrm{P})/\Gamma$ is approximately 100, the frequency shifts can reach $\Gamma/100$, constituting another advantage of moving to higher values of $n$.

There is however a major disadvantage of moving to higher values of $n$, which is the scaling of the dc-Stark shift with $n^7$. This is discussed in detail in Section 2.4.

### 2.2.2 The 2S-6P transition

The level scheme of the 2S-6P transition as relevant to the experiment discussed here is shown in Fig. 2.1. The atoms are first prepared in the metastable $2\mathrm{S}_{1/2}^{F=0}$, $m_F = 0$ level by optical excitation from the thermally populated $1\mathrm{S}_{1/2}^{F=0}$, $m_F = 0$ ground level, corresponding to a frequency difference of $\nu_{\text{1S-2S}} = 2.466\,\mathrm{PHz}$. To this end, the two-photon transition between the two levels is driven in a Doppler-free manner with the 1S-2S preparation laser at a wavelength of $\lambda_{\text{1S-2S}} = 243\,\mathrm{nm}$ (see Section 2.2.6).

The 2S-6P transition, with a transition frequency of $\nu_{\text{2S-6P}} = 730.7\,\mathrm{THz}$, is probed with a second laser. This 2S-6P spectroscopy laser, which is linearly polarized and has a wavelength of $\lambda_{\text{2S-6P}} = 410\,\mathrm{nm}$, is set to be resonant with either of the two transitions from the initial $2\mathrm{S}_{1/2}^{F=0}$, $m_F = 0$ level to the $6\mathrm{P}_{1/2}^{F=1}$, $m_F = 0$ or $6\mathrm{P}_{3/2}^{F=1}$, $m_F = 0$ excited level. Throughout this work, the two transitions are referred to as 2S-$6\mathrm{P}_{1/2}$ and 2S-$6\mathrm{P}_{3/2}$ transition, respectively, while the term 2S-6P transition is used when referring to both transitions. The $6\mathrm{P}_{1/2}^{F=1}$, $m_F = 0$ and $6\mathrm{P}_{3/2}^{F=1}$, $m_F = 0$ level belong to different fine-structure manifolds and are separated in energy by the 6P fine-structure splitting $\Delta\nu_{\mathrm{FS}}(6\mathrm{P}) \approx 405\,\mathrm{MHz}$ (see Section 6.2.4.6).

Counter-propagating laser beams are used to suppress the first-order Doppler shift of this one-photon transition (see Section 2.2.4). The excitation of all other 6P levels by the spectroscopy laser is forbidden by angular momentum conservation. In this way, only one hyperfine sublevel is addressed for each manifold. If this was not the case, e.g., if circular instead of linear laser polarization was used, the unresolved hyperfine structure would lead to overlapping resonances, since the hyperfine splitting of the 6P levels is smaller than their linewidth.

Table 2.1: Atomic properties of the 2S-6P transition, as used in the simulations and theory corrections in this work. Either one of the two dipole-allowed one-photon transitions from the initial metastable $2S_{1/2}^{F=0}$, $m_F = 0$ level ($|i\rangle$) to the excited $6P_J^{F=1}$, $m_F = 0$ level ($|e_1\rangle$ and $|e_2\rangle$ for $J = 1/2$ and $J = 3/2$, respectively) are probed with a linearly polarized laser. The hydrogen level energies, including fine and hyperfine structure, needed to calculate the properties were taken from [40] and assume the muonic value of the proton radius $r_p$ from [22]. The physical constants needed to convert from atomic units to SI units were taken from [46]. Dipole moments, Rabi frequencies, decay rates, and ac-Stark shift and photoionization coefficients are calculated using dipole moments derived from non-relativistic wave functions [38] (relativistic corrections to the dipole moments are of order $\alpha^2$ [47]). Only electric dipole (E1) transitions are considered, which is sufficient at the given level of accuracy [47]. The corresponding atomic properties of the 2S-4P transition are given in Table B.1.

| $J$ (excited level) | 1/2 ($\lvert e_1\rangle$) | 3/2 ($\lvert e_2\rangle$) |
|---|---|---|
| Transition frequency $\nu_{A,0}$ (kHz) | 730 690 111 486.4 | 730 690 516 650.9 |
| Transition wavenumber $K_L$ (1/m) | 15 314 132 | 15 314 141 |
| Dipole moment $\mu$ ($e\,a_0$) | $-\frac{m_e+m_p}{m_p}\frac{9}{512}\sqrt{105}$ | $\frac{m_e+m_p}{m_p}\frac{9}{512}\sqrt{2\times 105}$ |
| Dipole moment $\mu$ ($10^{-30}$ C m) | $-1.528$ | 2.161 |
| Rabi frequency $\Omega_0$ (krad/s (W/m$^2$)$^{-1/2}$) | $2\pi \times 63.30$ | $2\pi \times 89.52$ |
| Natural linewidth $\Gamma$ (kHz) | 3 899 | |
| Decay rates $A$ (kdcy/s) | | |
| $\gamma_{e\text{-}2S}$: $\lvert e_{1/2}\rangle \to$ 2S manifold | $2\pi \times 462.0$ | |
| $\gamma_{ei}$: $\lvert e_{1/2}\rangle \to \lvert i\rangle$ | $2\pi \times 153.6$ | $2\pi \times 306.4$ |
| $\Gamma_{e\text{-}1S}$: $\lvert e_{1/2}\rangle \to$ 1S manifold | $2\pi \times 3\,437$ | |
| $\Gamma_{det}$: Detected signal (Ly-$\epsilon$)[a] | $2\pi \times 3\,140$ | |
| Non-resonant ac-Stark shift coefficient[b] $\beta_{ac,0}$ (Hz/(W/m$^2$)) | $8.879 \times 10^{-5}$ | $9.275 \times 10^{-5}$ |
| Photoionization coefficient[c] $\beta_{ioni}$ (Hz/(W/m$^2$)) | $1.611 \times 10^{-5}$ | $1.887 \times 10^{-5}$ |
| Mass of hydrogen atom $m_H$ (kg) | $1.673\,533 \times 10^{-27}$ | |
| Recoil shift $\Delta\nu_{rec}$ (Hz) | 1 176 026 | 1 176 027 |
| Recoil velocity $v_{rec}$ (m/s) | 0.965 016 | 0.965 016 |

[a] In the experiment, only Lyman decays are detected, with Ly-$\epsilon$ photons accounting for ~97 % of the signal (see Tables 2.2 and 4.1).

[b] This coefficient is derived from a perturbative calculation (see Appendix C.1) and does not include near-resonant contributions.

[c] See Appendix C.2. Only the excited level can be photoionized by light of frequency $\nu_{A,0}$.

As described above, the excited levels rapidly decay, directly or through cascades, to the 1S and 2S manifold with rate $\Gamma_{e\text{-}1S}$ and $\gamma_{e\text{-}2S}$, respectively, resulting in a natural linewidth of $\Gamma = 3.90$ MHz. The fluorescence from these decays serve as the signal in the experi-ment[1]. Most decays (branching ratio $\Gamma_{\text{e-1S}}/\Gamma = 88.2\,\%$) are Lyman (Ly) decays leading to the 1S manifold, with direct Ly-$\epsilon$ decays (branching ratio $\Gamma_{\text{det}}/\Gamma = 80.5\,\%$, energy $h\nu = 13.22\,\text{eV}$) predominantly detected in the experiment. The remaining $\gamma_{\text{e-2S}}/\Gamma = 11.8\,\%$ of decays are Balmer (Ba) decays leading to the 2S manifold, with $\gamma_{\text{ei}}/\Gamma = 3.9\,\%$ and $\gamma_{\text{ei}}/\Gamma = 7.9\,\%$ leading back to the initial level from the $6P_{1/2}^{F=1}$, $m_F = 0$ and $6P_{3/2}^{F=1}$, $m_F = 0$ level, respectively. The rate of this latter back decay is a main difference between the 2S-6P$_{1/2}$ and 2S-6P$_{3/2}$ transition, and influences the observed line shape. Specifically, both the light force shift (see Chapter 3) and the quantum interference shifts (see Section 2.3.2) increase with increasing back decay.

The atomic properties of the 2S-6P$_{1/2}$ and 2S-6P$_{3/2}$ transition are summarized in Table 2.1. The 2S-6P transition has to our knowledge so far not been studied in detail using laser spectroscopy[2].

### 2.2.3 Resonance condition for one-photon absorption

Energy and momentum conservation requires the frequency $\nu_{\text{L},0}$, as determined in the laboratory frame of reference, of the laser light to be absorbed by the atom to be different from the atom's resonance frequency $\nu_{\text{A},0}$. For an atom with velocity $\boldsymbol{v}$ ($|\boldsymbol{v}| \equiv v$) interacting with a laser beam with wave vector $\boldsymbol{K}_{\mathbf{L}}$ ($|\boldsymbol{K}_{\mathbf{L}}| \equiv K_{\text{L}} = 2\pi\nu_{\text{L},0}/c$), with both $\boldsymbol{v}$ and $\boldsymbol{K}_{\mathbf{L}}$ measured in the laboratory frame, a relativistic energy and momentum balances gives [48]

$$\frac{\nu_{\text{L},0}}{\nu_{\text{A},0}} = \frac{\sqrt{1-\frac{v^2}{c^2}}}{1-\frac{\boldsymbol{K}_{\mathbf{L}}}{2\pi\nu_{\text{L},0}}\cdot\boldsymbol{v}}\left(\frac{1}{1-\frac{h\nu_{\text{A},0}}{2mc^2}}\right). \tag{2.2}$$

Here, $m$ is the mass of atom referenced halfway between the upper and lower atomic levels separated by $\nu_{\text{A},0}$, i.e., $m = m_{\text{H}} + h(\nu_l + \nu_{\text{A},0})/c^2$ where $m_{\text{H}}$ is the rest mass of the atom in its ground level and $h\nu_l$ the energy of the lower level.

In the experiment described in this work, atoms cross a laser beam at almost right angles $\alpha = \pi/2 - \delta\alpha$, such that $\boldsymbol{K}_{\mathbf{L}}\cdot\boldsymbol{v}/2\pi\nu_{\text{L},0} = \cos(\alpha)v/c = \sin(\delta\alpha)v/c$, where $\delta\alpha$ ($|\delta\alpha| \ll \pi/2$) is the small angle from the orthogonal towards the laser beam propagation direction. Expanding Eq. (2.2) in $v/c$ and $h\nu_{\text{A},0}/m_{\text{H}}c^2$ then gives

$$\nu_{\text{L},0} = \nu_{\text{A},0} + \nu_{\text{A},0}\frac{\sin(\delta\alpha)v}{c} + \frac{h\nu_{\text{A},0}^2}{2m_{\text{H}}c^2} - \frac{\nu_{\text{A},0}}{2}\left(\frac{v}{c}\right)^2, \tag{2.3}$$

where terms up to second order in $v/c$ and first-order in $hR_\infty/m_{\text{H}}c$ and $\sin(\delta\alpha)v/c$ have been included, using the upper limit for the binding energy of an hydrogen atom of $cR_\infty$ for $\nu_{\text{A},0}$ and $\nu_l$ in the latter expression. For the typical experimental speed of $v_{\text{typ}} = 200\,\text{m/s}$ and typical angle of $\delta\alpha_{\text{typ}} = 4\,\text{mrad}$, $v/c \approx 6.7\times10^{-7}$, $(v/c)^2 \approx 4.5\times10^{-13}$, $\sin(\delta\alpha)v/c \approx 2.7\times10^{-9}$, and $hR_\infty/m_{\text{H}}c \approx 1.4\times10^{-8}$. Thus, dropping higher-order terms is a good approximation for the relative accuracy of one part in $10^{-13}$ required here.

The second term in Eq. (2.3), linear in $v$, is the first-order Doppler shift,

$$\Delta\nu_{\text{D}} = \frac{\nu_{\text{A},0}}{c}\sin(\delta\alpha)v = \beta_{\text{D}}v_{\|}, \tag{2.4}$$

[1]One may also detect the remaining population in the 2S manifold, as, e.g., done for the measurement of the 2S-4P transition of [32].

[2]The 2S-6P transition was used in [33] to study the velocity distribution of an atomic beam of hydrogen, but no precision measurement of its transition frequency was reported.

with the velocity component along the laser beam direction $v_{\|} = \sin(\delta\alpha)v$ and the first-order Doppler shift coefficient $\beta_\text{D} = \nu_{\text{A},0}/c$. For the 2S-6P transition, $\beta_\text{D} \approx 2.437\,\text{MHz}/(\text{m/s})$, resulting in a shift of $\Delta\nu_\text{D} \approx 2.0\,\text{MHz}$ for $v_\text{typ}$ and a typical angle of $\delta\alpha_\text{typ} = 4\,\text{mrad}$. This shift is almost four orders of magnitude larger than the accuracy of the experiment, requiring an intricate suppression as shown in Fig. 2.2 and discussed in the following section.

Of special importance to the experiment discussed here is the fact that any first-order Doppler shift can be expressed as the product of a Doppler slope[1] $\kappa$ and the atom's velocity $v$,

$$\Delta\nu_\text{D} = \kappa v. \tag{2.5}$$

For the case described in Eq. (2.4), the Doppler slope is given by $\kappa = \beta_\text{D}\sin(\delta\alpha)$, which for $\delta\alpha_\text{typ}$ results in $\kappa = 9.7\,\text{kHz}/(\text{m/s})$. Thus, the experiment requires a suppression of the Doppler slope to the order of $\kappa = 1\,\text{Hz}/(\text{m/s})$. If the Doppler slope $\kappa$ is not known initially, it can be determined from the experiment by measuring $\nu_{\text{L},0}$ for different values of $v$. Since $\Delta\nu_\text{D}$ is the only term linear in $v$ in $\nu_{\text{L},0}$, the slope of the linear dependence of $\nu_{\text{L},0}$ on $v$ is equal to $\kappa$. Likewise, extrapolating this linear dependence to $v = 0\,\text{m/s}$ gives $\nu_{\text{L},0}$ free from the first-order Doppler shift. Here, this procedure is referred to as Doppler extrapolation. Note that the value of $\nu_{\text{L},0}$ found through the Doppler extrapolation is not identical to the value of $\nu_{\text{L},0}$ at $v = 0\,\text{m/s}$, as the second-order Doppler shift, discussed below and quadratic in $v$, is not removed.

The third term in Eq. (2.3) is the recoil shift,

$$\Delta\nu_\text{rec} = \frac{h\nu_{\text{A},0}^2}{2m_\text{H}c^2}, \tag{2.6}$$

which is independent of $v$. Within the approximations here and for an atom initially at rest, $h\Delta\nu_\text{rec}$ can be understood as the kinetic energy required for the atom to take up the photon's momentum $\hbar K_\text{L}$. For each absorbed momentum $\hbar K_\text{L}$, the atom's velocity increases by the recoil velocity $v_\text{rec}$ in the direction of the laser beam, $\boldsymbol{K}_\mathbf{L}/K_\text{L}$, with

$$v_\text{rec} = \frac{h\nu_{\text{A},0}}{m_\text{H}c} \tag{2.7}$$

using[2] $\nu_{\text{L},0} \approx \nu_{\text{A},0}$. The same change in velocity occurs upon emission of a photon by the atom, but with the atom's velocity decreasing in the direction of emission. For stimulated emission, the emission is along the stimulating laser beam, while for spontaneous emission the direction is given by the corresponding radiation pattern of the transition. For the 2S-6P transition, $\Delta\nu_\text{rec} \approx 1.176\,\text{MHz}$ and $v_\text{rec} = 0.965\,\text{m/s}$ (see Table 2.1). Note that the first-order Doppler shift of an atom with $v_{\|} = v_\text{rec}$ is $\Delta\nu_\text{D} = \beta_\text{D} v_\text{rec} = 2\Delta\nu_\text{rec}$, i.e., after absorbing a photon from rest the atom sees a first-order Doppler shift corresponding to twice the recoil frequency. The recoil shift has been directly observed in absorption spectroscopy in [48].

The final term in Eq. (2.3) is the second-order or relativistic Doppler shift,

$$\Delta\nu_\text{SOD} = -\frac{\nu_{\text{A},0}}{2}\left(\frac{v}{c}\right)^2, \tag{2.8}$$

[1] In [28], the symbol $\eta$ is used for the Doppler slope. However, since $\eta$ is used for the asymmetry parameter in this work, here $\kappa$ is used as symbol for the Doppler slope instead.

[2] The recoil velocity $v_\text{rec}$ as defined here does not enter any frequency corrections, but is a convenient quantity for qualitative descriptions. As such, the approximation made here is adequate.

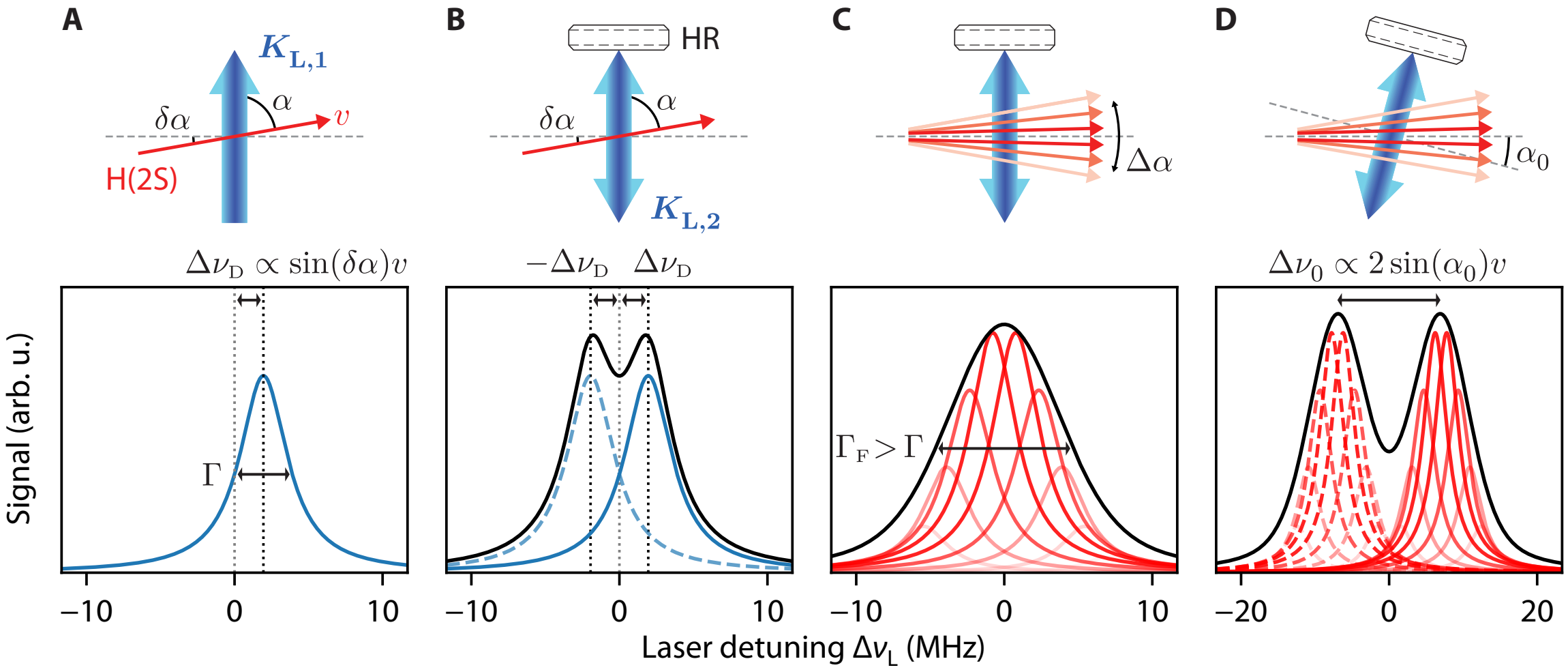

Figure 2.2: Scheme for the suppression of the first-order Doppler shift and the mechanism for Doppler broadening from atomic beam divergence, showing both the atom and laser beams (top) and the observed fluorescence signal versus laser detuning (bottom). (**A**) A hydrogen atom in the metastable 2S level (H(2S), red arrow) crossing a single laser beam (blue arrow) with wave vector $\boldsymbol{K}_{\mathbf{L},\mathbf{1}}$ at an angle $\alpha = \pi/2 - \delta\alpha$ and with speed $v$ experiences a first-order Doppler effect, shifting the observed atomic resonance (Lorentzian line shape, blue line) by $\Delta\nu_D \propto \sin(\delta\alpha)v$ from its unperturbed position (black and gray dotted lines). The line width $\Gamma$ of the resonance is unchanged from its unperturbed value. (**B**) A counter-propagating laser beam with wave vector $\boldsymbol{K}_{\mathbf{L},\mathbf{2}}$ that retraces the phase of the first beam ($\boldsymbol{K}_{\mathbf{L},\mathbf{2}} = -\boldsymbol{K}_{\mathbf{L},\mathbf{1}}$), e.g., generated by suitable retroreflection from a mirror (HR), adds to the signal a second resonance (dashed blue line). This resonance is Doppler-shifted by an equal, but opposite amount $-\Delta\nu_D$. The total signal (black line) from this doublet of resonances is not shifted in frequency. (**C**) Signal from multiple atoms (red arrows) forming an atomic beam with angular divergence $\Delta\alpha$. The total signal (Voigt line shape, black line) is made up of many resonance doublets with different Doppler shifts. As each doublet is unshifted in frequency, the total signal is also unshifted. However, its linewidth $\Gamma_F$ is broadened over the linewidth $\Gamma$ of each atomic resonance. (**D**) An offset angle $\alpha_0$ from the orthogonal between the atomic and laser beams splits the total signal into two resonances from the forward- (solid red lines) and backward-traveling (dashed red lines) laser beams, with the frequency splitting of this doublet given by $\Delta\nu_0 \propto 2\sin(\alpha_0)v$.

It is, as the name suggests, a consequence of relativistic time dilation, which leads to a higher laser frequency in the co-moving frame of the atom relative to the laser frequency measured in the laboratory frame. As such, and in contrast to the first-order Doppler shift, it does not depend on the direction of the atom's velocity, $\boldsymbol{v}/v$, but only on the square of the atom's speed $v$. For the 2S-6P transition and $v_{\mathrm{typ}}$, $\Delta\nu_{\mathrm{SOD}} \approx -163\,\mathrm{Hz}$. $\Delta\nu_{\mathrm{SOD}}$ has been tested within 6 % in a transversal configuration similar to the experiment discussed here in [49] and to much higher precision in a longitudinal configuration, e.g., in [50].

### 2.2.4 Suppression of the first-order Doppler shift

As detailed in the previous section, the first-order Doppler shift experience by atoms flying through the laser beam needs to be suppressed by many order of magnitude to allow for precision spectroscopy of the 2S-$n$P one-photon transitions. The basic idea followed here is to not use a single laser beam, but instead two counter-propagating laser beams to probe the atoms such that the Doppler shifts of the fluorescence signals from each beam cancel. This scheme is outlined in Fig. 2.2 and has been discussed at some length in publications focusing on the 2S-4P transition [28] and on the 2S-6P transition [35, 36].

#### 2.2.4.1 Suppression scheme

First, consider again the case of a hydrogen atom in the metastable 2S level with velocity $\boldsymbol{v}$ ($|\boldsymbol{v}| \equiv v$) crossing a single laser beam with wave vector $\boldsymbol{K}_{\mathbf{L,1}}$ and frequency $\nu_{\mathrm{L}} = \nu_{\mathrm{L,0}} + \Delta\nu_{\mathrm{L}}$ (see Fig. 2.2 (A)), where $\nu_{\mathrm{L,0}}$ here is the resonance frequency expected if no Doppler shift was present. The resulting fluorescence signal is shifted in frequency through the first-order Doppler effect by

$$\Delta\nu_{\mathrm{D}} = \frac{1}{2\pi}\boldsymbol{K}_{\mathbf{L,1}} \cdot \boldsymbol{v} = \beta_{\mathrm{D}} \sin(\delta\alpha) v, \tag{2.9}$$

where the crossing angle $\delta\alpha$ is again the small angle from the orthogonal between the atom's trajectory and the laser beam and $\beta_{\mathrm{D}}$ is the first-order Doppler shift coefficient introduced in Eq. (2.4). In the absence of homogeneous broadening, saturation, and quantum interference effects, the signal has a Lorentzian line shape with a full width at half maximum (FWHM) linewidth corresponding to the natural linewidth $\Gamma$ [51].

Adding a second, counter-propagating laser beam with wave vector $\boldsymbol{K}_{\mathbf{L,2}}$ such that $\boldsymbol{K}_{\mathbf{L,2}} = -\boldsymbol{K}_{\mathbf{L,1}}$ at any point causes the appearance of a second resonance in the signal (see Fig. 2.2 (B)). The condition $\boldsymbol{K}_{\mathbf{L,2}} = -\boldsymbol{K}_{\mathbf{L,1}}$ implies that the second beam exactly retraces the phase of the first laser beam, and that both beams have the same frequency. Hence, this second resonance is Doppler-shifted by an equal, but opposite amount $-\Delta\nu_{\mathrm{D}}$ as the first resonance. The total fluorescence signal is then a doublet of resonances separated in frequency by $2\Delta\nu_{\mathrm{D}}$. If the two beams additionally have exactly the same intensity at any point, the total signal is symmetric about the unperturbed resonance frequency and its center of mass is free from the first-order Doppler effect[1].

#### 2.2.4.2 Experimental realization and imperfections

While the Doppler-suppression scheme using two laser beams is conceptually straightforward, its experimental implementation is rather challenging for the level of suppression needed here [28, 35, 36]. The scheme used here is to retroreflect a single laser beam on a highly-reflective (HR) plane mirror to create a second, counter-propagating beam. If the beam waist of the first, forward-traveling beam is placed exactly on the mirror surface, and the mirror surface is perfectly plane and exactly orthogonal to the beam axis, the phase-retracing condition is fulfilled. If the mirror is also perfectly reflective, the forward- and backward-traveling beams will have the same intensity, and all conditions for Doppler suppression are fulfilled. The experimental realization and imperfections are discussed in Section 4.4.

Imperfections in the counter-propagating beams can degrade the suppression in two ways [28]. First, if the wave vectors of the two beams are, at a given point, not quite anti-parallel, but have a small angle of $\epsilon$ ($|\epsilon| \ll \pi$) between them, the total signal will have a residual

[1]In the experiment, the resonance frequency of the total signal is found by fitting a line shape function to the sampled data, which is a nonlinear procedure. Care has to be taken to use both a line shape function that matches the data well and an adequate sampling of the data in order to ensure that the determined resonance frequency approaches the Doppler-free center of mass.

Doppler shift of

$$\Delta\nu_{\mathrm{D}} = \beta_{\mathrm{D}}\frac{\sin(\delta\alpha)+\sin(-\delta\alpha+\epsilon)}{2}v \approx \frac{1}{2}\beta_{\mathrm{D}}\epsilon v = \kappa(\epsilon)v, \tag{2.10}$$

where in the approximation terms on the order of $\epsilon(\delta\alpha)^2$ have been neglected. The residual Doppler slope from this angular mismatch is thus $\kappa(\epsilon) = \frac{1}{2}\beta_{\mathrm{D}}\epsilon$, independent of the crossing angle $\delta\alpha$. A residual Doppler slope of $\kappa(\epsilon) = 1\,\mathrm{Hz/(m/s)}$ then corresponds to $\epsilon = 0.8\,\mu\mathrm{rad}$. In the experiment, the alignment of $\epsilon$ as given by the tip and tilt of the HR mirror from the beam axis is not expected to be on this level[1]. Additionally, the local angular mismatch seen by an atom flying through the laser beams might be much larger, as the phase-retracing condition is violated either by the beam waist not being placed exactly at the HR mirror or by residual aberrations present in the beam. However, $\epsilon$ is expected to randomly vary as vibrations affect the tip and tilt and as the experimental alignment to an as low as possible value of $\epsilon$, including the position of the beam waist, is redone many times during the measurement runs. Then, if $\epsilon$ varies, on average, about zero, $\kappa(\epsilon)$ is expected to average to zero as well.

Second, the two beams might also have a slightly different intensity. For example, if the reflectance of the HR mirror is $R = 1-\xi$, the power, and if the beams are phase-retracing also the intensity, of the backward-traveling beam is only $1-\xi$ of that of the forward-traveling beam. The residual Doppler shift in this case is

$$\Delta\nu_{\mathrm{D}} = \beta_{\mathrm{D}}\frac{\sin(\delta\alpha)+(1-\xi)\sin(-\delta\alpha)}{2}v \approx \frac{1}{2}\beta_{\mathrm{D}}\delta\alpha\xi v = \kappa(\xi)v. \tag{2.11}$$

The residual Doppler slope $\kappa(\xi) = \frac{1}{2}\beta_{\mathrm{D}}\delta\alpha\xi$ now depends on the crossing angle $\delta\alpha$, as opposed to the previous case. For the HR mirror used in the experiment, $\xi = 5\times10^{-5}$, resulting in $\kappa(\xi) = 0.24\,\mathrm{Hz/(m/s)}$ for the typical crossing angle $\delta\alpha_{\mathrm{typ}} = 4\,\mathrm{mrad}$ and sufficiently small for the accuracy reached in the experiment. However, just as for the local angular mismatch, the local intensity mismatch seen by the atoms might be much larger and therefore the reflectance of the HR mirror cannot serve as an upper limit of the intensity mismatch. As for the first case, if $\delta\alpha$, and the part of $\xi$ related to the position of the beam waist, varies about zero from alignment to alignment, $\kappa(\xi)$ is expected to average to zero as well.

For a symmetric atomic beam which is aligned such that, on average, the atoms cross the laser beams at a small offset angle $\alpha_0$ ($|\alpha_0| \ll \pi/2$) from the orthogonal, there is for each atom with a crossing angle of $\alpha_0+\delta\tilde{\alpha}$ another atom with a crossing angle $\alpha_0-\delta\tilde{\alpha}$, where $\delta\tilde{\alpha}$ is an angle within the beam divergence. For the first case of imperfections, the Doppler shift is then again given by Eq. (2.10) and is independent of both $\delta\tilde{\alpha}$ and $\alpha_0$. For the second case, $\delta\alpha$ is replaced with $\alpha_0$ in Eq. (2.11). Combining both cases then gives

$$\Delta\nu_{\mathrm{D}} \approx \frac{1}{2}\beta_{\mathrm{D}}\left(\epsilon+\alpha_0\xi\right)v = \kappa(\epsilon,\xi)v, \tag{2.12}$$

where a term proportional to $\epsilon\xi$ has been dropped. An alignment of the offset angle $\alpha_0$, together with the alignments affecting $\epsilon$ and $\xi$ described above, to a value as close to zero as possible is thus crucial to achieve a sufficient suppression of the Doppler shift. As detailed below, an offset angle of up to $\alpha_0 = 12\,\mathrm{mrad}$ is applied in some measurements to help with the characterization of some systematic effects. In order to still ensure that the Doppler shift averages to zero, the sign of this offset is then flipped every so often, e.g., the measurement is actually done at $\alpha_0 = \pm 12\,\mathrm{mrad}$.

[1] A misalignment of $\epsilon = 0.8\,\mu\mathrm{rad}$ only reduces the power coupled back into the fiber, an experimentally accessible value, by $5\times10^{-5}$, well below the detectable reduction of $\approx 0.1\,\%$, which corresponds to $\epsilon \approx 4\,\mu\mathrm{rad}$.

While the Doppler slope is thus ideally expected to average to zero, it is not clear on which time scale this averaging occurs and whether the realignments randomly vary about zero without systematic offsets. Additionally, beam aberrations in combination with the excitation dynamics of the atoms might lead to a Doppler slope that does not vary with the realignments. This is why in the experiment, as described in the previous section, the resonance frequency of atoms of different speeds $v$ is observed, allowing for the experimental determination of the residual Doppler slope $\kappa$ and thus an in-situ characterization of the suppression scheme. Importantly, even with this determination, an as-good-as-possible suppression is still crucial, as the modeling of the line shape and of systematic effects besides the Doppler shift relies on the presence of phase-retracing beams. Additionally, the speeds $v$ used in the experimental determination, also derived from simulations, have an associated uncertainty that is limiting for too large values of $\kappa$.

#### 2.2.4.3 Excitation with counter-propagating beams

The use of two counter-propagating beams instead of a single laser beam not only, on average, doubles the intensity seen by the atoms, but also creates an intensity standing wave with periodicity $\lambda_{2\mathrm{S}-6\mathrm{P}}/2 = 205\,\mathrm{nm}$, through which the atoms fly. This modifies the excitation dynamics, as the atoms can now absorb photons from either beam, with the associated recoil kick pointing in opposite directions. In turn, this gives rise to the light force shift, which is one the main systematic effects of the 2S-6P measurement and the topic of chapter Chapter 3.

#### 2.2.4.4 Doppler-free two-photon absorption

Finally, counter-propagating beams are also commonly employed to suppress the first-order Doppler shift of two-photon transitions [52]. Indeed, this technique is used in this work to prepare the atoms in the metastable 2S level through Doppler-free two-photon absorption on the 1S-2S transition, as was done in the 1S-2S measurement [23] and the 2S-4P measurement (see Appendix A). While the basic idea is the same for one- and two-photon transitions, i.e., the cancellation of the Doppler shifts of the two beams, the mechanism is different. In a two-photon transition, one photon from each of the beams is absorbed, resulting in a single resonance free from Doppler shifts, as opposed to two resonances that are only free from Doppler shifts when their combined signal is detected. Consequently, this Doppler-free component is also not subject to Doppler broadening (see next section). Additionally, there is a Doppler-shifted and -broadened component, corresponding to the absorption of two photons from either of the beams. However, the contribution of this component to the signal is usually negligible due to its large linewidth compared to the Doppler-free component.

### 2.2.5 Doppler broadening

While the use of two counter-propagating laser beams discussed in the previous section suppresses the first-order Doppler shift, it does not suppress the associated Doppler broadening of the observed linewidth. It is instructive to first consider two limiting cases: if the Doppler shift is much smaller than the natural linewidth, $|\Delta\nu_\mathrm{D}| \ll \Gamma$, the two resonances overlap almost completely and the linewidth of the total signal is close to $\Gamma$. On the other hand, for $|\Delta\nu_\mathrm{D}| \gg \Gamma$, the doublet will be well-separated in frequency and each resonance will again appear to be Lorentzian. However, if the Doppler shift $\Delta\nu_{\mathrm{D}}$ is on the order of the natural linewidth $\Gamma$, the two resonances will partly overlap and the total signal will appear to have an increased FWHM linewidth $\Gamma_{\mathrm{F}}$, as shown in Fig. 2.2 (B).

In an atomic beam experiment as described in this work, a distribution of atomic trajectories with different crossing angles $\delta\alpha$ is probed, with each trajectory contributing a doublet with frequency separation $2\Delta\nu_{\mathrm{D}}$ to the total signal (see Fig. 2.2 (C)). The total signal is then additionally inhomogeneously broadened, i.e., the broadening stems from the fact that the total signal is made up from the signal of many atoms, each with a different line shape. If the distribution of $\delta\alpha$ can be approximated by a Gaussian with a FWHM of $\Delta\alpha \ll \pi$, which corresponds to the atomic beam divergence, and centered at $\alpha_0 = 0\,\mathrm{mrad}$, the distribution of Doppler shifts $\Delta\nu_{\mathrm{D}}$ is also approximately Gaussian with a FWHM of $\Gamma_{\mathrm{G}} \approx \nu_{\mathrm{A},0}\Delta\alpha v/c$. The line shape of the total signal is given by the convolution of the doublet line shape from each atom, approximated by two Lorentzians, and the Gaussian distribution of Doppler shifts, resulting in a Voigt line shape [51]. The total FWHM linewidth of a Voigt line shape is approximately given by [53]

$$\Gamma_{\mathrm{F}} \approx 0.5346\Gamma_{\mathrm{L}} + \sqrt{0.2166\Gamma_{\mathrm{L}}^2 + \Gamma_{\mathrm{G}}^2} \tag{2.13}$$

and always equal or above $\Gamma_{\mathrm{L}}$ and $\Gamma_{\mathrm{G}}$. The atomic beam used for the spectroscopy of the 2S-6P transition has a divergence of $\Delta\alpha \approx 10\,\mathrm{mrad}$, which at the typical experimental speed of $v_{\mathrm{typ}} = 200\,\mathrm{m/s}$ results in $\Gamma_{\mathrm{G}} \approx 4.9\,\mathrm{MHz}$ and $\Gamma_{\mathrm{F}} \approx 7.3\,\mathrm{MHz}$, nearly twice the natural 2S-6P linewidth of $\Gamma = 3.9\,\mathrm{MHz}$.

This broadening substantially impacts the experiment. First, all else being equal, it increases the statistical uncertainty in the determination of the resonance frequency, which is approximately proportional to the total linewidth. Thus, in order to gain from the narrower natural linewidth of the 2S-$n$P transitions with increasing $n$, the divergence of the atomic beam must not be limiting the total linewidth. Second, it complicates and adds uncertainty to the theoretical modeling of the line shape, as it must include the properties of the atomic beam, which are only approximately described by Gaussian distributions. Importantly, however, the center of mass of the total line shape is still free from the first-order Doppler effect independent of the atomic beam properties, as long as the laser beams are phase-retracing.

So far, it has been assumed that $\alpha_0 = 0\,\mathrm{mrad}$, i.e., that the laser beams and the center of mass of the atomic beam cross at exactly right angles. In the experiment, this configuration is usually used to acquire spectroscopy data, and is ensured through an in-situ alignment in which the laser beams are rotated relative to the atomic beam. If the offset angle $\alpha_0$ is not zero, on the other hand, the total signal can again split into a doublet of two resonances (see Fig. 2.2 (D)) with a frequency splitting given by $\Delta\nu_0 = 2\nu_{\mathrm{A},0}\sin(\alpha_0)v/c$. Each resonance is associated with one of the laser beams, but contains the signal from all atoms. The total line shape is then the sum of two Voigts, which throughout this work is referred to as a Voigt doublet. While such a configuration is unfavorable in terms of statistical uncertainty, some systematic effects like the light force shift change their behavior, which can be exploited to test the theoretical modeling of these effects. Indeed, for this reason some spectroscopy data were acquired in the 2S-6P measurement for $\alpha_0 = 7.5\,\mathrm{mrad}$ and $\alpha_0 = 12.0\,\mathrm{mrad}$, corresponding to $\Delta\nu_0 = 7.3\,\mathrm{MHz}$ and $\Delta\nu_0 = 11.7\,\mathrm{MHz}$ for $v_{\mathrm{typ}}$, respectively. Note that in the experiment, limits on the accuracy of the determination and setting of $\alpha_0$ mean that even in the case where no offset angle is explicitly applied, a small residual, nonzero offset angle $\alpha_0$ will be present, which can lead to additional line broadening.

The discussion has up to now neglected saturation effects. Depending on their Doppler shift $\Delta\nu_{\mathrm{D}}$, atoms will resonantly interact, on the one extreme ($|\Delta\nu_{\mathrm{D}}| \gg \Gamma$), with only one or, on the other extreme ($|\Delta\nu_{\mathrm{D}}| \ll \Gamma$), with both laser beams at the same time. Thus, the intensity seen by the atoms depends on $\Delta\nu_{\mathrm{D}}$, which in turn implies that if saturation effects are non-negligible, the signal seen from the latter case will not be twice the signal as seen from the former case, as would be expected otherwise. This then leads to a smaller total signal for $\Delta\nu_{\mathrm{L}} \lesssim \Gamma$ than expected from the Voigt or Voigt doublet line shape, as both observed in the simulations and in the experiment. In theory, these saturation effects can then be included in a more sophisticated line shape model. However, experimentally they are hard to distinguish from deviations stemming from the non-Gaussian distribution of the atomic beam.

### 2.2.6 Preparation of metastable 2S atoms using the 1S-2S transition

The 1S-2S transition in atomic hydrogen has been extensively studied [23, 54, 55] as its narrow natural linewidth of $\Gamma_{2\mathrm{S}} = 1.31\,\mathrm{Hz}$ makes it an excellent choice for precision measurements. As opposed to the 2S-$n$P transitions, the 1S-2S transition is not dipole-allowed, but can be excited through the absorption of two photons at a wavelength of $\lambda_{1\mathrm{S}-2\mathrm{S}} = 243\,\mathrm{nm}$. This allows it to be studied free from first-order Doppler shifts through two-photon laser spectroscopy (see Section 2.2.4.4).

Here, this technique is used to prepare the hydrogen atoms, initially in the 1S ground level as they enter the experimental apparatus, in the metastable 2S level. Specifically, the transition from the $1\mathrm{S}_{1/2}^{F=0}$, $m_F = 0$ level ($|\mathrm{f}_1\rangle$) to the $2\mathrm{S}_{1/2}^{F=0}$, $m_F = 0$ level ($|\mathrm{i}\rangle$) is driven with the 1S-2S preparation laser, as shown in Fig. 2.1. The transition frequency is predicted to be

$$\begin{aligned}\nu_{\text{1S-2S}} &= \nu_{\text{1S-2S,cent}} + (3/4)\Delta\nu_{\text{HFS}}(1\mathrm{S}_{1/2}) - (3/4)\Delta\nu_{\text{HFS}}(2\mathrm{S}_{1/2}) \\ &= 2\,466\,062\,345\,323\,723(11)\,\mathrm{Hz},\end{aligned} \tag{2.14}$$

where $\nu_{\text{1S-2S,cent}}$ is the frequency of the 1S-2S hyperfine centroid, measured in [23], and $\Delta\nu_{\text{HFS}}(1\mathrm{S}_{1/2})$ and $\Delta\nu_{\text{HFS}}(2\mathrm{S}_{1/2})$ are 1S and 2S hyperfine splittings, measured in [56] and [57], respectively. This excitation scheme addresses one-quarter of the thermally-populated 1S atoms, as the three $F = 1$ ground levels are not resonantly coupled by the preparation laser. Likewise, the three $F = 1$ levels of the $2\mathrm{S}_{1/2}$ manifold are not populated.

The two-photon excitation of this transition is studied in detail in [58]. The time scale of the excitation is given by the two-photon Rabi frequency

$$\Omega_{\text{1S-2S}} = 2\left(\frac{m_{\mathrm{p}} + m_{\mathrm{e}}}{m_{\mathrm{p}}}\right)^3 2\left(2\pi\beta_{\mathrm{ge}}\right) I = I \times \left(463\,(\mathrm{rad/s})/(\mathrm{MW/m^2})\right), \tag{2.15}$$

where $I$ is the laser intensity per direction, i.e., the intensity of each of the counter-propagating beams. The coefficient $\beta_{\mathrm{ge}} = 3.681\,11 \times 10^{-5}\,\mathrm{Hz/(W/m^2)}$ is given in Table II of [58], and $m_{\mathrm{p}}$ and $m_{\mathrm{e}}$ are the proton and electron mass, respectively. For the typical parameters used here, the intensity at the center of the beams is $I = 7.2\,\mathrm{MW/m^2}$, resulting in $\Omega_{\text{1S-2S}} = 3.3 \times 10^3\,\mathrm{rad/s}$. Note that the two-photon Rabi frequency is proportional to the intensity and thus the square of the laser electric field, as opposed to the one-photon Rabi frequency introduced in Section 2.3.1, which is linear in the electric field. The 2S excitation probability for short interaction times $t \ll 1/\Omega_{\text{1S-2S}}$ is then approximately given by $P_{2\mathrm{S}} \approx (\Omega_{\text{1S-2S}}t)^2 \propto t^2 I^2$. In the experiment described here, $t$ can be on the order of 1 ms, and thus well outside the limit for short interaction times.

A single photon from the preparation laser, with an energy of $(3/8)hcR_\infty = 5.10\,\mathrm{eV}$, is energetic enough to photoionize the 2S level, which has an ionization energy of $(1/4)hcR_\infty = 3.40\,\mathrm{eV}$. The resulting ionization rate (see Appendix C.2) is

$$\begin{aligned}\gamma_{\mathrm{ioni}}(2\mathrm{S}) &= 2\left(\frac{m_\mathrm{p}+m_\mathrm{e}}{m_\mathrm{p}}\right)^3\left(-\sqrt{1/3}\right)2\pi\tilde{\beta}^{(0)}_{\mathrm{ioni}}(2\mathrm{S})I \\ &= I\times\left(1513\,(\mathrm{ionizations/s})/(\mathrm{MW/m^2})\right), \end{aligned} \tag{2.16}$$

with $\tilde{\beta}^{(0)}_{\mathrm{ioni}}(2\mathrm{S}) = -2.082\,06\times10^{-4}\,\mathrm{Hz/(W/m^2)}$ as given in Table C.4. For the parameters used here, the loss of 2S atoms from ionization dominates over the natural decay of the 2S level, leading to a decrease in lifetime of up to three orders of magnitude. Ionization limits the maximum achievable population in the 2S level, as with increasing 2S population, the loss rate due to ionization starts to dominate over the excitation rate to the 2S level. In fact, it can be shown [58] that the 2S population cannot exceed[1] $P_{2\mathrm{S}} = 0.175$, independent of the laser intensity used.

For linear laser polarization, as used here, the probability of a photoelectrons to be emitted into an infinitesimal solid angle $\mathrm{d}\Omega$ along direction $(\phi,\theta)$ is $p(\theta)\,\mathrm{d}\Omega \propto \cos^2(\theta)\,\mathrm{d}\Omega$, where $\theta$ is the angle to the polarization direction [38]. The photoelectron emerges with an energy of $1.70\,\mathrm{eV}$, taking up a fraction of $m_\mathrm{p}/(m_\mathrm{e}+m_\mathrm{p}) \approx 0.998$ of the excess energy. The proton, on the other hand, only gains $0.9\,\mathrm{meV}$ and recoils with a speed of $421\,\mathrm{m/s}$, which is however comparable to the thermal speed of the cold atoms in the atomic beam employed here.

While the excitation is free from the first-order Doppler shift, it is affected by both the second-order Doppler shift and the ac-Stark shift. The second-order Doppler shift is given through Eq. (2.8) as

$$\Delta\nu_{\mathrm{SOD,1S\text{-}2S}} = -\frac{\nu_{\mathrm{1S\text{-}2S}}}{2}\left(\frac{v}{c}\right)^2 = v^2\times\left(-13.7\,\mathrm{mHz/(m/s)^2}\right). \tag{2.17}$$

The mean speed of the atoms probed in the experiment described here ranges within $\bar{v} = 66\,\mathrm{m/s}\ldots256\,\mathrm{m/s}$ (see Table 5.1), resulting in $\Delta\nu_{\mathrm{SOD,1S\text{-}2S}} = -60\,\mathrm{Hz}\ldots-899\,\mathrm{Hz}$.

The ac-Stark shift, the shift of the energy levels due to the interaction with the oscillating laser field, is discussed in detail in Appendix C.1. For the 1S-2S transition, it is given by

$$\begin{aligned}\Delta\nu_{\mathrm{ac,1S\text{-}2S}} &= 2\left(\frac{m_\mathrm{p}+m_\mathrm{e}}{m_\mathrm{p}}\right)^3\left(-\sqrt{1/3}\right)\left(\tilde{\beta}^{(0)}_{\mathrm{ac}}(2\mathrm{S})-\tilde{\beta}^{(0)}_{\mathrm{ac}}(1\mathrm{S})\right)I \\ &= I\times\left(334\,\mathrm{Hz/(MW/m^2)}\right). \end{aligned} \tag{2.18}$$

The coefficients $\tilde{\beta}^{(0)}_{\mathrm{ac}}$ are given in Table C.2. For the typical intensity at the center of the beam, the ac-Stark shift is $\Delta\nu_{\mathrm{ac,1S\text{-}2S}} = 2.4\,\mathrm{kHz}$.

Both shifts are thus typically much larger than the natural linewidth. As the speed and the path through the laser beams vary from atom to atom, each atom sees different effective values for $\Delta\nu_{\mathrm{SOD,1S\text{-}2S}}$ and $\Delta\nu_{\mathrm{ac,1S\text{-}2S}}$. This leads to a substantial inhomogeneous broadening of the linewidth. Ionization, on the other hand, leads to a homogeneous broadening and line shape distortions. Additionally, the observed resonance is affected by time-of-flight broadening [59], as the time it takes the atoms to fly though the apparatus, 0.8 ms. . . 3.1 ms, is much shorter than the lifetime $\tau_{2S}$ of the 2S level. The amount of broadening again depends on the path the atoms take through the laser beams, but it is on the order of kHz as it is approximately given by the inverse of the interaction time. All in all, the four mechanism lead to an observed linewidth of $\Gamma_F \sim 3$ kHz, and the combination of the second-order Doppler and ac-Stark shifts, which are of opposite sign, lead to a shift of the observed transition frequency on the order of hundreds of Hz. To accurately take into account these mechanism in the formation of the atomic beam of metastable 2S atoms, a Monte Carlo simulation of many atomic trajectories is used as described in Section 5.2. A typical observation of the 1S-2S resonance with the apparatus presented in this thesis is shown in Fig. 7.2.

[1] The value given for the maximum 2S population was calculated neglecting the natural decay, and the second-order Doppler and ac-Stark shift, but including those effects does not change the value appreciably.

Static electric fields couple the 2S levels to the short-lived 2P levels, leading to dc-Stark shifts and quenching. The 1S levels are not affected, as there are no levels of opposite parity with the same principal quantum number. Specifically, the $2S_{1/2}^{F=0}$, $m_F = 0$ level couples to the $2P_{1/2}^{F=1}$ and $2P_{3/2}^{F=1}$ levels, separated in energy by $-910$ MHz and 10.0 GHz, respectively. This shifts the $2S_{1/2}^{F=0}$, $m_F = 0$ level[1], and thus the transition, by

$$\Delta\nu_{\mathrm{dc,1S\text{-}2S}} = \beta_{\mathrm{dc}}(2S_{1/2}^{F=0},\, m_F = 0)F^2 = F^2 \times \left(442\,\mathrm{mHz}/(\mathrm{V/m})^2\right), \tag{2.19}$$

where $F$ is the magnitude of the static electric field $\boldsymbol{F}$. This expression is valid for $\Delta\nu_{\mathrm{dc,1S\text{-}2S}} \ll$ 910 MHz.

The static electric fields quench the 2S level, i.e., reduce its effective lifetime, since the population transferred to the 2P level through the coupling quickly decays to the 1S ground level with a lifetime of $\tau_{2P} = 1/\Gamma_{2P} = 1.6$ ns. The quenched decay rate of the $2S_{1/2}^{F=0}$, $m_F = 0$ level[2] can be estimated as

$$\begin{aligned}\Gamma_{2S,qu} &= \Gamma_{2S} + \beta_{\mathrm{dc,qu}}(2S_{1/2}^{F=0},\, m_F = 0)\Gamma_{2P}F^2 \\ &= \Gamma_{2S} + \Gamma_{2P}F^2 \times (6.0 \times 10^{-10}\, 1/(\mathrm{V/m})^2) \approx F^2 \times \left(0.38\,(\mathrm{dcy/s})/(\mathrm{V/m})^2\right),\end{aligned} \tag{2.20}$$

where the approximation is valid within 1 % for $F > 47$ V/m. The approximate quenched 2S lifetime is then

$$\tau_{2S,qu} \approx 1/F^2 \times \left(2.6\,\mathrm{s}\,(\mathrm{V/m})^2\right). \tag{2.21}$$

Note that the minimum lifetime of the quenched 2S level is $2\tau_{2P}$.

## 2.3 Atom–light interaction

### 2.3.1 Master equation of atom–light interaction for a multi-level atom

The master equation of atom–light interaction is a description of the interaction of an atom with an electromagnetic (EM) field, modeled as a reservoir of photons. The EM field consist of both a coherent laser field and the vacuum electromagnetic field and thus lead to a description that includes absorption from and stimulated emission into the laser field, and spontaneous emission into free space modes, corresponding to emission stimulated by the vacuum field. The master equation gives the time evolution of the atom's density matrix $\varrho(t)$ as it evolves through the coupling to the EM field. The resulting coupled first-order differential equations for the elements of $\varrho(t)$ are known as the optical Bloch equations (OBEs). From $\varrho(t)$, the expectation values at time $t$ of operators, such as the population in an atomic level, can be calculated. The master equation reproduced here is derived in Chapter 2 in [61]. We will briefly outline the derivation and the assumptions made therein.

[1] The dc-Stark shift coefficients are found by solving the time-independent Schrödinger equation, using the level energies of [40] and thus including the fine and hyperfine structure. Unlike the ac-Stark shift given above, where the fine and hyperfine structure was ignored, this results in different coefficients for the different 2S hyperfine levels: $\beta_{\mathrm{dc}}$ is 352 mHz/(V/m)$^2$ and 329 mHz/(V/m)$^2$ for the $2S_{1/2}^{F=1}$, $m = 0$ and $m = \pm 1$ sublevels, respectively, with the electric field taken to be along the $z$-axis.

[2] As for the dc-Stark shift, the quenched decay rate depends on the 2S hyperfine level in question: $\beta_{\mathrm{dc,qu}}$ is $3.8 \times 10^{-10}$ 1/(V/m)$^2$ and $4.3 \times 10^{-10}$ 1/(V/m)$^2$ for the $2S_{1/2}^{F=1}$, $m = 0$ and $m = \pm 1$ sublevels, respectively. If the hyperfine structure is ignored, $\Gamma_{2S,qu} \approx F^2 \times \left(0.27\,(\mathrm{dcy/s})/(\mathrm{V/m})^2\right)$, comparable to the value of [60].

We consider an atom at rest with $M+1$ energy levels with energy eigenstates $|n\rangle, n = 0, 1, \dots, M$ and with corresponding energies $\hbar\tilde{\omega}_n$. The unperturbed atom is then described by the Hamiltonian

$$H_0 = \hbar \sum_{n=0}^{M} \tilde{\omega}_n |n\rangle\langle n|. \tag{2.22}$$

The population in each level is given by $a_n(t) = \mathrm{Tr}(\varrho(t)|n\rangle\langle n|)$. In the hydrogen atom, there are infinitely many bound states, but for levels not excited with a laser or otherwise populated by decays the population can be assumed to be zero at all times, allowing us to treat only a finite number of levels. Furthermore, the continuum states corresponding to an ionization of the atom are ignored. Both approximations hold for a sufficiently low enough laser intensity for a given precision goal, which is the case for the experimental situation considered here.

Only dipole-allowed transitions between the levels are considered here, since those are typically many orders of magnitude stronger than dipole-forbidden transitions. This is an adequate description for spectroscopy of the dipole-allowed 2S-$n$P transition, but this would not, e.g., describe the excitation of the 1S-2S transition. There are $K$ dipole-allowed transitions between the (lower) level $|m_i\rangle$ and the (upper) level $|n_i\rangle$, $i = 1, 2, \dots, K$, with a corresponding angular transition frequency $\omega_{n_i,m_i} = (E_{n_i} - E_{m_i})/\hbar$. The transitions are identified by the unique index $i$, with the corresponding projection operators

$$S_i^+ = |n_i\rangle\langle m_i|, \tag{2.23}$$

$$S_i^- = |m_i\rangle\langle n_i|, \tag{2.24}$$

and the angular transition frequency

$$\omega_i \equiv \omega_{n_i,m_i}. \tag{2.25}$$

Here, we limit the derivation through the approximations below to optical transitions, i.e., $\omega_i$ is on the order of hundreds of THz.

Each transition is associated with a dipole moment $\boldsymbol{\mu_i}$, given by the corresponding matrix element of the electric dipole operator $\boldsymbol{\mu}$, $\boldsymbol{\mu_i} = \langle n_i|\boldsymbol{\mu}|m_i\rangle$. The transition frequency and the magnitude of dipole moment, $\mu_i = |\boldsymbol{\mu_i}|$, determine the spontaneous emission rate $\Gamma_i$ of transition $i$ as

$$\Gamma_i = \frac{\omega_i^3 \mu_i^2}{3\pi\epsilon_0\hbar c^3}, \tag{2.26}$$

where $c$ is the speed of light in vacuum and $\epsilon_0$ is the permittivity of free space. $\Gamma_i$ is identical to the Einstein $A$ coefficient.

To describe the cross-damping through spontaneous emission between a pair of transitions $i$ and $j$, $i \neq j$, we define a parameter $\gamma_{ij}$ as

$$\gamma_{ij} = \sqrt{\Gamma_i \Gamma_j} \frac{\boldsymbol{\mu}_i \cdot \boldsymbol{\mu}_j}{\mu_i \mu_j} \quad (i \neq j). \tag{2.27}$$

$\gamma_{ij}$ thus depends on the relative orientation of the dipole moments $\boldsymbol{\mu}_i$ and $\boldsymbol{\mu}_j$, with $\gamma_{ij}$ vanishing for orthogonal dipole moments. For $i = j$, $\gamma_{ij}$ reduces to $\gamma_{ii} \equiv \Gamma_i$.

To simplify the equations, we work in the interaction picture, where the time evolution of the unperturbed atom is shifted from the density matrix $\varrho$ to the state vectors $|n\rangle$. Using the unitary transformation $U = e^{iH_0 t/\hbar}$, the density matrix in the interaction picture is given by

$$\varrho^{\mathrm{I}} = U \varrho U^\dagger. \tag{2.28}$$

Note that since $H_0$ is diagonal, $U$ is diagonal as well, and thus this transformation corresponds to a multiplication of each state vector with a phase factor. Then, $\mathrm{Tr}(\varrho^{\mathrm{I}}(t)|n\rangle\langle n|) = \mathrm{Tr}(\varrho(t)|n\rangle\langle n|) = a_n(t)$. Since we are here only interested in (the change of) the level populations $a_n(t)$, from which the fluorescence signal can be calculated, we do not need to transform back to $\varrho$ to interpret the results.

To derive the master equation, several approximations have to be made. The first approximation in [61] is the Born approximation, which assumes that the state of the EM field is not changed in time by the interaction with the atom. This is a valid approximation in our case, since only very few of the photons in the laser field are absorbed by the atoms, and the spontaneous decay of the atoms distributes photons in sufficiently many modes of the EM field, since there is no preferred mode as would, e.g., be the case with a cavity.

The second approximation made is the rotating wave approximation (RWA), in which rapidly oscillating terms of angular frequencies $\omega_i + \omega_j$ are ignored in the derivation. Neglecting these terms leads to a model error, the Bloch-Siegert shift, which tends to be very small for optical transitions. E.g., for the 2S-6P transitions, using the typical parameters given in Table 2.1, this shift is on the order of 1 µHz. We include the Bloch-Siegert shift in the non-resonant ac-Stark shift, as discussed in Appendix C.1.

The third approximation made is the Markov approximation. Under this approximation, it is assumed that $\varrho$ evolves on a time scale that is much shorter than the time scale of radiative processes in the atom, which in turn determine the typical correlation time of atom-field interaction. This approximation is justified because the time scale of radiative processes is given by the angular transition frequencies $\omega_i$, while $\varrho$ evolves at best with the spontaneous emission rates $\Gamma_i$ (here, $\Gamma_i \sim$ tens of Mdcy/s), which are many orders of magnitudes smaller.

In general, there can be $W$ laser fields interacting with the atoms with angular frequencies $\omega_{\mathrm{L},l}$ and Rabi frequencies $\Omega_i^{(l)}$ for $l = 1, \ldots, W$. This interaction is described by the atom-laser interaction Hamiltonian $H_{\mathrm{L}}$, with the laser fields described as coherent light fields[1], given by

$$H_{\mathrm{L}}(t) = -\frac{1}{2}\hbar \sum_{l=1}^{W} \sum_{i=1}^{K} \left[ \Omega_i^{(l)} S_i^+ e^{-i(\omega_{\mathrm{L},l} - \omega_i)t} + \mathrm{H.c.} \right]. \tag{2.29}$$

[1]A similar expression of $H_{\mathrm{L}}(t)$ is given in Eq. (2.65) of [61], but the expression here corresponds to a laser field phase-shifted by $\pi/2$ in order to recover the usual OBEs as given, e.g., in [62]. The factor of 1/2 in our Eq. (2.29) is a result of the rotating wave approximation, since only the co-rotating terms accounting for half the electric field were kept.

The Rabi frequency of transition $i$ and laser field $l$ is

$$\Omega_i^{(l)} = \frac{1}{\hbar}\boldsymbol{\mu_i} \cdot \boldsymbol{E_l}. \tag{2.30}$$

The electric field of the laser is

$$\boldsymbol{E_l} = \bar{\boldsymbol{e}}_l \sqrt{\frac{2I_l}{c\epsilon_0}}, \tag{2.31}$$

where $\bar{\boldsymbol{e}}_l$ is a unit vector giving the orientation of the laser polarization and $I_l$ is the intensity of the laser beam.

In the experiments discussed in this thesis, at any time only one single-frequency laser with angular frequency $\omega_\mathrm{L}$ is interacting with the atoms. However the laser beam is retroreflected to create a forward- ("+", $l = 1$) and a backward-traveling ("−", $l = 2$) beam with identical frequencies $\omega_\mathrm{L} = \omega_{\mathrm{L},1} = \omega_{\mathrm{L},2}$. For ideal retroreflection, both polarization and intensity will be the same for the two beams, $\bar{\boldsymbol{e}} = \bar{\boldsymbol{e}}_1 = \bar{\boldsymbol{e}}_2$ and $I = I_1 = I_2$. Taking the propagation direction to be along the $x$-axis and with the laser wavenumber $K_\mathrm{L} = \omega_\mathrm{L}/c$, the Rabi frequencies of the beams are

$$\Omega_i^{(1)} \equiv \Omega_i^+ = \Omega_{i,0} e^{iK_\mathrm{L}x}, \tag{2.32}$$

$$\Omega_i^{(2)} \equiv \Omega_i^- = \Omega_{i,0} e^{-iK_\mathrm{L}x}, \tag{2.33}$$

$$\Omega_{i,0} = \boldsymbol{\mu_i} \cdot \bar{\boldsymbol{e}} \sqrt{\frac{2I}{\hbar^2 c\epsilon_0}}. \tag{2.34}$$

With this, Eq. (2.29) becomes

$$\begin{aligned} H_\mathrm{L}(t) &= -\frac{1}{2}\hbar \sum_{i=1}^{K} \left[\Omega_{i,0}\left(e^{iK_\mathrm{L}x} + e^{-iK_\mathrm{L}x}\right) S_i^+ e^{-i(\omega_\mathrm{L}-\omega_i)t} + \mathrm{H.c.}\right] \\ &= -\hbar\cos(K_\mathrm{L}x) \sum_{i=1}^{K} \left[\Omega_{i,0} S_i^+ e^{-i(\omega_\mathrm{L}-\omega_i)t} + \mathrm{H.c.}\right], \end{aligned} \tag{2.35}$$

which corresponds to the atom encountering a standing light wave with local intensity $I_\mathrm{loc} = 4I\cos^2(K_\mathrm{L}x)$ and with a periodicity of half the laser wavelength $\lambda = c/\omega_\mathrm{L}$. We note that while the two laser beams have the same frequency in the laboratory frame, an atom moving with velocity $v_x$ along the $x$-axis will observe the frequencies $\omega_\mathrm{L} \pm K_\mathrm{L}v_x$ due to the first-order Doppler shift (see Eq. (2.4)).

We can now give an expression for the master equation in the interaction picture for atom–light interaction, including a coherent laser field (cf. Eq. (2.74) in [61], where we here assume zero thermal photons[1]):

$$\frac{\delta\varrho^\mathrm{I}}{\delta t} = \frac{1}{i\hbar}\left[H_\mathrm{L}(t), \varrho^\mathrm{I}\right] + \mathcal{L}^\mathrm{relax}(\varrho^\mathrm{I}), \tag{2.36}$$

$$\text{with} \quad \mathcal{L}^\mathrm{relax}(\varrho^\mathrm{I}) = \sum_{i,j=1}^{K} \mathcal{L}_{ij}^\mathrm{relax}(\varrho^\mathrm{I}), \tag{2.37}$$

$$\text{and} \quad \mathcal{L}_{ij}^\mathrm{relax}(\varrho^\mathrm{I}) = \gamma_{ij} e^{i(\omega_i-\omega_j)t}\left(-\frac{1}{2}\left(S_i^+ S_j^- \varrho^\mathrm{I} + \varrho^\mathrm{I} S_i^+ S_j^-\right) + S_j^- \varrho^\mathrm{I} S_i^+\right). \tag{2.38}$$

[1] This corresponds to $N = 0$ in [61].

Eq. (2.36) is the master equation in Lindblad form. The first term gives the time evolution as generated by the Hermitian Hamiltonian $H_\mathrm{L}$, as described by the quantum analog of the Liouville equation, the von Neumann equation. It describes the interaction of the atom with the coherent laser field, i.e., absorption and stimulated emission of laser photons. The second term is the relaxation superoperator $\mathcal{L}^\mathrm{relax}$ acting on the density matrix $\varrho^\mathrm{I}$, $\mathcal{L}^\mathrm{relax}(\varrho^\mathrm{I})$, corresponding to spontaneous emission of photons from the atom caused by the interaction with the vacuum electromagnetic field. The dynamics of the vacuum field itself are, by construction, not described by $\varrho^\mathrm{I}$. When we refer to the optical Bloch equations in this work, we generally mean the coupled first-order differential equations for the elements of $\varrho^\mathrm{I}$ as given in Eq. (2.36).

We note that $\mathcal{L}^\mathrm{relax}(\varrho^\mathrm{I})$ contains not only terms proportional to $\gamma_{ii} \equiv \Gamma_i$ and thus describing the spontaneous emission of a photon from transition $i$, but also terms proportional to $\gamma_{ij}$, $i \neq j$, describing the cross-damping through spontaneous emission between transitions $i$ and $j$.

It is instructive to look in detail at the contribution $\langle u|\mathcal{L}_{ij}^\mathrm{relax}(\varrho^\mathrm{I})|v\rangle$ to the matrix element $\langle u|\delta\varrho^\mathrm{I}/\delta t|v\rangle$ of $\delta\varrho^\mathrm{I}/\delta t$ from $\mathcal{L}^\mathrm{relax}(\varrho^\mathrm{I})$ for transitions $i$ and $j$. We distinguish two cases: first, $i$ and $j$ have the same lower level, $|m_i\rangle \equiv |m_j\rangle$ ($m_i = m_j$), and, second, $i$ and $j$ have different lower levels, $|m_i\rangle \neq |m_j\rangle$ ($m_i \neq m_j$). $\langle u|\mathcal{L}_{ij}^\mathrm{relax}(\varrho^\mathrm{I})|v\rangle$ is then given by

$$\begin{aligned}
&\langle u|\mathcal{L}_{ij}^\mathrm{relax}(\varrho^\mathrm{I})|v\rangle \\
&= \gamma_{ij} e^{i(\omega_i-\omega_j)t}\left(-\frac{1}{2}\delta_{m_i,m_j}\left(\delta_{u,n_i}\langle n_j|\varrho^\mathrm{I}|v\rangle + \delta_{v,n_j}\langle u|\varrho^\mathrm{I}|n_i\rangle\right) + \delta_{u,m_j}\delta_{v,m_i}\langle n_j|\varrho^\mathrm{I}|n_i\rangle\right) \\
&= \gamma_{ij} e^{i(\omega_i-\omega_j)t}\begin{cases} -\frac{1}{2}\left(\delta_{u,n_i}\langle n_j|\varrho^\mathrm{I}|v\rangle + \delta_{v,n_j}\langle u|\varrho^\mathrm{I}|n_i\rangle\right) + \delta_{u,m_j}\delta_{v,m_i}\langle n_j|\varrho^\mathrm{I}|n_i\rangle & m_i = m_j \\ \delta_{u,m_j}\delta_{v,m_i}\langle n_j|\varrho^\mathrm{I}|n_i\rangle & m_i \neq m_j. \end{cases}
\end{aligned} \tag{2.39}$$

For the first case, consider the change in the population of the (common) lower level from spontaneous decays of transitions $i$ and $j$, $\langle m_i|\mathcal{L}_{ij}^\mathrm{relax}(\varrho^\mathrm{I})|m_i\rangle \propto \gamma_{ij}\langle n_j|\varrho^\mathrm{I}|n_i\rangle$. This matrix element is of particular interest for experiments detecting fluorescence, as it is equal to the fluorescence signal of these decays. For $i = j$, this term just describes the spontaneous decay of the upper level $|n_i\rangle$. However, for $i \neq j$, this term becomes a cross-damping term and describes an additional contribution to the fluorescence signal proportional to both $\gamma_{ij}$ and the coherence between the two upper levels $\langle n_j|\varrho^\mathrm{I}|n_i\rangle$. Such a coherence is, for example, induced through the simultaneous (resonant or off-resonant) excitation of the upper levels from a common initial level. For the experiment discussed here, a term of this form allows the quantum interference between excitations and subsequent decays of two different fine-structure manifolds (see Section 2.3.2.3).

In the second case, the only nonzero matrix element is $\langle m_j|\mathcal{L}_{ij}^\mathrm{relax}(\varrho^\mathrm{I})|m_i\rangle \propto \gamma_{ij}\langle n_j|\varrho^\mathrm{I}|n_i\rangle$, which describes the evolution of the coherence between the distinct lower levels of transitions $i$ and $j$. Thus, this coherence is induced by cross-damping through spontaneous emission between the two transitions. In the special case where the upper levels are also distinct, this describes the transfer of the coherence between the upper levels to the lower levels, which will later play an important role when including the momentum exchange between atoms and photons in the atom–light interaction (see Section 3.4.1). However, as opposed to the first case, there is no quantum interference term contributing to the fluorescence signal from two decays with distinct lower levels (i.e., $\langle m_i|\mathcal{L}_{ij}^\mathrm{relax}(\varrho^\mathrm{I})|m_i\rangle = \langle m_j|\mathcal{L}_{ij}^\mathrm{relax}(\varrho^\mathrm{I})|m_j\rangle = 0$) independent on whether the decays have a common upper level or not.

In the corresponding expression for $\delta\varrho^{\mathrm{I}}/\delta t$ given in [61], there are also terms proportional to a coefficient $\delta_{ij}$, where the indices $i, j$ again identify atomic transitions. Terms proportional to $\delta_{ii}$ correspond to shifts of the energy levels caused by the coupling to the vacuum field. However, those energy shifts are already absorbed in the level energies $\omega_i$ used here, since they are included in the QED corrections discussed before. Terms proportional to $\delta_{ij}$, $i \neq j$ describe a dynamical coupling between atomic transitions driven by the vacuum field. However, for the hydrogen atom it can be shown that these contributions cancel when only (optical) transitions between manifolds of different principal quantum number $n$ are considered [63], as is the case here.

### 2.3.2 Quantum interference and the big model of the 2S-6P transition

Based on the procedure described in the previous section, a model for the spectroscopy of the 2S-6P transition as described in this work and shown in Fig. 2.1 is derived. Because this model includes all relevant atomic levels, it is here referred to as big model. A similar model was used for the 2S-4P measurement (see Appendix A) and the 1S-3S measurement of [26]. The big model is based on work by Arthur Matveev.

#### 2.3.2.1 Laser-driven excitations

The atom is assumed to be initially in the $2\mathrm{S}_{1/2}^{F=0}$, $m_F = 0$ level, which is taken to be stable as motivated in Section 2.2.1. The optical excitation to this initial level is taken into account with a separate model of the atomic beam, which is described in Section 5.2. Counter-propagating beams of the spectroscopy laser, with linear polarization along the $z$-axis, couple the initial level to the $6\mathrm{P}_{1/2}^{F=1}$, $m_F = 0$ and $6\mathrm{P}_{3/2}^{F=1}$, $m_F = 0$ level. As in the experiment, the spectroscopy laser frequency is scanned over either of the two corresponding transitions, i.e., the 2S-$6\mathrm{P}_{1/2}$ or 2S-$6\mathrm{P}_{3/2}$ transition. All other excitations to the 6P manifold are forbidden by angular momentum conservation for the chosen initial level and laser polarization. However, through decays, the $2\mathrm{S}_{1/2}^{F=1}$ levels may be populated, which are off-resonantly coupled to the 6P manifold. Therefore, all dipole-allowed excitations from the 2S manifold to the 6P manifold, and the corresponding levels, are included in the big model.

Non-resonant couplings by the laser light from the 2S levels (6P levels) to $n$P ($n$S and $n$D) levels with main quantum number $n \neq 6$ ($n \neq 2$) are not taken into account in the big model. Instead, the resulting non-resonant ac-Stark shift and photoionization are treated separately in Appendix C, but are found to be negligible at the current level of accuracy and for the laser intensities used here (see Table 2.1).

#### 2.3.2.2 Spontaneous decays

Dipole-allowed decays from the $6\mathrm{P}_{1/2}^{F=1}$, $m_F = 0$ and $6\mathrm{P}_{3/2}^{F=1}$, $m_F = 0$ level either directly, or through cascades, lead to either the 1S or 2S manifold, resulting in many different possible decay paths, with the final decay always a Lyman or Balmer decay, respectively. Upon a decay, the angular momentum of the atom changes by $\hbar$, as required by the conservation of momentum since the emitted photon has spin 1. The projection of the angular momentum of the photon onto the $z$-axis is $q\hbar$, where $q = -1, 0, 1$. Accordingly, the quantum number $m_F$ identifying the hyperfine sublevels of the atom changes by $\Delta m_F = -q$, i.e., $m_F \to m_F - q$.

Table 2.2: Probability $p$ and number $N$ of Lyman (Ly) and Balmer (Ba) decay paths, i.e., decay cascades with the final decay leading from the $n'$P level to the $n$S level for $n = 1$ and $n = 2$, respectively, for an atom initially in the $6P_{1/2}^{F=1}$, $m_F = 0$ level. The decay paths are grouped by the spherical component $q = -\Delta m_F = 0, \pm 1$ of the final decay, corresponding to $\pi, \sigma^{\pm}$ decays, respectively. $p$ is the ratio of the strengths of the considered decay paths to the total strength of all dipole-allowed decay paths leading to the 1S and 2S level. $N$ includes the number of possible paths leading to the $n'$P level from which the final decay starts. There are no decay paths to the 5P level and thus no Ly-$\delta$ or Ba-$\gamma$ decays.

| | $n$ | $n'$ | Energy | $\pi$ ($q = 0$) | | $\sigma^{\pm}$ ($q = \pm 1$) | | Sum ($\sigma^- + \pi + \sigma^+$) | |
|---|---|---|---|---|---|---|---|---|---|
| | | | (eV) | $p$ (%) | $N$ | $p$ (%) | $N$ | $p$ (%) | $N$ |
| Ly-$\alpha$ | 1 | 2 | 10.20 | 1.910 | 4479 | 2.188 | 4272 | 6.287 | 13 023 |
| Ly-$\beta$ | 1 | 3 | 12.09 | 0.339 | 108 | 0.387 | 106 | 1.113 | 320 |
| Ly-$\gamma$ | 1 | 4 | 12.75 | 0.068 | 54 | 0.078 | 53 | 0.223 | 160 |
| Ly-$\epsilon$ | 1 | 6 | 13.22 | 26.843 | 1 | 26.843 | 1 | 80.528 | 3 |
| Ly | 1 | | | 29.160 | 4642 | 29.496 | 4432 | 88.152 | 13 506 |
| Ba-$\alpha$ | 2 | 3 | 1.89 | 0.046 | 108 | 0.052 | 106 | 0.149 | 320 |
| Ba-$\beta$ | 2 | 4 | 2.55 | 0.010 | 54 | 0.011 | 53 | 0.032 | 160 |
| Ba-$\delta$ | 2 | 6 | 3.02 | 3.889 | 1 | 3.889 | 1 | 11.667 | 3 |
| Ba | 2 | | | 3.952 | 160 | 3.952 | 160 | 11.848 | 483 |
| Sum | | | | 33.448 | 4805 | 33.448 | 4592 | 100.000 | 13 989 |

Table 2.3: Same as Table 2.2, but for an atom initially in the $6P_{3/2}^{F=1}$, $m_F = 0$ level.

| | $n$ | $n'$ | Energy | $\pi$ ($q = 0$) | | $\sigma^{\pm}$ ($q = \pm 1$) | | Sum ($\sigma^- + \pi + \sigma^+$) | |
|---|---|---|---|---|---|---|---|---|---|
| | | | (eV) | $p$ (%) | $N$ | $p$ (%) | $N$ | $p$ (%) | $N$ |
| Ly-$\alpha$ | 1 | 2 | 10.20 | 2.087 | 6201 | 2.100 | 5898 | 6.287 | 17 997 |
| Ly-$\beta$ | 1 | 3 | 12.09 | 0.371 | 134 | 0.371 | 130 | 1.113 | 394 |
| Ly-$\gamma$ | 1 | 4 | 12.75 | 0.074 | 67 | 0.074 | 65 | 0.223 | 197 |
| Ly-$\epsilon$ | 1 | 6 | 13.22 | 53.685 | 1 | 13.421 | 1 | 80.528 | 3 |
| Ly | 1 | | | 56.218 | 6403 | 15.967 | 6094 | 88.151 | 18 591 |
| Ba-$\alpha$ | 2 | 3 | 1.89 | 0.005 | 134 | 0.005 | 130 | 0.149 | 394 |
| Ba-$\beta$ | 2 | 4 | 2.55 | 0.011 | 67 | 0.011 | 65 | 0.032 | 197 |
| Ba-$\delta$ | 2 | 6 | 3.02 | 7.778 | 1 | 1.945 | 1 | 11.667 | 3 |
| Ba | 2 | | | 7.839 | 202 | 2.005 | 196 | 11.849 | 594 |
| Sum | | | | 64.057 | 6605 | 17.972 | 6290 | 100.000 | 19 185 |

Decays with $q = -1, 0, 1$ are referred to as $\sigma^-$, $\pi$, $\sigma^+$ decays, respectively[1]. Identifying the decays by their value of $q$ corresponds to a decomposition of the electric dipole operator $\boldsymbol{\mu}$ into spherical harmonics with spherical components $\boldsymbol{\mu}_q$ [29]. For brevity, $q$ is here referred to as the spherical component. The angular distribution $p(\theta)$ of the emitted photon, where $\theta$ is the angle to the $z$-axis, is also governed by angular momentum conservation [64]. This results in

$$p(\theta) = \frac{3}{8\pi}\left(1 - \cos^2(\theta)\right) \tag{2.40}$$

for $\pi$ ($q = 0$) decays, and

$$p(\theta) = \frac{3}{16\pi}\left(1 + \cos^2(\theta)\right) \tag{2.41}$$

for $\sigma^{\pm}$ ($q = \pm 1$) decays.

[1]The definition used here corresponds to light emitted along the positive $z$-axis being right-handed (left-handed) circularly polarized about the $z$-axis for $\sigma^+$ ($\sigma^-$) decays. Along any other emission direction, the light is not purely circularly polarized.

Table 2.4: Probability $p$ and number $N$ of Lyman (Ly) and Balmer (Ba) decay paths, i.e., decay cascades with the final decay leading from the $n'$P level to the $n$S level for $n = 1$ and $n = 2$, respectively, for an atom initially in the $6\mathrm{P}_{1/2}^{F=1}$, $m_F = 0$ level. Similar to Table 2.2, but with the decay paths grouped by the final hyperfine level reached.

| | $n$ | $n'$ | $n\mathrm{S}_{1/2}^{F=0}$, $m_F = 0$ | | $n\mathrm{S}_{1/2}^{F=1}$, $m_F = 0$ | | $n\mathrm{S}_{1/2}^{F=1}$, $m_F = \pm 1$ | |
|---|---|---|---|---|---|---|---|---|
| | | | $p$ (%) | $N$ | $p$ (%) | $N$ | $p$ (%) | $N$ |
| Ly-$\alpha$ | 1 | 2 | 1.774 | 2623 | 1.177 | 3528 | 1.668 | 3436 |
| Ly-$\beta$ | 1 | 3 | 0.314 | 72 | 0.209 | 76 | 0.295 | 86 |
| Ly-$\gamma$ | 1 | 4 | 0.063 | 36 | 0.042 | 38 | 0.059 | 43 |
| Ly-$\epsilon$ | 1 | 6 | 26.843 | 1 | 0 | 0 | 26.843 | 1 |
| Ly | 1 | | 28.994 | 2732 | 1.427 | 3642 | 28.865 | 3566 |
| Ba-$\alpha$ | 2 | 3 | 0.042 | 72 | 0.028 | 76 | 0.004 | 86 |
| Ba-$\beta$ | 2 | 4 | 0.009 | 36 | 0.006 | 38 | 0.008 | 43 |
| Ba-$\delta$ | 2 | 6 | 3.889 | 1 | 0 | 0 | 3.889 | 1 |
| Ba | 2 | | 3.940 | 109 | 0.034 | 114 | 3.937 | 130 |

Table 2.5: Same as Table 2.4, but for an atom initially in the $6\mathrm{P}_{3/2}^{F=1}$, $m_F = 0$ level.

| | $n$ | $n'$ | $n\mathrm{S}_{1/2}^{F=0}$, $m_F = 0$ | | $n\mathrm{S}_{1/2}^{F=1}$, $m_F = 0$ | | $n\mathrm{S}_{1/2}^{F=1}$, $m_F = \pm 1$ | |
|---|---|---|---|---|---|---|---|---|
| | | | $p$ (%) | $N$ | $p$ (%) | $N$ | $p$ (%) | $N$ |
| Ly-$\alpha$ | 1 | 2 | 2.759 | 3557 | 1.011 | 4916 | 1.258 | 4762 |
| Ly-$\beta$ | 1 | 3 | 0.491 | 86 | 0.178 | 96 | 0.222 | 106 |
| Ly-$\gamma$ | 1 | 4 | 0.098 | 43 | 0.036 | 48 | 0.044 | 53 |
| Ly-$\epsilon$ | 1 | 6 | 53.685 | 1 | 0 | 0 | 13.421 | 1 |
| Ly | 1 | | 57.034 | 3687 | 1.225 | 5060 | 14.946 | 4922 |
| Ba-$\alpha$ | 2 | 3 | 0.066 | 86 | 0.024 | 96 | 0.03 | 106 |
| Ba-$\beta$ | 2 | 4 | 0.014 | 43 | 0.005 | 48 | 0.006 | 53 |
| Ba-$\delta$ | 2 | 6 | 7.778 | 1 | 0 | 0 | 1.945 | 1 |
| Ba | 2 | | 7.858 | 130 | 0.029 | 144 | 1.981 | 160 |

All dipole-allowed decay paths from the $6\mathrm{P}_{1/2}^{F=1}$, $m_F\!=\!0$ and $6\mathrm{P}_{1/2}^{F=1}$, $m_F\!=\!0$ level are listed according to their spherical component $q$ in Table 2.2 and Table 2.3, respectively. Likewise,

Table 2.4 and Table 2.5 list the same decay paths for the two excited levels, but grouped according to which final hyperfine level is reached. All 13 989 or, respectively, 19 185 listed decay paths and the involved intermediate and final levels are included in the big model. The corresponding Tables B.2 to B.5 for the 4P excited levels are given in Appendix B.

#### 2.3.2.3 Quantum interference

Quantum interference (QI) between two possible paths leading from the initial level to the detection of a fluorescence photon through either fine-structure manifold gives rise to distortions of the observed fluorescence signal [29, 45]. The first (second) path consists of a laser-driven excitation to the $6\mathrm{P}_{1/2}^{F=1}$, $m_F=0$ ($6\mathrm{P}_{3/2}^{F=1}$, $m_F=0$) level and a subsequent decay with spherical component $q$. Since the laser at any time is only close to resonance for one the excited levels, which are separated by in energy by $\Delta = \Delta\nu_{\mathrm{FS}}(6\mathrm{P}) \approx 100 \times \Gamma$, the off-resonant path essentially acts as a perturbation to the resonant path. The resulting distorted line shape of the 2S-$n$P transitions is discussed in detail in Appendix A and [29]. A corresponding line shape model, the Fano-Voigt line shape, is also derived therein, based on the Kramers-Heisenberg formula, which describes the scattering of photons by atomic electrons [65]. Here, we do not employ this line shape model, but instead use the big model to describe and correct for the line shape distortions from quantum interference.

The quantum interference is enabled by cross-damping through spontaneous emission between the decays $i$ and $j$ from the excited level, described by the corresponding cross-damping terms in the master equation (see Eqs. (2.36) and (2.39)), which are proportional to $\gamma_{ij}$ (see Eq. (2.27)). If the decays $i$ and $j$ correspond to different spherical components $q$, their dipole moments $\boldsymbol{\mu_i}$ and $\boldsymbol{\mu_j}$ are orthogonal and $\gamma_{ij}$ is zero, and thus there is no interference. In this way, separate interference effects are observed for decays with different $q$, and the resulting fluorescence signals with their characteristic angular emission distributions as given in Eqs. (2.40) and (2.41) can be added up incoherently. There is no interference in the fluorescence signal from decays leading to different final levels, as in this case the paths are distinguishable and the corresponding cross-damping terms, but not $\gamma_{ij}$, are zero (see Eq. (2.39) and discussion thereafter).

Quantum interference as described above is observable in the fluorescence signal from any of the decays of the excited levels. In principle, subsequent decays from intermediate levels, forming a decay cascade, are also subject to quantum interference. However, the interference tends to wash out as the decay cascade progresses, as many intermediate levels are involved.

Furthermore, in addition to the two paths through either fine-structure manifold with the same main quantum number $n$, there are also far-off-resonant paths leading through $n'$P levels with different main quantum numbers $n' \neq n$. However, since these levels are separated in energy from the resonant level by $\Delta > 4\,000\,000 \times \Gamma$, and the line shape distortions from quantum interference approximately scale as $\Gamma^2/\Delta$ [29], the effect from these paths is negligible.

In the experiment described here, the direct Ly-$\epsilon$ decay of the excited levels to the 1S ground levels constitute approximately 97 % of the total detected fluorescence signal (see Section 4.6.5), which is thus subject to almost the full QI distortions. Because both the QI distortion and the angular emission distribution depend on $q$, the observed QI distortion depends on the solid angle of the fluorescence detection [29]. Only in the case with a full detection coverage, i.e., a detection solid angle of $4\pi$, do the distortions disappear in the limit of negligible optical pumping effects. This is possible because the distortions of the fluorescence signal are of opposite sign for the $\pi$ and $\sigma^{\pm}$ decays in the excitation scheme used here. The dependence of the line shape distortions on the detection was investigated in detail in the 2S-4P measurement (see Appendix A). For the 2S-6P measurement discussed in this work, the detection solid angle, and the orientation of the linear laser polarization relative to the detector assembly, was chosen such that QI distortions and the associated line shifts are minimized.

#### 2.3.2.4 Derivation of the OBEs

Using the list of excitations and decays, and all 150 involved levels, the optical Bloch equations (OBEs) of the big model are derived using Eq. (2.36). Importantly, these equations include cross-damping terms proportional to $\gamma_{ij}$, $i \neq j$, which lead to quantum interference. The level energies are taken from [40], and the physical constants needed to convert from atomic units to SI units are taken from [46]. Some fast-rotating terms that do not affect the dynamics of the system, as their effect is either small or averages out, are removed from the OBEs to speed up the numerical integration. The details of this procedure will be given in the aforementioned upcoming publication.

To model the fluorescence observed in the experiment, separate signal equations for each Lyman or Balmer decay and each spherical component $q$ are added. Each signal equation is defined as the sum, over all contributing decays and their cross-damping terms, of the increase in population of the corresponding lower level due to spontaneous decay of the upper levels, as given by Eq. (2.39). In this way, the effective solid angle of the detection in the experiment can be included by weighting the signals with different $q$ by the product of their angular emission distribution (see Eqs. (2.40) and (2.41)) and the spatial detection efficiency of the fluorescence detection, found through simulations (see Section 4.6.6). Likewise, the signals corresponding to fluorescence at different frequencies can be weighted by the spectral sensitivity of the fluorescence detection (see Section 4.6.5).

Finally, the resulting equations are separated into real and imaginary parts, and equations that are always zero for the initial state given above are removed. In total, this procedure results in OBEs consisting of 732 real-valued coupled differential equations, including 42 signal equations. The OBEs are then numerically integrated for a given trajectory, i.e., the path an atom takes through the spectroscopy laser beams, and for different frequency detunings $\Delta\nu_{\text{2S-6P}}$ of the spectroscopy laser, as detailed in Section 5.3.

The big model as described here does not include external electric or magnetic fields, but either can be included in the Hamiltonian. This has been done for electric fields to model the dc-Stark shift, as discussed in Section 2.4.

## 2.4 dc-Stark shift of resonances observed through fluorescence

Static electric fields, just like the time-varying electric field of the laser, couple different atomic levels. This coupling leads to dc-Stark shifts of the level energies and new eigenstates which contain an admixture of other levels to the level of interest.

A coupling between the levels $|1\rangle$ and $|2\rangle$ with energies $E(|1\rangle)$ and $E(|2\rangle)$ leads to new eigenstates $|1'\rangle$ and $|2'\rangle$ with energies $E(|1'\rangle)$ and $E(|2'\rangle)$. The energy of level $|1'\rangle$, formed by an electric-field-induced coupling between levels $|1\rangle$ and $|2\rangle$, is in leading order given by time-independent second-order perturbation theory as [64]

$$E(|1'\rangle) = E(|1\rangle) - \frac{|\boldsymbol{\mu} \cdot \boldsymbol{F}|^2}{\Delta E}, \tag{2.42}$$

where $\Delta E = E(|2\rangle) - E(|1\rangle)$. $\boldsymbol{\mu}$ is the dipole moment of the transition between $|1\rangle$ and $|2\rangle$ (see Section 2.3.1), and $\boldsymbol{F}$ is the applied static electric field with field strength $F = |\boldsymbol{F}|$. An analogous expression applies to the energies of level $|2'\rangle$. The energy difference, in frequency units, between the unperturbed level $|1\rangle$ and the perturbed level $|1'\rangle$ is known as the dc-Stark shift of level $|1\rangle$,

$$\Delta\nu_{\mathrm{dc}}(|1\rangle) = \frac{E(|1'\rangle) - E(|1\rangle)}{h} = -\frac{|\boldsymbol{\mu} \cdot \boldsymbol{F}|^2}{h\Delta E} = \beta_{\mathrm{dc}}(|1\rangle)F^2. \tag{2.43}$$

Likewise, $\Delta\nu_{\mathrm{dc}}(|2\rangle) = -\Delta\nu_{\mathrm{dc}}(|1\rangle)$. $\beta_{\mathrm{dc}}(|1\rangle)$ is the quadratic dc-Stark shift coefficient of level $|1\rangle$, which in general depends on the relative orientation of $\boldsymbol{\mu}$ and $\boldsymbol{F}$. The shift is proportional to the inverse energy difference between the coupled levels and the square of the electric field strength $F$, and $\Delta\nu_{\mathrm{dc}}(|1\rangle)$ is negative if $\Delta E > 0$ and positive otherwise. That is, the energy difference between $|1'\rangle$ and $|2'\rangle$ increases with increasing $F$.

Likewise using time-independent second-order perturbation theory, the new eigenstates are given by

$$|1'\rangle = \left(1 - \frac{1}{2}\frac{|\boldsymbol{\mu} \cdot \boldsymbol{F}|^2}{\Delta E^2}\right)|1\rangle - \frac{\boldsymbol{\mu} \cdot \boldsymbol{F}}{\Delta E}|2\rangle = a|1\rangle - b|2\rangle, \tag{2.44}$$

$$|2'\rangle = \left(1 - \frac{1}{2}\frac{|\boldsymbol{\mu} \cdot \boldsymbol{F}|^2}{\Delta E^2}\right)|2\rangle + \frac{\boldsymbol{\mu} \cdot \boldsymbol{F}}{\Delta E}|1\rangle = a|2\rangle + b|1\rangle, \tag{2.45}$$

where $|a|^2 + |b|^2 = 1 + \mathcal{O}(|\boldsymbol{\mu} \cdot \boldsymbol{F}|^4/\Delta E^4)$.

Now, consider the case where $|1\rangle$ is the excited level of a dipole-allowed transition from an initial level $|\mathrm{i}\rangle$ with energy difference $E(|1\rangle) - E(|\mathrm{i}\rangle) \gg |\Delta E|$. There is no dipole moment between $|\mathrm{i}\rangle$ and $|2\rangle$, as they are by definition of same parity, and thus if no electric field is applied only the transition $|\mathrm{i}\rangle \to |1\rangle$ is dipole-allowed. However, if an electric field is applied, the new eigenstates $|1'\rangle$ and $|2'\rangle$ both contain a contribution from $|1\rangle$. Therefore, both the transition $|\mathrm{i}\rangle \to |1'\rangle$ and the transition $|\mathrm{i}\rangle \to |2'\rangle$ are allowed, and, in the absence of saturation effects, the relative excitation probability of $|2'\rangle$ as compared to $|1'\rangle$ is $|b|^2/|a|^2 = |\boldsymbol{\mu} \cdot \boldsymbol{F}|^2/\Delta E^2 + \mathcal{O}(|\boldsymbol{\mu} \cdot \boldsymbol{F}|^4/\Delta E^4)$.

The effective dc-Stark shift of a transition frequency depends on the how the frequency is determined. If the resonances of the $|\mathrm{i}\rangle \to |1'\rangle$ and $|\mathrm{i}\rangle \to |2'\rangle$ transition are well-separated in frequency, their transition frequencies can be determined independently, if incoherent and coherent line shifts from the far-reaching slopes of the resonances and quantum interference, respectively, are neglected. Then, the determined frequency of the $|\mathrm{i}\rangle \to |1'\rangle$ ($|\mathrm{i}\rangle \to |2'\rangle$) transition will be shifted by $\Delta\nu_{\mathrm{dc}}(|1\rangle)$ ($\Delta\nu_{\mathrm{dc}}(|2\rangle) = -\Delta\nu_{\mathrm{dc}}(|1\rangle)$). That is, the frequency shift of the transition frequencies corresponds to the frequency shift of the level energies.

If the resonances are however not well-separated in frequency, a measurement will yield some combination of the transition frequencies and the associated dc-Stark shifts. For example, for the case that the measurement determines the center of mass of both resonances, the determined transition frequency will be shifted from the unperturbed frequency of the

$|\mathrm{i}\rangle \to |1\rangle$ transition by

$$\begin{aligned}\Delta\nu_{\mathrm{dc}}(\mathrm{COM}) &= |a|^2\Delta\nu_{\mathrm{dc}}(|1\rangle) + |b|^2\left(\frac{\Delta E}{h} + \Delta\nu_{\mathrm{dc}}(|2\rangle)\right)\\ &= -\left(1 - \frac{1}{2}\frac{|\boldsymbol{\mu}\cdot\boldsymbol{F}|^2}{\Delta E^2}\right)^2\frac{|\boldsymbol{\mu}\cdot\boldsymbol{F}|^2}{h\Delta E} + \frac{|\boldsymbol{\mu}\cdot\boldsymbol{F}|^2}{\Delta E^2}\left(\frac{\Delta E}{h} + \frac{|\boldsymbol{\mu}\cdot\boldsymbol{F}|^2}{h\Delta E}\right)\\ &= \mathcal{O}\left(\frac{|\boldsymbol{\mu}\cdot\boldsymbol{F}|^4}{\Delta E^4}\right). \end{aligned} \tag{2.46}$$

That is, to leading order the dc-Stark shifts cancel out and the unperturbed transition frequency is recovered. The frequency shift of the observed resonance is thus not given by the frequency shift of the level energies, but can be much smaller.

In a real atomic system, the levels $|1\rangle$ and $|2\rangle$ will have limited lifetimes and thus nonzero linewidths, which are also not necessarily identical. The lifetime cannot be included within time-independent perturbation theory, and the expressions derived above are not valid if the frequency difference between the coupled levels approaches their linewidths. A treatment taking into account the time evolution of the system is then necessary, e.g., using optical Bloch equations (OBEs).

The excited levels will also in general not decay to the same final levels and the emitted fluorescence will likewise in general not be detected with the same efficiency in a given experimental apparatus, i.e., the relative weight of the resonances depends on the experimental realization. Moreover, there might be more than one perturbing level $|2\rangle$ that needs to be taken into account. To complicate things further, the shift of a combination of resonances is not necessarily given by the center of mass. In the experiment discussed here, the transition frequency is found by fitting a line shape function, which is a nonlinear procedure, to a resonance sampled at discrete frequency detunings. However, the two limiting cases discussed above are still instructive, and indeed the dc-Stark shift of the 2S-6P$_{1/2}$ transition corresponds approximately to the dc-Stark shift of the 2S-6P$_{1/2}$ level, while the 2S-6P$_{3/2}$ transition experiences a much smaller dc-Stark shift than would be expected from the shift of the 6P$^{F=1}_{3/2}$ level.

The dc-Stark shift of the 6P$^{F=1}_{1/2}$, $m_F = 0$ level results mostly from the coupling to the 6S$^{F=0}_{1/2}$ and 6S$^{F=1}_{1/2}$ levels, which are separated in frequency from the 6P$^{F=1}_{1/2}$ level by 34 MHz and 41 MHz, respectively, and thus outside both the natural linewidth of $\Gamma = 3.90$ MHz and the observed linewidth of $\Gamma_{\mathrm{F}} \lesssim 10$ MHz. The observed dc-Stark shift $\Delta\nu_{\mathrm{dc}}$(2S-6P$_{1/2}$) of the 2S-6P$_{1/2}$ transition then approximately matches that of the 6P$^{F=1}_{1/2}$, $m_F = 0$ level, minus the much smaller dc-Stark shift of the initial 2S$^{F=0}_{1/2}$, $m_F = 0$ level (see Section 2.2.6). This results in

$$\Delta\nu_{\mathrm{dc}}(\text{2S-6P}_{1/2}) = \left(\beta_{\mathrm{dc}}(\text{6P}^{F=1}_{1/2}, m_F = 0)F^2 - \beta_{\mathrm{dc}}(\text{2S}^{F=0}_{1/2}, m_F = 0)\right)F^2, \tag{2.47}$$

with the values of $\beta_{\mathrm{dc}}$ given in Table 2.6. An example of the experimentally observed shift of the 2S-6P$_{1/2}$ transition is shown in Fig. 6.10 (A, C).

The dc-Stark shift of the 6P$^{F=1}_{3/2}$, $m_F = 0$ level, on the other hand, results mostly from the coupling to the 6D$^{F=2}_{3/2}$ and 6D$^{F=1}_{3/2}$ levels, which are only separated in frequency from the 6P$^{F=1}_{3/2}$ level by 57 kHz and −469 kHz, respectively. That is, they are well within the linewidth of the 6P$^{F=1}_{3/2}$ level, and the observed shift of the 2S-6P$_{3/2}$ transition is not expected to correspond to the shift of the $6\mathrm{P}_{3/2}^{F=1}$, $m_F=0$ level. This also means that if a strong enough electric field is applied the observed resonance splits into multiple components, with the contribution from the $6\mathrm{D}_{3/2}^{F=2}$ level dominating as the excitation probability scales as $1/\Delta E^2$.

Table 2.6: Quadratic dc-Stark shift coefficients $\beta_{\mathrm{dc}}$, $\tilde{\beta}_{\mathrm{dc},\pi}$, and $\tilde{\beta}_{\mathrm{dc},\sigma^\pm}$ of the relevant 2S and 6P levels. $\beta_{\mathrm{dc}}$ is calculated using second-order perturbation theory and gives the shift of the level energies, applicable for the $2\mathrm{S}_{1/2}^{F=0}$, $m_F=0$ and $6\mathrm{P}_{1/2}^{F=1}$, $m_F=0$ level. $\tilde{\beta}_{\mathrm{dc},\pi}$ and $\tilde{\beta}_{\mathrm{dc},\sigma^\pm}$, on the other hand, are derived from preliminary simulations of the shift of the observed resonance for the 2S-$6\mathrm{P}_{3/2}$ transition (using the big model; $|\boldsymbol{F}| = 0\,\mathrm{V/m}\ldots1.2\,\mathrm{V/m}$). $\tilde{\beta}_{\mathrm{dc},\pi}$ and $\tilde{\beta}_{\mathrm{dc},\sigma^\pm}$ give the shift observed for the fluorescence signal (here the simple sum over all Lyman decays) from $\pi$ and $\sigma^\pm$ decays, respectively. If all decays are detected with equal efficiency, the observed shift is given by $\pi + 2\sigma^\pm$. The static electric field $\boldsymbol{F}$ is taken to be either parallel ($\|$) or perpendicular ($\perp$) to the electric field $\boldsymbol{E}$ of the linearly polarized laser. See text for details.

| Level | Orientation of $\boldsymbol{F}$ | $\beta_{\mathrm{dc}}$ | $\tilde{\beta}_{\mathrm{dc},\pi}$ | $\tilde{\beta}_{\mathrm{dc},\sigma^\pm}$ | $\tilde{\beta}_{\mathrm{dc},\pi}+2\tilde{\beta}_{\mathrm{dc},\sigma^\pm}$ |
|---|---|---|---|---|---|
| | | (Hz/(V/m)$^2$) | | | |
| $2\mathrm{S}_{1/2}^{F=0}$, $m_F=0$ | — | 0.442 | — | — | — |
| $6\mathrm{P}_{1/2}^{F=1}$, $m_F=0$ | $\boldsymbol{F} \parallel \boldsymbol{E}$ | −1751 | — | — | — |
| | $\boldsymbol{F} \perp \boldsymbol{E}$ | −1505 | — | — | — |
| $6\mathrm{P}_{3/2}^{F=1}$, $m_F=0$ | $\boldsymbol{F} \parallel \boldsymbol{E}$ | −33 602 | −370 | −441 | −411 |
| | $\boldsymbol{F} \perp \boldsymbol{E}$ | −167 | 1215 | −2165 | −115 |

To find the dc-Stark shift of the 2S-$6\mathrm{P}_{3/2}$ transition as observed in the fluorescence signal, the couplings caused by the static electric field are included in the OBEs of the big model (see Section 2.3.2). Note that to this end, levels which otherwise would not be coupled, i.e., the 6S and 6D levels, need to be added, together with the levels they decay to, which may not already be taken into account. Then, the OBEs are solved for various electric field strength $F$ and with the electric field applied either along or perpendicular to the electric field vector $E$ of the laser field. For field strengths $F \gtrsim 5\,\mathrm{V/m}$, a splitting of the resonance into two components, which shift into opposite directions as $F$ increases, is visible. This is also clearly observed in the experiment, as shown in Fig. 6.10 (B).

The modeling is complicated by the fact that the resulting line shape will in general be different for distinct Lyman decays and the different spherical components of the decays, i.e., $\pi$ and $\sigma^\pm$ decays. As is the case for line shifts from quantum interference discussed in Section 2.3.2, the observed dc-Stark shift thus depends on the spectral detection efficiency across Lyman decays, and on the spatial detection efficiency and the orientation of the laser electric field $\boldsymbol{E}$ relative to the detector assembly through the different radiation patterns of the decays. For the preliminary simulations discussed here, the simple sum over all Lyman decays is used as signal for each spherical component, with the effect of the spectral detection efficiency to be investigated. The line shapes corresponding to $\pi$ and $\sigma^\pm$ decays are separately fit with a line shape function to extract the resonance frequencies $\nu_0$. $\nu_0$ versus $F$ then gives the dc-Stark shift of the 2S-$6\mathrm{P}_{3/2}$ transition for a given spherical component. For the experimental parameters used here and the investigated range of field strengths $F = 0\,\mathrm{V/m}\ldots1.2\,\mathrm{V/m}$, $\nu_0$ shifts approximately quadratically with $F$, and $\nu_0$ versus $F$ is fit with a parabola to extract the effective dc-Stark shift coefficients $\tilde{\beta}_{\mathrm{dc},\pi}$ and $\tilde{\beta}_{\mathrm{dc},\sigma^\pm}$ for $\pi$ and $\sigma^\pm$ decays, respectively.

The coefficients also depend on the laser power, the frequency sampling of the resonance, and the line shape function. An experimental example of this quadratic behavior is shown in Fig. 6.10 (D).

Table 2.6 gives the result for an atom crossing the 2S-6P spectroscopy laser beams with a speed of $v_{\text{typ}} = 200\,\text{m/s}$ and at an angle $\delta\alpha = 2\,\text{mrad}$. A spectroscopy laser power of $P_{\text{2S-6P}} = 10\,\mu\text{W}$, and a uniform frequency sampling and Voigt line shapes are used. $\tilde{\beta}_{\text{dc},\pi}$ and $\tilde{\beta}_{\text{dc},\sigma^\pm}$ are two orders of magnitude smaller than $\beta_{\text{dc}}$ for the case that the electric field is oriented parallel to the laser field ($\boldsymbol{F} \parallel \boldsymbol{E}$). Without this partial cancellation, the dc-Stark shift as given by $\beta_{\text{dc}}$ would certainly be limiting to the experiment. For perpendicular orientation ($\boldsymbol{F} \perp \boldsymbol{E}$), the situation is reversed, and $\tilde{\beta}_{\text{dc},\pi}$ and $\tilde{\beta}_{\text{dc},\sigma^\pm}$ are an order of magnitude larger than $\beta_{\text{dc}}$, which however is comparatively small to begin with. The coefficients $\tilde{\beta}_{\text{dc},\pi}$ and $\tilde{\beta}_{\text{dc},\sigma^\pm}$ change by less than 15 % when the spectroscopy laser power is tripled. All in all, the dc-Stark shift of the 2S-6P$_{3/2}$ transition as observed through fluorescence is similar to that of the 2S-6P$_{1/2}$ transition.

In the experiment discussed here, stray electric fields and the resulting dc-Stark shift are determined in situ. To this end, bias electric fields of strength $F = 10\,\text{V/m} \ldots 45\,\text{V/m}$ are applied along a given direction (see Section 4.6.7) and the resonance frequency is measured as a function of $F$. This data is then fit with a parabola of the form

$$\nu_0(F) = \tilde{\beta}_{\text{dc}}(F - \Delta F)^2 + \nu_0(F = 0\,\text{V/m}), \tag{2.48}$$

where $\Delta F$ is the strength of the stray electric field, and $\tilde{\beta}_{\text{dc}}$ the effective quadratic dc-Stark shift coefficient, along the direction of the applied bias field. From this, the dc-Stark shift caused by the stray electric field along the given bias field direction is found through

$$\Delta\nu_{\text{dc,2S-6P}} = \tilde{\beta}_{\text{dc}} \Delta F^2. \tag{2.49}$$

This procedure is done for all three directions and the three dc-Stark shifts are added to find the total dc-Stark shift. Fig. 6.10 (C, D) show examples of this procedure from the 2S-6P measurement.

From such determinations during the 2S-6P measurement, $\tilde{\beta}_{\text{dc}}$ is found to be approximately within $-1700\,\text{Hz/(V/m)}^2$ and $-1470\,\text{Hz/(V/m)}^2$ for the different directions and for the 2S-6P$_{1/2}$ transition, in good agreement with the values given above. For the 2S-6P$_{3/2}$ transition, $\tilde{\beta}_{\text{dc}}$ is found to be approximately $-400\,\text{Hz/(V/m)}^2$ for all directions, demonstrating the much lower effective dc-Stark shift coefficient of the observed resonance as opposed to the shift of the level energies. A detailed investigation including the spectral and spatial detection efficiency and different spectroscopy laser powers used is still outstanding.

Finally, the dipole moments $\boldsymbol{\mu}$ and energy separations $\Delta E$ between $nl$ levels with the same $n$ scale as $n^2$ and $n^{-3}$, respectively [38]. Thus, the dc-Stark shift of the 2S-$n$P transitions scales approximately as $n^7$. For $n = 4 \rightarrow n = 6$, this corresponds to a 17-fold increased sensitivity to stray electric fields, which is why for the 2S-6P measurement an in-situ determination of these fields was necessary, while an estimation of the shift was sufficient for the 2S-4P measurement.

# Chapter 3

# Light force shift

The basic idea of the experiment treated in this thesis is to cross an atomic beam of hydrogen atoms (H) with two counter-propagating laser beams, resonant with the 2S-6P transition. The counter-propagating beams suppress the first-order Doppler shift, while also forming a standing light wave. This situation is shown in Fig. 3.1. The standing light wave can be thought of as a light grating causing diffraction of an atom or matter wave, analogous to a mechanical grating causing diffraction of a light wave. This chapter addresses the question of how to describe and model the light grating, the matter wave, and their interaction, ultimately leading to an analysis of the resulting shift of the resonance observed in fluorescence. This shift is referred to as the light force shift (LFS).

Since the experimental geometry is crucial in determining how to model the light force shift, Section 3.1 briefly describes the geometry and important parameters. The coherence properties of the atomic beam, and thus the question whether the atom indeed has to be described as a matter wave undergoing interference at the light grating, are discussed in Section 3.2 and an analogy to classic optics is drawn. In Section 3.3, the quantum mechanical description of the atomic beam is derived, using the link between classical phase-space density and its quantum analogue, the Wigner function. In Section 3.4, the knowledge gained from the quantum mechanical description is then used to develop a model of the light force shift that takes into account the momentum exchange of the atom–light interaction.

The analysis of the light force shift presented here was made possible by theory support from Yue Chang and Tao Shi, who started working on this problem, originally in the context of the 2S-4P measurement, during their time at the Theory division of the Max Planck Institute of Quantum Optics. A joint publication detailing these efforts is in preparation.

## 3.1 Experimental geometry and parameters

As shown in Fig. 3.1, hydrogen atoms at room temperature are fed to a cold ($T_{\mathrm{N}} = 4.8\,\mathrm{K}$) nozzle, where they thermalize by sticking to and escaping again from the $H_2$-coated walls. They leave the nozzle through orifices with radius $r_1 = 1\,\mathrm{mm}$, forming a beam along the $z$-axis. Each atom leaving the nozzle can be assigned a classical trajectory with momentum $\boldsymbol{p} = (p_x, p_y, p_z)$, describing its center-of-mass motion. The atoms can also undergo collisions in the nozzle, producing trajectories with the lowest transverse momentum ($|p_x| \ll |p_z|$). For a detailed drawing of the nozzle, see Fig. 4.25.

The 1S-2S preparation laser ($\lambda_{\text{1S-2S}} = 243\,\mathrm{nm}$), propagating along the $z$-axis, excites the ground level (1S, $|\mathrm{f}\rangle$) atoms to the initial level (2S, $|\mathrm{i}\rangle$). At a distance of $L_1 = 154\,\mathrm{mm}$ from the nozzle orifice, an aperture of width $2r_2 = 1.2\,\mathrm{mm}$ narrows the atomic beam's transverse (along $x$-axis) velocity distribution by blocking some atoms (the aperture height in the $y$-direction is $2r_{2,y} = 2\,\mathrm{mm}$). A further distance $L_2 = 50\,\mathrm{mm}$ from the aperture, a standing light wave along the $x$-axis is formed by the two counter-propagating beams of the active fiber-based retroreflector, used to suppress the Doppler shift of the 2S-6P transition. The beams have a transverse $1/e^2$ intensity radius $W_0 = 2.2\,\mathrm{mm}$ and are derived from the 2S-6P spectroscopy laser ($\lambda_{\text{2S-6P}} = 410\,\mathrm{nm}$). In this way, the 2S-6P ($|\mathrm{i}\rangle \to |\mathrm{e}\rangle$) transition is probed, with Lyman photons emitted upon decay to the ground level ($|\mathrm{e}\rangle \to |\mathrm{f}\rangle$) serving as signal. The standing light wave acts as a grating with periodicity $\lambda_{\text{2S-6P}}/2 = 205\,\mathrm{nm}$ for the atoms, leading to a light force shift (LFS) of the observed 2S-6P resonance.

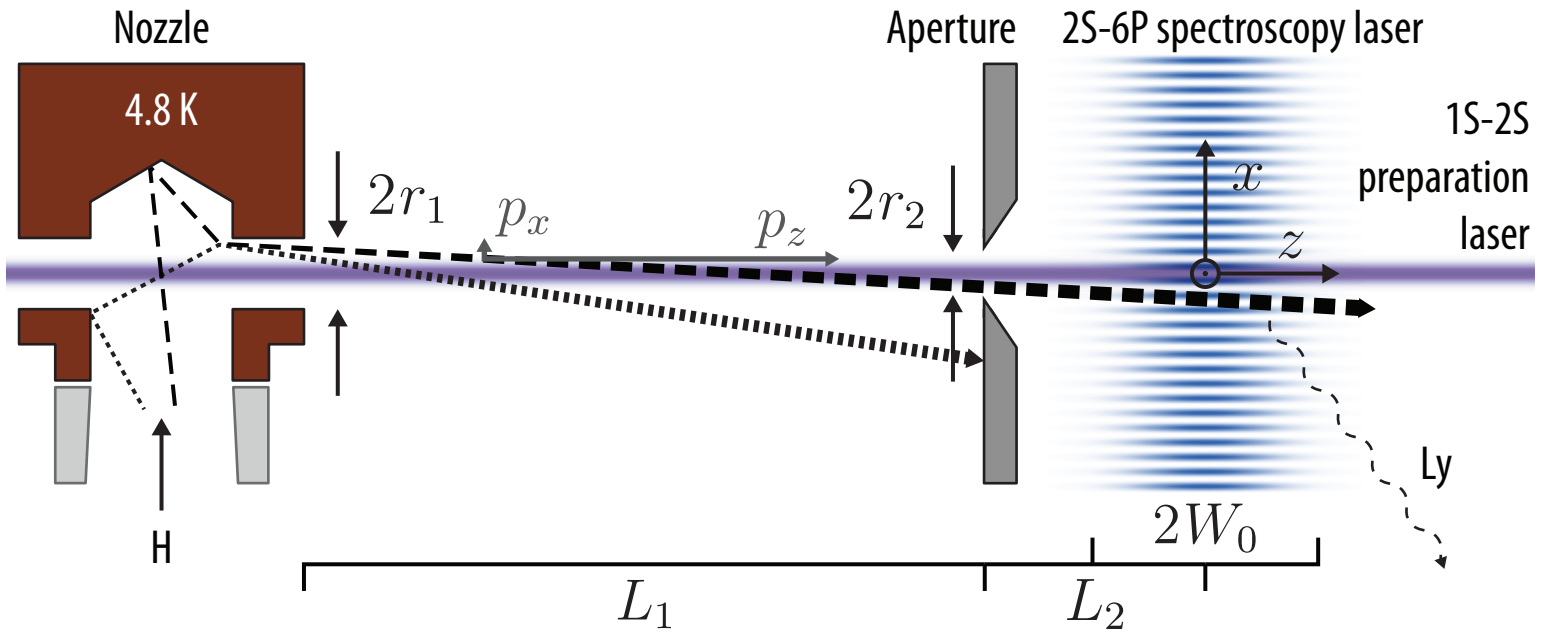

Figure 3.1: The geometry of the spectroscopy of the 2S-6P transition on a cryogenic atomic beam. Hydrogen atoms (H, dashed and dotted lines) at room temperature are fed to a cold ($T_N = 4.8\,\mathrm{K}$) nozzle (orifice radius $r_1 = 1\,\mathrm{mm}$), where they thermalize by sticking to and escaping again from the $H_2$-coated walls, forming a beam along the $z$-axis. Each atom leaving the nozzle can be assigned a classical trajectory with momentum $\boldsymbol{p} = (p_x, p_y, p_z)$, describing its center-of-mass motion. The atoms can also undergo collisions in the nozzle, producing trajectories with the lowest transverse momentum ($|p_x| \ll |p_z|$). The 1S-2S preparation laser ($\lambda_{\text{1S-2S}} = 243\,\mathrm{nm}$, purple), propagating along the $z$-axis, excites the ground level (1S, $|\mathrm{f}\rangle$) atoms to the initial level (2S, $|\mathrm{i}\rangle$). At a distance of $L_1 = 154\,\mathrm{mm}$ from the nozzle orifice, an aperture of width $2r_2 = d_2 = 1.2\,\mathrm{mm}$ narrows the atomic beam's transverse (along $x$-axis) velocity distribution by blocking some atoms (the aperture height in the $y$-direction is $2r_{2,y} = d_{2,y} = 2\,\mathrm{mm}$). A further distance $L_2 = 50\,\mathrm{mm}$ from the aperture, a standing light wave along the $x$-axis (blue) is formed by two counter-propagating beams with transverse $1/e^2$ intensity radius $W_0 = 2.2\,\mathrm{mm}$. The beams, derived from the 2S-6P spectroscopy laser ($\lambda_{\text{2S-6P}} = 410\,\mathrm{nm}$), probe the 2S-6P ($|\mathrm{i}\rangle \to |\mathrm{e}\rangle$) transition, with Lyman photons (wiggly line) emitted upon decay to the ground level ($|\mathrm{e}\rangle \to |\mathrm{f}\rangle$) serving as signal. The standing light wave acts as a grating with periodicity $\lambda_{\text{2S-6P}}/2 = 205\,\mathrm{nm}$ for the atoms, leading to a light force shift of the observed 2S-6P resonance. The transverse coherence length of the atomic wave increases during propagation, as symbolized by the widening line, and is comparable to the light grating's periodicity.

## 3.2 Coherence properties of the atomic beam

Two limiting regimes of the interaction between the light grating and the atoms can be distinguished. In the first limit, the atoms behave like classical well-localized particles, experiencing a force from the light grating acting at their center of mass. This limit is treated, e.g., in [62, 66], and was used to describe the observed channeling of atoms in a light grating [67], and line shifts in saturation spectroscopy of helium atoms [68]. In the other limit, the atoms are completely delocalized and behave like a plane wave, simultaneously interacting at all points with the light grating. This leads to a diffraction of the matter waves on the light grating [69], completely analogous to diffraction of light waves on a solid grating. In which regime, i.e., in either one of the limits or somewhere in between, an experiment operates depends on the coherence length $l_c$ of the atoms, which is the length scale over which interference can be observed. If the coherence length is much smaller than the periodicity of the light grating, the atoms can be treated as localized particles, but otherwise diffraction effects need to be taken into account.

Quantum mechanically, the nozzle can thought of as a source of matter waves. A matter wave of a particle with momentum $\boldsymbol{p}$ has a wavelength corresponding to de Broglie wavelength

$$\lambda_{\mathrm{dB}} = \frac{h}{p}, \tag{3.1}$$

where $p = |\mathbf{p}|$. For a hydrogen atom moving at $v_{\mathrm{typ}} = 200\,\mathrm{m/s}$, the de Broglie wavelength is $2\,\mathrm{nm}$. The de Broglie wavelength determines the size of the smallest wave packet that can constructed for the given particle [66], but is in general not identical to the coherence length, which can be much larger as we will see below. In this sense, a matter wave behaves identically to a light wave.

In a thermal gas of atoms at temperature $T_{\mathrm{N}}$, as is the case inside the nozzle, the average de Broglie wavelength is approximated by the thermal de Broglie wavelength

$$\lambda_{\mathrm{dB,th}} = \sqrt{\frac{2\pi\hbar^2 k_{\mathrm{B}} T_{\mathrm{N}}}{m_{\mathrm{H}}}}, \tag{3.2}$$

where $m_{\mathrm{H}}$ is the mass of the hydrogen atoms, and $k_{\mathrm{B}}$ is the Boltzmann constant. For $T_{\mathrm{N}} = 4.8\,\mathrm{K}$, as used in the experiment, $\lambda_{\mathrm{dB,th}} = 0.8\,\mathrm{nm}$. For the special case of a thermal gas of atoms, the coherence length is identical to its thermal de Broglie wavelength [70].

If we naively assume that the coherence length at the light grating is given by the coherence length of the thermal gas inside the nozzle, it would appear that interference effects from the light grating, which has a periodicity of $\lambda_{\text{2S-6P}}/2 \gg l_c \equiv \lambda_{\mathrm{dB,th}}$, are be negligible. However, as is well known from the Van Cittert-Zernike theorem in classical optics discussed below, interference patterns can be observed from gratings placed in beams even if the source is incoherent, i.e., the source's coherence length is negligible compared to the grating periodicity. This particular type of coherence is determined by the geometry of the experiment and is characterized by a transverse coherence length $l_{\mathrm{c,t}}$. As we will see in the quantum mechanical treatment given in Section 3.3, an analogous effect applies in the description of the atomic beam, and the transverse coherence length is found to be comparable to the periodicity of the light grating. Thus, the diffraction at the grating has to be taken into account in the theoretical description of the experiment.

### 3.2.1 The Van Cittert-Zernike theorem

From classical optics, it is well known that light from an extended incoherent light source[1] can produce interference patterns, i.e., have some degree of spatial coherence and thus coherently illuminate a certain area. The degree of coherence is contained in the function $g^{(1)}(\mathbf{P_1}, \mathbf{P_2})$, the absolute value of which gives the visibility of the interference pattern between light from spatially separated points $\mathbf{P_1}$ and $\mathbf{P_2}$. We here have dropped the time dependence of $g^{(1)}$, since we are only interested in spatial and not temporal coherence.

[1]Incoherence here is taken to mean that the phase of waves emerging from different sections of the source are uncorrelated, i.e., the time average of the product of the fields emerging from different sections is zero. Note that the incoherence of the source implies that there is some spread in the wavelength, i.e., the source cannot be strictly monochromatic [71].

The degree of coherence of an extended incoherent light source is described by the Van Cittert-Zernike theorem, which we will briefly outline here following section 10.4.2 in [71]. An incoherent source of light $A$ with constant intensity is assumed to occupy some part of a source plane, with the extent of the source given by $I(\xi, \eta)$, where $(\xi, \eta, 0)$ are the Cartesian coordinates of point $\mathbf{S}$ in the source plane. We assume $I$ to be normalized, i.e., $\iint_A I(\xi, \eta) d\xi d\eta = 1$. The light is observed at points $\mathbf{P_1}$ and $\mathbf{P_2}$ in an observation plane parallel to the source plane and at a distance $L$. The Cartesian coordinates of points $\mathbf{P_1}$ and $\mathbf{P_2}$ are $(X_1, Y_1, L)$ and $(X_2, Y_2, L)$.

We further assume that the extent of the source and the observation region, i.e., the region containing $\mathbf{P_i}$, are small compared to the distance between $\mathbf{P_i}$ and the source, and that the source and observation region are centered on an axis normal to the planes. This corresponds to the situation of a well-collimated light or atomic beam propagating along the $z$-axis.

We then find that the absolute value of the degree of coherence $|g^{(1)}|$ in the observation plane (referred to as $|\mu_{12}|$ in [71]) is given by

$$|g^{(1)}(\mathbf{P_1}, \mathbf{P_2})| = \left| \iint_A I(\xi, \eta) e^{-ik_{\mathrm{dB}}(p\xi + q\eta))} \, \mathrm{d}\xi \, \mathrm{d}\eta \right|, \tag{3.3}$$

with $p = (X_1 - X_2)/L$, $q = (Y_1 - Y_2)/L$, and $k_{\mathrm{dB}} = 2\pi/\lambda_{\mathrm{dB}}$. Thus, $|g^{(1)}|$ is equal to the absolute value of the Fourier transform of the source intensity $I$. Correspondingly, $|g^{(1)}|^2$ is the intensity distribution of the far-field diffraction pattern at the observation plane of a uniformly illuminated diffraction aperture with the same shape as the source placed in the source plane.

For the special case of a circular source with radius $r_1$, Eq. (3.3) evaluates to

$$|g^{(1)}| = \frac{2J_1(v)}{v}, \tag{3.4}$$

with the Bessel function of first kind and first order $J_1$, and $v = 2\pi r_1 \sqrt{p^2 + q^2}/\lambda_{\mathrm{dB}} = 2\pi r_1 \sqrt{(X_1 - X_2)^2 + (Y_1 - Y_2)^2}/\lambda_{\mathrm{dB}} L$. The expression $2J_1(v)/v$ has a central lobe of amplitude 1, dropping to $1/e$ at $\Delta v \approx 2.584$. After a zero crossing, where no interference is observed, some degree of coherence is reached with another lobe at higher $v$, but only with a maximum amplitude of 0.132. The amplitude further decreases for larger $v$. We can thus think of $\Delta v$ as the typical scale of $v$ for which coherence is observed. This translates to a typical length scale over which points $\mathbf{P_1}$ and $\mathbf{P_2}$ can be separated transverse to the axis of the system and still be illuminated coherently, given by

$$l_{\mathrm{c,t}} = \frac{\Delta v}{2\pi} \frac{L}{r_1} \lambda_{\mathrm{dB}} = \frac{1.29}{\pi} \frac{L}{r_1} \lambda_{\mathrm{dB}}. \tag{3.5}$$

We identify this typical length scale as the transverse coherence length $l_{\mathrm{c,t}}$, which corresponds to the distance of double slits placed transverse to the beam for which the visibility of the observed interference fringes is $1/e$.

For simplicity and to be able to compare with results to be derived below, we replace the circular source with a Gaussian source with a $1/e$ width of $r_1$, i.e., $I(\xi, \eta) = e^{-(\xi^2 + \eta^2)/r_1^2}/\pi r_1^2$.

Then, (the absolute value of) the degree of coherence (Eq. (3.3)) itself is also a Gaussian,

$$|g^{(1)}| = \exp\left(-\frac{k_{\mathrm{dB}}^2 r_1^2}{4L^2}\left((X_1 - X_2)^2 + (Y_1 - Y_2)^2\right)\right), \tag{3.6}$$

with the corresponding transverse coherence length

$$l_{\mathrm{c,t}} = \frac{1}{\pi}\frac{L}{r_1}\lambda_{\mathrm{dB}}. \tag{3.7}$$

Thus, with this choice of the size of the Gaussian source the result is comparable with the one of the circular source.

## 3.3 Quantum mechanical description of the atomic beam

### 3.3.1 Definition of Wigner function and first-order correlation function

So far, we have used a classical description that neglects quantum features such as the uncertainty principle. A more complete description is the Wigner function[1] $W(x, p_x)$, a quasiprobability distribution in position $x$ and momentum $p_x$, forming the quantum analog to the classical phase-space distribution [72, 73]. It is connected to the density matrix $\rho$ by a Wigner transform, defined as

$$W(x, p_x) = \frac{1}{\pi\hbar}\int_{-\infty}^{\infty} \langle x + x'|\rho|x - x'\rangle e^{-2ip_x x'/\hbar}\, \mathrm{d}x', \tag{3.8}$$

with the normalization property $\iint W(x, p_x)\mathrm{d}x\mathrm{d}p_x = 1$.

The marginals of $W(x, p_x)$ give the $x$ and $p_x$ probability distributions, $f_x(x) = \int_{-\infty}^{\infty} W(x, p_x)\mathrm{d}p_x$ and $f_p(p_x) = \int_{-\infty}^{\infty} W(x, p_x)\mathrm{d}x$.

$\rho$ can be recovered from $W(x, p_x)$ by a Weyl transform, which gives $\rho$ in position space as

$$\langle x|\rho|x'\rangle = \int_{-\infty}^{\infty} W(\frac{x + x'}{2}, p)e^{i(x-x')p_x/\hbar}\, \mathrm{d}p_x. \tag{3.9}$$

Likewise, the momentum space representation of $\rho$ can be retrieved through

$$\langle p_x|\rho|p_x'\rangle = \int_{-\infty}^{\infty} W(x, \frac{p_x + p_x'}{2})e^{i(p_x - p_x')x/\hbar}\, \mathrm{d}x. \tag{3.10}$$

The coherence properties in position and momentum space for a translation of $\delta x$ and momentum offset $\delta p_x$, respectively, are given by the first-order correlation functions

$$g^{(1)}(\delta x) = \frac{\langle x + \frac{\delta x}{2}|\rho|x - \frac{\delta x}{2}\rangle}{\sqrt{\langle x + \frac{\delta x}{2}|\rho|x + \frac{\delta x}{2}\rangle\langle x - \frac{\delta x}{2}|\rho|x - \frac{\delta x}{2}\rangle}}, \tag{3.11}$$

$$g^{(1)}(\delta p_x) = \frac{\langle p_x + \frac{\delta p_x}{2}|\rho|p_x - \frac{\delta p_x}{2}\rangle}{\sqrt{\langle p_x + \frac{\delta p_x}{2}|\rho|p_x + \frac{\delta p_x}{2}\rangle\langle p_x - \frac{\delta p_x}{2}|\rho|p_x - \frac{\delta p_x}{2}\rangle}}. \tag{3.12}$$

This definition is analogous to the degree of coherence in classical optics as used in the derivation of the Van Cittert-Zernike theorem above.

[1] Note that in this work the symbols $W$ and $W_0$ are used for the beam radius $W$ and waist radius $W_0$ of the 2S-6P spectroscopy laser as well as the Wigner functions $W(x, p_x)$ and $W_0(x, p_x)$.

### 3.3.2 Wigner function at the nozzle

Describing our experiment with a Wigner function then however requires the knowledge of the density matrix of the atoms leaving the nozzle, which a priori is unknown. We reduce the problem to two dimensions, transverse and along the atomic beam, chosen to be along the $x$-axis and $z$-axis, respectively. Because we are interested in the transverse coherence of the beam, we limit the Wigner function to the transverse dimension $(x, p_x)$, while the movement along the beam axis $(z, p_z)$ is treated classically, i.e., it enters the Wigner function only as a parameter.

First, let us find the classical phase-space distribution $f_0(x, p_x)$ of the atoms as they are leaving the nozzle. The atoms are assumed to be in thermal equilibrium and thus their distribution of transverse momenta follows the Maxwell-Boltzmann distribution. The probability to find an atom with momentum $p_x$ is $f_p(p_x) = \exp\left(-p_x^2/2\sigma_{p_x}^2\right)/\sqrt{2\pi}\sigma_{p_x}$, with the transverse momentum distribution width $\sigma_{p_x} = \sqrt{m_\mathrm{H}/\beta}$. Here, $m_\mathrm{H}$ is the mass of the hydrogen atoms, $\beta = k_\mathrm{B}T_\mathrm{N}$, and $k_\mathrm{B}$ is the Boltzmann constant. To model the spatial confinement of the nozzle with radius $r_1$, we again choose to work with a Gaussian distribution in space to simplify the derived expressions, i.e., the probability to find an atom at position $x$ is $f_x(x) = \exp\left(-x^2/r_1^2\right)/\sqrt{\pi}r_1$. Thus, the classical phase-space distribution at the nozzle is

$$f_0(x, p_x) = \frac{1}{\sqrt{2}\pi\sigma_{p_x} r_1} \exp\left(-\frac{p_x^2}{2\sigma_{p_x}}\right) \exp\left(-\frac{x^2}{r_1^2}\right). \tag{3.13}$$

It is tempting to assume that the initial Wigner function is identical to $f_0(x, p_x)$, but in general this is not the case. Such a direct replacement would, e.g., not exclude the case of decreasing both $\sigma_{p_x}$ and $r_1$ towards zero, leading to a violation of the Heisenberg uncertainty principle.

To see whether in our case $f_0(x, p_x)$ is a good approximation to the corresponding Wigner function $W_0(x, p_x)$, it is instructive to compare the classical phase-space distribution $f_\mathrm{HO}(x, p_x)$ of thermally excited one-dimensional classical harmonic oscillator with the Wigner function $W_\mathrm{HO}(x, p_x)$ of its quantum equivalent [72, 74]. The Hamiltonian of these systems is given by $H = (p_x^2/2m_\mathrm{H}) + \frac{1}{2}m_\mathrm{H}\omega^2 x^2$, $\omega$ being the angular frequency of the oscillator, and with $H, x, p_x$ replaced by their operator equivalents in the quantum case. We find

$$f_\mathrm{HO}(x, p_x) = \frac{\beta\omega}{2\pi} \exp\left(-\beta\left(\frac{p_x^2}{2m_\mathrm{H}} + \frac{1}{2}m_\mathrm{H}\omega^2 x^2\right)\right) \tag{3.14}$$

for the classical case and

$$W_\mathrm{HO}(x, p_x) = \frac{1}{\pi\hbar} \tanh\left(\hbar\omega\beta/2\right) \exp\left(-\beta\frac{\tanh\left(\hbar\omega\beta/2\right)}{\hbar\omega\beta/2}\left(\frac{p_x^2}{2m_\mathrm{H}} + \frac{1}{2}m_\mathrm{H}\omega^2 x^2\right)\right) \tag{3.15}$$

for the quantum mechanical case.

The difference between the classical and quantum case is the factor $\kappa = \frac{\tanh(\hbar\omega\beta/2)}{\hbar\omega\beta/2}$ in the exponent, which ensures that the uncertainty principle is not violated as $T_\mathrm{N} \to 0$. In our case, $f_\mathrm{HO}(x, p_x)$ is identical with $f_0(x, p_x)$ when $\omega = \sqrt{2/\beta m_\mathrm{H}} r_1$, i.e., we can model that atoms at the nozzle exit as a harmonic oscillator. For typical values of our experiment (see Table 2.1 for $m_\mathrm{H}$), $T_\mathrm{N} = 4.8\,\mathrm{K}$, $r_1 = 1\,\mathrm{mm}$, we find $1 - \kappa \approx 2 \times 10^{-14}$. Thus, the differences between the classical and quantum case are very small in our case and we can reasonably assume that $W(x, p_x)$ is well approximated by $f(x, p_x)$.

Now that the initial Wigner function $W_0(x, p_x)$ is known, $W_0(x, p_x) \equiv f_0(x, p_x)$, we can give expressions for the initial density matrix $\rho_0$ using Eqs. (3.9) and (3.10):

$$\langle x + \frac{x'}{2}|\rho_0|x - \frac{x'}{2}\rangle = \frac{1}{\sqrt{\pi}}\frac{1}{r_1} \exp\left(-\frac{1}{r_1^2}x^2 - \frac{\sigma_{p_x}^2}{2\hbar^2}x'^2\right), \tag{3.16}$$

$$\langle p_x + \frac{p_x'}{2}|\rho_0|p_x - \frac{p_x'}{2}\rangle = \frac{1}{\sqrt{2\pi}\sigma_{p_x}} \exp\left(-\frac{1}{2\sigma_{p_x}^2}p_x^2 - \frac{r_1^2}{4\hbar^2}p_x'^2\right). \tag{3.17}$$

From Eqs. (3.11) and (3.16), the transverse coherence length at the nozzle, $l_{\mathrm{c,t,0}}$, is found to be (assuming $\delta x \ll r_1$)

$$l_{\mathrm{c,t,0}} = \frac{\sqrt{2}\hbar}{\sigma_{p_x}} = \frac{\lambda_{\mathrm{dB,th}}}{\sqrt{\pi}}. \tag{3.18}$$

As expected, it corresponds to the thermal de Broglie wavelength, with a prefactor of $1/\sqrt{\pi}$ owing to differing definitions of coherence length in the literature.

Analogous to a coherence length, a typical momentum can be given over which the atomic state is coherent. Such a property will determine the visibility of an interferometer that creates a momentum difference between the two interfering paths. From Eqs. (3.12) and (3.17) we find this transverse momentum coherence scale $p_{\mathrm{c,t,0}}$ to be (assuming $\delta p_x \ll \sigma_{p_x}$)

$$p_{\mathrm{c,t,0}} = \frac{2\hbar}{r_1}. \tag{3.19}$$

For the experimental parameters used so far, $l_{\mathrm{c,t,0}} = 0.45\,\mathrm{nm}$ and $p_{\mathrm{c,t,0}}/m_{\mathrm{H}} = 1.3 \times 10^{-4}\,\mathrm{m/s}$.

### 3.3.3 Wigner function at the light grating without intermediate aperture

Knowing the Wigner function $W_0(x, p_x)$ at the exit of the nozzle, the Wigner function further downstream of the atomic beam along the $z$-axis can readily be found when assuming that atom-atom interactions are negligible. In this case, the atoms propagate force-free and the time evolution of the Wigner function is given by the classical Liouville equation [72]. At the position of the light grating, $z = L$, the Wigner function is

$$\begin{aligned} W_{2'}(x, p_x) &= W_0(x - p_x\frac{L}{p_z}, p_x) \\ &= \frac{1}{\sqrt{2\pi}\sigma_{p_x} r_1} \exp\left(-\frac{p_x^2}{2\sigma_{p_x}}\right) \exp\left(-\frac{\left(x - p_x\frac{L}{p_z}\right)^2}{r_1^2}\right). \end{aligned} \tag{3.20}$$

The free evolution of the atoms thus causes a shearing of the Wigner function in phase-space, mapping the initial momentum distribution onto the position distribution. A measurement of the position distribution for various evolution distances $L$ can then be used to reconstruct the Wigner function, as has, e.g., been done for a beam of He atoms [75].

The density matrix at $z = L$ is found by substituting $W_{2'}(x, p_x)$ into Eqs. (3.9) and (3.10), giving

$$\langle x + \frac{x'}{2}|\rho_{2'}|x - \frac{x'}{2}\rangle = \frac{1}{\sqrt{\pi}}\frac{1}{r_{\mathrm{b},2'}}\exp\left(-\frac{1}{r_{\mathrm{b},2'}^2}x^2 - \frac{r_1^2}{r_{\mathrm{b},2'}^2}\frac{\sigma_{p_x}^2}{2\hbar^2}x'^2 + i\frac{2\sigma_{p_x}^2 L}{\hbar r_{\mathrm{b},2'}^2 p_z}xx'\right), \qquad (3.21)$$

$$\begin{aligned}\langle p_x + \frac{p'_x}{2}|\rho_{2'}|p_x - \frac{p'_x}{2}\rangle &= \frac{1}{\sqrt{2\pi}\sigma_{p_x}}\exp\left(-\frac{1}{2\sigma_{p_x}^2}p_x^2 - \frac{r_1^2}{4\hbar^2}p_x'^2 + i\frac{L}{\hbar p_z}p_x p'_x\right)\\ &= \langle p_x + \frac{p'_x}{2}|\rho_0|p_x - \frac{p'_x}{2}\rangle\exp\left(i\frac{L}{\hbar p_z}p_x p'_x\right), \qquad (3.22)\end{aligned}$$

where $r_{\mathrm{b},2'} = \sqrt{r_1^2 + 2\sigma_{p_x}^2\left(\frac{L}{p_z}\right)^2}$ is the resulting width of the distribution in position space.

The increasing width $r_{\mathrm{b},2'}$ can be understood as resulting from the convolution of two Gaussians, the initial position distribution, and the spreading in space for each point in this distribution given by the initial momentum distribution. The phase factor $\exp\left(i(2\sigma_{p_x}^2 L/\hbar r_{\mathrm{b},2'}^2 p_z)xx'\right)$ in Eq. (3.21) accounts for the curvature of the wavefront emerging from the nozzle.

From Eqs. (3.11) and (3.21), the transverse coherence length at $z = L$ is found to be (assuming $\delta x \ll r_{\mathrm{b},2'}$)

$$\begin{aligned}l_{\mathrm{c,t},2'} &= \frac{r_{\mathrm{b},2'}}{r_1}\frac{\sqrt{2}\hbar}{\sigma_{p_x}} = \frac{r_{\mathrm{b},2'}}{r_1}l_{\mathrm{c,t},0}\\ &= \sqrt{l_{\mathrm{c,t},0}^2 + \frac{4\hbar^2 L^2}{p_z^2 r_1^2}}. \qquad (3.23)\end{aligned}$$

Thus, the transverse coherence length is enhanced by a geometrical factor $r_{\mathrm{b},2'}/r_1$, a behavior expected from the Van Cittert-Zernike theorem. $l_{\mathrm{c,t},2'}$ can also be interpreted as the root of the sum of squares of the initial transverse coherence length $l_{\mathrm{c,t},0}$ and a contribution growing linear with $L$, which dominates for $L \gg r_1(p_z^2/\sigma_{p_x}^2)$.

For the case of a well-collimated atomic beam as assumed in the derivation of the Van Cittert-Zernike theorem above, $p_x \ll p_z$ and $L \gg r_1$, and thus $p \approx p_z$ and $\lambda_{\mathrm{dB}} \approx h/p_z$. With these approximations, we find

$$l_{\mathrm{c,t},2'} \approx \sqrt{l_{\mathrm{c,t},0}^2 + \left(\frac{1}{\pi}\frac{L}{r_1}\lambda_{\mathrm{dB}}\right)^2} \approx \frac{1}{\pi}\frac{L}{r_1}\lambda_{\mathrm{dB}}, \qquad (3.24)$$

which is identical to the result of the Van Cittert-Zernike theorem as given in Eq. (3.7). Under these assumptions, the phase factor in Eq. (3.21) reduces to $\exp\left(i(\pi/L\lambda_{\mathrm{dB}})r^2\right)$, with $r$ the transverse distance from the beam axis ($x = r/2$, $x' = -r$).

The transverse momentum coherence scale, on the other hand, is not enhanced during the propagation and remains unchanged from its initial value, $p_{\mathrm{c,t},2'} \equiv p_{\mathrm{c,t},0}$. Indeed, the density matrix in momentum space only acquires a phase factor, corresponding to a phase shift between different momentum components $p_x$.

For the experimental parameters used so far, $l_{\mathrm{c,t},2'} = 129\,\mathrm{nm}$ and thus $l_{\mathrm{c,t},2'} \gg l_{\mathrm{c,t},0}$, owing to the geometrical enhancement during free propagation.

### 3.3.4 Wigner function at the light grating with intermediate aperture

In the experiment, there is an aperture between the nozzle and the light grating to reduce the transverse momentum width of the atomic beam and thus the Doppler broadening of the observed hydrogen resonance. The aperture has a radius $r_2$ and is placed at $z = L_1 < L$. The light grating is then a further distance $L_2 = L - L_1$ from this aperture.

As done when deriving the Wigner function at the nozzle, $W_0(x, p_x)$, we will assume that we can treat the Wigner function like a classical phase-space density. Then, the action of the aperture can taken into account by multiplying the Wigner function at $z = L_1$ with a mask in position space with a transmission proportional to $\exp(-x^2/r_2^2)$, thus modeling the aperture as a soft Gaussian aperture as has been done for the nozzle. The resulting Wigner function at $z = L_1$ is

$$W_1(x, p_x) = \frac{r_\mathrm{m}}{r_2}\sqrt{1 + \frac{2L_1^2\sigma_{p_x}^2}{p_z^2 r_\mathrm{m}^2}} \exp\left(-\frac{x^2}{r_2^2}\right) W_0(x - p_x\frac{L_1}{p_z}, p_x), \tag{3.25}$$

where $r_\mathrm{m} = \sqrt{r_1^2 + r_2^2}$ and the prefactor ensure proper normalization.

Propagating $W_1(x, p_x)$ a further distance $L_2$ along the $z$-axis, we arrive at the Wigner function after the aperture and at the position of the light grating, $z = L$,

$$\begin{aligned} W_2(x, p_x) &= W_1(x - p_x\frac{L_2}{p_z}, p_x) \\ &= \frac{r_\mathrm{m}}{\sqrt{2}\pi\sigma_{p_x} r_1 r_2}\sqrt{1 + \frac{2L_1^2\sigma_{p_x}^2}{p_z^2 r_\mathrm{m}^2}} \exp\left(-\frac{p_x^2}{2\sigma_{p_x}}\right) \exp\left(-\frac{\left(x - p_x\frac{L}{p_z}\right)^2}{r_1^2} - \frac{\left(x - p_x\frac{L_2}{p_z}\right)^2}{r_2^2}\right). \end{aligned} \tag{3.26}$$

As before, plugging $W_2(x, p_x)$ in Eq. (3.9), we find the density matrix in position space

$$\langle x + \frac{x'}{2}|\rho_2|x - \frac{x'}{2}\rangle = \frac{1}{\sqrt{\pi}}\frac{1}{r_{\mathrm{b},2}} \exp\left(-\frac{1}{r_{\mathrm{b},2}^2}x^2 - \frac{1}{l_{\mathrm{c,t},2}^2}x'^2 + i\frac{\phi_2}{r_1^2}xx'\right), \tag{3.27}$$

$$\text{with} \quad r_{\mathrm{b},2} = \frac{r_1 r_2}{r_\mathrm{m}}\sqrt{\frac{\frac{p_z^2}{\sigma_{p_x}^2} + 2\left(\frac{L^2}{r_1^2} + \frac{L_2^2}{r_2^2}\right)}{\frac{p_z^2}{\sigma_{p_x}^2} + 2\frac{L_1^2}{r_\mathrm{m}^2}}}, \tag{3.28}$$

$$\mathfrak{l}_{\mathrm{c,t},2} = \sqrt{l_{\mathrm{c,t},0}^2 + \frac{4\hbar^2}{p_z^2}\left(\frac{L^2}{r_1^2} + \frac{L_2^2}{r_2^2}\right)}, \tag{3.29}$$

$$\phi_2 = \frac{p_z}{\hbar}\frac{2\left(L + \frac{r_1^2}{r_2^2}L_2\right)}{\frac{p_z^2}{\sigma_{p_x}^2} + 2\left(\frac{L^2}{r_1^2} + \frac{L_2^2}{r_2^2}\right)}. \tag{3.30}$$

The width of the distribution in position space $r_{\mathrm{b},2}$ has two limiting regimes, determined by the ratio between $r_\mathrm{m}/L_1$ and $\sigma_{p_x}/p_z$, which are related to the beam divergences purely from the geometry and purely from the initial transverse momentum width, respectively. For $L_2 \gg L_1, L \approx L_2$, nozzle and aperture act as a single source and the width can be written as $r_{\mathrm{b},2} \approx L\alpha_\mathrm{far}$ with the far-field $1/e$ beam divergence angle $\alpha_\mathrm{far}$. Then, in the case of $r_\mathrm{m}/L_1 \gg \sigma_{p_x}/p_z$, the divergence becomes $\alpha_\mathrm{far} = \sqrt{2}\sigma_{p_x}/p_z$, while for $r_\mathrm{m}/L_1 \ll \sigma_{p_x}/p_z$ it is given by $\alpha_\mathrm{far} = r_\mathrm{m}/L_1$. Our experiment operates in the latter limit.

As before, using Eq. (3.11) and assuming $\delta x \ll r_{\mathrm{b},2}$, we can identify $l_{\mathrm{c,t},2}$ as the transverse coherence length. Comparing it to the case without aperture, $l_{\mathrm{c,t},2'}$ (Eq. (3.23)), we find

$$\frac{l_{\mathrm{c,t},2}}{l_{\mathrm{c,t},2'}} = \sqrt{1 + \left(\frac{L_2}{L}\frac{r_1}{r_2}\right)^2}. \tag{3.31}$$

Thus, the effect of the aperture on the transverse coherence length dominates only when $\frac{L_2}{r_2} \gg \frac{L}{r_1}$, in which case the aperture effectively acts as source.

The phase $\phi_2$ now includes the effects from diffraction on the aperture on the curvature of the wavefront, and as for the transverse coherence length these effects only become dominant for $\frac{L_2}{r_2} \gg \frac{L}{r_1}$.

The density matrix in momentum space derived from $W_2(x, p_x)$ using Eq. (3.9) is

$$\begin{aligned}\langle p_x + \frac{p'_x}{2}|\rho_2|p_x - \frac{p'_x}{2}\rangle = \frac{1}{\sqrt{2\pi}\sigma_{p_x}}\sqrt{1 + \frac{2L_1^2\sigma_{p_x}^2}{p_z^2 r_\mathrm{m}^2}}\exp\left(-\frac{1}{2\sigma_{p_x,2}^2}p_x^2 - \frac{r_2^2}{r_\mathrm{m}^2}\frac{r_1^2}{4\hbar^2}p_x'^2\right) \\ \times \exp\left(i\frac{1}{\hbar p_z}\left(\frac{r_2^2}{r_\mathrm{m}^2}L_1 + L_2\right)p_x p'_x\right),\end{aligned} \tag{3.32}$$

$$\text{with} \quad \sigma_{p_x,2} = \sigma_{p_x}\frac{1}{\sqrt{1 + \frac{2\sigma_{p_x}^2 L_1^2}{p_z^2 r_\mathrm{m}^2}}} = p_z\frac{1}{\sqrt{\frac{p_z^2}{\sigma_{p_x}^2} + \frac{2L_1^2}{r_\mathrm{m}^2}}}, \tag{3.33}$$

$$p_{\mathrm{c,t},2} = \frac{2\hbar r_\mathrm{m}}{r_1 r_2} = \frac{r_\mathrm{m}}{r_2}p_{\mathrm{c,t},0}. \tag{3.34}$$

Unsurprisingly, the width of the distribution in momentum space $\sigma_{p_x,2}$ has again two limiting regimes. For $r_\mathrm{m}/L_1 \gg \sigma_{p_x}/p_z$, the influence of the aperture is negligible and $\sigma_{p_x,2} \approx \sigma_{p_x,2'}$, while for $r_\mathrm{m}/L_1 \ll \sigma_{p_x}/p_z$ the momentum width is dominated by the nozzle and aperture radii and not the initial momentum distribution width, giving $\sigma_{p_x,2} \approx (p_z/\sqrt{2})(r_\mathrm{m}/L_1) \ll \sigma_{p_x}$.

We recall that so far, the transverse momentum coherence scale $p_{\mathrm{c,t},0}$ (Eq. (3.19)) was given by the inverse source size and, unlike the transverse coherence length, not geometrically enhanced. This is also the case for the transverse momentum coherence scale $p_{\mathrm{c,t},2}$ (Eq. (3.34)), found using Eq. (3.12) and assuming $\delta p_x \ll \sigma_{p_x}$. $p_{\mathrm{c,t},2}$ now reflects source and aperture size, but not the propagation distance, and only changes by a factor of $r_\mathrm{m}/r_2$ from the previous value $p_{\mathrm{c,t},0}$. The geometric effect on the transverse coherence length is instead again reflected in a phase factor of the density matrix, now taking into account both $L_1$ and $L_2$.

Having derived these results, it is instructive to again look at our experimental situation, where the aperture is placed close to the light grating ($L_2/L_1 = 0.33$) and has approximately the same size as the nozzle ($r_2/r_1 = 0.6$), giving $r_\mathrm{m} \approx 1.17\,\mathrm{mm}$. Since the purpose of the aperture is to reduce the momentum width of the beam, the parameters are chosen such that $r_\mathrm{m}/L_1 \ll \sigma_{p_x}/p_z$, resulting in a reduction from $\sigma_{p_x}/m_\mathrm{H} = 199\,\mathrm{m/s}$ to $\sigma_{p_x,2} = 1.07\,\mathrm{m/s}$, corresponding to divergence angle of $\alpha_\mathrm{far} = r_\mathrm{m}/L_1 = 7.6\,\mathrm{mrad}$. At the same time, the transverse coherence length $l_{\mathrm{c,t},2}$ only increases (Eq. (3.31)) by a small factor of 1.08 to as compared to the situation without the aperture. Put another way, there is no direct relationship between the transverse coherence length and the transverse momentum width.

Thus, for the experimental parameters used so far, $l_{\mathrm{c,t},2} = 139\,\mathrm{nm}$ and $p_{\mathrm{c,t},2}/m_\mathrm{H} = 2.5 \times 10^{-4}\,\mathrm{m/s}$.

### 3.3.5 Comparison with fully coherent Gaussian wave packet

It is instructive to compare the results obtained above, where we assumed the atoms to be in a thermal state with essentially classical behavior, with a fully coherent Gaussian wave packet described by a pure state. The wave function $\psi_{\mathrm{G}}$ of such a wave packet in 1D with an initial $1/e$ probability radius of $w_x(t=0) \equiv w_{x,0}$ and centered at position $x = x_0$ is given in position and momentum space at time $t = 0$ by

$$\langle x|\psi_{\mathrm{G}}\rangle = \frac{1}{\sqrt{\sqrt{\pi}w_{x,0}}} \exp\left(-\frac{(x-x_0)^2}{2w_{x,0}^2}\right), \tag{3.35}$$

$$\langle p_x|\psi_{\mathrm{G}}\rangle = \sqrt{\frac{w_{x,0}}{\sqrt{\pi}\hbar}} \exp\left(-\frac{w_{x,0}^2 p_x^2}{2\hbar^2} + i\frac{x_0 p_x}{\hbar}\right). \tag{3.36}$$

Thus, in momentum space its standard deviation probability width is given by $\sigma_{p_x,\mathrm{G}} = \hbar/w_{x,0}\sqrt{2}$. This wave packet corresponds to a minimum uncertainty case with $\sigma_{p_x}\sigma_{x,0} = \sigma_{p_x} w_{x,0}/\sqrt{2} = \hbar/2$.

For a free particle of mass $m_{\mathrm{H}}$, the solution to the time-dependent Schrödinger equation with Hamiltonian $H_{\mathrm{FP}} = p_x^2/2m_{\mathrm{H}}$ can be directly given in momentum space as

$$\langle p_x|\psi_{\mathrm{G}}(t)\rangle = \langle p_x|\psi_{\mathrm{G}}\rangle \exp\left(-i\frac{tp_x^2}{2\hbar m_{\mathrm{H}}}\right). \tag{3.37}$$

As expected for a free particle, its momentum width does not change in time. In position space, we find

$$\langle x|\psi_{\mathrm{G}}(t)\rangle = \frac{1}{\sqrt{\sqrt{\pi}}}\sqrt{\frac{1}{w_{x,0} + i\frac{\hbar t}{w_{x,0}m_{\mathrm{M}}}}} \exp\left(-\frac{(x-x_0)^2}{2w_x^2(t)} + i\frac{\hbar t(x-x_0)^2}{2m_{\mathrm{H}}w_{x,0}^2 w_x^2(t)}\right), \tag{3.38}$$

with the $1/e$ probability width at time $t$

$$w_x(t) = \sqrt{w_{x,0}^2 + \frac{\hbar^2 t^2}{m_{\mathrm{H}}^2 w_{x,0}^2}}. \tag{3.39}$$

The size of the wave packet increases over time, its probability to be observed at any point in space is said to be spreading. The spreading is not a quantum mechanical feature in itself, but just the probability density expected of an ensemble of freely expanding classical particles initially found in a region with size $w_{x,0}$ and with momentum distribution of width $\sigma_{p_x,\mathrm{G}}$. However, in the classical case we can chose the initial size and momentum distribution independently, while here it is dictated by the uncertainty principle.

The density matrix at time $t$ is

$$\langle x + \frac{x'}{2}|\rho_G|x - \frac{x'}{2}\rangle = \frac{1}{\sqrt{\pi}w_x(t)} \exp\left(-\frac{(x-x_0)^2}{w_x^2(t)} - \frac{x'^2}{4w_x^2(t)} + i\frac{\hbar t(x-x_0)x'}{m_{\mathrm{H}}w_{x,0}^2 w_x^2(t)}\right), \tag{3.40}$$

$$\langle p_x + \frac{p_x'}{2}|\rho_G|p_x - \frac{p_x'}{2}\rangle = \frac{w_{x,0}}{\sqrt{\pi}\hbar} \exp\left(-\frac{w_{x,0}^2 p_x^2}{\hbar^2} - \frac{w_{x,0}^2 p_x'^2}{4\hbar^2} + i\frac{x_0 p_x'}{\hbar} + i\frac{tp_x p_x'}{\hbar m_{\mathrm{H}}}\right). \tag{3.41}$$

Plugging the density matrix into the expressions for the degree of coherence, Eqs. (3.11) and (3.12), gives

$$g^{(1)}(\delta x) = \exp\left(i\frac{\hbar t(x-x_0)\delta x}{m_{\mathrm{H}} w_{x,0}^2 w_x^2(t)}\right), \tag{3.42}$$

$$g^{(1)}(\delta p_x) = \exp\left(-i\frac{x_0 \delta p_x}{\hbar} + i\frac{t p_x \delta p_x}{\hbar m_{\mathrm{H}}}\right). \tag{3.43}$$

The degree of coherence in both position and momentum space is a pure phase factor with an absolute value of unity for all times, and thus the coherence length and the coherence momentum scale are infinite.

It can be shown that averaging the initial position $x_0$ over the periodicity of the light grating will result in a signal identical to that from an (incoherent) sum over the signals from fully delocalized atoms, as long as the sum covers the same momentum content as the wave packet (see Section 3.4.2). Thus, for our experimental parameters, an initial-position-averaged wave packet will give the same result as the Wigner function. This is an especially import result in light of some authors using wave packets as input states for experimental situations that might be more accurately described by a thermal state as given by a Wigner function [76].

## 3.4 Light force shift of the 2S-6P transition

In the previous section, the density matrix $\rho_2$ (Eqs. (3.27) and (3.32)) of the atomic beam at the position of the light grating has been derived, using the Wigner function. The resulting transverse coherence length is comparable to the periodicity of the light grating formed by the 2S-6P spectroscopy laser, and thus the atoms cannot be described as localized particle as they cross the grating, but instead must be treated quantum mechanically. In the following, a model of the 2S-6P excitation taking into account this delocalization is developed. This model is then used to simulate the light force shift (LFS) of the 2S-6P transition as observed for the atomic beam.

### 3.4.1 Atom–light interaction including momentum exchange

To this end, the optical Bloch equations (OBEs) as introduced in Section 2.3.1 must be modified to not only take the internal degrees of freedom of the atom, the energy eigenstates or levels $|n\rangle$, into account, but also its external degrees of freedom, i.e., its momentum $\boldsymbol{p}$. The external and internal degrees of freedom are coupled through the atom–light interaction, since each absorption of a laser photon by the atom not only changes the internal level of the atom, but the atom also takes up the momentum $\hbar\boldsymbol{K}_{\mathbf{L}}$ of the photon, where $\boldsymbol{K}_{\mathbf{L}}$ is the wave vector of the laser beam. Likewise, the atom both loses momentum and changes its internal level during an emission event. Therefore, the state of the atom can be described in the combined basis $|n\rangle|\boldsymbol{p}\rangle$, where $|\boldsymbol{p}\rangle$ is a momentum eigenstate. In position space, a momentum eigenstate $|\boldsymbol{p}\rangle$ corresponds to a plane wave state with wave vector $\boldsymbol{p}/\hbar$.

Absorption and stimulated emission of photons only changes the atom's momentum by discrete amounts and in the direction of the laser beams, i.e., by $\pm\hbar\boldsymbol{K}_{\mathbf{L}}$, corresponding to a change in velocity by $v_{\mathrm{rec}} = \hbar|\boldsymbol{K}_{\mathbf{L}}|/m_{\mathrm{H}} \approx 0.97\,\mathrm{m/s}$. Spontaneous emission, on the other hand, leads to momentum change in a random direction, as given by the radiation pattern of the transition, and the momentum change given by the energy bridged by the decay. However, there are only at most a few emission events as the atoms cross the laser beams, since the atoms predominantly decay to the 1S ground level, which is not coupled by the spectroscopy laser. Along the axis of the atomic beam ($z$-axis), the typical velocity is much larger than the recoil velocity, $v_z \gg v_{\text{rec}}$, and thus spontaneous emission will not change this velocity and the interaction time $T = W_0/v_z$ appreciably. The velocity along the $y$-axis is comparable to the $x$-axis, $v_y \approx v_x$, but in contrast to the $x$-axis there is no light grating along the $y$-axis. An emission of $N$ photons along the $y$-axis will lead to a shift in $y$-position after crossing the beam of $\Delta y = N v_{\text{rec}} T = N W_0 v_{\text{rec}}/v_z$. Since $v_z \gg v_{\text{rec}}$, this shift is much smaller than the beam size and thus does not change the dynamics of the atom–light interaction. We thus here restrict our quantum mechanical treatment of the momentum to the dimension ($x$-axis) along the laser beams. This restriction reduces the basis of momentum eigenstates to $|p\rangle_x$, where $p$ is the transverse momentum along the $x$-axis, and the combined basis becomes $|n\rangle|p\rangle_x$.

The level scheme for 2S-6P spectroscopy (see Fig. 2.1) is simplified to model the light force shift (LFS). To describe the excitation of the atom by the spectroscopy laser, we assume that the laser field only couples to a single transition[1], the $2\text{S}_{1/2}^{F=0}$, $m_F = 0$–$6\text{P}_J^{F=1}$, $m_F = 0$ (here abbreviated as 2S-6P) transition, with fine-structure component $J$ either 1/2 or 3/2. This transition with transition frequency $\nu_{\text{A},0}$ couples the metastable initial level $|\text{i}\rangle = |2\text{S}_{1/2}^{F=0}, m_F = 0\rangle$ with the excited level $|\text{e}\rangle = |6\text{P}_J^{F=1}, m_F = 0\rangle$. The atom is assumed to be initially in the $|\text{i}\rangle$ level, as is the case in the experiment. The laser field is, as derived in Section 2.3.1, assumed to be composed of two laser beams counter-propagating along the $x$-axis with identical frequency $\nu_{\text{L}}$ and identical polarization and intensity, resulting in identical Rabi frequencies $\Omega_\pm$ for to forward- (+) and backward-traveling (−) beam. The absorption and stimulated emission of a photon from and into the forward-traveling (backward-traveling) beam changes the momentum of the atom by $\hbar K_{\text{L}}$ ($-\hbar K_{\text{L}}$) and $-\hbar K_{\text{L}}$ ($\hbar K_{\text{L}}$), respectively, where $K_{\text{L}} = |\boldsymbol{K}_{\mathbf{L}}|$ is the wavenumber of the laser beams.

As a further simplification, the decay channels of the excited level $|\text{e}\rangle$ are restricted to a Ba-$\delta$ decay back to the initial level $|\text{i}\rangle$ with rate $\gamma_{\text{ei}}$ ("back decay"), and a Ly-$\epsilon$ decay to the 1S ground level $|\text{f}\rangle$ with rate $\Gamma_{\text{ef}} \gg \gamma_{\text{ei}}$. The internal dynamics of the atom are in this way reduced to a three-level system, with the $2\text{S}_{1/2}^{F=0}$ levels, and the decay of the excited level to those levels with rate $\gamma_{\text{e-2S}} - \gamma_{\text{ei}}$, not explicitly taken into account. $\gamma_{\text{ei}}$ is a factor of two larger for the 2S-6P$_{3/2}$ transition ($\gamma_{\text{ei}}/\Gamma \approx 8\,\%$) than for the 2S-6P$_{1/2}$ transition ($\gamma_{\text{ei}}/\Gamma \approx 4\,\%$), which leads to a larger LFS for the former transition as seen below. In order to match the linewidth of the excited level to the value of the natural linewidth $\Gamma$, the decays to the $2\text{S}_{1/2}^{F=0}$ levels are assumed to lead to the ground level, i.e., $\Gamma_{\text{ef}} = \Gamma_{\text{e-1S}} + \gamma_{\text{e-2S}} - \gamma_{\text{ei}}$. This is a reasonable approximation since the $2\text{S}_{1/2}^{F=0}$ are only coupled off-resonantly to the 6P manifold by the spectroscopy laser. The Ly-$\epsilon$ decays constitute the fluorescence signal in the experiment. To retrieve this signal in the model, the signal seen on the decay back to the ground level is scaled by $\Gamma_{\text{det}}/\Gamma$. Note that this does not change the dynamics of the system. The ground level is not coupled by the spectroscopy laser, i.e., atoms that have reached this level are of no interest in the description of the experiment. The values of the decay rates are given in Table 2.1.

[1] This implies that this model of the light force shift does not take into account the quantum interference between the fine-structure components of the 2S-6P transition, which is instead treated with the big model including both fine-structure components and other coupled levels, but not the momentum of the atom (see Section 2.3.2).

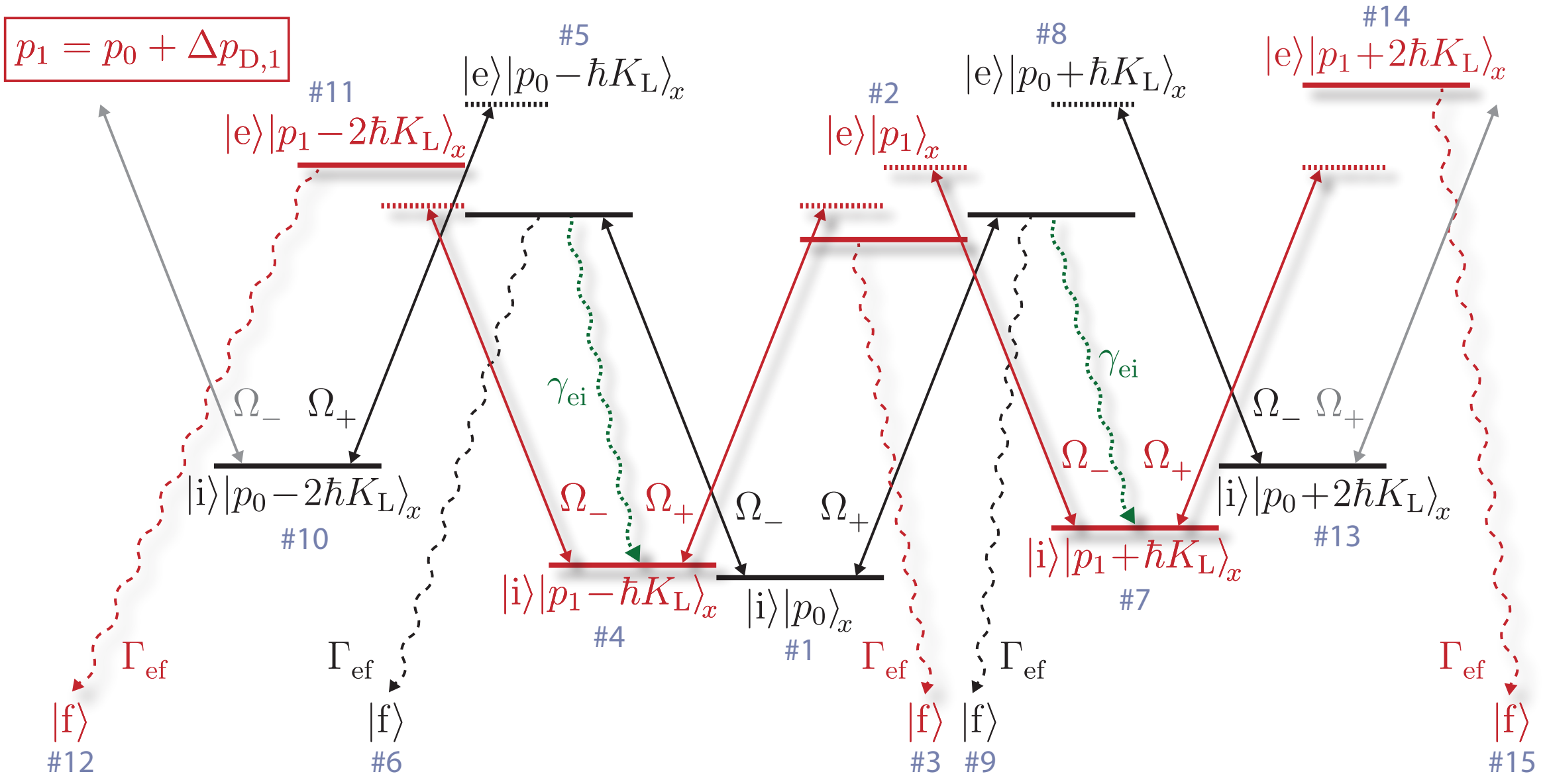

Figure 3.2: The coupling of internal atomic levels and the atom's momentum by the transfer of momentum $\hbar K_{\mathrm{L}}$ between the counter-propagating laser beams (Rabi frequencies $\Omega_{\pm}$) and atom during absorption and stimulated emission events (straight arrows) leads to a ladder-like level scheme (black). The coupled internal atomic levels are the initial (2S, $|\mathrm{i}\rangle$), exci.e., (6P, $|\mathrm{e}\rangle$), and ground (1S, $|\mathrm{f}\rangle$) levels. Using the basis of momentum eigenstates $|p\rangle_x$, where $p$ is the atom's transverse momentum along the $x$-axis, the state of the atom can be described in a combined basis $|\mathrm{i/e/f}\rangle|p\rangle_x$ (solid horizontal lines). The ladder continues indefinitely for negative and positive momenta (gray arrows), and is here shown in the laboratory frame for the case of zero $p_0$. A spontaneous back decay from $|\mathrm{e}\rangle$ to $|\mathrm{i}\rangle$ (green dotted wiggly lines), occurring with rate $\gamma_{\mathrm{ei}}$, randomly changes the momentum along the $x$-axis by $\Delta p_{\mathrm{D},1} \in [-\hbar K_{\mathrm{L}}, \hbar K_{\mathrm{L}}]$, i.e., $p_0 \to p_1 = p_0 + \Delta p_{\mathrm{D},1}$, leading to a ladder-like level scheme (red) shifted in momentum. Decays to $|\mathrm{f}\rangle$ (rate $\Gamma_{\mathrm{ef}}$, black and red dashed wiggly lines) also change the momentum, but this is of no consequence to the evolution of the system as $|\mathrm{f}\rangle$ is not coupled to other levels by the laser beams. The dotted horizontal lines mark the laser detuning from the excited states (not to scale) with the laser on-resonance with the $|\mathrm{i}\rangle|0\hbar K_{\mathrm{L}}\rangle_x$ to $|\mathrm{e}\rangle|\pm\hbar K_{\mathrm{L}}\rangle_x$ transitions. The light blue digits number the shown states.

A spontaneous emission event with a defined frequency, but random direction, randomly changes the momentum along the $x$-axis by $\Delta p_{\mathrm{D}}$. For the back decay to the initial level, $\Delta p_{\mathrm{D}} \in [-\hbar K_{\mathrm{L}}, \hbar K_{\mathrm{L}}]$ with a normalized probability density of [77]

$$\mathcal{N}(\Delta p_{\mathrm{D}}) = (3/8)\left(1 + \left(\frac{\Delta p_{\mathrm{D}}}{\hbar K_{\mathrm{L}}}\right)^2\right), \tag{3.44}$$

corresponding to the projection of the radiation pattern of a $\pi$ decay onto the $x$-axis (see Eq. (2.40)).

The coupling of internal atomic levels and the atom's momentum leads to a ladder-like level scheme, shown in Fig. 3.2 in the laboratory frame of reference, where both laser beams have the same frequency. The states are coupled by the laser beams through absorption and stimulated emission (straight arrows in Fig. 3.2), and by spontaneous emission (wiggly lines). An atom initially completely delocalized and with transverse momentum $p_0$ along the $x$-axis is in the state $|\mathrm{i}\rangle|p_0\rangle_x$ (#1). By absorbing a photon with momentum $\pm\hbar K_{\mathrm{L}}$ from either the forward- or the backward-traveling laser beam, the atom is excited to the level $|\mathrm{e}\rangle$ while gaining the photon's momentum, i.e., it is now in the state $|\mathrm{e}\rangle|p_0 \pm \hbar K_\mathrm{L}\rangle_x$. Generally, the atom interacts with the combined field of both laser beams simultaneously, leading to the atom being a superposition of multiple states. The probability density of such a superposition is modulated at a spatial frequency of $\lambda_{\mathrm{2S\text{-}6P}}/2$, corresponding to a localization of the atom through the interaction with the light grating.

Because there are two counter-propagating laser beams, the level scheme and thus the LFS is symmetric with respect to the incoming transverse momentum, i.e., invariant for $p_0 \to -p_0$. Furthermore, by taking into account the momentum exchange, the recoil shift and first-order Doppler shift have been naturally included in the model. That is, the energy difference between the states $|\mathrm{i}\rangle|p_0\rangle_x$ and $|\mathrm{e}\rangle|p_0 \pm \hbar K_\mathrm{L}\rangle_x$ is $h\nu_{\mathrm{A},0} + \hbar^2 K_\mathrm{L}^2/2m_\mathrm{H} + p_0 \hbar K_\mathrm{L}/m_\mathrm{H}$, with the three terms corresponding to the transition frequency $\nu_{\mathrm{A},0}$, the recoil shift $\Delta\nu_\mathrm{rec}$, and the first-order Doppler shift $\Delta\nu_\mathrm{D}$ (see Section 2.2.3). The frequency detuning $\Delta\nu_{\mathrm{2S\text{-}6P}}$ of the spectroscopy laser is here defined to take the recoil shift into account, such that zero detuning corresponds to a laser frequency of $\nu_{\mathrm{A},0} + \Delta\nu_\mathrm{rec}$.

A stimulated emission event into the laser beams can further increase the atom's momentum (states #10 and #13), leading to a total momentum transfer of $\pm 2\hbar K_\mathrm{L}$. This two-photon process is resonant for the special case of $p_0 = \pm\hbar K_\mathrm{L}$, and corresponds to first-order Bragg scattering of the atom on the light grating [78]. However, since the laser frequency is kept close to the transition frequency between initial and excited levels, the process is near-resonant with the intermediate excited state $|\mathrm{e}\rangle|0\hbar K_\mathrm{L}\rangle_x$, leading to a loss into the ground level by spontaneous decay. Higher-order resonant processes for even higher momentum transfer are also possible, i.e., transition from $|\mathrm{i}\rangle|\pm N\hbar K_\mathrm{L}\rangle_x$ to $|\mathrm{i}\rangle|\mp N\hbar K_\mathrm{L}\rangle_x$, where $N$ is an integer. The losses due to spontaneous emission increase accordingly, as $N$ intermediate near-resonant states are involved, which limits the efficiency of the process. Thus, depending on the level of accuracy needed, the maximum momentum change through absorption and stimulated emission can be constrained to allow for a numerical solution. To this end, the integer $N_{k,\mathrm{max}}$ is introduced, with the maximum momentum change given by $\pm N_{k,\mathrm{max}}\hbar K_\mathrm{L}$, resulting in $2N_{k,\mathrm{max}} + 1$ states for each value of $p$.

A spontaneous back decay from $|\mathrm{e}\rangle$ to $|\mathrm{i}\rangle$ (green dotted wiggly lines in Fig. 3.2) changes the momentum along the $x$-axis by the random amount $\Delta p_{\mathrm{D},1}$, i.e., $p_0 \to p_1 = p_0 + \Delta p_{\mathrm{D},1}$, leading to a ladder-like level scheme (red) shifted in momentum. If the atom is again excited to $|\mathrm{e}\rangle$, another back decay with $p_1 \to p_2 = p_1 + \Delta p_{\mathrm{D},2}$ can lead to another level scheme shifted in momentum by the random amount $\Delta p_{\mathrm{D},2}$. In principle, this process can continue indefinitely. However, since $\Gamma_\mathrm{ef} \gg \gamma_\mathrm{ei}$, the atom in most case decays to the ground level $|\mathrm{f}\rangle$, from where it cannot be excited again. Thus, for a given level of accuracy, the maximum number of back decays can be constrained to $N_{\mathrm{BD,max}}$. After $N_{\mathrm{BD,max}}$ back decays, the atom can thus only decay to the 1S ground level. In order to keep the linewidth of the excited level at its natural linewidth $\Gamma$, the decay rate for this final decay is set to $\Gamma$.

As mentioned above, the atom is in general in a superposition of excited states. A spontaneous decay then transfers this superposition from the excited to the initial or ground level, and shifts the momentum of all contained states by the same amount. That is, the coherence between the states with different momenta, but the same internal level, is preserved. It is only through this transfer of coherence that the spatial modulation of the probability density of the excited levels is likewise transferred to the initial or ground level. This transfer of coherence is contained in the cross-damping between the decays linking the respective excited and initial or ground levels, as described in Section 2.3.1. Importantly, this also means that if the system is to be described with OBEs, the corresponding cross-damping terms of the master equation (see Eq. (2.36)) must not be neglected.

Since $\Delta p_{\mathrm{D}}$ and thus $\Delta p_{\mathrm{D},1}$, $\Delta p_{\mathrm{D},2}$, ... are random, the ladder-like level scheme in principle consists of an infinite number of states, even if finite $N_{k,\mathrm{max}}$ and $N_{\mathrm{BD,max}}$ are introduced. To allow the translation of the system into OBEs, the following approximations are used: $N_{\mathrm{BD,max}}$ and $N_{k,\mathrm{max}}$ are set to 1 and 4, respectively, and the value of $\Delta p_{\mathrm{D},1}$ is fixed. This results in OBEs including 27 states and consisting of 207 real-valued coupled differential equations. Four signal equations are included, which contain, respectively, the signal of the Ba-$\delta$ back decays, the Ly-$\epsilon$ decays before a back decay has occurred, the Ly-$\epsilon$ decays after a back decay has occurred, and the sum of the latter two[1], where the Ly-$\epsilon$ signals have been scaled as mentioned above. For each set of input parameters, the OBEs are then numerically integrated $N_{\Delta p_{\mathrm{D}}}$ times for $N_{\Delta p_{\mathrm{D}}}$ different values of $\Delta p_{\mathrm{D},1}$, and the results are averaged according to Eq. (3.44). Specifically, we here use the Gaussian quadrature rule with $N_{\Delta p_{\mathrm{D}}} = 4$ points to average over $\Delta p_{\mathrm{D},1}$.

The LFS is then found by following this averaging procedure for a given input state and for a range of laser detunings $\Delta\nu_{\text{2S-6P}}$. From these results, a line scan, i.e., the signal as a function of $\Delta\nu_{\text{2S-6P}}$, can be constructed, with all Ly-$\epsilon$ decays constituting the signal as in the experiment. This line scan is then treated like an experimental line scan (see Chapter 5), and the LFS $\nu_{0,\mathrm{LFS}}$ is found by fitting a line shape function.

The results of a numerical integration of the LFS model for the 2S-6P$_{1/2}$ and 2S-6P$_{3/2}$ transition are shown in Fig. 3.3. The input state corresponds to a completely delocalized atom with transverse momentum $p_0$, i.e., at the start of the numerical integration only the state $|\mathrm{i}\rangle|p_0\rangle_x$ is populated. The speed of the atom is $v = 200\,\mathrm{m/s}$, and it thus crosses the light grating at an angle $\delta\alpha \approx p_0/m_{\mathrm{H}}v$ from the orthogonal. The resonance frequency $\nu_0$ of the simulated line scans is here determined by fitting Voigt line shapes, with $\nu_{0,\mathrm{LFS}} \equiv \nu_0$ constituting the value of the LFS. The power of the spectroscopy laser is $P_{\text{2S-6P}} = 30\,\mu\mathrm{W}$ ($15\,\mu\mathrm{W}$) for the 2S-6P$_{1/2}$ (2S-6P$_{3/2}$) transition, and the transverse beam profile is Gaussian with a $1/e^2$ intensity radius of $W_0 = 2.2\,\mathrm{mm}$, as used in the experiment. Note that the dipole moment of the 2S-6P$_{3/2}$ transition is a factor of $\sqrt{2}$ larger than that of the 2S-6P$_{1/2}$ transition, and thus using a power of $P_{\text{2S-6P}}/2$ for the 2S-6P$_{3/2}$ transition is equivalent to using $P_{\text{2S-6P}}$ for the 2S-6P$_{1/2}$ transition. $\nu_{0,\mathrm{LFS}}$ shows small wiggles at $p_0 \approx 0.3\hbar K_{\mathrm{L}}$, $0.85\hbar K_{\mathrm{L}}$, ..., which are numerical artifacts from the averaging over the back decay momentum change $\Delta p_{\mathrm{D},1}$.

As shown in Fig. 3.3 (A), for zero $p_0$ the LFS is found to be approximately $-600\,\mathrm{Hz}$ for both the 2S-6P$_{1/2}$ transition (solid green line) and the 2S-6P$_{3/2}$ transition (dashed red line). With increasing $p_0$, the LFS becomes more negative, till at $p_0 = \hbar K_{\mathrm{L}}$ a resonance-like feature is reached. Above this resonance, the light force shift is always positive. The LFS is generally larger for the 2S-6P$_{3/2}$ transition, which is a result of its two times larger back decay rate $\gamma_{\mathrm{ei}}$. This is demonstrated by setting $\gamma_{\mathrm{ei}}$ artificially to zero (black dotted line).

The resonance at $p_0 = \hbar K_{\mathrm{L}}$ is not fully shown in Fig. 3.3 (A), but highlighted in Fig. 3.3 (B). Around this point, the LFS reaches up to 230 kHz. This is a result of the two-photon transition from $|\mathrm{i}\rangle|\hbar K_{\mathrm{L}}\rangle_x$ to $|\mathrm{i}\rangle|-\hbar K_{\mathrm{L}}\rangle_x$, which becomes resonant for $p_0 = \hbar K_{\mathrm{L}}$, as discussed above. The LFS here is identical for both transitions, because the resonance is not a consequence of the back decay, with the same LFS observed for a zero $\gamma_{\mathrm{ei}}$. The width of this two-photon resonance is given by the limited interaction time between atom and the laser beams. Since the resonance always occurs at $p_0 = \hbar K_\mathrm{L}$, the corresponding transverse angle $\delta\alpha \approx \hbar K_\mathrm{L}/m_\mathrm{H}v$ is inversely proportional to $v$.

[1]No additional signal equation is needed for this incoherent sum of two other signals, but is included for convenience.

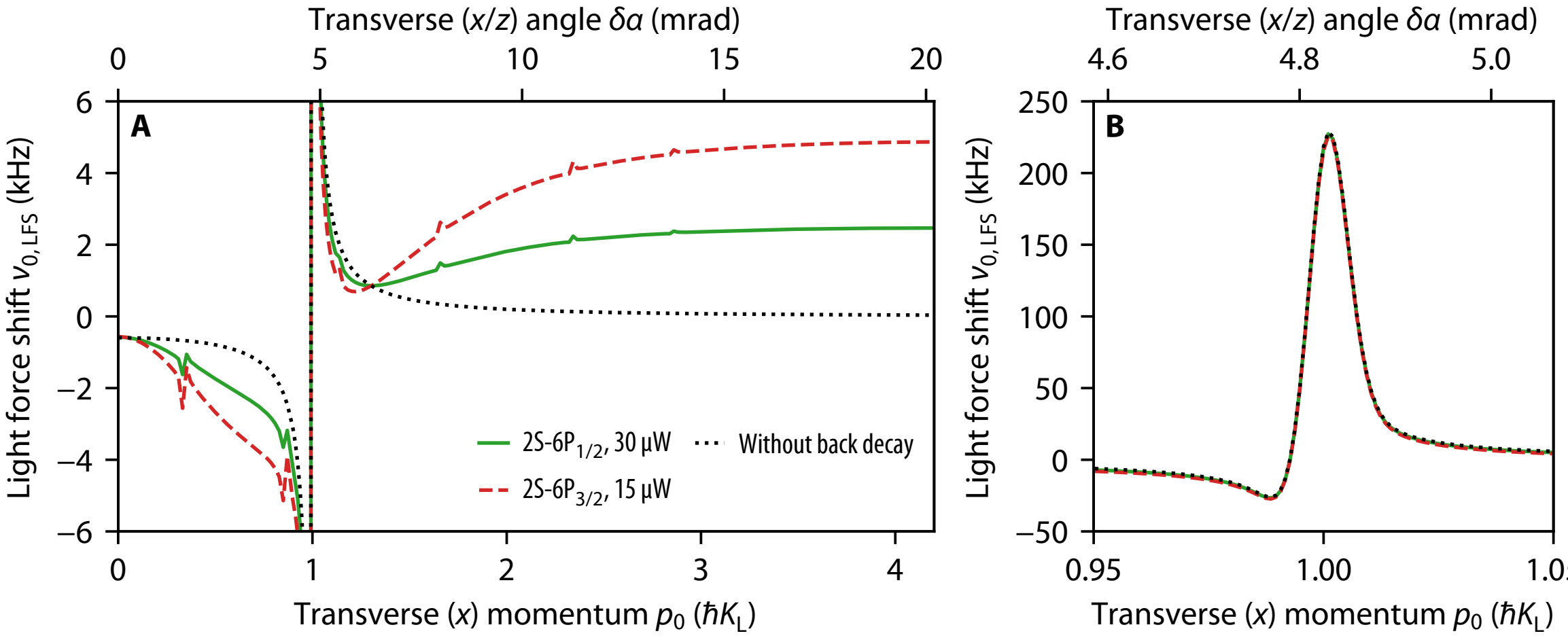

Figure 3.3: (**A**) Simulation of the light force shift (LFS) $\nu_{0,\mathrm{LFS}}$ versus transverse momentum $p_0$ (bottom axis) for the 2S-6P$_{1/2}$ transition (solid green line) and 2S-6P$_{3/2}$ transition (dashed red line). $p_0$, given in units of the photon momentum $\hbar K_\mathrm{L}$, is the momentum along the spectroscopy laser beams forming the light grating. The incoming atom is in the 2S level and delocalized along the transverse direction ($|\mathrm{e}\rangle|p_0\rangle_x$). The speed of the atom is $v = 200\,\mathrm{m/s}$, and it thus crosses the light grating at an angle $\delta\alpha \approx p_0/m_\mathrm{H}v$ (top axis) from the orthogonal. $\nu_{0,\mathrm{LFS}}$ is determined by fitting Voigt line shapes to simulated line scans, found by integrating the optical Bloch equations of the LFS model. The corresponding level scheme is shown in Fig. 3.2. The power of the spectroscopy laser is $P_{\text{2S-6P}} = 30\,\mu\mathrm{W}$ ($15\,\mu\mathrm{W}$) for the 2S-6P$_{1/2}$ (2S-6P$_{3/2}$) transition, giving the same Rabi frequency for both transitions because of the larger dipole moment for the 2S-6P$_{3/2}$ transition. The larger LFS of the 2S-6P$_{3/2}$ transition is caused by its two times higher back decay rate $\gamma_\mathrm{ei}$ to the 2S initial level. The LFS without back decay, i.e., $\gamma_\mathrm{ei}$ set to zero, is also shown (black dotted line), using the same Rabi frequency as with nonzero $\gamma_\mathrm{ei}$. The wiggles at $p_0 \approx 0.3\hbar K_\mathrm{L}$, $0.85\hbar K_\mathrm{L}$, ... are numerical artifacts from the averaging over the back decay momentum change $\Delta p_{\mathrm{D},1}$. (**B**) Same data as (A), but showing the region around $p_0 = \hbar K_\mathrm{L}$, where a two-photon resonance leads to an especially large LFS independent of the back decay rate.

Above the resonance, the now positive LFS trends towards a value of 2.7 kHz and 4.9 kHz for the 2S-6P$_{1/2}$ and 2S-6P$_{3/2}$ transition, respectively. If the back decay is not included in the description, however, the LFS trends towards zero above the resonance, again demonstrating the importance of including the back decay in the description of the LFS.

Fig. 3.4 shows the LFS for the 2S-6P$_{1/2}$ transition for three different values of the spectroscopy laser power used in the 2S-6P measurement, $P_{\text{2S-6P}} = 10\,\mu\mathrm{W}$, $20\,\mu\mathrm{W}$, and $30\,\mu\mathrm{W}$. The LFS scales approximately linearly with laser power, independent of the value of the transverse momentum, and including at the resonance.

The accuracy of the OBE model was verified using simulations based on the Monte Carlo wave function (MCWF) (or quantum jump) method [77, 79, 80]. In this method, the number of back decays need not be limited and the back decay momentum change $\Delta p_\mathrm{D}$ need not be discretized. This is possible because in this method, the wave function is evolved in time using a pseudo-Hamiltonian, and at each time step a decay with a random momentum change may take place. This procedure is then repeated for many of these quantum trajectories[1] to find the average evolution of the system, from which the density matrix, which is the object calculated with the OBEs, can be found. While the MCWF method is thus a very versatile method to solve problems involved spontaneous decay, it is for the problem discussed here much more computationally expensive than the OBE approach, as a large number quantum trajectories are needed to accurately model the LFS.

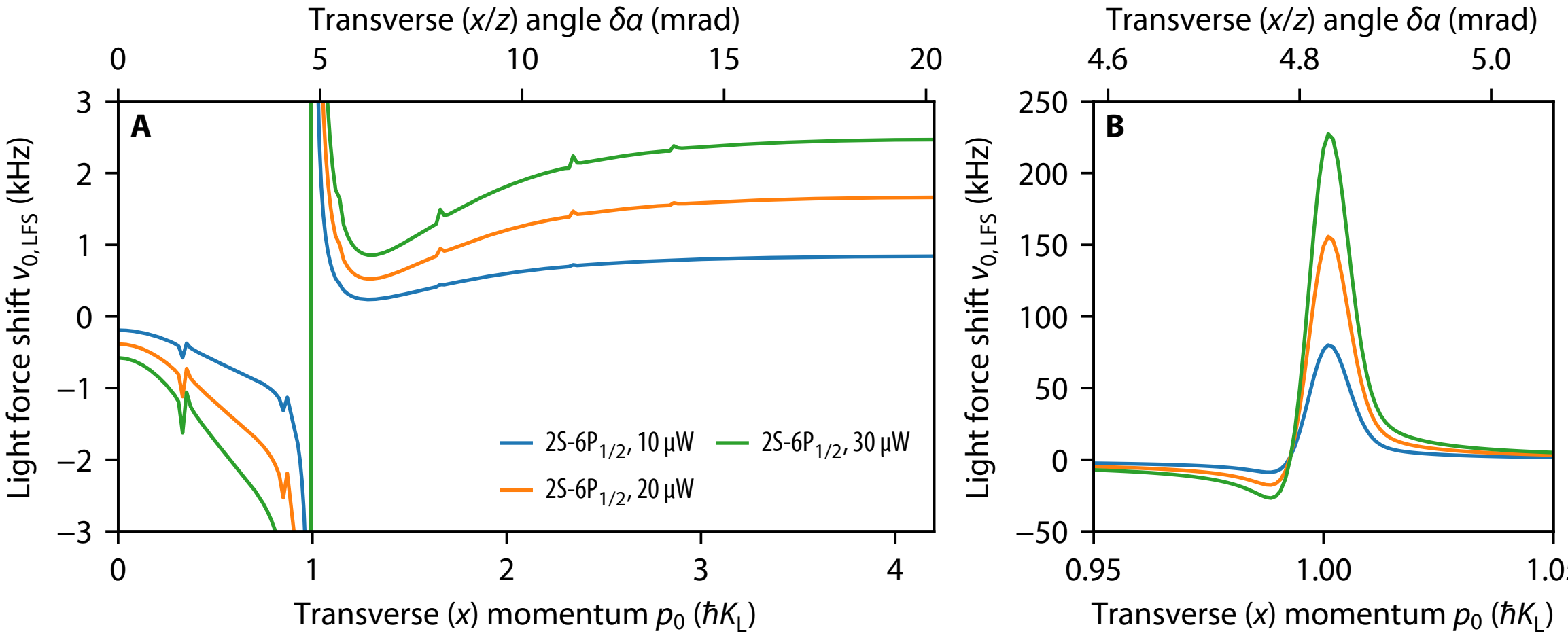

Figure 3.4: (**A**) Simulation of the light force shift (LFS) $\nu_{0,\mathrm{LFS}}$ versus transverse momentum $p_0$ and the spectroscopy laser power $P_{2\mathrm{S}\text{-}6\mathrm{P}}$. Similar to Fig. 3.3, but only results for the 2S-6P$_{1/2}$ transition are included. The LFS scales approximately linearly with $P_{2\mathrm{S}\text{-}6\mathrm{P}}$, independent of the value of the transverse momentum. (**B**) Same data as (A), but showing the region around $p_0 = \hbar K_\mathrm{L}$, where a two-photon resonance leads to an especially large LFS, which also scales approximately linearly with $P_{2\mathrm{S}\text{-}6\mathrm{P}}$.

The LFS was calculated with the MCWM method for atoms with a speed of $v = 200\,\mathrm{m/s}$ and transverse momenta of $p_0/\hbar K_\mathrm{L} = 0, 1, 2$, using $2 \times 10^8$ quantum trajectories each. The 2S-6P$_{3/2}$ transition was used and the laser power was set to $P_{2\mathrm{S}\text{-}6\mathrm{P}} = 15\,\mu\mathrm{W}$. First, the number of back decays is not limited and $N_{k,\mathrm{max}}$, that is the maximum momentum change, is varied. $\nu_{0,\mathrm{LFS}}$ changes by less than 5 Hz, even at the resonance for $p_0/\hbar K_\mathrm{L} = 1$, from $N_{k,\mathrm{max}} = 3$ to $N_{k,\mathrm{max}} = 4$, and by less than 100 mHz when increasing $N_{k,\mathrm{max}}$ further to 5. Second, using $N_{k,\mathrm{max}} = 4$, the maximum number of back decays $N_{\mathrm{BD,max}}$ is varied and compared to the situation where no such restriction is imposed. For $N_{\mathrm{BD,max}} = 1$ ($N_{\mathrm{BD,max}} = 2$), the difference to the latter case is below 30 Hz (0.4 Hz), corresponding to less than 4 % ($5 \times 10^{-4}$) of the corresponding LFS. Thus, using $N_{k,\mathrm{max}} = 4$ and $N_{\mathrm{BD,max}} = 1$ in the OBE model is a good approximation at the current level of accuracy, where the LFS needs to be corrected to within ∼30 % of its size. This OBE model is referred as LFS model throughout this work and is used to correct the results of the 2S-6P measurement, as discussed below.

Finally, the error made by using the Gaussian quadrature rule to average over the back decay momentum change $\Delta p_{\mathrm{D},1}$ in the LFS model was estimated. To this end, the average LFS for $\delta\alpha = 0$ mrad to 20 mrad was evaluated using, on the one hand, the Gaussian quadrature rule with $N_{\Delta p_\mathrm{D}} = 4$, and, on the other hand, an evenly spaced sampling of $\Delta p_{\mathrm{D},1}$ with

[1] Not to be confused with the (classical) trajectory of an atom, i.e., its position as a function of time. Here, the term trajectory is used for such classical trajectories, especially in the context of the Monte Carlo simulation of the atomic beam, while the quantum trajectories of the MCWF method are referred to as such.

$N_{\Delta p_D} = 101$. No numerical artifacts are visible when using the latter, but the computation time is 25 times larger. The results were found to agree within ≈1 %, which we here assume to be the error associated with the use of the Gaussian quadrature rule.

### 3.4.2 Light force shift for the atomic beam

In the previous section, we have shown how the light force shift (LFS) of the 2S-6P transition can be modeled for a single delocalized atom with defined transverse momentum $p_0$ and speed $v$, using optical Bloch equations (OBEs). We now return to the question of how to simulate the light force shift as observed for the atomic beam. In principle, the momentum space representation of the density matrix $\rho_2$ of the atomic beam (Eq. (3.32)) can be directly used as input state for the LFS model. There are however two caveats: first, $\rho_2$ is continuous, while the model uses discrete momentum states. Second, the excitation of the atoms to the 2S initial level by two-photon absorption from the 1S-2S preparation laser has so far been neglected, and the circular nozzle and the rectangular aperture have been approximated by Gaussians.

These points can be addressed by comparing the transverse momentum coherence scale $p_{c,t,2} = m_H \times 2.5 \times 10^{-4}$ m/s of the atomic beam, derived in Section 3.3.4, with the typical transverse momentum scale of the atom–light interaction, given by $\hbar K_L = m_H v_{rec} \approx m_H \times 0.97$ m/s. Since $p_{c,t,2} \ll \hbar K_L$, the coherence between momentum states can be neglected in the description of the light force shift. In other words, the density matrix in momentum space can be treated as diagonal, since the coherence between the occupied momentum states drops of very rapidly and is effectively zero for momentum states coupled by the laser. Likewise, momentum states that have some mutual coherence are separated by such a small amount of momentum compared to the momentum scale of the interaction that they can treated as a single input state.

We can thus model the atomic beam as an incoherent sum of delocalized atoms, i.e., plane waves in position space. This closely matches the situation of the extended incoherent light source treated above (see Section 3.2.1), where fully coherent plane waves are emitted from each point in the light source. It is only through the incoherent sum, i.e., the addition of the resulting intensity interference patterns from each plane wave, that the coherence is limited to the value given by the spatial coherence length.

In this way, the LFS model can be connected to the modeling of the atomic beam as a set of $N_{traj}$ classical atom trajectories with defined position and momentum, introduced in Section 5.2. This Monte Carlo modeling takes into account the 1S-2S excitation and the time-resolved detection of the experiment through an appropriate weighting of the trajectories. For each trajectory, the LFS model is then solved for a delocalized atom with a transverse momentum matching that of the trajectory. The signal found in this way is then summed up over all $N_{traj}$ trajectories, corresponding to an incoherent sum of the interference patterns observed in the signals. The sum of the signals as a function of laser detuning is then fit with the appropriate line shape function, with the determined resonance frequency $\nu_{0,LFS}$ corresponding to the light force shift for this set of trajectories. The results of such a procedure for trajectory sets describing the 2S-6P measurement are shown in Fig. 6.6 (A).

Note that using such classical trajectories instead of the Wigner function corresponds to neglecting the diffraction of the atomic beam on the aperture, but importantly not the light grating. The diffraction at the aperture however is negligible, since it is much larger than the transverse coherence length.

# Chapter 4

# Hydrogen spectrometer

This chapter discusses the experimental setup, or hydrogen spectrometer, used for laser spectroscopy on the 2S-$n$P transitions in atomic hydrogen (H) at the Laser Spectroscopy Division of the Max Planck Institute of Quantum Optics (MPQ). First, the core component of the spectrometer, the atomic beam apparatus, is introduced in Section 4.1. The vacuum system containing the atomic beam apparatus is the subject of Section 4.2. Details of the laser system used to optically excite hydrogen atoms from the 1S ground level to the metastable 2S level are given in Section 4.3. The spectroscopy laser that drives the 2S-6P transition and the active fiber-based retroreflector (AFR), necessary to suppress the first-order Doppler shift of the atoms moving through the spectroscopy laser beams, are detailed in Section 4.4. The formation of the cryogenic atomic beam of hydrogen atoms is described in Section 4.5. Section 4.6 focuses on the detector assembly that collects the fluorescence from the atoms as they decay from the 6P level. The data acquisition hardware and software that controls the experiment and collects and stores the measurement data are discussed in Section 4.7. Finally, Section 4.8 covers the determination of the laser frequencies in SI units.

## 4.1 Atomic beam apparatus

A 3D view of the core part of the hydrogen spectrometer, the atomic beam apparatus, is shown in Fig. 4.1. Here, the various components shown in the figure are introduced, each labeled with a two-letter acronym, which will then be discussed in more detail in the next sections.

The atomic beam apparatus sits inside a vacuum chamber, which is split into two differentially pumped regions (see Section 4.2). The outer vacuum region, pumped with a turbopump, is connected to the inner high-vacuum region only through the high-vacuum entrance (**EA**) and output (**OA**) apertures. The high-vacuum region is pumped with a cryopump through holes (**CC**) at the bottom of the high-vacuum enclosure (**HV**) and through a wire mesh at the bottom of the detector cylinder (**DC**). A single-layer magnetic shield (**MS**) encloses the high-vacuum region.

Atomic hydrogen (H) at room temperature, generated by dissociating molecular hydrogen (see Section 4.5.1), is fed through tubing (**TT**) made from PTFE[1] to a copper nozzle (**NZ**) (see Section 4.5.2). The nozzle is attached to a helium continuous-flow cryostat (**CS**) and cooled down to $T_{\mathrm{N}} = 4.8\,\mathrm{K}$. H thermalizes to the nozzle temperature and exits the nozzle channel through two orifices, forming a cold beam (**AB**). Only atoms passing through the high-vacuum entrance aperture, i.e., to the right in Fig. 4.1, and into the high-vacuum region are probed in the experiment. Together with the nozzle channel, an aperture of fixed height and variable width (**VA**) further downstream limits the divergence of the atomic beam. The width of the aperture is remote-controlled using two actuators (**AM**) (see Section 4.5.3).

[1]PTFE: polytetrafluoroethylene, commonly known under its brand name Teflon.

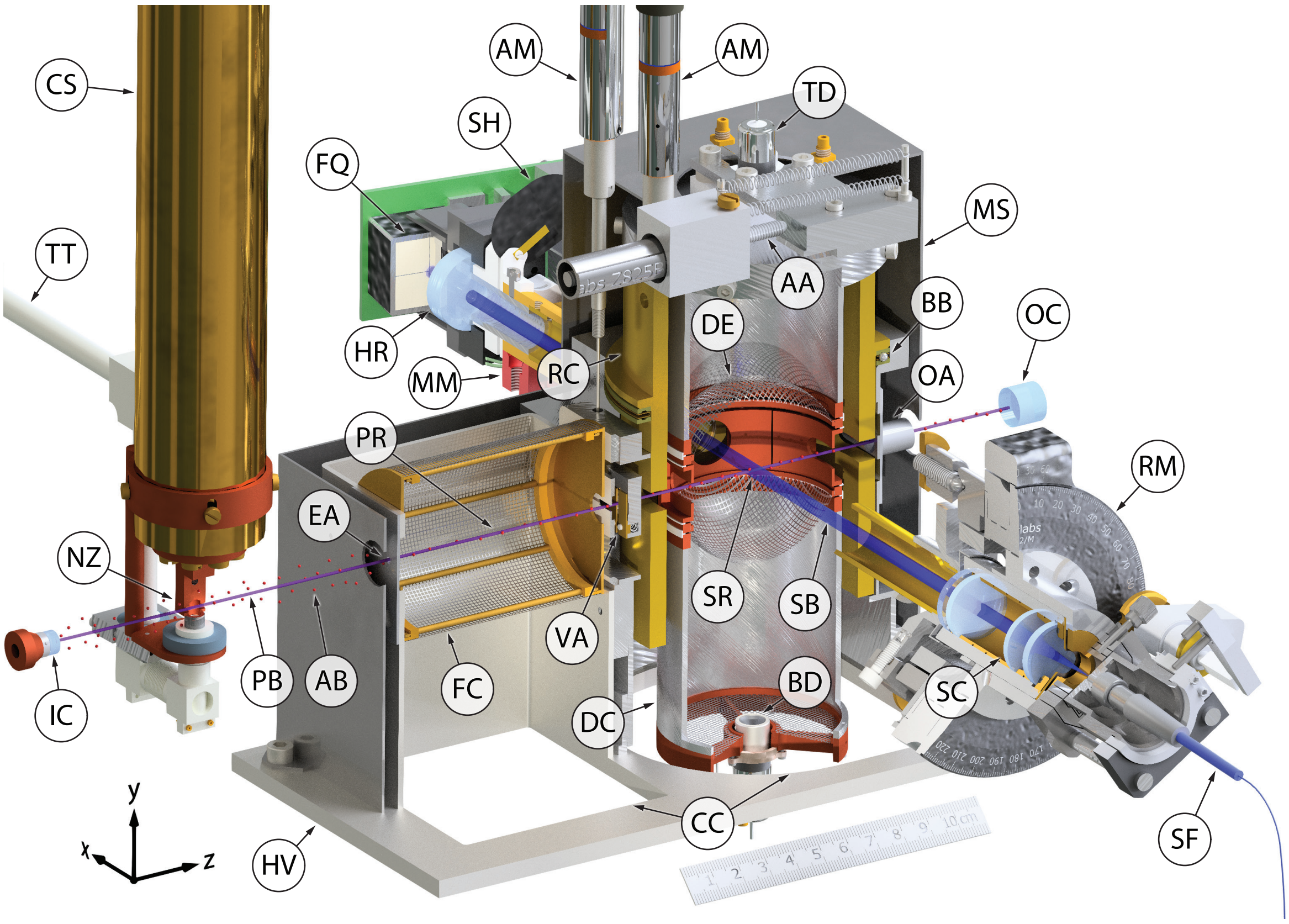

Figure 4.1: 3D view (orthographic projection) of the atomic beam apparatus, as used to determine the transition frequency of the 2S-6P transition. Some parts are shown cut open to reveal otherwise hidden details. See Section 4.1 for details. **AA**: $\alpha_0$ alignment actuator, **AB**: atomic beam, **AM**: variable aperture actuator, **BB**: base cylinder ball bearing, **BD**: bottom detector, **CC**: connection to cryopump, **CS**: cryostat, **DC**: detector cylinder, **DE**: detector electrodes, **EA**: high-vacuum entrance aperture, **FC**: 1S-2S Faraday cage, **FQ**: AFR four-quadrant photomultiplier, **HR**: AFR high-reflectivity mirror, **HV**: high-vacuum enclosure, **IC**: piezo-actuated 243 nm incoupling mirror, **MM**: piezo-actuated AFR mirror mount, **MS**: magnetic shield, **NZ**: copper nozzle, **OA**: high-vacuum output aperture, **OC**: 243 nm outcoupling mirror, **PB**: atomic beam and 1S-2S preparation laser beam, **PR**: 1S-2S preparation region, **RC**: rotatable base cylinder, **RM**: collimator rotation mount, **SB**: 2S-6P spectroscopy laser beams, **SC**: four-lens fiber collimator, **SF**: polarization-maintaining fiber, **SH**: AFR shutter, **SR**: 2S-6P spectroscopy region, **TD**: top detector, **TT**: PTFE tubing, **VA**: variable aperture; AFR: active fiber-based retroreflector, PTFE: polytetrafluoroethylene (Teflon).

The atoms emerging from the nozzle are initially in the 1S electronic ground level. To generate a cold beam of atoms in the metastable 2S level (2S atoms), the atomic beam is overlapped with a collinear preparation laser beam (**PB**) with a wavelength of 243 nm, resonant with the two-photon 1S-2S transition (see Section 4.3). A linear two-mirror enhancement cavity with an $1/e^2$ intensity radius of 0.3 mm (**IC**: 243 nm incoupling mirror with reflectance $R_{IC} = 98.4\,\%$, **OC**: 243 nm outcoupling mirror with reflectance $R_{OC} = 99.8\,\%$) is used to enhance the 1S-2S laser intensity seen by the atoms (see Section 4.3.3). In this way, the atoms are excited to the 2S level while flying through the apparatus, shielded from stray electric fields inside the 1S-2S preparation region (**PR**) by the 1S-2S Faraday cage (**FC**).

After passing through the variable aperture (**VA**), this atomic beam then interacts in the 2S-6P spectroscopy region (**SR**) with the two counter-propagating 2S-6P spectroscopy laser beams (**SB**) with a wavelength of 410 nm. The laser beams are derived from the 2S-6P spectroscopy laser (see Section 4.4.1), which is guided into the vacuum chamber using a polarization-maintaining optical fiber (**SF**). The centerpiece of the apparatus is the active fiber-based retroreflector (AFR, see Section 4.4.2), consisting of a four-lens fiber collimator (**SC**), a high-reflectivity (HR) mirror (**HR**) mounted in a piezo-actuated mirror mount (**MM**), a remotely-operated shutter (**SH**), and a four-quadrant photomultiplier (PMT) (**FQ**). The collimator produces a high-quality Gaussian beam with an $1/e^2$ intensity radius of $W_0 = 2.2$ mm, which is retroreflected at the HR mirror. The tip and tilt of the mirror is actively controlled, using the actuated mirror mount, such that the beam is coupled back into the optical fiber. In this way, two counter-propagating, wavefront-retracing[1] laser beams are produced. The forward-traveling beam can also be blocked right before the mirror with the shutter, such that no returning laser beam is created. A small of part of the laser light leaks through the HR mirror and is detected on the four-quadrant PMT. The collimator and optical fiber can be rotated about the beam axis using a manual rotation mount (**RM**), allowing for a rotation of the linear polarization direction as determined by the fiber.

The 2S atoms are excited to the short-lived 6P level while flying through the 2S-6P spectroscopy laser beams. They decay while still in the spectroscopy region, primarily emitting fluorescence photons at a wavelength of 93.8 nm. These photons are energetic enough to eject photoelectrons from the walls of the detector cylinder (**DC**) (see Section 4.6), which are then detected using electron multipliers at the top (**TD**) and bottom (**BD**) of the detector cylinder. The input face of the electron multipliers is held at a positive bias voltage relative to the grounded detector cylinder to attract the photoelectrons. In order not to disturb the atoms, the resulting electric fields are shielded with wire meshes (**DE**) above and below the spectroscopy region. Bias voltages to measure and offset remaining stray electric fields can be applied to these meshes and four copper electrodes forming the walls of the detector cylinder within the spectroscopy region (see Fig. 4.34).

The detector cylinder, which is mounted on a rotatable base cylinder (**RC**) itself connected to the high-vacuum enclosure through a ball bearing (**BB**), can be remotely rotated about its axis, using the $\alpha_0$ alignment actuator[2] (**AA**). Since the components of the AFR, defining the propagation axis of the spectroscopy laser, are attached to the rotation mount, this allows an adjustment of the offset angle $\alpha_0$ between the spectroscopy laser and the atomic beam, as defined by the apertures and the propagation axis of the preparation laser beam. Note that the offset angle $\alpha_0$ is defined as the angle from the orthogonal between the atomic and laser

[1]Wavefront-retracing is here taken to mean that the wave vectors of the two beams are antiparallel at any point. This property was referred to as phase-retracing in our previous publications [24, 28].

[2]Thorlabs Z825BV dc servo motor actuator with rotary encoder. Absolute on-axis accuracy is specified as 130 µm, and the bi-directional repeatability as $<1.5$ µm. A linear movement of 1 mm corresponds to a rotation of the cylinder by 16.7 mrad at $\alpha_0 \approx 0$ mrad. In the experiment, the position of the actuator (as given by its encoder) that corresponds to $\alpha_0 = 0$ mrad is determined in situ [28], and all subsequent rotations are performed relative to this position.

beams, which cross at close to right angles and thus $\alpha_0$ is on the order of a few mrad.

The experiment itself is performed by repeating a fixed measurement cycle, a chopper cycle, with a rate of $f_{\text{chop}} = 160\,\text{Hz}$. First, the 1S-2S preparation laser beam is unblocked and the length of the enhancement cavity stabilized to it. Atoms traveling through the apparatus are thus excited to the 2S metastable level. After $(1/f_{\text{chop}})/2 = 3.125\,\text{ms}$, the 1S-2S laser beam is blocked again. The 2S-6P spectroscopy laser beams are not blocked and thus atoms flying through them can be excited from the 2S to the 6P level. The blocking of the 1S-2S laser beam triggers the start at $\tau = 0\,\mu\text{s}$ of the time-resolved detection of fluorescence from atoms decaying from the 6P level. Each pulse from either electron multiplier is recorded as a single count. Since the 1S-2S laser beam is blocked during detection, it cannot photoionize the 2S atoms, which would otherwise lead to a large background count rate. Since also no 1S-2S excitation takes place anymore, atoms detected at a delay time $\tau$ must have been flying through the apparatus for at least a time $\tau$. This limits the maximum speed of the detected atoms to[1] $v_{\text{max}} \approx L/\tau$, where $L$ is the distance from the nozzle to the 2S-6P laser beam. Thus, the time-resolved detection gives way to a velocity-resolved detection. After again 3.125 ms of detection, the 1S-2S laser beam is unblocked again and the cycle begins anew.

In total, 160 chopper cycles are repeated within 1 s and the number of counts as a function of delay time $\tau$ is summed up over the cycles. Then, experimental parameters such as the frequency of the 2S-6P spectroscopy laser are changed and the experiment is repeated (see Section 4.7). The essential data set is then the number of counts as a function of delay time $\tau$ and laser frequency, referred to as a line scan through this work. From this, the transition frequency of the 2S-6P transition can be determined if combined with the determination of the laser frequency (see Section 4.8).

Another, longer experimental cycle is imposed by the nozzle being slowly clogged by freezing hydrogen. Because of this, the nozzle channel reduces in diameter over time, till the 1S-2S laser beam which passes through the channel is attenuated to such a degree that operation of the experiment is not possible anymore. The nozzle is then heated up to room temperature to remove the frozen hydrogen and other trace gases ("unfreezing the nozzle"). This experimental cycle is here referred to as freezing cycle (FC), with each measurement day separated into multiple freezing cycles. The duration of a freezing cycle for the nozzle geometry, nozzle temperature, and hydrogen flow used here is approximately $\Delta t_{\text{FC}} = 120\,\text{min}$. The FCs that make up the 2S-6P measurement are each assigned a name consisting of the measurement run they belong to (see Table 6.1) and a consecutive number within that run, e.g., B29 is the 29$^{\text{th}}$ FC within run B.

## 4.2 Vacuum system

The atomic beam apparatus introduced in Section 4.1 sits inside a vacuum chamber in order to probe the hydrogen atoms free from collisions with background gas particles. A 3D view of the vacuum system is shown in Fig. 4.2. As in Fig. 4.1, key components are labeled with a two-letter acronym, with components shown in both figures sharing the same acronym.

[1]Strictly speaking, the maximum longitudinal velocity $v_z$ along the atomic beam is constricted to $L/\tau$, but here $v \approx v_z$ since the atomic beam is well-collimated.

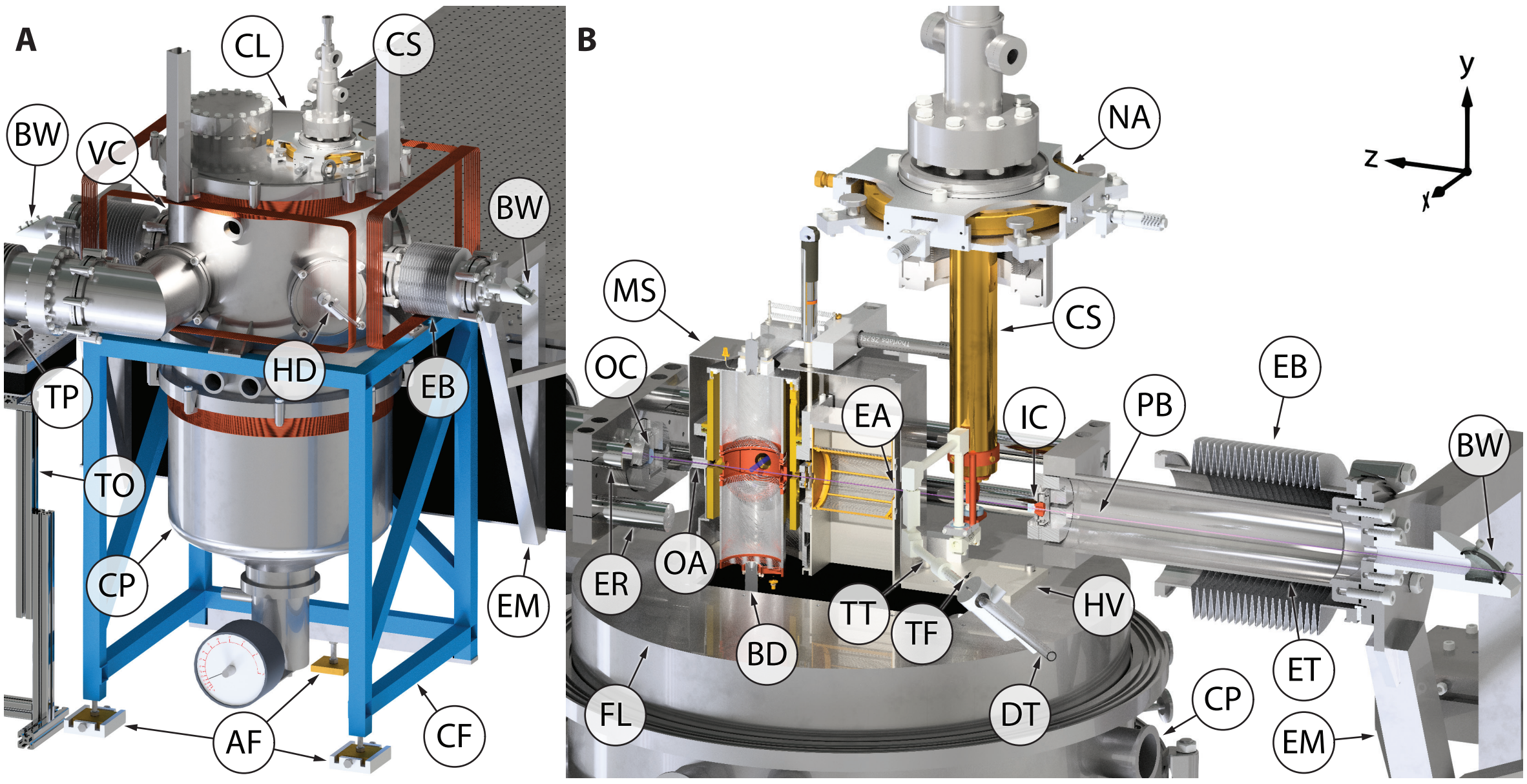

Figure 4.2: 3D view (orthographic projection) of the vacuum system containing the atomic beam apparatus (see Fig. 4.1): (**A**) Complete system including support frames and pumps, and (**B**) upper part of system with the circular vacuum chamber and its lid removed, and the enhancement cavity spacer and atomic beam apparatus cut open to reveal the mounting and differential pumping scheme. Coordinate axes for both views are shown in the upper right corner. See Section 4.2 for details. **AF**: adjustable feet, **BD**: bottom detector, **BW**: 243 nm Brewster's window, **CF**: vacuum chamber frame, **CL**: vacuum chamber lid, **CP**: cryopump, **CS**: cryostat, **DT**: dissociator discharge tube, **EA**: high-vacuum entrance aperture, **EB**: EC spacer bellows, **EM**: EC mounting brackets, **ER**: EC spacer Invar rods, **ET**: EC spacer tubes, **FL**: cylindrical vacuum chamber floor, **HD**: hydrogen dissociator, **HV**: high-vacuum enclosure, **IC**: piezo-actuated 243 nm incoupling mirror, **MS**: magnetic shield, **MW**: dissociator microwave cavity, **NA**: nozzle alignment stage, **OA**: high-vacuum output aperture, **PB**: atomic beam and 1S-2S preparation laser beam, **TF**: temperature sensor vacuum chamber floor, **TO**: turbopump optical table, **TP**: turbopump, **TT**: PTFE tubing, **VC**: cylindrical vacuum chamber; EC: 243 nm enhancement cavity, PTFE: polytetrafluoroethylene (Teflon).

### 4.2.1 Vacuum chamber

The vacuum chamber (**VC**) containing the atomic beam apparatus is a stainless steel cylinder of 498 mm inner diameter and 332 mm usable height, as measured from the chamber floor (**FL**) to the chamber lid (**CL**). The vacuum chamber is split into two differentially pumped regions, the outer vacuum region chiefly containing the cryostat (**CS**) with the hydrogen nozzle and the 243 nm enhancement cavity, and the inner high-vacuum region containing the 1S-2S preparation region and the 2S-6P spectroscopy region. The two vacuum regions are separated by the chamber floor and the high-vacuum enclosure (**HV**), and are only connected through two apertures, an entrance aperture (**EA**) of 2.4 mm diameter and 1.0 mm length, and an output aperture (**OA**) of 7 mm diameter and 13 mm length. Additionally, a bypass controlled by a valve can be used to temporarily connect the two regions with much higher conductance.

### 4.2.2 Outer vacuum region

The outer vacuum region is pumped with a hybrid-bearing turbopump[1] (**TP**) with a pumping speed of 555 l/s for $H_2$. The required fore-vacuum is generated by an oil-free roughing pump[2]. The turbopump is attached to the vacuum chamber through a vibration isolator and mounted on a separate optical table (**TO**), standing on the laboratory floor. When no hydrogen is introduced into the vacuum system, the background pressure in the outer vacuum region ($P_{\mathrm{OV}}$, calibrated for $N_2$ gas, see below) is typically $2 \times 10^{-6}$ mbar.

### 4.2.3 High-vacuum region and cryopump

The inner high-vacuum region is pumped by a cryopump[3] (**CP**) with a pumping speed of 10 000 l/s for $H_2$, attached to the bottom of the vacuum chamber. The cryopump is based on the Gifford-McMahon cycle [81], using helium as working fluid and an external compressor. The cryopump is a major design constraint because of its large size, the strong vibrations caused by the movement of the displacer inside the cryopump, and the large temperature difference between vacuum chamber and cryopump. In fact, spectroscopy data can only be acquired when the cryopump is not running, forcing an experimental cycle of switching, every few minutes, between running the cryopump and taking data. However, the cryopump offers a pumping speed and capacity for hydrogen unmatched by other types of oil-free vacuum pumps. The cryopump cools down to a temperature of 19 K, which is below the freezing point of common residual gases such as water vapor, nitrogen, and oxygen, but above the freezing point of hydrogen. Instead, hydrogen is removed from the environment by cryosorption in activated charcoal coating the inside of the cryopump [81]. The cryopump's very large pumping speeds of 10 000 l/s and 29 000 l/s for $N_2$ and $H_2O$, respectively, allow for a fast cycling time of breaking vacuum for maintenance and then pumping down to pressures on the order of $1 \times 10^{-7}$ mbar within approximately 8 h. When no hydrogen is introduced into the vacuum system, the background pressure in the high-vacuum region ($P_{\mathrm{HV2}}$, calibrated for $N_2$ gas, see below) is typically $5 \times 10^{-8}$ mbar.

### 4.2.4 Vacuum diagnostics

The pressures in the two vacuum regions are measured with hybrid pressure gauges[4], combining a gauge based on a thermal conductivity measurement of the gas (Pirani gauge) and a hot-filament ionization gauge (Bayard-Alpert gauge), used above and below pressures of $5 \times 10^{-3}$ mbar, respectively. To monitor the pressure $P_{\mathrm{OV}}$ in the outer vacuum region, a gauge is placed after a right-angle bend attached to the vacuum chamber. The high-vacuum region is monitored with two pressure gauges. One gauge is directly attached to a port on the vacuum chamber below the chamber floor, thus probing the pressure $P_{\mathrm{HV1}}$ close to the spectroscopy region. The hot filament of this gauge however creates charged particles that are detected mainly in the bottom electron multiplier (**BD**), leading to a background count rate on the order of 100 counts/s, hundred times higher than when the hot filament is switched off. For this reason, a second gauge, measuring pressure $P_{\mathrm{HV2}}$, is placed after two right-angle bends, attached to another port below the chamber floor. This gauge does not lead to an increase in the background count rate, and during a measurement only this gauge (and the outer vacuum gauge) are switched on. The pressure ratio between the two high-vacuum gauges in general depends on the pressure, temperature, and gas types present. During measurement conditions, i.e., with both the cryopump and nozzle cold and hydrogen streaming into the apparatus, $P_{\mathrm{HV1}}/P_{\mathrm{HV2}} \approx 0.65$. Furthermore, for these conditions the background gas is dominated by $H_2$, while the gauges are calibrated for $N_2$ gas, which leads to a correction factor of 2.4, determined by the manufacturer. These two factors are used to determine the background pressure in the high-vacuum region from the measured value of $P_{\mathrm{HV2}}$.

[1] Pfeiffer HiPace 700, compression ratio of $4 \times 10^5$ for $H_2$, with static magnetic bearing on the high-vacuum side and mechanical bearing on the fore-vacuum side. After the end of the measurement, the turbopump was replaced with a magnetically levitated version with similar specifications to further reduce mechanical vibrations.

[2] Pfeiffer ACP 40, 37 $m^3$/h nominal pumping speed.

[3] Leybold RPK 10000.

[4] Leybold Ionivac ITR 90 and Ionivac ITR 200 S.

A residual gas analyzer[1] (RGA) using a quadrupole mass spectrometer is also attached to the high-vacuum region through a right-angle bend. The RGA is mainly used for vacuum diagnostics such as identifying small leaks.

### 4.2.5 Mounting of in-vacuum 243 nm enhancement cavity

In order to keep the 243 nm enhancement cavity on resonance with the laser even when the cryopump is running, the enhancement cavity is decoupled from the vacuum chamber. The vacuum chamber is mounted on a aluminum frame (**CF**), which is fixed to the laboratory floor. The enhancement cavity, however, is mounted on a spacer that in turn is mounted on an optical table using brackets (**EM**). The laser system itself is also placed on this optical table. The spacer (see Fig. 4.2 (B)) consists of two steel tubes (**ET**) and four Invar rods (**ER**). The vacuum chamber and the spacer are connected through flexible bellows (**EB**) to seal the vacuum connection while suppressing the transmission of vibrations. The incoupling (**IC**) and outcoupling (**OC**) mirrors of the enhancement cavity are mounted in manually-actuated, top-adjusted mirror mounts. In order to adjust the position of the high-vacuum region relative to the beam axis of the enhancement cavity, the whole vacuum chamber can be moved using the three adjustable feet (**AF**), leaving the enhancement cavity essentially unaffected. The height of each of the three feet can be adjusted independently, and the two front feet, i.e., on the far end from the optical table, can be moved towards and away from the optical table. To allow the latter motion, the foot close to the optical table can slide on a metal plate.

During the 2S-4P measurement (see Appendix A), a second spacer holding the mirror mounts was mounted inside the first spacer, using needle-like rods and O-rings, in an effort to further reduce the amount of vibrations reaching the enhancement cavity from the vacuum chamber. However, this lead to frequent shifts of the enhancement cavity alignment when pumping down to vacuum, and, ultimately, was found to be more sensitive to vibrations of the vacuum chamber than the current arrangement.

### 4.2.6 Mounting of hydrogen nozzle cryostat

The cryostat (**CS**) which cools down the hydrogen nozzle to $T_{\mathrm{N}} = 4.8\,\mathrm{K}$ is mounted in an upright position in the lid of the vacuum chamber. It is a continuous-flow cryostat using liquid helium from a storage dewar. The evaporated helium is captured in a recycling system. Upon cooling down from room temperature, the cryostat contracts[2] such that the nozzle channel is not aligned with the 1S-2S laser beam passing through it. To compensate for this and to align the nozzle to the laser beam, the cryostat can be positioned relative to the chamber lid in all three dimensions using the nozzle alignment stage (**NA**). A flexible bellow protruding into the vacuum chamber is used to seal the vacuum while allowing the cryostat to be pushed across the chamber lid, using micrometer screws, and to be moved in the vertical direction using a translation screw.

[1]Pfeiffer PrismaPlus QMG 220 F1.

[2]The contraction predominantly results in a vertical movement of the nozzle, but there is also some movement in the horizontal plane due to a bending motion.

The cryostat also acts as a cryopump for the outer vacuum region. When the nozzle is cooled down, the heat shield of the cryostat (visible as the gold-colored tube in Fig. 4.2 (B)) is held at an intermediate temperature. With its large surface area, it freezes out water vapor and other contaminants, such as residual organic solvents and hydrocarbons. This mechanism can possibly increase the life time of the 243 nm enhancement cavity mirrors, which are believed to degrade through the accumulation of hydrocarbons on their surfaces facilitated by molecular-bond-cracking UV light. For this reason, the enhancement cavity is only operated with the nozzle, and thus the cryostat, cooled down.

### 4.2.7 Cryopump-induced temperature drifts and gradients

Unfortunately, the cryopump also cools down the vacuum chamber and its components through radiative cooling and direct thermal contact. An in-vacuum temperature sensor[1] (**TF**) taped to the high-vacuum enclosure directly above the cryopump measures the temperature of the vacuum chamber floor. Another sensor measures the temperature of the enhancement cavity spacer. With only the cryopump running, but the nozzle still at room temperature, the vacuum chamber floor is cooled down from room temperature to approximately 13 °C over a course of approximately 10 h. During the experimental cycle, when the nozzle is also cooled down, this temperature further decreases and reaches values of down to 10 °C. This is attributed to the nozzle's cryostat cooling down the vacuum chamber through the lid on which it is mounted[2].

This cooling adversely affects the experiment in several ways. First, the alignment of the various components, contracting at different rates, drifts as the temperature changes. This effect is most crucial for the optics of the 2S-6P spectroscopy laser beam, where especially the distance between the fiber and the fiber collimator is highly sensitive to shifts on the µm level. Second, components itself may be directly affected by the temperature change. In particular, the bottom electron multiplier (**BD**), which sits directly above the cryopump without thermal shielding, cools down by such a degree that its internal resistance and gain properties change substantially (see Section 4.6.2). Third, thermal gradients throughout the apparatus are inevitable as the chamber floor cools down but the rest of the vacuum chamber is held at room temperature. Thus, gradients are expected to be mainly along the vertical direction. A stray electric field in the spectroscopy region (this measurement is based on spectroscopy of the 2S-6P transition and was therefore only performed with the cryopump cooled down) along this direction was repeatedly observed for various detector designs, which is suspected to be caused by such a temperature gradient, especially since the detector cylinder is only weakly thermally coupled to the vacuum chamber.

To mitigate these adverse effects, a temperature state close to the steady state is favorable.

[1]Analog Devices AD590.

[2]No effort was made to shield the in-vacuum temperature sensors from thermal radiation and at least some of the temperature decrease (seen by both sensors) when the cryostat is cooled down could be attributed to thermal radiation.

This is achieved by running the cryopump till the temperature in the chamber stabilizes before starting the measurement, a process that takes about half a day. This is much longer than the times it takes for the cryopump itself to reach its target temperature, which is about 2 h. It is still necessary to stop the cryopump after a day of measurement to release the cryopumped particles to avoid a saturation of the pump. However, this only takes about 5 h, after which the cryopump can be restarted. During this time, the components in the chamber only warm up slightly, and the experiment can be resumed after running the cryopump for about 6 h. This procedure reduces the remaining temperature drifts and gradients substantially. The overall temperature gradient from the cold cryopump to the warm vacuum chamber, however, is of course still present. Note that this procedure deviates considerably from how the experiment was operated previously during the 1S-2S measurement [23] and the 2S-4P measurement (see Appendix A), where the cryopump was only switched on a few hours before starting the measurement, leading to strong temperature drifts requiring constant adjustment of various components during the measurement.

### 4.2.8 Cryopump-induced vibrations

The atomic beam apparatus discussed in Section 4.2, apart from the 243 nm enhancement cavity and the cryostat, is mounted on the vacuum chamber floor. It is thus mechanically directly coupled to the floor of the laboratory and subjected to the vibrations from the cryopump. This especially affects the active fiber-based retroreflector (AFR), since it relies on the precise alignment of its high-reflectivity mirror creating the retroreflected beam. In fact, disturbances on the laboratory floor, such as walking, are easily visible as fluctuations in the amount of light being coupled back into the fiber of the AFR. When the cryopump is running, this signal is strongly disturbed. Thus, when spectroscopy data are taken, the cryopump is switched off and any disturbances in the laboratory are avoided.

An improvement to this situation could be to mount the high-vacuum enclosure and its contents to the spacer holding the 243 nm enhancement cavity, with the resulting gap between the enclosure and the vacuum chamber floor closed with a flexible seal. This mounting needs to be adjustable, since the enclosure can now no longer be aligned relative to the enhancement cavity by adjusting the vacuum chamber position. Such a mounting is currently being evaluated. It remains to be seen if this does not adversely affect the performance of the enhancement cavity.

### 4.2.9 Access to the inside of vacuum chamber

To access the inside of the vacuum chamber, the chamber lid, including the cryostat, can be lifted up using a pulley attached to an overhead crane. First, however, the PTFE tubing (**TT**) leading from the discharge tube (**DT**) of the hydrogen dissociator (**HD**) to the nozzle has to be disconnected. This is achieved by removing the right-angle bend leading to the turbopump and then disconnecting the PTFE tubing through the now accessible flange. Unfortunately, this also requires removing the four-quadrant PMT ((**FQ**) in Fig. 4.1).

To break the vacuum, the vacuum chamber is filled with dry nitrogen gas, which is captured from evaporating liquid nitrogen stored in a dewar. Care was taken to have a constant stream of nitrogen even when the chamber lid was open in order to reduce contamination of the vacuum chamber and to protect the electron multipliers from moisture.

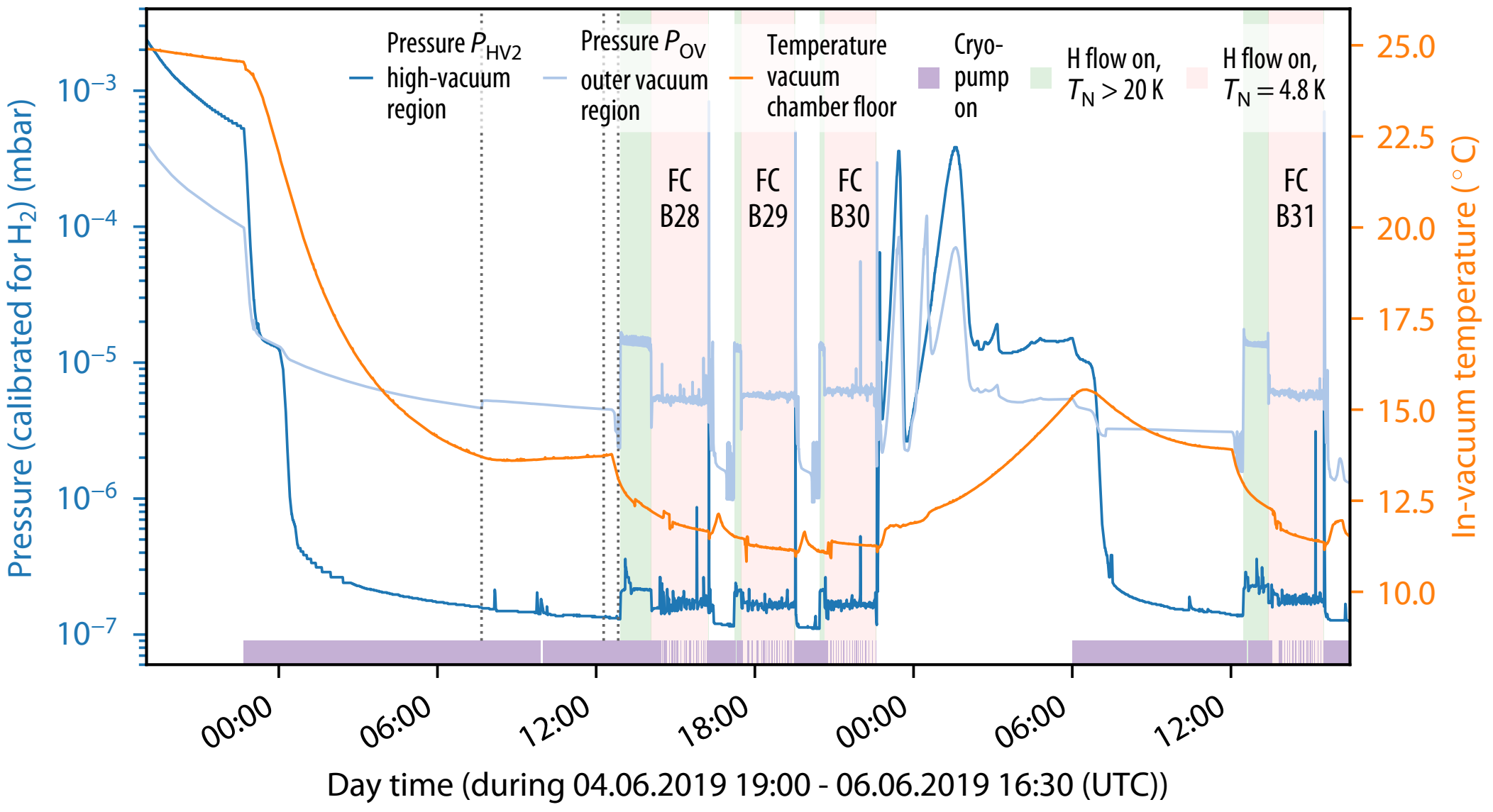

Figure 4.3: Typical pressures and temperatures in the vacuum chamber during an experimental run. The pressures (left scale) in the high-vacuum (dark blue line) and outer vacuum (light blue line) region are shown together with the temperature (right scale) of the vacuum chamber floor (orange line). Also marked are the times when the cryopump was running (purple shading), and when hydrogen was introduced into the apparatus through the nozzle at temperature $T_N > 20$ K (green shading) or $T_N = 4.8$ K (light pink shading). During the latter times, experimental data were taken, split into freezing cycles (FCs). Part of the hydrogen freezes on the nozzle, ultimately clogging it, requiring an intermediate unfreezing between FCs. The turbopump was started two hours before the beginning of the shown data (at 17:00), after the vacuum chamber was opened for maintenance and exposed to dry nitrogen at ambient pressure for 7 h. The pressure readings have been calibrated for $H_2$. See text for details and for special times (dashed gray vertical lines).

### 4.2.10 Typical pressures and temperatures

Fig. 4.3 shows the pressures and the temperature in the vacuum chamber during the course of a measurement day. The pressure readings in the figure have been calibrated for $H_2$ (increased by a factor of 2.4 with respect to standard $N_2$ calibration), leading to an overestimation of the pressure when no hydrogen is flowed into the system. Here, the pressure values given are calibrated for $H_2$ when hydrogen is introduced into the system, and calibrated for $N_2$ otherwise. The procedure and timing discussed here is similar on all other measurements days.

The vacuum chamber was opened for maintenance and exposed to dry nitrogen at ambient pressure for 7 h on the previous day. In the course of this, the hydrogen nozzle was removed from the cryostat, cleaned as described in Section 4.5.2.1, and re-installed. The vacuum chamber was closed and the turbopump started two hours before the beginning of the data shown in Fig. 4.3. After approximately 5 h, the pressures in the outer vacuum (light blue line in Fig. 4.3) and high-vacuum (dark blue line in Fig. 4.3) regions reach $P_{OV} = 4 \times 10^{-5}$ mbar and $P_{HV2} = 2 \times 10^{-4}$ mbar, respectively, and the cryopump is switched on (purple shading). After approximately 2 h, the cryopump reaches its target temperature of 19 K, with the two steep drops in pressure visible during this time thought to correspond to the cryopump reaching the freezing point of first water vapor and then nitrogen. Over the next 14 h, the pressures decrease to $P_{OV} = 2 \times 10^{-6}$ mbar and $P_{HV2} = 6 \times 10^{-8}$ mbar. During the same time, the temperature of the vacuum chamber floor (orange line) decreases from room temperature to 14 °C, with the initial decrease taking about 10 h. However, the temperature of other components further removed from the cryopump is still changing after this initial decrease, as, e.g., observed in the collimation of the 2S-6P spectroscopy laser beam. At 7:40, the bypass between the vacuum regions is closed, leading to a small pressure increase in the outer vacuum region as the cryopump stops pumping substantially on this region.

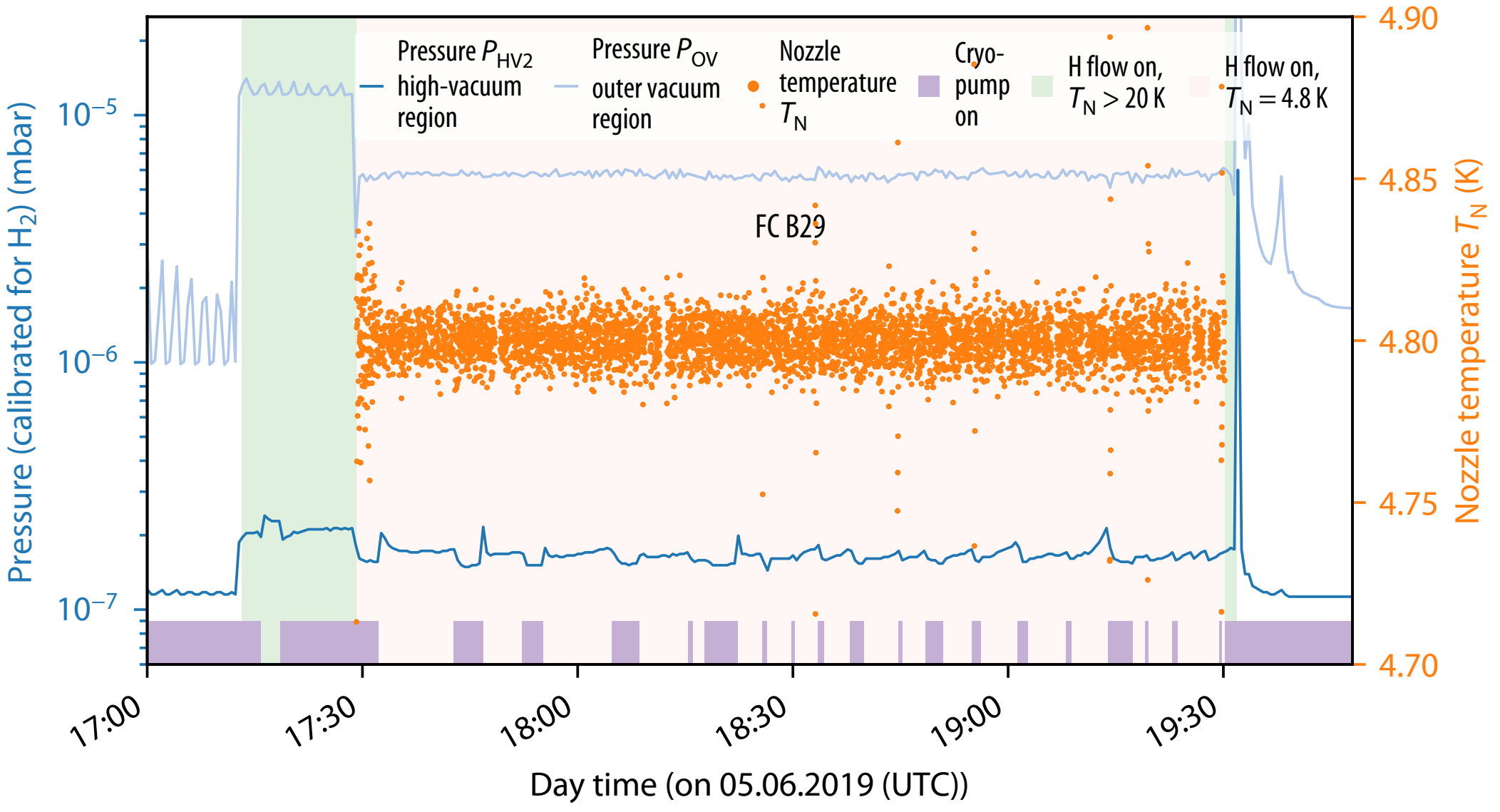

Figure 4.4: Similar to Fig. 4.3, but showing only the situation during the freezing cycle FC B29. Instead of the temperature of the vacuum chamber floor, the temperature $T_N$ of the nozzle is shown (orange points).

At 12:17, the cooldown of the cryostat is started, leading to both a slight drop in pressure in the outer vacuum region (to $P_{OV} = 1.3 \times 10^{-6}$ mbar) as the cryostat's heat shield acts as a cryopump and a further drop in chamber floor temperature as the cryostat cools down the chamber lid. By 12:50, the nozzle temperature has reached $T_N = 30$ K, where it is kept using a heating wire wrapped around the cryostat. At this temperature, hydrogen does not freeze out on the nozzle. The actual temperature oscillates by several K, as the feedback loop is optimized for lower temperatures, leading to the pressure oscillations visible in Fig. 4.4. Next, the hydrogen dissociator is put into operation by flowing $Q_{H_2} = 0.35$ ml/min (at 0 °C and 1013.25 mbar) of $H_2$ into the system and starting the discharge (green shading). This increases the pressure in the two vacuum regions to $P_{OV} = 1.4 \times 10^{-5}$ mbar and $P_{HV2} = 2.2 \times 10^{-7}$ mbar, with the values now calibrated for $H_2$ as it dominates over the background gas. The flow corresponds to a leak rate of $6.3 \times 10^{-3}$ mbar l/s, and thus from the value of $P_{OV}$ the effective pumping speed of hydrogen out of the outer vacuum region can be deduced to be 450 l/s. Since the nozzle had be cleaned, the discharge is kept running for approximately 1 h to flush out residues from the cleaning process (see Section 4.5.2.1). At the same time, the 243 nm enhancement cavity is aligned.

To start the data taking, the nozzle is cooled down to $T_N = 4.8$ K at 14:05. This corresponds to the start of a freezing cycle (FC), as at this temperature hydrogen freezes to the nozzle. Thus, less hydrogen is flowing into the vacuum chamber, reducing the pressures[1] to $P_{\mathrm{OV}} = 5.5 \times 10^{-6}$ mbar and $P_{\mathrm{HV2}} = 1.6 \times 10^{-7}$ mbar. From this drop in pressure, and the pressures without hydrogen, the fraction of hydrogen $Q_{\mathrm{cold}}/Q_{\mathrm{warm}}$ that does not freeze out on the nozzle can be estimated. Taking the pressures in the outer vacuum region, where the pressure with hydrogen flowing into chamber is much larger than the background pressure, $Q_{\mathrm{cold}}/Q_{\mathrm{warm}}$ is found to be approximately 1/3. This measurement is reproducible and is used to characterize the hydrogen transport and nozzle properties, as discussed in Section 4.5.2.

During the freezing cycle, the cryopump needs to be switched off intermittently in order to take spectroscopy data. In Fig. 4.3, the pressures and the nozzle temperature during FC B29 are shown, illustrating the operation of the cryopump (FC B29 instead of FC B28 is shown, because the alignment done during FC B28 makes it a somewhat atypical example). Typically, the cryopump is switched on for approximately a quarter of the time during an FC, keeping the pressure in the high-vacuum region below $P_{\mathrm{HV2}} = 2 \times 10^{-7}$ mbar. However, this procedure consumes more than a quarter of the available measurement time, as the tip-tilt stabilization of the HR mirror of the AFR needs some time to settle down before the acquisition of spectroscopy data can be continued.

After about 2 h, the hydrogen frozen in the nozzle starts to disturb the operation of the 243 nm enhancement cavity, and the nozzle needs to be heated up to melt the hydrogen ice. As the nozzle temperature passes the hydrogen freezing point, the pressure spikes up to $P_{\mathrm{OV}} = 1 \times 10^{-2}$ mbar for a few seconds caused by the large amounts of hydrogen suddenly released. The nozzle is heated up to room temperature to remove contaminants frozen to it. During the 2S-4P measurement (see Appendix A), the nozzle was only heated up to approximately 50 K, with the fluorescence signal varying substantially between the different FCs. It was found that heating up to room temperature, on the other hand, allows for a quite reproducible fluorescence signal. It is conceivable that water vapor is produced in the hydrogen dissociator, freezing to the nozzle and influencing the atomic beam formation if not removed between FCs.

Heating the nozzle up to room temperature and then again cooling it down to 30 K takes approximately 45 min. Then, the next FC can be started, following the routine outlined above. At the end of the last FC of the measurement day, here after the end of FC B30 at 22:35, the nozzle is heated up, the cryopump is switched off, and the bypass from the high-vacuum to the outer vacuum region is opened. As the cryopump warms up, it releases the cryopumped gases, causing multiple peaks in pressure, with the pressures stabilizing at $P_{\mathrm{OV}} = 2 \times 10^{-6}$ mbar and $P_{\mathrm{HV2}} = 5 \times 10^{-6}$ mbar after about 5 h. In total, the cryopump is switched off for 7.5 h before being started again at 6:00 at the start of the next measurement day. This ensures that the temperature of the atomic beam apparatus does not drift too much, with the vacuum chamber floor peaking at 16 °C before decreasing again. After about 7 h of running the cryopump, the first FC of the next day starts. This measurement schedule continues till experimental problems, such as a degradation of the mirrors of the 243 nm enhancement cavity, force the measurement run to be interrupted. For the data acquired for this thesis, the longest such run lasted 8 d.

[1] If no hydrogen is flowing into the vacuum chamber, the decrease in pressures when the nozzle temperature is reduced from 30 K to 4.8 K is much smaller, i.e., the additional cryopumping is negligible compared to the loss of hydrogen on the nozzle.

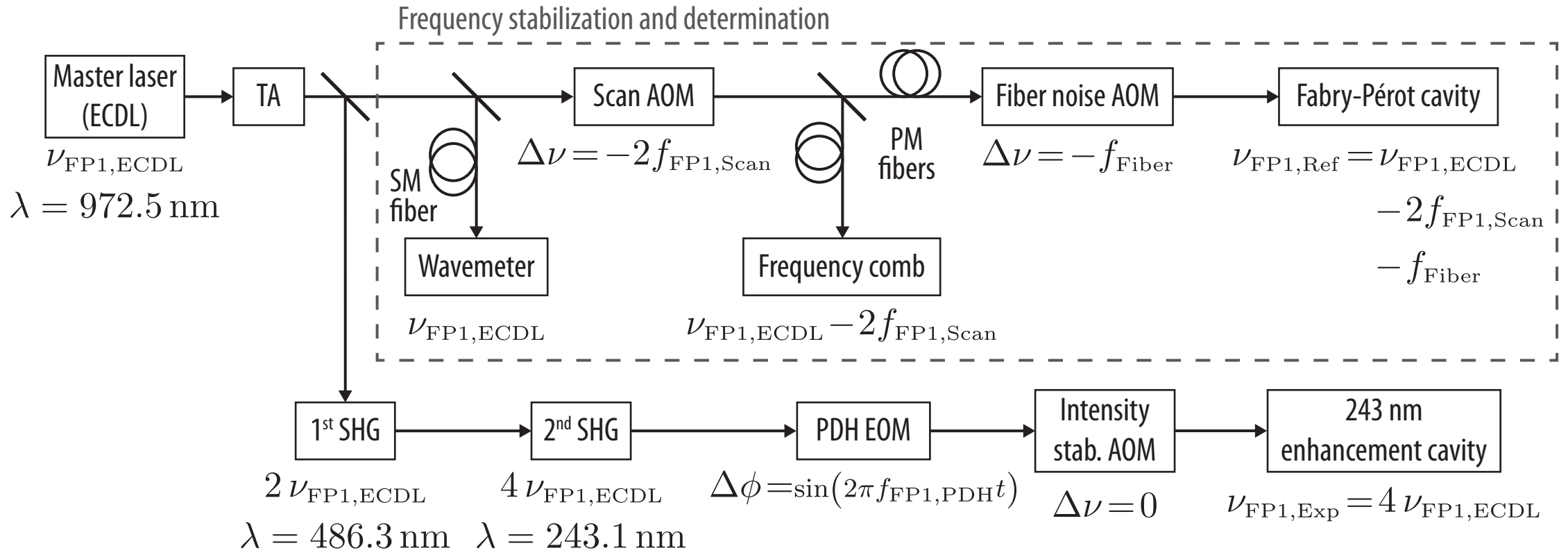

Figure 4.5: Sketch of relevant components of the 243 nm laser system ("FP1"), part of the 1S-2S preparation laser, with a focus on components that change the frequency, phase, or power of the laser light. See Section 4.3.1 for details. AOM: acousto-optic modulator, ECDL: external-cavity diode laser, EOM: electro-optic modulator, PDH: Pound-Drever-Hall technique, PM fiber: polarization-maintaining fiber, SHG: second-harmonic generation, SM fiber: single-mode fiber, TA: tapered amplifier.

## 4.3 1S-2S preparation laser

The 1S-2S preparation laser prepares the hydrogen atoms in the metastable 2S level by Doppler-free two-photon excitation from the 1S ground level. Specifically, as detailed in Section 2.2.6, the $1\mathrm{S}_{1/2}^{F=0} - 2\mathrm{S}_{1/2}^{F=0}$ transition, which has a transition frequency of $\nu_{\mathrm{1S\text{-}2S}} = 2.466\,\mathrm{PHz}$, is excited, with the atomic detuning $\Delta\nu_{\mathrm{1S\text{-}2S}}$ from the transition frequency measured relative to $\nu_{\mathrm{1S\text{-}2S}}$, as opposed to the half-as-large UV laser frequency. The generation of the required ultraviolet (UV) laser light at wavelength $\lambda_{\mathrm{1S\text{-}2S}} = 243.1\,\mathrm{nm}$ is described in Section 4.3.1. To enable the velocity-resolved detection employed in the experiment, this light is periodically blocked and unblocked by an optical chopper, discussed in Section 4.3.2. The UV laser power from the laser system is not sufficient to excite a large enough fraction of the atoms to the metastable 2S level, and thus the intensity seen by the atoms is further increased using an in-vacuum enhancement cavity. This cavity and its operation during the experiment are discussed in detail in Section 4.3.3.

### 4.3.1 243 nm laser system

Fig. 4.5 sketches the relevant components of the 243 nm laser system (known as "FP1" in the hydrogen laboratory), generating the laser light necessary to drive the 1S-2S transition. The following discussion mainly focuses on components that change the frequency, phase, or power of the laser light.

The laser system starts with a home-built master laser in the infrared at frequency $\nu_{\mathrm{FP1,ECDL}}$ ($\lambda \approx 972.5\,\mathrm{nm}$). The laser is an external-cavity (semiconductor) diode laser (ECDL, cavity length of 23 cm) in the Littrow configuration, described in detail in [82]. The 30 mW of power from this laser is amplified with a semiconductor tapered amplifier (TA), delivering up to 2.7 W when supplied with a current of 4 A.

Part of this infrared light (≈65 mW) is split off to measure and stabilize the laser frequency (dashed box in Fig. 4.5). To coarsely determine the master laser frequency, some light is sent to a wavemeter[1] with a frequency accuracy of 60 MHz through a single-mode (SM) optical fiber. The rest of the light is shifted in frequency by an acousto-optic modulator (scan AOM) operating in a double-pass configuration using the $-1^{\text{st}}$ diffraction order, i.e., the total frequency shift $\Delta\nu$ corresponds to twice the (radio) frequency sent to the AOM, $f_{\text{FP1,Scan}} \approx 435\,\text{MHz}$, giving $\Delta\nu = -2f_{\text{FP1,Scan}}$. This AOM is used to scan and set the frequency of the light interacting with the atoms in the apparatus. Some light, now of frequency $\nu_{\text{FP1,ECDL}} - 2f_{\text{FP1,Scan}}$, is split off and sent through a polarization-maintaining (PM) optical fiber to the frequency comb described in Section 4.8 to determine the laser frequency in Hz, and to compare the optical spectrum of the master laser to that of the 2S-6P spectroscopy laser.

The remaining light is sent through another PM fiber to an enclosure with passive acoustic and active vibration isolation. Here, another AOM (fiber noise AOM) operating in the $-1^{\text{st}}$ diffraction order shifts the light frequency by $\Delta\nu = -f_{\text{Fiber}} = 39\,337\,184\,\text{Hz}$. This AOM is part of a scheme to cancel the phase noise introduced by thermal expansion and acoustic noise in the PM fiber connecting the enclosure and the master laser [83]. The remaining light (approximately 6 µW) is then sent to a high-finesse Fabry-Pérot cavity (finesse of ≈400 000 and free spectral range of 1.933 GHz), consisting of two high-reflecting mirrors and a spacer made from ultra-low expansion glass (ULE) sitting inside a thermally stabilized vacuum enclosure, described in detail in [84]. The master laser frequency is stabilized to the resonance frequency of this reference cavity ("locked") using the Pound-Drever-Hall (PDH) technique [85] (modulation frequency of ≈ 20 MHz). To this end, an electro-optic modulator (EOM) inside the master laser cavity is used as an actuator, with the output of a fast (bandwidth ∼1 MHz) and a slow (bandwidth ∼100 kHz) feedback loop applied separately to the two electrodes of the EOM's crystal [82, 84]. An additional feedback loop (bandwidth ∼1 Hz) compensates drifts of the cavity length by moving the grating which forms the cavity end mirror.

The main part of the infrared light is converted to the UV and used in the experiment (lower branch in Fig. 4.5). To this end, the light is frequency doubled twice, using two resonant second-harmonic generation (SHG) ring cavities [86] in a four-mirror bow-tie configuration. These two SHG cavities and the TA described before are part of a commercial laser system[2]. The first SHG cavity uses a lithium triborate (LBO) crystal as nonlinear material, giving up to 1.2 W of power at $\lambda \approx 486.3\,\text{nm}$. The second SHG cavity uses a β-barium borate (BBO) crystal as nonlinear material, generating up to 90 mW of power at $\lambda_{\text{1S-2S}} \approx 243.1\,\text{nm}$, that is light with the frequency $\nu_{\text{FP1,Exp}} = 4\nu_{\text{FP1,ECDL}}$. Both SHG cavities use critical phase matching [87] and their optical length is stabilized using the PDH technique and piezo actuators. It was found that while the output power of the first SHG cavity is quite stable and only needs to be optimized once a day, the output power of the second SHG cavity is susceptible to both slow downward drifts and sudden jumps. The original output power can then usually be recovered by moving the BBO crystal such that a different part of it is used in the SHG process.

The intensity and phase of the generated UV light are then further modified before being coupled into the in-vacuum 243 nm enhancement cavity, where it interacts with the H atoms at frequency $\nu_{\text{FP1,Exp}}$. The various involved optical elements are shown in Fig. 4.6 and will be discussed in the next sections.

[1]HighFinesse WS7.

[2]Toptica TA-FHG pro.

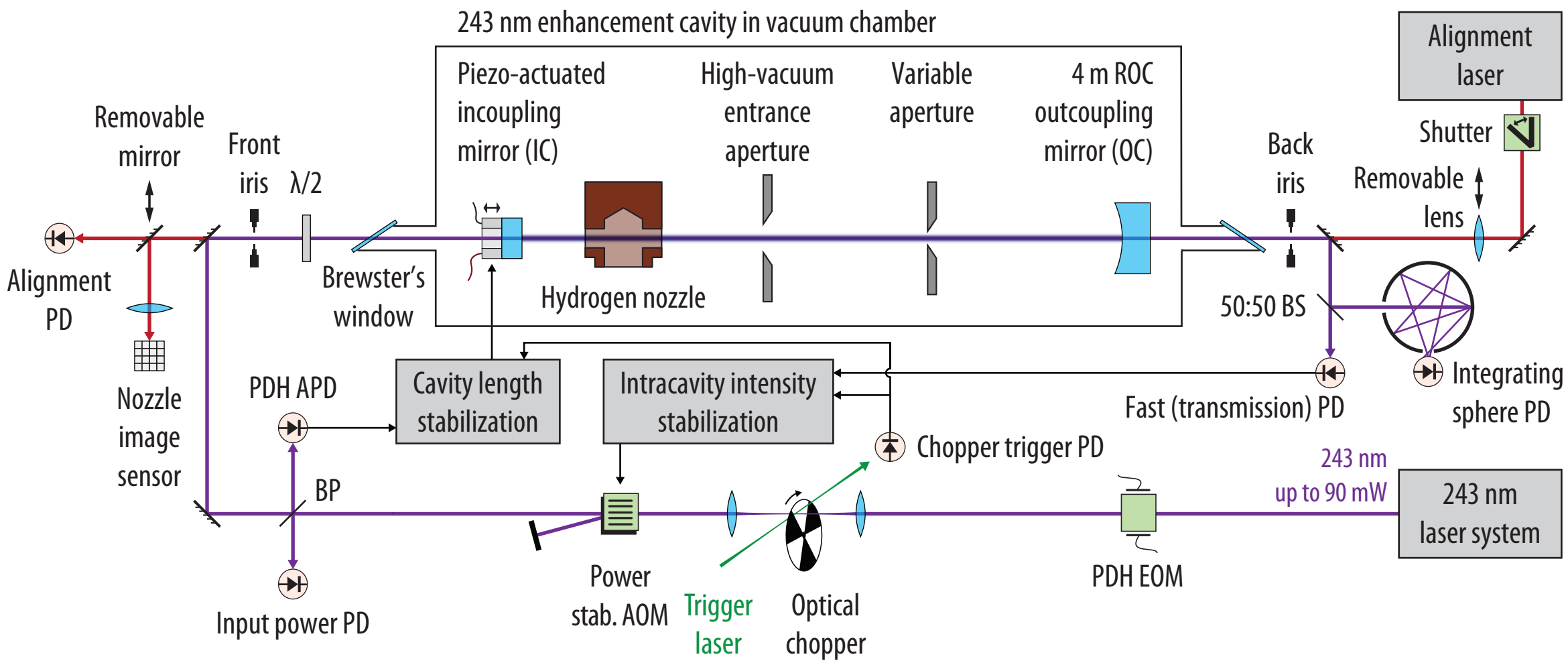

Figure 4.6: Optical layout of the 1S-2S preparation laser beam and the 243 nm enhancement cavity. See Section 4.3.3 for details. $\lambda/2$: half-wave plate, AOM: acousto-optic modulator, APD: avalanche photodiode detector, BP: beam sampler, BS: beamsplitter, EOM: electro-optic modulator, PD: photodetector, PDH: Pound-Drever-Hall technique, ROC: radius of curvature.

### 4.3.2 Optical chopper

Our measurement scheme relies on the velocity-resolved detection of metastable H, as discussed in Section 4.1. This detection is enabled by periodically switching off and on the excitation of H to the metastable level by intermittently blocking the 1S-2S preparation laser beam. To this end, the laser beam is focused through an optical chopper[1] with an equal-width slotted wheel. The chopper runs at 160 Hz, i.e., switching the transmission on or off every 3.125 ms, with the resulting phases here referred to as bright and dark phase, respectively. The rise and fall times are 3 µs, limited by the focus size and slot width. Since large parts of the experiments are triggered on this chopping of the beam, it is advantageous to generate an electric trigger using an auxiliary trigger laser[2] as shown in Fig. 4.6. The two laser beams cross at a slight angle in the plane of the chopper wheel, with the trigger laser detected on a photodetector. The overlap is checked routinely and the jitter in the switching time the two beams is typically found to be less than 3 µs. Possibly due to imperfections of the chopper wheel, there is also a variation of approximately $\pm 5$ µs in the duration for which the laser is switched on or off. From the photodetector signal, a TTL-compatible[3] chopper trigger is derived and distributed[4] to various devices.

[1] Stanford Research Systems SR540 (O540RCH head and O5402530 blade) with New Focus 3501 controller.

[2] Thorlabs CPS532-C2 (532 nm, 0.9 mW) or CPS405 (405 nm, 4.5 mW).

[3] TTL: transistor–transistor logic, which here is used synonymous with requiring a logic level for the true state of 5 V.

[4] The TTL trigger is distributed using a fan-out line driver (Pulse Research Labs PRL-414C-BNC) to drive the 50 Ω-terminated trigger inputs of some devices. Delayed triggers are generated with a pulse generator (Quantum Composers 9300) or various function generators (Rigol DG1022Z, DG1062Z), which were found to be very versatile.

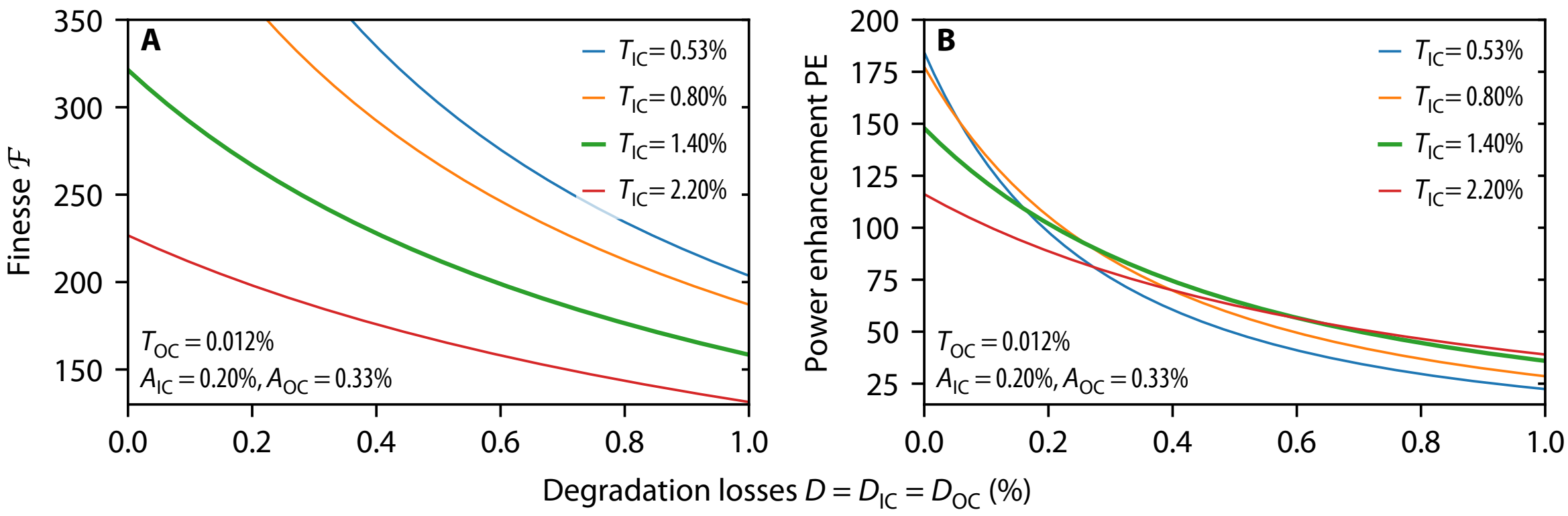

Figure 4.7: Model of the effect of mirror degradation $D$ on (**A**) finesse $\mathcal{F}$ and (**B**) power enhancement PE of the 243 nm enhancement cavity. $D$ is the fraction of light that is lost due to degradation on each reflection of the incoupling (IC) and outcoupling (OC) mirror, i.e., $D = D_{\mathrm{IC}} = D_{\mathrm{OC}}$. The OC transmittance is $T_{\mathrm{OC}} = 0.012\,\%$ and the absorption and scattering losses of the IC and OC are $A_{\mathrm{IC}} = 0.20\,\%$ and $A_{\mathrm{OC}} = 0.33\,\%$. Different values (colored lines) of the IC transmittance $T_{\mathrm{IC}}$ are shown, with $T_{\mathrm{OC}} = 1.40\,\%$ (green line) corresponding to the experimental situation.

### 4.3.3 In-vacuum 243 nm enhancement cavity

The UV laser power from the laser system is not sufficient to excite a large enough fraction of the hydrogen atoms to the metastable 2S level during the limited time available as the atoms travel from the nozzle to the spectroscopy region. To increase the amplitude of the electromagnetic field seen by the atoms, and thus the excitation fraction, an in-vacuum two-mirror enhancement cavity is used in which the UV light circulates resonantly [88]. However, the cavity mirrors degrade in the presence of strong UV light, limiting the time of operation of the experiment, and requiring some design compromises. In the followng, the design of the cavity (see Sections 4.3.3.1 and 4.3.3.2), the mirror degradation (see Section 4.3.3.3), the determination of intracavity power (see Section 4.3.3.4), the stabilization of the cavity length and intracavity power (see Section 4.3.3.5), and the auxiliary alignment laser (see Section 4.3.3.6) are discussed.

#### 4.3.3.1 Cavity layout, finesse and power enhancement

The cavity consists of a flat ($r_{\mathrm{IC}} = \infty$) incoupling mirror (IC) and a concave outcoupling mirror (OC) with a nominal radius of curvature (ROC) of $r_{\mathrm{OC}} = 4\,\mathrm{m}$, spaced apart by $L_{\mathrm{EC}} = 355\,\mathrm{mm}$ (see Fig. 4.6). This corresponds to a running-wave half-symmetric cavity [88] with a free spectral range of $\mathrm{FSR} = c/2L_{\mathrm{EC}} = 422\,\mathrm{MHz}$. The beam waist, which is located on the flat IC, has a $1/e^2$ intensity radius of $w_{\text{1S-2S}} = 297\,\mu\mathrm{m}$, with the beam radius only increasing to 311 µm on the OC, as the Rayleigh range $z_{\mathrm{r,1S-2S}} = 1.14\,\mathrm{m}$ is more than three times larger than $L_{\mathrm{EC}}$. The linear polarization of the laser field inside the cavity is oriented along the (horizontal) $x$-axis, i.e., coming out of the page in Fig. 4.6.

The resonance frequency $\nu_{q,n,m}$ of a cavity for an $nm$-th order Hermite-Gaussian mode is given by (see Eq. (23) of Chapter 19 of [88])

$$\nu_{q,n,m} = \left( q + (n+m+1)\frac{\arccos(\sqrt{g_1 g_2})}{\pi} \right) \mathrm{FSR}, \tag{4.1}$$

where $q$ is an integer and $g_1, g_2 > 0$. For the enhancement cavity, $g_1 = 1 - L_{\mathrm{EC}}/r_{\mathrm{IC}}$ and $g_2 = 1 - L_{\mathrm{EC}}/r_{\mathrm{OC}}$. Thus, the resonance frequency of higher-order modes is shifted from that of the fundamental (Gaussian) mode ($n = m = 0$) by the transverse mode spacing

$$\Delta\nu_{\mathrm{TMS}} = \frac{\arccos(\sqrt{g_1 g_2})}{\pi}\,\mathrm{FSR}. \tag{4.2}$$

For the enhancement cavity, $\Delta\nu_{\mathrm{TMS}} = 40.7\,\mathrm{MHz}$, giving $\mathrm{FSR}/\Delta\nu_{\mathrm{TMS}} \approx 10.4$.

The reflectances $R_{\mathrm{IC}}$ and $R_{\mathrm{OC}}$ of the IC and OC, respectively, can be written as

$$R_{\mathrm{IC}} = 1 - T_{\mathrm{IC}} - A_{\mathrm{IC}} - D_{\mathrm{IC}} = R_{\mathrm{IC},0} - D_{\mathrm{IC}}, \tag{4.3}$$

$$R_{\mathrm{OC}} = 1 - T_{\mathrm{OC}} - A_{\mathrm{OC}} - D_{\mathrm{OC}} = R_{\mathrm{OC},0} - D_{\mathrm{OC}}, \tag{4.4}$$

where $T_{\mathrm{IC}}$ and $T_{\mathrm{OC}}$ are the transmittances, and $A_{\mathrm{IC}}$ and $A_{\mathrm{OC}}$ are losses due to absorption and scattering. $D_{\mathrm{IC}}$ and $D_{\mathrm{OC}}$ are additional losses due to mirror degradation, i.e., they are assumed to be zero initially, when the mirrors have reflectances $R_{\mathrm{IC},0}$ and $R_{\mathrm{OC},0}$, but increase as the mirrors degrade.

The finesse $\mathcal{F}$ of the enhancement cavity is given by [88]

$$\mathcal{F} = \frac{\pi\,(R_{\mathrm{IC}}R_{\mathrm{OC}})^{1/4}}{1 - \sqrt{R_{\mathrm{IC}}R_{\mathrm{OC}}}}. \tag{4.5}$$

The finesse is a particularly useful quantity, because it is experimentally accessible through a measurement of the spectral width $\delta\nu$ of the cavity resonances, from which the finesse is given by $\mathcal{F} = \mathrm{FSR}/\delta\nu$.

The power enhancement PE is defined as the ratio of the optical power impinging on the IC, $P_{\mathrm{in}}$, to the optical power per direction circulating inside the cavity, $P_{\mathrm{circ}}$. Including the effect of mirror degradation, it is given by

$$\mathrm{PE} = \frac{P_{\mathrm{circ}}}{P_{\mathrm{in}}} = \eta_{\mathrm{mode}} \frac{(1 - D_{\mathrm{IC}})\,T_{\mathrm{IC}}}{\left(1 - \sqrt{R_{\mathrm{IC}}R_{\mathrm{OC}}}\right)^2}, \tag{4.6}$$

where $\eta_{\mathrm{mode}}$ is the spatial overlap or mode matching between the free space mode and the resonant cavity mode. The factor $(1 - D_{\mathrm{IC}})$ in the numerator accounts for the fact that mirror degradation also lowers the power transmitted into the cavity. For given values of all losses besides the IC transmittance, $L_{\mathrm{total}} = T_{\mathrm{OC}} + A_{\mathrm{IC}} + A_{\mathrm{OC}} + D_{\mathrm{IC}} + D_{\mathrm{OC}}$, the power enhancement is maximal for the impedance-matched case where $T_{\mathrm{IC}} = L_{\mathrm{total}}$.

In the experiment, the cavity is kept on resonance with the fundamental Gaussian mode by appropriately stabilizing ("locking") the cavity length to the preparation laser. Beam shaping optics are used to optimize the overlap of the incoming laser beam with this mode, resulting on average in a mode matching of $\eta_{\mathrm{mode}} \approx 85\,\%$. $\eta_{\mathrm{mode}}$ is here determined by measuring the cavity transmission for the fundamental mode and for the visible higher-order modes.

The finesse $\mathcal{F}$ and power enhancement PE as a function of the degradation losses $D = D_{\mathrm{IC}} = D_{\mathrm{OC}}$ for different values of $T_{\mathrm{IC}}$ are shown in Fig. 4.7, assuming perfect mode matching, i.e., $\eta_{\mathrm{mode}} = 1$. The OC transmittance $T_{\mathrm{OC}}$, and the absorption and scattering losses of the IC and OC, given by $A_{\mathrm{IC}}$ and $A_{\mathrm{OC}}$, respectively, are set to the values determined for the mirrors used here (see Section 4.3.3.2). Note that instead assuming $D_{\mathrm{IC}} = 2D, D_{\mathrm{OC}} = 0$ or vice versa leaves the values of $\mathcal{F}$ unchanged and only changes PE by a few percent. For the case with no degradation, the impedance-matched case ($T_{\mathrm{IC}} = 0.53\,\%$, blue line) shows a power enhancement of 184. However, as the degradation losses increase, the originally impedance-matched case turns into an undercoupled case ($T_{IC} < L_{total}$) and the power enhancement quickly decreases. If the cavity however is initially overcoupled ($T_{IC} > L_{total}$), the maximal power enhancement is lower initially, but can stay over a certain threshold for a wider range of degradation losses. This is the regime used here, with $T_{IC}$ chosen to be 1.40 % (green line), for which an power enhancement of up to 148 is expected, which reduces to 126 when taking the typical experimental mode matching into account.

#### 4.3.3.2 Cavity incoupling and outcoupling mirrors

The IC[1] has a multi-layer dielectric coating with a specified reflectance of 98.3 % and transmittance of 1.4 % at 0° angle of incidence (AOI), as determined by the manufacturer[2]. The coating consists of alternating layers of hafnium dioxide ($HfO_2$, $n \approx 2.1$ [89]) and silica ($SiO_2$, $n \approx 1.51$ [90]), deposited to the substrate using ion beam sputtering (IBS), with a calculated absorption of 0.18 %. To avoid etalons between the two mirror surfaces, the backside of the IC both has an anti-reflection coating (reflectance $< 2 \times 10^{-4}$) and is wedged by 30′. While it is straightforward to confirm the transmittance measurement in the laboratory, giving the same value of $T_{IC} = 1.4\,\%$ as determined by the manufacturer, a reliable reflectance measurement is more challenging. To this end, an OC[3] with the identical coating applied in the same run as the IC was used. The finesse $\mathcal{F}$ of a cavity consisting of the IC and this special OC was found to be 195, corresponding to total losses of 3.2 % per round trip. Using the measured transmittance, the remaining losses due to absorption and scattering are then deduced to be $A_{IC} = 0.20\,\%$ per mirror. This value closely matches the calculated coating absorption, which thus seems to be the limiting factor, with scattering losses playing a negligible role. From these measurements, the non-degraded reflectance of the IC is thus found to be $R_{IC,0} = 98.4\,\%$.

The regularly used OC[4] has a multi-layer dielectric high-reflectivity coating with a specified reflectance of > 99.9 %. The coating materials and deposition technique are the same as for the IC and a similar reflectance bandwidth is achieved. Unfortunately, the backside is neither anti-reflection coated nor wedged, giving rise to etalon effects on the transmitted light on the order of 4 % under 0° AOI, which is the configuration used in the cavity. The transmittance of the OC was measured to be $T_{OC} = 1.2 \times 10^{-4}$ at 8° AOI, using a photodetector at different, calibrated gain settings. The reflectance again is deduced from the finesse of the cavity, now using the OC with the IC described above. A maximum finesse of $\mathcal{F} = 320$ is observed (see Fig. 4.8) in this configuration, corresponding to total losses of 1.94 % per round trip. Using the measured transmittance, and absorption and scattering losses of the IC, the total losses of the OC are found to be 0.34 %. Subtracting the measured transmittance, the losses due to absorption and scattering are then $A_{OC} = 0.33\,\%$. While no calculated coating absorption is supplied by the manufacturer, it is reasonable to assume that theses losses are also predominantly caused by coating absorption, since the coating and substrate polishing are very similar to the IC. Furthermore, a slightly higher coating absorption is expected from the OC because of its higher reflectance and thus the need for more coating layers. Finally, the non-degraded reflectance of the OC is then $R_{\mathrm{OC},0} = 99.7\,\%$, below the specified value.

[1]Substrate and custom coating from LayerTec (purchased 2016, article number 132659, coating run numbers 131914 (frontside) and 131916 (backside)). Substrate material is excimer-grade fused silica, dimensions are 7.75 mm diameter, 4 mm thickness, and 6 mm diameter clear aperture. Substrate has been superpolished to a root-mean-square surface roughness below 0.15 nm.

[2]The determination adjusts process parameters to match a theoretical model of the coating to a broad-band transmission measurement (230 nm. . . 1100 nm). The reflectance and transmittance at the wavelength of interest are then taken from the adjusted model.

[3]LayerTec (article number 132666), same substrate material and polishing as IC, dimensions are 12.7 mm diameter and 6.35 mm thickness.

[4]Substrate and custom coating from Research Electro-Optics (purchased 2002, lot number 602-0617-02, coating run number C2512). Substrate material is fused silica, dimensions are 12.7 mm diameter and 10 mm thickness. Substrate has been superpolished to a root-mean-square surface roughness below 0.1 nm.

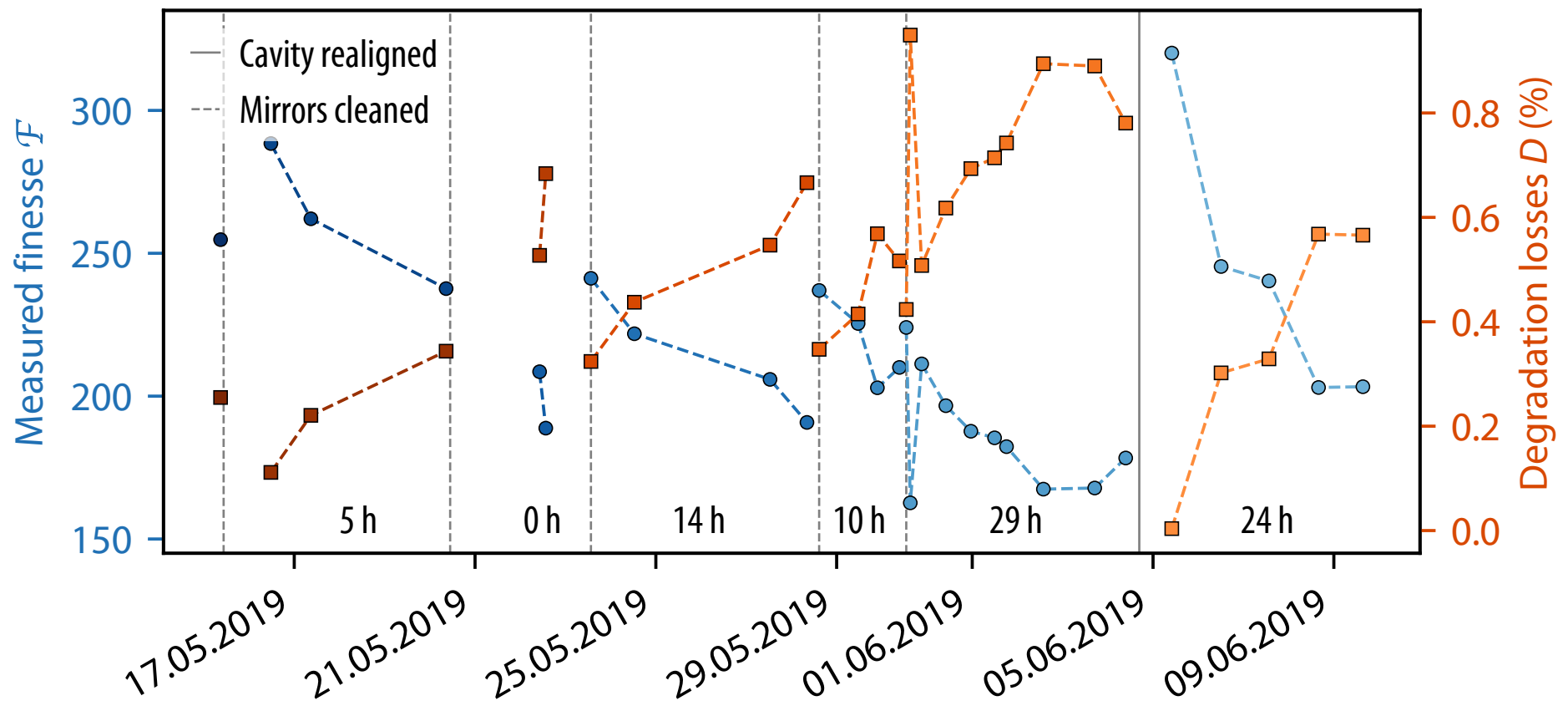

Figure 4.8: Measured finesse (blue circles) and deduced degradation losses $D$ (red squares) of the 243 nm enhancement cavity over time. $D$, the fraction of light that is lost due to degradation on each reflection of the incoupling (IC) and outcoupling (OC) mirror, is determined by assuming that any decrease in finesse is caused by an increase of $D$. The gray vertical lines mark times when the cavity was realigned (solid line), leading to the light inside the cavity hitting the mirrors at different spots, and when the mirrors were cleaned (dashed lines). Not marked is a previous cavity realignment on 18.04.2019, after which the cavity was not operated until the start of the data shown here. The runtime of the cavity, i.e., the time it was resonant with the preparation laser, during the time span from the first to the last finesse measurement within each section, as marked by the vertical lines, is given at the bottom of the plot.

The IC is specified with a irregularity and spherical bending power of less than[1] 55 nm over its clear aperture, corresponding to an ROC of less than $\pm 80$ m. The tolerance of the ROC of the OC is not documented, but similar mirrors from the same manufacturer have an specified tolerance of 1 %. Fortunately, the waist radius is not very sensitive to changes in the ROC of the mirrors. Varying $r_{\mathrm{IC}}$ by its tolerance and $r_{\mathrm{OC}}$ within 5 %, the cavity waist radius varies only between $w_{\text{1S-2S}} = 290\,\mu\text{m} \ldots 304\,\mu\text{m}$. This range does not limit the accuracy of the intensity determination (see Section 4.3.3.4) nor significantly influences the result of the atomic beam simulations, and the nominal value of $w_{\text{1S-2S}} = 297\,\mu\text{m}$ is used, with the beam waist placed at the position of the IC.

A measurement of the transverse mode spacing $\Delta\nu_{\mathrm{TMS}}$ can also be used to determine the ROC of either of the mirrors if the ROC of one mirror, or the ratio of the ROCs, is known. It was however found that the piezo actuator (see Section 4.3.3.5) on which the IC is glued deforms the IC, leading to shift in $\Delta\nu_{\mathrm{TMS}}$ as the voltage applied to the piezo actuator is varied[1]. Assuming $r_{\mathrm{OC}} = 4\,\mathrm{m}$, the deformation corresponds to an $r_{\mathrm{IC}}$, for the IC used during the measurement, between $-1000\,\mathrm{m} \ldots -50\,\mathrm{m}$, i.e., the IC becomes convex as seen from inside the cavity.

[1]This corresponds to $\lambda/10$ irregularity and spherical bending power at the measurement wavelength of $\lambda = 546$ nm.

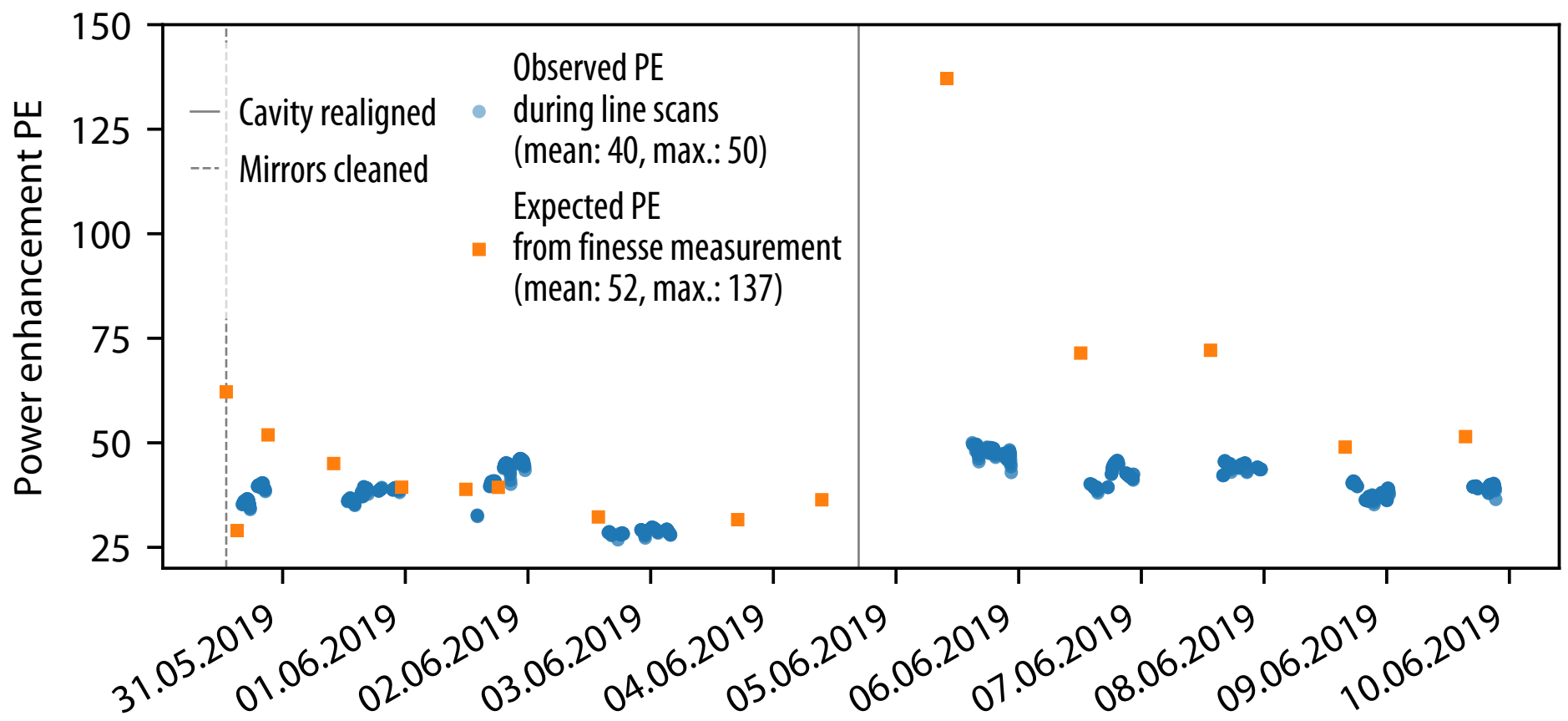

Figure 4.9: Observed (blue circles) and expected (orange squares) power enhancement PE of the 243 nm enhancement cavity over time. The observed PE is determined from measurements of the impinging power $P_{\mathrm{in}}$, the circulating power $P_{\mathrm{circ}}$, and the mode matching $\eta_{\mathrm{mode}}$, with only the value of PE during line scans shown. The expected PE is deduced from a subset of the finesse measurements shown in Fig. 4.8 and using Eq. (4.6). The gray vertical lines mark times when the cavity was realigned (solid line), leading to the light inside the cavity hitting the mirrors at different spots, and when the mirrors were cleaned (dashed lines).

#### 4.3.3.3 Mirror degradation and power enhancement

The degradation of mirrors illuminated with intense UV light has been observed in various experiment under different conditions and for different types of mirror coatings [91–95]. Several, possibly related degradation mechanisms have been identified: first, the contamination of the mirror surfaces through the UV-assisted photodeposition of hydrocarbons present in the background gas [91, 96]. Second, the volume degradation through defect formation inside the coatings and the mirror substrate itself [91]. Third, the depletion of oxygen from the top coating layer for certain coating materials [97].

The degradation of the mirrors of the 243 nm enhancement cavity severely limits the time the cavity can be operated before the power enhancement is reduced below an acceptable level. The degradation is clearly visible as a decrease in finesse over time, as shown in Fig. 4.8 for the time period of run B of the 2S-6P measurement (see Table 6.1). The degradation can be slowed somewhat by intermediately cleaning the cavity mirrors[2] (dashed lines in Fig. 4.8), which however requires breaking the vacuum and risks a misalignment of the cavity. At some point, the power enhancement cannot be restored through cleaning and the mirrors either need to be exchanged or, alternatively, the cavity needs to be realigned such that the different spots on the mirrors are illuminated by the cavity mode. For the data shown in Fig. 4.8, the runtime between such two realignments was 60 h, during which the mirrors were cleaned four times. Within this time, the finesse decreases from ≈300 to ≈180. To prolong the life time of the mirrors and thus increase the time of operation of the experiment as much as possible, the cavity is only locked during the measurement itself, with all other alignments done while the length of the cavity is scanned. This also ensures that the amounts of residual trace gases other than hydrogen are reduced as much as possible during operation, since both the cryopump and the heat shield of the cryostat should efficiently remove traces gases and especially hydrocarbons from the background gas.

[1] One might assume that when no piezo voltage is applied, the IC is not deformed, leading to $r_{\mathrm{OC}} = 4.03(17)\,\mathrm{m}$ when taking into account the IC ROC tolerance. However, the IC is glued while the piezo actuator, mirror, and piezo holder are heated up to $\approx 50\,^\circ\mathrm{C}$ and thus might be permanently deformed at room temperature.

[2] Either methanol or acetone were used as solvents for cleaning, with no notable difference found in the cleaning efficiency.

The degraded mirrors showed localized spots, especially visible when breathing upon the mirror, at the position of the cavity mode. These spots could not be removed with common organic solvents. Interestingly, both when cleaning and replacing the mirrors, it was found that the incoupling mirror (IC) was substantially more affected by degradation than the outcoupling mirror (OC), to the point where it was sufficient to only clean the OC during the whole 2S-6P measurement, while it was necessary to realign the beam position on the IC multiple times. Note that the reduction in finesse and in power enhancement is expected to be not or only weakly dependent, respectively, on which mirror degrades. All occasions during the 2S-6P measurement on which the mirrors were cleaned, the cavity was realigned, or the IC was replaced are marked in Fig. 4.10. A similar behavior was also observed with a previous batch of ICs from the same manufacturer as the current OC and most likely coated using electron beam evaporation instead of the IBS used for the current batch of ICs. A notable difference between the IC and OC is the distance to the nozzle, and thus the flux of atomic hydrogen, with the IC sittings six times closer to it than the OC. However, the IC was also observed to be more affected by degradation[1] in the experiment of [95] at the same wavelength and with a similar cavity configuration, even without the presence of hydrogen. This hints at volume degradation playing a role, since for the IC the deeper coating layers and the substrate are exposed to the intensity of the impinging light. For the OC, on the other hand, these parts are only exposed to the transmitted light, whose intensity is smaller by more than two orders of magnitude. Furthermore, mirror degradation was also found to be an issue in the study of antihydrogen [93]. There, an enhancement cavity very similar to the one discussed here was used, but placed inside a cryostat and thus in an environment with a much lower background pressure, which should drastically reduce surface contamination. They also tested their cavity within a high vacuum environment such as the one used here and found a much faster rate of degradation. The oxygen depletion of the top layer found in [97] was observed to be preventable by using $SiO_2$ as material for the top layer, which is the case for the IC, while the material of the top layer of the OC is unknown. However, both mirrors used in [95], where as mentioned a similar degradation as found here was observed, had a $SiO_2$ top layer. All in all, it is likely that multiple degradation mechanism are at play, with different mechanism dominating for different intensities and vacuum contaminations.

In [91] it was found that degraded mirrors could be partially restored using a combination of oxygen plasma treatment to clean the surfaces, and thermal annealing to heal defects in the bulk. The experiments of [92, 95] instead placed the mirrors in an oxygen atmosphere, which was found to prevent degradation at sufficient pressures on the order of 1 mbar. Inspired by this, it was here attempted to undo the degradation by admitting ≈10 mbar of oxygen into the vacuum chamber and locking the cavity. This indeed was found to increase the circulating power on the timescale of minutes. Interestingly, the same behavior was observed when using nitrogen instead of oxygen. However, when then locking the cavity again during the measurement, the power enhancement and thus finesse quickly dropped to the level seen before the cleaning procedure. Interestingly, a similar procedure was successfully used to undo some degradation in [93]. After the completion of the 2S-6P measurement, work was started to implement an oxygen atmosphere for the mirrors of the 243 nm enhancement cavity. This however requires multiple differential pumping stages to keep the partial pressure of oxygen at a sufficiently low level in the spectroscopy region.

[1] D. C. Yost, private communication.

The observed reduction in finesse $\mathcal{F}$ can be translated into degradation losses $D$ as detailed in Section 4.3.3.1. The resulting values, assuming that each of the mirrors has additional losses $D$, are shown in Fig. 4.8 and are within $0\,\%\ldots 1\,\%$, with the latter value too large to continue the experiment. An almost equivalent graph is obtained when assuming losses $2D$ on either the IC or OC. Using these values and Eq. (4.6), the expected power enhancement can be calculated, which is shown together with the observed power enhancement in Fig. 4.9 for a subset of the measurements days of measurement run B. The observed and expected PE are in reasonable agreement for the first half of the shown data, while for the second half, after a cavity realignment, the expected PE is larger than the observed PE. Especially the large PE of 137 deduced from a finesse measurement right after the realignment could not be observed during the measurement, where the maximum achieved[1] PE is 50. In fact, a rapid degradation within tens of seconds was commonly observed upon locking the cavity for initially very high finesse configurations, with the further degradation proceeding at a much slower pace as shown in Fig. 4.8.

#### 4.3.3.4 Determination of intracavity power

The power $P_{\text{1S-2S}} = P_{\text{circ}}$ circulating inside the enhancement cavity is deduced by measuring the power $P_{\text{tr}}$ of the light transmitted through the cavity's outcoupling mirror (OC), which is on the order of 100 µW. Here, $P_{\text{1S-2S}}$ is defined as the power per direction inside the cavity during the bright phase of the optical chopper, i.e., the atoms see two counter-propagating beams with a power of $P_{\text{1S-2S}}$ each while the preparation laser is unblocked. As the power here is referenced to the bright phase, the circulating power averaged over many chopper cycles is $\eta_{\text{circ}} P_{\text{1S-2S}}$, where $\eta_{\text{circ}}$ is the fraction of time the light is admitted to the cavity. For the equal slit width chopper used here, $\eta_{\text{circ}}$ is nominally 1/2. A 50:50 non-polarizing beamsplitter[2] allows the transmission to be simultaneously monitored on a high-gain, low-bandwidth photodetector[3] attached to an integrating sphere[4], and a low-gain, high-bandwidth photodector[5]. The output of both photodetectors is further amplified and filtered with low-noise voltage preamplifiers[6], and in total the detection bandwidths are 5 kHz and 1 MHz for the integrating sphere and fast photodetectors, respectively. The beam is focused on the fast photodetector, but left collimated for the integrating sphere. Both photodetectors are shielded from stray light using blackout fabric and beam tubes, and electronic offsets were subtracted at the preamplifier. Using an integrating sphere reduces the effect of beam pointing fluctuations and drifts on the detector signal, as the attached photodetector is illuminated from many directions and at many positions. However, this reduces the power seen by the detector, necessitating a large gain and thus reduced bandwidth. Therefore, the higher-bandwidth detector is used to deduce the finesse of the cavity while scanning its length, and to remove intensity noise, as discussed in the Section 4.3.3.5. Of course, the latter application will in principle then add any beam pointing dependencies of the fast detector to the intensity noise. However, these beam pointing effects are seen on a timescale longer than a typical frequency scan.

[1] [95] reported a PE of 80 with similar cavity mirrors and at the same wavelength, but with the mirrors placed in an oxygen atmosphere as mentioned above.

[2] Thorlabs BSW20.

[3] Home-built with 5 kHz bandwidth.

[4] LabSphere LBS-3P-GPS with a diameter of 5.3 in. The reflectance of the inside of the sphere is specified as 97 % at a wavelength of 250 nm and an angle of incidence of 8°.

[5] Thorlabs PDA25K2, using a gallium phosphide (GaP)-based photodiode with responsivity of 0.024 A/W at 243 nm. It has a built-in, gain-adjustable transimpedance amplifier, used at the 20 dB gain setting offering 15 kV/A gain and 1 MHz bandwidth.

[6] Stanford Research Systems SR560. For the integrating sphere and fast photodetectors, voltage gains of 200 and 50 (before 28.05.2019) or 20 (from 28.05.2019 on) and low-pass filters (6 dB/decade) with corner frequencies 30 kHz and 1 MHz are used, respectively.

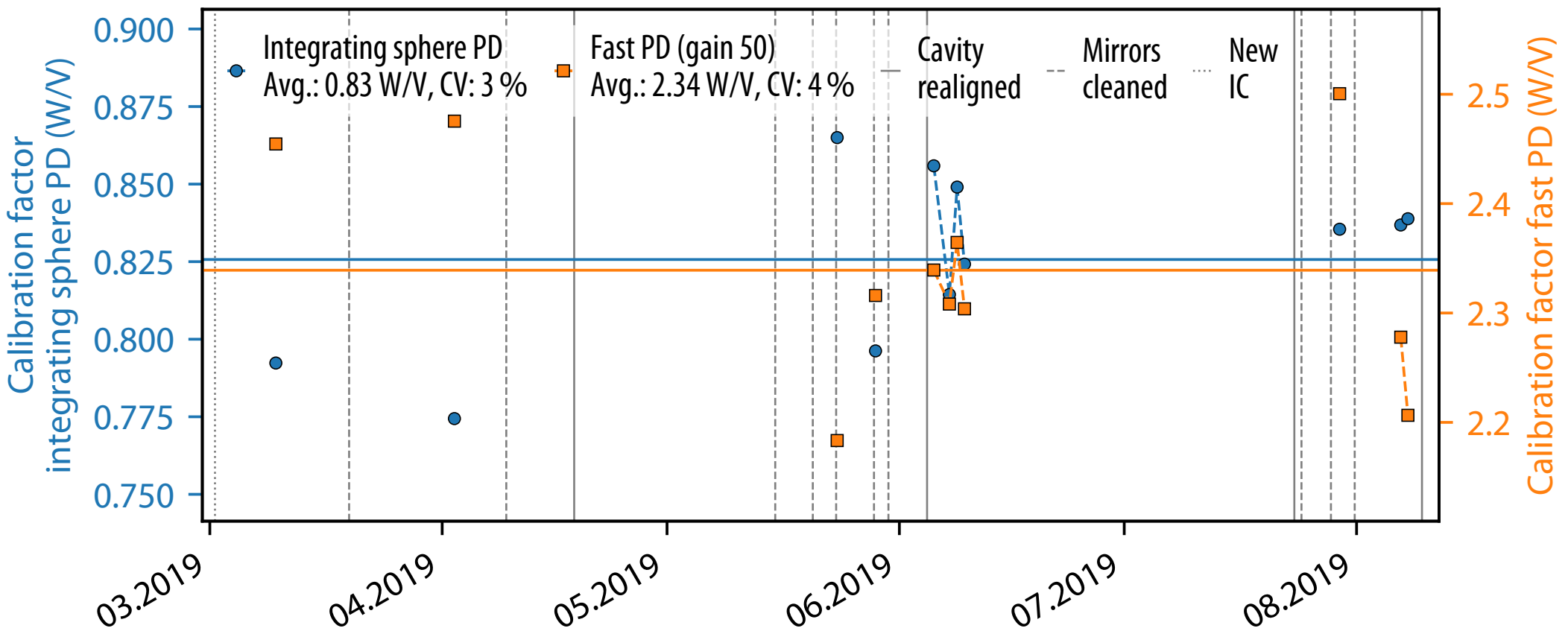

Figure 4.10: The calibration factors for the integrating sphere photodetector (blue points, left scale) and the fast photodetector (orange points, right scale). Both detectors monitoring the light transmitted through the 243 nm enhancement cavity, from which the intracavity power $P_{1S\text{-}2S}$ is deduced. The corresponding solid horizontal lines show the average values. As in Fig. 4.8, the gray vertical lines mark times when the cavity was realigned (solid line), the cavity mirrors were cleaned (dashed lines), and a new IC was installed (dotted line). The calibration factor for the fast photodetector is shown for gain of 50 of its preamplifier by scaling the measurements taken at a gain of 20 accordingly. CV: coefficient of variation, IC: incoupling mirror.

To convert the voltage readings of the photodetectors to optical power, the transmitted power is measured with a calibrated photodiode power meter[1] while the cavity is length-stabilized to the laser. The reading of the power meter, $\tilde{P}_{\mathrm{tr,PM}}$, is the transmitted power averaged over many seconds, and the average circulating power when the laser is not blocked by the chopper is given by $P_{\mathrm{tr,PM}} = \tilde{P}_{\mathrm{tr,PM}}/\eta_{\mathrm{circ}}$. The power meter is placed directly before the beamsplitter (see Fig. 4.6), and thus the light transmitted through the OC additionally passes through a Brewster's window and is reflected on a steering mirror before reaching the power meter. This additional path leads to a reduction in power to 92 % of its level at the position of the OC and thus the transmitted power is $P_{\mathrm{tr}} = P_{\mathrm{tr,PM}}/0.92$. Using the measured transmittance $T_{\mathrm{OC}}$ of the OC, the power circulating inside the cavity is then given by $P_{1S\text{-}2S} = P_{\mathrm{tr,PM}}/(0.92 \times T_{\mathrm{OC}}) = P_{\mathrm{tr,PM}}/1.1 \times 10^{-4}$. Finally, $P_{\mathrm{tr,PM}}$ is compared to the voltage readings of the two photodetectors directly before and after placing the power meter before the beamsplitter, from which two calibration factors corresponding to the ratio of the intracavity power $P_{\text{1S-2S}}$ to the voltage readings are derived. The voltage readings are derived from the time-resolved signal of the photodetectors as detailed in Section 4.3.3.5. The calibration factors were measured whenever the enhancement cavity was re-aligned and at additional intermediate times, with a total of 11 calibration factors for each photodetector available during the precision measurement time period. These calibration factors are shown in Fig. 4.10, with the calibration factor for the fast photodetector shown for a gain of 50 of its preamplifier by scaling the measurements taken at a gain of 20 accordingly. For the integrating sphere photodetector, the average calibration factor is found to be 0.83 W/V with a coefficient of variation (CV) of 3 % and a maximum deviation from the mean of 6 %. For the fast photodetector, the average calibration factor is found to be 2.34 W/V with a CV of 4 % and a maximum deviation from the mean of 7 %. Interestingly, the reproducibility of the fast photodetector is thus comparable to that of the integrating sphere photodetector.

[1]Newport 918D-UV-OD3R, based on a UV-enhanced silicon photodiode. The calibration uncertainty is specified as ±2 % at $\lambda = 243$ nm.

An intracavity power $P_{\text{1S-2S}}$ is found for each line scan in the following way: the appropriate calibration factor is retrieved, either using the value measured on the same or an earlier day. Then, the average voltage reading of the integrating sphere photodetector (see Fig. 4.11 and Section 4.3.3.5) for the data points within the line scan is multiplied with the calibration factor, resulting in a value of $P_{\text{1S-2S}}$ for each data point. Finally, the average value and the standard deviation of $P_{\text{1S-2S}}$ over these data points is evaluated and assigned to the corresponding line scan. Fig. 4.12 shows the resulting values of $P_{\text{1S-2S}}$ for the 2S-6P measurement discussed in this work.

#### 4.3.3.5 Stabilization of cavity length and intracavity power

The length of the 243 nm enhancement cavity is stabilized such that it is resonant with the light of the preparation laser using a servo loop. The length of the cavity can be varied through the voltage applied to a circular piezo actuator[1] to which the incoupling mirror is glued. A correction or error signal is generated using the Pound-Drever-Hall (PDH) technique [85, 98] and fed back to the piezo actuator using a proportional–integral (PI) controller[2]. The capture range of this error signal and thus of the servo loop, i.e., the maximum instantaneous change in length for which the error signal can still be used to return the cavity to resonance, is critical for the stabilization. This is because during the 3 ms-long dark phase where the chopper blocks the preparation laser, no error signal is available and no feedback can be applied to the piezo actuator. Thus, any length change during this time larger than the capture range will disturb the stabilization, with large disturbances especially likely when the cryopump is running. The capture range of the PDH error signal is given by $f_{\text{PDH}}\lambda_{\text{1S-2S}}/\text{FSR}$, where $f_{\text{PDH}}$ is the frequency of the sinusoidal phase modulation on the laser light necessary in the PDH technique. Here, $f_{\text{PDH}} \approx 47.3\,\text{MHz}$ is used, corresponding to a capture range of $\approx$27 nm, for which the cavity length can be reliably stabilized even when the cryopump is running[3].

The phase modulation is applied to the laser light using an EOM[4] (PDH EOM in Fig. 4.6). The nonlinear material of the EOM is a BBO crystal, since this material has a high damage threshold in the UV region [99]. The crystal is anti-reflection coated and initially (in 2014) showed a transmission of 95 %, which had decreased to 85 % by 2019. Unfortunately, some unwanted static birefringence[1] was observed, thought to originate from strain induced in the crystal by its mount[2] or an improperly cut crystal. This birefringence can spoil the linear polarization of the UV light. It is possible to minimize this effect by adjusting the polarization direction. However, the direction found this way must not coincide with the modulation axis of the crystal, leading to residual amplitude modulation in combination with polarizing elements, such as the Brewster's windows. An intermediate adjustment aiming to minimize both the residual circular polarization and amplitude modulation is used. Since only linearly polarized light can drive the $1S_{1/2}^{F=0} - 2S_{1/2}^{F=0}$ transition, the presence of circular polarization lowers the effective excitation rate.

[1]PI Ceramic PD080.31, 8.0 mm outer diameter, 4.5 mm inner diameter, and 2 µm range. Agilent Torr Seal is used to glue the mirror to the actuator and the actuator to its copper mount.

[2]Vescent Photonics D2-125.

[3]This was not the case for the previously used modulation frequency of 20 MHz in combination with the previously used second cavity spacer as described in Section 4.2.5.

[4]Qubig EO-47B3-UV, with an integrated resonant circuit with a specified quality factor of 125 at 47.2 MHz. Note that the exact resonance frequency depends on the amount of RF power sent to the EOM.

Before the laser light enters the cavity, a small part is split off, using an AR-coated plane window as beam sampler, and sent to the input power photodetector (see Fig. 4.6), whose signal is thus proportional to $P_{\text{in}}$. Likewise, the counter-propagating reflection off of the cavity is sent to an avalanche photodiode detector[3], and the PDH error signal is derived from its signal. The length stabilization is switched off during the dark phase to prevent noise and offsets on the error signal from affecting the servo loop, which otherwise could move the cavity length outside the loop's capture range. Specifically, the stabilization is switched on 10 µs after the start of the bright phase, as heralded by the trigger signal derived from the optical chopper (see Section 4.3.2), for a duration of 2995 µs, and thus is typically switched off ≈20 µs before the end of the bright phase. This safety margin ensures that the stabilization is always switched off before the light itself is blocked. It however also effectively increases the delay time $\tau$, nominally measured from the start of the dark phase as determined by the chopper trigger, by ≈20 µs.

The performance of the cavity length stabilization during a typical freezing cycle as seen on the various photodetector signals is shown in Fig. 4.11 (A), with Fig. 4.11 (B) and (C) showing the start and end of the bright phase in detail. The optical chopper unblocks the preparation laser at $\tau \approx -3130$ µs, as heralded by the chopper trigger (green line and shading), and seen as a step-like increase on the input power photodetector (red line and shading). The cavity transmission as measured by the fast photodetector (purple line and shading) is initially at a low value, increasing to its steady-state value after the length stabilization is switched on (dashed vertical lines). The stabilization typically settles within ≈150 µs after the start of the bright phase, with the signal approximately rising with a time constant of 50 µs, corresponding to a stabilization bandwidth of ≈7 kHz, after an initial drift within the capture range. The integrating sphere photodetector (blue line and shading) sees a slower rise time of 90 µs, limited by its detection bandwidth. At the end of the bright phase, the transmission quickly decreases as the length stabilization is switched off, shortly before the laser itself is blocked. Note that the bright phase here lasts ≈3130 µs instead of the expected 3125 µs, probably caused by imperfections of the optical chopper wheel. This is also the reason why here the length stabilization is switched off ≈25 µs instead of ≈20 µs before the end of the bright phase. To determine the delay-time-averaged impinging and intracavity power, the corresponding signals are averaged over a window between 500 µs. . . 3000 µs after the start of the bright phase (dotted lines in Fig. 4.11) (see also Section 4.7.3). As discussed in the Section 4.3.3.4, the intracavity power $P_{1S\text{-}2S}$ is found from the signal of the integrating sphere photodetector, with the start time of the averaging window matched to the rise time of this photodetector.

[1]BBO is an uniaxial crystal, with the dominant electro-optic coefficient being $r_{11} = 2.5$ pm/V. Thus, for phase modulation, the light is sent along the optical axis and a sinusoidal electric field is applied transverse to the optical axis. There should then be no static birefringence in the ideal case.

[2]The birefringence increases as the RF power sent to the EOM is increased. Judging from the time scale of the increase, the effect seems to be of thermal origin.

[3]Thorlabs APD430A2/M.

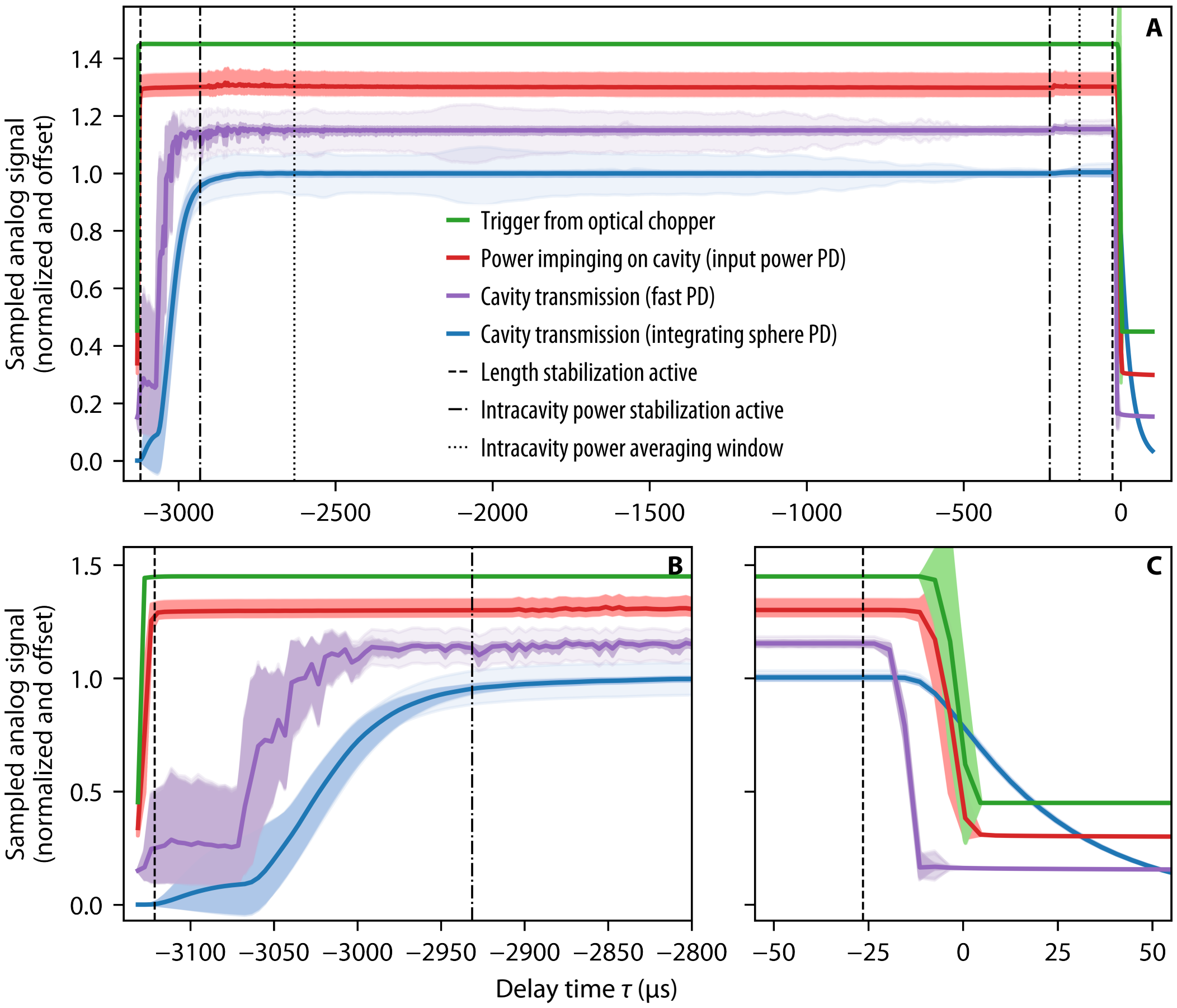

Figure 4.11: Performance of the 243 nm enhancement cavity length and intracavity power stabilization. Shown are delay-time-resolved sampled analog signals, normalized and offset for clarity, of 1920 data points, belonging to 64 line scans during a single freezing cycle (FC B35). The signal of each data point has been averaged over 160 chopper cycles, with the standard deviation over these cycles shown as superimposed shaded regions. The solid lines mark the average over all data points and chopper cycles. (**A**) The signals during the bright phase of the chopper cycle. Shown are the trigger from the optical chopper (green), the power impinging on the cavity as measured by the input power photodetector (red), and the cavity transmission measured with the fast (purple) and integrating sphere (blue) photodetectors. The faint shaded region with a large standard deviation visible on the transmission signals is caused by a single data point. The vertical lines window the times during which the length (dashed lines) and intracavity power (dash-dotted lines) stabilizations are active, and the portion of the transmission signals used to determine the average intracavity power $P_{1S\text{-}2S}$ (dotted lines). The lower row shows in detail the same signals during (**B**) the start and (**C**) the end of the bright phase.

The transient behavior when switching the length stabilization on and off also affects the determination of the intracavity power. In Section 4.3.3.4, it was assumed that for a fraction $\eta_{\text{circ}} = 1/2$ of the time, the cavity is stabilized at a constant circulating power $P_{1S\text{-}2S}$, while for the rest of the time no power is circulating. Because of the transient behavior, the atoms in the experiment actually see a circulating power below $P_{\text{1S-2S}}$ at the start of the bright phase, but a circulating power above $P_{\text{1S-2S}}$ when the cavity has stabilized. By comparing the integrals of the transmission photodiode signal over the averaging window, on the one hand, and over the whole chopper cycle, on the other hand, the instantaneous power during the averaging window can be related to the power averaged over many chopper cycles as measured with the power meter. For the data of the fast photodetector shown in Fig. 4.11, the ratio between the integrals is 2.054. That is, the instantaneous power is $2.054/2 - 1 = 2.7\,\%$ higher than the average power assuming the cavity is fully stabilized exactly half the time. In any case, the atoms in the experiment sample the complete bright phase and not only the averaging window. Thus, for a complete picture, the simulations of the experiment (see Section 5.2) ideally should use the time-dependent power as determined from the combination of the full photodiode signal and the power meter measurements. As an approximation, here the cavity is assumed to be fully stabilized exactly half the time with an instantaneous power equal to the average power $P_{\text{1S-2S}}$.

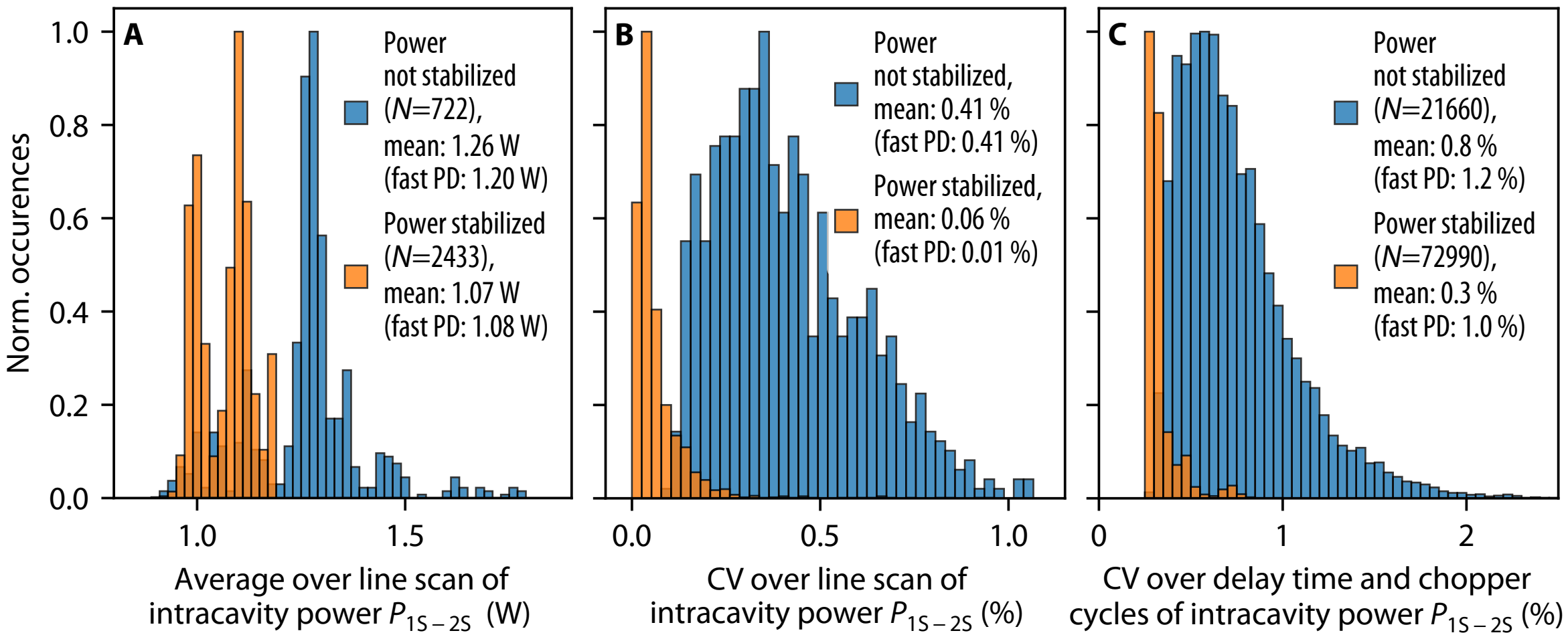

Figure 4.12: Stability of the 1S-2S intracavity laser power $P_{\text{1S-2S}}$ during the 2S-6P measurement, separated into line scans where the intracavity power was not (722 line scans, blue bars) or was (2433 line scans, orange bars) stabilized (see Table 6.1). (**A**) Histogram of $P_{\text{1S-2S}}$ assigned to the line scans by averaging over the values of $P_{\text{1S-2S}}$ of the 30 data points each scan contains. For each data point, $P_{\text{1S-2S}}$ in turn is found by averaging the time-resolved cavity transmission signal over a window of delay times and chopper cycles as shown in Fig. 4.11, and converting the signal to a power reading with the calibration factors of Fig. 4.10. The signal of the integrating sphere photodetector is used, but the mean value from the fast photodetector is given in the legend. (**B**) Coefficient of variation (CV) of $P_{\text{1S-2S}}$ over the points within each line scan, with the points on average separated in time by 2.3 s. (**C**) CV over delay times and chopper cycles within each data point. See text for details.

During the 2S-4P measurement (see Appendix A), the intracavity power during the bright phase was not stabilized, but left free-running and thus subject to drifts of both the input power and the power enhancement of the cavity. This was also the case for some of the data taken during the 2S-6P measurement. However, an intracavity power stabilization was added on 28.05.2019 and used during the bulk of measurement run B and the complete run C (see

Table 6.1). To this end, an AOM[1] is used to control the power impinging on the cavity by diffracting some of the light to a beam dump. The amount of light dumped is controlled[2] with the amount of RF power sent to the AOM. The fast transmission photodetector is used to generate the error signal for this stabilization, which uses an identical PI controller as the length stabilization. The integrating sphere photodetector provides an out-of-loop cavity transmission signal. The power stabilization is only active during part of the bright phase to minimize interference with the length stabilization. It is switched on 200 µs after the start of the bright phase to ensure that the length stabilization has settled, and switched off again after a further 2705 µs (dash-dotted lines in Fig. 4.11), i.e., 200 µs before the length stabilization is switched off[3]. The effect of the intracavity power stabilization is clearly visible in Fig. 4.11, with the steady-state cavity transmission (purple shading) varying less over time than the input power (red shading), which is measured after the AOM and thus shows the compensation applied by the feedback loop.

Fig. 4.12 shows the behavior of the intracavity power $P_{\text{1S-2S}}$ during the line scans of the 2S-6P measurement, separated into sets where the power was not stabilized (blue bars) and was stabilized (orange bars). When the power is not stabilized, the coefficient of variation (CV) over the data points within each line scan (see Fig. 4.12 (B)) is on average 0.41 %, but reduces to 0.06 % for the data with active power stabilization. Note that the fast photodetector shows a lower average CV of 0.01 % for the latter case, but the same CV for the former case, which is to be expected since it is part of the feedback loop. Likewise, the average CV over delay time and chopper cycles for each data point (see Fig. 4.12 (C)) drops from 0.8 % to 0.3 %. Interestingly, the fast photodetector shows a larger average CV of 1.0 % and only a small effect of the stabilization in this case. This is thought to be caused by the detection bandwidth of the fast photodiode, limited to 250 kHz by the sampling of the signal, exceeding the ≈50 kHz feedback bandwidth of the stabilization.

#### 4.3.3.6 Alignment laser

An auxiliary alignment laser that follows the beam path of the 1S-2S preparation laser through the vacuum chamber, as shown in Fig. 4.6, is used to adjust the variable aperture and image the hydrogen nozzle. A helium–neon laser at $\lambda = 632.8$ nm is used for this purpose, as the 243 nm mirrors used in the enhancement cavity and for steering are only weakly reflective at this wavelength and because of its good beam quality. The beam paths of the alignment laser and the 1S-2S laser, which are counter-propagating, are combined using the steering mirrors right before and after the two Brewster's windows of the vacuum chamber as beamsplitters. Two irises, placed close to either of the two mirrors and referred to as front and back iris, respectively, are used to fine-adjust the overlap of the beams. This overlap is adjusted at least once every measurement day before the start of the measurement itself. The alignment laser can be blocked with a remote-controlled shutter, with the laser always blocked when spectroscopy data are acquired.

A removable mirror sends the beam of the alignment laser, after passing through the vac-uum chamber, either to a photodetector (alignment PD) or to an image sensor (nozzle image sensor). When using the laser to adjust the width of the variable aperture, its collimated beam, which has an $1/e^2$ intensity radius of ≈450 µm, is sent through the chamber and detected on the alignment photodetector to enable a knife-edge measurement as described in Section 4.5.3. To suppress intensity fluctuations during these measurements, the measurement signal is normalized to the output power of the alignment laser using an additional photodetector.

[1] IntraAction ASM-1101M3, 110 MHz center frequency. The active medium is UV-grade fused silica with a broadband UV anti-reflection coating and a transmission of 98 % at 243 nm.

[2] To adjust the RF power, the amplitude modulation feature (3 dB bandwidth of 50 kHz) of the RF synthesizer (Rohde & Schwarz SMC100A) is employed.

[3] The intracavity power stabilization is thus switched off ≈100 µs before the end of the averaging window. This is unintended, but was only realized after the end of the measurement. In fact, there is no reason the power stabilization should not be kept on till right before the length stabilization is switched off.

To image the nozzle, the alignment laser beam is made to diverge using a removable lens creating an intermediate focus, such that the nozzle channel is approximately uniformly illuminated. The nozzle channel is then imaged onto the nozzle image sensor using an additional lens in front of the sensor. This image is used to adjust the nozzle position and observe the nozzle during the measurement, as discussed in Section 4.5.2.5.

#### 4.3.3.7 Apertures inside the cavity and cavity alignment

There are multiple apertures inside the 243 nm enhancement cavity as shown in Fig. 4.1 and Fig. 4.6. Of those, the circular nozzle channel (diameter $2r_1 = 2\,\mathrm{mm}$), the circular high-vacuum entrance aperture (diameter 2.4 mm), and the rectangular variable aperture (size 1.2 mm×2 mm) are relevant to the operation of the cavity. Each pass of the laser beam through an aperture causes losses $A_{\mathrm{AL}}$, such that the total losses per round trip inside the cavity are increased by $2A_{\mathrm{AL}}$. The losses $A_{\mathrm{AL}}$ for circular, square, and rectangular apertures for the beam parameters of the cavity are shown in Fig. 4.13, with the losses from the three aforementioned apertures all below $6 \times 10^{-5}$. However, this is only the case when the laser beam is exactly centered on the apertures. An offset of 200 µm from the center of the variable aperture along its narrower width increases the apertures losses to $A_{\mathrm{AL}} = 0.35\,\%$, leading to almost 40 % higher total round-trip losses compared to the non-degraded case with perfect alignment. For this reason, the width of the variable aperture can be remotely adjusted along the horizontal dimension ($x$-axis) to compensate for drifts in the alignment (see Section 4.5.3). Furthermore, as the nozzle channel diameter decreases due to the accumulation of solid hydrogen during the freezing cycle, the losses quickly increase below a certain threshold, with $A_{\mathrm{AL}} = 0.35\,\%$ and $A_{\mathrm{AL}} = 1.0\,\%$ reached for $2r_1 = 1\,\mathrm{mm}$ and $2r_1 = 0.9\,\mathrm{mm}$, respectively. When this threshold is reached, the experiment has to be stopped as the power enhancement of the cavity is insufficient.

The alignment of the apertures relative to the optical axis of the cavity is done in multiple steps. For the initial alignment of the enhancement cavity, done while the vacuum chamber lid is removed, the entrance and output aperture of the high-vacuum enclosure are removed, the cryostat and attached nozzle are moved out of the cavity beam, and the variable aperture is fully opened. The beam is then centered on both the IC and OC using the steering mirrors before the cavity, before closing the beam path inside the cavity by adjusting first the tip and tilt of the OC and then of the IC. After aligning the enhancement cavity, the entrance and output apertures are installed again. Using the adjustable feet of the vacuum chamber as shown in Fig. 4.2, the chamber, to which the apertures are fixed, is positioned such that the 1S-2S laser beam is centered on the entrance aperture and the variable aperture, which is still fully open. A beam profiling camera and lens tissue, which scatters the laser beam such that the shadow of the apertures are visible, are used during this alignment procedure, with the camera alternately placed after the entrance and variable apertures. This alignment procedure is estimated to be accurate within 100 µm. When the chamber is evacuated, this alignment might drift, since the cavity is fixed to the optical table and not the vacuum chamber itself. This drift is monitored by observing the transmission of higher-order modes of the cavity, which are especially susceptible to aperture losses due to their larger spatial extent. Typically, no change in transmission is observed and the alignment is thought to be intact after evacuation.

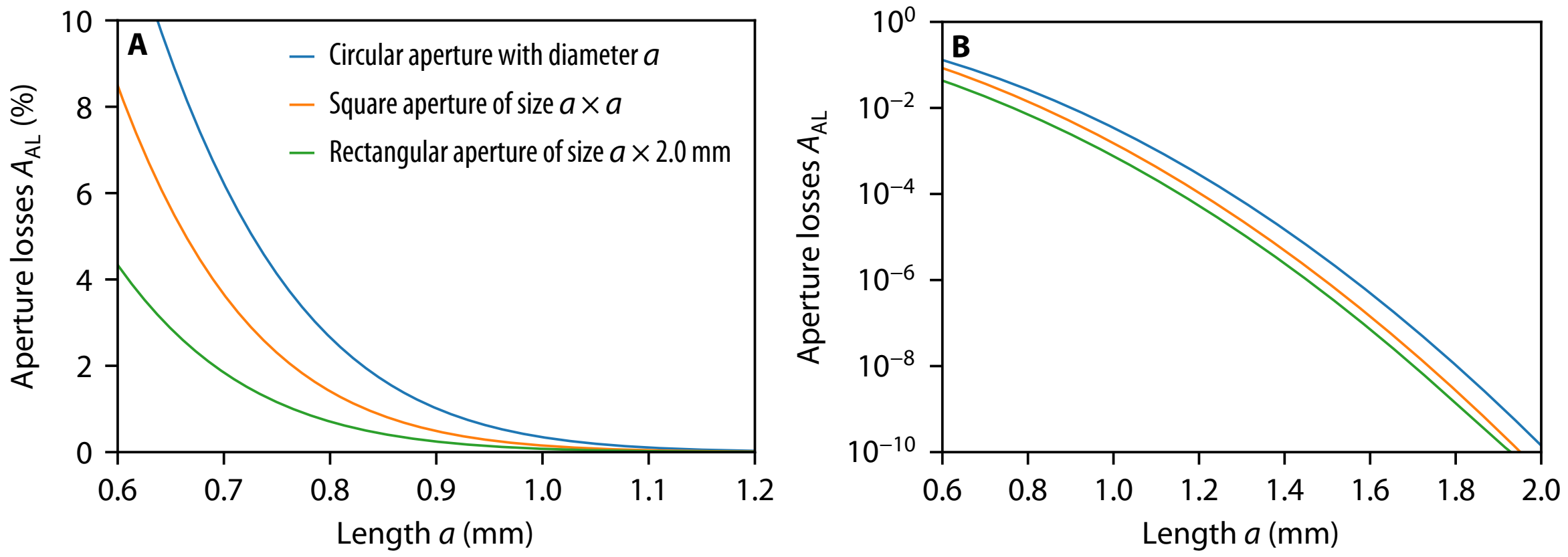

Figure 4.13: (**A**) Calculated losses $A_{AL}$ introduced by different apertures for a Gaussian laser beam with waist radius $w_{1S\text{-}2S} = 297\,\mu m$ as used in the 243 nm enhancement cavity: circular aperture with diameter $a$ (blue line), square aperture of size $a \times a$ (orange line), and rectangular aperture of $a \times 2\,mm$ (green line). (**B**) shows the same results as (A), but on a logarithmic scale.

The nozzle position and the horizontal width of the variable aperture can be adjusted independently of the other apertures and while the chamber is under vacuum. These alignments are described in Section 4.5.2.5 and Section 4.5.3, respectively.

From the aperture losses, the number of photons that hit the inside of the fluorescence detector assembly can be estimated. The by far largest losses of $5.3 \times 10^{-5}$ occur at the variable aperture. From a Monte Carlo integration, it is estimated that 5 % of the photons scattered from the edges of this aperture, as limited by the geometry, reach the interior walls of the detector. For an intracavity power of $P_{1S\text{-}2S} = 1\,W$, this corresponds to $3.3 \times 10^{12}$ photons/s hitting the detector walls during the bright phase of the optical chopper. The illuminated sections of the detector walls are coated with graphite, resulting in the emission of $2 \times 10^7$ photonelectrons/s, using the photoelectron yield given in Table 4.1. Note that the 2S-6P fluorescence signal is only detected during the dark phase of the optical chopper, i.e., when the preparation laser is blocked, and thus these scattered photons and resulting photoelectrons do not affect the signal.

## 4.4 2S-6P spectroscopy laser

The 2S-6P spectroscopy laser excites hydrogen atoms from the metastable 2S level to the 6P level, using light at a frequency of $\nu_{2S\text{-}6P} = 730.7\,THz$ or, correspondingly, a wavelength of $\lambda_{2S\text{-}6P} = 410.3\,nm$, with its detuning $\Delta\nu_{2S\text{-}6P}$ from the transition frequency measured relative to $\nu_{2S\text{-}6P}$. The laser light is generated by frequency-doubling an infrared laser, as outlined in Section 4.4.1. A crucial component in the spectroscopy of the 2S-$n$P transitions is the active fiber-based retroreflector (AFR), which uses counter-propagating laser beams to suppress the first-order Doppler shift the hydrogen atoms moving through the beams would otherwise experience. To this end, a low-aberration collimator is required, which had to be specially designed for use with the 2S-6P transition, as detailed in Section 4.4.3. The control and stabilization of the retroreflection used to generate the counter-propagating beams are discussed in Section 4.4.4, while the stabilization of the intensity of these beams is the topic of Section 4.4.5. Finally, the polarization of the laser beams is also monitored, as shown in Section 4.4.6. The content of this section is largely identical to a publication co-authored with Vitaly Wirthl [35, 36], with more details found therein.

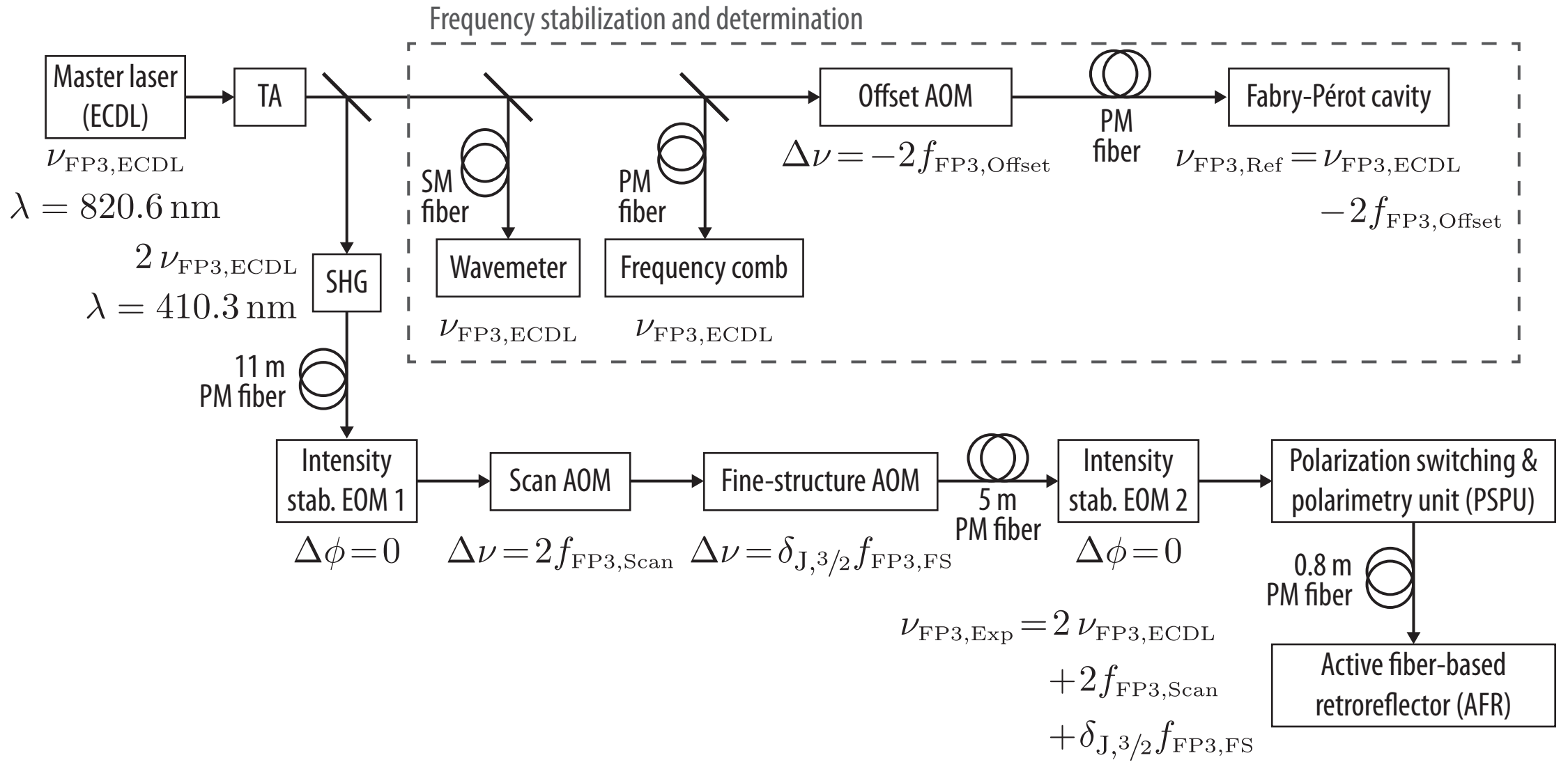

Figure 4.14: Sketch of relevant components of the 410 nm laser system ("FP3"), part of the 2S-6P spectroscopy laser, with a focus on components that change the frequency, phase, or power of the laser light. $\delta_{\mathrm{J},3/2}$ is 0 (1) when the 2S-6P$_{1/2}$ (2S-6P$_{3/2}$) transition is probed. See Section 4.4.1 for details. AOM: acousto-optic modulator, ECDL: external-cavity diode laser, EOM: electro-optic modulator, PM fiber: polarization-maintaining fiber, SHG: second-harmonic generation, SM fiber: single-mode fiber, TA: tapered amplifier.

### 4.4.1 410 nm laser system

The 410 nm laser system (known as "FP3" in the hydrogen laboratory), shown in Fig. 4.14, generates the laser light necessary to drive the 2S-6P transition. The system is very similar to the 243 nm laser system described in Section 4.3.1. The master laser, operating in the infrared at frequency $\nu_{\mathrm{FP3,ECDL}}$ ($\lambda \approx 820.6\,\mathrm{nm}$), is a home-built external-cavity diode laser (ECDL, cavity length of 20 cm) in the Littrow configuration [82]. The output of the master laser is amplified to an optical power of 1.1 W using a tapered amplifier (TA), supplied with a current of 2.2 A. As opposed to the 243 nm laser system, the infrared light is doubled in frequency only once. This is achieved using a lithium triborate (LBO) crystal, placed in a home-built bow-tie ring cavity, generating 100 mW... 150 mW of 410 nm laser light. The light is then sent through an 11 m-long polarization-maintaining (PM) fiber to the optical table of the hydrogen spectrometer.

The frequency stabilization and determination of the 410 nm system is analogous to that of the 243 nm system. The frequency of the master laser is stabilized to an identically constructed high-finesse Fabry-Pérot reference cavity (free spectral range of 1.932 GHz). Some light at frequency $\nu_{\mathrm{FP3,ECDL}}$ is sent through a PM fiber to the frequency comb to determine the laser frequency in Hz, and to compare the optical spectrum of the master laser to that of the 1S-2S preparation laser (see Section 4.8).

However, as opposed to the 243 nm system, no cancellation for the phase noise picked up by the fiber between the cavity and the master laser is implemented because of the lower requirements on the laser spectrum. The double-passed acousto-optic modulator (offset AOM) placed between the master laser and the reference cavity shifts the laser frequency by $-2f_{\mathrm{FP3,Offset}}$ and into resonance with the reference cavity. Unlike for the 243 nm system, this AOM is not used to scan the frequency of the light seen by the atoms, as the required range of 100 MHz at the second harmonic exceeds the tuning range of the master laser. Instead, a double-passed AOM[1] (scan AOM) operating at $f_{\mathrm{FP3,Scan}} = 325\,\mathrm{MHz} \ldots 375\,\mathrm{MHz}$ is used to shift the frequency of the frequency-doubled light. An additional AOM[2] (fine-structure AOM) further shifts the frequency by $f_{\mathrm{FP3,FS}} \approx 406\,\mathrm{MHz}$ when addressing the 2S-6P$_{3/2}$ transition, but is bypassed when the 2S-6P$_{1/2}$ transition is probed.

The remaining components are discussed as part of the active fiber-based retroreflector in the next section.

### 4.4.2 Active fiber-based retroreflector

The general concept of an active fiber-based retroreflector (AFR) is discussed in detail in [28]. The main idea is to collimate a laser beam from a single-mode fiber and use a highly-reflective (HR) mirror to couple the light back through the collimator into the same fiber. Using a beamsplitter before the fiber, the backcoupled light fraction can be monitored and maximized by adjusting the distance between the collimator and the fiber as well as tip-tilt alignment of the HR mirror. This tip-tilt alignment is actively stabilized with piezoelectric actuators on the mirror mount[3]. Maximizing the backcoupled light fraction corresponds to optimizing the wavefront-retracing property of the beam: for a perfectly flat, fully reflective HR mirror and an aberration-free beam, the wavefront-retracing and amplitude matching conditions are satisfied if the waist of the collimated beam is located on the mirror, resulting in all of the optical power being coupled back into the fiber.

The optical setup of our hydrogen spectrometer relevant to the AFR is shown in Fig. 4.15. We start with around 100 mW of 410 nm laser light which is sent to the experiment through a 11 m-long polarization-maintaining (PM) fiber from the 410 nm laser system. An acousto-optic modulator (AOM, same as scan AOM in Fig. 4.14) is used to scan the optical frequency across the resonance and an electro-optical modulator (EOM 1) in sequence with a polarizing beamsplitter (PBS) and half-wave plates is used for intensity control. The light is transferred via another 5 m-long PM fiber to a polarization switching and polarimetry unit (PSPU). The intensity after this fiber is stabilized using the signal of photodetector PD 1 in combination with EOM 1 as an actuator (1$^{\mathrm{st}}$ intensity stabilization). Additional photodetectors (PD 2, PD 3 and PD 4) monitor the intensity out-of-loop. In order to switch between horizontal and vertical linear polarizations, either of the two paths are opened by mechanical shutters. Glan-Thompson polarizers with PER $> 50\,\mathrm{dB}$ improve the polarization extinction ratio PER $= P_{\mathrm{max}}/P_{\mathrm{min}}$ of light coupled into the last fiber, where $P_{\mathrm{max}}$ and $P_{\mathrm{min}}$ are the maximal and minimal transmitted optical powers in the two orthogonal polarization directions. Isolators in each arm suppress optical etalons, with lowest number of optical surfaces as possible placed after the isolators, and shutters closing the unused arm. After the PSPU, the light passes a non-polarizing beamsplitter (BS 2) and is sent through another beamsplitter (BS 3).

[1]Brimrose TEF-350-100-400, 350 MHz center frequency and 100 MHz bandwidth.

[2]TEF-350-200-380/500, 350 MHz center frequency and 150 MHz. . . 200 MHz bandwidth.

[3]Radiant Dyes MDI-H with piezo drive. Two Newport 8301-UHV Picomotor precision motors have been added to increase the alignment range.

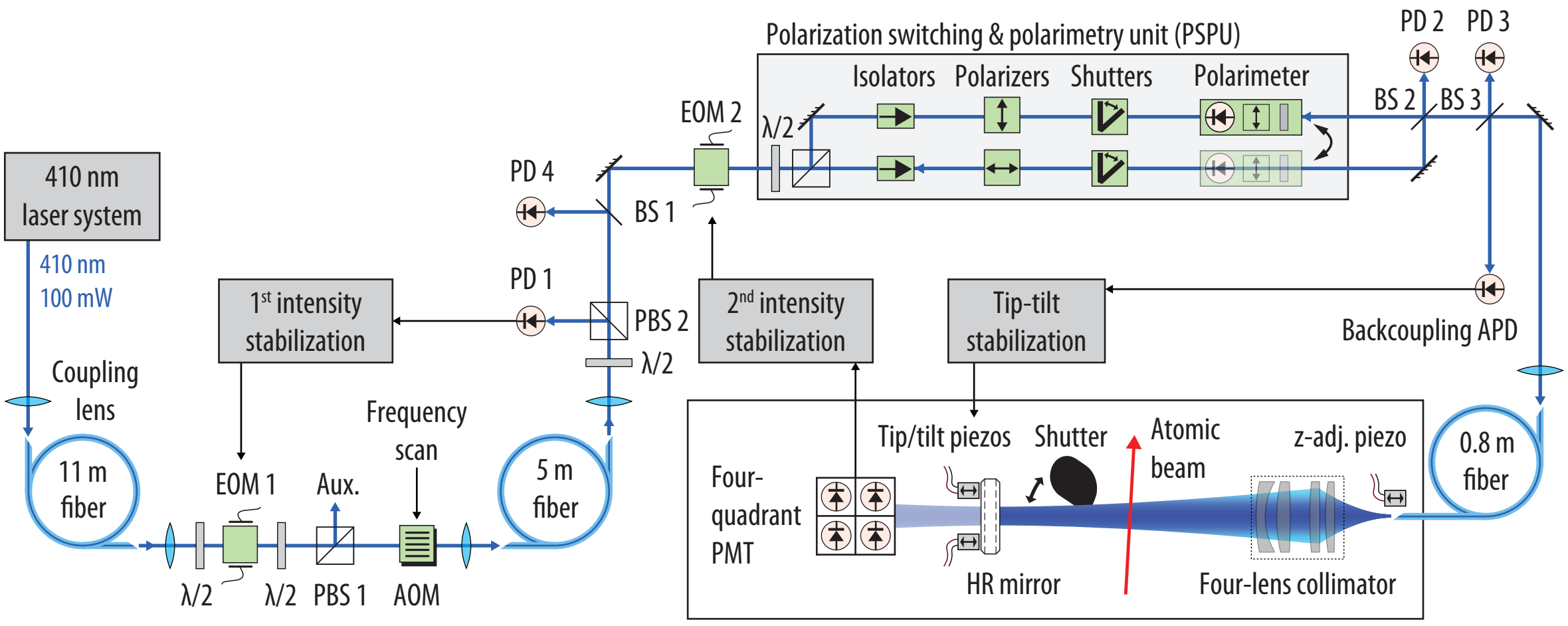

Figure 4.15: Optical layout of the 2S-6P spectroscopy laser and the active fiber-based retroreflector. Note that the fine-structure AOM included in Fig. 4.14 is not shown here, but is present in the experiment. $\lambda/2$: half-wave plate, AFR: active fiber-based retroreflector, AOM: acousto-optic modulator, APD: avalanche photodiode detector, BS: beamsplitter, EOM: electro-optic modulator, HR mirror: high-reflectivity mirror, PBS: polarizing beamsplitter, PD: photodetector, PMT: photomultiplier, z-adj. piezo: adjustment of the fiber–collimator distance.

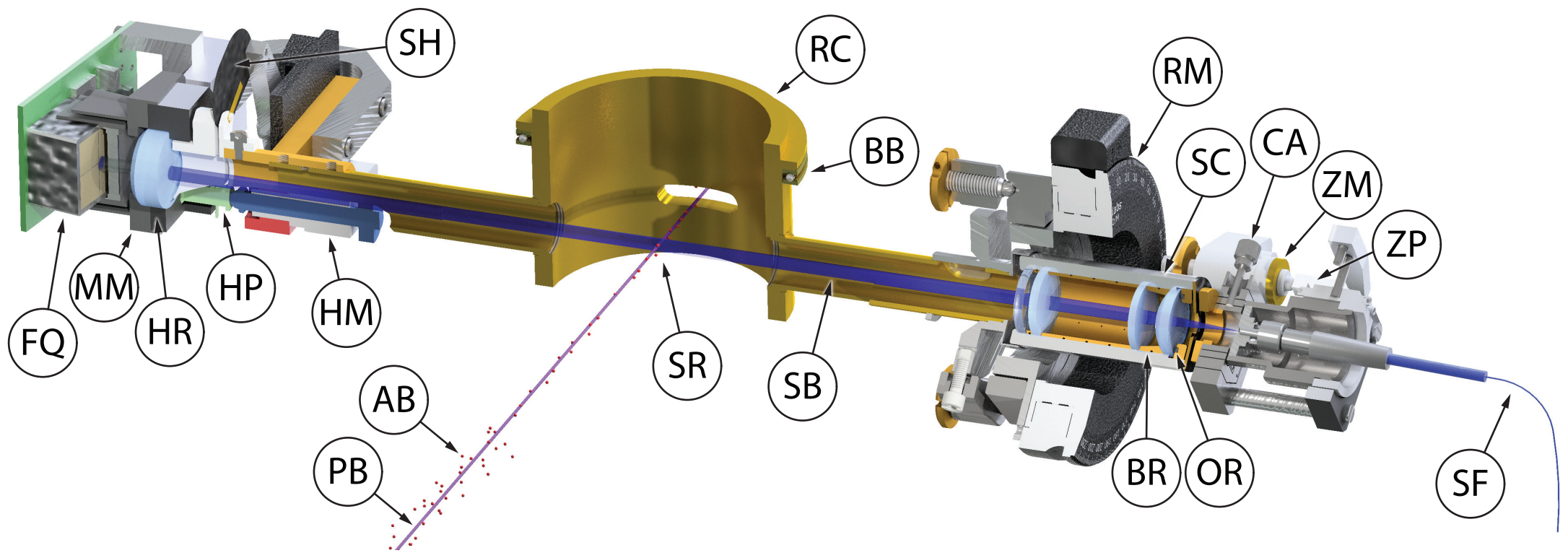

Figure 4.16: 3D view (orthographic projection) of the in-vacuum active fiber-based retroreflector (AFR) setup, which is part of the atomic beam apparatus shown in Fig. 4.1. **AB**: atomic beam, **BB**: base cylinder ball bearing, **BR**: brass ring spacer between collimator lenses, **CA**: collimator–fiber alignment, **FQ**: AFR four-quadrant photomultiplier, **HM/HP**: horizontal (tip) precision motor (HM) and piezo actuator (HP), **HR**: AFR high-reflectivity mirror, **MM**: piezo-actuated AFR mirror mount, **OR**: O-ring holding collimator lens system in place, **PB**: 1S-2S preparation laser beam, **RC**: rotatable base cylinder, **RM**: collimator rotation mount, **SB**: 2S-6P spectroscopy laser beams, **SC**: four-lens collimator, **SF**: polarization-maintaining fiber, **SH**: AFR shutter, **SR**: 2S-6P spectroscopy region, **ZM/ZP**: precision motor (ZM) and piezo actuator (ZP) for fiber–collimator distance adjustment.

Finally, the laser light (5 µW…30 µW laser power) reaches the in-vacuum AFR setup,

shown in Fig. 4.16, via an 80 cm-long polarization-maintaining (PM) fiber[1] (**SF**). The whole in-vacuum AFR setup is mounted on the rotatable base cylinder (**RC**), which sits on a ball bearing (**BB**) such that the angle between the spectroscopy laser beams (**SB**) and the atomic beam of metastable hydrogen (**AB**), which is collinear to the 1S-2S preparation laser beam (**PB**), can be aligned close to 90°. The four-lens collimator (**SC**) is mounted onto a mirror mount combined with a manual precision rotation stage[2] (**RM**). The distances from the fiber tip and the last lens surface of the collimator to the center of the cylinder are 20 cm and 12 cm, respectively. The collimation is adjusted with the help of a commercial fiber translation mount[3] which we rebuilt for accurate distance control using a precision motor[4] (**ZM**) combined with a piezo actuator[5] (z-adj. piezo, **ZP**). This part is placed onto a cage system mounted to a flexure adjustment plate[6] (**CA**) needed for precise centering between the fiber and the collimator. The light is retroreflected by the HR mirror[7] (**HR**), which is mounted in a piezo-actuated mirror mount (**MM**) at a distance of 16 cm from the center of the cylinder. The reflected light passes back through the fiber, where approximately one-half of the backcoupled light is detected after BS 3 on the avalanche photodiode (backcoupling APD) whose signal is used for the tip-tilt stabilization using the piezo actuators of the mirror mount (only the horizontal piezo actuator (**HP**) is shown in Fig. 4.16). Another fraction of the backcoupled light passes BS 2 where the polarimeter is placed in the unused beam path. The home-built remotely controlled shutter (**SH**) makes it possible to block the reflected beam to acquire spectroscopy data without the Doppler suppression provided by the counter-propagating beams.

The sum signal of the four-quadrant photomultiplier[8] (PMT, **FQ**) after the HR mirror is used for a second intensity stabilization of the wavefront-retracing beams, with EOM 2 serving as actuator. The PMT is mounted under an angle of ≈10° with an interference bandpass filter[9] and scattering disk[10]. The use of a position-sensitive PMT has the practical advantage of misalignment monitoring. An angular misalignment of the HR mirror or the collimator-fiber system by 200 µrad leads to a complete loss of the backcoupled signal [28]. If the optimal orientation of AFR is lost (e.g., during work on the apparatus), the horizontal and vertical position signals of the PMT help to retrieve the alignment.

During the 2S-6P measurement, the 410 nm laser power was measured, on each measurement day, right before the fiber to the in-vacuum AFR setup, using a calibrated power meter[11]. The 2S-6P spectroscopy laser power $P_{\text{2S-6P}}$ is then given by the fiber coupling efficiency multiplied by the power meter reading. The power meter is specified to have a calibration uncertainty of ±5 %, a power linearity of ±1 %, and a active area uniformity of ±1 %. By placing the power meter in a designated mount for each power measurement, we find that the relative measurement uncertainty limited by the active area uniformity can be reduced to ≈0.5 %. The fiber coupling efficiencies were measured[1] to be 0.788(7) and 0.807(7) for vertical and horizontal polarization, respectively, with the two polarization directions used for the linear laser polarization angle setting of, respectively, $\theta_{\mathrm{L}} = 56.5°$ and $\theta_{\mathrm{L}} = 146.5°$ during the 2S-6P measurement. The fiber coupling was optimized on each measurement day by optimizing the sum signal on the PMT (with the intensity stabilization switched off). The uncertainty for the absolute value of $P_{\mathrm{2S\text{-}6P}}$ is estimated, from the specifications of the power meter, to be ≈7 %, while the relative uncertainty between measurement days is estimated to be ≈1.3 % (quadrature sum of uncertainties for fiber coupling efficiency, power meter linearity, and active area uniformity).

[1] Vacuum-compatible Nufern PM-S405-XP fiber in a 900 µm-diameter PEEK jacket, produced and AR-coated by Diamond GmbH.

[2] Thorlabs POLARIS-K2S3, PRM2/M.

[3] Thorlabs SM1.

[4] Newport 8301-V.

[5] Thorlabs PK4FQP2.

[6] Thorlabs CP1XY.

[7] Custom order from LayerTec: reflectivity $R > 99.995\,\%$, transmission $T \approx 2 \times 10^{-5}$, substrate with $\lambda/30$ @ 633 nm irregularity and <1.5 Å RMS roughness.

[8] Hamamatsu R11265-200-M4.

[9] Edmund Optics 34-494, 10 nm-wide passband (FWHM), centered at 413 nm.

[10] DG10-1500-A.

[11] Thorlabs PM160.

### 4.4.3 Low-aberration collimator

In the AFR, the collimator plays a central role since aberrations may be imprinted on the wavefronts of the spectroscopy laser beams. These aberrations distort the wavefronts such that there may be no position in the collimated beam with a plane wavefront, and thus the backward-traveling beam will not retrace the wavefronts of the forward-traveling beam. The backcoupled light fraction is a quantity which characterizes how well the wavefront-retracing property is maintained, because this quantity is directly linked to the overlap integral of the forward- and backward-traveling beams. In our previous setup of the AFR at 486 nm, a collimator design based on two achromatic lens doublets was used to minimize aberrations and achieve a backcoupled fraction consistent with 100 % within 1 % [28]. Apart from correcting chromatic aberrations, which are irrelevant for our single-wavelength application, achromatic lens doublets have the advantage of reducing spherical aberration compared to a single lens.

With shorter wavelengths, designing suitable optics becomes more challenging since fewer glass types are sufficiently transparent and can be combined into achromatic lens doublets. In theory, aberration-free collimation can be achieved with a single aspheric lens of the desired shape. To this end, we tested custom-made aspheres[2] machined with the advanced technique of magnetorheological finishing (MRF) [100, 101]. Unfortunately, imperfections from polishing were still clearly visible on the collimated beam and only around 80 % of backcoupled light fraction could be achieved.

Therefore, we chose to only work with spherical lenses which are available with small surface roughness. First, at 410 nm we tested a three-lens collimator based on spherical fused-silica lenses[3]. A design with minimized aberrations was found by following conventional ray-tracing techniques such as optimizing the point-spread function and minimizing the optical path difference of rays, similar to the previous two-achromats design at 486 nm [28]. However, when testing the assembled custom-made collimator we found that residual spherical aberrations limit the backcoupled fraction to 94(1.2) %. Contrary to our previous experience where collimator imperfections were clearly visible as distortions in the collimated beam [28], aber-rations of the three-lens collimator were not visible in the intensity profiles of the collimated beam and were revealed only by a caustic measurement, i.e., beam profile measurements at different position within a caustic, shown below in Fig. 4.19. For our application, the usual ray tracing design process had to be extended by wave optics propagation tools. Finally, together with the manufacturing company[1] we arrived at a four-lens design whose optical performance was confirmed with a caustic measurement and showed no aberrations above our detection limit. With this collimator, we achieved a measured backcoupled light fraction of 99.3(1.2) %, consistent with 100 %. In the following, the design and characterization process of the four-lens collimator is summarized.

[1] Note that when measuring the power before the fiber, one has to either take into account or block the second-order reflection from the beamsplitter BS 3, which propagates at an angle to the main beam and is not coupled into the fiber, but contains 1.7 % of the total power of both beams.

[2] Thorlabs MRF-polished diffraction-limited, high-precision aspheres AL1225H (stock item) and AL1225H-50URAD-SP (custom order, best possible surface quality with 50 µrad peak-valley slope, optimized for performance between 380–410 nm.)

[3] Note that this is not the three-lens collimator mentioned in [28], which was an even earlier version of poor manufacturing quality and an unfortunate design, since the last surface of the last lens was chosen to be flat, forming an etalon with the HR mirror.

#### 4.4.3.1 Design process

For our previous collimator at 486 nm, each of the commercial achromatic doublets exhibited little spherical aberrations by itself, and the combination of two doublets found by ray tracing turned out to satisfy the requirements of the AFR without further investigation. For shorter wavelengths, after we found that the three-lens collimator designed by ray tracing optics alone showed residual aberrations revealed through the caustic measurement (see Fig. 4.19 introduced below), the design process was extended by wave optic propagation tools[2]. Ray tracing allows the minimization of wavefront aberrations within a given aperture width, with the Gaussian beam profile not easily accounted for. For a given number of lenses a compromise has to be made between the width of the aperture employed for minimizing the wavefront deviations and the magnitude of acceptable deviations from the aberration-free wavefront.

The aberration-free focusing phase and the leading spherical aberration term can be written as

$$\phi_{\text{foc}}(r) = -\frac{k\,r^2}{2\,f}, \quad \phi_{\text{ab}}(r) \approx S\left(\frac{r^4}{W^4} - 2\frac{r^2}{W^2}\right), \tag{4.7}$$

with $f$ being the focal distance, $k$ the wave-number and $W$ the beam radius at the position where the aberration is imprinted. The focusing effect of the $r^4$-term in $\phi_{\text{ab}}(r)$ is compensated by the $r^2$-term to isolate the contribution of the aberration alone. The parameter $S$ characterizes the strength of spherical aberration. If in our case ($\lambda = 410\,\text{nm}$, $f = 30\,\text{mm}$, $W = 2.2\,\text{mm}$) we would use a single thin plano-convex collimating spherical lens, spherical aberrations would be as large as $S \approx 3$ following the analytical expression of [102]. In the case of spherical aberrations, deviations from the aberration-free wavefront increase as $\propto r^4$. Hence, for larger radial distances $r$ it becomes progressively more difficult to meet the ray tracing compromise between the width of the aperture employed for minimizing the wavefront deviations and the magnitude of acceptable deviations, especially at shorter wavelength. Because the wings of a laser beam extend to large $r$, it is a priori unclear which ray tracing criteria should be used.

Therefore, we followed an iterative design procedure together with the manufacturing company. Ray tracing was used as a guidance based on the manufacturability of lenses and the requirement of effective focal length of $f \approx 30\,\text{mm}$. Using wave optics propagation, the designs found in this way were evaluated with simulated intensity profiles in the caustic measurement simulation. Furthermore, the electric field phase and amplitude were extracted for simulations of residual Doppler shift with optical Bloch equations. Another important design criterion is the consideration of residual reflections from lens surfaces back to the fiber and to the spectroscopy region. After several iterations we found that three lenses are not enough to meet our requirements and a four-lens design was needed. In order to be able to use the same collimator for spectroscopy of all 2S-$n$P transitions with $n \geq 4$, the collimator was designed for the wavelength range from 380 nm to 486 nm. The final collimator design with an effective focal length of $f = 31.02$ mm is shown in Fig. 4.17. This focal length, in combination with the fiber used here, results in a collimated beam with a waist radius of $W_0 = 2.2$ mm [35, 36].

[1] Bernhard Halle Nachfl. GmbH, Germany.

[2] Zemax OpticStudio 15.5 Professional, which includes the Physical Optics Propagation (POP) module.

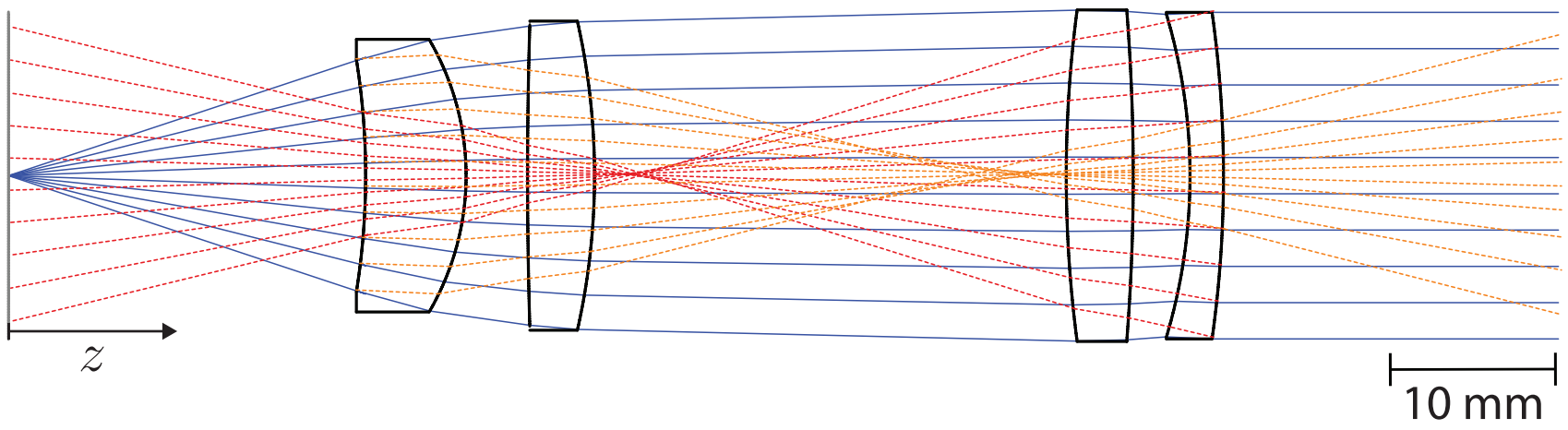

Figure 4.17: Four-lens collimator design with blue solid rays illustrating the collimated beam from the fiber. Shown lens radius corresponds to the open lens apertures. Dashed rays exemplary show reflections from the last surface back into the fiber (red) and from the first surface back to the HR mirror (orange). The analysis of these spurious reflections is important for the collimator design process, since they may lead to a performance loss of the active fiber-based retroreflector or induce a residual Doppler shift.

#### 4.4.3.2 Residual reflections from lens surfaces

Though all the collimator lenses are AR-coated for the desired wavelength[1], residual reflections can lead to performance loss of the AFR. Here, we only consider single reflections from lens surfaces, since multiple reflections are strongly suppressed through the AR coating. Then, two types of reflections need to be considered. First, reflections of the forward-traveling wave from lens surfaces back towards the fiber. The part of these reflections that is coupled back into the fiber can disturb the tip-tilt stabilization. To reduce the influence of these reflections, efficient coupling into the fiber needs to be avoided which can be evaluated by calculating the spatial overlap $\eta_{\text{fiber}}$ of the reflected beam with the fiber mode.

The second type of reflections are reflections of the backward-traveling beam back toward the atomic spectroscopy region and the HR mirror. These reflections can influence the spectroscopy in two ways. Just like the reflection toward the fiber, an optical etalon is formed with the collimated beam, with the spatial overlap given by $\eta_{\text{coll}}$. The presence of this etalon will lead to intensity variations in the spectroscopy region, depending on the distance between the reflecting surface and the HR mirror as well as the laser frequency, which is varied to record the atomic resonance. However, since the laser intensity is stabilized to the signal of the PMT behind the HR mirror, these intensity modulations are suppressed and will instead influence the backcoupled light, possibly disturbing the tip-tilt stabilization. The reflections can also give rise to a residual Doppler effect. The more the reflections are focused near the spectroscopy region, either before or after the reflection of the HR mirror, the greater the intensity imbalance between the beams as seen by the atoms. To minimize the influence of these reflections, the design was chosen such that all reflections are diverging with a beam radius $W_{\text{refl}} \gg W$ at the first pass through the spectroscopy region.

[1]Custom coating from LayerTec, reflectivity $R < 0.15\,\%$ for 0–10° angle of incidence for 400–490 nm. A different AR coating would be necessary if the collimator is to be used at shorter wavelengths, e.g., for the 2S-8/9/10P transitions. An uncoated version of the collimator is available for this purpose.

The values of $\eta_{\text{fiber}}$ and $\eta_{\text{coll}}$ for all 8 possible first-order reflections are smaller than $\eta < 1 \times 10^{-5}$, such that with an additional suppression from the AR coating the overlaps are $<1 \times 10^{-8}$. The beam radius $W_{\text{refl}}$ of the second type of reflections is always above 20 mm. An example of a reflection back to the fiber from the last collimator surface is illustrated with red dashed rays in Fig. 4.17. Orange dashed rays illustrate the reflection back to the HR mirror and the spectroscopy region from the first surface, demonstrating that not only surfaces with negative curvature radii may focus the reflected beam towards the atoms. The highly suppressed 28 combinations of reflections from two lens surfaces of the forward-traveling beam lead to 26 strongly diverging and two nearly-collimated beams at the spectroscopy region.

#### 4.4.3.3 Doppler shift simulations with optical Bloch equations

Simulations of the Doppler shift with optical Bloch equations played a decisive role in the collimator design process. The collimator designs were evaluated using the wave optics propagation tool of our optics design software, with the Gaussian beam profile (beam waist radius of $w_0 = 1.9\,\mu\text{m}$ representing the fiber mode) as an input beam. The actual fiber mode deviates slightly from a Gaussian beam profile, which is treated in detail in [35, 36]. We extracted the electric field amplitude and phase after the collimator to perform simulations of the residual Doppler shift in the AFR. For this purpose, optical Bloch equations were numerically solved for our configuration of the experiment, using the atomic system for the 2S-6P transitions in hydrogen with 10 µW of laser power. We simulated individual trajectories of atoms moving through the center of the laser beam at different angles $\alpha = 90° + \delta\alpha$, and determine the Doppler shift $\Delta\nu_{\text{D}}$ by fitting a Voigt function to the resulting fluorescence line shape. The atomic velocity is set to $v = 200\,\text{m/s}$ which would result in an unsuppressed collinear ($\alpha = 0°$) Doppler shift of $\Delta\nu_{\text{D}} = 490\,\text{MHz}$ and $\Delta\nu_{\text{D}} = 2\,\text{MHz}$ for $\delta\alpha = 4\,\text{mrad}$. In the simulations, we evaluate the Doppler shift as a function of the fiber–collimator distance $\delta d_{\text{fc}}$ defined such that zero $\delta d_{\text{fc}}$ corresponds to the collimation with maximized backcoupled light fraction. No tip-tilt misalignment of the reflected beam from the HR mirror is assumed here.

Fig. 4.18 (A) compares simulations of perfect paraxial collimation to the four-lens collimator. Both cases are evaluated for different angles $\delta\alpha$. The top graph shows the resulting Doppler shift $\Delta\nu_{\text{D}}$. The bottom graphs show two AFR beam properties: the intensity mismatch of forward- and backward-traveling beams at their beam centers, $\xi_{\text{cent}}$, and the backcoupled light fraction $P_{\text{bc}}$. In the perfectly orthogonal case (zero $\delta\alpha$, gray points and lines), the Doppler shift is strongly suppressed and here found to be zero within the numerical uncertainty independent of $\delta d_{\text{fc}}$. For $\delta\alpha \neq 0$, there is only a single distinguished value of $\delta d_{\text{fc}}$ where the Doppler shift vanishes independent of $\delta\alpha$. For an aberration-free Gaussian beam, this value corresponds to the maximum backcoupled light fraction ($\delta d_{\text{fc}} = 0\,\mu\text{m}$).

Here, only single atomic trajectories are evaluated, though in the experiment a finite atomic beam divergence of 8–10 mrad (FWHM) is present (see Fig. 5.4). However, as Fig. 4.18 (A) demonstrates, the Doppler shift is approximately linear in $\delta\alpha$ within the range of interest. For a symmetric atomic beam which is aligned such that, on average, the atoms cross the laser beams at an small offset angle $\alpha_0$ ($|\alpha_0| \ll \pi/2$) from the orthogonal, there is for each atom with a crossing angle of $\alpha_0 + \delta\tilde{\alpha}$ another atom with a crossing angle $\alpha_0 - \delta\tilde{\alpha}$, where $\delta\tilde{\alpha}$ is an angle within the beam divergence. This results in a partial cancellation of the overall Doppler shift, with the remaining residual Doppler shift corresponding to that of a single trajectory with angle $\delta\alpha = \alpha_0$. As a figure of merit for the AFR performance, we evaluate the Doppler shift at the point of maximum backcoupling (zero $\delta d_{fc}$) for an angular displacement of $\delta\alpha = 4$ mrad, corresponding to the typical alignment accuracy in the experiment (see blue points and lines in Fig. 4.18 (A)). The residual Doppler shift for the aberration-free collimation with the Gaussian beam is exactly zero. For the four-lens collimator with the Gaussian beam as the input beam, we find almost no difference to the aberration-free Gaussian beam such that $\Delta\nu_D = -0.01$ kHz.

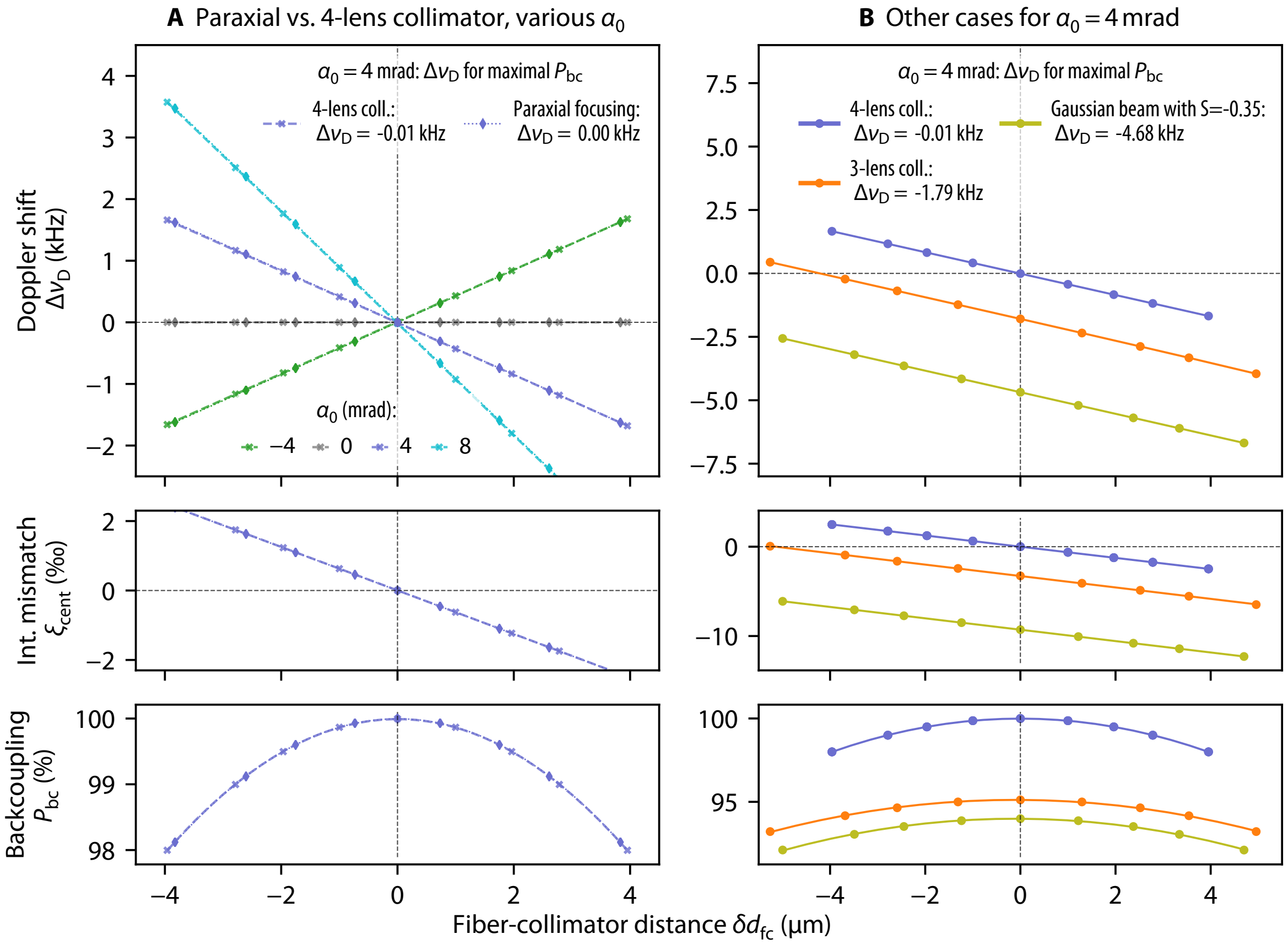

Figure 4.18: The top graphs show the simulated Doppler shift $\Delta\nu_D$ for the 2S-6P transition for different active fiber-based retroreflector (AFR) configurations versus the fiber–collimator distance $\delta d_{fc}$. The bottom graphs show the intensity mismatch $\xi_{cent}$ and the backcoupled light fraction $P_{bc}$, also versus $\delta d_{fc}$. (**A**) Aberration-free paraxial collimation compared to the four-lens collimator for different angles $\delta\alpha$. As a figure of merit, the Doppler shift is evaluated at the maximal backcoupling ($\delta d_{fc} = 0\,\mu m$) for $\delta\alpha = 4$ mrad (blue points and lines), demonstrating that the four-lens collimator performs almost as well as the aberration-free collimation, with no substantial difference observed in the simulations. (**B**) Simulations for $\delta\alpha = 4$ mrad. The simulation for the four-lens collimator is shown in blue as a reference to (A). The three-lens collimator (orange) as well as a Gaussian beam with spherical aberrations (green) show a residual Doppler shift of $\Delta\nu_D \approx 2$–5 kHz, thereby limiting the performance of the AFR. This residual Doppler shift is mainly caused by the intensity mismatch $\xi_{cent}$ which vanishes for approximately the same value of $\delta d_{fc}$ as the Doppler shift.

In Fig. 4.18 (B) several simulations are shown for $\delta\alpha = 4$ mrad. For the three-lens collimator with the Gaussian beam as input (orange line) we find $\Delta\nu_D = -1.79$ kHz. Though this residual Doppler shift still corresponds to a suppression factor $>10^5$ of the full collinear shift, its value is comparable to the uncertainty of the previous 2S-4P result [24]. The calculated backcoupled light fraction for the three-lens collimator approximately agrees with the measured value of 94.0(1.2) %. The Gaussian beam with spherical aberrations of $S \approx -0.35$ from Eq. (4.7) (an approximate value according to the measured beam quality factor from Fig. 4.19) results in $\Delta\nu_{\mathrm{D}} = -4.68\,\mathrm{kHz}$ (green line). We find again that the residual Doppler shift is mainly caused by the intensity mismatch which vanishes approximately for the same value of $\delta d_{\mathrm{fc}}$. Therefore, in principle, for an aberrated beam in the AFR one could adjust $\delta d_{\mathrm{fc}}$ such that the Doppler shift vanishes. However, in practice, reliable identification and adjustment of this position is challenging unless it is the point of maximized backcoupled light fraction.

Two effects may lead to a residual Doppler shift in the AFR (see Section 2.2.4): first, non-matching wavefronts of the forward- and backward-traveling beams, and second, imbalances of their intensities. Surprisingly, we find in our simulations, that the second effect dominates the induced Doppler shift for imperfections caused by aberrations. We do not observe significant deviations from our simulation results if, after propagation of forward- and backward-traveling beam, the wavefront mismatch but not the intensity mismatch of both beams is fully neglected in the spectroscopy region. Therefore, though aberrations are initially imprinted in the wavefront of the beam, after propagation, they are effectively manifested in an intensity mismatch in terms of their influence on the Doppler shift. For an aberration-free beam, the fiber–collimator distance with optimal backcoupled fraction ($\delta d_{\mathrm{fc}} = 0\,\mu\mathrm{m}$) corresponds to the same distance with balanced intensities of forward- and backward-traveling beams. In the presence of aberrations, those distances are not the same, such that for $\delta d_{\mathrm{fc}} = 0\,\mu\mathrm{m}$ there is a residual Doppler-shift mainly due to the intensity imbalance.

#### 4.4.3.4 Measurement of collimator performance

We measured the collimator performance at 410 nm by analyzing intensity profiles in the caustic measurement shown in Fig. 4.19. If the radius $W(z)$ of a beam with an arbitrary mode decomposition is defined through the second-moments of the transverse intensity distribution (“D4$\sigma$-method” according to the ISO standard [103]), any beam radius $W(z)$ follows the hyperbolic propagation law [88, 104, 105],

$$W(z) = W_0\sqrt{1 + (z - z_0)^2/z_{\mathrm{R}}^2}, \quad \text{with} \quad z_{\mathrm{R}} = \frac{\pi W_0^2}{M^2\lambda}, \tag{4.8}$$

where $z_0$ is the waist position, $z_{\mathrm{R}}$ the Rayleigh length, and $W_0$ the beam radius at the waist (or simply waist radius). The factor $M^2$ entering the above equation through $z_{\mathrm{R}}$ is denoted as the beam quality factor. This factor relates the waist radius $W_0$ as defined above to the waist radius $w_0$ of a Gaussian beam with the same Rayleigh length $z_{\mathrm{R}}$ through $W_0 = M\,w_0$. As we will find below, the collimator used here has a value of $M^2$ very close to one, and thus the values of $W_0$ and $w_0$ are identical within the measurement uncertainty in this special case.

For an impinging beam with a beam quality factor $M_0^2$ passing through optics with spherical aberrations of strength $S$ as defined in Eq. (4.7), the beam quality factor $M^2$ of the transmitted beam is modified by an additional contribution $M_S^2$ as [102]

$$M^2 = \sqrt{\left(M_0^2\right)^2 + \left(M_S^2\right)^2}, \quad \text{with} \quad M_S^2 \approx \sqrt{2}S. \tag{4.9}$$

The beam radius $W(z)$ is determined for the orthogonal $x$ and $y$ transverse directions according to the second-moment definition with a self-convergent-width factor [106] of $F_s = 3$.

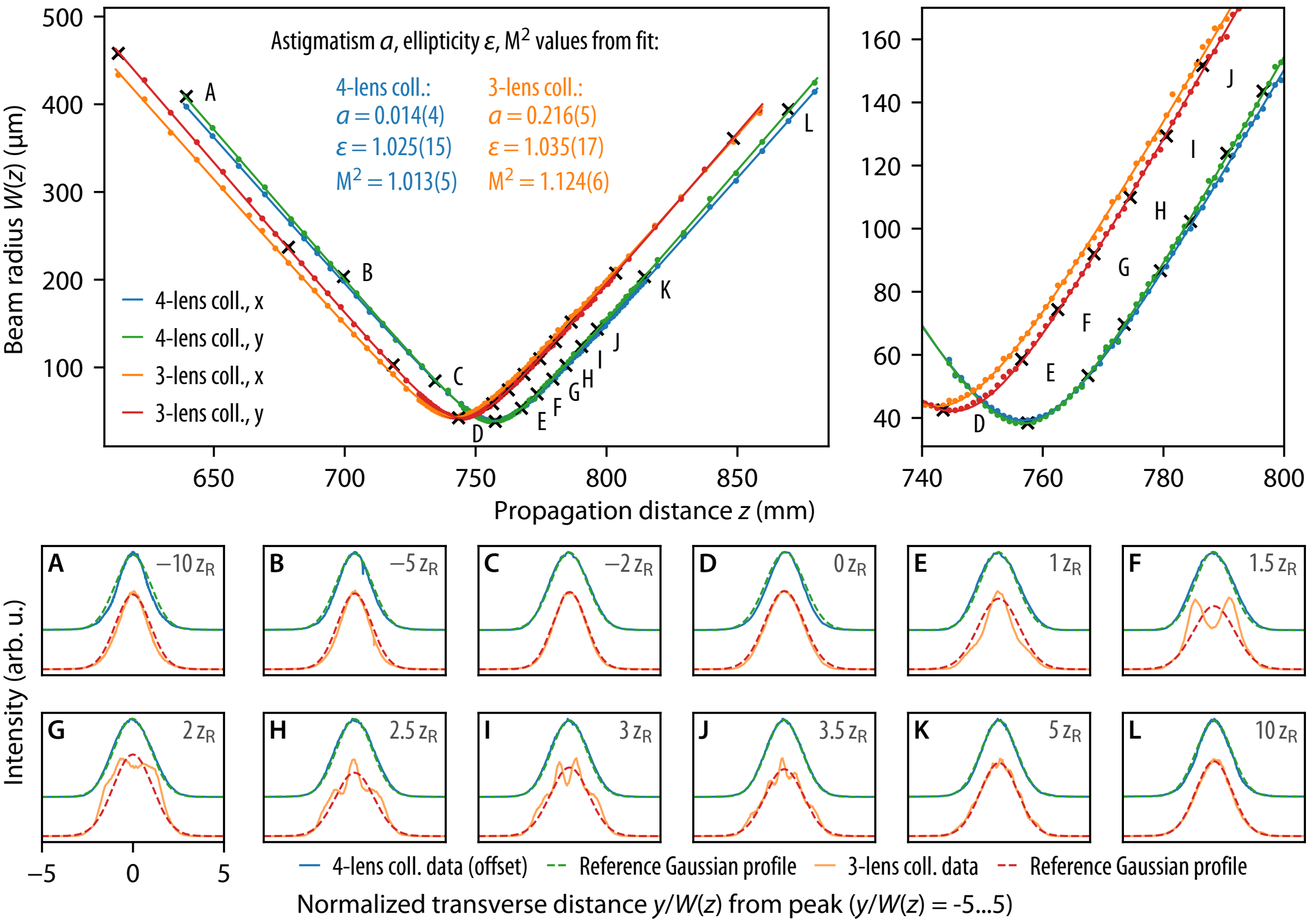

Figure 4.19: Caustic measurement of the three-lens and four-lens collimators at 410 nm. **Top:** The beam radius $W(z)$, obtained from second-order moments in the $x$ and $y$ directions, is shown versus the propagation distance $z$ (with different offset for the two collimators) after the $f = 750$ mm focusing lens. The plot on the right shows the same data in detail in the region of up to four Rayleigh lengths $z_{\mathrm{R}}$ after the focus, where distortions due to spherical aberrations are expected to be most pronounced. The fitted beam quality value for the three-lens collimator is $M^2 = 1.124(6)$, caused by spherical aberrations with $|S| \approx 0.34$, whereas for the four-lens collimator the beam quality of $M^2 = 1.013(5)$ is not limited by spherical aberrations, but only by the fiber mode profile. **Bottom:** Normalized intensity profiles at selected propagation distances **A** to **L** (from $-10\,z_{\mathrm{R}}$ to $+10\,z_{\mathrm{R}}$, marked on top plots) are depicted, along with reference Gaussian beams of $1/\mathrm{e}^2$ intensity radius $W(z)$.

This beam radius determination is performed at different positions around the focus of the $M^2$ lens ($f = 750$ mm), corresponding to propagation distances from $-10\,z_{\mathrm{R}}$ to $+10\,z_{\mathrm{R}}$ around the waist position. From the fit according to Eq. (4.8) the beam waist radii $W_{0,x}$ and $W_{0,y}$, waist positions $z_{0,x}$ and $z_{0,y}$, and beam quality factors $M_x^2$ and $M_y^2$ are extracted. These values determine the ellipticity $\varepsilon$, astigmatism $a$, and combined beam quality factors $M^2$ as

$$\varepsilon = \max\left(\frac{W_{0,x}}{W_{0,y}}, \frac{W_{0,y}}{W_{0,x}}\right), \quad a = \frac{z_{0,x} - z_{0,y}}{(z_{\mathrm{R},x} + z_{\mathrm{R},y})/2}, \quad M^2 = \sqrt{M_x^2\, M_y^2}. \tag{4.10}$$

The fits and the determined parameters are shown at the upper part of Fig. 4.19. The measured ellipticity of a few percent is in agreement with the slightly elliptical beam from our polarization-maintaining fiber. The reduced astigmatism of the four-lens collimator, $\alpha = 0.014(4)$, was achieved with the help of alignment using the caustic measurement, compared to the three-lens collimator with $\alpha = 0.216(5)$ where the collimator was aligned with the help of centering the observed distortions in the intensity profiles. For the three-lens collimator we find $M^2 = 1.124(6)$ which corresponds to $|S| = 0.33(1)$ according to Eq. (4.9). Aberrations are also revealed as distortions in intensity around the focus after the $M^2$ lens, see bottom of Fig. 4.19 where the intensity profiles at selected positions A to L, as marked on the top, are shown. For the three-lens collimator we observe the characteristic intensity profiles for a beam with spherical aberrations [107]. Note that distortions in intensity appear only in the region between $z_\mathrm{R}$ and $4\,z_\mathrm{R}$ after the focus, which demonstrates that the manifestation of aberrations cannot be readily observed in intensity of collimated beam without the caustic measurement. In the simulations of caustic intensity profiles for the given three-lens collimator design, we find a larger dip depth and more distorted intensity profiles than observed, which we attribute to high sensitivity to single-mode fiber input parameters. For the four-lens collimator, no beam profile distortions are observed in agreement with simulations using the given lens design, and the beam quality factor $M^2 = 1.013(5)$ is limited only by the not exactly Gaussian profile from the single-mode fiber [35, 36].

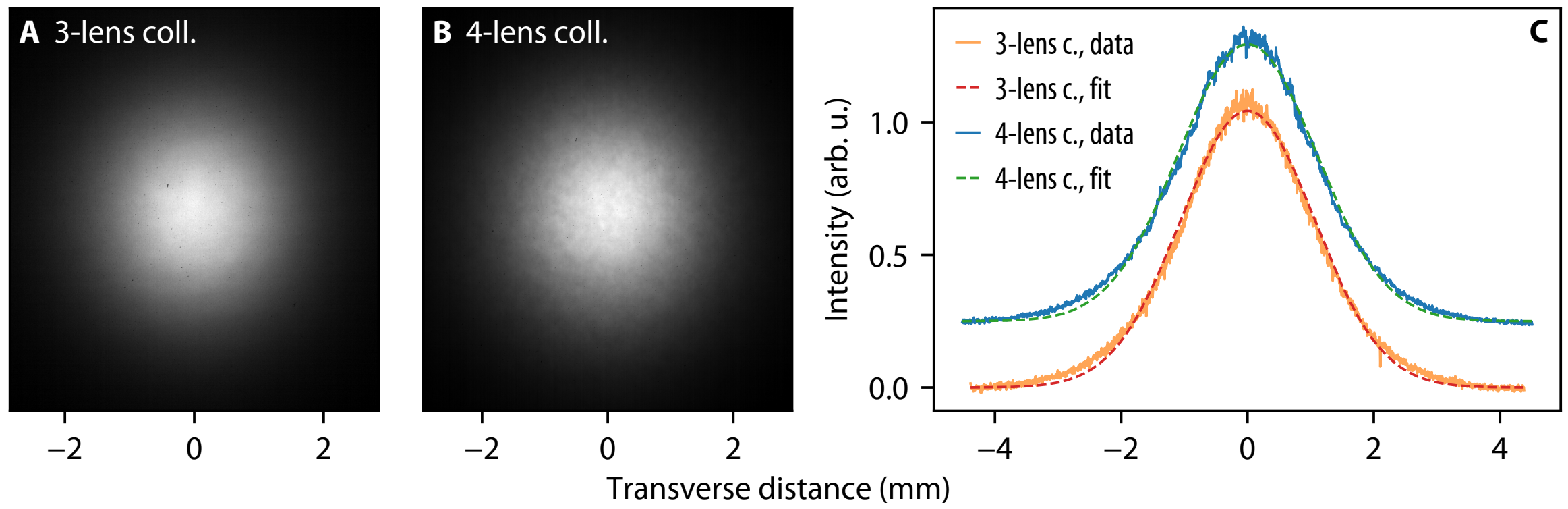

Figure 4.20: Measured beam profile of the collimated beam in the spectroscopy region of the active fiber-based retroreflector (12 cm after the collimator) for: (**A**) 3-lens collimator, (**B**) 4-lens collimator. A weak "orange-peel" structure is visible in both beam profiles, but more pronounced on the 4-lens collimator beam profile. Cuts through the beam profiles are shown in (**C**) along with a Gaussian fit to the data (offset for 4-lens collimator).

The collimated beam intensity profiles in the spectroscopy region (corresponding to 12 cm propagation distance after the collimator) are shown in Fig. 4.20 (A) and (B) for the three-lens and four-lens collimator, respectively. Though no performance shortcomings of the four-lens collimator are observed in the caustic measurement, a weak residual "orange-peel" structure is observed on the collimated beam of both collimators, which is more pronounced for the four-lens collimator. This structure is barely observed on cuts through intensity profiles as shown in Fig. 4.20 (C). We observed similar but much stronger deviations in beam profiles from aspheric lenses where these mid-spatial-frequency errors are more pronounced, as well as for some other collimators. It is important to note that these lens imperfections imprint phase distortions which may disturb the wavefront-retracing property of the AFR. After propagation, these phase distortions transform into intensity distortions and may introduce a residual Doppler shift. Therefore, it is important to ensure best possible lens polishing quality, minimizing mid-spatial frequency errors. In our case, the lens surfaces and polishing processes responsible for deviations observed in Fig. 4.20 (B) could not yet be identified and remain under investigation. Ultimately, for our experiment only the velocity-resolved spectroscopy measurement provides certainty on the suppression factor of the Doppler shift. In the preliminary data analysis discussed in this thesis, we see no evidence for a residual Doppler shift within the uncertainty of the measurement[1].

As discussed at the beginning of this section, the backcoupled light fraction gives an important figure of merit of the collimator performance. Compared to the three-lens collimator with a backcoupled fraction of 94.0(1.2) %, the backcoupled light fraction of the four-lens collimator is measured to be 99.3(1.2) %. Note that this number gives the spatial overlap of the forward- and backward-traveling beams in the AFR, with the known transmission losses of the beam path from the spectroscopy region to the backcoupling APD are taken into account. The uncertainty of 1.2 % is deduced from the quadrature sum of uncertainties for position-dependent photodiode sensitivity, beamsplitter transmission, fiber attenuation, as well as AR coating uncertainties of the fiber coupling lens, fiber tips, and collimator lenses.

### 4.4.4 Retroreflection control and stabilization

In order to achieve the wavefront-retracing retroreflection in the AFR, it is necessary to adjust the distance between the fiber and the collimator (collimation distance) such that the position of the flat wavefront of the collimated beam is at the HR mirror. Moreover, the horizontal and vertical (tip and tilt) directions of the HR mirror need to be oriented such that the wave vectors of the forward- and backward-propagating waves are antiparallel to each other. If both conditions (collimation and tip-tilt alignment) are optimized, the backcoupled light fraction is maximized.

The stabilization of the HR mirror tip-tilt orientation is described in detail in the previous work [28]. In short, modulating the two piezo actuators for the tip-tilt movement of the HR mirror mount with weak signals of different frequencies (producing maximal angular misalignment of approximately ±1 µrad) and detecting this modulation in the backcoupled light with two lock-in amplifiers, two error signals are generated which are used for the tip and tilt stabilization feedback loops. We tried improving the bandwidth of the feedback by using higher modulation frequencies. Due to the entanglement between horizontal and vertical movement of the mirror mount, as well as due to mechanical resonances, a crosstalk between the horizontal and vertical piezo actuators is present. In order to find frequencies with minimal crosstalk between horizontal and vertical modulation, we measure the corresponding transfer functions of the HR mirror assembly. For this measurement we use an auxiliary laser beam hitting the HR mirror from the back side to allow for an in-situ measurement (with PMT being removed) under a small (≈5°) angle, and detect the reflection with a position-sensitive detector while sweeping the frequency of horizontal and vertical piezo actuators. Though higher modulation frequencies of 1.52 kHz (vertical) and 2.09 kHz (horizontal) with minimal crosstalk (typically 10–30 % amplitude ratio of horizontal to vertical error signals) could be identified, the feedback bandwidth could not be improved due to large mechanical resonances around 30 Hz caused by the rotatable geometry of the whole AFR setup.

Fig. 4.21 shows the performance of the tip-tilt stabilization, where in (A) and (B) the in-loop error signals are plotted. The bandwidth of stabilization as deduced from the in-loop error signals is around 10 Hz. However, no significant noise suppression is observed on the spectrum of the backcoupled light shown in (C), with even a slight increase of noise visible for low frequencies when the tip-tilt stabilization is switched on. Only a small decrease of noise is visible on the backcoupled light for resonances around 30 Hz. Apart from Rayleigh scattering from fiber discussed below, the unobserved noise suppression in the backcoupled light could originate from the large amount of noise common to both directions which cannot be suppressed due to large cross-talk. Though for inspection of this issue no out-of-loop measurement of the tip-tilt stabilization is available in our setup, the performance of the tip-tilt feedback is clearly observed in the backcoupled light when scanning the piezo actuator controlling the fiber–collimator distance as described below and shown in Fig. 4.22. Likewise, during a typical hour-long precision spectroscopy measurement, the tip-tilt feedback maintains the retroreflecting condition.

[1] The preliminary analysis of the 2S-6P measurement (see Chapter 6) results in a residual Doppler slope $|\kappa| < 3.5\,\mathrm{Hz/(m/s)}$, corresponding to a Doppler shift of $< 700\,\mathrm{Hz}$ for an atom with a speed of $v = 200\,\mathrm{m/s}$.

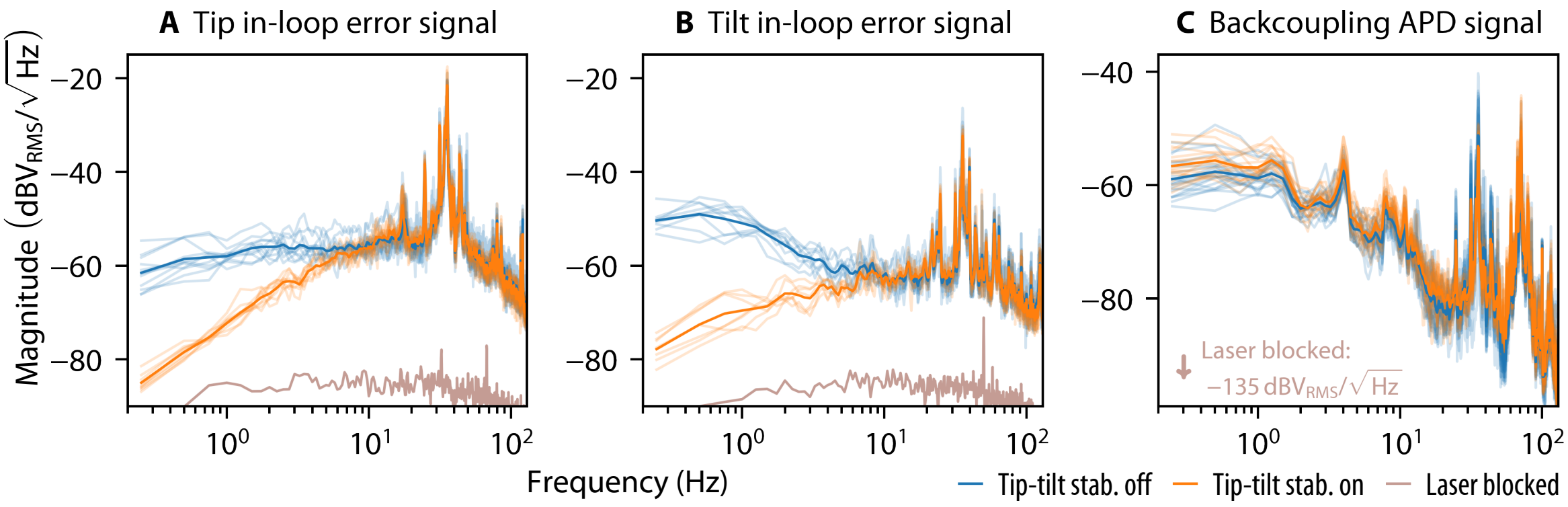

Figure 4.21: Performance of the tip-tilt stabilization of the active fiber-based retroreflector. Spectra of the in-loop error signals produced by the lock-in amplifiers (1 ms time constant) for the horizontal and vertical (tip and tilt) feedbacks of the HR mirror are shown in (**A**) and (**B**). The spectrum of the backcoupled light on the backcoupling APD (see Fig. 4.15) is shown in (**C**). All spectra are shown with tip-tilt stabilization switched on (orange line), off (blue line) and with the laser blocked (gray line). Each faint line is a 15 s average with a resolution bandwidth of 0.25 Hz, whereas heavy lines are the average of all data. Large resonances around 30 Hz are observed on all signals, limiting the feedback bandwidth to around 10 Hz as deduced from the in-loop error signals.

In the previous setup [28] the collimation distance was adjusted by using a remote-controlled motor and maximizing the observed signal on the backcoupling APD. With this procedure the backcoupled signal could typically be optimized within ≈1 %. We improved the distance control by adding a piezo actuator to the precision motor, which is now used for pre-alignment only. In order to determine the optimal piezo voltage, we typically scan the applied voltage with a frequency of 1 Hz over a period of 30 s such that the fluctuations on the backcoupling APD are averaged out, see Fig. 4.22. Due to the not exactly on axis translation of the fiber, we observe tip-tilt misalignment on the in-loop error signals shown in (C) and (D), resulting in a large drop of the backcoupled light fraction (3 %) shown in (A) with tip-tilt stabilization switched off (brown lines). With the tip-tilt stabilization switched on (blue and orange lines), the drop of the backcoupled light fraction by ±0.5 % for ±1.5 µm around $\delta d_{\mathrm{fc}} = 0$ µm agrees with simulations from Fig. 4.18. From the fitted dashed line the optimal piezo voltage is determined, allowing to set the fiber–collimator distance to within approximately ±0.2 µm of the optimal value corresponding to ±0.1 % of the maximum back-coupled light fraction value. In our setup we observe only slow drifts of the optimal collimation distance on the order of 0.2 µm per hour (correlated with temperature) such that typically an adjustment is performed every 1-2 hours with no need of active stabilization.

The observed modulation on the backcoupled light fraction in Fig. 4.22 (A) demonstrates

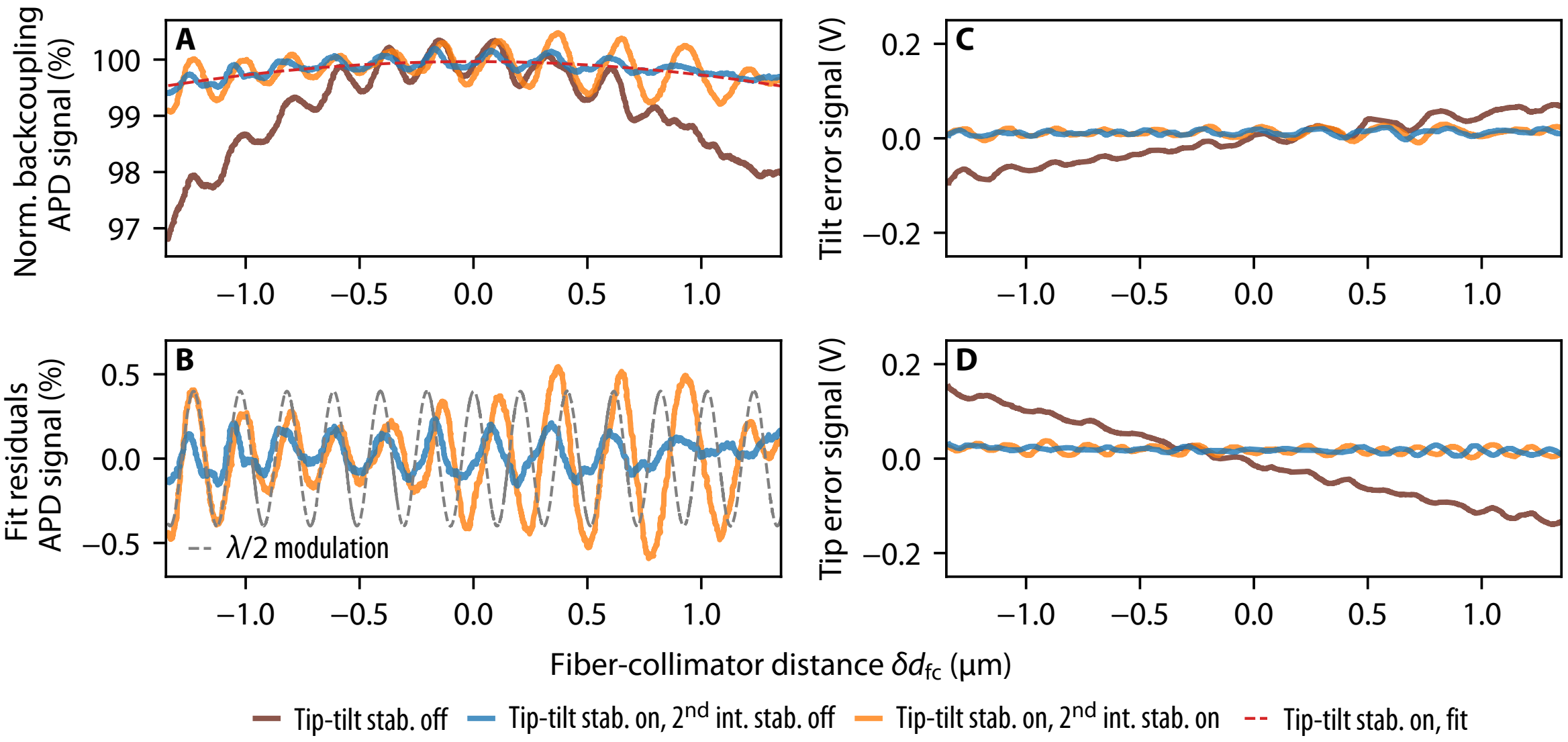

Figure 4.22: Demonstration of the improved fiber–collimator distance control using a piezo actuator. All plots show the recorded data when scanning the voltage of the piezo actuator with a frequency of 1 Hz over an averaging time of 30 s. The common $x$-axis has been converted from the applied voltage and leverage factor to the fiber–collimator distance $\delta d_{\text{fc}}$, with zero $\delta d_{\text{fc}}$ corresponding to maximum backcoupling. In (**A**) the normalized signal of the backcoupled light is shown. If the tip-tilt stabilization is switched off (brown lines), clear tip-tilt misalignment is observed on the in-loop error signals of the tip-tilt stabilization shown in (**C**) and (**D**), leading to a large drop in the backcoupled light fraction of ≈3 % in (A) over the full distance range. With tip-tilt being stabilized (blue and orange lines), this drop is reduced to ≈0.5 % in agreement with simulations from Fig. 4.18. The dashed line in (A) shows a quadratic fit to the data with tip-tilt stabilization switched on. A clearly visible modulation is observed on all of the signals, originating from Rayleigh scattering inside the fiber leading to an etalon-like effect. Along with the fit residuals of the backcoupled fraction in (**B**), the expected $\lambda/2$ modulation is drawn (dashed gray line) which also reveals the nonlinearity of the piezo actuator. As expected, the modulation is stronger when the 2nd intensity stabilization (to the PMT after the HR mirror) is switched on (orange line), as compared to the case with intensity not stabilized (blue line).

how in addition to the precise distance control, the piezo actuator provides the possibility to inspect the AFR for optical etalons. The corresponding $\lambda/2$ modulation is depicted along with fit residuals in (B), not exactly matching all the data due to expected nonlinearities of the piezo actuator of ∼20 % over the full range. In order to avoid etalons, we use only AR-coated optics and place all photo-detectors at large angles ($\gtrsim 10°$). The fiber tips are angle-cleaved under 8° and AR-coated. After a thorough investigation, we found that the modulation we observe originates from Rayleigh scattering inside the fiber. A small fraction on the order of $10^{-3}$ of the scattered light from randomly distributed scattering points inside the fused silica of the fiber is guided forward and backward in the fiber mode [108–110], interfering with the strong reflection from the HR mirror. As expected, stabilizing the intensity after the fiber leads to an increased modulation (orange lines) as compared to the case with the stabilization switched off (blue lines). Since the amplitude of fluctuations from Rayleigh scattering in the backcoupled light increases with the square root of the fiber length, we use an as short as possible fiber. The detailed report of our investigation of the etalon-like effect from Rayleigh

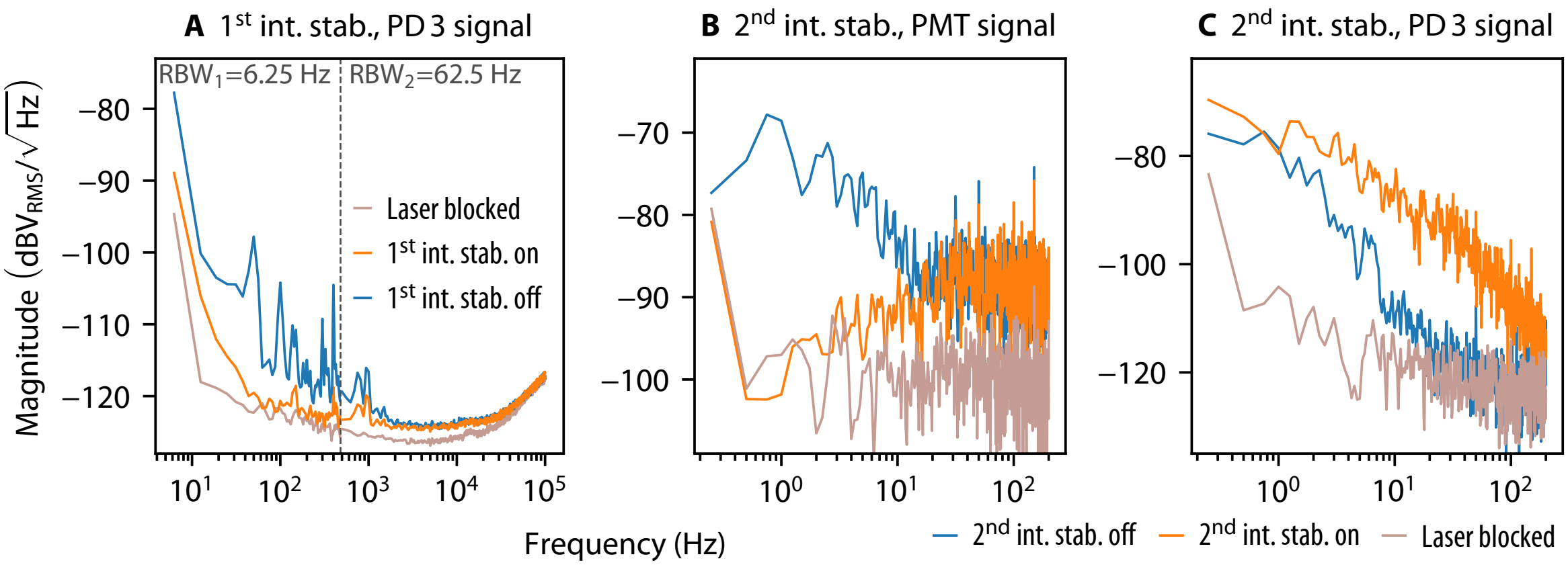

Figure 4.23: Spectra of the PMT and PD 3 photodetector signals showing the performance of the two intensity stabilizations of the active fiber-based retroreflector, shown for active stabilization on (orange line) or off (blue line) along with the background noise (gray line). (**A**) Spectrum on the out-of-loop PD 3 detector demonstrating the performance of the first high-bandwidth (∼30 kHz) intensity stabilization to PD 1 (with second intensity stabilization switched off). The frequency interval up to 500 Hz is plotted with a resolution bandwidth ($RBW_1$) of 6.25 Hz and with $RBW_2 = 62.5$ Hz for higher frequencies. (**B**, **C**) Second intensity stabilization of wavefront-retracing beams using the PMT signal after the HR mirror (with first intensity stabilization switched on). The in-loop PMT spectrum in (B) demonstrates the bandwidth of around 10 Hz while additional noise is imprinted on PD 3 as shown in (C).

scattering in fiber will be subject of a future publication.

### 4.4.5 Intensity stabilization

Since the fluorescence signal depends on the intensity of the exciting laser beams, it is advantageous to implement an intensity stabilization in the AFR. Various sources like pointing fluctuations, polarization drifts, electronic noise in the laser system or frequency-dependent AOM efficiencies lead to intensity fluctuations already before the light reaches the AFR fiber. In the previous setup only the intensity of this light before the fiber was stabilized [28]. However, the intensity of the wavefront-retracing beams in the AFR is then still affected by the coupling efficiency of the AFR fiber, subject to pointing fluctuations, as well as by the frequency-dependent interference from the Rayleigh scattering mentioned above. In order to stabilize the intensity of the wavefront-retracing beams, we use the summed signal of all four quadrants of the PMT behind the HR mirror (see Fig. 4.15). Due to the low power reaching the PMT (70–400 pW) and the resulting large shot noise, only a low-bandwidth stabilization can be achieved such that the high-bandwidth first intensity stabilization (with PD 1 in Fig. 4.15 serving as detector) is still needed to suppress other noise. Two different electro-optic modulators (EOM 1 and EOM 2 in Fig. 4.15) are being used as actuators, with a low-pass filter placed before EOM 2 to adapt for lower feedback bandwidth.

Fig. 4.23 shows the performance of the two intensity stabilizations by comparing the spectra for the corresponding stabilization switched on (orange line) and off (blue line). In (A) the spectrum of the out-of-loop detector PD 3 demonstrates the (∼30 kHz bandwidth of the first intensity stabilization (with the second intensity stabilization switched off). The spectrum of the PMT signal which is the in-loop detector of the second intensity stabilization is shown in (B), where the feedback bandwidth of ∼10 Hz is observed from the merging point of the blue and orange data lines. The second intensity stabilization imprints the noise of the stabilized signal on the PMT to the light before fiber observed on the spectrum of PD 3 signal in (C). Note that also shot noise from the PMT signal is imprinted which cannot be fully avoided. Using a low-pass filter with a corner frequency of 80 Hz in the second intensity stabilization we found a compromise between a sufficient suppression of noise and an acceptable additional imprinting of shot noise at lower frequencies.

When scanning the atomic resonance, at each point the frequency applied to the AOM in Fig. 4.15 is switched, thereby causing a short dead time (in our case ∼400 µs using the signal generator[1]) where no light passes through the AOM. The error signal generated from noise during that dead time perturbs the feedback loops. To avoid this, we use a pulse generator triggered to frequency switching to place the feedback loops on hold during and after the dead time, adapted to each of the feedback loops (20 ms for 1st intensity stabilization, 22 ms for 2nd intensity stabilization, 50 ms for tip-tilt stabilization).

Furthermore, we implemented the possibility to automatically switch the power of the spectroscopy laser beams by using digital step attenuators[2] connected in series at the output of the photodiodes. In this way, the signal levels of the feedback loops remain unchanged without the need to modify the feedback loop parameters. The power switching allows us to perform simultaneous spectroscopy measurements with different laser powers, thereby investigating the light force shift discussed in Chapter 3.

### 4.4.6 Polarization monitor

For the 2S-6P transition measurement, we try to achieve the best possible linear polarization of the wavefront-retracing beams. Any residual circularly polarized light leads to a first-order Zeeman shift which vanishes for fully linearly polarized light (see Section 6.2.4.3). Furthermore, the linear polarization rotation angle at the position of the atoms, $\theta_{\mathrm{L}}$, strongly influences the quantum interference line shift (see Section 6.2.3). One way to achieve a well-controlled polarization in the AFR would be to place a polarizer with high polarization extinction ratio (PER) after the collimator. However, such a polarizer might lead to optical etalons, wavefront distortions, and residual intensity fluctuations, and requires additional space currently not available in our setup. Therefore, we choose to work only with a well-characterized polarization-maintaining (PM) fiber.

Here, we use the Stokes formalism to describe the evolution of the state of polarization, which is at any point given by a Stokes vector $\boldsymbol{S} = (S_0, S_1, S_2, S_3)$ [111]. In particular, the residual circularly polarized light fraction is given by $S_3/S_0$, where $S_0$ is the total intensity and $S_3$ the intensity difference between right and left circularly polarized light. The orientation of the polarization ellipse, or linear polarization rotation angle, $\theta$, is given by the other two Stokes parameters as $\tan 2\theta = S_2/S_1$. For fully polarized light, as is the case here, $S_1^2 + S_2^2 + S_3^2 = S_0^2$, and the polarization extinction ratio PER is related to the residual circularly polarized light fraction as $|S_3/S_0| \approx 2\sqrt{1/\mathrm{PER}}$ for $\mathrm{PER} \gg 1$.

In order to achieve a high PER after a PM fiber, it is important to use incoming light with a high PER, and to align the linear polarization rotation angle of the incoming light to the polarization-maintaining axis of the fiber. For our PM fiber, this alignment has to be better than 1°, which we achieve by placing the polarizers in the PSPU (see Fig. 4.15) on rotation mounts. Furthermore, we find that the coupling lens as well as mirrors and beamsplitters after the polarizers may distort the input polarization due to stress-induced birefringence. We minimize this effect by also placing the fiber onto a rotation mount, and systematically varying the orientation of both the polarizers and the fiber mount. For their optimal orientations, the polarization is aligned to both the stress-induced birefringence axis of optical components after the polarizers (such that the resulting effect from their birefringence is minimized), and the polarization-maintaining axis of the fiber.

[1] Rohde & Schwarz SMC100A.

[2] Mini-Circuits ZX76-31R5A-PPS+.

Typically, for PM fibers in the near UV, the specified PER of output polarization (for optimal alignment of input polarization) is around 20 dB, corresponding to a circularly polarized fraction of $|S_3/S_0| < 20\,\%$. For our PM fibers, we find that stress-induced birefringence at the fiber connectors mostly limits the achieved polarization extinction ratio and that in a sample of commercial, connectorized fibers some have a higher PER of >26 dB, corresponding to $|S_3/S_0| < 10\,\%$. Likewise, the fluctuations over time of the orientation $\theta$ of the polarization ellipse at the fiber output are found to be below 3°. Such a specially selected fiber is used in the AFR. Furthermore, we observe that by thermally isolating the part of the fiber outside of vacuum, only slow polarization drifts occur on the time scale of hours.

To monitor the polarization in the AFR, we implemented polarimetry of the backcoupled light by placing a polarimeter[1] to measure the polarization state $\boldsymbol{S}_{\text{meas}}$ in the unused beam path before BS 2 as depicted in Fig. 4.15. This allows a partial reconstruction of the polarization state $\boldsymbol{S}_{\text{atom}}$ after the collimator, and thus the polarization state of the light interacting with the atoms. The full polarization state could in principle be obtained from polarimetry of the light leaking through the HR mirror, assuming the effects of the HR mirror on the transmitted polarization state are sufficiently characterized and are constant over time. Since this would further increase the complexity of the setup and require an in-vacuum polarimeter for <100 pW of laser power, this approach was not implemented here.

In the following, we summarize how the measured polarization state $\boldsymbol{S}_{\text{meas}}$ of the backcoupled light relates to the polarization state $\boldsymbol{S}_{\text{atom}}$ of the light interacting with the atoms. The light starts with a well-characterized polarization state $\boldsymbol{S}_{\text{in}}$, which in our setup is taken to be between beamsplitters BS 2 and BS 3 (see Fig. 4.15). It then passes various optical components, including the fiber and the collimator, in the forward direction, resulting in polarization state $\boldsymbol{S}_{\text{atom}}$. It is then retroreflected at the HR mirror and passes through the same components in the backward direction, resulting in a polarization state $\boldsymbol{S}_{\text{back}}$ at the same position as where $\boldsymbol{S}_{\text{in}}$ is defined. Finally, the backcoupled light passes through additional components before reaching the polarimeter, where the polarization state $\boldsymbol{S}_{\text{meas}}$ is measured. We separate the problem into two parts. First, the reconstruction of $\boldsymbol{S}_{\text{atom}}$ from $\boldsymbol{S}_{\text{back}}$ is treated, which involves only components common to both the forward- and backward-traveling direction. This part includes the fiber, which here is the dominant source of polarization drifts, caused mainly by temperature fluctuations in the laboratory. Second, the polarization evolution from $\boldsymbol{S}_{\text{back}}$ to $\boldsymbol{S}_{\text{meas}}$ is considered, which involves components not common to the forward- and backward-traveling direction, and not subject to significant polarization drifts in our setup.

The total birefringence effect of any number of non-polarizing components can be described by the combined Mueller matrix $\boldsymbol{R}(\phi)\,\boldsymbol{\Gamma}(\delta,\beta)$ [112, 113]. $\boldsymbol{R}(\phi)$ is the Mueller matrix of a rotator (such as an element with circular birefringence) with rotation angle $\phi$, and $\boldsymbol{\Gamma}(\delta,\beta)$ is the Mueller matrix of a linear retarder (such as a wave plate) with retardance $\delta$ and the birefringence axis oriented at an angle $\beta$. The matrix $\boldsymbol{R}(\phi)$ does not change the circularly polarized fraction $S_3/S_0$, but only rotates the orientation of the polarization ellipse. Note that the parameters $\phi$, $\delta$, and $\beta$ are only effective parameters describing the overall polarization behavior of the system, and are not necessarily linked to the circular or linear birefringence of each element (e.g., the fiber).

[1] Schäfter+Kirchhoff SK010PA-UV.

Importantly, the parameter $\phi$, which describes the effective circular birefrigence, can be nonzero even if none of the physical objects in the system exhibit circular birefringence, e.g., when cascading multiple linear retarders. This can be easily understood when visualized on the Poincaré sphere: The action of a wave plate corresponds to a rotation about an axis in the equatorial plane, while the action of a rotator corresponds to a rotation about the north–south axis. The combination of two or more wave plates can be described as a single rotation, with the axis of this rotation however not generally lying in the equatorial plane. Thus, the resulting rotation cannot be described by a single wave plate, and the combination of wave plate and rotator is necessary.

We can therefore describe the forward propagation through the common optical system by the Mueller matrix

$$\boldsymbol{F}(\delta, \beta, \phi) = \boldsymbol{R}(\phi)\boldsymbol{\Gamma}(\delta, \beta), \tag{4.11}$$

such that

$$\boldsymbol{S}_{\mathrm{atom}} = \boldsymbol{F}(\delta, \beta, \phi)\boldsymbol{S}_{\mathrm{in}}. \tag{4.12}$$

Because the coordinate system is different for the return path, the common optical system is not described by the same Mueller matrix as in the forward direction, but by its representation in the coordinate system of the returning beam. For the linear retarder this corresponds to flipping the sign of the orientation to the vertical, $\beta \to -\beta$, while the retardation stays unchanged. On the other hand, because for a reciprocal medium the direction of rotation is always the same when referenced to the direction of propagation, the Mueller matrix for the rotator for the return path is identical to that of the forward path. Thus, the combined Mueller matrix for forward propagation through the system, retroreflection, and backward propagation through the system reads

$$\boldsymbol{B}(\delta, \beta, \phi) = \boldsymbol{\Gamma}(\delta, -\beta)\boldsymbol{R}(\phi)\boldsymbol{M}\boldsymbol{R}(\phi)\boldsymbol{\Gamma}(\delta, \beta), \tag{4.13}$$

and thus

$$\boldsymbol{S}_{\mathrm{back}} = \boldsymbol{B}(\delta, \beta, \phi)\boldsymbol{S}_{\mathrm{in}}. \tag{4.14}$$

Here, $\boldsymbol{M} = \boldsymbol{\Gamma}(\pi, 0)$ is the Mueller matrix of a mirror [111]. We note that

$$\boldsymbol{R}(\phi)\boldsymbol{M}\boldsymbol{R}(\phi) = \boldsymbol{M}, \tag{4.15}$$

that is, the orientation of the polarization ellipse after forward propagation cannot be determined from the backcoupled light, since its rotation in the forward direction is exactly reversed in the backward direction. However, since the shape of the ellipse is not affected, information about the circularly polarized fraction after forward propagation is still contained in the backcoupled light.

Consider first a perfect incoming horizontal or linear polarization, $(S_2/S_0)_{\text{in}} = \pm 1$, $(S_2/S_0)_{\text{in}} = (S_3/S_0)_{\text{in}} = 0$, and, for simplicity, $(S_0)_{\text{in}} = 1$. This choice results in a polarization state after forward propagation of

$$\begin{aligned}\boldsymbol{S}_{\text{atom}} &= \boldsymbol{F}(\delta,\beta,\phi)\boldsymbol{S}_{\text{in}} \\ &= \begin{pmatrix} 1 \\ \pm\cos(2\beta)\cos(2\beta+2\phi) \pm \sin(2\beta)\cos(\delta)\sin(2\beta+2\phi) \\ \pm\cos(2\beta)\sin(2\beta+2\phi) \mp \sin(2\beta)\cos(\delta)\cos(2\beta+2\phi) \\ \pm\sin(2\beta)\sin(\delta) \end{pmatrix}, \end{aligned} \tag{4.16}$$

and a backcoupled polarization state of

$$\begin{aligned}\boldsymbol{S}_{\text{back}} &= \boldsymbol{B}(\delta,\beta,\phi)\boldsymbol{S}_{\text{in}} \\ &= \begin{pmatrix} 1 \\ \pm\sin^2(2\beta)\cos(2\delta) \pm \cos^2(2\beta) \\ \mp\sin(4\beta)\sin^2(\delta) \\ \mp\sin(2\beta)\sin(2\delta) \end{pmatrix}. \end{aligned} \tag{4.17}$$

As expected, the second and third components of $\boldsymbol{S}_{\text{atom}}$, describing the orientation $\theta_{\text{atom}} \equiv \theta_{\text{L}}$ of the polarization ellipse after forward propagation, depend on the effective circular birefringence rotation angle $\phi$, while $\boldsymbol{S}_{\text{back}}$ is independent of $\phi$ and only depends on $\delta$ and $\beta$. Thus, $\theta_{\text{L}}$ cannot be determined from $\boldsymbol{S}_{\text{back}}$. Even for the case of $\phi = 0$, $\theta_{\text{L}}$ generally cannot be reconstructed, as can be seen by, e.g., considering the case of a half-wave plate ($\delta = \pi$).

The fourth component of $\boldsymbol{S}_{\text{atom}}$, $(S_3/S_0)_{\text{atom}}$, describing the circularly polarized fraction, only depends on $\delta$ and $\beta$, and we can solve for $(S_3/S_0)_{\text{atom}}$ given $\boldsymbol{S}_{\text{back}} = (1, s_1, s_2, s_3)$. It turns out that the absolute value of $(S_3/S_0)_{\text{atom}}$, but not its sign, can be determined in this way. Furthermore, there are two solutions for $|(S_3/S_0)_{\text{atom}}|$, given by

$$|(S_3/S_0)_{\text{atom}}|_\pm = A_\pm |s_3| \sqrt{s_3^4 + s_2^2\left(2 + s_3^2 \pm 2\sqrt{1 - s_2^2 - s_3^2}\right)}, \tag{4.18}$$

$$\text{with} \quad A_\pm = \sqrt{\frac{2s_2^2 + s_3^2 \pm s_3^2\sqrt{1 - s_2^2 - s_3^2}}{2s_3^2(s_2^2 + s_3^2)(4s_2^2 + s_3^4)}}. \tag{4.19}$$

We find that $|(S_3/S_0)_{\text{atom}}|_- \leq 1/\sqrt{2}$ and $|(S_3/S_0)_{\text{atom}}|_+ \geq 1/\sqrt{2}$. If we can constrain the value of $|(S_3/S_0)_{\text{atom}}|$, e.g., through an auxiliary measurement, the ambiguity between $|(S_3/S_0)_{\text{atom}}|_+$ and $|(S_3/S_0)_{\text{atom}}|_-$ can be resolved. This is the case for the AFR, where $|(S_3/S_0)_{\text{atom}}| < 20\,\%$, and we will use the solution $|(S_3/S_0)_{\text{atom}}|_-$ for $|(S_3/S_0)_{\text{atom}}|$ in the following.

Keeping only the lowest-order terms of $s_2 \equiv (S_2/S_0)_{\text{back}}$ and $s_3 \equiv (S_3/S_0)_{\text{back}}$, $|(S_3/S_0)_{\text{atom}}|_-$ reduces to

$$|(S_3/S_0)_{\text{atom}}|_- \simeq \frac{1}{2}\sqrt{(S_2/S_0)^2_{\text{back}} + (S_3/S_0)^2_{\text{back}}}. \tag{4.20}$$

This approximation deviates by less than $2\,\%$, corresponding to our typical measurement uncertainty, from the full solution for $|(S_3/S_0)_{\text{atom}}| < 0.3$. The same approximation can also be directly retrieved by assuming $|\beta| \ll 1$ in Eqs. (4.16) and (4.17). The latter derivation

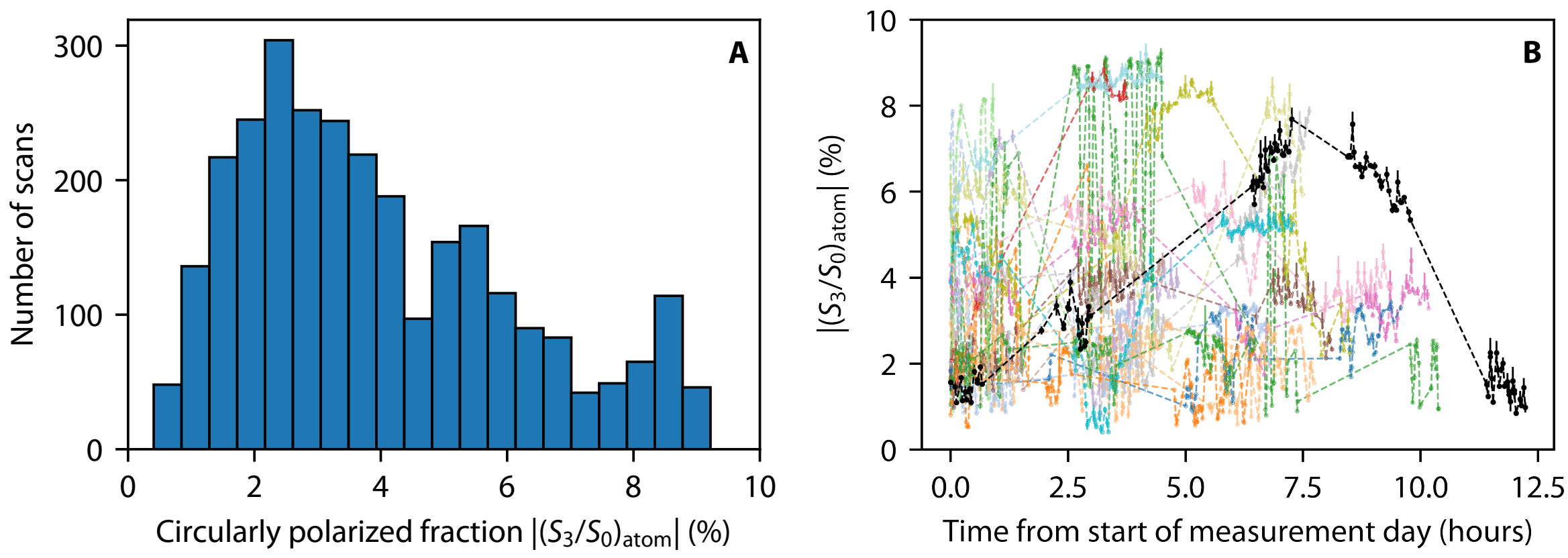

Figure 4.24: The absolute value of the circularly polarized fraction $|(S_3/S_0)_{\mathrm{atom}}|$ of the light seen by the atoms, as deduced from polarimetry of light backcoupled through the active fiber-based retroreflector. $|(S_3/S_0)_{\mathrm{atom}}|$ is found using Eqs. (4.21) and (4.22). (**A**) Histogram summarizing the polarization data from 2875 2S-6P line scans (22 measurement days). $|(S_3/S_0)_{\mathrm{atom}}|$ is always less than 10 %, corresponding to a polarization extinction ratio of PER > 26 dB. (**B**) $|(S_3/S_0)_{\mathrm{atom}}|$ for the individual line scans versus time (faint lines) for each measurement day, with one exemplary day (199 spectroscopy line scans) highlighted (black line).

agrees with the intuitive picture that for a given wave plate with an arbitrary value of $\delta$, a linear input polarization is only (approximately) preserved if $\beta \ll 1$.

Finally, small imperfections of the incoming polarization state $\boldsymbol{S}_{\mathrm{in}}$, that is, a small circularly polarized fraction, $(S_3/S_0)_{\mathrm{in}} \ll 1$, and a small rotation from the vertical or horizontal, $(S_2/S_0)_{\mathrm{in}} \ll 1$, can be taken into account with

$$|(S_3/S_0)_{\mathrm{atom}}| \simeq \frac{1}{2}\sqrt{\left((S_3/S_0)_{\mathrm{back}} - (S_3/S_0)_{\mathrm{in}}\right)^2 + \left((S_2/S_0)_{\mathrm{back}} + (S_2/S_0)_{\mathrm{in}}\right)^2}. \tag{4.21}$$

In our setup, $(S_3/S_0)_{\mathrm{in}} \simeq -0.024\,(\simeq 0.012)$ and $(S_2/S_0)_{\mathrm{in}} \simeq -0.036\,(\simeq 0.039)$ for vertical (horizontal) input polarization.

Optical elements, including non-polarizing beamsplitters, can be subject to polarization-dependent transmission and reflectivity, thereby acting as partial polarizers. For the components of the common optical system, we here find that the latter are unbalanced between orthogonal polarizations on the level of 2 % and thus negligible. If this were not the case, the above treatment would be insufficient and the system would instead need to be modeled as a combination of a partial polarizer in between two different wave plates and an additional rotator [112].

We now consider the polarization evolution of the backcoupled light on the non-common path from $\boldsymbol{S}_{\mathrm{back}}$ to $\boldsymbol{S}_{\mathrm{meas}}$, which can be described by a Mueller matrix $\boldsymbol{M}_{\mathrm{BS}}$, such that

$$\boldsymbol{S}_{\mathrm{back}} = \boldsymbol{M}_{\mathrm{BS}}^{-1}\,\boldsymbol{S}_{\mathrm{meas}}. \tag{4.22}$$

$\boldsymbol{M}_{\mathrm{BS}}$ is experimentally determined and is found to be mainly determined by the properties of beam splitter BS 2 (see Fig. 4.15), which shows a polarizing effect on the order of 20 %. Because the backreflected light is transmitted through or reflected from BS 2 for vertical or horizontal polarization, respectively, $\boldsymbol{M}_{\mathrm{BS}}$ is found to be different for the two input polarizations.

Eqs. (4.21) and (4.22) allow the monitoring of the absolute value of the circularly polarized fraction of the light seen by atoms from polarimetry of the backcoupled light, given a characterization of the polarization effect of the optical components before the polarimeter. We confirmed this method in a test measurement by measuring the polarization after the collimator with a second polarimeter placed behind the HR mirror (with the PMT of Fig. 4.15 removed), and comparing it to the derived result from the simultaneous measurement of the backcoupled polarization. Based on the uncertainties for Stokes parameter measurements ($\sim$1 %), combined with the uncertainty in the determination of $\boldsymbol{M}_{\mathrm{BS}}$, we estimate the absolute accuracy on deducing $|(S_3/S_0)_{\mathrm{atom}}|$ to be 2 %. The same test measurement was also used to confirm that fluctuations over time of the linear polarization rotation angle at the position of the atoms, $\theta_{\mathrm{L}}$, are below 3°.

On most measurement days of the 2S-6P measurement, we continuously took polarimetry data. In Fig. 4.24, the absolute value of the circularly polarized fraction $|(S_3/S_0)_{\mathrm{atom}}|$ of the light seen by the atoms is shown, as deduced from polarimetry of the backcoupled light according to the above equations. The histogram of Fig. 4.24 (A) shows the values for $|(S_3/S_0)_{\mathrm{atom}}|$ for 22 measurement days with a total number of 2875 spectroscopy line scans. $|(S_3/S_0)_{\mathrm{atom}}|$ is always less than 10 %, corresponding to a polarization extinction ratio of PER $>$ 26 dB. Note that the distribution of $|(S_3/S_0)_{\mathrm{atom}}|$ is skewed, which is partly caused by an offset in $(S_3/S_0)_{\mathrm{atom}}$ introduced by the stress-induced birefringence of the fiber connectors and the collimator. The time variation of the circularly polarized fraction is shown in Fig. 4.24 (B), showing slow drifts on the time scale of hours attributed to thermal fluctuations in the laboratory. Because of the in situ monitoring of $|(S_3/S_0)_{\mathrm{atom}}|$, a subset of line scans with $|(S_3/S_0)_{\mathrm{atom}}|$ well below 10 % can be selected, if needed.

## 4.5 Cryogenic atomic beam

To reach the required accuracy in the spectroscopy of the 2S-$n$P transitions, a cryogenic atomic beam with a high flux of atomic hydrogen is required. This is because the lower the temperature $T_{\mathrm{N}}$ of the atoms in the beam, the lower their speed $v \propto \sqrt{T_{\mathrm{N}}}$, in turn leading to reduced systematic effects, foremost the first-order Doppler effect, which scales as $\propto v$. The high flux is needed to reach high enough statistics to accurately determine the line center. Here, a temperature of $T_{\mathrm{N}} \approx 5\,\mathrm{K}$ is used, which can be reached with relatively simple liquid helium cryostats. The main challenge in producing such a beam is the recombination of atomic hydrogen atoms (H), initially produced in a dissociator (see Section 4.5.1) from molecular hydrogen ($H_2$), back into energetically preferred hydrogen molecules. This recombination tends to occur as the atoms collide with walls and is typically only negligible for certain materials at certain temperatures. On the other hand, the atoms need to be guided, i.e., through collisions with walls, from the dissociator to a cold nozzle at $T_{\mathrm{N}}$ (see Section 4.5.2), where they thermalize again through collisions, before escaping into vacuum and forming the atomic beam. The challenge is then to find a design with as little recombination as possible while also thermalizing the atoms to a low temperature, which here is achieved by accumulating solid $H_2$ on the cold walls of the nozzle. Finally, the atomic beam created in this way needs to be collimated to limit the Doppler broadening of the atomic resonance, which is achieved through a variable beam aperture (see Section 4.5.3).

### 4.5.1 Hydrogen dissociator

The hydrogen atoms probed in the experiment are produced in a dissociator, in which hydrogen molecules ($H_2$) are dissociated into two hydrogen atoms (H). The $H_2$ gas is taken from a gas cylinder and has a purity of $\geq 99.999\,\%$. It is fed through a palladium hydrogen purifier[1] to further increase the purity.

The hydrogen dissociator itself consists of an $H_2$-filled discharge tube inserted into a microwave (MW) cavity. The MW fields sustain an electrodeless discharge inside the tube, producing a plasma, and, finally, hydrogen atoms. The design of both the discharge tube assembly and the microwave cavity are very similar to that described in detail in [114] (see especially Figs. 2 and 3 therein). The MW cavity extends approximately 28 mm along the discharge tube and is resonant at 2.45 GHz. The input MW is derived from a solid-state MW generator[2]. 40 W of MW power is sent to the cavity, and, after adjusting the MW cavity length, typically no reflection back to the generator is observed when the discharge is running, except when the discharge tube is degraded (see below).

The discharge tube has an inner diameter of 7.25 mm (outer diameter 9.3 mm, length 250 mm) and is made of crystalline sapphire ($Al_2O_3$), which has a superior thermal conductivity compared to fused quartz as used in [114]. As opposed to the discharge tube used in [114], the sapphire tube itself has no built-in small orifice to limit to flow of H into the vacuum chamber. Instead, at the end of the discharge tube towards the vacuum chamber, a PTFE tube with an inner diameter of 5.3 mm and extending 35 mm into the discharge tube is inserted. The PTFE tube itself has a small, exchangeable[3] orifice with a diameter of 300 µm and a length of 2.07 mm, corresponding to a conductance of $6.0 \times 10^{-6}\,\mathrm{m^3/s}$ for H at a temperature of $-20\,°\mathrm{C}$ using the equations given in [115]. The orifice sits at a distance of 60 mm from the center of the MW cavity. The discharge tube is cooled at the position of the MW cavity with gaseous nitrogen, which itself is cooled by passing through tubing immersed in liquid nitrogen. The temperature of the nitrogen gas after the discharge tube is typically kept at $-30\,°\mathrm{C} \ldots -10\,°\mathrm{C}$, controlled by varying the flow of nitrogen gas.

The flow of $H_2$ into the discharge tube, $Q_{\mathrm{H_2}}$, is controlled with a gas dosing valve. The flow is measured right before this valve with a thermal mass flow meter[4], which gives readings of the volumetric flow in units of ml/min referenced to a temperature of 0 °C and a pressure of 1013.25 mbar. Throughout this work, these reference conditions are used for flows given in units of ml/min. Additionally, the pressure of $H_2$ is monitored directly after the valve and right before the discharge tube, using a pressure gauge[5] which itself measures the thermal conductivity of the gas (Pirani gauge). Typically, the discharge is operated with a flow of $Q_{\mathrm{H_2}} = 0.35\,\mathrm{ml/min}$, corresponding to $1.57 \times 10^{17}$ molecules/s, for which an $H_2$ pressure of approximately 1.6 mbar is measured. Using the conductance of the orifice for H and assuming a gas temperature of $-20\,^{\circ}$C, the flow for this pressure is calculated to be 0.31 ml/min, in reasonable agreement with the measured flow.

[1]The purifier was removed during the light force shift measurements at the end of July 2019, as the purity of the molecular hydrogen was found to be most likely limited by air leaks after the purifier.

[2]SAIREM GMS, max. output power 200 W.

[3]From February to May 2019, an orifice of 500 µm diameter and 3.5 mm length was used (conductance $1.7 \times 10^{-5}\,\mathrm{m^3/s}$ for H at and at $-20\,°\mathrm{C}$). To achieve the same flow of H with the newer, smaller orifice, an approximately three times higher pressure of $H_2$ was used. The smaller orifice allows lower H flows without degrading the discharge tube.

[4]Bronkhorst F-111B, calibrated for molecular hydrogen for volumetric flows between 0.18 ml/min…9 ml/min. The flow measurement has an uncertainty of ±0.5 % of the reading plus ±0.009 ml/min, i.e 3 % at 0.35 ml/min.

[5]Leybold TTR 101 N. The gauge reading is calibrated for $N_2$ gas, which is converted to a $H_2$ reading using the calibration curve supplied by the manufacturer.

The dissociator is operated by first filling the discharge tube with $H_2$, then switching on both the MW power and the discharge tube cooling, and finally starting the discharge with a high-voltage pulse applied to the outside of the discharge tube[1]. The pressure reading drops slightly after the discharge is started, which is attributed to a rise in the temperature of the gas mixture inside the discharge tube. At the same time, the flow reading remains constant, indicating that the cooled orifice limiting the flow stays at a constant temperature. No attempt was made at measuring the degree of dissociation of the gas leaving the discharge. However, in the similar design of [114] a degree of dissociation[2] $\alpha_{\mathrm{dis}}$ on the order of 90 % was routinely achieved, and it seems reasonable to assume a similar performance in our system. Furthermore, the limiting factor for the degree of dissociation is most likely the transport of the gas to the nozzle.

The output end of the discharge tube protrudes into the vacuum chamber. PTFE tubing of 6 mm inner diameter, attached to the PTFE tube inserted into the discharge tube, is used to guide the produced H to the nozzle. In total, a distance of 435 mm is covered, including five right-angle bends owing to the geometry of the atomic beam apparatus (the second half of the tubing is visible on the left in Fig. 4.1, labeled (**TT**)). The straight sections use commercially available PTFE tubes, while the bends were machined from bulk PTFE. The recombination probability per collision of H on PTFE surfaces was found to range between $\gamma_{\mathrm{PTFE}} = 1 \times 10^{-4} \ldots 4.5 \times 10^{-4}$ for commercial PTFE tubing in[3] [114]. Neglecting the bends in our tubing and assuming free molecular flow, the number of wall collisions during the transport of H through the tubing is found to be approximately 8000 [114]. The actual number of collisions is certainly higher due to the bends, but no estimation of this correction was attempted. Using this number of collisions, the degree of dissociation $\alpha_{\mathrm{dis}}$ after the tubing is expected to range between 14 %. . . 53 % for initially pure atomic hydrogen. Using $\alpha_{\mathrm{dis}} = 90\,\%$ at the input of the tubing, this further decreases to $\alpha_{\mathrm{dis}} = 12\,\% \ldots 47\,\%$ at the nozzle. The signal observed in the 2S-6P measurement hints at a substantially lower $\alpha_{\mathrm{dis}}$ than estimated here, with the underestimation of the collisions in the tubing a likely cause for this discrepancy.

While it may be tempting to use the residual gas analyzer (RGA) attached to the chamber to measure $\alpha_{\mathrm{dis}}$, this is not easily possible since the hydrogen atoms have to undergo many wall collisions, each associated with a high probability of recombination, before reaching the RGA (see Section 4.2.4). Furthermore, a strong signal corresponding to the mass of hydrogen atoms is always present from the dissociation of heavier hydrogen-containing compounds by the mass spectrometer. However, one might interfere some measure of the degree of dissociation by monitoring the signal corresponding to molecular hydrogen. Previously, the RGA was attached to the cylindrical vacuum chamber, i.e., probing the outer vacuum region. In this configuration, a drop of ≈6 % in the molecular hydrogen signal was observed upon starting the discharge in the dissociator and with the nozzle at room temperature. Note that for this measurement, $Q_{\mathrm{H}_2}$ was set to ≈2.3 ml/min, much higher than used in the experiment described here.

[1] Electro-Technic Products BD-10ASV high-frequency generator, generating 20 kV. . . 50 kV pulses.

[2] Throughout this work, the definition the degree of dissociation as $\alpha_{\mathrm{dis}} = N_{\mathrm{H}}/(N_{\mathrm{H}} + 2N_{\mathrm{H}_2})$ is adopted, also used in [114], where $N_{\mathrm{H}}$ and $N_{\mathrm{H}_2}$ are the number of hydrogen atoms and hydrogen molecules, respectively.

[3] Interestingly, [114] also cite, before discussing their own results, a much lower value of $2 \times 10^{-5}$ for $\gamma_{\mathrm{PTFE}}$ from an experiment described in [116]. However, [116] do not give a value for the recombination probability per collision $\gamma_{\mathrm{PTFE}}$, but instead measure a loss rate of 2.1/s due to recombination on PTFE-covered walls held at room temperature. To convert this loss rate to a value of $\gamma_{\mathrm{PTFE}}$, the rate of wall collisions in the experiment of [116] needs to be estimated, which is not done in [116]. This nontrivial conversion is apparently done, but not mentioned, in [114], resulting in $\gamma_{\mathrm{PTFE}} = 2 \times 10^{-5}$. Since the conversion thus cannot be retraced, and because the value of $\gamma_{\mathrm{PTFE}}$ found in this way is much lower than values measured in [114], this result is not used here. On the other hand, the values of $\gamma_{\mathrm{PTFE}}$ measured in [114], and given here in the main text, are derived from the measured degree of dissociation $\alpha_{\mathrm{dis}}$ after a straight PTFE tube combined with a detailed estimation of the number of collisions during transport in this tube.

During the experiment, it was sometimes observed that the reflection of MW power from the cavity could not be removed anymore by adjusting the MW cavity (and with the purple glow of the discharge dimmed noticeably), worsening over time and ultimately leading to the point where the discharge could not be started anymore. This behavior coincides with the degradation of the discharge tube, with a metallic layer forming on the inside at the location of the MW cavity. This layer is suspected to consist of aluminum, formed by a reaction of sapphire with atomic hydrogen. The intervals in which the layer formed are not clearly linked to the operating time of the discharge. In one particular case, occurring after the discharge was operated with a lower pressure of hydrogen (<1 mbar) than previously used, the layer formed more quickly than usual. It was found that the layer can be quickly removed by immersing the discharge tube in a solution of 5 % NaOH in deionized water. Afterwards, the discharge tube is cleaned in isopropyl alcohol using an ultrasonic cleaner. After re-assembling the discharge, the discharge is run for a few hours to flush out remaining solvent or other contaminants. After this procedure, the experiment can be continued.

## 4.5.2 Cryogenic nozzle

### 4.5.2.1 Nozzle design and temperature stabilization

After being transported through the PTFE tube, the room temperature hydrogen atoms reach the cryogenic nozzle, where they ideally thermalize to the nozzle's temperature $T_{\mathrm{N}}$ before forming an atomic beam. The nozzle, shown in Fig. 4.25, is a copper block with a through-hole with a diameter of 2.0 mm and a length of 8.0 mm as nozzle channel and a 4.0 mm diameter blind hole as input channel, with the channels thus forming a t-shape. This through-hole design is necessary since the 1S-2S preparation laser beam is collinear with the atomic beam and thus needs to propagate through the nozzle channel. The nozzle block is attached to the cold finger of the helium continuous-flow cryostat[1]. A silicon diode temperature sensor[2] is mounted to the nozzle block to measure its temperature $T_{\mathrm{N}}$. A thin layer of thermal grease[3] is applied to the polished interfaces between nozzle and cryostat, and nozzle and temperature sensor. Brass screws are used for mounting, as brass is non-magnetic and its thermal expansion is slightly higher than that of the copper parts. An additional sensor[4] measures the temperature inside the cold finger of the cryostat.

The PTFE tubing carrying the H atoms from the discharge tapers toward the nozzle input channel to efficiently transfer the atoms into the nozzle. A PTFE spacer[5] with a wall thickness of only 0.45 mm creates a small gap of 0.2 mm width between the tubing and the cryogenic input channel of the nozzle while reducing the thermal conduction between the two. The tubing is supported through a copper holder attached to the thermal shield of the cryostat, with the tubing and holder thermally isolated using a polyurea aerogel spacer[1]. This design is meant to keep the PTFE tubing, which is held at room temperature at its other end near the dissociator, from cooling down to the nozzle temperature, where the recombination probability on PTFE increases substantially [117, 118]. In an earlier version of the nozzle, used to acquire the data shown in Fig. 4.28 and Fig. 4.30 but not during the 2S-6P measurement, the PTFE tubing was directly attached to the nozzle without a spacer.

[1]ICEoxford ICICLE, 3.6 W (4.6 W) cooling power at 4.2 K (4.7 K).

[2]LakeShore DT-670-BO-1.4L, accuracy of temperature reading is ±12 mK between 1 K and 10 K.

[3]Apiezon N.

[4]LakeShore CX-1030-AA-1.4L thin film resistance cryogenic temperature sensor.

[5]Before the PTFE spacer was installed on 28.05.2019, a polyurea aerogel spacer was used. This spacer however could easily be flattened when mounting the nozzle due to its softness, making it challenging to maintain a controlled, small gap between the nozzle and the PTFE tubings.

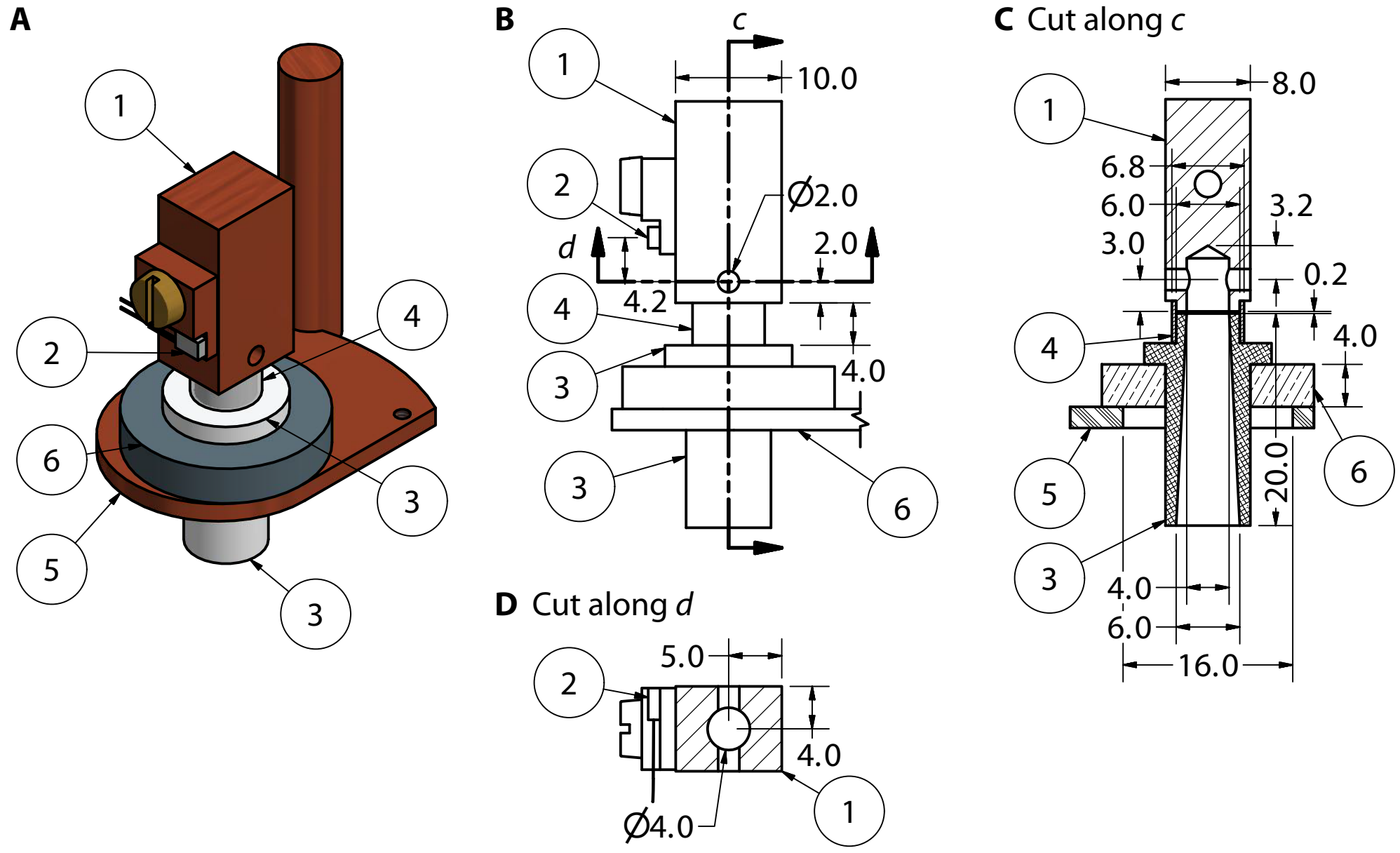

Figure 4.25: Drawing of the hydrogen nozzle: (**A**) orthographic projection, (**B**) drawing in $x$-$y$-plane as seen from 2S-6P spectroscopy region, and section view cuts along (**C**) (vertical) $y$-$z$-plane $c$ and (**D**) (horizontal) $x$-$z$-plane $d$. All dimensions are in mm. The (1) t-shaped nozzle is a copper block with a through-hole with a diameter of 2.0 mm and a length of 8.0 mm as nozzle channel and a 4.0 mm diameter blind hole as input channel. The nozzle block is attached to the cold finger of the cryostat (upper part of nozzle block and cryostat not shown). A (2) temperature sensor is attached to the nozzle block. Hydrogen atoms flow to the nozzle through (3) PTFE tubing, which tapers toward the nozzle input channel. A thin-walled (4) PTFE spacer creates a small gap between the tubing and input channel and reduces the thermal conduction. The tubing is supported through a (5) copper holder attached to the thermal shield of the cryostat (upper part of holder and thermal shield not shown). Tubing and holder are thermally isolated using a (6) polyurea aerogel spacer. PTFE: polytetrafluoroethylene (Teflon).

As detailed below, the fluorescence signal is sensitive to the nozzle temperature $T_\mathrm{N}$ and thus the temperature fluctuations, especially during a line scan, should be kept as small as possible to avoid adding excess noise on the signal. To this end, the nozzle temperature is actively stabilized[2] using a heating wire inside the cryostat as actuator. The bandwidth of this feedback loop is limited by the controller update rate of 10 Hz, and the temperature is read out and saved once per data point, approximately every 2 s. Fig. 4.4 shows the nozzle temperature during a typical freezing cycle, while Fig. 4.26 (A) shows a histogram of the nozzle temperature for the line scans and data points of the 2S-6P measurement, where $T_N$ was stabilized to 4.8 K. For the latter, $T_N$ fluctuates from point-to-point with a standard deviation of 7.5 mK. The fluctuations of the scan-averaged temperature from scan to scan, which are separated by at least ≈1 min, are substantially smaller, with a standard deviation of 0.8 mK. Thus, the time scale of the visible fluctuations is on the order seconds, as may be expected from the feedback loop. Fluctuations on a shorter time scale, not resolved here, seem unlikely due to the macroscopic size of the cold head and nozzle. As can be seen from the histogram of the standard deviations of $T_N$ over each line scan, shown in Fig. 4.26 (B), the fluctuations are fairly similar for each line scan. The exact choice of feedback loop parameters seemed to have little impact on the size of the fluctuations, hinting at the fact that the feedback bandwidth is insufficient to suppress them. They could, however, be somewhat reduced by finding a suitable combination of helium flow and power sent to the heating wire, with lower flows tending to produce lower fluctuations. Care however has to be taken that enough cooling power, i.e., enough helium flow, is available at any time to allow the operation of the feedback loop. Otherwise, large temperature spikes can occur, which can release frozen hydrogen from the nozzle and in turn lead to pressure spikes in the vacuum chamber.

[1] Aerogel Technologies Airloy X103 M, thermal conductivity $k = 29\,\mathrm{mW/(m\,K)}$.

[2] Implemented with the Cryogenic Control Systems (Cryo-con) Model 32B temperature controller and its built-in a PID controller.

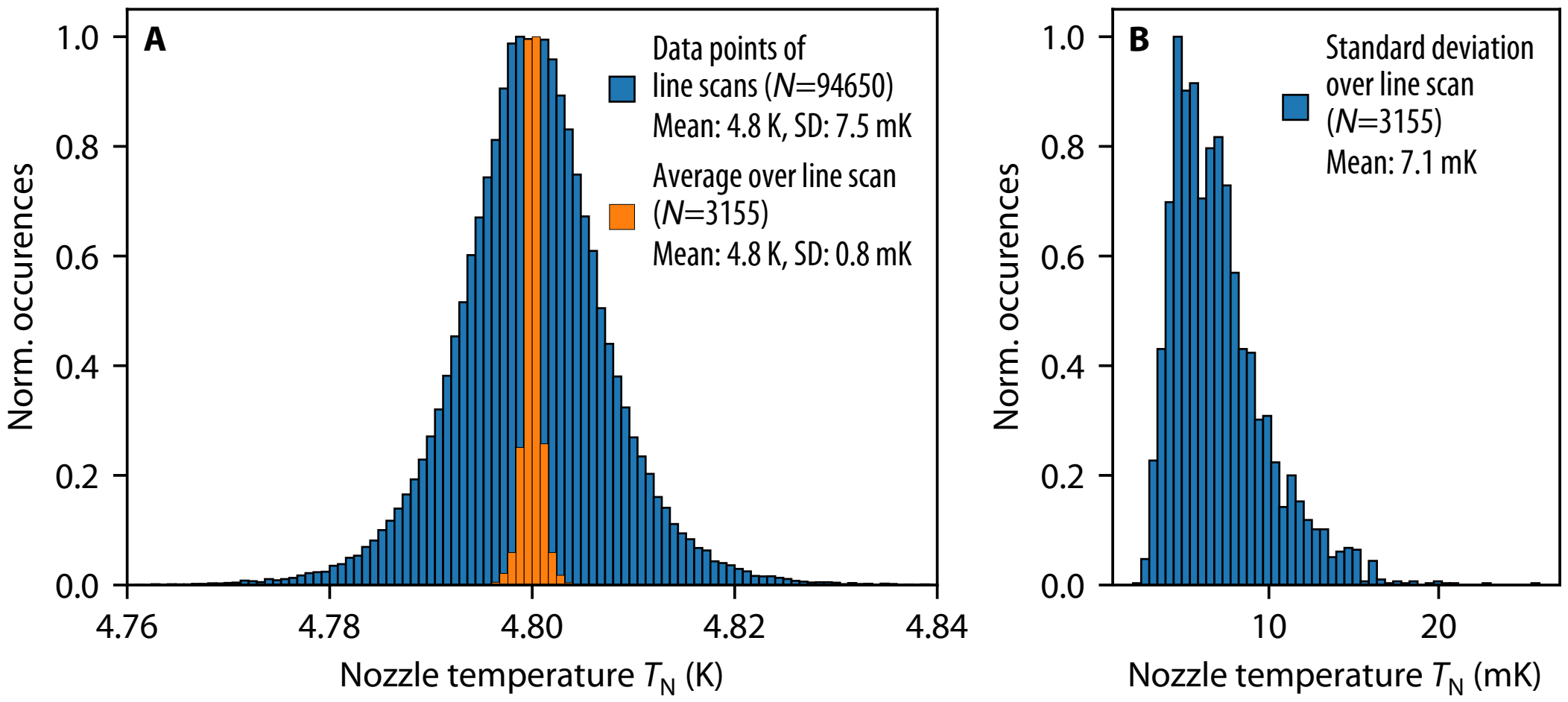

Figure 4.26: Stability of the nozzle temperature $T_N$ during the 2S-6P measurement. (**A**) Histogram of $T_N$ during the measurement's 3155 line scans (see Table 6.1). Each line scan contains 30 data points, which are on average separated in time by 2.3 s, with the nozzle temperature read out once per data point. $T_N$, which is stabilized to 4.8 K, shows point-to-point fluctuations (blue bars) with a standard deviation (SD) of 7.5 mK. The scan-averaged nozzle temperature (orange bars) fluctuates substantially less from scan to scan, with a SD of 0.8 mK. (**B**) Histogram of the standard deviation of $T_N$ over each line scan. See text for details.

At some points, yellowish deposits were found on the nozzle near the inlet. These deposits could be removed using industrial soap[1], with the nozzle afterwards cleaned in isopropyl alcohol, and with an ultrasonic cleaner used for both steps. One suspicion is that the deposits are remains of the cleaning of the discharge tube as discussed in Section 4.5.1. When installing a newly machined nozzle, the same cleaning produce was followed. The nozzle temperature was found to exhibit intermittent temperature spikes for the first few freezing cycles after such a cleaning procedure, possibly caused by soap and solvent residues interfering with the formation of the frozen layer of $H_2$.

[1]Edisonite Super.

#### 4.5.2.2 Trajectory distribution in the atomic beam

For an ideal gas of atomic hydrogen contained in a closed container and at thermal equilibrium with the container walls at temperature $T_\mathrm{N}$, the probability distribution $p_\mathrm{eq}(v)$ of the speeds[1] $v$ of the atoms is given by the Maxwell-Boltzmann distribution [119]

$$p_\mathrm{eq}(v)\,\mathrm{d}v = \left(\frac{m_\mathrm{H}}{2\pi k_\mathrm{B} T_\mathrm{N}}\right)^{3/2} 4\pi v^2 e^{-\frac{m_\mathrm{H} v^2}{2k_\mathrm{B} T_\mathrm{N}}}\,\mathrm{d}v, \tag{4.23}$$

where $k_\mathrm{B}$ is the Boltzmann constant and $p_\mathrm{eq}(v)\,\mathrm{d}v$ is the probability to find an atom within the speed interval $v$ to $v + \mathrm{d}v$. The mean speed of this distribution is

$$v_\mathrm{th} = \sqrt{\frac{8k_\mathrm{B} T_\mathrm{N}}{\pi m_\mathrm{H}}}, \tag{4.24}$$

which is here used as a measure of the thermal speed. For room temperature, $v_\mathrm{th}$ is 2482 m/s, while for the typical nozzle temperature of $T_\mathrm{N} = 4.8\,\mathrm{K}$, $v_\mathrm{th}$ is reduced to 318 m/s.

An effusive atomic beam is created by adding a small source orifice to the container, such that the speed distribution inside the container remains unchanged. The probability for atoms to escape the container through the orifice is proportional to $v$, and thus the speed distribution $p_\mathrm{eff}(v)$ of the atoms contributing to the signal is proportional to $p_\mathrm{eq}(v)$ multiplied with $v$ [119], i.e.,

$$p_\mathrm{eff}(v)\,\mathrm{d}v \propto v^3 e^{-\frac{m_\mathrm{H} v^2}{2k_\mathrm{B} T_\mathrm{N}}}\,\mathrm{d}v. \tag{4.25}$$

The mean speed of this distribution is given by [119]

$$v_\mathrm{eff} = \frac{3}{4}\sqrt{\frac{2\pi k_\mathrm{B} T_\mathrm{N}}{m_\mathrm{H}}}, \tag{4.26}$$

corresponding to 2924 m/s and 374 m/s for room temperature and $T_\mathrm{N} = 4.8\,\mathrm{K}$, respectively. To distinguish the two distributions, Eq. (4.23) and Eq. (4.25) are sometimes referred to as the volume or number density distribution and flux distribution, respectively.

The nozzle used here only vaguely resembles this ideal situation of an effusive beam. The thermalization and beam formation are based on collisions of the room-temperature atoms coming from the dissociator with the cold walls of the nozzle and with other atoms and molecules. Most atoms leave the nozzle after only a few collisions, and in [114, 120] cryogenic atomic beams of hydrogen with a speed distribution given by Eq. (4.25) were observed for nozzle designs where the atoms underwent only a few collisions with the cold walls. However, [120] also found evidence of deviations from Eq. (4.25) when collisions between the particles in the nozzle are thought to play an important role. Specifically, they found that if the source orifices are actually part of a through-hole, as is the case here, the speed distribution of atoms emerging under a small angle to the beam axis showed fewer slow atoms as expected from Eq. (4.25). These atom trajectories can only be the product of collisions with other particles, as there is no direct line of sight to a wall. Similar deviations in the speed distribution have also been observed in room-temperature beams of atomic hydrogen [27, 94, 121].

[1] The usual definition of the speed $v$ as the magnitude of the velocity $\boldsymbol{v}$ is used here. However, often times the term velocity distribution instead of speed distribution is used when referring to the probability distribution of $v$.

Such a depletion of slower atoms associated with intra-beam collisions between the particles inside the beam is sometimes referred to as the Zacharias effect, after the failed attempt to construct the first cesium fountain clock relying on the presence of these atoms in a thermal beam [122, 123]. This depletion was clearly observed in thermal beams of potassium and cesium and shown to be the product of collisions inside the beam right after the source orifice in [124]. The difference between these collisions and the collisions leading to thermalization inside a closed container is that the former can deflect atoms from the beam and thus from the region of high particle density, such that additional collisions returning the atoms to the beam are unlikely. These atoms are then effectively removed from the beam. In the experiment here, only atoms with an divergence angle of less than $\approx 10\,\mathrm{mrad}$ to the beam axis are probed, and thus a small deflection is sufficient for this removal. The slower the relative speed of the colliding particles, the larger the interaction time and therefore the deflection angle, leading to a preferred removal of slow atoms [124, 125]. Inside the nozzle, deflected atoms can return to the beam through wall collisions. However, fast atoms traveling at a small divergence angle to the beam axis have a higher probability to escape the nozzle than slower atoms, again leading to a depletion of slow atoms for the low-divergence part of the beam [126]. Various attempts have been made to find an analytical description of the influence of these collisions on the resulting speed distribution, focusing on collisions inside the nozzle [126], collisions right after the source orifice [124], and collisions with particles from the residual background gas as the atoms propagate through the apparatus [125]. The models generally rely on simplifying assumptions that are not entirely justified here, and the input parameters such as the composition of the colliding particles are not well known here. Additionally, all three effects most likely contribute here.

To estimate how important a role collisions play inside and right after the nozzle, one can compare the mean free path $\lambda_{\mathrm{mfp}}$ of the atoms to the dimensions of the nozzle. The situation is here complicated by the fact that both collisions between hydrogen atoms and between hydrogen atoms and molecules are thought to be important (see below). If the collision cross section $\sigma$ of the colliding particles is known, $\lambda_{\mathrm{mfp}}$ can be calculated through the relation $\lambda_{\mathrm{mfp}} = 1/(\sqrt{2}\sigma n)$, where $n$ is the number density of collision partners [119]. For a thermal gas, often the viscosity $\eta$ instead of $\sigma$ is given, which is related through $\sigma = (1/\eta)\sqrt{m k_{\mathrm{B}} T/\pi}$ and where $m$ and $T$ are the particle mass and temperature of the gas, respectively [127]. The viscosity of a gas of hydrogen atoms has been calculated at $T = 5\,\mathrm{K}$ as $0.365\,\mathrm{Pa\,s}$ in [128], giving[1] $\sigma = 5.2 \times 10^{-19}\,\mathrm{m}^2$. The author is not aware of calculations of $\eta$ for a mixture of H and $H_2$ at cryogenic temperatures, but [129] found that $\eta$ of this mixture is less than $50\,\%$ larger than for a gas of H between $100\,\mathrm{K}$ and room temperature. As an estimate, here the cross section found for H is also used for H–$H_2$ collisions. For the typical pressure of $3.6 \times 10^{-5}\,\mathrm{mbar}$ inside the nozzle used in the experiment (see below), the particle density is $n = 5.4 \times 10^{19}/\mathrm{m}^3$, and the mean free path is $\lambda_{\mathrm{mfp}} = 25\,\mathrm{mm}$. Thus, $\lambda_{\mathrm{mfp}}$ is about three times larger than the channel length of the nozzle, $l = 8\,\mathrm{mm}$, with this ratio known as the Knudsen number $Kn = \lambda_{\mathrm{mfp}}/l \approx 3.1$ [126]. In the model of [126], the onset of modifications on the speed distribution is found to be for $Kn$ below 10. For comparison, $Kn$ is estimated to be 0.01 inside the nozzle of the aforementioned room temperature hydrogen beam [94].

[1] This cross section is thus almost two orders of magnitudes larger than the sometimes used coarse estimation of $\sigma$ as $\pi a_0^2 = 8.8 \times 10^{-21}\,\mathrm{m}^2$, where $a_0$ is the Bohr radius.

In any case, the depletion of slow atoms is clearly observed in the experiment discussed here, as the amplitude of the 2S-6P resonance decreases with increasing delay time $\tau$, and thus decreasing mean atom speed $\bar{v}$, more rapidly than expected for an effusive beam. Furthermore, the depletion increases with larger flows of hydrogen into the system, leading to a larger particle density inside the nozzle, inside the beam, and in the vacuum chamber itself due to the higher gas load (see Section 4.5.2.4). Since the particle density approximately drops of with the square of the inverse distance to the nozzle, collisions within or shortly after the nozzle are more likely than during the propagation to the 2S-6P spectroscopy region. Importantly, the magnitude of the depletion typically increases during the course of the freezing cycle as the volume inside the nozzle shrinks due to the freezing, implying that collisions in this region are largely responsible for the depletion. However, the background counts on the fluorescence signal, detected far away from the nozzle and only present when 2S atoms are produced, also significantly increases with larger flows. This suggests that intra-beam collisions and collisions with the residual gas inside the vacuum chamber, kicking 2S atoms out of the beam, are responsible for the background. Thus, both collisions close to the nozzle and during the propagation to the spectroscopy region must be assumed to be important in the description of the experiment.

Here, a phenomenological model of the depletion of slow atoms is used. To this end, Eq. (4.25) is multiplied with an exponential suppression term, giving

$$p(v)\,\mathrm{d}v \propto v^3 e^{-\frac{m_{\mathrm{H}}v^2}{2k_{\mathrm{B}}T_{\mathrm{N}}}} e^{-\frac{v_{\mathrm{cutoff}}}{v}}\,\mathrm{d}v. \tag{4.27}$$

The cutoff speed $v_{\mathrm{cutoff}}$ is the characteristic speed below which atoms are removed from the beam and is treated as an adjustable parameter. This approach is motivated by the analytical results of [124], where the depletion of slow atoms through intra-beam collisions was found to be approximately of this functional form (see Eq. (8) and Fig. 13 of [124]). It is determined from the spectroscopy data by matching the observed behavior of the resonance amplitude with the delay time $\tau$ with the prediction of simulations using Eq. (4.27). For the 2S-6P measurement, $v_{\mathrm{cutoff}}$ ranges between $-3\,\mathrm{m/s}\ldots112\,\mathrm{m/s}$ (see Fig. 6.1). A similar approach was used in [27, 121].

To model the atomic beam and thus the experiment, not the speed distribution, but the joint spatial, angular, and speed distribution of the atoms emerging from the nozzle is needed, here referred to as the distribution of atom trajectories. For an effusive beam emerging through an source orifice such that no collisions occur as the atoms escape the container, the distribution $p(\theta)\,\mathrm{d}\Omega$ of the angle $\theta$ of the atoms in the beam, measured to the normal of the plane in which the orifice lies, is [119]

$$p(\theta)\,\mathrm{d}\Omega \propto \cos\theta\,\mathrm{d}\Omega, \tag{4.28}$$

where $\Omega$ is the solid angle and $\mathrm{d}\Omega = 2\pi\sin\theta\,\mathrm{d}\theta$ is the surface element of a unit sphere in spherical coordinates. Eq. (4.28) is often referred to as cosine law or, in optics, as Lambert's cosine law, with the factor $\cos\theta$ leads to a peaking of the otherwise isotropic emission in the direction of the normal of the orifice[1]. This distribution is modified if collisions occur as the atoms pass through the orifice, and for a nozzle channel a peaking of the beam is expected [114]. Again, however, these descriptions rely on insufficiently determined parameters, and with the most commonly used approximation of a long channel, i.e., a channel much longer than its radius, not applicable here. In any case, here only atoms flying with an angle $\theta \leq 10.4\,\mathrm{mrad}$ can reach the spectroscopy region due to geometrical constraints (for the typical width of the variable aperture of 1.2 mm). On this scale, the beam peaking from the nozzle channel does not change the shape of the angular distribution in a relevant way[2].

However, a possibly much larger influence on the distribution of atom trajectories is that trajectories with an angle smaller than $\theta = 2.5\,\mathrm{mrad}$ can only be the product of collisions with other particles, as there is no direct line of sight to the walls of the nozzle. Thus, if those collisions produce substantially less trajectories compared to wall collisions, the atomic beam at the spectroscopy region might actually be best described as a hollow tube[3]. It is thus of great interest to model the conditions and the collisions inside the nozzle and the beam with good accuracy. While some preliminary studies have been made, this is beyond the scope of this thesis, but will be revisited for the final analysis of the 2S-6P measurement. For the analysis presented here, a spatially-independent emission is assumed, with the angular and speed distributions given by Eq. (4.28) and Eq. (4.27), respectively.

So far, the discussion has not included the excitation of the ground state (1S) atoms emerging from the nozzle to the metastable 2S level. In the description of the experiment, however, ultimately the distribution of 2S atoms is needed, since only those atoms are probed in the spectroscopy of the 2S-$n$P transitions. The excitation takes place as the atoms fly through the laser beam of the 1S-2S preparation laser, with the probability to find an atom in the 2S level now itself depending on the exact trajectory as, e.g., some atoms spend most of time inside the laser beam and others just cross it for a short time. A Monte Carlo simulation is used to model the trajectory distribution of the 2S atoms, as described in detail in Section 5.2.

#### 4.5.2.3 Hydrogen recombination, freezing, and degree of dissociation

The atoms undergo only a few collisions with the cold nozzle walls before forming the atomic beam. However, for most materials, including PTFE, copper, and aluminum, the recombination probability increases strongly below 60 K and reaches up to 20 % per collision [117].

[1] The occurrence of the cosine law can be explained as follows: in the container, there is no preferred direction, and each combination of position and direction is equally likely. For each position and direction that leads through the orifice, i.e., the trajectory after the last collision with a wall or another particle, the probability to leave is proportional to $v$. However, the effective size of the orifice is proportional to $\cos\theta$, where $\theta$ is the angle between the direction and the normal of the orifice. Thus, the probability to find trajectories with an angle $\theta$ outside the orifice is proportional to $\cos\theta$.

[2] The fraction of atoms that reach the spectroscopy region, however, is substantially influenced by the angular distribution, with, e.g., the cosine law increasing the on-axis flux of atoms by a factor of two over isotropic emission. This results in a large uncertainty when attempting to estimate the number of atoms probed in the spectroscopy region.

[3] Note that this does not imply that there are no atoms crossing the 2S-6P spectroscopy laser beams with angles smaller than $\delta\alpha = 2.5\,\mathrm{mrad}$ to the beam axis, as $\delta\alpha$ is the projection of the angle $\theta$ from the atomic beam axis on the laser beam axis. For the typical atom speed $v_{\mathrm{typ}} = 200\,\mathrm{m/s}$, a trajectory crossing the 2S-6P spectroscopy laser beams with an angle $\delta\alpha = 2.5\,\mathrm{mrad}$ experiences a Doppler shift of $\Delta\nu_{\mathrm{D}} \approx 1.2\,\mathrm{MHz}$ (see Section 2.2.3). This corresponds to ≈30 % of the natural linewidth of the 2S-6P transition. Thus, the line shape resulting from such a hollow atomic beam might not be immediately distinguishable from the case of a uniform beam.

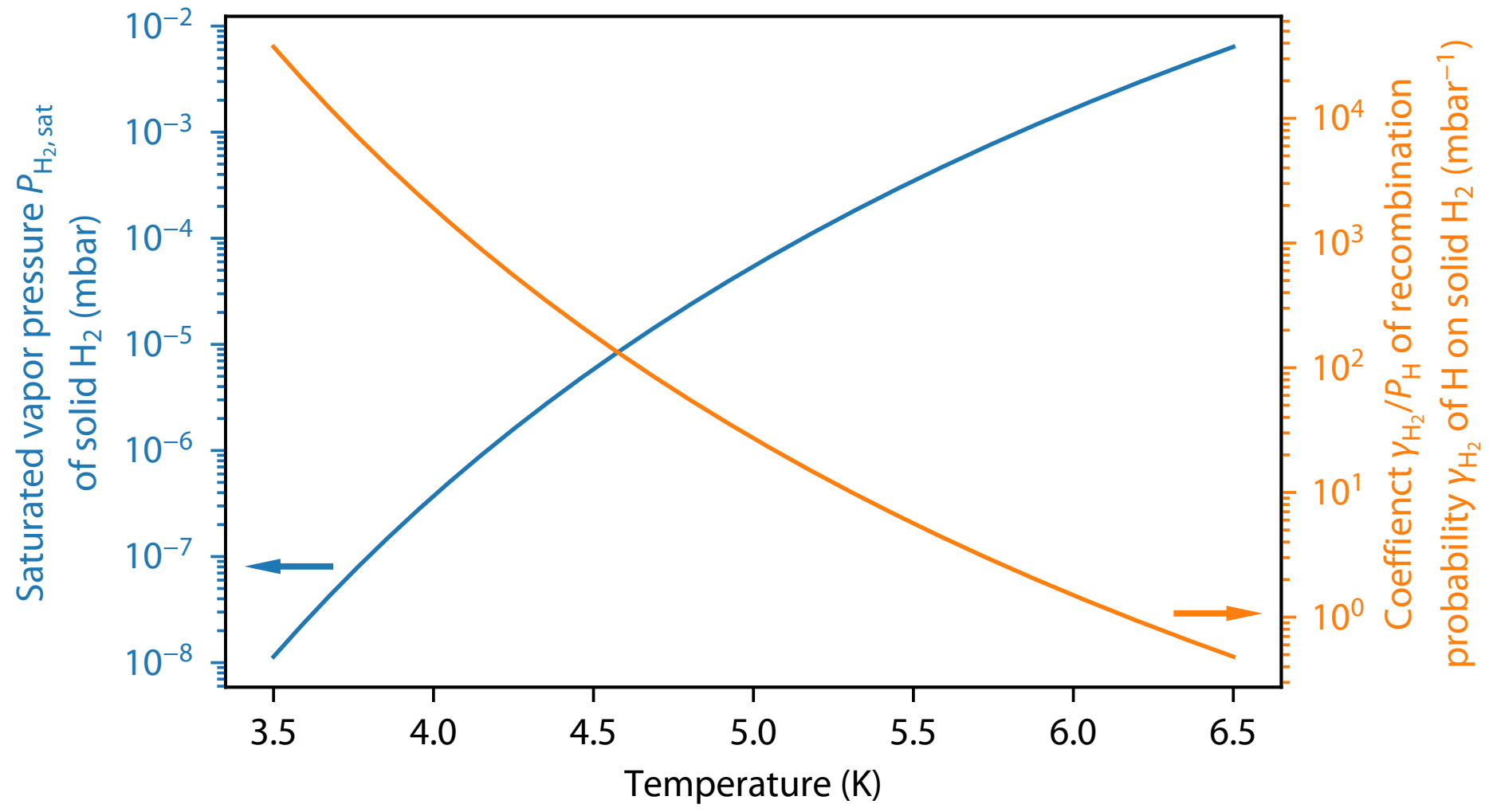

Figure 4.27: Saturated vapor pressure $P_{H_2,sat}$ of solid $H_2$ (blue line, left scale) and the recombination probability $\gamma_{H_2}$ of H per collision with solid $H_2$ as a function of temperature. $\gamma_{H_2}$ is given by multiplying $\gamma_{H_2}/P_H$ (orange line, right scale) with the partial pressure of H, $P_H$, which is estimated to be $4.8 \times 10^{-6}$ mbar inside the nozzle for the experimental conditions discussed here. At this partial pressure and a temperature of 4.8 K, $P_{H_2,sat} = 2.3 \times 10^{-5}$ mbar, $\gamma_{H_2}/P_H = 56\,\text{mbar}^{-1}$, and $\gamma_{H_2} = 2.7 \times 10^{-4}$. The underlying measurements are taken from [114, 130, 131], see text for details.

Thus, even a few collisions lead to large decrease in the degree of dissociation $\alpha_{dis}$. On the other hand, the recombination probability for collisions with solid $H_2$, while also strongly increasing with decreasing temperature, can be much lower in that temperature range (see orange line in Fig. 4.27). It is thus advantageous to cover the inside of the nozzle with solid $H_2$, which implies that the partial pressure of $H_2$ inside the nozzle must not be below the saturated vapor pressure of solid $H_2$. As the vapor pressure strongly increases with temperature (see blue line in Fig. 4.27), this limits the temperature range to below about 6 K if the pressure inside the nozzle and the vacuum system are to be kept within a reasonable range. An even lower temperature yet is advantageous, as higher pressures inside the nozzle also disturb the velocity distribution of the atomic beam as discussed above. The $H_2$ freezing out on the nozzle is in this way also largely removed from the atomic beam and from the vacuum chamber, where it otherwise can lead to collisions removing atoms from the beam and pressure shifts of the observed resonance. However, this freezing will also lead to the nozzle clogging over time as more and more solid $H_2$ accumulates. Thus, a balance between sufficient coverage of the nozzle walls with $H_2$, the time it takes the nozzle to freeze[1], and the recombination probability has to be found. Here, this is achieved by experimentally finding a suitable combination of nozzle temperature $T_N$ and flow $Q_{H_2}$ of hydrogen into the system, with $\alpha_{dis}$ at the nozzle is fixed by the properties of the PTFE tubing. For the 2S-6P measurement, the combination $T_N = 4.8$ K and $Q_{H_2} = 0.35$ ml/min was used.

H adsorbs to $H_2$-coated surfaces with an absorption energy of 38 K [131]. The adsorbed atoms are thought to be mobile on the surface and can recombine to $H_2$ with other adsorbed atoms. Since the number of absorbed atoms is proportional to the volume density of atoms $n_H$, the recombination probability $\gamma_{\mathrm{H}_2}$ per collision with the $\mathrm{H}_2$-coated wall itself is proportional to $n_\mathrm{H}$ as $\gamma_{\mathrm{H}_2} = (4/v_\mathrm{th})K_{\mathrm{S}2}n_\mathrm{H}$ [114], where $v_\mathrm{th}$ is the thermal speed of the atoms as given in Eq. (4.24). The measurements of [131] give $K_{\mathrm{S}2} = 8.5 \times 10^{-28}\,\mathrm{m}^4\,\mathrm{K}^{1/2}\,\mathrm{s}^{-1}$, resulting in $\gamma_{\mathrm{H}_2} = n_\mathrm{H} \times (3.7 \times 10^{-23}\,\mathrm{m}^3) = P_\mathrm{H} \times (56\,\mathrm{mbar}^{-1})$ at a temperature of 4.8 K, where $P_\mathrm{H}$ is the partial pressure of H. The resulting temperature dependence of $\gamma_{\mathrm{H}_2}/P_\mathrm{H}$ is shown as the orange line in Fig. 4.27.

[1] If $\alpha_{dis}$ at the nozzle is so large such that almost no $H_2$ can freeze out, one might also cover the walls of the nozzle with $H_2$ by initially keeping the discharge switched off. Then, after starting the discharge, both the recombination probability and the growth of $H_2$ are reduced.

The saturated vapor pressure $P_{\mathrm{H}_2,\mathrm{sat}}$ of solid $\mathrm{H}_2$ increases from $1 \times 10^{-8}$ mbar to $6 \times 10^{-3}$ mbar between 3.5 K and 6.5 K [130] (see blue line in Fig. 4.27). At a temperature of $T_\mathrm{N} = 4.8\,\mathrm{K}$, $P_{\mathrm{H}_2,\mathrm{sat}} = 2.3 \times 10^{-5}$ mbar, which is above the typical pressure of $P_\mathrm{OV} = 5.5 \times 10^{-6}$ mbar in the outer vacuum region (see Section 4.2.10), in which the nozzle sits. However, inside the nozzle the pressure is higher due to the constant influx of hydrogen. Indeed, the freezing of hydrogen in the nozzle is observed in the experiment, constituting a major limiting factor for the measurement time as the frozen hydrogen blocks the 1S-2S preparation laser beam.

The gas dynamics inside the nozzle are rather complicated as $\mathrm{H}_2$ freezes to and sublimates from the nozzle walls and H adsorbs to the $\mathrm{H}_2$-coated walls and recombines, depending on and at the same time changing the local density of H and $\mathrm{H}_2$. Furthermore, no direct measurement of the degree of dissociation $\alpha_\mathrm{dis}$ after the PTFE tubing or after the nozzle is available. Nevertheless, some coarse estimates regarding the gas dynamics can be made. The typical flow of $Q_{\mathrm{H}_2} = 0.35$ ml/min of hydrogen molecules into the dissociator corresponds to $1.57 \times 10^{17}$ molecules/s, as detailed in Section 4.5.1. Using an FEM simulation[1] and neglecting the adsorption of $\mathrm{H}_2$ to the walls as well as the recombination of H, the partial pressures of H and $\mathrm{H}_2$ inside the nozzle for different values of the degree of dissociation $\alpha_\mathrm{dis}$ at the nozzle input can be found. Note that the partial pressure of $\mathrm{H}_2$ can be above the solid phase vapor pressure in the simulation, which is not the case in equilibrium. Here, this is interpreted as a sign that freezing of $\mathrm{H}_2$ will occur. For $T_\mathrm{N} = 4.8\,\mathrm{K}$ and $\alpha_\mathrm{dis} = 50\,\%$, 25 %, 10 %, the partial pressures at the center of the nozzle are found to be $P_\mathrm{H} = 2.4 \times 10^{-5}$ mbar, $1.2 \times 10^{-5}$ mbar, $4.8 \times 10^{-6}$ mbar and $P_{\mathrm{H}_2} = 1.7 \times 10^{-5}$ mbar, $2.6 \times 10^{-5}$ mbar, $3.1 \times 10^{-5}$ mbar for H and $\mathrm{H}_2$, respectively. The corresponding recombination probabilities per collision for these H partial pressures are $\gamma_{\mathrm{H}_2} = 1.3 \times 10^{-3}$, $6.7 \times 10^{-4}$, $2.7 \times 10^{-4}$, at least a factor of ten lower than for PTFE, copper, or aluminum at the same temperature [117]. However, for lower temperatures $\gamma_{\mathrm{H}_2}$ rapidly increases, reaching $5 \times 10^{-2}$ for $T_\mathrm{N} = 3.7\,\mathrm{K}$ and $P_\mathrm{H} = 4.8 \times 10^{-6}$ mbar. At $T_\mathrm{N} = 4.8\,\mathrm{K}$, $Q_{\mathrm{H}_2} = 0.35$ ml/min, and $\alpha_\mathrm{dis} \lesssim 25\,\%$, $P_{\mathrm{H}_2}$ is thus slightly above $P_{\mathrm{H}_2,\mathrm{sat}} = 2.3 \times 10^{-5}$ mbar, allowing the formation of solid $\mathrm{H}_2$. However, since $P_{\mathrm{H}_2}$, $P_{\mathrm{H}_2,\mathrm{sat}}$, and $P_\mathrm{H}$ are all of comparable magnitude, the atomic beam at least partly consists not only of H, but also of $\mathrm{H}_2$.

The number of cold wall collisions of particles leaving the nozzle is estimated with a particle tracing simulation[2], in which the particles are released 28 mm below the nozzle inlet inside the PTFE tubing, which is assumed to be at a temperature such that no solid $\mathrm{H}_2$ is formed there. Collisions between the particles are neglected. In this way, two distributions are determined: first, the total number $N_\mathrm{recomb}$ of cold wall collisions, relevant to the recombination dynamics. Second, the number $N_\mathrm{th}$ of cold wall collisions after the last PTFE (hot) wall collision, relevant for the thermalization dynamics. Both distributions posses a steep initial rise with a long exponential tail. $N_\mathrm{recomb}$ has a mode of 7 and a mean of 26 collisions, with 95 % of particles leaving the nozzle after less than 75 collisions. $N_\mathrm{th}$ has a mode of 3 and a mean of 7 collisions, with 95 % of particles undergoing less than 19 collisions and 91 % of particles undergoing more than one collision. As H is adsorbed to $H_2$, it is reasonable to assume that a single cold wall collision is enough to thermalize H to the wall temperature, with at least a single collision necessary for particles to leave the t-shaped nozzle.

[1]Finite element method, implemented using the Free Molecular Flow module of the commercial simulation software package COMSOL.

[2]Implemented using the Mathematical Particle Tracing module of COMSOL. Adding collisions between the particles does not change the results substantially.

Using the mean value of 26 cold collisions and the values of $\gamma_{H_2}$ given above, the degree of dissociation is only negligibly reduced to 97 % . . . 99 % of its level at the nozzle input. Thus, recombination of H on the $H_2$-coated walls of the nozzle plays a minor role compared to the recombination during transport through the PTFE tubing. With this, the result found in Section 4.2.10 that at $T_N = 4.8\,\mathrm{K}$ the hydrogen flow into the outer vacuum region is only $Q_{\mathrm{cold}}/Q_{\mathrm{warm}} = 1/3$ of its value at $T_N = 30\,\mathrm{K}$ can be interpreted as an upper bound of $\alpha_{\mathrm{dis}}$ at the nozzle input. This is because, to good approximation, no new $H_2$ is formed in the nozzle and all hydrogen not leaving the nozzle must freeze out on its walls as $H_2$. The real value of $\alpha_{\mathrm{dis}}$ might be much lower as some of the $H_2$ still leaves the nozzle, as discussed above.

From these coarse estimates, including those made in Section 4.5.1, it is not unreasonable to assume that the degree of dissociation at the nozzle input is only on the order of $\alpha_{\mathrm{dis}} = 10\,\%$. Together with the observed flow reduction $Q_{\mathrm{cold}}/Q_{\mathrm{warm}}$, this results in[1] $\alpha_{\mathrm{dis}} = 30\,\%$ after the cold nozzle ($T_N = 4.8\,\mathrm{K}$). In other words, for each H atom leaving the nozzle, on average 1.17 $H_2$ molecules are also emitted, and only $3.13 \times 10^{16}$ atoms/s leave the nozzle as $1.57 \times 10^{17}$ molecules/s stream into the dissociator. Note that the nozzle has two orifices, and only half of the particles leaving the nozzle are emitted towards the 2S-6P spectroscopy region and thus are of interest in the experiment here, while the other half hits the incoupling mirror of the 243 nm enhancement cavity. That is, the number of atoms and molecules leaving the nozzle per second in the direction of the spectroscopy region is estimated to be $N'_{1S} = 1.6 \times 10^{16}$ atoms/s and $N'_{H_2} = 1.8 \times 10^{16}$ molecules/s, respectively.

At the same time, $1.04 \times 10^{17}$ molecules/s freeze out on the nozzle, corresponding to a growth of solid hydrogen[2] of 14.1 mm$^3$/h. This volume is sufficient to cover the whole interior of the nozzle with an approximately 0.3 mm thick layer of solid hydrogen after two hours, shrinking the diameter of the nozzle channel to 1.4 mm. Two hours corresponds to the time after which the losses of the 1S-2S preparation laser beam propagating through the nozzle become too large to continue the experiment and the solid hydrogen needs to be removed by heating the nozzle. For these losses to occur, however, the reduction in diameter needs to be ≈1.0 mm (see Section 4.3.3.7). Indeed, this reduction is close to the actual value observed in the imaging of the nozzle as shown in Fig. 4.33, with the difference between the naively expected and the observed reduction attributed to uneven freezing inside the nozzle, as detailed in Section 4.5.2.5.

#### 4.5.2.4 Influence of nozzle temperature and hydrogen flow on atomic beam

To study the influence of the nozzle temperature $T_N$ and hydrogen flow $Q_{H_2}$ on the properties of the atomic beam, test measurements were conducted where both parameters were varied while recording line scans of the 2S-6P transition. The resulting time-resolved count rates of the top detector, $N_{\text{on-res}}$, with the spectroscopy laser on resonance, and $N_{\text{off-res}}$, with the laser detuned by 50 MHz from the resonance where almost no excitation takes place, are combined to give the amplitude, $N_{\text{on-res}} - N_{\text{off-res}}$, and the background-to-amplitude ratio (BAR), $N_{\text{off-res}}/(N_{\text{on-res}} - N_{\text{off-res}})$. This procedure mimics the analysis of the 2S-6P resonances by fitting line shapes (see Fig. 5.1), but allows for a finer time resolution. Fig. 4.28 shows the result of the first test measurement, where three freezing cycles (FC) were acquired in sequence in a single day, with $Q_{H_2}$ set to 1.00 ml/min, 0.50 ml/min, and 0.35 ml/min, respectively, while $T_N = 4.6\,K$ was kept constant. As done during the 2S-6P measurement, the nozzle was heated up to room temperature between the second and third FC to remove frozen trace gases. Between the first and second FC, however, the nozzle was only heated up to ≈35 K, as was previously done in the 2S-4P measurement (see Appendix A), which removes frozen hydrogen, but not necessarily other trace gases present in the vacuum system. The residual pressure in the outer vacuum region during the freezing cycles is dominated by $H_2$ and is found to be $P_{OV} = 1.1 \times 10^{-5}$ mbar, $6.2 \times 10^{-6}$ mbar, $4.8 \times 10^{-6}$ mbar for the three values of $Q_{H_2}$.

[1] For a degree of dissociation at the nozzle input of $\alpha_{\mathrm{dis}}$ and a ratio $Q_{\mathrm{cold}}/Q_{\mathrm{warm}}$ between the flow for a cold nozzle, where some of the $H_2$ freezes out, but H recombination is negligible, and the flow for a warm nozzle, where no $H_2$ freezes out, the degree of dissociation after the nozzle is $\alpha_{\mathrm{dis}}/(Q_{\mathrm{cold}}/Q_{\mathrm{warm}})$.

[2] The density of solid hydrogen is 0.089 g/cm$^3$ at a temperature of 4.8 K [130].

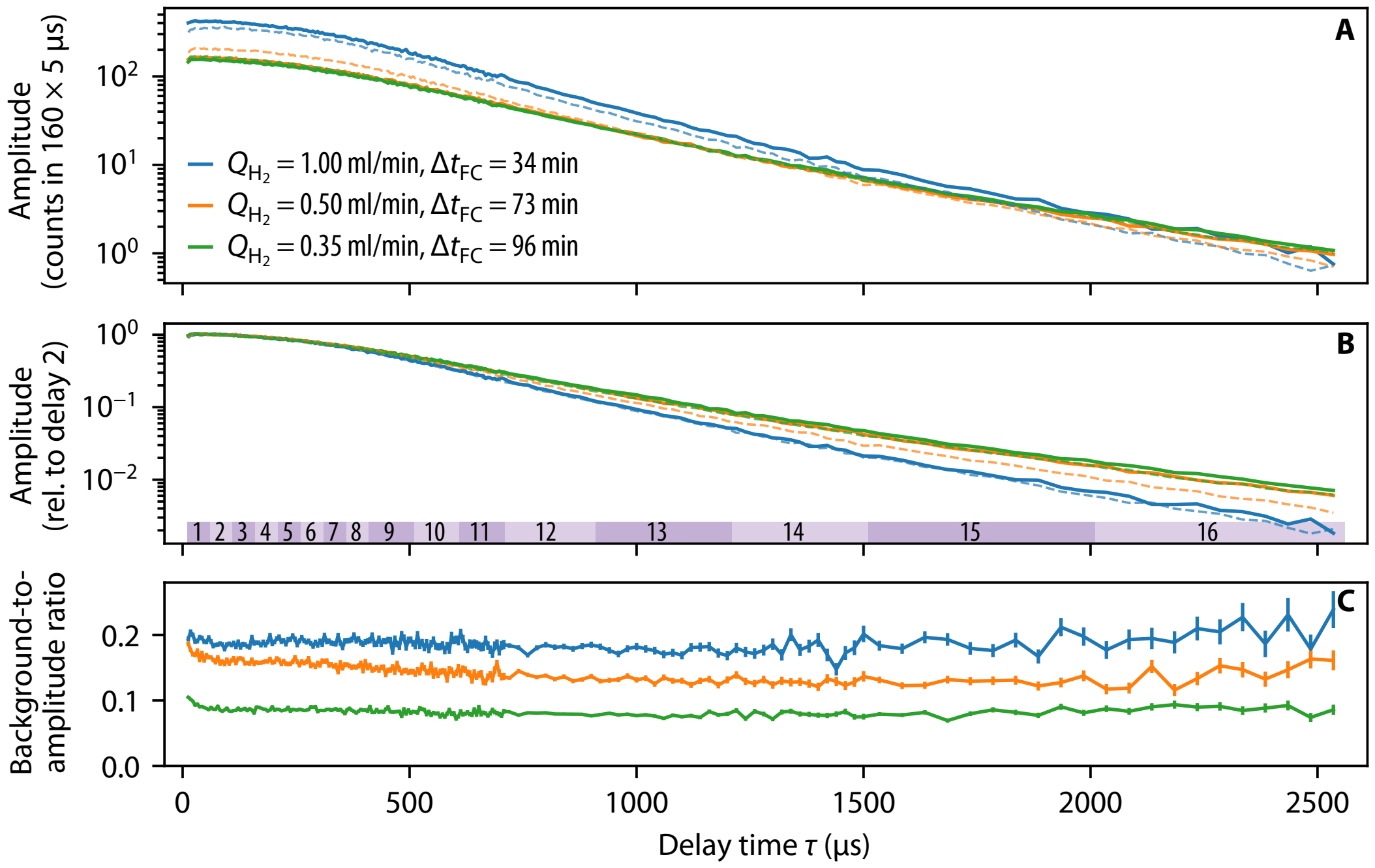

Figure 4.28: Average number of time-resolved fluorescence counts detected at the top detector for different flows $Q_{H_2}$ of $H_2$ into the hydrogen dissociator. The nozzle is held at a temperature of $T_N = 4.6\,K$, slightly lower than the typical 4.8 K, and, different to the design used in the precision measurement, the PTFE input tube is directly attached to the nozzle without a spacer. For each flow, data are recorded for one freezing cycle (FC), limited by the freezing of the nozzle after a time $\Delta t_{FC}$. (**A**) 2S-6P amplitude $N_{\text{on-res}} - N_{\text{off-res}}$, where $N_{\text{on-res}}$ and $N_{\text{off-res}}$ are the counts with the spectroscopy laser on resonance and detuned from the resonance, respectively. The counts are sorted into 5 µs-long bins at delay time $\tau$, summed over 160 chopper cycles, and averaged of over the first half (solid lines) and second half (dashed lines) of the FCs. (**B**) Same data as in (A), but the amplitude is normalized to the average of delay 2. The purple-shaded regions indicate the duration of the delays 1-16 used in the analysis (see Table 5.1). (**C**) The background-to-amplitude ratio $N_{\text{off-res}}/(N_{\text{on-res}} - N_{\text{off-res}})$ averaged over the complete FC. The data are additionally averaged over 4 and 10 bins for delays 13 to 15 and delay 16, respectively, to reduce the noise level. The powers of the 2S-6P spectroscopy and 1S-2S preparation lasers were $P_{2S\text{-}6P} = 30\,\mu W$ and $P_{1S\text{-}2S} = 0.85\,W$, respectively.

Assuming again $\alpha_{\text{dis}} = 10\,\%$ at the nozzle input and using the FEM simulation described above, the three hydrogen flows are found to correspond to $P_\mathrm{H} = 1.3 \times 10^{-5}$ mbar, $6.6 \times 10^{-6}$ mbar, $4.7 \times 10^{-6}$ mbar ($P_{\mathrm{H}_2} = 8.5 \times 10^{-5}$ mbar, $4.3 \times 10^{-5}$ mbar, $3.0 \times 10^{-5}$ mbar) and $\gamma_{\mathrm{H}_2} = 1.6 \times 10^{-3}$, $7.9 \times 10^{-4}$, $5.7 \times 10^{-4}$. For 26 collisions with $H_2$-coated walls inside the nozzle, this results in a maximum reduction of $\alpha_\mathrm{dis}$ to 96 % of the value at the input. The duration $\Delta t_\mathrm{FC}$ of the three FCs, i.e., the time it takes till the growing layer of solid $H_2$ interferes with the 1S-2S spectroscopy laser, was found to be 34 min, 73 min, and 93 min. Thus, $\Delta t_\mathrm{FC}$ is inversely proportional to $Q_{\mathrm{H}_2}$, which implies that the $H_2$ growth rate is independent of $Q_{\mathrm{H}_2}$ and thus $P_\mathrm{H}$. This is consistent with the estimation above that indeed the recombination inside the nozzle is negligible in comparison to the recombination on the PTFE tubings, which does not depend on $P_\mathrm{H}$. Likewise, the prompt amplitude (delay time $\tau \lesssim 500\,\mu\mathrm{s}$) increases approximately linearly when increasing the flow from 0.35 ml/min to 1.00 ml/min for both the first and the second half of the FCs (solid and dashed lines in Fig. 4.28, respectively). For 0.50 ml/min, the amplitude expected from this scaling only is observed during the second half of the FC, which might be related to the low temperature used to unfreeze the nozzle. This linear behavior also suggests that the formation of the $H_2$ layer progressed in a similar way, compatible with the result of the FEM simulation that the partial pressure of $H_2$ is always well above the saturated vapor pressure of solid $H_2$ of $P_{\mathrm{H}_2,\mathrm{sat}} = 9.5 \times 10^{-6}$ mbar. Note that for $T_\mathrm{N} = 4.8$ K and $Q_{\mathrm{H}_2} = 0.35$ ml/min as used in the 2S-6P measurement, the mode of the FC durations[1] $\Delta t_\mathrm{FC}$ is 120 min, about 30 % longer than in the test measurements here at $T_\mathrm{N} = 4.6$ K. If this difference is indeed significant, it could be caused by the absence of a thermally isolating spacer between the PTFE tubings and the nozzle in the test measurements, as discussed in Section 4.5.2.1.

Importantly, for larger $\tau$, the scaling of the amplitude with $Q_{\mathrm{H}_2}$ as seen for the prompt amplitude reduces, with all three values of $Q_{\mathrm{H}_2}$ resulting in the same number of counts for the maximum delay time. This implies a relative decrease of slower atoms, which dominate the amplitude at large $\tau$, for increasing flow. This behavior is highlighted in Fig. 4.28 (B), which shows the same amplitudes, but normalized to the average number of counts in delay 2 ($\tau = 60\,\mu\mathrm{s} \ldots 110\,\mu\mathrm{s}$) to show their relative scaling with $\tau$. As detailed in Section 4.5.2.2, this behavior is expected as the increasing pressure inside the nozzle and the atomic beam leads to more collisions, removing primarily slow atoms from the beam. For the maximum delay time $\tau = 2.5$ ms, there are a factor of 3.5 less atoms for $Q_{\mathrm{H}_2} = 1.00$ ml/min as compared to three times lower hydrogen flow of $Q_{\mathrm{H}_2} = 0.35$ ml/min.

Fig. 4.28 (C) shows the background-to-amplitude ratio. The background is known to be related to the presence of 2S atoms, as the background is observed only if, simultaneously, the preparation laser is on-resonance with the 1S-2S transition, hydrogen is introduced into the system, and the atomic beam is overlapped with the preparation laser. The BAR increases with increasing hydrogen flow, i.e., the background counts increase more than the on-resonance counts shown in Fig. 4.28 (A, B). This behavior is expected if the background counts are caused by 2S atoms quenched or kicked out of the beam by intra-beam collisions with other atoms and molecules, and by collisions with residual hydrogen molecules in the

[1] FCs stopped for other reasons than the nozzle freezing are not included in this statistic. The coefficient of variation of $\Delta t_\mathrm{FC}$ is 10 % over the 56 FCs considered.

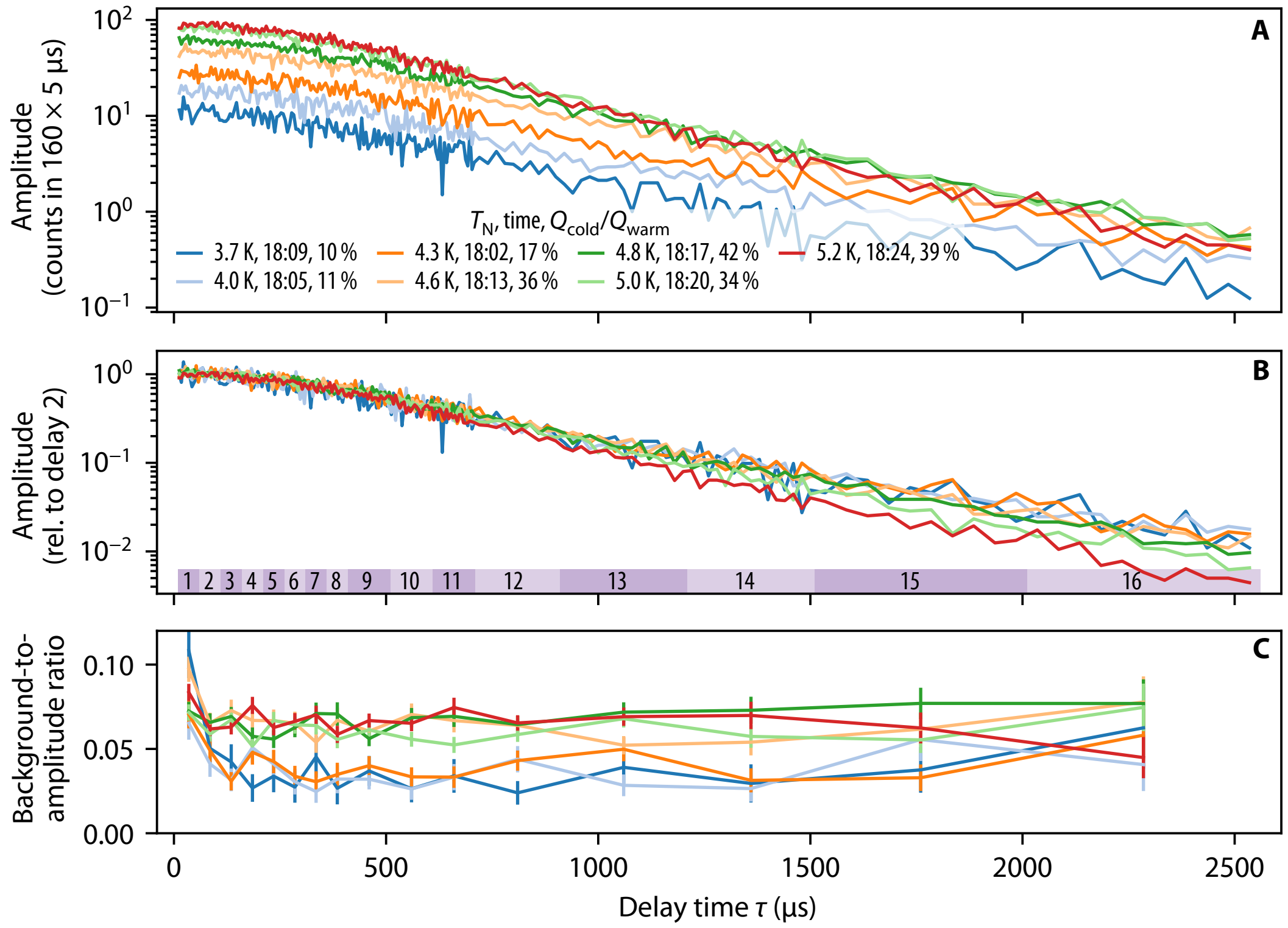

Figure 4.29: Average number of time-resolved fluorescence counts detected at the top detector, similar to Fig. 4.28, but for different nozzle temperatures $T_N = 3.7\,\mathrm{K} \ldots 5.2\,\mathrm{K}$ and a constant flow of $Q_{H_2} = 0.35\,\mathrm{ml/min}$ of $H_2$ into the dissociator. For each temperature, data are recorded for a few minutes, with the temperature, the time of the measurement, and the fraction $Q_{cold}/Q_{warm}$ of hydrogen not freezing out on the nozzle shown in the legend of (A). For (A, B), the data are averaged over 4 and 10 bins for delays 13 to 15 and delay 16, respectively, while for (C) all bins within each delay are averaged. The powers of the 2S-6P spectroscopy and 1S-2S preparation lasers were $P_{2S\text{-}6P} = 30\,\mu\mathrm{W}$ and $P_{1S\text{-}2S} = 0.65\,\mathrm{W}$, respectively.

vacuum chamber, as the number density of both increases with increasing hydrogen flow[1].

In a second test measurement, shown in Fig. 4.29 and Fig. 4.30, the hydrogen flow was kept constant at $Q_{H_2} = 0.35\,\mathrm{ml/min}$ while the temperature $T_N$ of the nozzle was varied between 3.7 K and 6.0 K. This measurement consists of a single FC, with the temperature changed every few minutes and a few line scans recorded at each temperature. The prompt amplitude (see Fig. 4.29 (A)) increases as the temperature is increased from initially 3.7 K (at 18:09), with the maximum amplitude reached for approximately 5.2 K (at 18:24). At the same time, the ratio $Q_{cold}/Q_{warm}$ of hydrogen freezing out on the nozzle increases from 10 % to 39 %. These observations are consistent with an increase in recombination of H on the solid $H_2$ inside the nozzle with lower temperature, as expected from the rapid increase of the recombination probability per collision $\gamma_{\mathrm{H}_2}$ (see Fig. 4.27). For $T_{\mathrm{N}} = 3.7\,\mathrm{K}$, $\gamma_{\mathrm{H}_2}$ is estimated to be 4.3 % and $\alpha_{\mathrm{dis}}$ is consequently reduced to 42 % of its value at the nozzle input.

[1] Intra-beam collisions and collisions with residual hydrogen molecules could possibly be distinguished by their different delay dependence. This is because the probability for collisions that quench or remove 2S atoms from the beam is expected to depend on the relative speed of the collision partners. With increasing $\tau$, the speed of the 2S atoms decreases, while the speed of other 1S atoms and molecules in both the beam and in the residual gas stays is independent of $\tau$, as only a small fraction of atoms is excited to the 2S level. However, since the residual gas is at room temperature, the relative speed of collisions with these residual particles is dominated by the speed of the room-temperature particles. Thus, it changes less with $\tau$ as compared to intra-beam collisions. This situation is complicated by the fact that the 6P excitation probability, which gives the amplitude, also increases with decreasing speed of the 2S atoms.

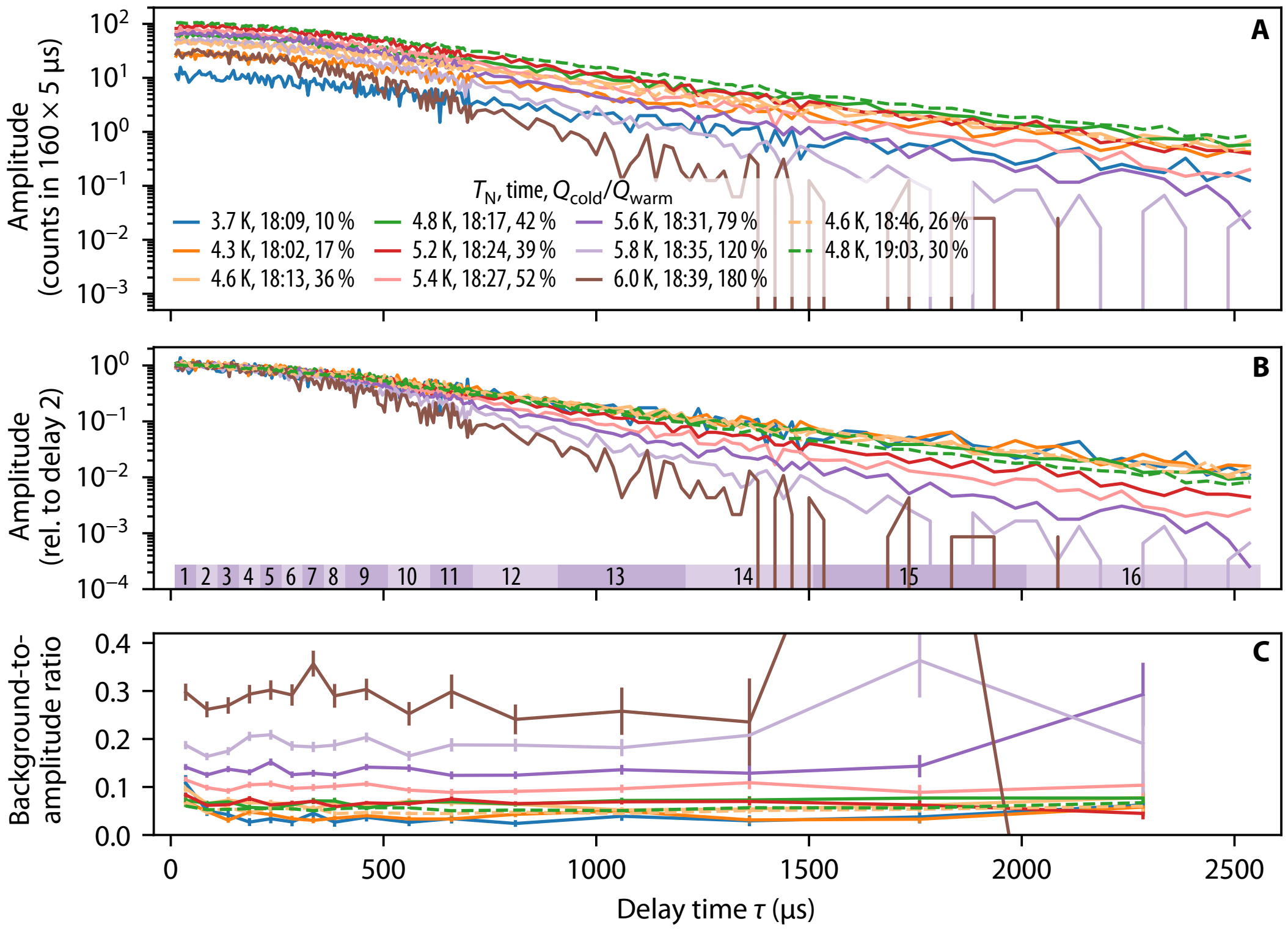

Figure 4.30: Similar to Fig. 4.29, but additionally including data for a larger nozzle temperature range $T_{\mathrm{N}} = 3.7\,\mathrm{K} \ldots 6.0\,\mathrm{K}$ (solid lines), and data acquired after the temperature was cycled to check the reproducibility of the observed signal (dashed lines).

For higher temperatures than 5.2 K, shown in Fig. 4.30, the prompt amplitude decreases again, while $Q_{\mathrm{cold}}/Q_{\mathrm{warm}}$ keeps increasing. At 5.8 K, $Q_{\mathrm{cold}}/Q_{\mathrm{warm}}$ even surpasses 100 %, implying that more hydrogen leaves the nozzle than enters it. This could be a transient effect caused by hydrogen frozen to the nozzle walls at lower temperatures sublimating into the gas phase again at higher temperatures. Indeed, for the parameters used so far, the saturated vapor pressure of solid $H_2$ is expected to surpass the partial pressure of $H_2$ inside the nozzle at approximately 4.9 K (see Fig. 4.27). At around this point, $H_2$ will then stop freezing on the nozzle and the pressure inside the nozzle, the beam, and the vacuum system increases. The latter pressure increases more than four-fold from $4.8 \times 10^{-6}$ mbar at 5.2 K to $2.2 \times 10^{-5}$ mbar at 6.0 K. The removal of frozen $H_2$ at some points during FC is also supported by the unusually long duration of this FC of $\Delta t_{\mathrm{FC}} \approx 160\,\mathrm{min}$. Because of the transient nature of this test measurement, the inside of the nozzle is still at least partially covered by solid $H_2$ at all times, as confirmed by the imaging of the nozzle. Thus, the reduction of the amplitude could be caused both by recombination on exposed copper walls, where the recombination probability per collisions could be as high as 20 % [117], and by the increased pressure interfering with the formation of the atomic beam. However, right after the maximum temperature at 6.0 K, the temperature was again reduced to 4.6 K at 18:46 (dashed orange line in Fig. 4.30), and a very similar amplitude as at 18:13 was found, suggesting that either no walls were exposed during at higher temperatures or the exposed sections were quickly covered by solid $H_2$ again. The amplitude was also found to be reproducible when returning, after the temperature cycling was completed, to 4.8 K, ≈45 min after first measuring at this temperature (dashed dark green line in Fig. 4.30).

Fig. 4.29 (B) and Fig. 4.30 (B) again show the same amplitudes, but normalized to the average number of counts in delay 2. For temperatures up to 4.6 K, no change in the normalized amplitude is observed. Starting from 4.8 K, again a depletion of slow atoms is visible, with the amplitude at the maximum delay time decreasing by almost two orders of magnitude from 4.6 K to 6.0 K and by one order of magnitude for intermediate delay times of $\tau \approx 1000\,\mu\text{s}$, corresponding to mean atom speeds of $\bar{v} \approx 60\,\text{m/s}$ and $\bar{v} \approx 120\,\text{m/s}$, respectively (see Table 5.1). This depletion is attributed to the increased pressure in the system, removing slower atoms from the beam as detailed in Section 4.5.2.2.

The BAR, shown in Fig. 4.29 (C) and Fig. 4.30 (C), starts to increase strongly at 5.4 K, with a five-fold increase observed between 4.8 K and 6.0 K. This further supports the assumption that intra-beam collisions and collisions with residual hydrogen molecules cause the background, as both increase as more molecular hydrogen leaves the nozzle with increasing temperature.

For all temperatures, the BAR increases slightly with increasing $\tau$. As discussed before, this is to be expected as the mean speed of the 2S atoms decreases with $\tau$, but the velocity distributions of 1S atoms and molecules in the beam, and residual particles in the background gas, stay unchanged. However, for $T_\text{N} \lesssim 4.8\,\text{K}$, the prompt BAR within the first ≈50 µs is observed to be larger, at ≈10 %, than for the next 1 ms or so. For 3.7 K, where this effect is most pronounced, the prompt BAR reaches ≈11 % before dropping to ≈3 %. As the nozzle temperature, and with it the overall BAR, increases, the effect is less and less visible. The effect is also visible in the BAR of the first measurement (see Fig. 4.28 (C)). A possible mechanism could be that some of the atoms leaving the nozzle are, conversely to the arguments made before, not thermally accommodated to its temperature, e.g., because they only underwent one or more elastic collision inside the nozzle. These atoms can then also be excited to the metastable 2S level and consequently contribute to both the background and on-resonance signal. The latter contribution is strongly suppressed by the low 2S-6P excitation probability for room temperature atoms due to the $1/v$ scaling for short interaction times. The contribution to the background from collisions with residual hydrogen molecules, on the other hand, is only expected to be reduced by a factor of $\approx \sqrt{2}$ as compared to cold atoms. Thus, the presence of room-temperature atoms can lead to an increased BAR. For the thermal velocity of hydrogen atoms at room temperature, 2924 m/s, the maximum possible delay time for 2S excitation is 70 µs, which matches the observed time scale. The disappearance of the effect for increasing temperature, on the other hand, could then be related to the increased pressure in the nozzle, leading to thermalizing collisions between particles, and changes in the temperature accommodation properties of the nozzle walls.

It is instructive to compare the measured amplitudes to the behavior expected for an atomic beam with a speed distribution of the form of Eq. (4.27), which takes the depletion of slow atoms into account with an exponential suppression term characterized by the cutoff speed $v_\text{cutoff}$. The expected behavior is extracted from Monte Carlo simulations, as detailed in Section 5.2, using the corresponding nozzle temperatures and laser powers. The experimentally observed and simulated amplitudes for the first measurement at different hydrogen flows $Q_{H_2}$ are shown in Fig. 4.31 (A). For all flows, the amplitude for large delay times $\tau$ and thus slower atoms is below the simulated amplitude if no exponential suppression is assumed (crosses), i.e., $v_{cutoff} = 0\,m/s$. However, the observed amplitude, where here the data from the second half of the FCs are used (dashed lines), can be described by the simulations when assuming $v_{cutoff}$ to be 130 m/s (diamonds), 65 m/s (right-pointing triangles), and 30 m/s (left-pointing triangles) for $Q_{H_2} = 1.00\,ml/min$, 0.50 ml/min, and 0.35 ml/min, respectively.

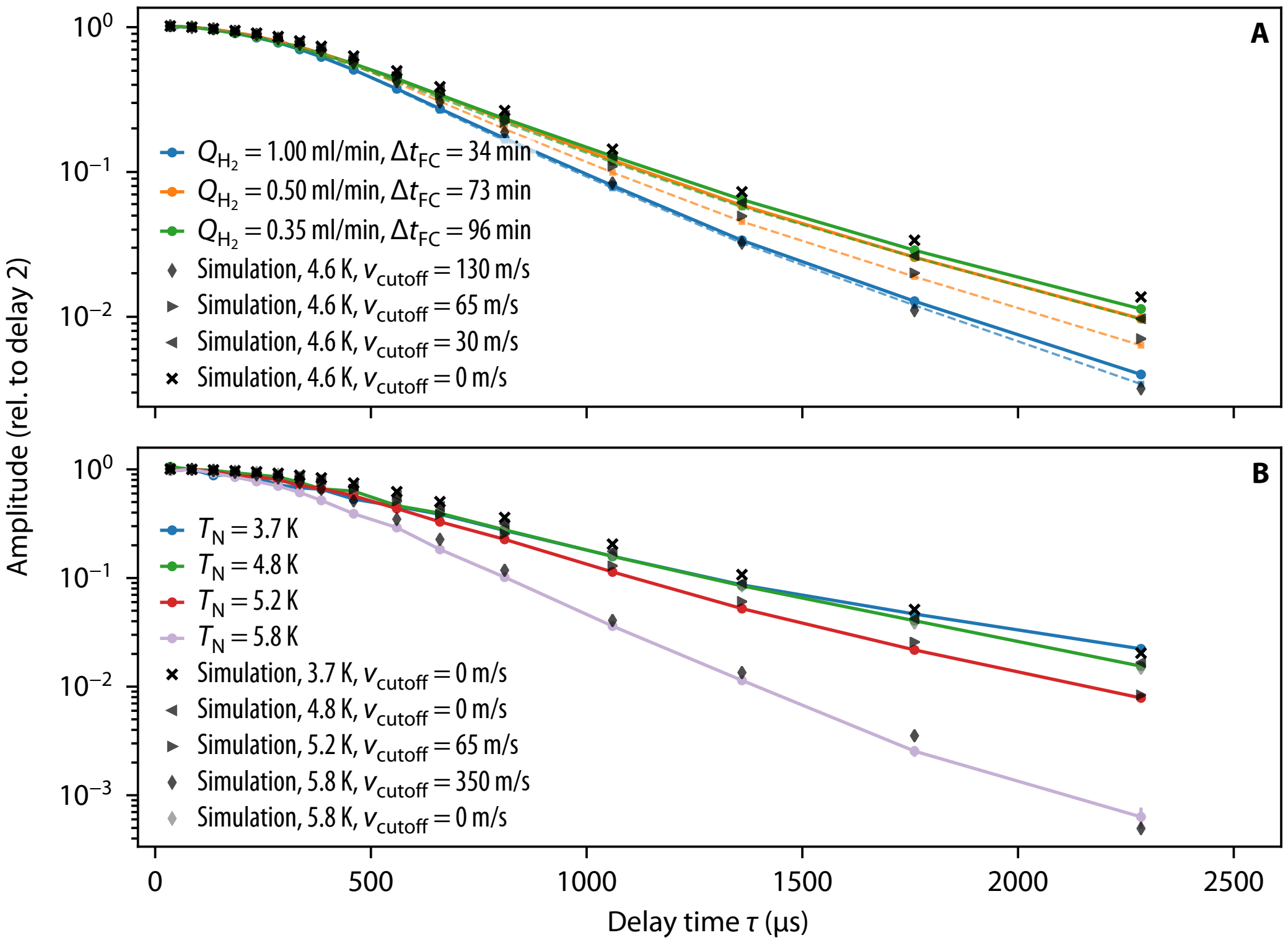

Figure 4.31: Comparison with simulation results for the experimental time-resolved fluorescence count rate for (**A**) the various hydrogen flows $Q_{H_2}$, at a nozzle temperature of $T_N = 4.6\,K$, of Fig. 4.28 and (**B**) the various nozzle temperatures $T_N$, at a hydrogen flow of $Q_{H_2} = 0.35\,ml/min$, of Fig. 4.30. The underlying speed distribution of the simulations is given by Eq. (4.27). The bins within each delay are averaged to match the simulation results.

Fig. 4.31 (B) shows the corresponding situation for the second measurement for four values of the nozzle temperature $T_N$. Within the experimental uncertainty, the experimental data for $T_N = 3.7\,K$ (blue line) and $T_N = 4.8\,K$ (green line) are approximately described by unsuppressed speed distributions (crosses and left-pointing triangles). For increasing temperatures, however, a nonzero cutoff speed $v_{cutoff}$ is needed to describe the data, found to be 65 m/s and 350 m/s for $T_N = 5.2\,K$ (red line and right-point triangles) and 5.8 K (purple line and diamonds). For reference, the expected behavior for $v_{cutoff} = 0\,m/s$ at $T_N = 5.8\,K$ is also shown (gray diamonds), demonstrating that the observed suppression of slow atoms is much larger than to be expected from the increased nozzle temperature alone.

The nozzle temperature and hydrogen flow at which to acquire spectroscopy data are thus a compromise between the size of the prompt amplitude, the depletion of slow atoms, the background-to-amplitude ratio, pressure shifts, and the time it takes the nozzle to freeze. For the 2S-6P measurement, a temperature of $T_N = 4.8\,K$ and a flow of $Q_{H_2} = 0.35\,ml/min$ was chosen. The prompt amplitude is close to the maximum at this temperature, while the depletion of slow atoms is only just visible, and the BAR is still reasonably low. At the same time, the FC duration of ≈2 h is long enough to include alignment procedures and auxiliary measurements in each FC without reducing the amount of acquired line scans by too much.

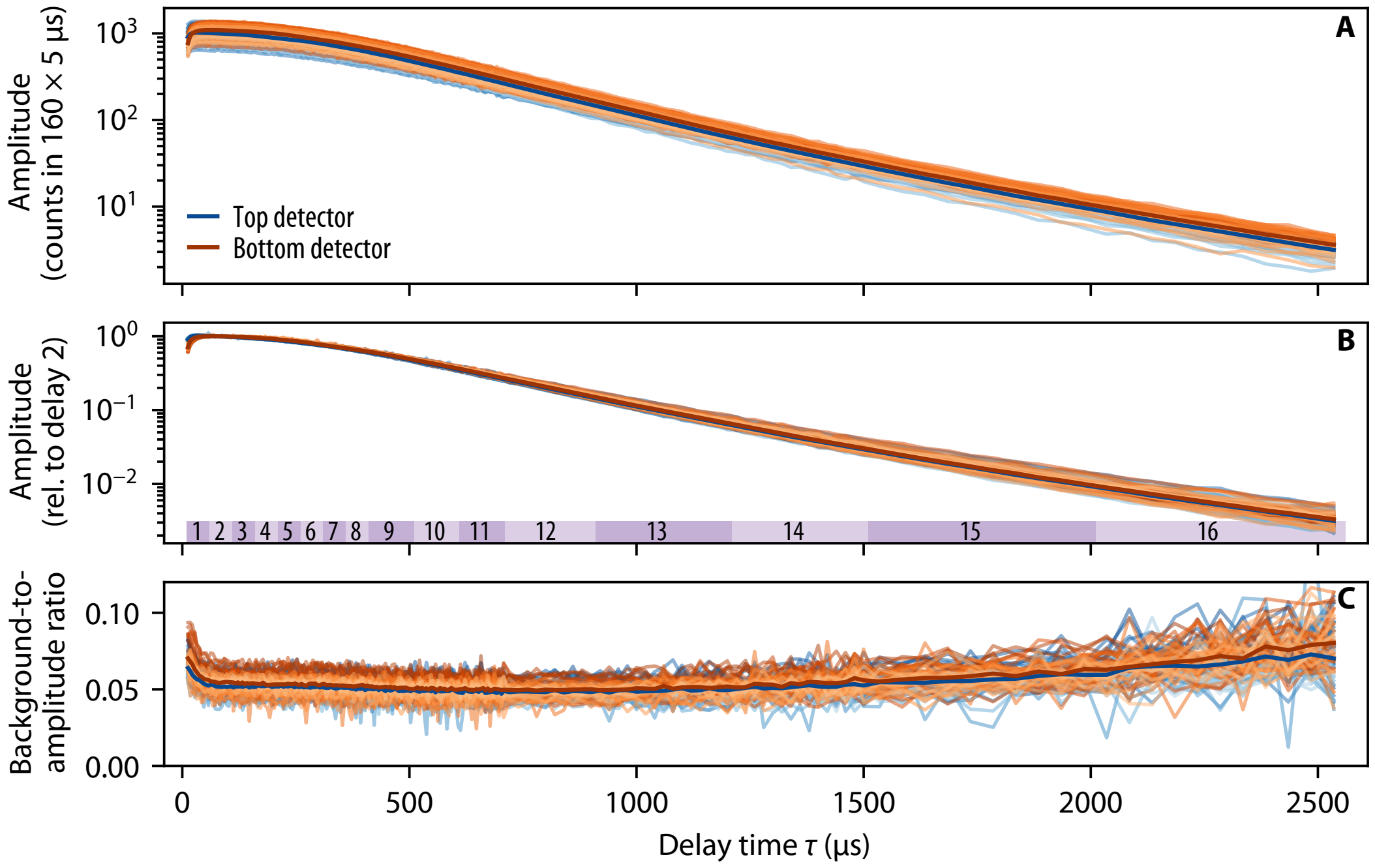

Figure 4.32: Time-resolved fluorescence count rate of the top (blue lines) and bottom (orange lines) detector for all 41 freezing cycles (FCs) during run B of the 2S-6P measurement (see Table 6.1). Only data contributing to the measurement of the $2S\text{-}6P_{1/2}$ ($2S\text{-}6P_{3/2}$) transition with a spectroscopy laser power of $P_{2S\text{-}6P} = 30\,\mu W$ ($P_{2S\text{-}6P} = 15\,\mu W$) are shown (data groups G1B, G2–3 and G7B, G8–9, see Table 6.2). The nozzle temperature is $T_N = 4.8\,K$ and the hydrogen flow is $Q_{H_2} = 0.35\,ml/min$, except for one and two FCs for which $Q_{H_2} = 0.25\,ml/min$ and $0.30\,ml/min$, respectively. As in Fig. 4.28, (**A**) the 2S-6P amplitude, (**B**) the amplitude normalized to the average of delay 2, and (**C**) the background-to-amplitude ratio are shown. Each line corresponds to the average over a single FC, with the bold line the average over all FCs. The data are additionally averaged over 4 and 10 delay time bins for delays 13 to 15 and delay 16, respectively, to reduce the noise level.

Fig. 4.32 shows the fluorescence count rate on both the top and bottom detector during all 41 FCs of run B of the 2S-6P measurement (see Table 6.1). The chosen nozzle parameters were used for all FCs, except for one and two FCs, where $Q_{H_2}$ was slightly lower at 0.25 ml/min and 0.30 ml/min, respectively.

#### 4.5.2.5 Imaging and alignment of nozzle

In the experiment, the cryogenic nozzle is observed on an imaging sensor. To this end, as described in Section 4.3.3.6 and shown in Fig. 4.6, the transmission of the alignment laser propagating through the vacuum chamber, including the variable aperture and the nozzle channel, is detected. Some of the resulting nozzle images are shown in Fig. 4.33, with the visible interference fringes caused by the coherent illumination. The perspective corresponds to the view along the atomic beam towards the spectroscopy region.

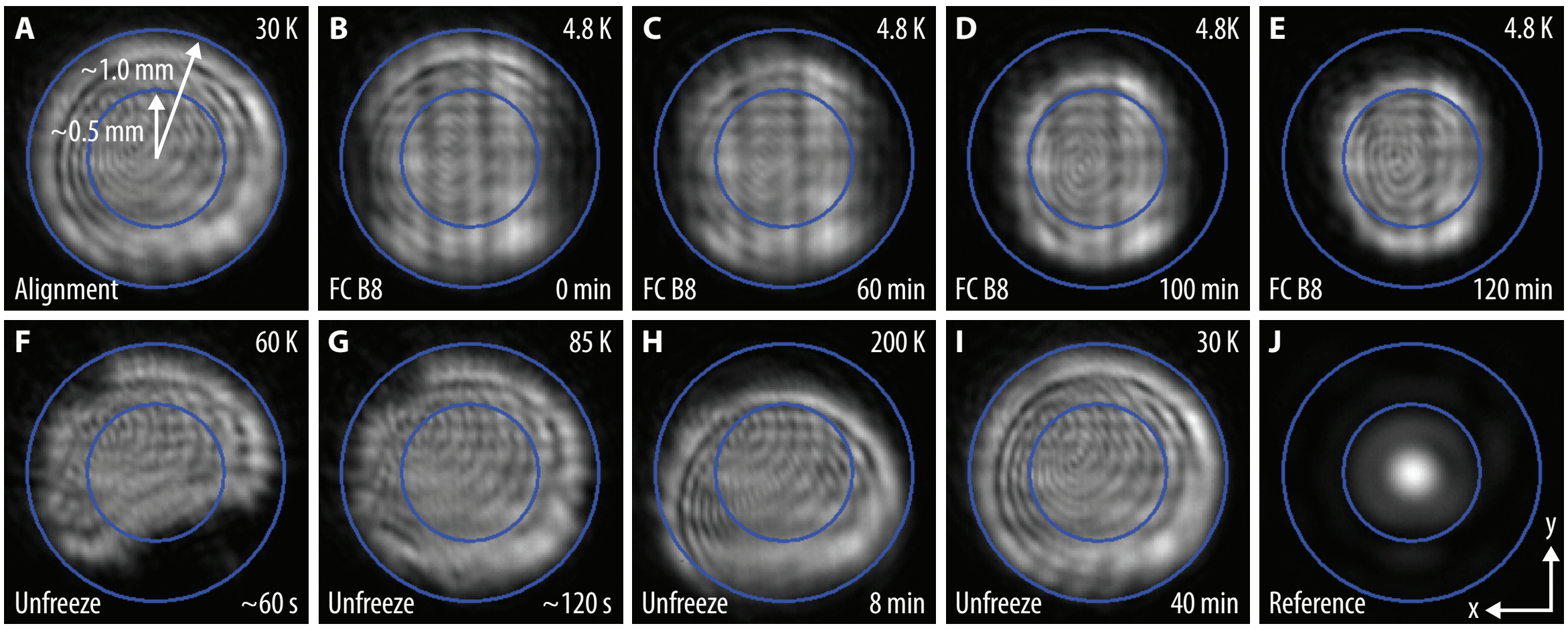

Figure 4.33: Imaging and alignment of the cryogenic hydrogen nozzle. The nozzle temperature $T_N$ is shown in the top right corner. (**A**–**E**) The clogging through the accumulation of solid $H_2$ inside the nozzle during a typical freezing cycle (FC; shown is FC B8), with the images taken right before, and 0 min, 60 min, 100 min, 120 min after the start of the FC. The time since the start of the FC is shown in the bottom right corner. From (A) to (B), the width of the variable aperture is reduced, blocking part of transmission laser used for illumination. (**F**–**H**) The unfreezing of the nozzle after the FC, with the time passed since starting the heating of the nozzle to room temperature shown in the bottom right corner. (**I**) The result of the unfreezing, with nozzle ready for the next FC. (**J**) Reference circles centered on alignment laser, with the nozzle moved out of the laser beam and the front iris almost fully closed. The same reference circles are shown in all images and have an radius of ≈1.0 mm and ≈0.5 mm.

To ensure the symmetry of the atomic beam, the nozzle channel needs to be centered on the apertures of the beam apparatus as well as on the optical axis of the 243 nm enhancement cavity. As described in Section 4.3.3.7, the latter is itself centered on the apertures, such that centering the nozzle channel on the cavity's optical axis is sufficient. Since the alignment laser in turn is overlapped with the optical axis of the cavity, the problem is reduced to centering the nozzle channel on the beam axis of the alignment laser. The nozzle can be moved in the three dimensions by moving the cryostat relative to the vacuum chamber (see Section 4.2.6). The nozzle alignment proceeds as follows: first, the nozzle is moved out of beam of the alignment laser and the variable aperture is fully opened. Next, the front iris (see Fig. 4.6), which is centered on the alignment laser, is almost fully closed and the resulting diffraction spot on the imaging sensor is marked with reference circles (see Fig. 4.33 (J)). The nozzle is then moved back into the beam of the alignment laser such that the image of the nozzle channel is centered within the reference circles (see Fig. 4.33 (A)). Finally, the width of the variable aperture is reduced to the measurement setting, blocking part of the alignment laser and thus part of the transmission through the nozzle channel (see Fig. 4.33 (B), where the aperture width was set to 1.2 mm). This procedure is done at least once per measurement day before the first freezing cycle (FC) and whenever deemed necessary, as judged from the images, before the subsequent FCs.

The imaging of the nozzle is also useful to monitor and study the clogging of the nozzle with solid $H_2$ during the course of an FC. Fig. 4.33 (A-E) show the nozzle right before, and 0 min, 60 min, 100 min, 120 min after the nozzle was cooled down to $T_N = 4.8$ K from 30 K, corresponding to the start of an FC. For the FC shown, FC B8, the nozzle including the PTFE spacer between the PTFE tubings and the nozzle input was used (see Section 4.5.2.1). The nozzle was cleaned on the same day, with one test FC done before FC 8B to remove residues from the cleaning procedure. The reduction of the radius of the nozzle channel from initially 1 mm to ≈0.5 mm is clearly visible. The channel does not freeze symmetrically (note that the clipping on the left and right edges already visible in Fig. 4.33 (B) is from the variable aperture, and not freezing), with less $H_2$ accumulating at the bottom towards the inlet. Nevertheless, some freezing does occur at the bottom. This freezing pattern could correspond to the situation where material mainly accumulates at top and the sides of inner region of the nozzle, where the input (radius of 2 mm) and nozzle channel cross, with the nozzle channel itself being uniformly covered with a comparatively thin layer of $H_2$. This is also compatible with the observation from the particle tracing simulation described above that most wall collisions occur at this inner region.

After the end of the FC, the nozzle is heated up to room temperature to remove the accumulated solid $H_2$ and other frozen trace gases by stopping the flow of liquid helium to the cryostat and switching on the heating wire inside it to maximum power. The nozzle during this unfreezing is shown in Fig. 4.33 (F–H), with the nozzle temperature having reached 60 K, 85 K, 200 K within 60 s, 120 s, 8 min, respectively. The variable aperture has been fully opened again for these images and at 200 K the thermal expansion of the cryostat has visibly shifted the nozzle downward. The frozen material still present at 60 K and 85 K is thought to be mainly remaining solid $H_2$, with the unfreezing process commonly observed to last a few minutes. It was also sometimes observed that pieces of material disappeared from the field of view only to reappear after a few seconds, which could correspond to these pieces moving within the inner region of the nozzle with the view blocked by the smaller nozzle channel. At 200 K, no large pieces of material remain visible. However, the outline of the nozzle channel frequently appeared fuzzy as compared to the start of the measurement, with the fuzziness only disappearing close to room temperature. This could be caused by trace gases with a high melting point such as water. After 40 min, as shown in Fig. 4.33 (I), the cycle of heating the nozzle up to room temperature and back down to 30 K is complete and the next FC can begin. In this case, the nozzle has returned to its initial position and no alignment of its position is necessary.

### 4.5.3 Variable beam aperture

A variable beam aperture is used to limit the divergence of the atomic beam. The divergence along the propagation direction ($x$-axis) of the 2S-6P spectroscopy laser is especially important, since it limits the observed linewidth of the 2S-6P resonance through Doppler broadening (see Section 2.2.5). However, the minimal size of the aperture is limited by the beam size of the 1S-2S preparation laser, as its beam path leads through the aperture and a too small aperture size will lead to excessive transmission losses (see Section 4.3.3.7). In order to achieve an as small as possible aperture width $d_2$ along the $x$-axis, the aperture consists of two vertical blades that can be moved along $x$-axis using remote-controlled actuators[1]. This adjustment capability, as opposed to using a fixed-width aperture as done during the 2S-4P measurement (see Appendix A), is necessary to compensate for drifts in the alignment of the apparatus (see also Section 4.3.3.7). The width can be varied between 0 mm. . . 8 mm, with $d_2 = 1.2$ mm used during the 2S-6P measurement. The height of the aperture (i.e., along the $y$-axis), on the other hand, is fixed to $d_{2,y} = 2.0$ mm. The variable aperture (**VA**), along with its actuators (**AM**), is shown in Fig. 4.1.

[1]Thorlabs Z806V dc servo motor actuators with rotary encoders. Absolute on-axis accuracy is specified as 42 µm.

To both calibrate the width of the aperture, and to center it on the 1S-2S preparation laser, the alignment laser, which is overlapped with the preparation laser, is used (see Section 4.3.3.6). To this end, first a knife-edge measurement for each blade is performed, from which the position of the blade edge, as given by the rotary encoders of the actuators, relative to the alignment laser can be extracted. Then, the blades are moved, again using the rotary encoders, to the positions corresponding to the desired aperture width. To this end, a conversion factor between the movement of the actuators and the blades needs to be taken into account, as the blades are driven through a 90° bellcrank. This factor was determined from the technical drawings of the aperture and confirmed by test measurements. Finally, the centering on the 1S-2S preparation laser is checked by observing the transmission of the 243 nm enhancement cavity while shifting the position of the aperture with the width held constant. Typically, a small offset on the order of 100 µm is found and corrected, which is attributed to imperfect overlap between the alignment and preparation laser. This procedure is repeated on each measurement day. A proof-of-principle test of the calibration scheme, using an early version of the aperture, was carried out by Clarissa Kroll as part of her bachelor's thesis [132].

## 4.6 Fluorescence detector assembly

The purpose of the fluorescence detector assembly is two-fold: first, it should detect as many of the fluorescence photons emitted by the decay of the 6P level as possible. Second, it should provide a spectroscopy region free from electric and magnetic fields, and with a sufficiently low background pressure.

The first requirement is met by basing the design on the detection of Ly-$\epsilon$ photons, which constitute over 80 % of the fluorescence (see Tables 2.2 and 2.3). These photons, with a wavelength of 94 nm, are quickly absorbed by common optical materials, making an efficient detection, covering a large solid angle, using refractive of reflective optics rather challenging. However, they are energetic enough ($h\nu = 13.22$ eV) to eject photoelectrons from many materials, including graphite and aluminum, used here, which have work functions of ≈4.6 eV and ≈4.0 eV, respectively [133]. The basic design idea is then to cover an as large solid angle as possible with these materials, while efficiently collecting and detecting the photoelectrons. The collection of the photoelectrons is achieved using applied electric fields, which guide the electrons to two channel electron multipliers (CEM). Finally, these CEMs convert each electron into an electric pulse, which is subsequently counted in a time-resolved fashion.

To fulfill the second requirement, the 2S-6P excitation and the decay take place inside a shielded inner region of the detector assembly, constituting the 2S-6P spectroscopy region. Here, the atomic beam and the 2S-6P spectroscopy laser beams cross. Grounded electrodes surround the region and shield it from stray electric fields. A magnetic shield around the inner vacuum region, together with coils outside the vacuum chamber, reduce the magnetic fields in the spectroscopy region to an acceptable level. A low background pressure inside the region is achieved by a direct connection to the cryopump.

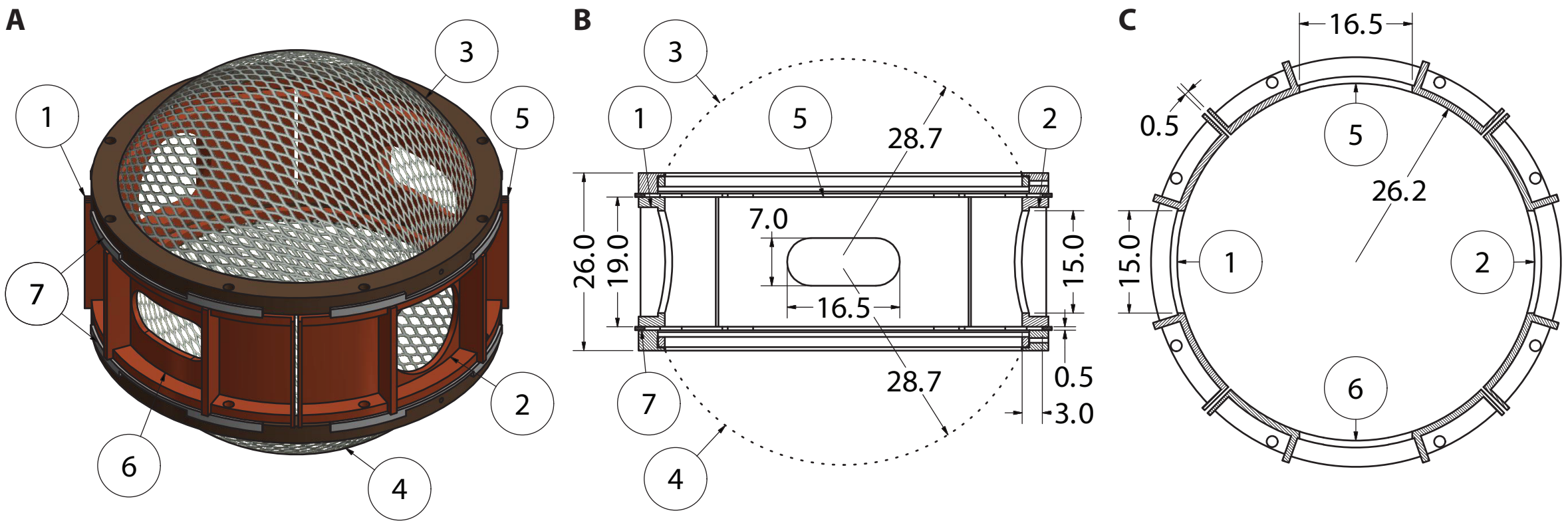

Figure 4.34: Drawing of the inner region of the fluorescence detector assembly, enclosing the 2S-6P spectroscopy region: (**A**) Orthographic projection, and section view cuts along (**B**) vertical plane and (**C**) horizontal plane through center. All dimensions are in mm. There are four ring electrodes that are quarter segments of a ring: Two electrodes with circular apertures (15.0 mm diameter) for the 2S-6P spectroscopy laser beams, (1) towards HR mirror ($+x$), and (2) towards fiber collimator ($-x$); two electrodes with rectangular apertures ($16.5 \times 7.0\,\mathrm{mm}^2$) for the atomic beam and the 1S-2S preparation laser beam, (5) downstream ($+z$), and (6) upstream ($-z$). The two remaining electrodes are spherical segments of wire mesh, (3) top ($+y$), and (4) bottom ($-y$); they are electrically isolated from (1-2, 5-6) with a (7) polyimide spacer.

### 4.6.1 Detector design

The detector assembly is shown cut-open and sitting at its position inside the beam apparatus in[1] Fig. 4.1. It is built around the detector cylinder (**DC** in Fig. 4.1), which has an inner radius of 28 mm and a length of 174.5 mm, and is oriented along the vertical $y$-axis. The atomic beam (**PB**) and 2S-6P spectroscopy laser beams (**SB**), which propagate in the horizontal ($x$-$z$-) plane through the cylinder's center, enter and exit through apertures in the cylinder's wall. The beams cross in the center of the cylinder, the interaction point, which is the point of origin of the fluorescence. The cylinder is split into three sections along its length, the top and bottom sections, housing the channel electron multipliers (CEMs) and made from aluminum, and the center section, containing the interaction point and made from copper. The two CEMs, the top (**TD**) and bottom detector (**BD**), sit at the center of each respective end of the cylinder. The CEMs slightly protrude into the cylinder by 2.3 mm and 2.8 mm, respectively, such that both their input surfaces are at a distance of 84.7 mm from the interaction point. The top of the cylinder is closed with an aluminum cap, while the bottom of the cylinder is only separated from the cryopump right below it by a fine graphite-coated wire mesh, with a graphite-coated copper frame holding the bottom detector. During usual operation, all parts of the detector assembly are held at the same potential, except the CEMs, whose input surfaces are held at 270 V to create electric fields to collect the photoelectrons emitted from the walls. The detector cylinder is mounted coaxially to the rotatable base cylinder (**RC**), which has an inner radius of 36 mm. The rotatable base cylinder is made from brass and is connected to the high-vacuum enclosure (**HV**) through a ball bearing, allowing it to rotate, together with the detector cylinder, about its axis.

The 26 mm-long center section of the cylinder forms the inner region of the detector assembly, enclosing the spectroscopy region, with the interaction point at its center. It is shown in detail in Fig. 4.34. The inner radius of this section of 26.2 mm is slightly smaller than that of the rest of the cylinder. The section's side wall is split into four electrically isolated ring electrodes, with each electrode forming a quarter section of a ring and containing an aperture[1]: two electrodes with circular apertures with a diameter of 15.0 mm for the 2S-6P spectroscopy laser beams ((1, 2) in Fig. 4.34), and two electrodes with rectangular apertures with a width of 16.5 mm and a height of 7.0 mm for the atomic beam and the 1S-2S preparation laser beam ((3, 4) in Fig. 4.34). The electrodes are made from copper and separated from each other by a small gap of ≈0.5 mm. To fully enclose the inner region with electrodes, two spherical segments of wire mesh sit above and below the ring electrodes ((5, 6) in Fig. 4.34). The spherical segments are not quite hemispherical, as their radius of 28.7 mm is slightly larger than the inner radius of the ring electrodes. The wire mesh[2] is made from 30.5 µm-diameter stainless steel wires spaced 0.508 mm apart, resulting in an open area of 88 %. The wire mesh, which is initially flat, is bent into shape and then kept in shape by clamping its rim into copper rings while under tension. The same wire mesh is used in the bottom of the detector cylinder. The ring and spherical electrodes, and the spherical electrodes and the rest of the detector assembly, are electrically isolated using 0.5 mm-thick spacers machined from polyimide blocks[3] ((7) in Fig. 4.34). These spacers are hidden from the inner region by recessing from the inner walls by 3.0 mm (see (7) in Fig. 4.34 (B)). The electrode assembly is held in place and fixed to the outer sections of the cylinder by titanium screws, which are isolated by polyimide sleeves to prevent a short-circuiting of the electrodes.

[1]Note that both Fig. 4.1 and Fig. 4.34 do not show the graphite coating applied to some parts, in order to reveal the material of the parts.

The six electrodes form a Faraday cage, shielding the spectroscopy region from stray electric fields and the electric fields used to collect the photoelectrons at the CEMs. The electrodes, including the wire meshes, are coated with colloidal graphite, serving multiple purposes. First, the metals used can form isolating oxide layers, which can accumulate charge, e.g., from photoelectrons, and thus cause stray electric fields. Second, the use of both copper and stainless steel electrodes gives rise to contact potentials between dissimilar electrodes[4], and thus electric fields, as the work functions of both materials are different on the order of 0.1 eV [133, 134]. Third, surface contamination, e.g., from residues of the solvents used for cleaning, oxidation, and the crystal structure of the metals affect the local work function [133–135], likewise leading to contact potentials and stray electric fields. By covering the electrodes with colloidal graphite, which does not form oxide layers and is applied right before the parts are installed in the vacuum chamber, a uniform conductive layer with a uniform work function is created and these effects are suppressed. Such a coating has been successfully used to suppress stray electric fields in previous experiments in the same apparatus as used here [136], and in other hydrogen atomic beam experiments [33, 137]. It was also found to prevent stray electric fields from contact potentials from dissimilar materials in [138]. The graphite coating was also applied to all parts in close proximity to the atomic beam, including the high-vacuum entrance aperture (**EA** in Fig. 4.1), the 1S-2S Faraday cage (**FC**), made from the same wire mesh as used for the electrodes, the variable aperture (**VA**), the rotatable base cylinder (**RC**), and the high-vacuum output aperture (**OA**).

[1]The corresponding apertures in the rotation mount are slightly larger: the circular apertures have a diameter of 17.0 mm, and the rectangular apertures have a width of 27.0 mm and a height of 7.0 mm.

[2]TWP, part number 050X050T0012.

[3]DuPont Vespel is used, which is a polyimide-based plastic.

[4]One might intuitively understand the electric fields arising from contact potentials in the following way: when the electrodes are electrically connected with negligible resistance, as is the case here when they are grounded, the Fermi level is identical in all electrodes. The work function is the energy required to remove an electron from a metal to the surrounding vacuum. Removing an electron from one electrode and into another, dissimilar electrode, would then result in an energy gain or loss, given by the difference in work functions, if no electric field exists between the electrodes. As the Fermi levels are however identical, this violates energy conservation, unless an electric field exists between the electrodes in which the electron losses or gains the difference in work functions.

Colloidal graphite is known to consist of 1 µm-diameter plates [139, 140], with various commercial suspensions in water or organic solvents available[1]. Here, colloidal graphite suspended in isopropyl alcohol[2] is used, which is spray-applied to the surfaces. The open area of the wire mesh, i.e., its transparency, reduces as graphite accumulates on the wires. It was found through evaluation of camera and optical microscope images to vary between 72 %. . . 81 %, while for uncoated case the expected open area of 88 % was recovered. This corresponds to a 23 µm. . . 10 µm-thick graphite layer covering the wires. To achieve the lower layer thickness while still covering the complete mesh from all sides, an iterative approach of repeatedly spray-painting and checking the graphite coverage under an optical microscope was necessary. As this process was improved during the course of the 2S-6P measurement, the mesh transparency is estimated to be 70 % during measurement run A, and 80 % during runs B and C (see Table 6.1).

Moreover, during run A, additional blocking meshes were installed in the detector assembly. These meshes are identical to the spherical wire mesh electrodes, and are placed 5 mm above and below the top and bottom electrodes, respectively. They are electrically isolated from both the top and bottom sections, and the top and bottom electrodes, using identical spacers as used between the latter and the ring electrodes. The original purpose of the blocking meshes was to prevent charged particles from entering the top and bottom sections by applying appropriate voltages. Ultimately, this was not done and the blocking meshes were left grounded. However, the additional meshes result in a reduced effective transparency of the combination of mesh electrode and blocking mesh of $0.7^2 \approx 50\,\%$.

As opposed to the inner region, the top and bottom sections of the detector cylinder are not coated with graphite, but the aluminum is left exposed. This is because, as discussed in Section 4.6.5, the photoelectron yield of aluminum is substantially higher than for graphite. At the same time, the suppression of stray electric fields is of no concern inside these sections, as they are expected to be much smaller than the applied electric field. Importantly, the aluminum is not protected from oxidation, but machined, cleaned with industrial soap and organic solvent, and installed in the vacuum chamber, during which it is exposed to ambient air for at least a few days. Thus, the aluminum is expected to be covered by an aluminum oxide layer [141]. Furthermore, the aluminum parts were left as is and not polished after machining, leading to a surface with poor optical properties, even for visible light. The stainless steel wire mesh at the bottom of the cylinder is coated with graphite to keep that number of materials to be considered in the simulation of the detector at two. Note that for the photon energies relevant here, the photoelectron yield of colloidal graphite is higher than for copper, but lower than for stainless steel, which is comparable to aluminum [133, 134].

[1]Commonly used and experimentally studied are Aquadag and DAG 580, available from Agar Scientific, which are suspensions of colloidal graphite in distilled water and ethanol, respectively.

[2]Kontakt Chemie Graphit 33 (manufacturer reference 76009-AC), spray can. It is not a scientific product, and variations in the resulting coating between batches were observed. For future experiments, a more well-controlled scientific product is recommended.

### 4.6.2 Channel electron multipliers

Channel electron multipliers (CEMs) allow the detection of single electrons and photons. To this end, a bias voltage $V_{\text{bias}}$ is applied to a channel of suitable geometry and made from appropriately processed glass, such that electrons entering the channel release secondary electrons upon striking the channel walls, which in turn release more secondary electrons, ultimately resulting in an electronic pulse consisting of many millions of electrons that can be easily detected [142, 143]. The detectors here are operated in the pulse counting mode, in which the amplitude of the output pulses is independent of the exact conditions in the early stages of electron multiplication. This is achieved by using a sufficiently high bias voltage, such that the number of electrons in each pulse, and thus the gain, eventually saturates during the multiplication process through space charge effects. In this mode, the output pulses can then be converted with a fixed-threshold discriminator into logic pulses, which are subsequently counted with a multichannel scaler.

The electron pulses leaving the channel correspond to a current, given by $I_{\text{cts}} = NeG_{\text{CEM}}$, where $N$ is the number of pulses per second, or count rate, $e$ is the elementary charge, and $G_{\text{CEM}}$ the gain of the CEM. A bias current $I_{\text{bias}} = V_{\text{bias}}/R_{\text{chan}}$, where $R_{\text{chan}}$ is the channel resistance, flows in the same direction as $I_{\text{cts}}$ and replenishes the electrons in the channel walls. This gives an upper limit on the count rate, as $I_{\text{cts}}$ cannot exceed $I_{\text{bias}}$. However, the gain starts to decrease around $I_{\text{cts}} \gtrsim 0.1\, I_{\text{bias}}$ [143], which in an application where the count rate exhibits a high dynamic range further limits the maximum count rate. This is because the gain has to be adjusted such that the detector operates in pulse counting mode within the whole dynamic range, which is not possible if the gain itself changes too much with the count rate. This also applies to the experiment discussed here, and the maximum count rate should thus not be much higher than $N \sim 0.1 I_{\text{bias}}/eG_{\text{CEM}}$.

To detect electrons, the input end of the CEM must be at a potential $V_{\text{in}}$ that is higher than its surroundings. The output end of the channel, in turn, must be at a higher potential $V_{\text{out}}$, such that $V_{\text{out}} - V_{\text{in}} = V_{\text{bias}}$. The pulses exiting the channel are collected on an electrically isolated collector, which has to be placed at a yet even higher potential $V_{\text{coll}}$. To extract and count the pulses from the CEM, the collector is capacitively coupled to a signal cable.

The CEMs used here[1] have six spiral channels, twisted around a solid center, with their inputs and outputs, respectively, at the same potential. The input surface of the CEMs conically tapers from a diameter of 10.2 mm to a flat surface of 3.0 mm diameter, which contains the channel openings. A titanium cap with an outer diameter of 11.1 mm and an aperture of 8.5 mm is slipped onto the input end, serving both as a holder and as an electric contact, held at[2] $V_{\text{in}} = 270\,\text{V}$. The titanium cap replaces the stainless steel cap installed by the manufacturer, which protects the rim of input surface from chipping. The cap is clamped at its outer edges between two rings made from PTFE, which electrically isolate it from the detector assembly. To prevent the accumulation of charges on the insulating rings, they are hidden from the inside of the detector cylinder behind grounded conductors.

[1] Photonis MAGNUM Electron Multiplier 5901 EDR (specification number PS36892). Total resistance specified as $R_{\text{CEM}} = 10.8\,\text{M}\Omega \ldots 16.9\,\text{M}\Omega$, maximum operating voltage is 3 kV. See [144] for specifications.

[2] None of the various high-voltage sources tested could maintain an arbitrary voltage at the channel input when a high voltage was applied to the channel output. This problem could be circumvented by adding a $500\,\text{k}\Omega \ll R_{\text{CEM}}$ resistance to ground in parallel at the output of the supplies. This effectively flips the (conventional) current direction from into the voltage supply to out of the voltage supply. FUG MCN 35-350 and FUG MCN 140-1250 were used as voltage supplies for the input ends of the top and bottom detector, respectively.

The CEMs have an integrated bias resistor with resistance $R_{\text{bias}}$ between the channel output and a flap to which a voltage $V_{\text{flap}}$ is applied. The total resistance between the flap and the input is then $R_{\text{CEM}} = R_{\text{chan}} + R_{\text{bias}}$, with $R_{\text{bias}}$ specified as ranging within $10\,\%\ldots20\,\%$ of $R_{\text{CEM}}$. In the following, it is assumed that $R_{\text{bias}} = 0.15R_{\text{CEM}}$. The CEMs are special extended dynamic range (EDR) versions, which were found to have, on average and at room temperature, a total resistance of $R_{\text{CEM}} \approx 15\,\text{M}\Omega$ and thus a channel resistance of $R_{\text{chan}} \approx 13\,\text{M}\Omega$. This is about a factor of two lower resistance compared to the previously used standard version, increasing the typical bias current to $I_{\text{bias}} \approx 120\,\mu\text{A}$. This, in turn, corresponds to a maximum count rate of $N = 1.5 \times 10^6$ counts/s for a typical gain of $G_{\text{CEM}} = 5 \times 10^7$. The use of a low-resistance CEM is especially necessary for the bottom detector, which sits directly above the cryopump and is cooled well below room temperature. As the temperature coefficient of the CEM resistance is negative, this cooling leads to an increase of $R_{\text{CEM}}$ by $50\,\%\ldots70\,\%$ and a corresponding decrease in the maximum count rate. Using the temperature coefficients measured in [145], this increase corresponds to the bottom detector cooling down to ≈220 K. The same two CEMs were used during the 2S-6P measurement, and their resistances when the vacuum chamber was not cooled down was found to be $R_{\text{CEM}} = 17.6\,\text{M}\Omega$ and $R_{\text{CEM}} = 18.0\,\text{M}\Omega$ for the top and bottom detector, respectively. Cooling down the vacuum chamber through the cryopump leads to an average increase to $R_{\text{CEM}} = 18.3\,\text{M}\Omega$ and $R_{\text{CEM}} = 27.2\,\text{M}\Omega$, respectively.

The voltage at the channel output is $V_{\text{out}} = (V_{\text{flap}} - V_{\text{in}})R_{\text{bias}}/R_{\text{CEM}} + V_{\text{in}} = 0.85V_{\text{flap}} + 0.15V_{\text{in}} \approx 0.85V_{\text{flap}}$, and the bias voltage is $V_{\text{bias}} = 0.85(V_{\text{flap}} - V_{\text{in}})$. The purpose of this bias resistor is that the same high-voltage supply can be used to for both the channel output and the collector. Here, separate high-voltage supplies[1] are used for the channel output and the collector in order to be able to adjust the voltage across the gap between the channel output and the collector. To this end, a small electrical circuit, made from standard high-voltage-compatible components and enclosed in vacuum-compatible epoxy, is directly attached to the output pin of the collector. The high-frequency pulses are split off with a 470 pF capacitor and are transmitted through a 50 Ω resistor, defining the output impedance, to the coaxial signal cable. The high voltage $V_{\text{coll}}$ is applied to the collector through a current-limiting 270 kΩ resistor. Finally, the output side of the capacitor is connected to the grounded shield of the high-voltage coaxial cable through a 1 MΩ resistor to avoid high voltage building up through capacitor leakage currents. Here, $V_{\text{coll}}$ is held at $V_{\text{flap}} + 200\,\text{V}$, resulting in a voltage across the channel output–collector gap of approximately $0.15V_{\text{flap}} + 200\,\text{V} \approx 500\,\text{V}$.

Outside the vacuum chamber, preamplifiers[2] with a voltage gain of 10 further increase the pulse height. If necessary, the pulses are attenuated before the preamplifiers. Resistive dc-coupled 1:2 splitters are used to send equal power to the discriminators and to an oscilloscope to monitor the pulses. All electrical components involved have an impedance of 50 Ω to avoid reflections and ringing, except the low-impedance output of the preamplifiers. Discriminators[3] convert the pulses into logic pulses if they cross a given voltage threshold, here typically set to −100 mV, and these logic pulses are counted by the data acquisition as detailed in Section 4.7.3.

The pulses from the CEMs have a full width at half maximum (FWHM) of approximately ≈5 ns, with no ringing or reflections visible. The distribution of the (absolute value of the) amplitude of the pulses, known as the pulse height distribution (PHD), is recorded with the oscilloscope. Fig. 4.35 (A) shows the PHD for various bias voltages $V_{\text{bias}}$ for the CEM used as bottom detector during the 2S-6P measurement. For low values of $V_{\text{bias}}$ ($\lesssim 1.4\,\text{kV}$ in Fig. 4.35), the CEM gain is not saturated and the PHD is approximately an exponential distribution (black dashed lines in Fig. 4.35 (A)). As $V_{\text{bias}}$ is increased, the gain increases, with the distribution starting to resemble a Gaussian distribution (dashed lines in Fig. 4.35 (A)) as the gain starts to saturate ($\approx 1.6\,\text{kV}$ in Fig. 4.35). This is the onset of the pulse counting mode [143]. The resulting distribution is typically characterized by its FWHM divided by its mean, FWHM/$V_0$, ranging between 40 % to 110 % for the CEMs used here, and found to be ≈80 % for the CEM of Fig. 4.35. The count rate versus $V_{\text{bias}}$, shown in Fig. 4.35 (B), is also an important indicator of whether the right combination of $V_{\text{bias}}$ and discriminator threshold has been found, since the count rate should be insensitive to the exact value of both parameters around this point[1]. Further increasing $V_{\text{bias}}$ beyond this point eventually leads to a strong increase in count rate as ion feedback disturbs the gain mechanism [143]. In the experiment, both the PHD and the behavior of the count rate are used to find the value of $V_{\text{bias}}$ where the CEM operates in pulse counting mode, which for data of Fig. 4.35 was determined to be $V_{\text{bias}} = 1.68\,\text{kV}$. This procedure is here, and in Fig. 4.35, typically done using charged particles stemming from an ionization pressure gauge as input particles (see Section 4.2.4). Usually, no substantial difference in the PHD was found when using the fluorescence of the 2S-6P transition instead, for which of course the whole experiment needs to be operated. However, because of the much higher count rate from the fluorescence, which can lower the gain of the CEM as discussed above, $V_{\text{bias}}$ sometimes needed to be further increased by 40 V. . . 80 V. The gain $G_{\text{CEM}}$ of the CEM can also be estimated from the data of Fig. 4.35 (A) by deducing the mean charge in the pulses from $V_0$ and the pulse length. Taking into account the additional voltage gain of 2 from the combination of preamplifier, attenuator, and splitter, this estimation gives $G_{\text{CEM}} = 7.6 \times 10^7$ at $V_{\text{bias}} = 1.68\,\text{kV}$. The dark count rate of the CEMs used in the experiment in this configuration is typically below 10 cts/s.

[1] Heinzinger LNC3000-20 pos high-voltage supplies, max. voltage 3 kV.
[2] FAST ComTec TA1000B-10, dc-coupled, 3 dB bandwidth of 710 MHz, maximum output voltage ±1.3 V.
[3] Phillips Scientific 704 with four channels, max. continuous repetition rate of 300 MHz.

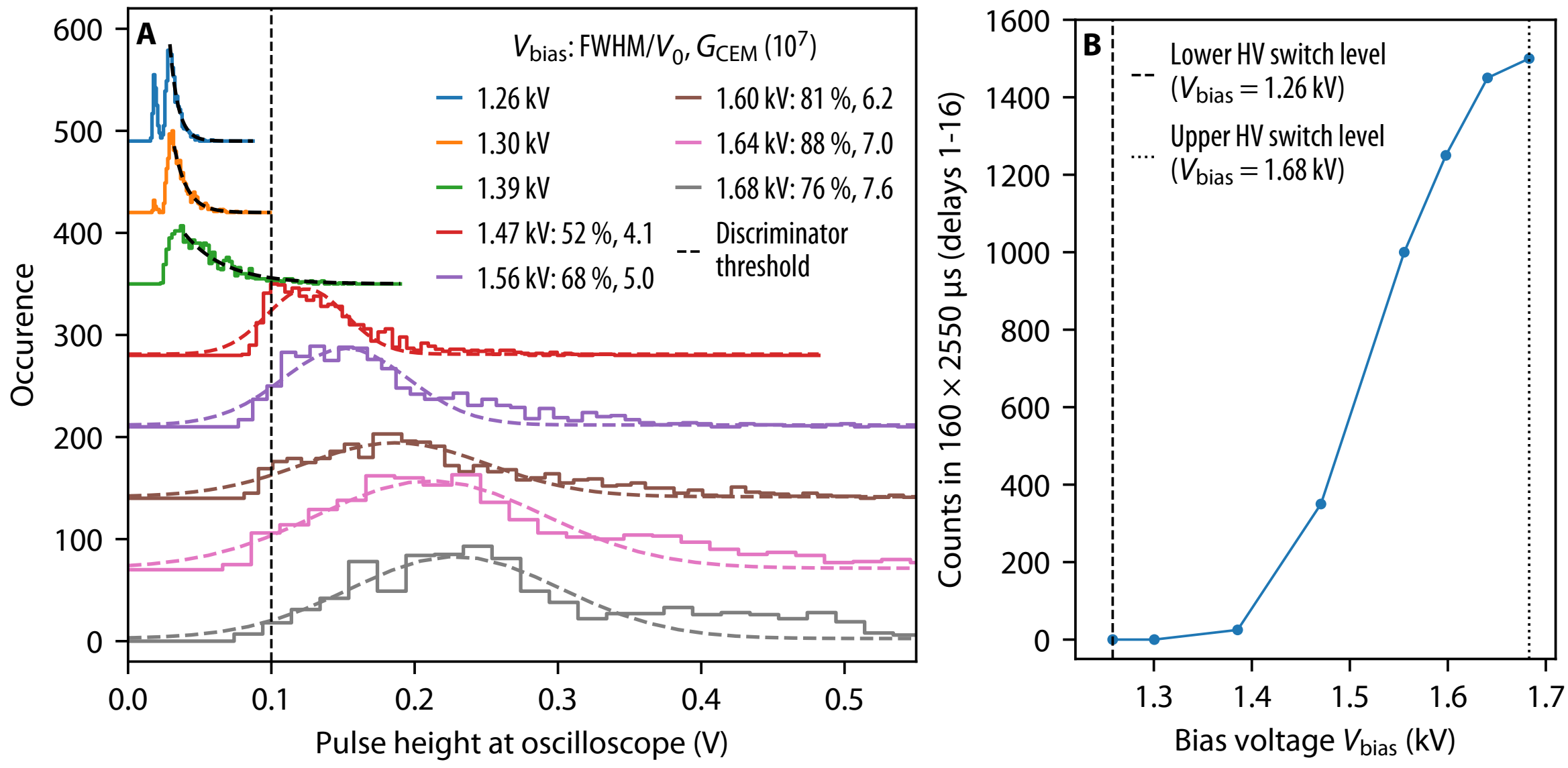

Figure 4.35: (**A**) Pulse height distribution for different bias voltages $V_{\text{bias}}$ of the channel electron multiplier (CEM) used as bottom detector during the 2S-6P measurement (solid lines). For $V_{\text{bias}} \lesssim 1.4\,\text{kV}$, the distribution is approximately exponential (black dashed lines). For increasing $V_{\text{bias}}$, it is better described by a Gaussian distribution (dashed lines), characterized by FWHM/$V_0$, where FWHM is the full width at half maximum, and $V_0$ the mean of the distribution. (**B**) Counts in delays 1-16 versus $V_{\text{bias}}$. Charged particles from an ionization pressure gauge are used as input particles. The pulse height at the output of the CEM is a factor of two lower than as detected at the oscilloscope. See text for details.

[1]Here, increasing $V_{\text{bias}}$ much beyond the onset of pulse counting mode was avoided to avoid damaging the CEMs.

The more charge has been extracted from a CEM, the lower its gain tends to be for a given bias voltage. This aging is thought to stem from degradation of the channel surfaces near the output [143, 146]. Therefore, the bias voltage $V_{\text{bias}}$ for which the CEMs operate in pulse counting mode needs to be checked regularly, which was done as detailed above on each measurement day during the 2S-6P measurement. Due to aging, $V_{\text{bias}}$ needed to be increased during this measurement from 1.51 kV to 1.64 kV for the top detector and from 1.64 kV to 1.9 kV for the bottom detector.

### 4.6.3 Saturation of channel electron multipliers through photoionization of 2S atoms

As discussed in the previous section, the CEMs can only sustain a certain maximum count rate if the pulse height distribution is to remain constant, while at the same time operation at high count rates can limit the lifetime of the CEMs. Unfortunately, the photoionization of metastable 2S atoms (see Section 2.2.6) through the 1S-2S preparation laser leads to a high number of photoelectrons inside the detector during the bright phase of the optical chopper (scattered photons, as discussed below, only play a minor role). These electrons are then readily detected by the CEMs, and can lead to count rates in excess of $1 \times 10^7$ counts/s, limited only by the available bias current and thus well above the count rate sustainable in pulse counting mode. This overwhelming of the CEMs also affects the operation during the dark phase, when the preparation laser is blocked, while at the same time considerably decreasing the lifetime of the CEMs.

Fig. 4.36 shows the result of a test measurement of the CEM performance under these conditions, for both the bright and the dark phase. The data were taken using a hydrogen nozzle of an earlier design (the same one as used during the 2S-4P measurement), which behaves differently from the nozzle used here otherwise and described above, and was operated at a slightly higher temperature of $T_{\text{N}} = 5.8\,\text{K}$, and at a much higher hydrogen flow of $Q_{\text{H}_2} = 1.95\,\text{ml/min}$, than the latter. Additionally, all active surfaces of the detector were graphite-coated, and the variable aperture width was set to $d_2 = 2\,\text{mm}$. The test measurement used the top detector, which had a different, but of the same type, CEM ($R_{\text{CEM}} = 13.5\,\text{M}\Omega$) installed as during the 2S-6P measurement. Fig. 4.36 (A) shows the count rate of the top detector versus delay time $\tau$, averaged over about 8 min, with $\tau < 3125\,\mu\text{s}$ and $\tau \geq 3125\,\mu\text{s}$ corresponding to the dark and bright phases, respectively. Note that not the complete bright phase is captured, which continues to $\tau = 6250\,\mu\text{s}$. The spectroscopy laser is on-resonance with the 2S-6P transition at all times. While the count rate during the bright phase follows the familiar pattern discussed before and stays below 1 Mcts/s, during the bright phase it saturates at ≈18 Mcts/s, close the maximum limit allowed by the bias current. The transient oscillations of the enhancement cavity stabilization (not shown here, see Section 4.3.3.5) are reflected in the count rate at the start of the bright phase. Interestingly, the count rate initially rises over a few µs at the start of the dark phase, right after the preparation laser is blocked, as shown in detail in Fig. 4.36 (B). Such a behavior is not expected from the fluorescence itself, because the 2S-6P spectroscopy laser is always on and the flux of 2S atoms does not change within the time scale of the initial rise. Instead, this behavior is here attributed to the CEM itself as it adjusts to the sudden decrease in count rate, with the time scale given by the effective RC time constant $\tau_{\text{rise}}$ of the CEM and the high-voltage supplies. An exponential fit (red line) results in $\tau_{\text{rise}} = 5.8\,\mu\text{s}$.

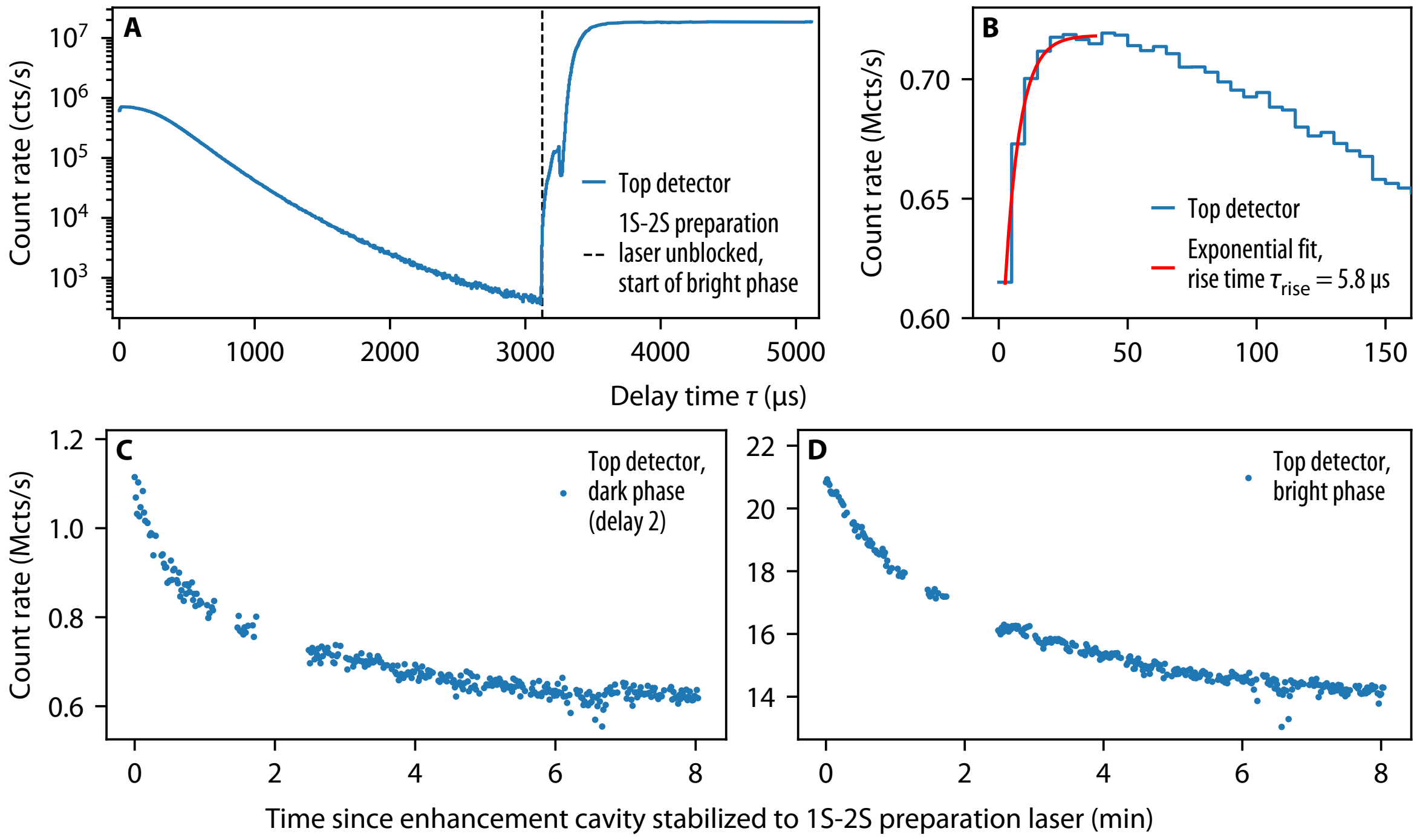

Figure 4.36: Count rate of the top detector (blue lines and circles) during the dark and bright phase, i.e., when the 1S-2S preparation laser is blocked and unblocked, respectively, by the optical chopper. (**A**, **B**) Count rate versus delay time, and count rate during (**C**) dark and (**D**) bright phase versus measurement time. An exponential fit (red line) of the form $a(1 - b\exp((t - t_0)/\tau_{\text{rise}}))$, where $a$, $b$, $t_0$, and $\tau_{\text{rise}}$ are free parameters, to the initial rise of the count rate ($\tau < 40\,\mu\text{s}$) reveals a rise time of $\tau_{\text{rise}} = 5.8\,\mu\text{s}$. The data were taken using a hydrogen nozzle of an earlier design, which behaves differently from the nozzle used here otherwise, and was operated at a temperature of $T_{\text{N}} = 5.8\,\text{K}$ and a hydrogen flow of $Q_{\text{H}_2} = 1.95\,\text{ml/min}$. All active surfaces of the detector were graphite-coated. The powers of the 2S-6P spectroscopy and 1S-2S preparation lasers were $P_{\text{2S-6P}} = 30\,\mu\text{W}$ and $P_{\text{1S-2S}} = 0.95\,\text{W}$, respectively, and the width of the variable aperture was set to $d_2 = 2\,\text{mm}$. See text for details.

A decrease in the count rate over the course of minutes after starting the test measurement is also observed, affecting both the dark and bright phase, as shown in Fig. 4.36 (C) and (D), respectively. The power of both preparation and spectroscopy laser was constant during this time, as was the hydrogen flow and nozzle temperature. Furthermore, the hydrogen discharge was started and the nozzle cooled down 15 min before the enhancement cavity was stabilized to the preparation laser, corresponding to the start of the test measurement. At the same time, the pulse height distribution (PHD) of the CEM changes from a Gaussian distribution, which at that point was stable for two hours of operation at low count rate, to an exponential distribution, while the high voltages applied to the CEM were kept constant. Therefore, the high count rate seems to reduce the gain of the CEM, which in turn does not operate in pulse counting mode anymore, leading to a decrease in the number of pulses counted also during the dark phase. Just as the change in the PHD happens on the course of a few minutes, it here does not return to its previous state immediately after stopping the test measurement. However, after switching off the CEM for a few minutes, the PHD returns to its original state.

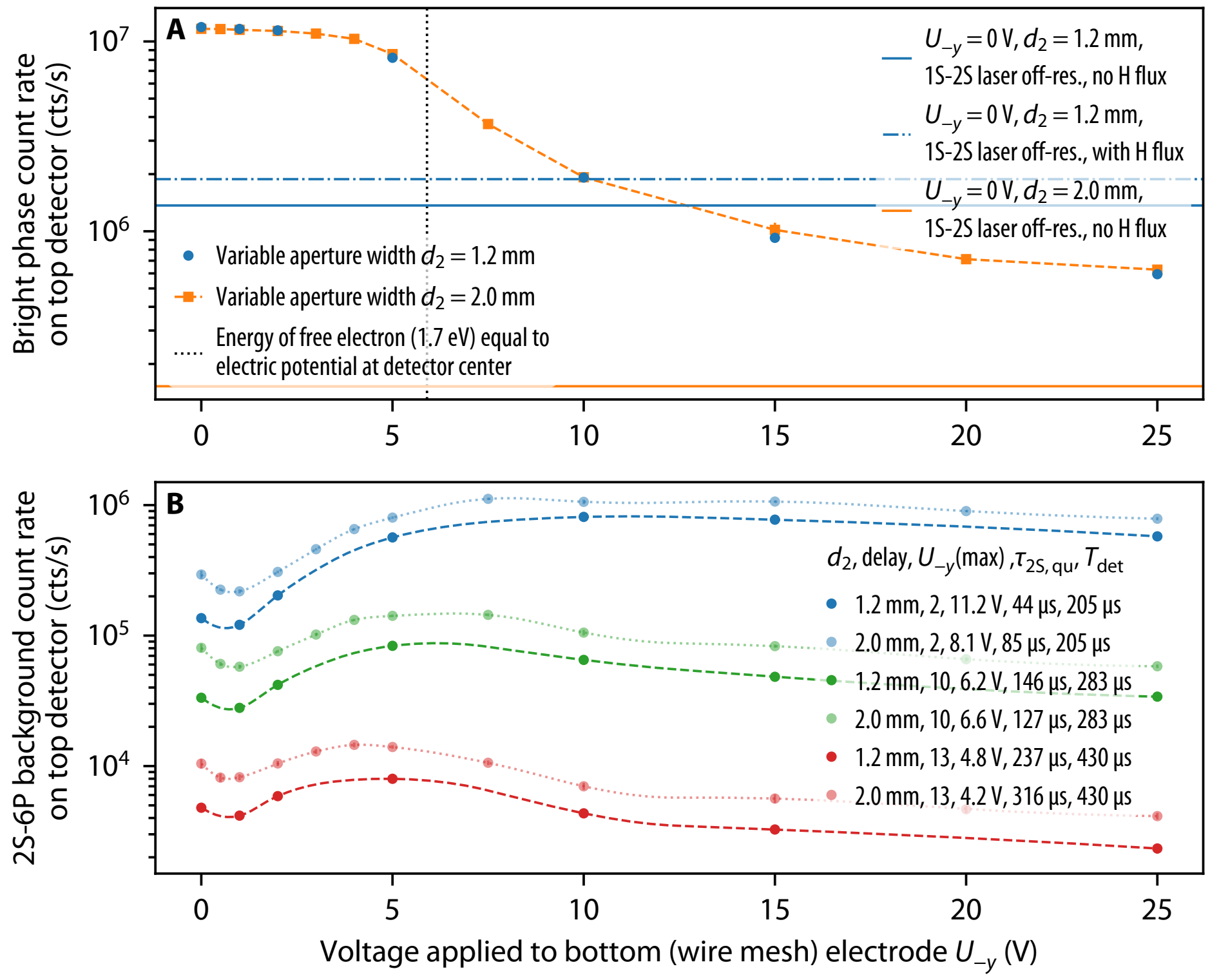

Figure 4.37: Count rate of the top detector for the (**A**) bright and (**B**) dark phase, versus the positive voltage $U_{-y}$ applied to the bottom electrode. The experimental conditions are the same as for Fig. 4.36, but with the 2S-6P preparation laser kept off-resonance. See text for details.

A second test measurement was carried out, two weeks after the first test measurement, to pinpoint the source of the counts during the bright phase. The parameters were the same as for the first measurement, except that the 2S-6P spectroscopy laser is kept off-resonance and the variable aperture width is initially set to $d_2 = 1.2\,\text{mm}$. Fig. 4.37 (A) shows the count rate of the top detector during the bright phase versus the voltage $U_{-y}$ applied to the bottom electrode of the detector assembly's inner region. For $U_{-y} = 0\,\text{V}$, with the no hydrogen flowing into the system and the preparation laser detuned from the 1S-2S transition by $\Delta\nu_{\text{1S-2S}} = 200\,\text{kHz}$, but circulating with $P_{\text{1S-2S}} = 0.95\,\text{W}$ inside the enhancement cavity, a count rate of 1.4 Mcts/s is observed (solid blue line). These counts are attributed to photons scattered out of the enhancement cavity at the apertures along the beam path. Starting the flow atomic hydrogen into the chamber, with the preparation laser still off-resonant, increases the count rate to 1.9 Mcts/s (dash-dotted blue line). This increase could possibly stem from the Doppler-broadened excitation of 2S atoms.

Setting the frequency of the preparation laser on-resonance with the Doppler-free 1S-

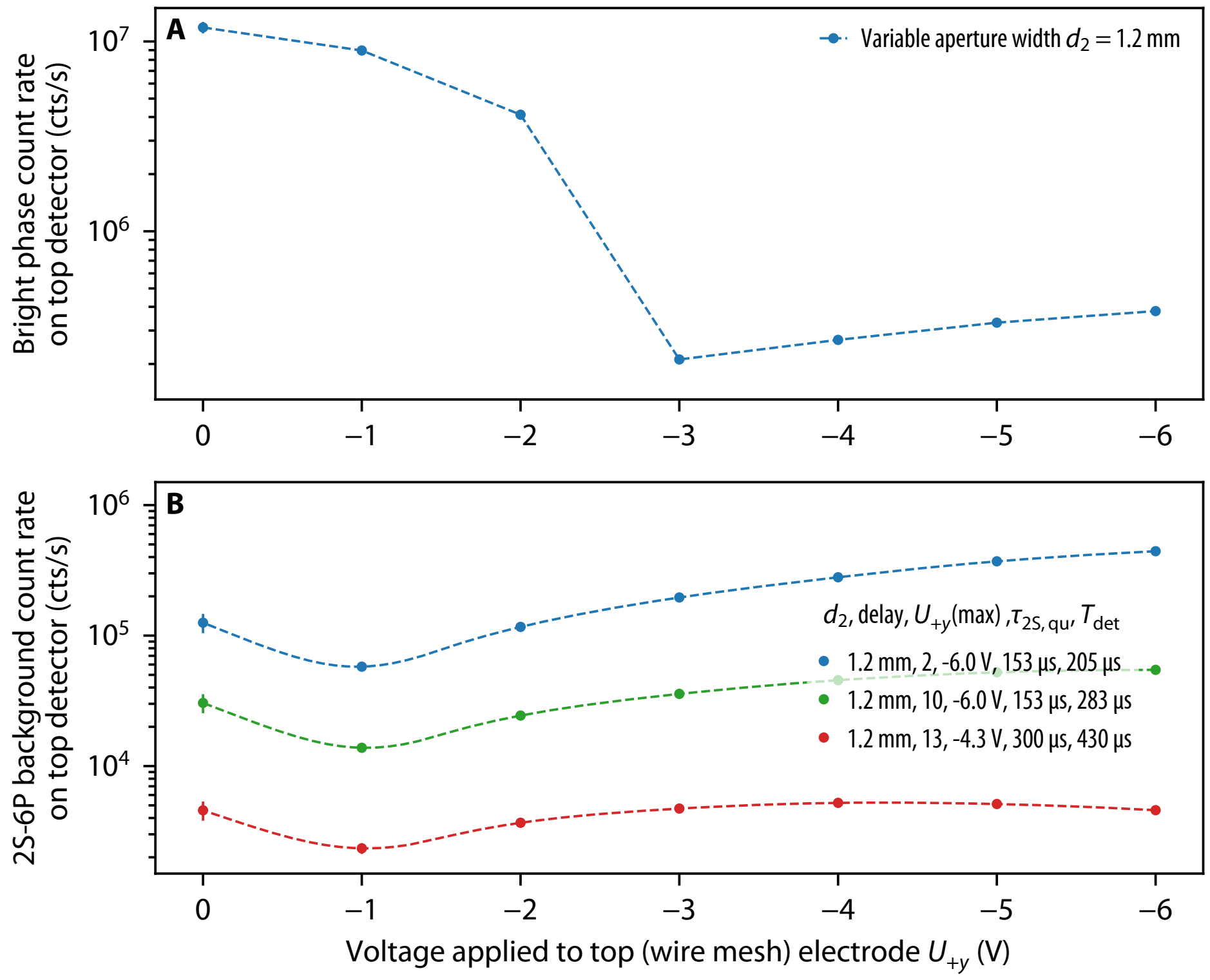

Figure 4.38: Similar to Fig. 4.37, but with a negative voltage $U_{+y}$ applied to the top electrode instead of the bottom electrode. See text for details.

2S transition increases the count rate by almost an order of magnitude to 11.9 Mcts/s (blue circles). If these additional counts stem from electrons created by the photoionization of 2S atoms, it should be possible to reduce the number of counts by pulling the electrons towards the bottom detector with an electric field. To this end, $U_{-y}$ is increased, while all other electrodes are kept grounded. This results in an electric field and potential at the center of the detector of $F_0 = U_{-y} \times (21.7\,1/\text{m})$ and $U_0 = 0.288\,U_{-y}$, respectively, as determined by simulations (see Section 4.6.7). As the photoelectrons have an energy of 1.70 eV (see Section 2.2.6), they should not be able to escape to the top detector for a voltage of $U_{-y} = 1.70/0.288 = 5.9\,\text{V}$ (dotted black line). In reality, this condition is less stringent, as the 2S atoms are photoionized at all positions along the atomic beam, while the electric potential drops off towards to apertures in the electrodes (at the apertures, it is only $U \approx 0.03\,U_{-y}$). The expected behavior is indeed observed as $U_{-y}$ is increased, with the count rate starting to drop around $U_{-y} = 5\,\text{V}$, and reaching 1.9 Mcts/s for $U_{-y} = 10\,\text{V}$. For even higher voltages, the count rate drops below the value seen with the preparation laser off-resonant. This is because the scattered photons are expected to be mainly detected through the emission of photoelectrons from the detector walls, of which those emitted from inside the inner region are likewise pulled towards the bottom detector.

The measurement is repeated with the variable aperture width increased to $d_2 = 2.0$ mm (orange points and dashed line). The increased aperture width does not change the count rate appreciably. This behavior is expected if the majority of the counts stem from the photoionization of 2S atoms, since most atoms that would otherwise be blocked by a narrower aperture do not cross the beam of the preparation laser again. For increasing $U_{-y}$, a behavior very similar for the earlier case is observed. The number of counts without hydrogen, on the other hand, is strongly reduced to 0.15 Mcts/s (orange solid line), as less photons are scattered on the aperture.

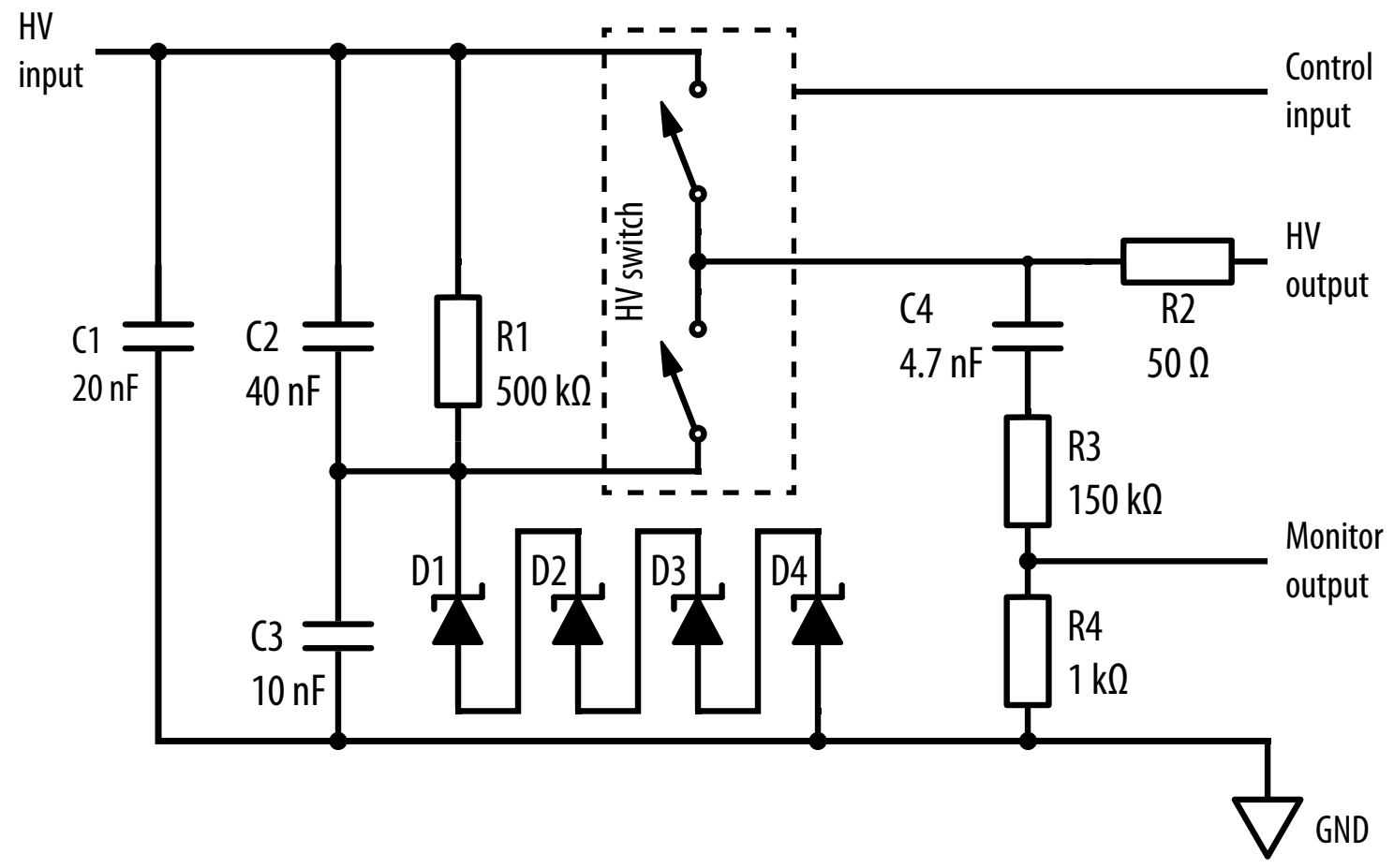

Figure 4.39: Circuit diagram of the high-voltage switches. The circuit shown integrates the Behlke HTS 41-03-GSM switch (dashed box), used for the top detector. The circuit used for the bottom detector integrates the HTS 31-01-GSM switch and is identical, except that the resistance of R1 is 1 MΩ. The number and breakdown voltage of the Zener diodes (here, four diodes D1–4 are shown) is adjusted as needed. See text for details.

A complementary measurement, instead of applying positive voltage to the lower electrode, is to apply a negative voltage $U_{+y}$ to the upper electrode instead. For large enough values of $U_{+y}$, this also prevents the photoelectrons from reaching the top detector. The results of the measurement, now again for $d_2 = 1.2\,\mathrm{mm}$, are shown in Fig. 4.38. While here also a suppression of the count rate is observed, a voltage about a factor two lower is needed. This could be because the electric field inside the top section, which guides the electrons to the CEM, is now also disturbed. Indeed, the field lines directly above the top electrode point towards the walls of the detector cylinder, as the electric field from the CEM is weak in this region.

### 4.6.4 Gain switching of channel electron multipliers

As detailed in the preceding section, the large count rate on the CEMs during the bright phase, much above the sustainable limit, alters the pulse height distribution (PHD), leads to a transient nonlinear detector response at the onset of the dark phase, and reduces the lifetime of the CEMs. While in principle the decrease in gain can be countered by an increase of the bias voltage, it is not clear how the CEMs behave during the strongly different count rates of the dark and bright phase[1]. For example, it could be possible that this reduces the dynamic range in the dark phase, distorting the line shape of the 2S-6P resonance.

A way to circumvent this problem is to substantially reduce the gain during the bright phase, such that the CEMs are effectively switched off and almost no current from ampli-fied pulses is flowing[1]. This is here achieved by reducing the bias voltage $V_{\text{bias}}$ using high-voltage (HV) switches. To not disturb the fluorescence count rate at the start of the dark phase, this gain switching needs to happen within a few µs. To this end, push-pull transistor switches[2] are used to vary the voltage $V_{\text{flap}}$ applied to the flap of the CEMs, resulting in a lower gain voltage through $V_{\text{bias}} = 0.85(V_{\text{flap}} - V_{\text{in}})$. During the dark phase, the switches set $V_{\text{flap}}$ to $V_{\text{flap,high}}$, for which the CEMs are in pulse counting mode, while during the bright phase, $V_{\text{flap}}$ is set to $V_{\text{flap,low}} < V_{\text{flap,high}}$, such that close to no output pulses are detected (see Fig. 4.35 (B)). $V_{\text{flap,high}}$ is derived directly from the HV supplies, while $V_{\text{flap,low}}$ is derived from $V_{\text{flap,high}}$ using Zener diodes to ground in a circuit resembling a voltage regulator[3]. The corresponding circuit diagram is shown in Fig. 4.39, with $V_{\text{flap,high}}$ applied at the HV input terminal and the CEM flap connected to the HV output terminal. By choosing and adding in series Zener diodes with different breakdown voltages, $V_{\text{flap,low}}$ can be picked almost arbitrarily and thus matched to the gain characteristics of the CEM. Note that setting $V_{\text{flap,low}}$ to ground would lead to a large potential difference across the channel output–collector gap, and reverse the field inside the channel as the input is still held at a positive voltage, both of which should be avoided. During the 2S-6P measurement, $V_{\text{flap,low}}$ was set to 1.76 kV and 1.85 kV for the top and bottom detector, respectively.

[1]In principle, this could be investigated by observing the PHD of dark and bright phase separately, using an RF switch.

A TTL signal, derived from a function generator that is triggered on the chopper trigger, is applied to the control input of the HV switches, where the high signal level corresponds to the HV output of the switches set to the high level. The rising edge of the TTL signal is delayed by 2 µs from the falling edge of the chopper trigger, heralding the beginning of the dark phase, and the signal is held at high level for 2.6 ms. Thus, the complete delay time $\tau = 10\,\mu\text{s} \ldots 2560\,\mu\text{s}$ during which the 2S-6P fluorescence is counted is covered. The rise and fall time of the HV output, as determined from the ac-coupled monitor output (see Fig. 4.39), was found to be ≈200 ns, with a delay of likewise ≈200 ns between the TTL signal and the HV output, well within the requirement stated above. However, the internal power driver of the HV switches produces pulses every few hundred µs on the HV output. These pulses mimic the output pulses of the CEMs and are counted by the data acquisition. A low-pass filter with an RC time constant of 680 ns, placed between the HV output and the CEM, is sufficient to reduce the amplitude of these pulses to a negligible level, while at the same time keeping the rise and fall time within the acceptable range. There are also pulses on the HV output during switching, which are of larger amplitude and not completely suppressed. However, as the switching occurs before and after the fluorescence is counted, the resulting spurious counts are excluded. As these spurious counts also disturb the measurement of the PHD, RF switches installed before the oscilloscope used for this measurement (see Section 4.6.2) block any signal from the CEMs during the gain switching.

The gain switching of the CEMs was first tested using charged particles from an ionization pressure gauge, as used to measure the PHD in Section 4.6.2, since they result in a constant, time-independent count rate on the order of 10 kcts/s. The test used the same top detector as used in the test measurement discussed in Section 4.6.3. When switching $V_{\text{flap}}$ from $V_{\text{flap,low}}$ to $V_{\text{flap,high}}$, the count rate stabilizes with a time constant of $\tau_{\text{rise}} = 2\,\mu s \ldots 3\,\mu s$, thought to be limited by the capacitance of the CEM itself, similar to the behavior observed due to saturation without the HV switches (see Section 4.6.3 and Fig. 4.36 (B)). There is also an increase with a much longer time constant of ≈300 μs after the initial rise, during which the count rate increases by approximately 1.7 %, also observed on the HV output voltage. The behavior is identical for both HV switches. The count rate when the HV output is at its low level, i.e., $V_{\text{flap,low}}$, is zero.

[1]During the 2S-4P measurement, an RF switch was used to block the counts from the CEMs from reaching the discriminator during the bright phase. This however does not prevent the adverse effects on the CEMs themselves.

[2]Behlke HTS 31-01-GSM and HTS 41-03-GSM, max. operating voltage 3 kV and 4 kV, respectively.

[3]Initially, both $V_{\text{flap,high}}$ and $V_{\text{flap,low}}$ were derived from independent HV supplies. However, this was found to limit the switching speed as the charge stored in the CEM has to flow across the supplies upon switching. Using Zener diodes, on the other hand, allows the charges to flow to ground across a much lower resistance.

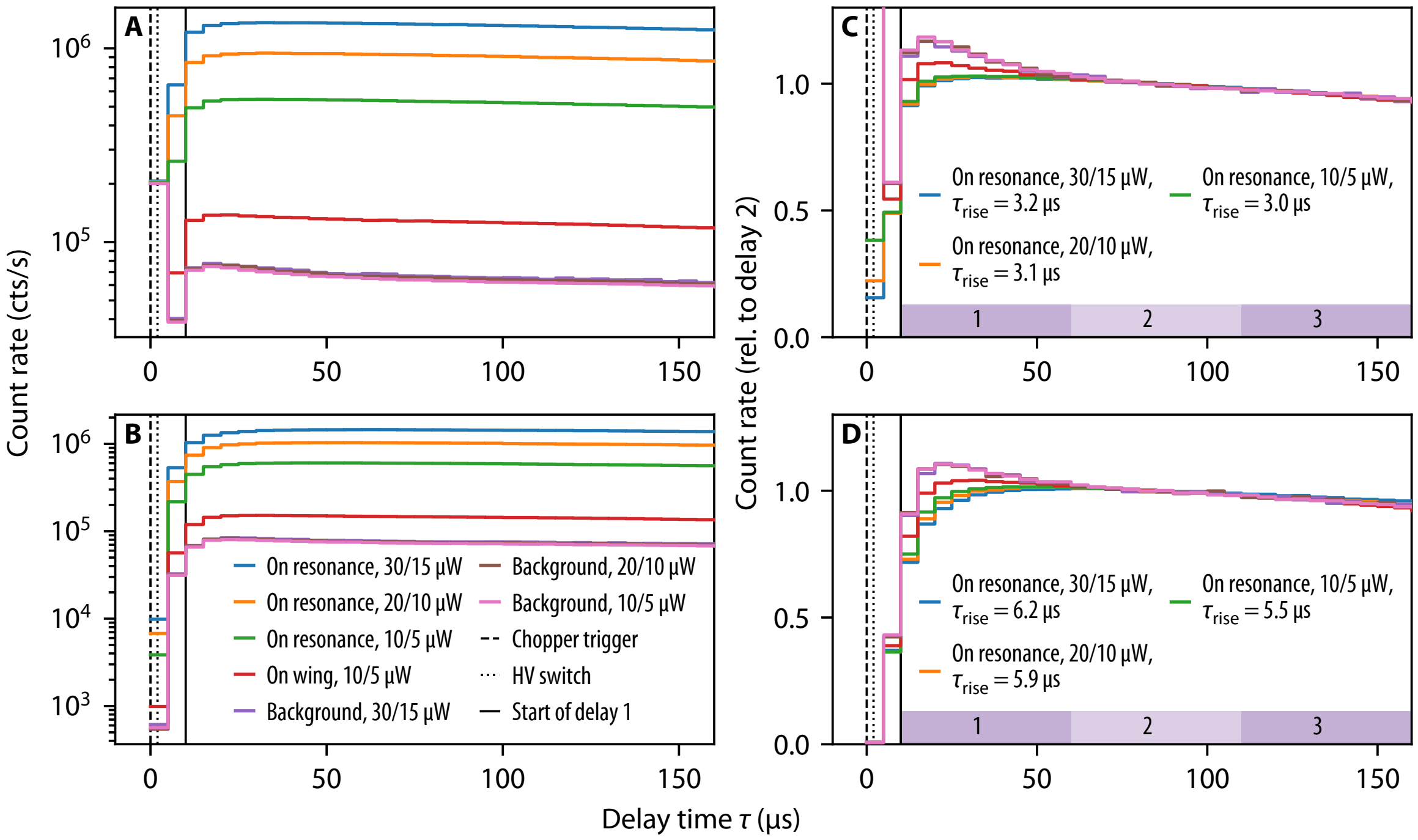

Figure 4.40: Time-resolved fluorescence count rate of the (**A**, **C**) top and (**B**, **D**) bottom detector during run B of the 2S-6P measurement (see Table 6.1). Only the data for $\tau < 160\,\mu s$ are shown (see Fig. 4.32 for the full data). (C, D) show the same data as (A, B), but normalized to the average count rate in delay 2. The legend of (B) is applicable to all plots, with on resonance, on wing, and background corresponding to a detuning of the 2S-6P spectroscopy laser from the resonance of $|\Delta\nu_{2S\text{-}6P}| = 0\,$MHz, 8 MHz. . . 10 MHz, and 50 MHz, respectively. The data are further grouped by the spectroscopy laser power $P_{2S\text{-}6P} = 30/15\,\mu W$, 20/10 μW, and 10/5 μW for the 2S-6P$_{1/2}$/2S-6P$_{3/2}$ transition (data groups G1B/G7B, G2/G8, G3/G9, see Table 6.2). The delay times of the falling chopper trigger ($\tau = 0\,\mu s$, dashed black line), the rising HV switch ($\tau = 2\,\mu s$, dotted black line), and the start of delay 1 ($\tau = 10\,\mu s$, solid black line) are marked. The legends in (C) and (D) show the rise time $\tau_{\text{rise}}$ of the on-resonance fluorescence count rate, as determined from an exponential fit within $5\,\mu s \leq \tau < 40\,\mu s$, similar as done in Fig. 4.36.

The HV switches were used during the complete 2S-6P measurement. Fig. 4.40 shows the fluorescence count rate for delay times $\tau < 160\,\mu s$ during run B (see Table 6.1) of the measurement. The data are grouped into on resonance, on wing, and background, corresponding to a detuning of the 2S-6P spectroscopy laser from the resonance of $|\Delta\nu_{2S\text{-}6P}| = 0\,$MHz, 8 MHz. . . 10 MHz, and 50 MHz, respectively, and into spectroscopy laser powers $P_{2S\text{-}6P} = 30/15\,\mu W$, 20/10 μW, and 10/5 μW for the 2S-6P$_{1/2}$/2S-6P$_{3/2}$ transition. The maximum on-resonance count rate reaches ≈1.4 Mcts/s for $P_{2S\text{-}6P} = 30/15\,\mu W$. Spurious counts as the HV output switches to its high level (dotted black line) are visible, with the top detector affected more strongly. The initial rise of the fluorescence count rate caused by the switching of the gain is clearly visible in the normalized count rate shown in Fig. 4.40 (C) and (D). Exponential fits (not shown) to the on-resonance count rate within $5\,\mu s \leq \tau < 40\,\mu s$ reveal rise times of 3.0 µs. . . 3.2 µs and 5.5 µs. . . 6.2 µs for the top and bottom detector, respectively. The difference between the detectors is attributed to their different resistances $R_{\mathrm{CEM}}$, owing to the cold temperature of the bottom detector (see Section 4.6.2). A slight, but statistically significant, decrease in the rise time with decreasing laser power is present, hinting at a slight dependence of the rise time on the count rate, which depends linearly on the laser power.

The background count rate, as expected, is nearly identical for all laser powers, i.e., the background-to-amplitude ratio increases with decreasing laser power. However, it shows a clear excess of counts compared to the on-resonance count rate within the first 50 µs, superimposed on the rise from the gain switching. As the on-resonance count rate is at least a factor of 7 larger than the background count rate, the excess is not expected to be visible in the latter. It is thus unclear whether this excess is also present with the laser on resonance with the 2S-6P transition, as the background signal, coming from 2S atoms, cannot be readily disentangled from the much larger fluorescence from the decay of the 6P level. However, placing the laser not on resonance, but on the wing of the resonance such that the 6P fluorescence contributes the same count rate as the background (red line), the excess should be half as strong if it indeed only is present on the background signal. This is indeed clearly the case. As discussed in Section 4.5.2.4, this excess could be caused by atoms that have not or only partially thermalized to the temperature of the nozzle, but further experiments, e.g., using different nozzle designs, are needed to study this hypothesis.

During the bright phase, less than 0.03 cts/s are detected on average for the top detector. The reduction of the gain of the bottom detector, on the other, is not quite as strong, and on average 51 cts/s are still detected in the bright phase. Using the ratio of the residual bright phase count rate when the laser is on- and off-resonance, and the corresponding count rates at the start of the dark phase, the count rate the detectors would be exposed to during the bright phase without gain switching can be estimated to be close to 10 Mcts/s. This count rate would by far exceed the sustainable count rate in pulse counting mode and lead to a modified PHD as seen in Section 4.6.3. Furthermore, this bright phase count rate is 30 times higher than the average count rate during the dark phase, and thus the gain switching is expected to increase the lifetime of the CEMs, as given by the accumulated charge, by a factor of more than 30.

### 4.6.5 Detector properties pertaining to detection efficiency

The detection efficiency of CEMs was found to be above 90 % for electrons with kinetic energies within 200 eV. . . 700 eV in [143, 147]. Here, the input surface of the CEMs is held at $V_{\mathrm{in}} = 270\,\mathrm{V}$ above the rest of the grounded detector surfaces, and thus the photoelectrons emitted or reflected from these surfaces will have an energy close to 270 eV, as their initial kinetic energy is negligible in comparison with the energy gained from the electric field (see below). Therefore, the detection efficiency of photoelectrons is here assumed to be ≈90 %. The quantum efficiency of photon detection was measured in [148] to be ≈10 % for Ly-$\epsilon$ photons ($h\nu = 13.22\,\mathrm{eV}$) and ≈2 % for Ly-$\alpha$ photons ($h\nu = 10.20\,\mathrm{eV}$). The photons from the 1S-2S preparation laser ($h\nu = 5.10\,\mathrm{eV}$), on the other hand, are only detected with an quantum efficiency of $\sim 1 \times 10^{-8}$ according to [149]. The small titanium cap covering the rim of the CEM input cone is held at the same potential as the input surface itself. The secondary electron yield of titanium is ~2 [150] for 270 eV electrons and thus here it is assumed that any electron hitting the small cap is equivalent to hitting the input surface of the CEM, i.e., $P_{\mathrm{det,Ti}} = 100\,\%$.

Table 4.1: Properties of the materials used in the detector relevant to modeling the detection efficiency. See Section 4.6.5 for details and Fig. B.1 for the measurements from which the properties were extracted. Only the photon energies relevant to the modeling of the 2S-6P and 2S-4P measurements are included.

| Material | Photon | $h\nu$ (eV) | $Y_{\mathrm{el}}$ (%) | $P_{\mathrm{ph}}$ (%) | $n_{\mathrm{r}}$ | $E_{\mathrm{el},0}$ (eV) | $\Gamma_{\mathrm{el}}$ (eV) | $P_{\mathrm{el}}$ (%) |
|---|---|---|---|---|---|---|---|---|
| Colloidal graphite (gr) | 1S-2S | 5.10 | 0.0006 | — | — | — | — | — |
| | Ly-$\alpha$ | 10.20 | 0.7 | ≈7 | 1.7 | — | — | — |
| | Ly-$\gamma$ | 12.75 | 4.1 | ≈15 | 2.2 | 2.4 | 2.5 | ≈60 |
| | Ly-$\epsilon$ | 13.22 | 4.8 | ≈15 | 2.2 | 2.5 | 2.8 | ≈60 |
| Oxidized aluminum (Al) | 1S-2S | 5.10 | 0.0009 | — | — | — | — | — |
| | Ly-$\alpha$ | 10.20 | 3.9 | ≈40 | 4.4 | — | — | — |
| | Ly-$\gamma$ | 12.75 | 18 | ≈15 | 2.2 | 1.9 | 2.4 | ≈10 |
| | Ly-$\epsilon$ | 13.22 | 20 | ≈15 | 2.2 | 1.9 | 2.4 | ≈10 |

The active surfaces of the detector are either coated with colloidal graphite or consist of oxidized aluminum. These materials have work functions of 4.6 eV and 4.0 eV, respectively [133], and thus only photons stemming from Lyman decays, for which $h\nu \geq 10.20$ eV, are energetic enough to eject photoelectrons. As can be seen from Tables 2.2 and 2.3, for the 2S-6P transition 91 % of these decays are Ly-$\epsilon$ ($h\nu = 13.22$ eV) and 7 % are Ly-$\alpha$ ($h\nu = 10.20$ eV), while Ly-$\beta$ ($h\nu = 12.09$ eV) and Ly-$\gamma$ ($h\nu = 12.75$ eV) only contribute 1.3 % and 0.3 % of Lyman photons, respectively. The properties of the two materials relevant to modeling the spatial detection efficiency and spectral sensitivity of the fluorescence detection are given in Table 4.1 and discussed in the following.

The photoelectron yield $Y_{\mathrm{el}}$, photon reflectance $P_{\mathrm{ph}}$, and photoelectron energy distribution were measured for both materials[1] in [133] at near-normal incidence at an angle of $\alpha = 7.5°$. The measurements are shown in Fig. B.1 (A–C) for reference. Importantly, the aluminum sample used in [133] was left exposed to atmosphere for at least several days, and thus should be covered with an oxide layer just as the aluminum parts used here. Note that $Y_{\mathrm{el}}$ is here defined as the number of photoelectrons ejected per incoming photon, not per absorbed photon, with the latter given by $Y_{\mathrm{el}}/(1-P_{\mathrm{ph}})$. Using the Fresnel equations, the refractive index $n_{\mathrm{r}}$ of the material can be calculated from $P_{\mathrm{ph}}$, since at near-normal incidence the reflectance of s- and p-polarized light are sufficiently similar. The photoelectron energy distribution was not measured for graphite in its colloidal form, but for an atomically clean sample; it is here assumed that the distributions are identical. The distributions can be described by a normal distribution centered at energy $E_{\mathrm{el},0}$ and with a full width at half maximum (FWHM) of $\Gamma_{\mathrm{el}}$ for both materials, as given in Table 4.1.

The electron reflectance $P_{\mathrm{el}}$ was measured for colloidal graphite[2] in [140], and for non-oxidized aluminum in [151] as reproduced in [152]. Both measurements are shown for reference in Fig. B.1 (D). For the values of $P_{\mathrm{el}}$ given in Table 4.1, the incoming electron energy was assumed to correspond to $E_{\mathrm{el},0}$. The emission of secondary electrons is negligible here, as only the most energetic photoelectrons have an energy slightly above the work function of the materials considered. The scattering of both electrons and photons is assumed to be diffuse, since the surface irregularity is thought to be much larger than the wavelength of the particles (an electron with an energy of 1 eV has a de Broglie wavelength of 1.2 nm).

[1] In [133], the properties of colloidal graphite are measured for a coating made with Aquadag, which is a suspension of colloidal graphite in distilled water. Here, colloidal graphite suspended in isopropyl alcohol is used. It is assumed that the properties of the dried coatings are comparable.

[2] In [140], DAG 580, which is suspension of colloidal graphite in ethanol, was used to apply the graphite coating, which is assumed to have similar properties as the coatings prepared here.

Since the photoelectron yield for Ly-$\alpha$ photons is only 15 %. . . 20 % of the yield of Ly-$\epsilon$ photons, Ly-$\alpha$ photons only account for ≈1 % of the detected signal, similar to the contribution from Ly-$\beta$ and Ly-$\gamma$ decays. Thus, only Ly-$\epsilon$ photons need to be taken into account to model the detector with sufficient accuracy. Furthermore, Ly-$\epsilon$ photons have a 4.2 times higher photoelectron yield higher on oxidized aluminum than on colloidal graphite. The corresponding increase in signal has indeed been observed upon exchanging the detector material from graphite-coated copper to aluminum.

### 4.6.6 Simulation of spatial detection efficiency

To estimate the spatial detection efficiency, the detection of Ly-$\epsilon$ photons is modeled using a Monte Carlo particle tracing simulation[1]. To this end, the geometry of the detector cylinder is approximated, with the surfaces either corresponding to colloidal graphite or aluminum. The apertures in the electrodes are modeled as graphite-coated tubes, closed at the far end, with a length of ≈23 mm and the appropriate cross section. The wire meshes of the electrodes and the bottom of the detector cylinder are modeled as graphite-coated surfaces with a transparency of $T_{\mathrm{WM}}$ and $T_{\mathrm{WM,BD}}$, respectively, for both photons and electrons. For the electrodes, transmitted particles enter the top or bottom sections of the detector cylinder, while particles crossing the bottom of the detector cylinder disappear from the simulation. Each surface type is assigned a photoelectron yield $Y_{\mathrm{el}}$, a photon reflectance $P_{\mathrm{ph}}$, an electron reflectance $P_{\mathrm{el}}$, and a Gaussian photoelectron energy distribution centered at $E_{\mathrm{el},0}$ and with a FWHM of $\Gamma_{\mathrm{el}}$, with the corresponding values taken from Table 4.1.

All surfaces are grounded, except the CEM input surface, which is held at 270 V. The resulting electric field inside the top and bottom sections is found by solving Laplace's equation using the finite difference method. The step size of this calculation is set to 0.2 mm and $2\times10^5$ iterations are used. The center section, or spectroscopy region, is assumed to be free from electric fields.

The fluorescence photons are emitted from the interaction point. Spherical coordinates are used with the polar angle $\theta = -90° \ldots 90°$ measured against the vertical $y$-axis, i.e., the axis of the detector cylinder, and the azimuthal angle $\phi = -180° \ldots 180°$ measured against the atomic beam ($z$-) axis. For each run of the simulation, $M$ emission directions are picked at random, where here $M = 2 \times 10^5$ is used. For each photon emitted, or trajectory, in turn one of the emission directions is randomly selected, and the photons flies through the vacuum until a surface is encountered. Here, as determined by the properties of the surface, the photon is either reflected, absorbed and a photoelectron emitted, or absorbed (or enters the cryopump) and the trajectory is stopped. The polarization of the photon and the angle of incidence are not taken into account, i.e., $Y_{\mathrm{el}}$, $P_{\mathrm{ph}}$, and $P_{\mathrm{el}}$ are assumed to be independent of both[1]. This assumption is partly motivated by the large surface irregularity, which should reduce possible dependencies on the angle of incidence.

[1]The simulation is written in the Object Pascal programming language, using Embarcadero Delphi 10.3. The original version was created by Arthur Matveev, and was subsequently adapted and modified by the author.

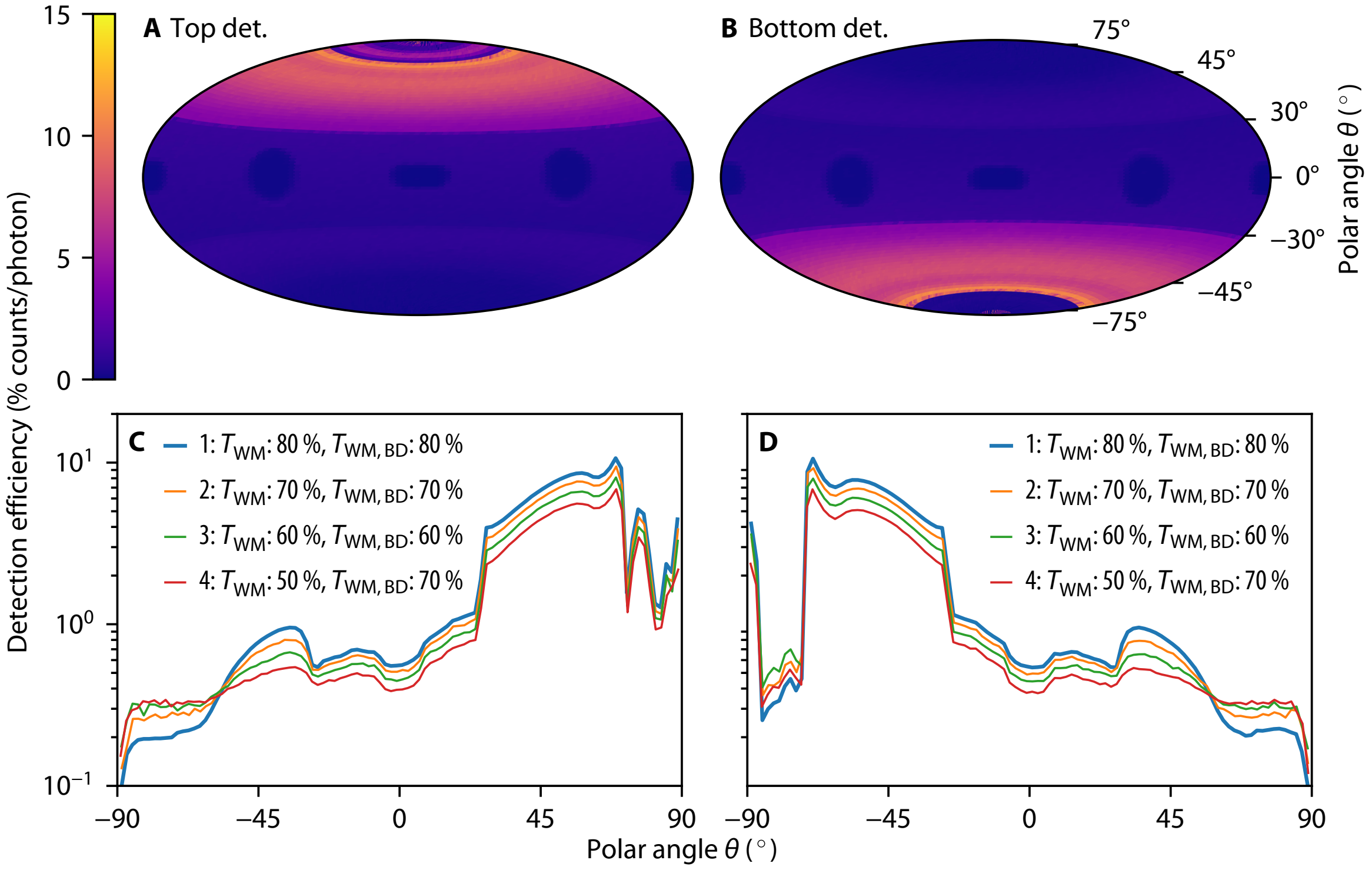

Figure 4.41: Spatial detection efficiencies of the (**A**) top and (**B**) bottom detector versus fluorescence emission direction, using the material properties given in Table 4.1 and wire mesh transparencies of $T_{\mathrm{WM}} = T_{\mathrm{WM,BD}} = 80\,\%$. Spherical coordinates are used with the polar angle $\theta = -90° \ldots 90°$ measured against the vertical $y$-axis, i.e., the axis of the detector cylinder, and the azimuthal angle $\phi = -180° \ldots 180°$ measured against the atomic beam ($z$-) axis. The projection of the resulting spherical surface to 2D uses the equal-area Hammer projection. (**C**, **D**) Detection efficiency versus polar angle $\theta$, averaged over the azimuthal angle $\phi$, for different simulation parameters for the (C) top and (D) bottom detector. The data from (A, B) correspond to the bold blue lines in (C, D).

The angular distribution of both the diffuse reflection of photons and electrons as well as the emission of photoelectrons, is assumed to be given by (Lambert's) cosine law (see Eq. (4.28)), according to which the emission angles are randomly found. If an photoelectron is emitted, its energy is randomly drawn from the Gaussian distribution for the emitting surface. The trajectory of electrons is affected by the local electric field through the Lorentz force, and upon striking a surface it is likewise reflected or absorbed. This procedure continues till the photon or electron is either absorbed or hits the input surface of one the CEMs. In the latter case, it is counted as a click on the corresponding detector with a probability of 90 % and 10 % for electrons and photons, respectively. In total, $N_{\mathrm{traj,ph}}$ trajectories are calculated in this way, where $N_{\mathrm{traj,ph}}$ here is typically on the order of $1 \times 10^8$.

[1]While implementing a simulation taking these effects into account is beyond the scope of this work, not least because experimental data on the polarization dependence of photoemission are scarce in the parameter region needed here, it should be estimated in future work if polarization-dependent effects like quantum interference need to be corrected for using only simulation data. For the 2S-6P measurement, the effects of quantum interference are reduced to a negligible level using the fine-structure centroid without having to rely on the accuracy of the detection efficiency simulation to a large degree.

Table 4.2: Overview of parameters and results of detection efficiency simulations as shown in Fig. 4.41. See text for details.

| Set | $T_{\mathrm{WM}}$ (%) | $T_{\mathrm{WM,BD}}$ (%) | $P_{\mathrm{ph}}$ (%) gr | Al | $P_{\mathrm{el}}$ (%) gr | Al | $P_{\mathrm{det,Ti}}$ (%) | $P_{\mathrm{DE,TD}}$ (%) | $P_{\mathrm{DE,BD}}$ (%) | $\eta_{\mathrm{DE,TD}}$ (%) | $\eta_{\mathrm{DE,BD}}$ (%) |
|---|---|---|---|---|---|---|---|---|---|---|---|
| 1 | 80 | 80 | 15 | 15 | 60 | 10 | 100 | 2.22 | 2.09 | 30.5 | 26.1 |
| 2 | 70 | 70 | 15 | 15 | 60 | 10 | 100 | 1.99 | 1.86 | 30.2 | 25.8 |
| 3 | 60 | 60 | 15 | 15 | 60 | 10 | 100 | 1.75 | 1.63 | 30.1 | 26.1 |
| 4 | 50 | 70 | 15 | 15 | 60 | 10 | 100 | 1.47 | 1.38 | 30.1 | 26.0 |

The resulting spatial detection efficiencies of the two detectors as a function of fluorescence emission direction is shown in Fig. 4.41 (A, B), using the material properties given in Table 4.1 and wire mesh transparencies of $T_{\mathrm{WM}} = T_{\mathrm{WM,BD}} = 80\,\%$. Spherical coordinates are used with the polar angle $\theta = -90° \ldots 90°$ measured against the vertical $y$-axis, i.e., the axis of the detector cylinder, and the azimuthal angle $\phi = -180° \ldots 180°$ measured against the atomic beam ($z$-) axis. The same data, averaged over $\phi$, are shown as bold blue lines in Fig. 4.41 (C) and (D) for the top and bottom detector, respectively. The detection efficiency is highest within the respective hemisphere, reaching ≈13 % for fluorescence photons hitting the detector cylinder walls close to the top or bottom ($\theta \approx \pm 70°$). It then decreases towards the center of the cylinder, with the steep drop at $\theta \approx \pm 28°$ corresponding to the boundary between the top/bottom sections and the inner region. Here, the material changes from aluminum to colloidal graphite. The boundary back to aluminum in the opposing hemisphere again is visible at $\theta \approx \mp 28°$. For the top detector, photons hitting the aluminum top of the cylinder lead to a peak around $\theta \approx 77°$, but the detection efficiency is lower than for photons hitting the wall as the photoelectrons need to be deflected by 180° to hit the CEM. On the other hand, for the bottom detector most photons for $\theta \gtrsim 70°$ escape into the cryopump, with only few photoelectrons ejected from the graphite-coated wire mesh. Photons directly hitting the CEMs lead to peaks at $\theta \approx \pm 90°$. The apertures in the electrodes are faintly visible as regions of reduced detection efficiency near the equator. The total detection efficiency is found to be $P_{\mathrm{DE,TD}} = 2.2\,\%$ and $P_{\mathrm{DE,BD}} = 2.1\,\%$ for the top and bottom detector, respectively. Note that $P_{\mathrm{DE,TD}} + P_{\mathrm{DE,BD}} \leq 100\,\%$, as each fluorescence photon can only be detected by either CEM, meaning that here 4.3 % of all photons emitted by the atoms are collected. It is also instructive to estimate the solid angle covered by the detection. Since the detection efficiency is not a step-like function, one way to do this is to pick a detection efficiency threshold above which the solid angle is thought to contribute. Here, a threshold of 20 % of the peak detection efficiency is picked, resulting in the top and bottom detector covering a fraction of $\eta_{\mathrm{DE,TD}} = 31\,\%$ and $\eta_{\mathrm{DE,BD}} = 26\,\%$ of the total solid angle, respectively.

Fig. 4.41 (C, D) additionally shows the detection efficiencies for wire mesh transparencies $T_{\mathrm{WM}} = T_{\mathrm{WM,BD}}$ of 70 % (orange line) and 60 % (green line). Between $\theta \approx -50° \ldots 90°$ and $\theta \approx -50° \ldots 70°$, the detection efficiency of the top and bottom detector, respectively, scales linearly with $T_{\mathrm{WM}}$, as all particles, photons or photoelectrons, need to cross the wire mesh electrodes to reach the CEMs from the inner region. For $\theta \lesssim -50°$ and $\theta \gtrsim 50°$, on the other hand, an inverse scaling with $T_{\mathrm{WM}}$ is observed, as photons hitting the wire mesh electrodes

in one hemisphere are predominantly detected by the CEM in the opposite hemisphere. The same behavior is visible for $\theta \lesssim 70°$ for the bottom detector, since a lower $T_{\mathrm{WM}}$ of the wire mesh at the bottom of the cylinder also implies less particles escaping into the cryopump.

To simulate the detection efficiency during run A, where the additional blocking meshes were installed in the detector assembly, $T_{\mathrm{WM}}$ and $T_{\mathrm{WM,BD}}$ are assumed to be 50 % and 70 %, respectively (red line in Fig. 4.41 (C, D)). The behavior of the detection efficiency is similar to that of $T_{\mathrm{WM}} = T_{\mathrm{WM,BD}} = 70\,\%$, but the overall detection efficiency is only ≈66 % of the latter. This approximately matches the differences seen in the fluorescence signal between run A, on the one hand, and run B and C, on the other hand.

The resulting total detection efficiencies and solid angle coverages of the simulations are given in Table 4.2.

### 4.6.7 Detector electrodes and bias electric fields

Stray static electric fields inside the inner region of the detector assembly can cause a systematic offset of the measured 2S-6P transition frequency through the dc-Stark shift. Such stray fields may originate from multiple sources. First, fields present outside the inner region, e.g., the electric fields used to collect electrons at the electron multipliers, fields from charges accumulating on poorly conducting surfaces, or fields from contact potentials, may extend into the inner region through the wire mesh or the apertures. Second, an imperfect graphite coating or foreign particles on the electrodes may lead to charges building up on the electrodes themselves. Third, the fields from charges on the electrically-isolating spacers between the electrodes might not be shielded sufficiently. Fourth, temperature differences between the electrodes might lead to a thermoelectric voltage difference through the Seebeck effect or cause a contact potential through the temperature-dependence of the work function. Fifth, an inhomogeneous distribution of charged particles[1] inside the inner region itself could cause electric fields.

For all but the last possibility, the stray electric fields can then be measured using the six electrodes surrounding the inner region, and the atoms themselves as field sensors. To this end, bias voltages are applied to opposing electrodes to create a bias electric field along the given direction, which adds to or subtracts from any present stray electric fields. By determining the voltage for which dc-Stark shifts of the 2S-6P transition from the stray electric fields are canceled in all three directions, the field strength and direction of the stray electric fields can be deduced.

The electric fields resulting from bias voltages applied to the three electrode pairs are determined using a FEM simulation[2] based on the geometry of the inner region as shown in Fig. 4.34. Only the electric field in the region where the atoms are probed by the 2S-

[1] The photoelectrons here have an energy of ≈2 eV, corresponding to a speed of $8.4 \times 10^5$ m/s. They thus cross the full diameter of the inner region of the detector within 60 ns. At a count rate of $1 \times 10^6$ cts/s, reached for the short delays for $P_{\mathrm{2S\text{-}6P}} = 30\,\mu\mathrm{W}$, and a detection efficiency of 2 % on each detector, ∼3 photoelectrons are then expected at any time within the inner region. Electrons created by photoionization of 2S atoms by the preparation laser (see Section 2.2.6) are expected to have left the inner region by the time the 2S-6P transition is probed in the dark phase. On the other hand, the protons produced by photoionization have a speed of 421 m/s in the rest frame of the initial atom–photon system. Thus, they are not necessarily expected to leave the inner region by the time the transition is probed. However, the stray electric fields invariably present inside the inner region are expected to remove these particles within tens of µs.

[2] Finite element method, implemented using the AC/DC module of the commercial simulation software package COMSOL.

Table 4.3: Electric field strength $F$ within a sphere of 5 mm radius, centered within the inner region of the detector assembly, when applying a voltage of ±1 V on opposing electrodes along axis $i$. The values were obtained with a FEM simulation using geometry of the inner region of the detector assembly as shown in Fig. 4.34. The field component $F_i$ along direction $i$ when moving from the center $((x, y, z) = (x_1, x_2, x_3) = (0, 0, 0))$ along the $j$-axis is well approximated by $F_i = F_{i,0}(1 + b_{i,j}x_j^2)$. See text for details.

| Axis | $F_{i,0}$ (V/m) | $b_{i,1}$ | $b_{i,2}$ ($10^{-3}$/mm$^2$) | $b_{i,3}$ | $\max\lvert(F_i-F)/F\rvert$ | $\max\lvert(F_i-\bar{F}_i)/\bar{F}_i\rvert$ |
|---|---|---|---|---|---|---|
| $x$ ($i=1$) | 18.99 | 4.04 | −3.09 | −0.95 | $3.02 \times 10^{-3}$ | $9.33 \times 10^{-2}$ |
| $y$ ($i=2$) | 43.47 | 0.23 | −0.54 | 0.22 | $2.21 \times 10^{-5}$ | $1.25 \times 10^{-2}$ |
| $z$ ($i=3$) | 19.65 | −1.06 | −3.11 | 4.20 | $3.05 \times 10^{-3}$ | $9.70 \times 10^{-2}$ |

6P spectroscopy laser beams is of interest, here taken to be a sphere of 5 mm radius centered within the inner region. To increase the uniformity of the fields, voltages of opposite sign are applied to the opposing electrodes along direction $i$, where $i = 1, 2, 3$ are taken to be along the $x$-,$y$-,$z$-axes, respectively. By symmetry, at the center of the sphere $((x, y, z) = (x_1, x_2, x_3) = (0, 0, 0))$, the electric field points along the $i$-axis with strength $F_{i,0}$. When moving from the center along the $j$-axis, the field component along $i$ is found to be well-described by a parabolic dependence, $F_i = F_{i,0}(1 + b_{i,j}x_j^2)$. Table 4.3 gives $F_{i,0}$ and $b_{i,j}$ for the case where a voltage of ±1 V is applied to the opposing electrodes along axis $i$. The non-uniformity within the sphere is here characterized by $\max\lvert(F_i-F)/F\rvert$ and $\max\left\lvert(F_i-\bar{F}_i)/\bar{F}_i\right\rvert$, also given in Table 4.3, and reaching up to 0.3 % and 9.7 %, respectively.

## 4.7 Data acquisition

The data acquisition (DAQ) of the spectrometer is a combination of software and hardware components responsible for controlling and reading out the various measurement devices, and saving and displaying the acquired data. The design of the DAQ follows the periodic nature of the experiment, which the repeating cycle of producing metastable 2S atoms, and then exciting them to and observing their decay from the 6P level in a time-resolved fashion. This requires the acquisition devices to be synchronized with each cycle, which is achieved through direct triggering of the involved hardware. Furthermore, it also requires large amounts of data to be quickly processed and stored to keep the experimental duty cycle as high as possible. In the following, first, the structure of the experimental data is explained, followed by a description of the software and hardware involved in the DAQ. Finally, the time-resolved sampling of analog signals and counting of fluorescence from the atoms is discussed in more detail. The DAQ described here was set-up during the course of this thesis, replacing all previous hard- and software.

### 4.7.1 Data structure

The experimental data are structured into chopper cycles, data points, and line scans, as shown in Fig. 4.42. The 6.25 ms long chopper cycle is the fundamental block, consisting of a bright and dark phase of an equal duration of 3.125 ms. During the bright phase, the 1S-

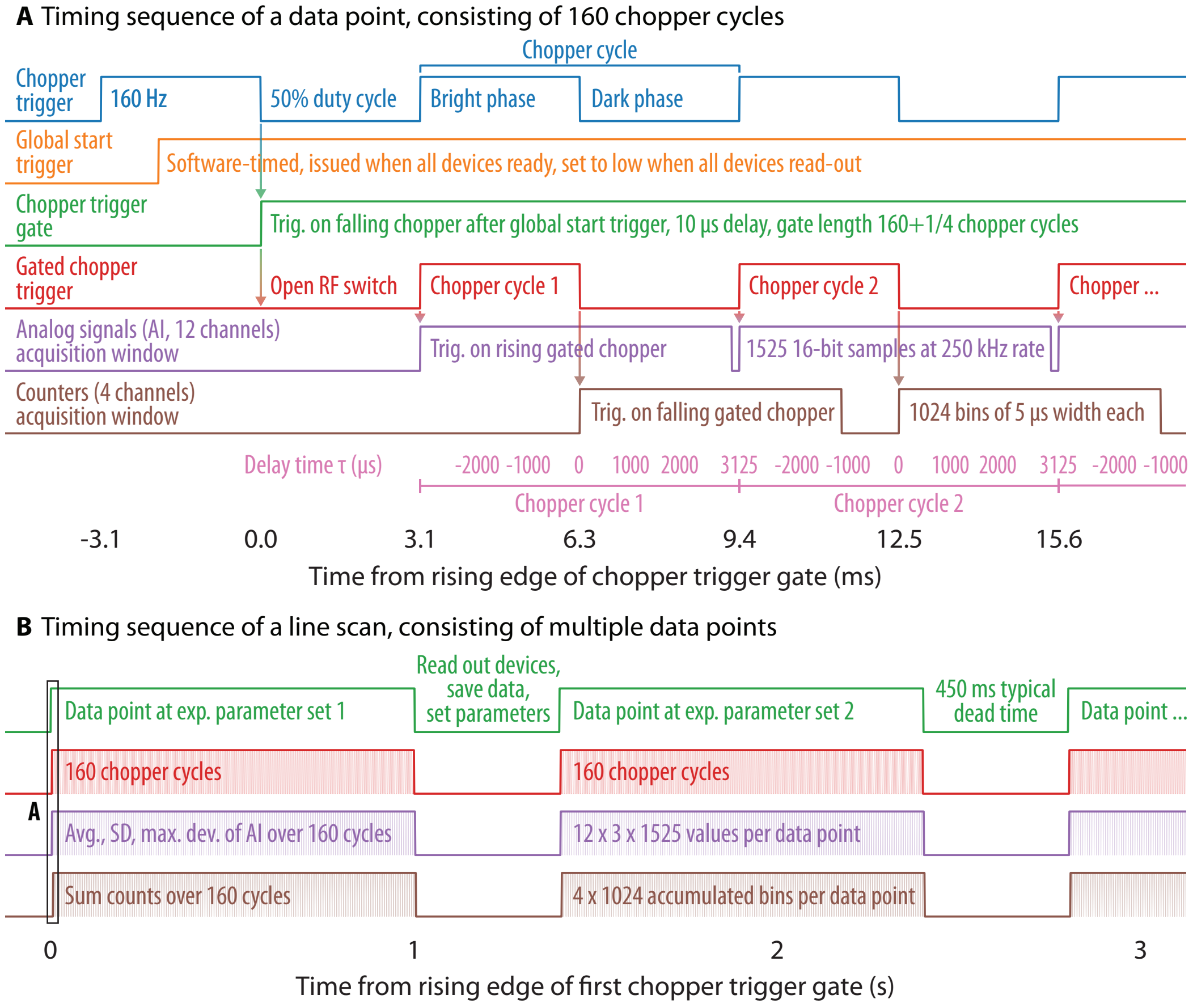

Figure 4.42: Timing sequence for the acquisition of (**A**) a single data point, consisting of 160 chopper cycles, and (**B**) a line scan, consisting of multiple data points. The signals shown in (B) correspond to those of the same color in (A). See text for details. SD: standard deviation, max. dev.: maximum absolute deviation from average.

2S preparation laser is allowed into the beam apparatus and excites some of the atoms to the metastable 2S level. The preparation laser is then blocked by the optical chopper, heralding the start of the dark phase at delay time $\tau = 0\,\mu s$, during which the 2S atoms are excited to the 6P level by the 2S-6P spectroscopy laser and subsequently decay. The fluorescence of these decays is detected, with individual detector clicks counted in a time-resolved fashion. Likewise, samples of analog signals are recorded as a function of delay time. After the optical chopper unblocks the preparation laser again, the cycle begins anew.

In total, 160 chopper cycles are repeated in sequence and with the same experimental parameters, forming a data point that thus contains 1 s of data. The analog samples are averaged over the cycles, but not the delay times, and the fluorescence counts are accumulated over the cycles into delay time bins. Additional data, such as the various synthesizer frequencies, the nozzle temperature, and so on, are read out once per data point.

Finally, a line scan is a sequence of multiple data points, with the frequency of either the 2S-6P spectroscopy laser (for line scans of the 2S-6P resonance) or the 1S-2S preparation laser (for line scans of the 1S-2S resonance) changed, or scanned, in between the data points. Scan parameters such as the line sampling method used are saved for each line scan.

### 4.7.2 Soft- and hardware overview and line scans

The software of the DAQ, referred to as Pythonic hydrogen (pyh), is written in the Python programming language[1] and allows the control of the spectrometer through a graphical user interface. It controls and reads out various devices, that are connected using a variety of interfaces to a control computer running pyh. The software also saves and displays the acquired data, and fits the line scans with various line shape models. pyh is written in a modular fashion, allowing, e.g., the use of modules only concerned with reading and analyzing the data, but not acquiring it, to be used during the post-analysis of the data. In total, the software controls 12 analog input channels sampling each chopper cycle, two analog input channels only read out when needed, four counter channels, six analog output channels, 10 digital output channels, six frequency synthesizers, three mechanical actuators, two voltage sources, and one cryogenic temperature controller. For precise timing signals independent of the status of the control computer, a programmable multifunction DAQ device[2] is used.

Two classes of data are distinguished: first, purely numerical data with many values acquired per second from the sampling of analog signals (16-bit floating point values) and the counting of atomic fluorescence (32-bit integers). Second, a mixture of categorical and numerical data with only a single set of data acquired at intervals of one second or longer, i.e., attached to a data point or line scan. The first class is stored in arrays[3] of the appropriate data type. The second data type is stored in table-like date frames[4] that can contain a mixture of data types. Each data point and each line scan are assigned a 128-bit long universally unique identifier (UUID), which is used to cross-reference the data, and a coordinated universal time (UTC) timestamp. Fits to the line scans are also saved in data frames.

To acquire data, the software first sets the control parameters of the various devices. For some devices, settling times are enforced after a change in parameters, such as when the frequency of the 2S-6P spectroscopy laser is modified, to allow the various feedback loops to settle before data are acquired. When all devices are ready and the settling times have passed, a software-timed global start trigger is issued. Then, starting from the next bright phase of the chopper analog signals are sampled and fluorescence counts are accumulated over 160 chopper cycles. This sequence is timed in hardware, as described in the next section. After these chopper cycles, the data are read-out from the devices and stored in the corresponding arrays and data frames, with a new UUID assigned to this data point. Next, new parameters are set, and the next data point is acquired. The dead time, including settling times, between subsequent data points was on average 450 ms for the 2S-6P measurement, i.e., the time between two data points was 1.45 s. If the data points are part of a line scan, a new entry to the scan data frame is added after the scan is complete, and the UUID assigned to the line scan is also added to the entries of all data points involved. After each line scan or a fixed number of data point, the data are written to the hard disk of the control computer using the HDF5[5] file format. This results in approximately 250 kB of data written for each data point, with lossless compression used as implemented within the framework of HDF5.

For a line scan, the only parameter varied between subsequent data points is the laser frequency. Additionally, there is also the possibility to acquire dual (line) scans. The idea is to perform two line scans at almost the same time while varying one or more auxiliary experimental parameters between these scans. To this end, a single data point of the first of the two scans is acquired. Then, the chosen auxiliary parameters, but not the frequency of the scanned laser, are changed, and a single data point of the second scan is acquired. This procedure results in two line scans that are effectively recorded within seconds of each other, as opposed to two line scans taken in sequence, which would be separated by almost one minute. During measurement run B of the 2S-6P measurement, dual scans were used to acquire line scans at different powers of the 2S-6P spectroscopy laser (see Table 6.2), in an attempt to further distinguish drifts in the experimental apparatus from line shifts related to the change in laser power. In this case, the dead time of 370 ms between data points for dual scans was slightly less than for non-dual scans, while the time between data points for each of the two line scans within the dual scans increased to 2.7 s. This resulted in an average time between data points of 2.3 s for the combination of non-dual and dual scans in the 2S-6P measurement.

[1] Python Software Foundation, version 3.6. Available at https://www.python.org.

[2] National Instruments PCIe-6353.

[3] Implemented using the ndarray object of NumPy, a Python package for scientific computing. Available at https://www.numpy.org. See, e.g., [153].

[4] Implemented using the DataFrame objects of pandas, a Python package for data analysis. Available at https://pandas.pydata.org. See, e.g., [154].

[5] Hierarchical Data Format, version 5 [155].

### 4.7.3 Time-resolved detection

Fig. 4.42 shows the timing sequence during the acquisition of a data point and a line scan. The experimental scheme hinges on a precise determination of the delay time $\tau$ relative to the start of the dark phase of each chopper cycle. Hence, the data acquisition is referenced to the chopper trigger (shown in blue in Fig. 4.42), which is derived from optical chopper itself, as described in Section 4.3.2. As the optical chopper is always running, the chopper trigger is gated such that exactly 160 chopper cycles are covered for each data point. The start of each data point is signaled by the software with a global start trigger (shown in orange), which arrives at random times relative to the optical chopper cycle. From this trigger, the chopper trigger gate (shown in green) is generated with the multifunction DAQ device. This gate is triggered on the falling edge of the chopper trigger with a delay of 10 µs, such that the triggering falling edge is not contained with the gate anymore. The length of the gate is set to match the duration of $160 + 1/4$ chopper cycles. In this way, the gate covers 160 rising and 160 falling edges of the chopper trigger, with a safety margin of $\pm 1/4$ chopper cycles to account for variations in the actual length of the individual chopper cycles. The gate is used to open an RF switch[1], transmitting the chopper trigger and resulting in the gated chopper trigger (shown in red). Using more sophisticated hardware such as a custom-programmed FPGA (field-programmable gate array) could replace the RF switch while adding flexibility, but this approach was not pursued here.

The sampling of analog signals (shown in purple) and the counting of the fluorescence signals (shown in brown) are then referenced to the rising and the falling edge, respectively, of the gated chopper trigger. This triggering scheme is implemented because the acquisition needs to be synced with each individual chopper cycle, which is subject to some timing jitter stemming from imperfections of the optical chopper wheel (see Section 4.3.2). Furthermore, the acquisition hardware cannot be re-triggered while an acquisition is in progress, forcing a short dead time between the chopper cycles. Since the experimental scheme hinges on an precise determination of the delay time of the fluorescence counts from the start of the dark phase, and the detectors are switched off during the bright phase, the counting is triggered on the start of the dark phase. The analog signals, on the other hand, should cover the bright phase directly preceding the dark phase covered by the fluorescence counters, and the main part of the dark phase itself. In this way, e.g., intensity variations of the 1S-2S preparation laser during the bright phase can be captured. Since at the end of the dark phase there are no fluorescence counts observed anymore, the necessary dead time for re-triggering is placed there and the trigger is set to the start of the bright phase. Therefore, the acquisition window for the fluorescence signals then covers two sequential chopper cycles, as is shown at the bottom of Fig. 4.42 (A) (pink scale).

[1]Mini-Circuits ZX80-DR230+, frequency range is from DC to 3 GHz and the absolute max. input voltage is 5 V. The voltage of the TTL true state is thus too high for the switch, and NIM (Nuclear Instrumentation Module) logic levels are used, where the false and true states are at 0 V and −0.8 V, respectively. To this end, the chopper trigger is converted from TTL to NIM, and the gated chopper trigger from NIM to TTL where needed.

Multichannel scalers[1] count the number of fluorescence clicks from the two detectors as a function of the delay time $\tau$ from the falling edge of the gated chopper trigger. The counts are sorted into 1024 time bins with a width of 5 µs each. The bins cover the complete dark phase with delay times $\tau = 0\,\mu\text{s} \ldots 3125\,\mu\text{s}$ and part of the bright phase of the next chopper cycles, corresponding to approximately $\tau = -3125\,\mu\text{s} \ldots -1130\,\mu\text{s}$. The latter bins are used to checked whether the gain of the detectors is indeed reduced far enough such that no clicks are detected. The multichannel scalers internally accumulate the counts in these bins over the 160 chopper cycles before sending the data to the control computer.

Analog-to-digital converters[2] (ADCs) take 1525 16-bit samples of the analog signals with a rate of 250 kHz, starting from the rising edge of the chopper trigger. The covered delay times are approximately $\tau = -3125\,\mu\text{s} \ldots 2975\,\mu\text{s}$. The sampling times are defined using a trigger signal, generated with the multifunction DAQ device, and distributed to the ADCs. After acquiring data for 160 chopper cycles, the ADCs send $160 \times 1525$ samples to the control computer for each channel. There, the average over the chopper cycles, but not the delay time, is calculated and saved, along with the standard deviation and the maximum deviation from the mean for each delay time. The chopper trigger itself is also sampled. In this way, the exact time of its falling edge, which is within the sampling region, can be determined, and delay times relative to it can be assigned to the samples. Thus, the analog and fluorescence signals shared a common delay time reference.

Additionally, the average, the standard error of the mean, and the maximum deviation from the mean over both the chopper cycles and a given window of the delay times is calculated for each channel. This window is defined relative to either the rising or falling edge of the chopper trigger. In this way, e.g the average intensity of the 1S-2S preparation laser for each data point is determined using a window starting and ending 500 µs and 3000 µs, respectively, after the rising edge (see Fig. 4.11).

During a line scan, after the data acquired over the 160 chopper cycles are processed, a new set of experimental parameters is sent to the devices and the next 160 chopper cycles are acquired, as shown in Fig. 4.42 (B). This is repeated till all data points constituting the line scan are recorded, after which the next line scan starts, and so on.

[1]Two FAST ComTec MCS4 multichannel scalers with two stop channels each, maximum count rate of 400 MHz and a minimum bin length of 50 ns. Two of the channels are used for the detectors, and a third channel records pulses from a generator also triggered on the optical chopper for debugging purposes.

[2]National Instruments PCI-6143 and PXI-6143 in a PXIe-1073 chassis, each offering eight analog input channels with a maximum sampling rate of 250 kHz and a maximum vertical resolution of 16 bit.

### 4.7.4 Logging of additional parameters

Some additional parameters, such as the pressures inside the vacuum system, the laboratory temperature and humidity, the signals of various monitoring photodetectors, etc. are continuously logged at intervals of 30 s. . . 60 s. The underlying software is referred to as Pythonic hydrogen logger (pyhl). The logged data are stored, along with UTC timestamps, in a separate data base, and can be combined with the data acquired with pyh by matching the timestamps.

## 4.8 Determination of laser frequencies

An optical frequency comb[1] [156] is used to compare the optical frequencies of the 1S-2S preparation laser and the 2S-6P spectroscopy laser to the microwave frequency of a passive hydrogen maser[2]. All frequency synthesizer used to modify the optical frequencies of the lasers before interacting with the atoms are also referenced to the hydrogen maser. The frequency of the hydrogen maser, in turn, is continuously compared against the caesium standard, which is the basis of the unit of hertz in the International System of Units (SI), using a global navigation satellite system (GNSS) receiver. The fractional frequency offset between the maser and the caesium standard is kept below $1 \times 10^{-13}$ by manual adjustment, usually required every few months. This fractional offset translates into an absolute offset of 73 Hz for the 2S-6P transition and 247 Hz for the 1S-2S transition. By using the GNSS corrections and averaging the frequency comparison over a sufficiently long gating window, here typically 30 min, this offset can be further reduced during the data analysis.

However, the full GNSS corrections can only be applied after a delay of a few days, when the satellite ephemerides become available. This is not a problem for the measurement of the 2S-6P transition, as at the time of the measurement its line center only needs to be known on the order of 10 kHz to ensure correct sampling of the resonance. On the other hand, the typical offset limits the accuracy with which the optimal frequency of the 1S-2S preparation laser can be predicted during the measurement. This is why this optimal frequency, i.e., the frequency at which the highest number of 2S atoms is produced for a given delay, is found experimentally for each freezing cycle. Because of the length drift of their respective high-finesse Fabry-Pérot cavities, the fundamental frequencies (in the infrared) of the preparation and spectroscopy lasers drift with linear rates of approximately 2.7(2.6) kHz/day and 5.4(2.9) kHz/day, respectively (parentheses give standard deviation over measurement days). During the ≈1 min required to record a single line scan, this corresponds to a drift at the atomic transition frequencies of ≈15 Hz and ≈8 Hz, respectively, well below the level required here. During the measurement, the frequency drift of the preparation laser is compensated multiple times per freezing cycle, while the line sampling of the 2S-6P resonance is adjusted before each freezing cycle.

The frequency $\nu$ of any mode of the frequency comb can be fully determined by its mode number $N$, the repetition rate $f_{\text{rep}}$, and the carrier-envelope offset (CEO) frequency $f_{\text{CEO}}$ through the relation $\nu = N \times f_{\text{rep}} + f_{\text{CEO}}$. Both $f_{\text{rep}}$ and $f_{\text{CEO}}$ are radio frequencies that are straightforward to synthesize and count electronically. The fundamental comb, with its spectrum centered at a wavelength of ≈1.5 µm, is frequency-doubled to reach the visible spectrum. This leaves the repetition rate unchanged, but doubles the CEO frequency, such that the frequency of a mode of the frequency-doubled comb is given by $\nu = N \times f_{\mathrm{rep}} + 2f_{\mathrm{CEO}}$. Here, instead of stabilizing $f_{\mathrm{rep}}$ and $f_{\mathrm{CEO}}$, one mode of the frequency-doubled comb, with mode number $N_{\mathrm{FP3}} = 1\,461\,379$, is phase-coherently stabilized, with an offset frequency of $f_{\mathrm{FP3,FC}} \approx 47\,\mathrm{MHz}$, to the fundamental of the 2S-6P spectroscopy laser with frequency $\nu_{\mathrm{FP3,ECDL}} \approx 365.344\,708\,\mathrm{THz}$ (see Fig. 4.14). To this end, an optical beat note between the comb mode and the spectroscopy laser is detected with a photodetector. The CEO frequency of the frequency comb is phase-coherently stabilized, using an f-to-2f interferometer, to $f_{\mathrm{CEO}} = -45\,\mathrm{MHz}$. In this way, the frequency of the comb modes is fully determined, with the frequency of the spectroscopy laser given by

$$\nu_{\mathrm{FP3,ECDL}} = N_{\mathrm{FP3}} \times f_{\mathrm{rep}} + 2f_{\mathrm{CEO}} + f_{\mathrm{FP3,FC}}. \tag{4.29}$$

[1] Menlo Systems FC1500-250-ULN ultra-low-noise optical frequency comb. The integrated femtosecond laser system generating the frequency comb is based on polarization-maintaining erbium-doped fibers.

[2] T4Science pH Maser 1008, Allan deviation specified as $5 \times 10^{-13}$ and $9 \times 10^{-15}$ at 1 s and 1 h, respectively.

The offset frequency $f_{\mathrm{FP3,FC}}$ is chosen such that the repetition rate of the frequency comb is approximately $f_{\mathrm{rep}} = 250\,\mathrm{MHz}$.

Likewise, the frequency of the preparation laser (see Fig. 4.5) is given by

$$\nu_{\mathrm{FP1,ECDL}} = N_{\mathrm{FP1}} \times f_{\mathrm{rep}} + 2f_{\mathrm{CEO}} + f_{\mathrm{FP1,FC}} + 2f_{\mathrm{FP1,Scan}}, \tag{4.30}$$

where $f_{\mathrm{FP1,FC}} \approx 40\,\mathrm{MHz}$ is the beat frequency between the preparation laser and the mode of the frequency-doubled comb with mode number $N_{\mathrm{FP1}} = 1\,233\,028$. $f_{\mathrm{FP1,Scan}}$ is the radio frequency, referenced to the hydrogen maser, sent to an acousto-optic modulator (AOM) to shift the laser frequency relative to the resonance frequency of the high-finesse cavity (see Section 4.3.1).

The frequencies $f_{\mathrm{rep}}$, $f_{\mathrm{CEO}}$, $f_{\mathrm{FP1,FC}}$, and $f_{\mathrm{FP3,FC}}$ are continuously determined using Λ-type counters[1] [157, 158], referenced to the hydrogen maser. $f_{\mathrm{FP1,FC}}$ and $f_{\mathrm{FP3,FC}}$ are additionally counted using independent redundancy counters, used to diagnose counting issues.

The frequency of the laser light interacting with the atoms can then be found by appropriately taking into account the frequency shifting by AOMs and doubling or quadrupling in nonlinear crystals. For the 1S-2S preparation laser, the frequency inside the 243 nm enhancement cavity in the laboratory frame is (see Section 4.3.1 and Fig. 4.5)

$$\nu_{\mathrm{FP1,Exp}} = 4\nu_{\mathrm{FP1,ECDL}}. \tag{4.31}$$

For the 2S-6P spectroscopy laser, the frequency, likewise in the laboratory frame, of the light coupled into the active fiber-based retroreflector is (see Section 4.4.1 and Fig. 4.14)

$$\nu_{\mathrm{FP3,Exp}} = 2\nu_{\mathrm{FP3,ECDL}} + 2f_{\mathrm{FP3,Scan}} + \delta_{\mathrm{J},3/2} f_{\mathrm{FP3,FS}}. \tag{4.32}$$

$f_{\mathrm{FP3,Scan}}$ and $f_{\mathrm{FP3,FS}}$ are radio frequencies referenced to the hydrogen maser, and $\delta_{\mathrm{J},3/2}$ is 0 (1) when the 2S-6P$_{1/2}$ (2S-6P$_{3/2}$) transition is probed. $f_{\mathrm{FP3,FS}}$ is set to match the 6P fine-structure splitting $\Delta\nu_{\mathrm{FS}}^{\mathrm{pred}}(6\mathrm{P})$ (see Section 6.2.4.6 and Eq. (6.23)). Offsets, randomly generated[2] and unknown to the author, are added to $\nu_{\mathrm{FP3,Exp}}$, without modifying the original data, during the blind data analysis of 2S-6P transition frequency. Two separate offsets are used for the analysis of the 2S-6$P_{1/2}$ transition and of the 2S-6$P_{3/2}$ transition.

[1] K+K Messtechnik FX80.

[2] The random offset is drawn from a normal distribution centered at the expected 2S-6P transition frequency as taken from [40], i.e., using the muonic proton radius of [22], and with a standard deviation of 12 kHz. The same distribution is used to determine the center of the frequency sampling of the experimental resonance.

Because of the low intrinsic noise and the fast feedback bandwidth of the frequency comb, the beat note of the frequency comb with the 1S-2S preparation laser while the frequency comb is stabilized to the 2S-6P spectroscopy laser reveals the combined spectrum of the preparation laser and spectroscopy laser. Both laser systems use external-cavity diode lasers (ECDLs) similar to the one described in [82] as master lasers, and their combined spectrum is comparable to that expected from the spectra shown therein.

# Chapter 5

# Data analysis and numerical modeling

## 5.1 Data analysis procedure

The goal of the experiment described in this work is the determination of the resonance frequency of the 2S-6P transitions in atomic hydrogen. To this end, the line scans over these resonances are analyzed to extract the resonance frequency $\nu_0$. Each line scan consists of the number of fluorescence photons detected, or counts, as a function of the delay time $\tau$ and the frequency detuning $\Delta\nu_{\text{2S-6P}}$ of the 2S-6P spectroscopy laser, with the latter sampled at 30 points. Fig. 5.1 shows a typical line scan, in this case for the 2S-6P$_{1/2}$ transition and using $P_{\text{2S-6P}} = 10\,\mu\text{W}$ of spectroscopy laser power. The time-resolved counts of each line scan are binned into 16 delays ($N_{\text{dlys}} = 16$), as defined in Table 5.1. The resulting 16 signals each contain the contribution from a different velocity group of atoms, ranging in mean atom speed $\bar{v}$ from 256 m/s to 66 m/s for delay 1 and delay 16, respectively.

Each of these delays is analyzed independently by fitting a line shape function to the data, as described in Section 5.1.1. This results in a value for the resonance frequency $\nu_0$ for each delay, as shown in Fig. 5.3 for the same line scan as shown in Fig. 5.1. Since each delay corresponds to a different mean atom speed $\bar{v}$, any velocity-dependent frequency shift will show up as a modulation of $\nu_0$ as a function of $\bar{v}$. The dominant shift in this experiment is the first-order Doppler shift, linear in $\bar{v}$. This shift is removed by a linear extrapolation of $\nu_0$ to zero speed, resulting in a resonance frequency free from the first-order Doppler shift, $\nu_{0,\text{e}}$, as discussed in Section 5.1.2. The extrapolated frequencies $\nu_{0,\text{e}}$ form the basis of the determination of the 2S-6P transition frequencies. Thus, the frequency sampling, i.e., the frequency detuning of the 30 points ($N_\Delta = 30$) sampling the resonances, has been chosen such that the statistical uncertainty in $\nu_{0,\text{e}}$ is as low as possible, as detailed in Section 5.1.3.

### 5.1.1 Line shape functions

Ideally, the line shape function that is fit to the data is identical to the experimentally observed line shape. However, this exact function is rarely known, as is the case here, where, e.g., the properties of the atomic beam, which are only known to some degree, substantially influence the line shape. Instead, a line shape function $\text{LS}(\Delta\nu_\text{L}, \boldsymbol{p})$ is used, which for appropriate values of the free parameters $\boldsymbol{p}$ approximates the the experimental line shape sampled at frequency

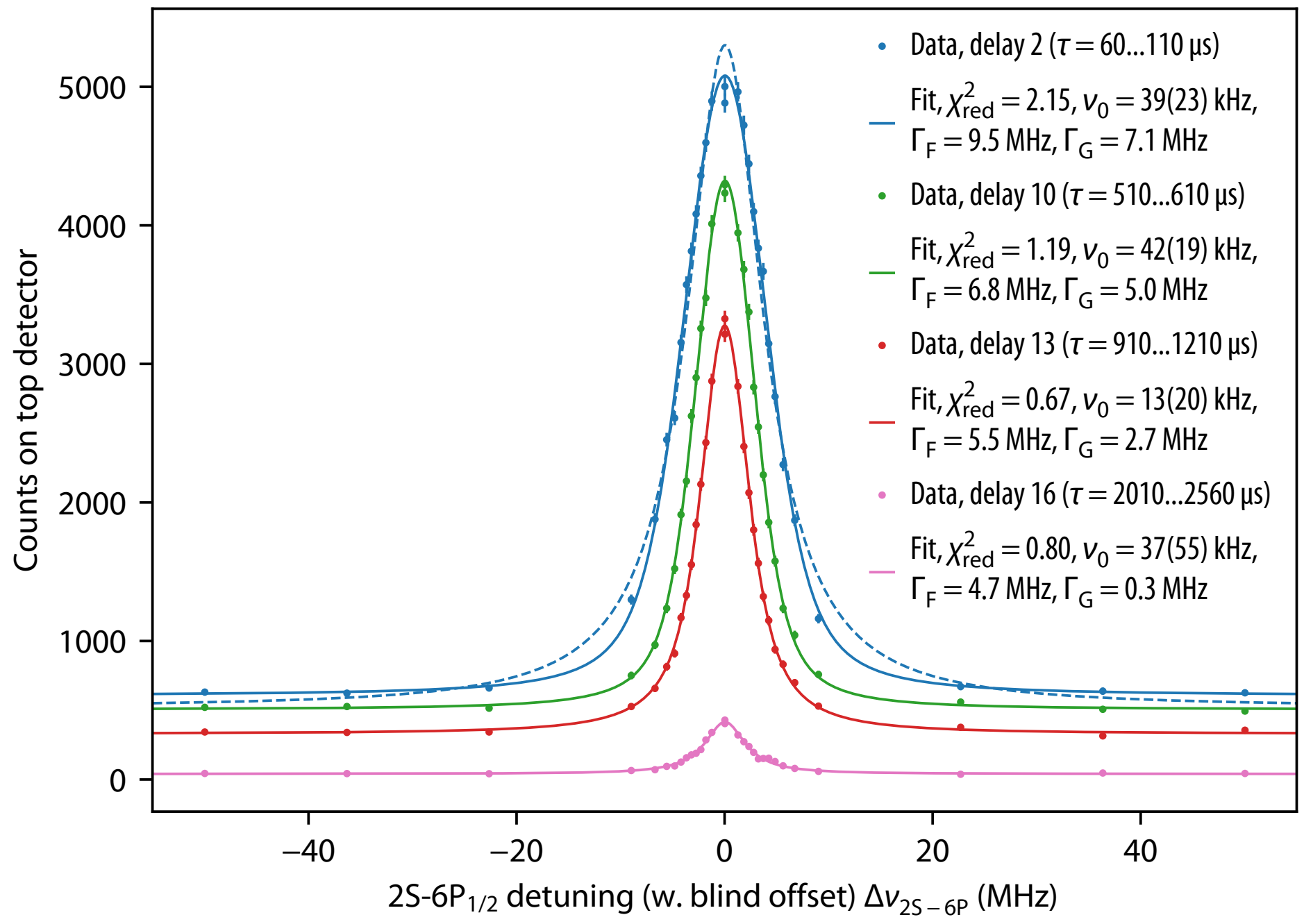

Figure 5.1: Typical line scan of the 2S-6P$_{1/2}$ transition, consisting of 30 data points of the counts on the top detector as a function of frequency detuning $\Delta\nu_{\text{2S-6P}}$ of the 2S-6P spectroscopy laser. The counts are binned into 16 delays covering a range of delay times $\tau$ (see Table 5.1), with four of the resulting signals for delays 2, 10, 13, and 16 shown here (points). The uncertainty on each data point is assumed to be shot noise only, i.e., the square root of the counts. Each delay is fit independently with a Voigt line shape (solid lines) to extract the resonance frequency $\nu_0$. For larger $\tau$, the linewidth $\Gamma_\text{F}$ decreases as slower atoms are probed and the Doppler broadening $\Gamma_\text{G}$ reduces, with $\Gamma_\text{F}$ only 20 % above the natural linewidth of $\Gamma = 3.90\,\text{MHz}$ for delay 16. The number of counts decreases with increasing $\tau$, leading to an increasing dominance of shot noise over other noise sources such as atomic beam fluctuations, which tends to improve the reduced chi-squared $\chi^2_\text{red}$ of the fit. A Lorentzian fit shown for delay 2 (blue dashed line), does not properly account for the Doppler broadening for short $\tau$, while for delay 16 a Voigt fit is indistinguishable from a Lorentzian. The counts on the bottom detector are comparable, but not shown here for clarity. The powers of the 2S-6P spectroscopy and 1S-2S preparation lasers were $P_{\text{2S-6P}} = 10\,\mu\text{W}$ and $P_{\text{1S-2S}} = 1.1\,\text{W}$, respectively, and the line scan is part of data group G3 (see Table 6.2).

detunings $\Delta\nu_\text{L}$.

#### 5.1.1.1 Lorentzian line shape

The fluorescence signal from a single atomic resonance has a Lorentzian line shape [51], which can be expressed as

$$\text{L}(\Delta\nu_\text{L}, \nu_0, A, \Gamma_\text{L}, y_0) = A\frac{(\Gamma_\text{L}/2)^2}{(\Delta\nu_\text{L}-\nu_0)^2 + (\Gamma_\text{L}/2)^2} + y_0. \tag{5.1}$$

$\Delta\nu_\text{L}$ is the frequency detuning, i.e., the frequency difference to some reference frequency, $\nu_0$ is the resonance frequency or line center, $A$ is the line amplitude, $\Gamma_\text{L}$ is the full width at half maximum (FWHM) linewidth, and $y_0$ is a detuning-independent offset of the signal. The latter four are used as free fit parameters in the data analysis. The total FWHM linewidth $\Gamma_{\mathrm{F}}$ is by definition identical to $\Gamma_{\mathrm{L}}$. The Lorentzian line shape is symmetric about $\Delta\nu_{\mathrm{L}} = \nu_0$ and the signals at $\Delta\nu_{\mathrm{L}} = \nu_0$ and $\Delta\nu_{\mathrm{L}} = \nu_0 \pm \Gamma_{\mathrm{L}}/2$ are $A + y_0$ and $A/2 + y_0$, respectively. The Lorentzian line shape however is not a good fit to the experimental data, except for the longest delay times and thus slowest atoms, as its shape does not match the observed Doppler broadening (see blue dashed line in Fig. 5.1).

Table 5.1: Overview of the $N_{\mathrm{dlys}} = 16$ delays into which the time-resolved fluorescence counts are binned for analysis. Each delay covers a range of delay times $\tau$, relative to the blocking of the 1S-2S preparation laser at $\tau = 0\,\mu\mathrm{s}$, and spans a total delay time $\Delta\tau$. Using a Monte Carlo simulation of atomic trajectories (see Section 5.2) and a model of the 2S-6P fluorescence signal (see Section 5.3), the velocity distribution of the atoms probed within each delay is determined. From this, the mean speed $\bar{v}$ and the FWHM $\Delta v$ of the 6P speed distribution of the atoms contributing to the fluorescence signal is determined. $A_i$ is the expected line amplitude for delay $i$, given relative to $A_2$. $p_{\mathrm{6P,sig}}$ is the average probability for an 2S atom to contribute to the fluorescence signal when the spectroscopy laser is on resonance, given separately for the three values of the spectroscopy laser power $P_{\mathrm{2S\text{-}6P}}$ used in the 2S-6P measurement. $P_{\mathrm{2S\text{-}6P}}$ is here referenced to the 2S-6P$_{1/2}$ transition, i.e., the actual power used for the 2S-6P$_{3/2}$ transition is a factor of two lower. The number in parentheses is the standard deviation over multiple simulations covering the parameters given in Table 5.3 and the parameters of data groups G1–G12 of the 2S-6P measurement as given in Table 6.2.

| Delay $i$ | $\tau$ (µs) | $\Delta\tau$ (µs) | $\bar{v}$ (m/s) | $\Delta v$ (m/s) | $\Delta v_x$(2S) (m/s) | $\frac{A_i}{A_2}$ | $p_{\mathrm{6P,sig}}$ (%) $P_{\mathrm{2S\text{-}6P}}$ (µW) 10 | 20 | 30 |
|---|---|---|---|---|---|---|---|---|---|
| 1 | 10...60 | 50 | 256.4(6.5) | 217(5) | 3.36(11) | 1.022(2) | 6 | 12 | 17 |
| 2 | 60...110 | 50 | 255.0(6.4) | 215(5) | 3.36(11) | 1.000(0) | 6 | 12 | 17 |
| 3 | 110...160 | 50 | 253.0(6.3) | 213(5) | 3.33(11) | 0.970(2) | 6 | 12 | 17 |
| 4 | 160...210 | 50 | 250.0(6.1) | 210(5) | 3.27(11) | 0.932(3) | 6 | 12 | 17 |
| 5 | 210...260 | 50 | 245.5(5.8) | 206(4) | 3.17(10) | 0.882(6) | 6 | 12 | 18 |
| 6 | 260...310 | 50 | 238.6(5.4) | 198(4) | 3.02(8) | 0.818(8) | 7 | 13 | 19 |
| 7 | 310...360 | 50 | 229.8(5.0) | 190(3) | 2.82(7) | 0.743(11) | 7 | 14 | 20 |
| 8 | 360...410 | 50 | 219.8(4.5) | 180(3) | 2.61(6) | 0.663(13) | 8 | 15 | 22 |
| 9 | 410...510 | 100 | 204.4(3.9) | 167(3) | 2.33(4) | 1.090(29) | 9 | 17 | 24 |
| 10 | 510...610 | 100 | 184.3(3.3) | 150(4) | 2.01(3) | 0.811(29) | 11 | 20 | 28 |
| 11 | 610...710 | 100 | 166.5(2.7) | 136(5) | 1.76(3) | 0.595(27) | 13 | 23 | 32 |
| 12 | 710...910 | 200 | 146.1(2.3) | 114(5) | 1.50(2) | 0.758(45) | 15 | 27 | 37 |
| 13 | 910...1210 | 300 | 120.8(1.8) | 89(4) | 1.20(2) | 0.558(44) | 19 | 34 | 45 |
| 14 | 1210...1510 | 300 | 99.0(1.3) | 71(4) | 0.94(2) | 0.252(26) | 24 | 41 | 54 |
| 15 | 1510...2010 | 500 | 81.2(1.2) | 54(6) | 0.76(2) | 0.175(22) | 29 | 48 | 62 |
| 16 | 2010...2560 | 550 | 65.5(0.9) | 34(7) | 0.60(2) | 0.068(12) | 35 | 56 | 70 |

#### 5.1.1.2 Voigt line shape

The transverse velocity distribution of the atomic beam causing the Doppler broadening can be approximated by a Gaussian with a FWHM linewidth of $\Gamma_{\mathrm{G}}$, as simulations such as those shown in Fig. 5.4 (A) reveal. Then, the resulting line shape can be approximated with a Voigt function [51], which is defined as the convolution of a Lorentzian line shape with a Gaussian.

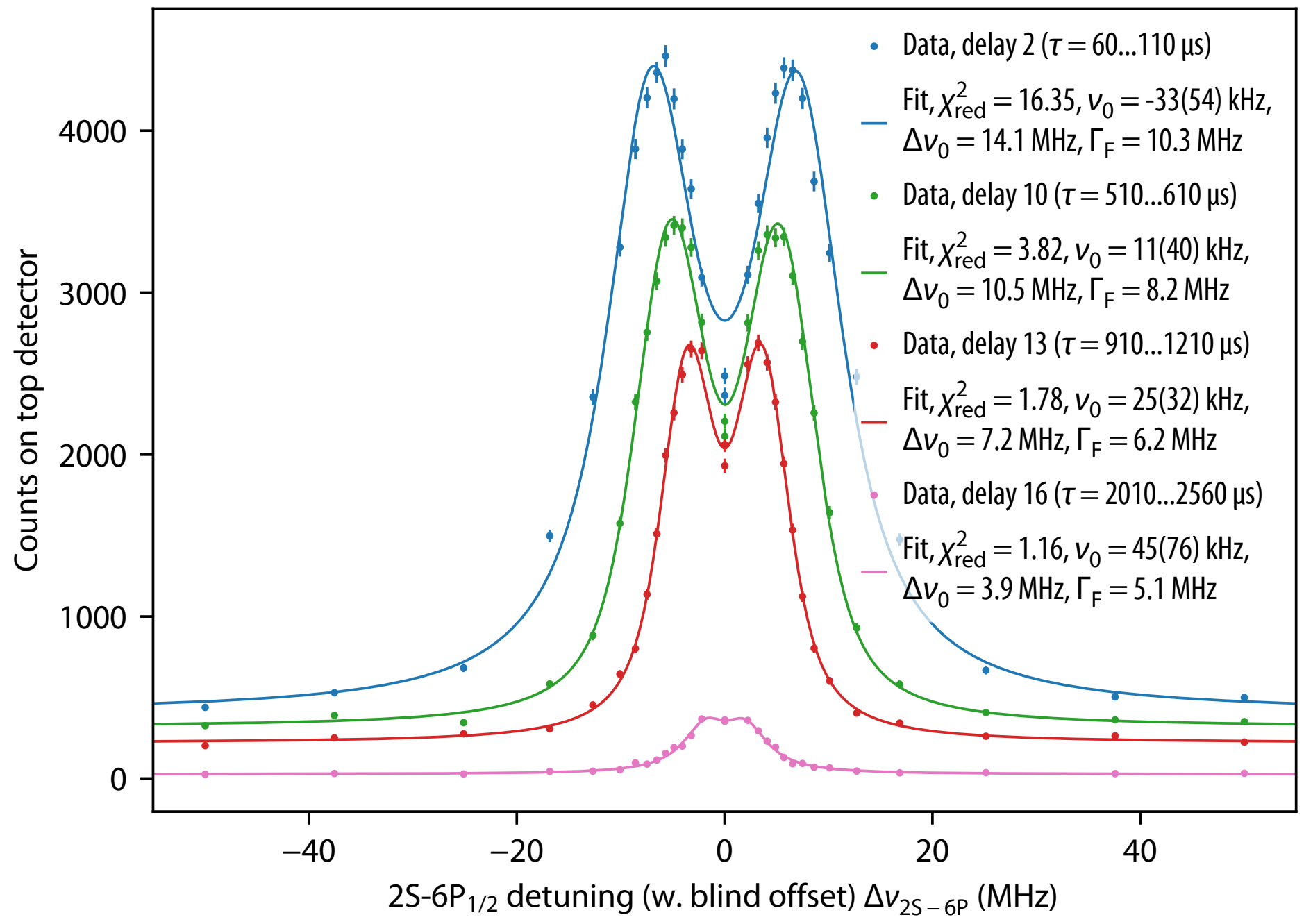

Figure 5.2: Line scan of the 2S-6P$_{1/2}$ transition with an offset angle of $\alpha_0 = 12.0\,\mathrm{mrad}$ from the orthogonal between the atomic and laser beams, leading to a splitting of the signal into a doublet of resonances separated in frequency by $\Delta\nu_0$. As in Fig. 5.1, four of the resulting signals for delays 2, 10, 13, and 16 are shown (points). Each delay is fit with a Voigt doublet line shape (solid lines), giving the resonance frequencies $\nu_1$ and $\nu_2$ of the two resonances, which are weighted with their amplitudes to find the resonance frequency $\nu_0$. The Voigt doublet does not include saturation effects, which lead to a lower-than-expected signal, as compared to the case of no saturation, for atoms interacting with both laser beams simultaneously ($\Delta\nu_{\text{2S-6P}} \lesssim \Gamma_{\text{F}}$). The fit minimizes the total deviation from the data, leading to the fitted line shape lying below the signal for atoms interacting with a single laser beam only ($\Delta\nu_{\text{2S-6P}} \gtrsim \Gamma_{\text{F}}$). This results in a large $\chi^2_{\text{red}}$, especially for short $\tau$ where $\Delta\nu_0 > \Gamma_{\text{F}}$. The powers of the 2S-6P spectroscopy and 1S-2S preparation lasers were $P_{\text{2S-6P}} = 30\,\mu\mathrm{W}$ and $P_{\text{1S-2S}} = 1.0\,\mathrm{W}$, respectively, and the line scan is part of data group G14 (see Table 6.2).

The Voigt line shape is given by

$$\mathrm{V}(\Delta\nu_{\mathrm{L}}, \nu_0, A, \Gamma_{\mathrm{L}}, \Gamma_{\mathrm{G}}, y_0) = \frac{A}{\mathrm{Re}[w(ib)]}\,\mathrm{Re}[w(z)] + y_0, \tag{5.2}$$

$$\text{with} \quad z = a + ib = \frac{2\sqrt{\ln 2}}{\Gamma_{\mathrm{G}}}\left((\Delta\nu_{\mathrm{L}} - \nu_0) + i\Gamma_{\mathrm{L}}/2\right). \tag{5.3}$$

$w(z) := e^{-z^2}\,\mathrm{erfc}(-iz)$ is the Faddeeva function, defined through the complex complementary error function erfc. $\Gamma_{\mathrm{L}}$ and $\Gamma_{\mathrm{G}}$ are the Lorentzian and Gaussian linewidths, respectively. $\Gamma_{\mathrm{G}}$ is treated as a free fit parameter in the data analysis, increasing the total number of free fit parameters to five. The total FWHM linewidth $\Gamma_{\mathrm{F}}$ is given by Eq. (2.13) and its uncertainty is found through propagation of the uncertainties of $\Gamma_{\mathrm{L}}$ and $\Gamma_{\mathrm{G}}$, including their correlation. The normalization is again such that the signal on resonance ($\Delta\nu_{\mathrm{L}} = \nu_0$) is $A + y_0$. The Voigt line shape describes the experimental line shape well for all delays, with $\Gamma_{\mathrm{F}}$ varying by more than a factor of two between the different delays (see solid lines in Fig. 5.1).

#### 5.1.1.3 Voigt doublet line shape

When an offset angle $\alpha_0$ between the atomic beam and the 2S-6P spectroscopy laser beams is set in the experiment, the observed signal splits into a doublet of two Doppler-broadened resonances (see Section 2.2.5). A line scan for such a configuration is shown in Fig. 5.2. In this case, the sum of two Voigts can be used to approximate the resulting line shape. This Voigt doublet is defined as

$$\begin{aligned} \mathrm{VD}(\Delta\nu_\mathrm{L}, \nu_1, \nu_2, A_1, A_2, \Gamma_\mathrm{L}, \Gamma_\mathrm{G}, y_0) = &\mathrm{V}(\Delta\nu_\mathrm{L}, \nu_1, A_1, \Gamma_\mathrm{L}, \Gamma_\mathrm{G}, y_0/2) \\ &+ \mathrm{V}(\Delta\nu_\mathrm{L}, \nu_2, A_2, \Gamma_\mathrm{L}, \Gamma_\mathrm{G}, y_0/2), \end{aligned} \tag{5.4}$$

where $\nu_1$ and $\nu_2$ are the resonance frequencies of the two resonances and $A_1$ and $A_2$ are their amplitudes. The underlying Voigt line shapes share the same Lorentzian, Gaussian, and total linewidths, since the natural linewidth and broadening mechanism are identical, and the detuning-independent offset $y_0$ is added to the total line shape. There are thus, compared to the Voigt line shape, two more free fit parameters, bringing the total to seven. In the data analysis, the resonance frequency $\nu_0$ of the total line shape is of interest, which here is defined as the average of $\nu_1$ and $\nu_2$, weighted by their amplitudes, resulting in

$$\nu_0 = \frac{A_1\nu_1 + A_2\nu_2}{A_1 + A_2}. \tag{5.5}$$

$\nu_0$ thus corresponds to the center of mass of the two resonances.

It is furthermore advantageous to consider additional combinations of the fit parameters: the frequency splitting of the doublet, given by $\Delta\nu_0 = |\nu_2 - \nu_1|$, and the amplitude ratio $A_1/A_2$ with $\nu_2 \geq \nu_1$, i.e., $A_1/A_2$ is always the ratio of the amplitude of the lower frequency resonance and the amplitude of the higher frequency resonance. The uncertainties of all these combinations are found through the propagation of uncertainties of the involved fit parameters, including their correlations as determined by the fit.

The Voigt doublet describes the experimental data only reasonably well (see solid lines in Fig. 5.2), which is mainly due to the fact that saturation effects, just like for the Voigt, are not included. These saturation effects lead to a lower-than-expected signal, as compared to the case of no saturation, for atoms interacting with both laser beams simultaneously ($\Delta\nu_{\text{2S-6P}} \lesssim \Gamma_\mathrm{F}$), while atoms interacting with a single laser beam only ($\Delta\nu_{\text{2S-6P}} \gtrsim \Gamma_\mathrm{F}$) are less affected by saturation effects. See Fig. 5.2 for details.

#### 5.1.1.4 Asymmetric line shapes

All the line shapes discussed so far are symmetric about the line center. The experimental line shape, on the other hand, is not expected to be symmetric, as both quantum interference effects and the light force shift introduce an asymmetry. Using a line shape that takes this asymmetry into account through an additional parameter allows to determine the asymmetry and remove the associated line shift from the experimental data. The analysis of the 2S-4P measurement (see Appendix A) used such a line shape, the Fano-Voigt line shape, to account for line shifts from quantum interference. However, the usually present correlation between the additional asymmetry parameter and the determined resonance frequency can increase the uncertainty in the latter. This is why, together with the overall much smaller asymmetry as compared to the 2S-4P measurement, a symmetric line shape is used in the data analysis presented here, with the line shifts from the remaining asymmetry accounted for through simulation corrections.

#### 5.1.1.5 Fitting of line shape functions to data

The line shape functions $\mathrm{LS}(\Delta\nu_\mathrm{L}, \boldsymbol{p})$ are fit to the data through a least-squares optimization. The data consist of $N_\Delta$ pairs of frequencies detunings $\Delta\nu_{\mathrm{L},i}$ and the number of detected counts $y_i$ for each delay. The square root of the number of counts $y_i$ is used as an estimate of the uncertainty in the number of counts, $\sigma_i = \sqrt{y_i}$. This corresponds to the assumption that the uncertainty is only given by the shot noise in the number of photons detected[1]. The least-square optimization minimizes the total squared deviations of the data from the model, weighted by the uncertainty, i.e., the quantity

$$\chi^2 = \sum_i \frac{\left(y_i - L(\Delta\nu_{\mathrm{L},i}, \boldsymbol{p})\right)^2}{\sigma_i^2}. \tag{5.6}$$

For each of the fit parameters $\boldsymbol{p}$, the optimization gives an estimate of its most likely value and an estimate of the one-standard-deviation uncertainty ($1\sigma$) of this value. The $1\sigma$ uncertainty of a parameter is defined as the change in the value of that parameter that leads to an increase in $\chi^2$ by one, all while the values of all other parameters are optimized to achieve the lowest $\chi^2$. Additionally, the optimization returns the estimated correlations between all fit parameters. For the assumption of Gaussian uncertainties, as used here, and further assuming that the measurement errors of the data points are uncorrelated, the least-square optimization corresponds to a maximum likelihood estimation [159, 160]. In this case, the distribution of $\chi^2$ follows the chi-squared distribution for $k$ degrees of freedom [159], where $k$ here is taken to be $N_\Delta - M$, with $M$ the number of fit parameters.

An useful quantity in this context is the reduced chi-squared statistic, given by

$$\chi^2_\mathrm{red} = \frac{\chi^2}{k}. \tag{5.7}$$

This is because, for the case of uncorrelated Gaussian uncertainties and large $k$, $\chi^2_\mathrm{red}$ is approximately normally distributed with a mean of 1 and a standard deviation of $\sqrt{2/k}$. Throughout this work, $\chi^2_\mathrm{red}$ is commonly given for fit results and weighted averages, with the value usually given in parentheses the standard deviation expected for the corresponding value of $k$.

When $k$ is not large, it is instructive to calculate the $p$-value, which is the probability of finding a $\chi^2$ at least as large as the observed $\chi^2$, assuming the chi-squared distribution with $k$ degrees of freedom describes the data (see Eq. (39.71) of [160]). It is furthermore sometimes helpful to convert this $p$-value into the single-sided significance $Z$ (see Eq. (39.46) of [160]). $Z$ is defined so that a $Z$ standard deviation upward fluctuation of a Gaussian random variable would have an upper tail area equal to $p$ [160]. The $p$-value is used to discuss the results of the 2S-6P measurement in Chapter 6, and the significance is given as $Z\sigma$ if $p < 0.5$.

### 5.1.2 Delay analysis and Doppler extrapolation

The fits of the line shape functions to each line scan results in a set of fit parameter estimates for each of the 16 delays. For example, the 16 resonance frequencies $\nu_0$ and

[1]Strictly speaking, the probability distribution of counts is given by the Poisson distribution, which however is well-approximated by a Gaussian centered at the expected number of counts $y_i$ with standard deviation of $\sqrt{y_i}$ for $y_i \gtrsim 10$ [159]. This condition is met for all experimental data of the 2S-6P transition. In this case, the expected number of counts $y_i$ can also be replaced with the observed number of counts, which is the procedure used here.

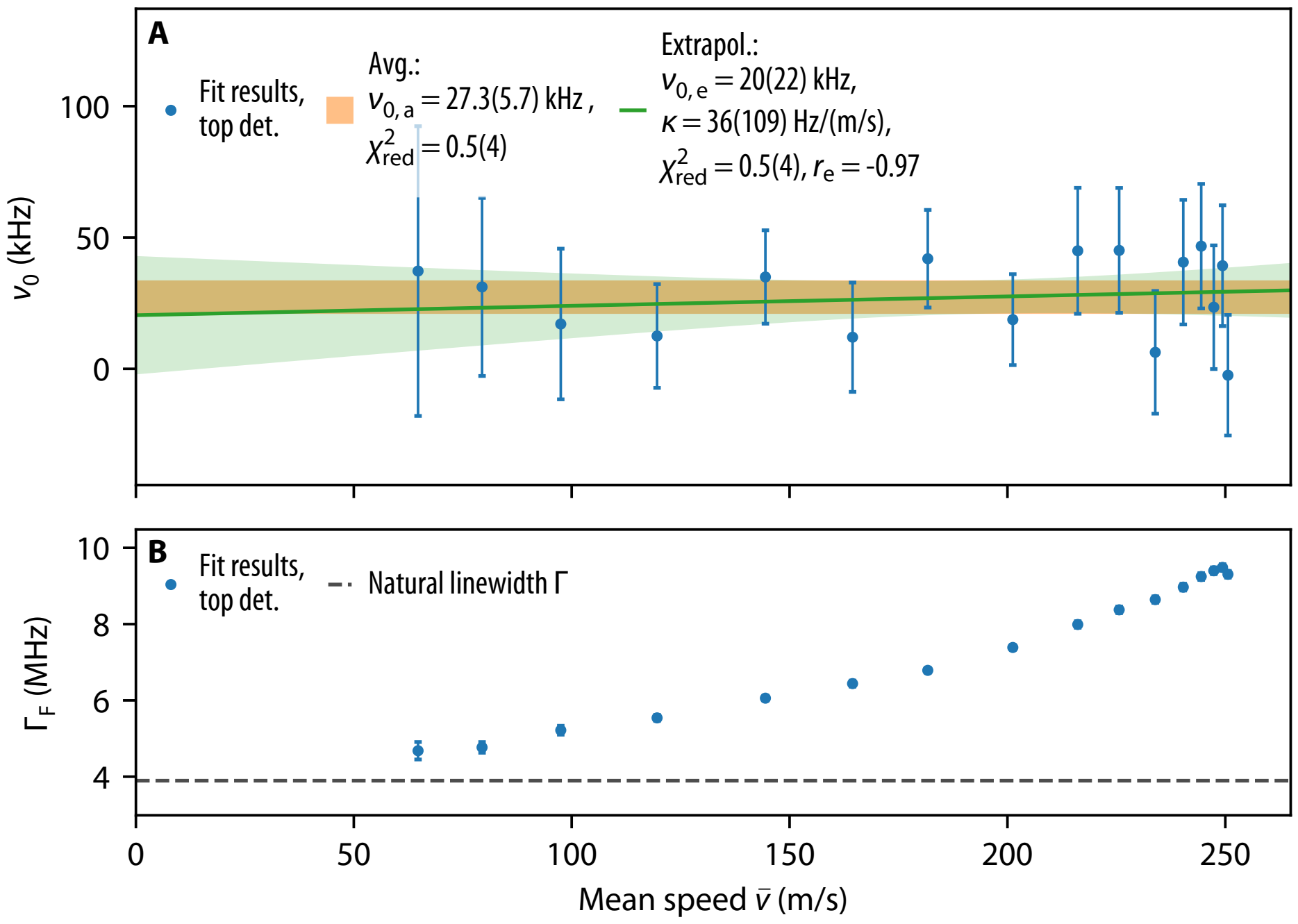

Figure 5.3: The (**A**) resonance frequencies $\nu_0$ and (**B**) FWHM linewidths $\Gamma_F$ of the delays of the line scan shown in Fig. 5.1 as a function of the delays' mean atom speeds $\bar{v}$ (see Table 5.1). The fit results and their uncertainties (blue points and error bars) are derived from fits of a Voigt line shape to the individual delays. (**A**) A linear fit (green line), the Doppler extrapolation, results in values for the Doppler-free resonance frequency $\nu_{0,e}$ and the Doppler slope $\kappa$. The uncertainty of the linear model (green shading) changes with $\bar{v}$ through the linear correlation coefficient $r_e$ (here, $-0.97$) between $\nu_{0,e}$ and $\kappa$. A weighted average gives the Doppler-averaged resonance frequency $\nu_{0,a}$ (orange shading, showing the region within $\nu_{0,a} \pm 1\sigma$). Here, the Doppler slope is in good agreement with zero and the values of $\nu_{0,e}$ of $\nu_{0,a}$ agree within their $1\sigma$ uncertainties. (**B**) The FWHM linewidth $\Gamma_F$ decreases with decreasing $\bar{v}$, as the Doppler broadening gets smaller and smaller. For the slowest delays, $\Gamma_F$ approaches the natural linewidth $\Gamma$ (dashed line).

16 FWHM linewidths $\Gamma_F$ of the line scan from Fig. 5.1 are shown in Fig. 5.3. The delay dependence of $\nu_0$ is of particular importance to this experiment, as it corresponds to a velocity dependence through the different mean speeds $\bar{v}$ of the delays (see Table 5.1). As discussed in Section 2.2.3 and Section 2.2.4, any residual, i.e., not removed by the Doppler suppression scheme, first-order Doppler shift will manifest itself as a linear dependence of $\nu_0$ on $\bar{v}$. That is, the velocity dependence of $\nu_0$ can be modeled as

$$\nu_0(\bar{v}) = \nu_{0,e} + \kappa\bar{v}, \tag{5.8}$$

where $\nu_{0,e}$ is the Doppler-free resonance frequency, i.e., the value of $\nu_0$ extrapolated to zero speed, and $\kappa$ is the (residual) Doppler slope. Fitting Eq. (5.8) to the data, referred to as Doppler extrapolation and equivalent to a linear regression [159], then results in estimates of $\nu_{0,e}$ and $\kappa$, their uncertainties as propagated from the uncertainties $\sigma_{\nu_0}$ in $\nu_0$, and the linear correlation coefficient $r_e$ between $\nu_{0,e}$ and $\kappa$. Note that no uncertainty[1] is assumed for the values of $\bar{v}$, which are derived from simulations. Instead, the underlying simulation parameters are varied, resulting in different values of $\bar{v}$, and the Doppler extrapolation is repeated. The result of such an Doppler extrapolation is shown (green line and shading) in Fig. 5.3 (A). The correlation coefficient $r_\mathrm{e}$ is completely determined by the behavior of the uncertainties $\sigma_{\nu_0}$ over the delays and the values of $\bar{v}$, independent of $\nu_0$ and the absolute values of $\sigma_{\nu_0}$. For the 2S-6P measurement, $r_\mathrm{e} = -0.95 \ldots -0.97$, and thus it is highly probable that random noise leading to a higher-than-average value of $\nu_{0,\mathrm{e}}$ will lead to a lower-than-average value of $\kappa$.

[1]Fitting a straight line with uncertainties in both coordinates is somewhat surprisingly a considerable statistical and numerical problem, see, e.g., Section 15.3 of [159].

Some insight might also be gained by an average over $\nu_0$, weighted by $\sigma_{\nu_0}$, giving the Doppler-averaged resonance frequency $\nu_{0,\mathrm{a}}$. $\nu_{0,\mathrm{a}}$ has a three to four times lower statistical uncertainty than $\nu_{0,\mathrm{e}}$, but is not free from the Doppler shift. However, while $\nu_{0,\mathrm{a}}$ thus cannot be used to determine the transition frequencies, it can be used to compare the resonance frequencies of different experimental configurations for which the Doppler shift can be assumed to be identical or to have averaged to a low enough level. This was used in the 2S-4P measurement (see Appendix A) to show the influence of quantum interference effects on $\nu_{0,\mathrm{a}}$, which otherwise would have not been visible for the larger uncertainty of $\nu_{0,\mathrm{e}}$. The Doppler average is also shown (orange shading) in Fig. 5.3 (A).

In the analysis of the 2S-6P measurement, the Doppler-free resonance frequencies $\nu_{0,\mathrm{e}}$ are used to determine the transition frequency. However, small simulation corrections as described in the following sections need to be taken into account. These corrections can also change with $\bar{v}$, as, e.g., atoms will spend more or less time interacting with the laser beams. To prevent that the corrections mimic a first-order Doppler shift and thus compromise the Doppler extrapolation, the simulation corrections are applied to $\nu_0$ of each delay before the extrapolation and averaging is done. The extrapolation and averaging is performed for each of the $M$ line scans, thus resulting in $M$ values for $\nu_{0,\mathrm{e}}$, $\kappa$, and $\nu_{0,\mathrm{a}}$. In general, the residual Doppler slope is not expected to be constant during the measurement, as the apparatus is repeatedly realigned. Thus, one might expect the values of $\kappa$ and $\nu_{0,\mathrm{a}}$ to show some excess scatter beyond what is expected from their uncertainties, while the Doppler-free values of $\nu_{0,\mathrm{e}}$ should be free from this scatter. In the experimental data, indeed a significant excess scatter is observed for $\kappa$ and $\nu_{0,\mathrm{a}}$. $\nu_{0,\mathrm{e}}$, however, also shows some, albeit lower, but significant, excess scatter. This is attributed to the fact that the uncertainties so far are all based on the assumption of pure shot noise on the observed signal, while technical noise has been neglected. Thus, this assumption is not entirely justified and the excess scatter needs to be taken into account, which is here done through an appropriate scaling of the uncertainties.

It should also be noted that this excess scatter is not clearly observed on the $\chi^2_\mathrm{red}$ distribution of the Doppler extrapolation. A possible explanation, as explored by Monte Carlo simulations, could be that technical noise, e.g., fluctuations of the nozzle temperature, is highly correlated between the delays, leading to an excess shift $\nu_0$ of the same sign and of similar size for all delays. Such a shift would then not be visible as excess scatter within a single Doppler extrapolation, but would show up as excess scatter between different Doppler extrapolations. The correlation between the delays comes about as the delays are separated by at most hundreds of µs, while the signal at each frequency point is accumulated for 1 s. Thus, technical noise on a time scale longer than hundreds of µs affects all delays in a similar way, which is the time scale expected for, e.g., temperature and atomic flux fluctuations. Further investigations into the influence of such fluctuations are in progress, but beyond the scope of this work.

The other fit parameters apart from $\nu_0$ (amplitude $A$, linewidths $\Gamma_\mathrm{L}$, $\Gamma_\mathrm{G}$, $\Gamma_\mathrm{F}$, offset $y_0$, doublet frequency splitting $\Delta\nu_0$) vary significantly over the delays and neither an average nor an extrapolation over the delays is instructive. The behavior of the FWHM linewidth $\Gamma_\mathrm{F}$ is shown as an example in Fig. 5.3 (B). Thus, the analysis has to take into account a large range of fit parameters, complicating the interpretation and visualization of the data. On the other hand, for some fit parameter combinations can be found that do not vary substantially over the delays, which here are the relative offset or background-to-amplitude ratio, $y_0/A$, and, for the Voigt doublet, the amplitude ratio $A_1/A_2$.

### 5.1.3 Frequency sampling of resonances

Table 5.2: The 15 unique absolute values of the frequency detuning $\Delta\nu_{2S\text{-}6P}$ used to sample the 2S-6P resonances during a line scan. The resonance is sampled symmetrically about zero detuning, including twice at zero detuning, resulting in a total of 30 frequency detuning points. Different frequency detunings are used depending on the offset angle $\alpha_0$ to account for the different line shapes. See Section 5.1.3 for details.

| | $\lvert\Delta\nu_{2S\text{-}6P}\rvert$ (MHz) | | |
|---|---|---|---|
| $\alpha_0$(mrad) | 0...6 | 6...10 | > 10 |
| | 0.0000 | 0.0000 | 0.0000 |
| | 1.2546 | 1.0702 | 2.2173 |
| | 1.8268 | 2.1403 | 3.2279 |
| | 2.3104 | 3.1276 | 4.0820 |
| | 2.7650 | 3.9738 | 4.8856 |
| | 3.2190 | 4.7845 | 5.6899 |
| | 3.6946 | 5.6168 | 6.5368 |
| | 4.2171 | 6.5261 | 7.4773 |
| | 4.8251 | 7.5970 | 8.5973 |
| | 5.5937 | 9.0174 | 10.0931 |
| | 6.7157 | 11.4855 | 12.6965 |
| | 8.9994 | 15.7649 | 16.8413 |
| | 22.6662 | 24.3237 | 25.1310 |
| | 36.3331 | 37.1618 | 37.5655 |
| | 50.0000 | 50.0000 | 50.0000 |

In order to gather enough statistics to accurately determine the 2S-6P transition frequency, the line scans over the resonance need to be repeated many times. It is thus crucial to select the frequency detuning $\Delta\nu_{2S\text{-}6P}$ of the $N_\Delta$ points at which the resonances are sampled such that the statistical uncertainty is as low as possible, while also sampling all parts of the experimental line shape. In other words, the frequency sampling should be chosen such that fitting the line shape function to the data results in an estimate of the true resonance frequency that is consistent, free from bias, efficient, and robust [160]. For a single resonance, these requirements are approximately fulfilled by placing most of the points on the slope of the resonance while also placing a few outer points at a large detuning and $L$ points at zero detuning. The frequency sampling is always symmetric about zero and thus $(N_\Delta - L)/2$ unique points with nonzero frequency detuning need to be found. To find the slope points, the derivative of the expected line shape with respect to frequency can be used as a guide to the local density of points. The problem of choosing the points is however complicated by the fact that all delays necessarily are subject to the same frequency sampling, while the linewidth and line shape changes greatly, as shown in Fig. 5.1. Additionally, in the end what should be minimized is the statistical uncertainty of the Doppler-free resonance frequency $\nu_{0,e}$, as determined by the Doppler extrapolation detailed in the previous chapter. Then, it might be advantageous to chose the points such that the frequency sampling of the slower delays is optimized at the cost of that of the faster delays to compensate for the decrease in signal.

To find a suitable frequency sampling, a Monte Carlo simulation using the approximate line shapes and amplitudes observed for the different delays in the experiment was carried out. A Voigt or Voigt doublet line shape with variable line width was used as a template to find the frequency detuning points. Then, the statistical uncertainty of $\nu_{0,e}$ was minimized by varying the template line width and the placement of the additional outer points. The number of points at zero detuning was chosen to be two to determine the drift in amplitude during the line scan. The total number of points, 30, and the integration time at each point, 1 s, is a compromise between, on the one hand, a comprehensive sampling of the line shape and a high enough number of counts to approximate the shot noise on the counts with a Gaussian distribution, and, on the other hand, a short enough measurement time per line scan to allow for many line scans during a freezing cycle. The latter helps in estimating the excess scatter seen in the experiment, which is mainly visible by comparing different line scans.

The 15 unique absolute values $|\Delta\nu_{\text{2S-6P}}|$ of frequency detuning points, including zero detuning, found in this way are shown in Table 5.2. To account for the different line shapes observed when using different offset angles $\alpha_0$ (see Fig. 5.1 and Fig. 5.2), three sets of $|\Delta\nu_{\text{2S-6P}}|$ are used for values of $\alpha_0$ close to 0 mrad, 8 mrad, and 12 mrad. The listed detunings were used to acquire all spectroscopy data except for some line scans from the data groups G1A, G7A, and G14, where slightly different detunings were used initially.

To suppress the influence of drifts in the signal during the line scan, caused by, e.g., drifts in atomic flux, the order in which the resonances are sampled is randomly picked. This is done again in a symmetric fashion, i.e., a detuning and a random sign is picked and data are acquired. Then, the sign of the detuning is flipped and again data are acquired, before moving on to the next randomly determined detuning. This randomization is done anew for each line scan. The points at zero detuning are exempt from this procedure, with data at zero detuning acquired at the start and at the end of a line scan. In this way, the changes in amplitude during the line scan can be determined and, assuming a linear drift, corrected for. This has not been done for the data presented in this thesis, but will be included in the final analysis.

By symmetry, for a perfectly symmetric line shape and a symmetric sampling about the usually a priori unknown line center, the resonance frequency determined through sampling is independent of the choice of frequency detunings. However, neither is the experimental line shape perfectly symmetric nor is the sampling symmetric about the line center, which additionally is not uniquely defined for an asymmetric line shape. Then, the choice of frequency detunings can influence the determined resonance frequency. The resulting sampling bias is here partly taken into account by also sampling the simulated line shapes at the same detunings as the experimental data, using the true resonance frequency, known in the simulations, as zero detuning. By correcting the experimental data with the simulations, the sampling bias is then also partially accounted for. To also take into account that zero detuning in the experiment might not correspond to zero detuning in the simulations, one can estimate this offset by comparing the frequency used for zero detuning for each line scan with the transition frequency determined in the measurement. Then, the simulations can be analyzed separately for each line scan using this offset. This was done for the 2S-4P measurement (see

Table 5.3: Range of parameters used for the Monte Carlo simulation of the trajectories of metastable 2S atoms. Each trajectory set uses a single value for each parameter, adjusted to the experimental conditions the trajectory set is describing.

| Parameter | Value |
|---|---|
| Geometry | |
| Nozzle orifice radius $r_1$ | 1.0 mm |
| Distance $L_1$ from nozzle orifice to variable aperture | 153.6 mm |
| Width $d_2 = 2r_2$ of variable aperture along $x$-axis | 1.2 mm |
| Height $d_{2,y} = 2r_{2,y}$ of variable aperture along $y$-axis | 2.0 mm |
| Distance $L$ from nozzle orifice to 2S-6P spectroscopy laser | 204.0 mm |
| Velocity distribution of ground state (1S) atoms | |
| Speed distribution $p(v)\,\mathrm{d}v$ | $\propto v^3 e^{-\frac{m_\mathrm{H} v^2}{2k_\mathrm{B} T_\mathrm{N}}} e^{-\frac{v_\mathrm{cutoff}}{v}}\,\mathrm{d}v$ |
| Angular distribution $p(\theta)\,\mathrm{d}\Omega$ | $\propto \cos(\theta)\,\mathrm{d}\Omega$ |
| Nozzle temperature $T_\mathrm{N}$ | 4.8 K |
| Cutoff speed $v_\mathrm{cutoff}$ | 30 m/s. . . 65 m/s |
| 1S-2S preparation laser | |
| Beam waist radius $w_\mathrm{1S\text{-}2S}$ ($1/e^2$ intensity radius) | 297 µm |
| Distance of waist to nozzle orifice | 56.2 mm |
| Intracavity power $P_\mathrm{1S\text{-}2S}$ (per direction) | 1.00 W. . . 1.50 W |
| (Atomic) detuning from $1\mathrm{S}_{1/2}^{F=0} - 2\mathrm{S}_{1/2}^{F=0}$ resonance $\Delta\nu_\mathrm{1S\text{-}2S}$ | 740 Hz. . . 1380 Hz |
| Chopper frequency $f_\mathrm{chop}$ (equal slit width) | 160 Hz |

Appendix A), where a large sampling bias on the order of the final uncertainty was found. For the 2S-6P measurement, care was taken to use a small offset on the order of 10 kHz, which should reduce the sampling bias well below the level seen in the 2S-4P measurement. However, for the final analysis this needs to be confirmed with the aforementioned procedure.

## 5.2 Modeling of the atomic beam of metastable hydrogen

The trajectory that the atoms take through the 2S-6P spectroscopy laser substantially influences both the observed line shape and systematic shifts of the line center, especially through the light force shift. Thus, an accurate model of the trajectories of metastable (2S) atoms is needed. The model used here is based on a Monte Carlo approach: first, a geometric trajectory is randomly picked, consisting of the origin and velocity of a ground state (1S) atom as it emerges from the nozzle into vacuum. Second, the probability to find this atom in the 2S level at the spectroscopy region and for a given delay time $\tau$ is determined. To this end, the interaction with the 1S-2S preparation laser beam as the atom flies from the nozzle to the spectroscopy region is simulated by numerically solving the appropriate optical Bloch equations (OBEs). This procedure is then repeated for $N_\mathrm{traj}$ trajectories. The range of parameters for this Monte Carlo simulation is listed in Table 5.3. A collection of $N_\mathrm{traj}$ for a given set of parameters is here referred to as a trajectory set.

### 5.2.1 Monte Carlo simulation of 2S trajectories

The speed distribution $p(v)\,\mathrm{d}v$ and angular distribution $p(\theta)\,\mathrm{d}\Omega$ of the atoms emerging from the nozzle have been discussed in Section 4.5.2.2. The atoms are assumed to originate from the nozzle orifice of radius $r_1$, and the speed and angular distribution are assumed to be identical over the orifice. For each trajectory, a random position within this orifice, a random speed $v$ according to $p(v)\,\mathrm{d}v$, and a random $\theta$ and $\phi$ according to $p(\theta)\,\mathrm{d}\Omega$ are picked. $\theta$ and $\phi$ are the polar and azimuthal angles with respect to the normal through the center of the orifice and parallel to the $z$-axis, and $\mathrm{d}\Omega = 2\pi \sin\theta\,\mathrm{d}\theta$ is the corresponding surface element. If the resulting trajectory passes through the variable aperture of width (height) $d_2 = 2r_2$ ($d_{2,y} = 2r_{2,y}$) along the $x$-axis ($y$-axis) and located at a distance $L_1$ along the $z$-axis from the orifice, the trajectory is kept. Otherwise, new trajectories are drawn till this condition is satisfied. For the parameters of Table 5.3, a fraction of $P_{\mathrm{geo}} = 3.2 \times 10^{-5}$ of the trajectories emerging from the nozzle in the direction of the 2S-6P spectroscopy region pass through the variable aperture into this region[1].

The dynamics of the 1S-2S excitation through two-photon absorption is discussed in Section 2.2.6. To model the excitation for each trajectory, here the OBEs given in Eqs. (10a–c) of [58] are used, taking into account the second-order Doppler shift, the ac-Stark shift, and photoionization of the 2S level. For the numerical integration of the OBEs, integration delays of constant length $\Delta\tau_{\mathrm{int}} < \min \Delta\tau_i$ are used, such that for each (experimental) delay $i$ of length $\Delta\tau_i$ there is an integer number $M_{\mathrm{int},i}$ of integration delays. Here, $\Delta\tau_{\mathrm{int}} = 10\,\mu\mathrm{s}$ is used, such that $M_{\mathrm{int},1} = 5$ and $M_{\mathrm{int},16} = 55$. The minimum and maximum delay time $\tau$ of these integration delays covers the full range of the experimental delays, resulting here in 255 integration delays. For each trajectory $k$, a random delay time $\tau_j$ is picked within each integration delay. The trajectory is then treated as if it had exited the nozzle orifice at a time such that it arrives at the center of the 2S-6P spectroscopy laser beams a time $\tau_j$ after the 1S-2S preparation laser has been switched off. Thus, $L/v_{z,k} - \tau_j$, where $v_{z,k}$ is the trajectory's longitudinal velocity and $L$ is the distance from the nozzle orifice to the spectroscopy laser beams, determines whether and for how long the trajectory can interact with the preparation laser. If $L/v_{z,k} - \tau_j \leq 0$, there is no interaction with the laser, i.e., the interaction time is $T_{\text{1S-2S},j,k} = 0$. For $0 < L/v_{z,k} - \tau_j \leq (1/f_{\mathrm{chop}})/2$, the trajectory starts interacting with the laser as it leaves the nozzle for a total interaction time of $T_{\text{1S-2S},j,k} = L/v_{z,k} - \tau_j$. The same interaction time results for $(1/f_{\mathrm{chop}})/2 < L/v_{z,k} - \tau_j \leq 1/f_{\mathrm{chop}}$, as the preparation laser is switched off by the optical chopper during $\tau_j - (1/f_{\mathrm{chop}}) \ldots \tau_j - (1/f_{\mathrm{chop}})/2$, but the position at which the trajectory starts interacting with the laser shifts from the nozzle towards the spectroscopy region. Finally, for $L/v_{z,k} - \tau_j > 1/f_{\mathrm{chop}}$, the trajectory interacts with more than one chopper cycle. For the parameters used here, this can only occur for trajectories with $v_{z,k} < 56\,\mathrm{m/s}$, and as an approximation only the first chopper cycle is taken into account. Furthermore, it is assumed here that the intracavity power of the preparation laser, as seen by the atoms, is $P_{\text{1S-2S}}$ for a duration of $1/2f_{\mathrm{chop}}$ when switched on by the optical chopper, and zero otherwise. As discussed in Section 4.3.3.5, due to the transient behavior of the cavity stabilization, this is not quite the case in the experiment and thus constitutes a further approximation. While beyond the scope of this thesis, in principle the measured time-resolved behavior of the intracavity power could be used in the simulations instead of this approximation.

[1] A fraction of $3 \times 10^{-4}$ of the trajectories pass through the upstream high-vacuum entrance aperture. Note that the nozzle has two orifices, and only half of the particles leaving the nozzle are emitted towards the 2S-6P spectroscopy region, while the other half hits the incoupling mirror of the 243 nm enhancement cavity.

Table 5.4: Average 2S excitation probabilities and related quantities as determined from the Monte Carlo simulation of 2S trajectories for the experimental delays as defined in Table 5.1 and the simulation parameters given in Table 5.3. See text for details.

| Delay $i$ | $P_{2S,i}$ | $\eta_{2S,i}$ | $\tilde{P}_{2S,i}$ | $\bar{T}_{1S\text{-}2S,i}$ (µs) | $N_{2S,i}/N'_{1S}$ (s) | $P_{2S,dcy}$ |
|---|---|---|---|---|---|---|
| 1 | $2.2 \times 10^{-2}$ | 1 | $2.2 \times 10^{-2}$ | 580 | $5.8 \times 10^{-9}$ | $1.6 \times 10^{-3}$ |
| 2 | $2.2 \times 10^{-2}$ | 1 | $2.2 \times 10^{-2}$ | 530 | $5.7 \times 10^{-9}$ | $1.6 \times 10^{-3}$ |
| 3 | $2.1 \times 10^{-2}$ | 1 | $2.1 \times 10^{-2}$ | 480 | $5.5 \times 10^{-9}$ | $1.6 \times 10^{-3}$ |
| 4 | $2.0 \times 10^{-2}$ | 1 | $2.0 \times 10^{-2}$ | 430 | $5.2 \times 10^{-9}$ | $1.6 \times 10^{-3}$ |
| 5 | $1.8 \times 10^{-2}$ | 1.0 | $1.8 \times 10^{-2}$ | 381 | $4.7 \times 10^{-9}$ | $1.7 \times 10^{-3}$ |
| 6 | $1.6 \times 10^{-2}$ | $9.9 \times 10^{-1}$ | $1.6 \times 10^{-2}$ | 335 | $4.2 \times 10^{-9}$ | $1.7 \times 10^{-3}$ |
| 7 | $1.4 \times 10^{-2}$ | $9.7 \times 10^{-1}$ | $1.4 \times 10^{-2}$ | 300 | $3.5 \times 10^{-9}$ | $1.8 \times 10^{-3}$ |
| 8 | $1.3 \times 10^{-2}$ | $9.0 \times 10^{-1}$ | $1.1 \times 10^{-2}$ | 277 | $2.9 \times 10^{-9}$ | $1.9 \times 10^{-3}$ |
| 9 | $1.0 \times 10^{-2}$ | $7.9 \times 10^{-1}$ | $8.2 \times 10^{-3}$ | 260 | $4.2 \times 10^{-9}$ | $2.1 \times 10^{-3}$ |
| 10 | $9.2 \times 10^{-3}$ | $5.6 \times 10^{-1}$ | $5.2 \times 10^{-3}$ | 258 | $2.7 \times 10^{-9}$ | $2.3 \times 10^{-3}$ |
| 11 | $8.9 \times 10^{-3}$ | $3.8 \times 10^{-1}$ | $3.3 \times 10^{-3}$ | 267 | $1.7 \times 10^{-9}$ | $2.6 \times 10^{-3}$ |
| 12 | $7.3 \times 10^{-3}$ | $2.5 \times 10^{-1}$ | $1.8 \times 10^{-3}$ | 289 | $1.9 \times 10^{-9}$ | $3.0 \times 10^{-3}$ |
| 13 | $6.5 \times 10^{-3}$ | $1.1 \times 10^{-1}$ | $7.2 \times 10^{-4}$ | 336 | $1.1 \times 10^{-9}$ | $3.6 \times 10^{-3}$ |
| 14 | $7.1 \times 10^{-3}$ | $3.8 \times 10^{-2}$ | $2.7 \times 10^{-4}$ | 396 | $4.2 \times 10^{-10}$ | $4.4 \times 10^{-3}$ |
| 15 | $6.1 \times 10^{-3}$ | $1.5 \times 10^{-2}$ | $9.5 \times 10^{-5}$ | 473 | $2.5 \times 10^{-10}$ | $5.4 \times 10^{-3}$ |
| 16 | $6.6 \times 10^{-3}$ | $4.4 \times 10^{-3}$ | $2.9 \times 10^{-5}$ | 569 | $8.4 \times 10^{-11}$ | $6.7 \times 10^{-3}$ |

For each trajectory and integration delay, the numerical integration of the OBEs is performed for the given interaction time $T_{1S\text{-}2S,j,k}$ and the given position at which the interaction starts[1]. The natural decay of the 2S population during the time $\tau_j$ after the preparation laser has been switched off but before the trajectory reaches the spectroscopy laser beams is also taken into account. This computation then gives the 2S excitation probability $P_{2S,int,j,l}$ of the trajectory, i.e., the probability of an atom described by this trajectory and for the delay time $\tau_j$ to be found in the metastable 2S level at the spectroscopy laser beams. $P_{2S,int,j,l}$ is set to zero for $T_{1S\text{-}2S,j,k} = 0$. The average effective 2S excitation probability $\tilde{P}_{2S,i}$ for the experimental delay $i$ is then found by the average of $P_{2S,int,j,k}$ over the $N_{traj}$ trajectories and the $M_{int,i}$ integration delays,

$$\tilde{P}_{2S,i} = \frac{1}{N_{traj} M_{int,i}} \sum_{j,k} P_{2S,int,j,k}. \tag{5.9}$$

Importantly, this procedure implies that every run of the simulation uses the same set of trajectories for all experimental delays. It is also instructive to find the average 2S excitation probability $P_{2S,i}$ and the average interaction time $\bar{T}_{1S\text{-}2S,i}$ of only those trajectories that have interacted with the laser, i.e., for which $T_{1S\text{-}2S,j,k} > 0$, as $P_{2S,i}$ is a measure of the excitation dynamics. Likewise, the fraction $\eta_{2S,i}$ of trajectories that have interacted with the laser is of interest, with $\tilde{P}_{2S,i} = \eta_{2S,i} P_{2S,i}$. The values of $P_{2S,i}$, $\eta_{2S,i}$, $\tilde{P}_{2S,i}$, and $\bar{T}_{1S\text{-}2S,i}$ for the preliminary analysis of the 2S-6P measurement are given in Table 5.4.

[1] If there are multiple integration delays for which the interaction starts at the nozzle, the computation can be reduced to a numerical integration with a single start position, but multiple end position, or, equivalently, interaction times. This speeds up the computation substantially, as many trajectories fall into this category.

Table 5.5: Speed distribution of the delays given in Table 5.1. The maximum speed $v_{\text{max}}$ of atoms that can contribute to the signal is determined by the delay time $\tau$ through $v_{\text{max}} \approx L/\tau$, where $L$ is the distance from the nozzle orifice to the 2S-6P spectroscopy laser beams. The mean speed of these atoms is different if no excitation is taken into account, giving the 1S distribution with mean speed $\bar{v}(1S)$, or if the weighting of the 1S-2S excitation or both the 1S-2S and 2S-6P excitation is taken into account, giving the 2S distribution with mean speed $\bar{v}(2S)$ and the signal distribution with mean speed $\bar{v}$, respectively. The latter distribution is relevant for the evaluation of systematic effects such as the Doppler shifts, with $\bar{v}$ also given Table 5.1. Likewise, the FWHM of the 1S and 2S distribution along the $x$- ($\Delta v_x$) and $y$-axis ($\Delta v_y$) is given. The number in parentheses is the standard deviation over multiple simulations covering the parameters given in Table 5.3 and the parameters of data groups G1–G12 as given in Table 6.2.

| Delay | $v_{\text{max}}$ (m/s) | $\bar{v}(1S)$ (m/s) | $\bar{v}(2S)$ (m/s) | $\bar{v}$ (m/s) | $\Delta v_x(1S)$ (m/s) | $\Delta v_x(2S)$ (m/s) | $\Delta v_y(1S)$ (m/s) | $\Delta v_y(2S)$ (m/s) |
|---|---|---|---|---|---|---|---|---|
| 1 | >1000 | 383.7(2.3) | 313.3(6.2) | 256.4(6.4) | 3.25(3) | 3.36(11) | 3.98(4) | 4.56(20) |
| 2 | >1000 | 383.7(2.3) | 312.4(6.2) | 255.0(6.3) | 3.25(3) | 3.36(11) | 3.98(4) | 4.60(20) |
| 3 | >1000 | 383.7(2.3) | 310.2(6.0) | 253.0(6.2) | 3.25(3) | 3.33(11) | 3.98(4) | 4.62(20) |
| 4 | >1000 | 383.7(2.3) | 305.2(5.8) | 250.0(6.1) | 3.25(3) | 3.27(10) | 3.98(4) | 4.61(20) |
| 5 | 971 | 383.7(2.3) | 296.3(5.3) | 245.5(5.8) | 3.25(3) | 3.17(10) | 3.98(4) | 4.55(20) |
| 6 | 785 | 381.4(2.3) | 283.5(4.9) | 238.6(5.4) | 3.25(3) | 3.02(8) | 3.97(4) | 4.44(18) |
| 7 | 658 | 371.9(2.1) | 268.2(4.4) | 229.8(4.9) | 3.23(3) | 2.82(7) | 3.95(4) | 4.25(16) |
| 8 | 567 | 354.4(1.9) | 252.1(4.0) | 219.8(4.5) | 3.16(3) | 2.61(6) | 3.89(4) | 4.01(14) |
| 9 | 498 | 332.6(1.7) | 229.9(3.5) | 204.4(3.9) | 3.05(2) | 2.33(4) | 3.76(3) | 3.65(11) |
| 10 | 400 | 288.5(1.3) | 203.0(2.9) | 184.3(3.2) | 2.73(2) | 2.01(3) | 3.39(3) | 3.21(8) |
| 11 | 334 | 251.0(1.2) | 180.6(2.5) | 166.5(2.7) | 2.41(2) | 1.76(3) | 3.00(2) | 2.84(6) |
| 12 | 287 | 220.7(1.1) | 156.5(2.0) | 146.1(2.2) | 2.14(2) | 1.50(2) | 2.66(2) | 2.43(4) |
| 13 | 224 | 176.8(0.9) | 127.5(1.6) | 120.8(1.8) | 1.72(1) | 1.20(2) | 2.15(2) | 1.96(3) |
| 14 | 169 | 135.5(0.8) | 103.0(1.1) | 99.0(1.3) | 1.33(1) | 0.94(2) | 1.66(2) | 1.56(2) |
| 15 | 135 | 109.8(0.7) | 83.8(1.1) | 81.2(1.2) | 1.08(1) | 0.76(2) | 1.35(2) | 1.26(2) |
| 16 | 101 | 83.5(0.7) | 66.8(0.9) | 65.5(0.9) | 0.82(1) | 0.60(2) | 1.03(2) | 1.00(3) |

In the experiment, 160 chopper cycles are acquired for each data point. Thus, for each delay $i$, data are acquired for a total time of $160 \times \tau_i$. The number of 2S atoms $N_{2S,i}$ that fly through the beams of the 2S-6P spectroscopy laser during this time is then

$$N_{2S,i} = \tilde{P}_{2S,i} \times 160 \times \Delta\tau_i \times P_{\text{geo}} \times N'_{1S}, \tag{5.10}$$

where $N'_{1S}$ is the number of 1S atoms leaving the nozzle per second in the direction of the 2S-6P spectroscopy region (see Section 4.5.2.3). $N_{2S,i}/N'_{1S}$ is also given in Table 5.4. In order to show the simulation results in a way that highlights the limitations of the Monte Carlo procedure used, the number of simulated 2S atoms defined by

$$N_{\text{2S,sim,i}} = \tilde{P}_{2S,i} \times \frac{\Delta\tau_i}{50\,\mu\text{s}} \times N_{\text{traj}} \tag{5.11}$$

is used, which is proportional to $N_{2S,i}$.

### 5.2.2 Velocity distribution of 2S trajectories

The results of a Monte Carlo simulation of 2S trajectories using $N_{\text{traj}} = 1 \times 10^7$ trajectories and for typical parameters of the 1S-2S preparation laser are shown in Fig. 5.4. A typical cutoff speed of $v_{\mathrm{cutoff}} = 65\,\mathrm{m/s}$ is used to model the initial speed distribution, with the values of the remaining input parameters given in Table 5.3. The distribution of the transverse velocity $v_x$ of the 2S atoms along the 2S-6P spectroscopy laser beams (or $x$-axis) is of particular importance for the experiment, as it determines the Doppler broadening and strongly influences the light force shift. Fig. 5.4 (A) shows the number of simulated 2S atoms $N_{2S,sim}$ versus $v_x$ for different delays (solid lines). The underlying distribution of the same trajectories, i.e., trajectories that have interacted with the preparation laser, but not weighted with the 2S excitation probability, is also shown for reference (dashed lines). For simplicity, this distribution is referred to as the 1S distribution throughout this work. Note that the distribution of all atoms flying through the spectroscopy region, i.e., including trajectories that have not interacted with the laser, is constant in time, with only the 2S excitation probability of these atoms being modulated with the chopper cycles. As all atoms up to delay 4 interact with the preparation laser, i.e., $P_{2S} = 1$, the 1S distribution shown for delay 2 in Fig. 5.4 corresponds to the unweighted distribution of all atoms. The 1S distribution has been scaled to envelope the 2S distribution, revealing that the latter is flattened and narrowed from the initially approximately Gaussian distribution of the 1S atoms. This is attributed to the following mechanism: atoms with low transverse velocity, depending on their initial position, tend to either not be excited to the 2S level as they miss the preparation laser beam, or already start to be ionized as they spend most of the time inside the laser beam. Then, the excitation probability is effectively lower for these atoms than for atoms crossing the laser beams such that ionization is not yet limiting. This results in a narrower FWHM $\Delta v_x(2S)$ of the 2S distribution as compared to the FWHM $\Delta v_x(1S)$ of the 1S distribution, with these values also given in Table 5.5.

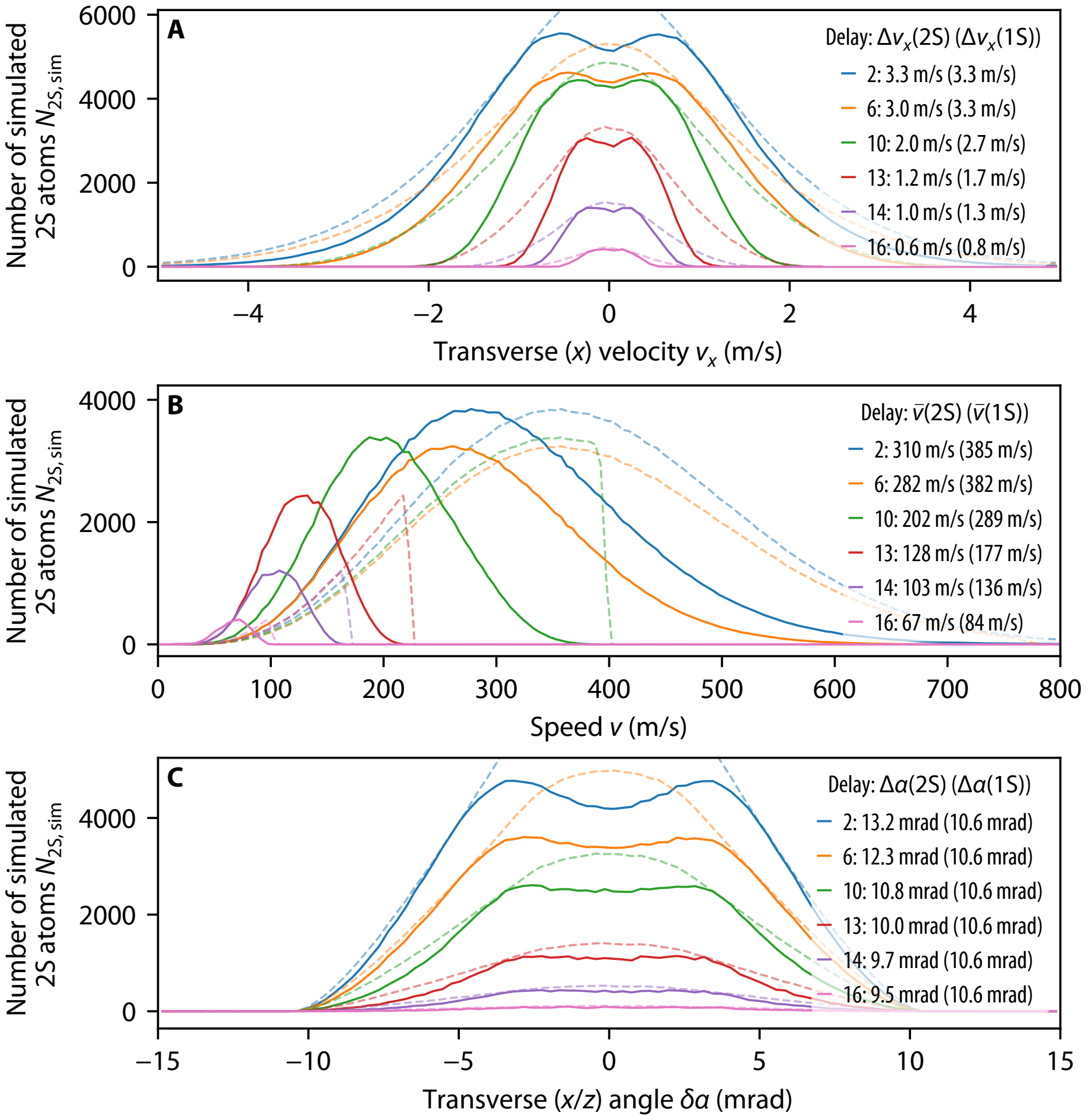

Figure 5.4: Monte Carlo simulation of the trajectories of metastable 2S atoms at the 2S-6P spectroscopy laser beams. The intracavity power and atomic detuning of the 1S-2S preparation laser are set to $P_{1S\text{-}2S} = 1.1\,\mathrm{W}$ and $\Delta\nu_{1S\text{-}2S} = 970\,\mathrm{Hz}$, respectively, and the speed distribution has a cutoff speed of $v_{\mathrm{cutoff}} = 65\,\mathrm{m/s}$. (**A**) Number of simulated 2S atoms $N_{2S,\mathrm{sim}}$ versus the transverse velocity $v_x$ along the 2S-6P spectroscopy laser beams for different delays (solid lines). The underlying distribution of 1S atoms is shown for reference (dashed lines), scaled to envelope the 2S distribution. The legend shows the FWHM $\Delta v_x$ of the 2S and 1S distributions. (**B**) Number of simulated 2S atoms $N_{2S,\mathrm{sim}}$ versus the speed $v$. The longer interaction time for slower atoms shifts the distribution to slower velocities as compared to the 1S distribution (dashed lines), shown normalized to the 2S distribution. The legend shows the mean speed $\bar{v}$ for the 2S and 1S distributions. (**C**) Number of simulated 2S atoms $N_{2S,\mathrm{sim}}$ versus the transverse angle $\delta\alpha$ from the $z-$ towards the $x$-axis (solid lines). The width of the 1S distribution (dashed lines), scaled to envelope the 2S distribution, is given by geometry and identical for all delays. The legend again shows the FWHM $\Delta\alpha$ of the 2S and 1S distributions. The simulation shown here contains $N_{\mathrm{traj}} = 1 \times 10^7$ random atomic trajectories. The remaining parameters of the simulation are listed in Table 5.3. See text for details.

The distribution for the transverse velocity $v_y$ along the $y$-axis, orthogonal to both the spectroscopy laser beams and the atomic beam axis, is not shown in Fig. 5.4, but the FWHM along this direction is given in Table 5.5. Geometrically, the divergence along this direction is expected to be larger by a factor of $(r_1 + r_{2,y})/(r_1 + r_2) = 1.25$, owing to the larger size of the variable aperture along the $y$-axis. The ratio $\Delta v_y(1S)/\Delta v_x(1S)$ indeed closely matches this expectation, while the ratio $\Delta v_y(2S)/\Delta v_x(2S)$ is somewhat larger for low delays and rises to almost 1.7 for high delays caused by the 2S excitation dynamics. This is an example of the complex interplay between geometric constraints and the 2S excitation, making the prediction of the beam properties without a trajectory simulation rather challenging[1].

The distribution of the speed $v$ is shown in Fig. 5.4 (B) for 2S atoms (solid lines) and 1S atoms (dashed lines), with the latter normalized to the 2S distribution. Since the atomic beam is well-collimated, $v_x, v_y \ll v$, it follows that $v_z \approx v$, where $v_z$ is the longitudinal velocity along the atomic beam ($z$-) axis. Importantly, $v$ determines the size of the first-order and second-order Doppler shifts. The maximum longitudinal velocity and thus speed of atoms interacting with the preparation laser is limited by the delay time $\tau$ to $v_{\max} = L/\tau$, substantially affecting the distribution starting with approx. delay 6, and visible as a sharp cut in the 1S distribution. The distribution of 2S atoms is shifted to slower velocities as compared to the 1S distribution, as slower atoms experience longer interaction times, leading to a larger excitation probability, as compared to faster atoms. In turn, the mean speed of 2S atoms, $\bar{v}(2S)$, is lower than that of the 1S atoms, $\bar{v}(1S)$. The values of $v_{\max}$, $\bar{v}(1S)$, and $\bar{v}(2S)$ are also given in Table 5.5. Note that $\bar{v}(1S)$ of delays $1 \ldots 6$ is slightly higher than the mean speed expected for an effusive beam (see Eq. (4.26)), which is caused by the effect of the cutoff speed on the speed distribution.

[1]Another example is that one might be tempted, in the expectation to reach a lower beam divergence, to decrease the beam waist radius of the preparation laser, allowing for a smaller width of the variable aperture. However, this was simulated to not necessarily lead to a lower divergence, as most of the low-divergence trajectories are ionized in the now higher peak intensity of the laser beam. Reducing the laser power, on the other hand, will lower the 2S excitation probability, and so on.

Lastly, Fig. 5.4 (C) shows $N_{2S,sim}$ versus the transverse angle $\delta\alpha$ from the $z-$ towards the $x$-axis (solid lines). The width of the 1S distribution (dashed lines), here shown scaled to envelope the 2S distribution, is given by geometry and identical for all delays with a FWHM of $\Delta\alpha(1S) = 10.6\,\text{mrad}$. However, the 2S distribution is modified by the aforementioned excitation dynamics, resulting in a narrower FWHM $\Delta\alpha(2\text{S})$ (see legend).

## 5.3 Modeling of the 2S-6P fluorescence signal

So far, the modeling of the atomic beam of metastable atoms as a set of trajectories has been discussed in Section 5.2. To complete the description of the experiment, the excitation of these 2S atoms to the 6P levels by the 2S-6P spectroscopy laser beams, and the resulting decay giving the fluorescence signal, has to be modeled. Two models are employed here: first, the big model, introduced in Section 2.3.2, which takes into account all atomic levels coupled through the spectroscopy laser or dipole-allowed decays, but does not include the exchange of momentum between the atoms and the laser beams. Second, the light force shift (LFS) model, introduced in Section 3.4, which uses a simplified three-level system, but explicitly includes the atom's momentum and the exchange of momentum.

### 5.3.1 Evaluating the big model and LFS model for the atomic beam

Both models consist of two sets of optical Bloch equations (OBEs), one for each of the two 2S-6P transitions. The frequency detuning $\Delta\nu_{\text{2S-6P}}$ of the spectroscopy laser is the frequency difference to the corresponding unperturbed transition frequency.

The OBEs are solved for one trajectory from the atomic beam model at a time. Each trajectory consists of an initial position $\boldsymbol{x_0}$, a velocity $\boldsymbol{v}$, and a set of excitation probabilities $\tilde{P}_{2S,i}$, with $i = 1 \ldots 16$, corresponding to the delays used in the experiment. Neglecting for now the momentum exchange with the laser beams, the atom's path through the apparatus and the laser beams versus time $t$ is a straight line given by $\boldsymbol{x}(t) = \boldsymbol{x_0} + \boldsymbol{v}t$.

The laser beams are assumed to have a Gaussian transverse intensity profile $I(y, z)$ with a $1/e^2$ intensity radius of $W_0$, with the peak intensity adjusted such that the power contained in each beam is $P_{\text{2S-6P}}$. The divergence of the laser beams along the $x$-axis is not taken into account, as the relative change of the intensity radius over the relevant region is less than $1 \times 10^{-7}$. For the 2S-6P measurement, $W_0 = 2.2\,\text{mm}$ is used. With this, the time-dependent intensity envelope $I(t) = I(y(t), z(t))$, i.e., not taking into account the standing-wave pattern of the intensity along the $x$-axis, as seen by the atom can be found. $I(t)$ in turn corresponds to a time-dependent Rabi frequency $\Omega_0(t) \propto \sqrt{I(t)}$ of the 2S-6P excitation (see Eqs. (2.30) and (2.31)). The integration boundaries of $t$ are chosen such that the integration starts and ends with the atom at $z - z' = -7.5\,\text{mm}$ and $z - z' = 7.5\,\text{mm}$, respectively, where $z$ is the coordinate along the atomic beam axis, and $z'$ is the position of the center of the spectroscopy laser beams.

For some experimental settings, an offset angle $\alpha_0$ between the atomic beam and the spectroscopy laser beams is applied. This angle is taken into account by rotating $\boldsymbol{v}$ about the $y$-axis by $\alpha_0$, with $\boldsymbol{x}(t)$ changing accordingly.

The big model does not depend explicitly on $\boldsymbol{x}(t)$, and its initial state is set such that all population is in the initial level $|\text{i}\rangle$ ($2\text{S}_{1/2}^{F=0}$, $m_F = 0$). The first-order Doppler shift is taken into account implicitly through the dependence of the (complex) Rabi frequencies $\Omega^{\pm}$ of the two counter-propagating beams on the position $x(t)$ of the atom along the beams (see Eqs. (2.32) to (2.34)). Note that through this, the standing-wave pattern formed by the beams, but not the diffraction on it, is included.

The LFS model explicitly includes the atom's momentum $p = m_\mathrm{H} v_x$ along the axis of the laser beams, and accordingly the initial state is set to $|\mathrm{i}\rangle|p\rangle_x$. This state corresponds to a completely delocalized atom, and therefore the initial position along the $x$-axis is of no consequence. Likewise, the Doppler shift and the standing-wave pattern are explicitly included through the antiparallel wave vectors of the counter-propagating beams. The momentum and position along the $y, z$-axes is only implicitly included through the intensity envelope $I(t)$, as done in the big model.

The OBEs of both models include signal equations that contain the expected fluorescence signal from each atom, i.e., the expected number of photons emitted by the decays included in the signal equation. For the big model, all Lyman decays for each spherical component $q$ are summed up with equal weights[1], resulting in three signals for the possible values of $q = -1, 0, 1$. These three signals are weighted with their detection efficiency, given by the simulations of the spatial detection efficiency for the two detectors (see Section 4.6.6), the angular distribution of the emitted photons, which depends on $q$ (see Eqs. (2.40) and (2.41)), and the linear laser polarization angle $\theta_\mathrm{L}$. $\theta_\mathrm{L}$ is the angle of the polarization vector of the 2S-6P spectroscopy laser beams, which lies in the $y$-$z$-plane, from the axis of the detector cylinder, oriented along the $y$-axis. It enters the detection efficiency of the signals as the photon angular distribution is given with respect to the polarization direction, while the spatial detection efficiency is given with respect to the axis of the detector cylinder. After weighting the three signals in this way, they are summed up, resulting in one signal for each detector and each value of $\theta_\mathrm{L}$.

For the LFS model, only the Ly-$\epsilon$ decays with $q = 0$ are used as signal, as there are no other Lyman decays in this model. The detection efficiency thus need not be included, and the signal is identical for both detectors and independent of $\theta_\mathrm{L}$.

The OBEs of both models are numerically integrated using a Runge-Kutta method[2]. This integration is repeated for the $N_\Delta$ values of the frequency detunings $\Delta\nu_\text{2S-6P}$ given in Table 5.2, depending on which experimental situation is simulated. To take into account the delay-dependent 1S-2S excitation, the signal is scaled with the excitation probabilities $\tilde{P}_{\mathrm{2S},i}$. In this way, the fluorescence signal from a single 2S trajectory for the $N_\text{dlys} = 16$ experimental delays and sampled at $N_\Delta$ detunings is found.

This procedure is repeated for all $N_\text{traj}$ trajectories in the given trajectory set. The signal for each delay and detuning is either summed up over all trajectories, or, as used below, for subsets of trajectories grouped by their speed $v = |\boldsymbol{v}|$. Thus, simulated line scans mimicking the experiment are found. The line scans are treated like their experimental counterparts, as detailed in Section 5.1. For clarity, the resonance frequencies determined from fits to the delays are identified by $\nu_0$, $\nu_\text{0,BM}$, and $\nu_\text{0,LFS}$ for experimental line scans, simulated line scans from the big model, and simulated line scans from the LFS model, respectively.

All in all, adding the 2S-6P excitation to a given set of 2S trajectories adds the fol-lowing parameters that should be matched to the experimental values: the probed transition ($2S\text{-}6P_{1/2}$ or $2S\text{-}6P_{3/2}$), the spectroscopy laser power $P_{\text{2S-6P}}$, the radius $W_0$ of the spectroscopy laser beams, the offset angle $\alpha_0$, and for the big model a specific detection efficiency simulation and the linear laser polarization angle $\theta_{\text{L}}$.

[1] This neglects the spectral sensitivity of the detection and overestimates the contribution from Ly-$\alpha$ decays relative to Ly-$\epsilon$ decays (see Table 4.1; other decays only contribute negligibly), which in turn can lead to an underestimation of quantum interference effects (see Section 2.3.2.3). In the final version of the analysis, the Lyman decays will be weighted with the spectral sensitivity of the fluorescence detection.

[2] The method Dopr853 of [159] is used, which is an eigth-order Dormand-Prince method. The absolute and relative tolerance are set to $1 \times 10^{-10}$ and $5 \times 10^{-11}$, respectively.

### 5.3.2 Speed distribution of atoms contributing to the fluorescence signal

The big model is used to determine the speed distribution of the atoms contributing to the 6P fluorescence signal. This 6P speed distribution is different from that of the underlying set of 2S trajectories, the 2S speed distribution, because the excitation probability to the 6P level, and thus the probability to emit a fluorescence photon, depends on the trajectory of the atom. This probability approximately scales, neglecting saturation effects, with the interaction time $T$ with the spectroscopy laser, $T \approx 2W_0/v$. Thus, the 6P speed distribution is shifted to lower speeds as compared to the 2S speed distribution.

To determine the 6P speed distribution for a given trajectory set, the trajectories are first grouped by their speed $v$ into 200 bins of width 5 m/s each, spanning $v = 0\,\text{m/s} \ldots 1000\,\text{m/s}$. Then, as described above, simulated line scans are calculated with the big model for each speed bin and fit with the appropriate line shape function. The 6P speed distribution is then given by $(v_j,\ A_{i,j})$, where $v_j$ is the mean speed of the $j^{\text{th}}$ speed bin and $A_{i,j}$ the line amplitude found by the fit to this bin and for delay $i = 1 \ldots 16$.

The mean speed $\bar{v}_i$ of the delay $i$ is found by the weighted average

$$\bar{v}_i = \frac{\sum_j A_{i,j} v_j}{\sum_j A_{i,j}}. \tag{5.12}$$

Likewise, the root mean square (RMS) speed $\bar{v}_{\text{RMS},i}$ is found by

$$\bar{v}_{\text{RMS},i} = \sqrt{\frac{\sum_j A_{i,j} v_j^2}{\sum_j A_{i,j}}}. \tag{5.13}$$

The full width at half maximum (FWHM) of the speed distribution, $\Delta v$, is determined by a Gaussian fit to $(v_j,\ A_{i,j})$ for each delay. The values of $\bar{v}$ and $\Delta v$ for the experimental parameters of the 2S-6P measurement are given in Table 5.1.

The line scans calculated for the speed bins are then summed up, giving line scans corresponding to the expected fluorescence signal from all atoms. Fits to these line scans then reveal the expected line amplitude $A_i$ for each delay $i$. The simulated line amplitudes cannot be directly compared to those seen in the experiment due to the large uncertainties such as the poorly known degree of dissociation of the atomic beam. However, the delay-dependence of the line amplitudes, i.e., $A_i/A_k$ as compared to a chosen reference delay $k$, can be compared, as factors common to all delays drop out. Here, $k = 2$ is used, and $A_i/A_2$ as found from simulations of the experiment is given in Table 5.1. The ratio of $A_i/A_2$ of the simulations to that of the experiment, i.e., $(A_i/A_2)_{\text{exp}}/(A_i/A_2)_{\text{sim}}$, is the experiment-to-simulation amplitude ratio. The cutoff speed $v_{\text{cutoff}}$ of the underlying 1S speed distribution is found by minimizing the experiment-to-simulation amplitude ratio. This corresponds to adjusting $v_{\text{cutoff}}$ such that the simulations best reproduce the delay-dependence of the line amplitudes observed in the experiment.

Finally, the ratio of $A_i$ to the number $N_{\text{2S,sim,i}}$ of simulated 2S atoms that contribute to delay $i$ is formed. This ratio corresponds to the average probability $p_{\text{6P,sig}}$ for an 2S atom to contribute to the fluorescence signal through excitation to the 6P level and subsequent emission of a Lyman photon. Table 5.1 gives $p_{6P,sig}$ separately for the three values of the spectroscopy laser power $P_{2S\text{-}6P}$ used in the 2S-6P measurement. As expected, $p_{6P,sig}$ increases with decreasing $\bar{v}$ and thus increasing interaction time $T$. It also approximately scales linearly with $P_{2S\text{-}6P}$, except for the slowest delays where saturation effects start to play a substantial role.

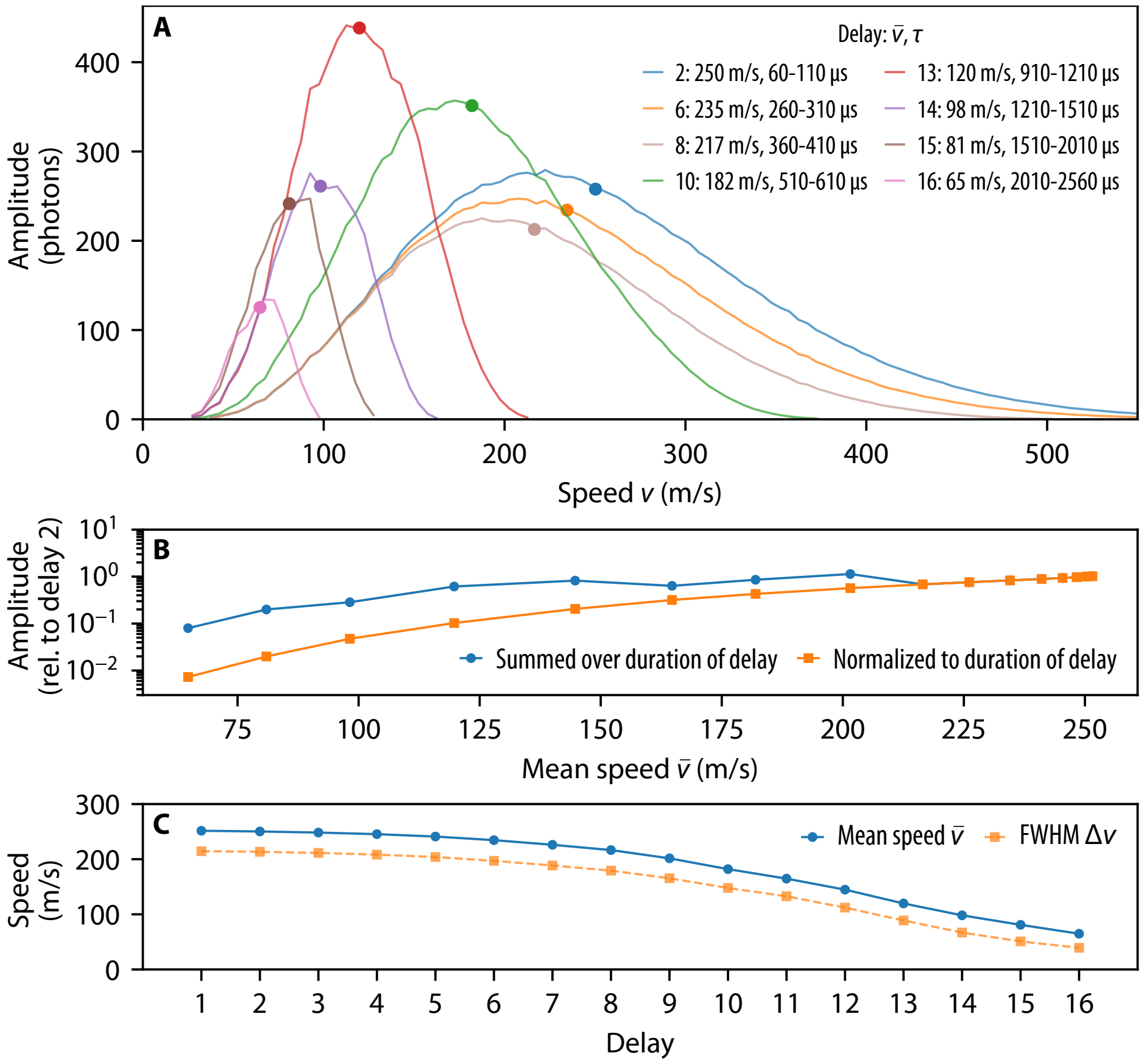

Figure 5.5: Simulation of the 6P speed distribution, i.e., the speed distribution of atoms contributing to the 6P fluorescence signal, based on the 2S trajectory set shown in Fig. 5.4 obtained from a Monte Carlo simulation. The trajectories are propagated through the 2S-6P spectroscopy laser using the big model to find the fluorescence signal, given by the expected number of photons emitted by Lyman decays versus detuning. The spectroscopy laser power is $P_{2S\text{-}6P} = 10\,\mu W$ and no offset angle $\alpha_0$ is applied. The trajectories are sorted by their speed $v$ into 200 bins of width 5 m/s each, and the resulting line scans are fit with a Voigt line shape for each bin and delay. (**A**) The number of fluorescence photons expected with the laser on the 2S-6P resonance, corresponding to the line amplitude $A$ found by the fits, versus the speed $v$. Selected delays are shown (colored lines). Within each delay, a mean atomic speed $\bar{v}$ (circles) is determined by an average of the binned speeds, using the amplitudes $A$ as weights. (**B**) The delay-dependence of the line amplitude, i.e., the line amplitude relative to that of delay 2, versus $\bar{v}$ (blue circles). The increasing duration of the delays for longer delay times partly compensates the drop in the number of emitted photons per unit time (orange squares). (**C**) The mean speed $\bar{v}$ (blue circles) and the FWHM $\Delta v$ (orange squares) of the 6P speed distribution for the 16 different delays.

Fig. 5.5 shows an example of such an analysis, using the 2S trajectory set of Fig. 5.4.

The spectroscopy laser power is set to $P_{\text{2S-6P}} = 10\,\mu\text{W}$ and no offset angle $\alpha_0$ is applied, and accordingly a Voigt line shape is used. The line amplitude $A_{i,j}$ versus the speed $v_j$ is shown for selected delays (colored lines) in Fig. 5.5 (A), with the resulting mean speed $\bar{v}_i$ marked (circles). Fig. 5.5 (B) shows the the delay-dependence of the line amplitude, $A_i/A_2$, versus $\bar{v}_i$ (blue circles). The same data, but normalized to the duration $\Delta\tau$ of the delays (orange squares), shows that the increasing duration of the delays for longer delay times partly compensates the drop in the number of emitted photons per unit time. Finally, Fig. 5.5 (C) shows the mean speed $\bar{v}$ (blue circles) and the FWHM $\Delta v$ (orange squares) of the 6P speed distribution for the 16 different delays.

### 5.3.3 Comparison of expected and experimentally observed fluorescence signal

Knowing $p_{\text{6P,sig}}$, the amplitude of the fluorescence signal expected in the experiment, i.e., the number of counts accumulated on each detector over 160 chopper cycles and with the spectroscopy laser resonant with the 2S-6P transition, can be estimated. For delay $i$, the number of counts is

$$N_{\text{6P},i} = N_{\text{2S},i} \times p_{\text{6P,sig},i} \times P_{\text{DE}}. \tag{5.14}$$

$N_{\text{2S},i}$ is the total number of 2S atoms that contribute to the signal, as given in Eq. (5.10). The ratio $N_{\text{2S},i}/N'_{\text{1S}}$ is given in Table 5.4 for the experimental parameters of the 2S-6P measurement, where $N'_{\text{1S}}$ is the number of 1S atoms leaving the nozzle per second in the direction of the 2S-6P spectroscopy region. $P_{\text{DE}}$ is the detection efficiency of fluorescence photons, which is derived from the simulations discussed in Section 4.6.6. Values of $P_{\text{DE}}$ for the top ($P_{\text{DE,TD}}$) and bottom ($P_{\text{DE,BD}}$) detector are given in Table 4.2 for the detection of Ly-$\epsilon$ photons.

Taking the corresponding values from Tables 4.2, 5.1 and 5.4 and using the estimate $N'_{\text{1S}} = 1.6 \times 10^{16}$ atoms/s of Section 4.5.2.3, the amplitude of the fluorescence signal of delay 13 for a spectroscopy laser power of $P_{\text{2S-6P}} = 10\,\mu\text{W}$ is expected to be $\approx$74 kcts. In the 2S-6P measurement, however, the maximum observed amplitude was only $\approx$4 kcts, that is, almost a factor of 20 lower than expected.

On the other hand, the simulations reproduce the observed linewidth $\Gamma_{\text{F}}$ for the different delays and spectroscopy laser powers within a few percent. Since $\Gamma_{\text{F}}$ is strongly affected by Doppler broadening and thus highly dependent on the distribution of 2S trajectories, this is a stringent test of the modeling of both the 2S trajectories and the 2S-6P fluorescence signal. The experimentally observed scaling of the signal with $P_{\text{2S-6P}}$ is also well reproduced by the simulations.

A more likely explanation for the discrepancy in the expected and observed signal is the overestimation of the degree of dissociation $\alpha_{\text{dis}}$, i.e., the fraction of hydrogen atoms as compared to hydrogen molecules in the beam. As discussed in Section 4.5.1, this value is poorly known, and it is possible that recombination losses, especially during the transport of the atoms from the dissociator to the nozzle, are much higher than estimated here. Further experimental studies, possibly including a direct measurement of $\alpha_{\text{dis}}$, are needed to resolve this discrepancy.

# Chapter 6

# 2S-6P transition frequency measurement

This chapter gives an overview over the preliminary results of the 2019 2S-6P transition frequency measurement. As the data analysis is still ongoing at the time of writing, all frequency results given are blinded by adding a random offset not known to the author. Furthermore, due to time constraints, and since the results are subject to change as the analysis progresses, the description is kept brief and should mainly serve as an outline of the data set.

## 6.1 Overview of acquired data and analysis procedure

### 6.1.1 2019 measurement runs

As described in the Introduction (see Chapter 1), in 2019 three measurement runs (A–C) to determine the 2S-6P transition frequency were conducted. Table 6.1 lists these runs and the corresponding experimental configurations, which are briefly discussed below. Each run consists of multiple measurement days, with each measurement day again separated into multiple freezing cycles (FCs). In total, 3155 line scans were acquired in 73 FCs.

For run A and B, the atomic beam and the 2S-6P spectroscopy laser beams were aligned to cross at right angles, i.e., $\alpha_0 = 0\,\mathrm{mrad}$, to minimize the observed linewidth, the residual Doppler shift, and the light force shift. During run C, on the other hand, the atomic beam and the laser beams were deliberately set to cross at a small offset angle $\alpha_0$ from the orthogonal. The primary purpose of this was to increase the size of the light force shift in order to check its theoretical modeling. Most of the data in run C were taken at $\alpha_0 = \pm 12.0\,\mathrm{mrad}$, but for some data $\alpha_0 = \pm 7.5\,\mathrm{mrad}$ and $\alpha_0 = 0\,\mathrm{mrad}$ were used.

Both the 2S-6P$_{1/2}$ and the 2S-6P$_{3/2}$ transitions were probed in run A and B, i.e., the transitions to the two fine-structure (FS) components 6P$_{1/2}$ and 6P$_{3/2}$, respectively. This allows the two respective transition frequencies $\nu_{1/2}$ and $\nu_{3/2}$ to be combined to form the 2S-6P centroid $\nu_{\text{2S-6P}}$, for which line shifts from quantum interference (QI) of the FS components are greatly suppressed (see Section 6.2.4.6). Run C, on the other hand, only contains data on the 2S-6P$_{1/2}$ transition. For each given FC, only one of the transitions was probed, with the number of FCs and valid line scans for each transition and run given in Table 6.1.

The power $P_{\text{2S-6P}}$ of the 2S-6P spectroscopy laser was set to, respectively, $30\,\mu\mathrm{W}$ and

Table 6.1: Overview of the three measurement runs A, B, and C during which the data of the 2S-6P transition frequency measurement was collected in 2019. $\alpha_0$: Offset of atomic–laser beam angle, $\theta_\mathrm{L}$: linear laser polarization angle, $P_\mathrm{1S\text{-}2S}$: intracavity power of 1S-2S preparation laser, detector blocking meshes: whether the additional blocking meshes were installed in the detector assembly, FS: probed fine-structure component, $P_\mathrm{2S\text{-}6P}$: power of 2S-6P spectroscopy laser, FCs: number of freezing cycles contained in run, valid line scans: number of valid line scans contained in run.

| Meas. run | Time period (2019) | $\alpha_0$ (mrad) | $\theta_\mathrm{L}$ (°) | $P_\mathrm{1S\text{-}2S}$ (W) | Detector blocking meshes | FS | $P_\mathrm{2S\text{-}6P}$ (µW) | FCs | Valid line scans |
|---|---|---|---|---|---|---|---|---|---|
| A | 24.3.–3.4. (6 days) | 0 | 56.5 | 1.0–1.7 | Installed | $6P_{1/2}$ | 30 | 13 | 285 |
| | | | | | | $6P_{3/2}$ | 15 | 3 | 162 |
| B | 23.5.–9.6. (14 days) | 0 | 56.5, 146.5 | 1.0–1.3 | Removed | $6P_{1/2}$ | $10, 20, 30$ | 21 | 1093 |
| | | | | | | $6P_{3/2}$ | $5, 10, 15$ | 20 | 992 |
| C | 29.7.–7.8. (5 days) | $0, \pm 7.5, \pm 12.0$ | 56.5 | 1.0–1.1 | Removed | $6P_{1/2}$ | 30 | 16 | 623 |
| Total | | | | | | | | 73 | 3155 |

15 µW for the 2S-$6P_{1/2}$ and 2S-$6P_{3/2}$ transitions during run A. For run B, data were taken at two-fold and three-fold reduced laser powers, with most of the data taken at the latter. This was done because using a lower laser power decreases the size of the light force shift and, to a lesser extend, line shifts from quantum interference (QI).

The majority of the data were acquired with the linear laser polarization of the 2S-6P spectroscopy laser aligned such that QI line shifts were approximately minimized (see Fig. 6.8). To this end, $\theta_\mathrm{L}$ was set to 56.5°, where $\theta_\mathrm{L}$ is the angle of the polarization vector from the axis of the detector cylinder. To confirm the theoretical modeling of the observed QI line shifts, which in turn critically depends on the modeling of the detection efficiency, some data during run B were taken at $\theta_\mathrm{L} = 146.5°$, where the QI line shifts are larger.

The intracavity power $P_\mathrm{1S\text{-}2S}$ of the 1S-2S preparation laser was kept between 1.0 W and 1.3 W during runs B and C. For run A, the power reached 1.7 W for the first FC, decreasing to 1.0 W for the final FC. For the bulk of run B and all of run C, the intracavity power was stabilized (see Section 4.3.3.5), but not for run A.

Finally, during run A additional blocking meshes were installed inside the detector assembly, as discussed in Section 4.6.1. The blocking meshes were removed prior to run B.

### 6.1.2 Data groups and their analysis

For data analysis purposes, the data are further grouped by measurement run, the probed transition, and the values of $\alpha_0$, $\theta_\mathrm{L}$, and $P_\mathrm{2S\text{-}6P}$. The resulting 17 data groups are listed in Table 6.2.

Table 6.2: Overview of the data groups G1–G14, with the line scans in each data group taken during the same measurement run and for the same fine-structure (FS) component, offset of atomic–laser beam angle $\alpha_0$, linear laser polarization angle $\theta_L$, and power of 2S-6P spectroscopy laser $P_{2S\text{-}6P}$. FCs: number of freezing cycles contained in group, valid line scans per detector: number of valid line scans contained in group for the top and bottom detector, line shape func.: line shape function used for analysis, dual scan: whether line scans were acquired as dual scans and which parameter was varied.

| Data group | Meas. run | FS | $\alpha_0$ (mrad) | $\theta_L$ (°) | $P_{2S\text{-}6P}$ (µW) | FCs | Valid line scans per detector: Top | Valid line scans per detector: Bottom | Line shape func. | Dual scan |
|---|---|---|---|---|---|---|---|---|---|---|
| G1A | A | $6P_{1/2}$ | 0 | 56.5 | 30 | 13 | 285 | 285 | Voigt | — |
| G1B | B | | | | 30 | 18 | 148 | 141 | | $P_{2S\text{-}6P}$ |
| G1C | C | | | | 30 | 16 | 77 | 68 | | — |
| G2 | B | | | | 20 | 18 | 147 | 138 | | $P_{2S\text{-}6P}$ |
| G3 | B | | | | 10 | 18 | 598 | 564 | | |
| G4 | B | $6P_{1/2}$ | 0 | 146.5 | 30 | 3 | 34 | 31 | Voigt | $P_{2S\text{-}6P}$ |
| G5 | B | | | | 20 | 3 | 34 | 31 | | |
| G6 | B | | | | 10 | 3 | 132 | 119 | | |
| G7A | A | $6P_{3/2}$ | 0 | 56.5 | 15 | 3 | 162 | 162 | Voigt | — |
| G7B | B | | | | 15 | 17 | 143 | 116 | | $P_{2S\text{-}6P}$ |
| G8 | B | | | | 10 | 18 | 151 | 124 | | |
| G9 | B | | | | 5 | 18 | 568 | 461 | | |
| G10 | B | $6P_{3/2}$ | 0 | 146.5 | 15 | 2 | 21 | 20 | Voigt | $P_{2S\text{-}6P}$ |
| G11 | B | | | | 10 | 2 | 22 | 21 | | |
| G12 | B | | | | 5 | 2 | 87 | 80 | | |
| G13 | C | $6P_{1/2}$ | 7.5 | 56.5 | 30 | 3 | 106 | 106 | Voigt | — |
| G14 | C | | 12.0 | | 30 | 11 | 440 | 394 | doublet | |

#### 6.1.2.1 Simulation corrections

For each data group, up to three sets of 2S trajectories are generated using a Monte Carlo simulation, as detailed in Section 5.2. For each set, the power $P_{1S\text{-}2S}$ and detuning $\Delta\nu_{1S\text{-}2S}$ of the 1S-2S spectroscopy laser are adjusted to match the experimental values. Multiple sets are used for data groups with a large spread of $P_{1S\text{-}2S}$ or $\Delta\nu_{1S\text{-}2S}$. Each line scan is then assigned the set most closely matching $P_{1S\text{-}2S}$ and $\Delta\nu_{1S\text{-}2S}$ during that line scan. The cutoff speed $v_{\text{cutoff}}$ of the velocity distribution is set[1] to approximately match the average $v_{\text{cutoff}}$ of the line scans of the data group, shown in Fig. 6.1. For all data groups of run A (run B and C), $v_{\text{cutoff}} = 30\,\text{m/s}$ ($v_{\text{cutoff}} = 65\,\text{m/s}$) is used. A total of $N_{\text{traj}} = 1 \times 10^6$ and $N_{\text{traj}} = 4 \times 10^6$ trajectories are used for data groups G1–G12 and G13–14, respectively. The latter data groups require a larger number of trajectories to reach a sufficient accuracy in the simulations of the light force shift.

[1] $v_{\text{cutoff}}$ is determined by comparing the delay dependencies of the experimental and simulated line amplitudes, as discussed in Section 5.3.2.

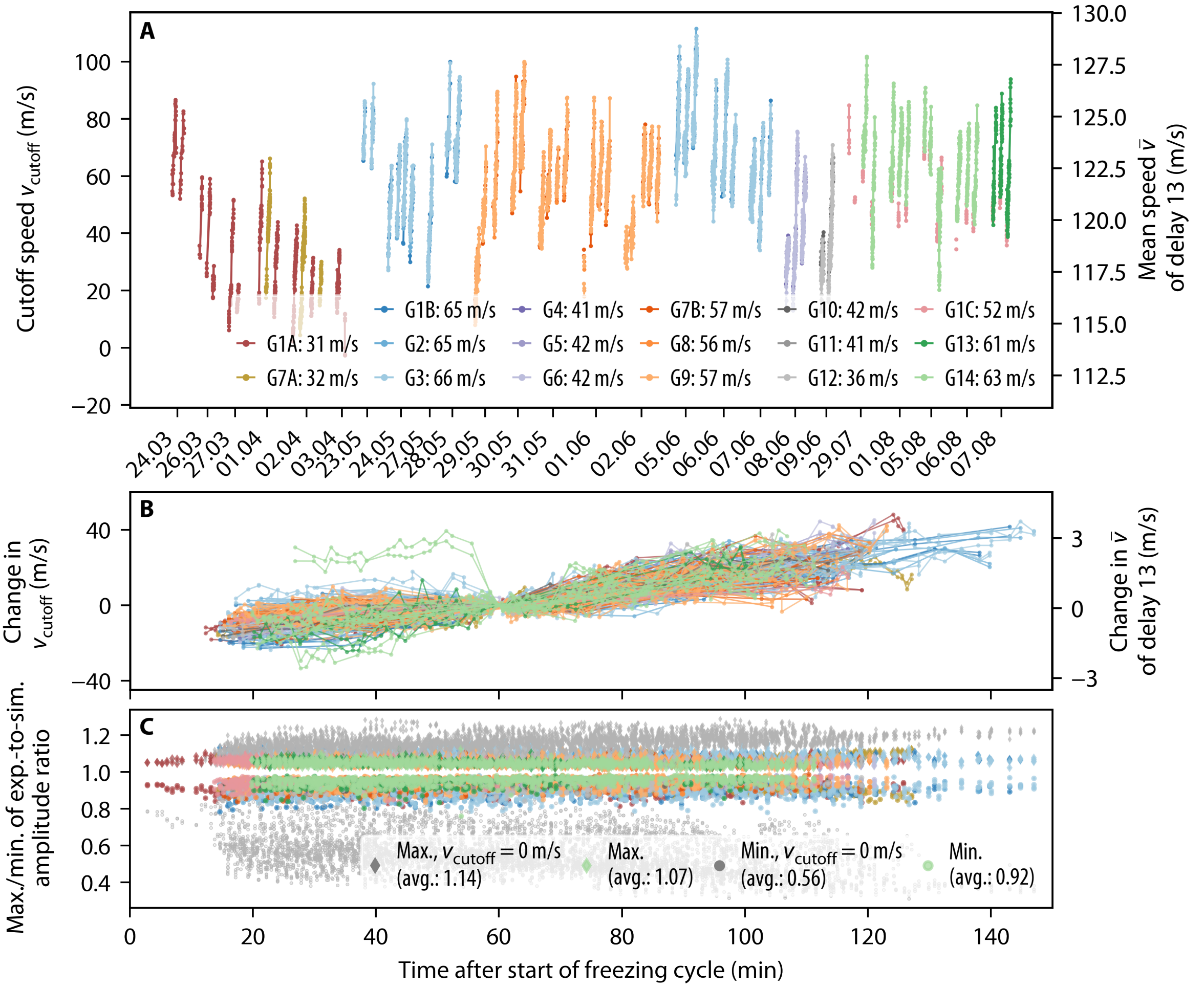

Figure 6.1: Variation of the cutoff speed $v_{\mathrm{cutoff}}$ during the 2S-6P measurement. $v_{\mathrm{cutoff}}$ is determined for each valid scan and each detector by minimizing the ratio of the experimental to the simulated amplitudes for all delays. (**A**) $v_{\mathrm{cutoff}}$ versus measurement time for the different data groups. Each almost vertical stripe (colored lines and points) corresponds to a freezing cycle (FC). (**B**) Same data as (A), but shown as change in $v_{\mathrm{cutoff}}$ during each FC (colored lines and points), relative to the value of $v_{\mathrm{cutoff}}$ at 60 min after the start of the FC. (**C**) The maximal and minimal experiment-to-simulation amplitude ratio for the data of (B) (colored diamonds and circles, respectively), and for reference when $v_{\mathrm{cutoff}}$ is set to zero (gray symbols).

The remaining parameters of the Monte Carlo simulation are identical for all data groups and their values are given in Table 5.3.

The generated trajectory sets serve as input to the big model and the light force shift model, as discussed in Section 5.3. For each trajectory set, line scans from both models are calculated, using the values of $\alpha_0$, $\theta_{\mathrm{L}}$, and $P_{\mathrm{2S\text{-}6P}}$ of the corresponding data group. For this preliminary analysis, the same detection efficiency simulation with mesh transparencies $T_{\mathrm{WM}} = T_{\mathrm{WM,BD}} = 80\,\%$ is used for all data groups (see Section 4.6.6), that is the blocking meshes during run A are neglected. The mean speed $\bar{v}$ and root mean square speed $\bar{v}_{\mathrm{RMS}}$ of each delay is determined for each trajectory set as described in Section 5.3.2.

Both the experimental and simulated line scans are fit using the same line shape function. For data groups with $\alpha_0 = 0\,\mathrm{mrad}$ a Voigt line shape is used (see Section 5.1.1.2 and Fig. 5.1), while for data groups G13 and G14, where $\alpha_0 = 7.5\,\mathrm{mrad}$ and $\alpha_0 = 12.0\,\mathrm{mrad}$, respectively, the Voigt doublet line shape is employed (see Section 5.1.1.3 and Fig. 5.2). The simulated line scans use the same frequency sampling of the spectroscopy laser detuning $\Delta\nu_{\text{2S-6P}}$ as used for the experimental line scans (see Table 5.2).

In this way, each delay of each experimental line scan is assigned a fit result from the big model (with resonance frequency $\nu_{0,\mathrm{BM}}$), a fit result from the light force shift model (with resonance frequency $\nu_{0,\mathrm{LFS}}$), and a value of $\bar{v}$ and $\bar{v}_{\mathrm{RMS}}$. From the latter, the second-order Doppler shift $\Delta\nu_{\mathrm{SOD}}$ can be calculated (see Section 6.2.4.1). The resonance frequency $\nu_0$ of each delay of each experimental line scan is then corrected for the SOD, shifts dominated by quantum interference (QI), and the light force shift by subtracting $\Delta\nu_{\mathrm{SOD}} + \nu_{0,\mathrm{BM}}+\nu_{0,\mathrm{LFS}}$, i.e., $\nu_0 \to \nu_0 - (\Delta\nu_{\mathrm{SOD}} + \nu_{0,\mathrm{BM}}+\nu_{0,\mathrm{LFS}})$. The complete analysis is also repeated for some or all simulation corrections omitted to find their impact on the determined resonance frequency of the 2S-6P transition (see Fig. 6.4). Note that, since both $\nu_{0,\mathrm{BM}}$ and $\nu_{0,\mathrm{LFS}}$ are determined from fits to line scans, systematic shifts from, e.g., the chosen frequency sampling might be accounted for twice. This remains to be studied in detail in the future. However, when the 2S-6P centroid is formed (see Section 6.2.4.6), the corrections from $\nu_{0,\mathrm{BM}}$ almost completely cancel out, leaving only those from $\nu_{0,\mathrm{LFS}}$, in which this particular problem is circumvented.

Using the mean speeds $\bar{v}$, a Doppler extrapolation (and averaging) as described in Section 5.1.2 is done for both the experimental and simulated line scans. This results in values of the Doppler-free resonance frequency $\nu_{0,\mathrm{e}}$, the Doppler slope $\kappa$, and the Doppler-averaged resonance frequency $\nu_{0,\mathrm{a}}$ for each line scan and detector. Note that because the simulation corrections have been applied to the resonance frequencies of the individual experimental delays, they influence the determined value of $\kappa$ and in general lead to corrections of different size and even sign for $\nu_{0,\mathrm{e}}$ and $\nu_{0,\mathrm{a}}$. $\nu_{0,\mathrm{e}}$, $\kappa$, and $\nu_{0,\mathrm{a}}$ from the simulations are not used directly in the analysis, but only to visualize the simulation corrections, as in, e.g., Fig. 6.8. Thus, in the following, $\nu_{0,\mathrm{e}}$, $\kappa$, and $\nu_{0,\mathrm{a}}$ refer to values derived from the experimental data unless otherwise mentioned.

#### 6.1.2.2 Detector correlations

The values of $\nu_{0,\mathrm{e}}$, $\kappa$, and $\nu_{0,\mathrm{a}}$ of each line scan are correlated for the two detectors because of noise on the fluorescence signal common to both detectors, i.e., noise other than shot noise or from the detectors themselves. Linear correlation coefficients $r$ are determined for each data group. They range between $0.29\ldots0.49$, $0.48\ldots0.70$, and $0.70\ldots0.80$, respectively, for the three quantities, with a larger correlation observed for data groups where a larger laser power $P_{\text{2S-6P}}$ was used. This is to be expected, as a larger $P_{\text{2S-6P}}$ leads to a higher fluorescence signal and thus a lower relative contribution of shot noise. The correlation coefficients are taken into account when averaging the results from the two detectors, with all data shown in this chapter corresponding to this detector average[1].

#### 6.1.2.3 Excess scatter

Finally, $\nu_{0,\mathrm{e}}$, $\kappa$, and $\nu_{0,\mathrm{a}}$ are averaged over each of the freezing cycles (FCs) contained in each data group, using the corresponding statistical uncertainties as weights. The reduced chi-squared $\chi^2_{\mathrm{red,FC}}$ of these weighted averages is typically found to be above one, indicating an excess scatter of the results from each line scan. For the 2S-6P measurement, the mean $\chi^2_{\text{red,FC}}$ is $\bar{\chi}^2_{\text{red,FC}} = 1.7$, 2.6, and 6.7 for the three quantities (see Figs. 6.2 and 6.3), with each FC containing a mean number of 18.8 scans. As discussed in detail in Section 5.1.2, this is expected for the averages of $\kappa$ and $\nu_{0,\text{a}}$ if the Doppler shift varies over time, e.g., from drifts of the experimental alignment. The Doppler extrapolation should remove those shifts and $\nu_{0,\text{e}}$ should ideally show no excess scatter. While the excess scatter is significantly lower for $\nu_{0,\text{e}}$ than for $\kappa$ and $\nu_{0,\text{a}}$, it still is significantly above one. This is attributed to additional noise on the fluorescence signal, e.g., from variations in the atomic flux from nozzle temperature fluctuations[1], which also leads to the aforementioned detector correlations.

[1]For some line scans, only data from the top detector are available (see Table 6.2), because of transient excess scatter and spikes on the bottom detector. In this case, only the data from the top detector, and not the detector average, are used.

To account for this excess scatter, the uncertainties of the FC-averaged values of $\nu_{0,\text{e}}$, $\kappa$, and $\nu_{0,\text{a}}$ are scaled by the corresponding $\sqrt{\chi^2_{\text{red,FC}}}$ (for $\chi^2_{\text{red,FC}} \leq 1$, no scaling is done, i.e., the uncertainty is never decreased). Whether and at what point this scaling is done is a somewhat arbitrary choice, and here the effect of the scaling on the value of $\nu_{0,\text{e}}$ found for the 2S-6P$_{1/2}$ and 2S-6P$_{3/2}$ transition is less than 30 Hz. The idea here is the following: first, before each FC, the nozzle is heated up to room temperature, cooled down, and centered in the preparation laser beam. The hydrogen dissociator, which is stopped during this process, is restarted. At the start of each FC, the optimal detuning of the preparation laser is found by spectroscopy of the 1S-2S laser and the offset angle $\alpha_0$ is set using the observed Doppler shift of the 2S-6P transition [28]. Second, it is assumed that there is additional noise on the fluorescence signal, and that this noise is approximately the same for each FC. Third, some variation of the residual Doppler shift from FC to FC is possible, as it should depend on the alignment at some level. Then, scaling the average over each FC by $\sqrt{\chi^2_{\text{red,FC}}}$ removes this noise. It does however not remove possible scatter from FC to FC and day to day from the preparation and alignment procedure, which manifests itself as a $\chi^2_{\text{red}}$ above one for the weighted average over the scaled, FC-averaged values of $\nu_{0,\text{e}}$. The results from this procedure for the 2S-6P measurement are presented in the next section.

## 6.2 Preliminary results of the 2S-6P measurement

### 6.2.1 Blinded transition frequencies

#### 6.2.1.1 The 2S-6P$_{1/2}$ transition

The blinded results from the measurement of the 2S-6P$_{1/2}$ transition are shown in Fig. 6.2. All data for this transition from the 2S-6P measurement is taken into account, including for a nonzero $\alpha_0$, i.e., data groups G1–G6, G13, and G14 (colored diamonds). The data within each group has been analyzed as described in Section 6.1.2, and corrections for the light force shift, quantum interference line shifts, and second-order Doppler shift have been included. This results in a value and uncertainty of the Doppler-free resonance frequency $\nu_{0,\text{e}}$, Doppler slope $\kappa$, and Doppler-averaged resonance frequency $\nu_{0,\text{a}}$ for each of the 106 freezing cycles (FC), shown in Fig. 6.2 (A), (C), and (E), respectively. The corresponding $\chi^2_{\text{red,FC}}$ of the weighted average over all line scans within each FC is given in (B), (D), and (F), respectively, and (G) shows the number of line scans in each FC. The uncertainties of $\nu_{0,\text{e}}$, $\kappa$, and $\nu_{0,\text{a}}$ have been scaled by $\sqrt{\chi^2_{\mathrm{red,FC}}}$, which results in a mean increase of $\sqrt{\bar{\chi}^2_{\mathrm{red,FC}}} = 1.30$, 1.63, and 2.66, respectively.

[1]The standard deviation of the nozzle temperature $T_\text{N}$ during the line scans and the $\chi^2_{\text{red}}$ of the fits to the line scan are found to be significantly positively correlated. This is currently, along with other noise source, being investigated further.

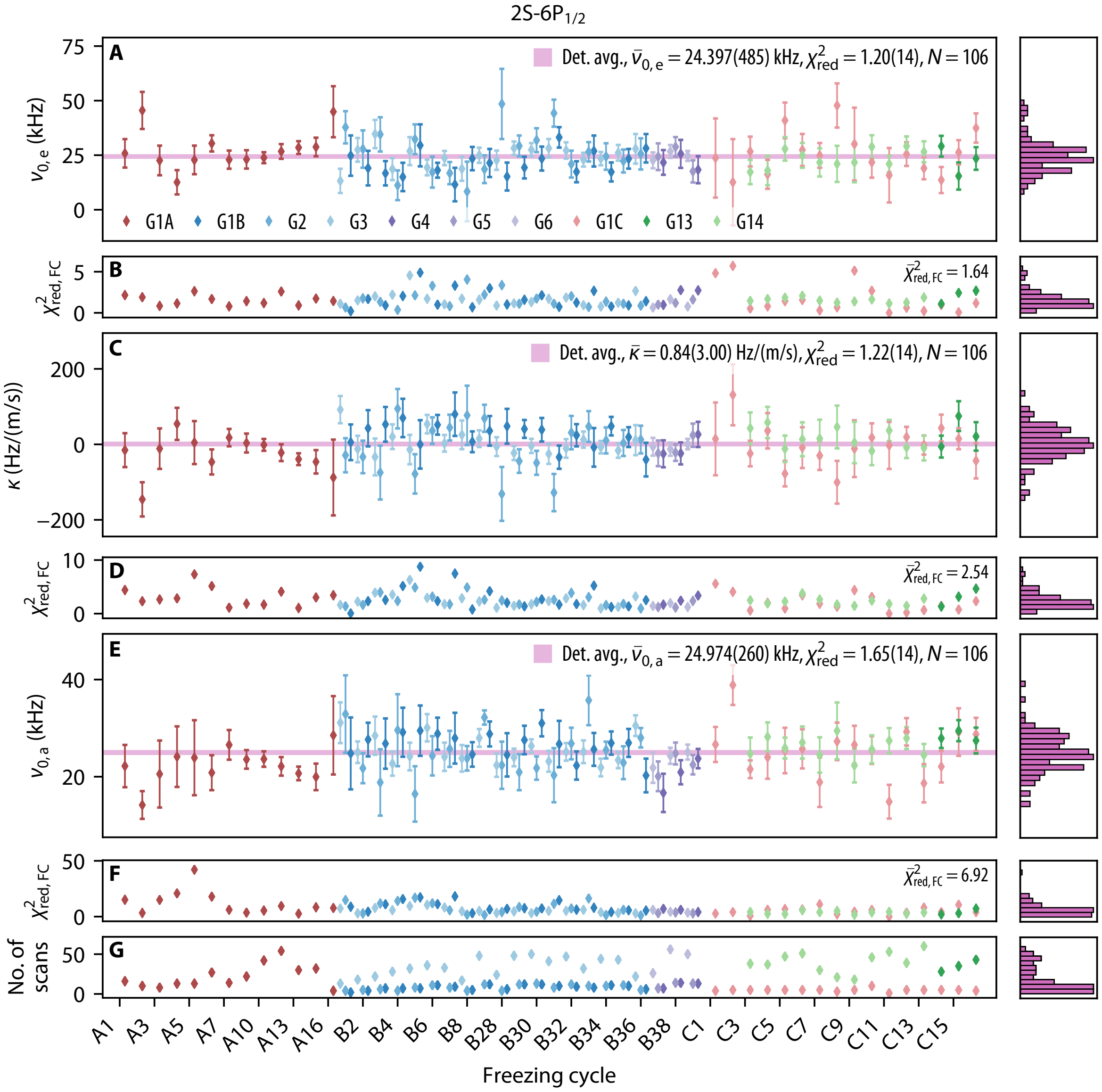

Figure 6.2: Blinded results of the measurement of the 2S-6P$_{1/2}$ transition for the different data groups and freezing cycles (FCs). All data, including for a nonzero $\alpha_0$, are considered (data groups G1–G6, G13, G14). Shown is the FC average of the (**A**) Doppler-free resonance frequency $\nu_{0,\mathrm{e}}$, (**C**) Doppler slope $\kappa$, and (**E**) Doppler-averaged resonance frequency $\nu_{0,\mathrm{a}}$. The corresponding $\chi^2_{\mathrm{red,FC}}$ of the weighted average over all line scans within each FC is given in (**B**), (**D**), and (**F**), respectively, and the uncertainties of $\nu_{0,\mathrm{e}}$, $\kappa$, and $\nu_{0,\mathrm{a}}$ have been scaled by $\sqrt{\chi^2_{\mathrm{red,FC}}}$. Purple bands indicate the $1\sigma$ uncertainty region of weighted averages over all 106 FCs. Corrections for the light force shift, quantum interference line shifts, and second-order Doppler shift have been included. Only the statistical uncertainty, including detector correlations, but not the uncertainty of the corrections, is taken into account. (**G**) The number of line scans within each FC. Histograms of the data within each plot are shown on the right.

The weighted average of the Doppler-free resonance frequencies $\nu_{0,\mathrm{e}}$ of all 106 FCs gives the blinded transition frequency $\tilde{\nu}_{1/2}$ of the 2S-6P$_{1/2}$ transition (purple band in Fig. 6.2 (A)),

$$\tilde{\nu}_{1/2} \equiv \bar{\nu}_{0,\mathrm{e}} = 24.397(485)\,\mathrm{kHz} \qquad \text{with} \quad \chi^2_{\mathrm{red}} = 1.20(14). \tag{6.1}$$

$\tilde{\nu}_{1/2}$ so far only includes the statistical uncertainty, but not the uncertainty of the corrections or other systematic uncertainties. In total, $\tilde{\nu}_{1/2}$ has been corrected by 760 Hz, consisting of a 696 Hz, 210 Hz, and −145 Hz correction for the light force shift (LFS), quantum interference (QI) line shifts, and second-order Doppler shift (SOD), respectively (see Fig. 6.4).

The $\chi^2_{\mathrm{red}}$ of the weighted average is in reasonable agreement with the assumption of no excess scatter between the FCs. If the uncertainties of each FC are not scaled up to account for the excess scatter within each FC, $\tilde{\nu}_{1/2}$ shifts by only 17 Hz, but the uncertainty decreases to 401 Hz, while $\chi^2_{\mathrm{red}}$ increases to $1.82 \approx (485\,\mathrm{Hz}/401\,\mathrm{Hz})^2 \times 1.20$. The statistical uncertainty of $\tilde{\nu}_{1/2}$ is a factor of 6.0 lower than the comparable uncertainty[1] of the 2S-4P$_{1/2}$ transition frequency of Appendix A.

Likewise, the weighted average of the Doppler slopes $\kappa$ of all FCs gives (purple band in Fig. 6.2 (C))

$$\bar{\kappa} = 0.84(3.00)\,\mathrm{Hz}/(\mathrm{m/s}) \qquad \text{with} \quad \chi^2_{\mathrm{red}} = 1.22(14), \tag{6.2}$$

$\bar{\kappa}$ is in good agreement with zero, and it corresponds to a Doppler shift of 168(600) Hz for an atom with $v_{\mathrm{typ}} = 200\,\mathrm{m/s}$. The corrections of the resonance frequencies of each line scan result in $\bar{\kappa}$ being corrected by −5.4 Hz/(m/s), consisting of a −4.7 Hz/(m/s), −2.4 Hz/(m/s), and 1.7 Hz/(m/s) correction for the LFS, QI line shifts, and SOD, respectively (see Fig. 6.4). Thus, were those corrections not applied, $\bar{\kappa}$ would only be reasonable compatible with zero by $2\sigma$. The $\chi^2_{\mathrm{red}}$ similar to that of $\tilde{\nu}_{1/2}$, indicates that the excess scatter between FCs and measurement days is not substantially larger than within each FC. This suggests that the alignment procedures performed for each FC result in a negligible, or at least reproducible, residual Doppler shift, compared to the uncertainty of the measurement and the change in residual Doppler shift during an FC. This was not the case for the 2S-4P measurement (see Appendix A), where excess day-to-day scatter was observed.

Finally, the weighted average of the Doppler-averaged resonance frequencies $\nu_{0,\mathrm{a}}$ of all FCs results in (purple band in Fig. 6.2 (E))

$$\bar{\nu}_{0,\mathrm{a}} = 24.974(260)\,\mathrm{kHz} \qquad \text{with} \quad \chi^2_{\mathrm{red}} = 1.65(14). \tag{6.3}$$

$\bar{\nu}_{0,\mathrm{a}}$ has been corrected by −274 Hz, consisting of a −253 Hz, −208 Hz, and 187 Hz correction for the LFS, QI line shifts, and SOD, respectively (see Fig. 6.4). With those corrections included, $\tilde{\nu}_{1/2}$ and $\bar{\nu}_{0,\mathrm{a}}$ agree within $1\sigma$ of their combined uncertainty. Without those corrections, however, there is a $3\sigma$ tension between $\tilde{\nu}_{1/2}$ and $\bar{\nu}_{0,\mathrm{a}}$.

The value of $\chi^2_{\mathrm{red}}$ indicates a significant excess scatter of $\bar{\nu}_{0,\mathrm{a}}$ between FCs, as opposed to $\tilde{\nu}_{1/2}$ and $\bar{\kappa}$. This is unsurprising, as $\nu_{0,\mathrm{a}}$ is more sensitive to residual Doppler shifts and additional noise on the fluorescence signal, as demonstrated by the much larger excess scatter within the FCs as compared $\nu_{0,\mathrm{e}}$ and $\kappa$. Thus, excess scatter between the FCs that is only visible as a slightly larger $\chi^2_{\mathrm{red}}$ for $\tilde{\nu}_{1/2}$ and $\bar{\kappa}$ can lead to a clearly too large value of $\chi^2_{\mathrm{red}}$ for $\bar{\nu}_{0,\mathrm{a}}$.

[1] For the 2S-4P measurement, the first-order Doppler shift was accounted for differently than here, and the statistical uncertainty of $\bar{\nu}_{0,\mathrm{e}}$ should be compared to the first-order Doppler shift uncertainty given in Table S2 of Appendix A.

However, if one is not interested in the absolute frequency, but only in frequency differences between subsets of the data, one may still use $\bar{\nu}_{0,\mathrm{a}}$ by taking the excess scatter into account. To this end, $\bar{\nu}_{0,\mathrm{a}}$ is again scaled by $\sqrt{\chi^2_{\mathrm{red}}}$, but this time using the $\chi^2_{\mathrm{red}}$ of the weighted average of the FCs in the subset of data in question. This gives the scaled Doppler-averaged resonance frequency $\hat{\nu}_{0,\mathrm{a}} = \sqrt{\chi^2_{\mathrm{red}}}\bar{\nu}_{0,\mathrm{a}} \approx 1.28\,\bar{\nu}_{0,\mathrm{a}}$. $\hat{\nu}_{0,\mathrm{a}}$ is used in the tests for the LFS model and the big model, which is used to calculate the QI line shifts. A similar procedure was used to observe the QI line shifts in the 2S-4P measurement (see Appendix A).

One such test is to look at the measurement results for the 2S-6P$_{1/2}$ transition versus the spectroscopy laser power $P_{\text{2S-6P}}$, shown in Fig. 6.5 (A–C). Only data for which $\alpha_0 = 0\,\mathrm{mrad}$ are considered, since data with nonzero $\alpha_0$ were only taken for $P_{\text{2S-6P}} = 30\,\mu\mathrm{W}$ and thus tend to complicate the interpretation of the test. With the power-dependent corrections for the light force shift and the QI line shifts taken into account, both $\bar{\nu}_{0,\mathrm{e}}$ and $\hat{\nu}_{0,\mathrm{a}}$ for different $P_{\text{2S-6P}}$ are found to be highly compatible.

#### 6.2.1.2 The 2S-6P$_{3/2}$ transition

Similarly, the blinded results from the measurement of the 2S-6P$_{3/2}$ transition are shown in Fig. 6.3. All 62 FCs of data for this transition from the 2S-6P measurement are taken into account, i.e., data groups G7–G12 (colored diamonds). Unlike for the 2S-6P$_{1/2}$ transition, no data were acquired for a nonzero $\alpha_0$. The weighted average of all FCs gives

$$\tilde{\nu}_{3/2} \equiv \bar{\nu}_{0,\mathrm{e}} = -3.315(601)\,\mathrm{kHz} \qquad \text{with} \quad \chi^2_{\mathrm{red}} = 1.23(18), \tag{6.4}$$

$$\bar{\kappa} = -3.14(3.75)\,\mathrm{Hz/(m/s)} \qquad \text{with} \quad \chi^2_{\mathrm{red}} = 1.22(18), \tag{6.5}$$

$$\bar{\nu}_{0,\mathrm{a}} = -3.695(312)\,\mathrm{kHz} \qquad \text{with} \quad \chi^2_{\mathrm{red}} = 1.40(18). \tag{6.6}$$

$\tilde{\nu}_{3/2}$ ($\bar{\nu}_{0,\mathrm{a}}$) has been corrected by 1149 Hz (729 Hz), consisting of a 1313 Hz (381 Hz), −22 Hz (163 Hz), and −142 Hz (186 Hz) correction for the LFS, QI line shifts, and SOD, respectively (see Fig. 6.4). As for the 2S-6P$_{1/2}$ transition, the LFS correction dominates the total correction. However, the correction is approximately a factor of two larger for $\tilde{\nu}_{3/2}$ than for $\tilde{\nu}_{1/2}$, which is mainly because there is some cancellation of the LFS for the data taken at zero $\alpha_0$ and at $\alpha_0 = 12.0\,\mathrm{mrad}$ for the latter.

The $\chi^2_{\mathrm{red}}$ for $\tilde{\nu}_{3/2}$ and $\bar{\kappa}$ are similar to those of the 2S-6P$_{1/2}$ transition and likewise in reasonable agreement with the assumption of no significant excess scatter between the FCs. If the uncertainties of each FC are not scaled up to account for the excess scatter within each FC, $\tilde{\nu}_{3/2}$ shifts by only −24 Hz, but the uncertainty decreases to 479 Hz, while $\chi^2_{\mathrm{red}}$ increases to $1.89 \approx (601\,\mathrm{Hz}/479\,\mathrm{Hz})^2 \times 1.23$. $\tilde{\nu}_{1/2}$ and $\bar{\nu}_{0,\mathrm{a}}$ agree within their combined uncertainty whether the corrections are included or not. The statistical uncertainty of $\tilde{\nu}_{3/2}$ is a factor of 4.7 lower than the comparable uncertainty of the 2S-4P$_{3/2}$ transition frequency of Appendix A.

The $\chi^2_{\mathrm{red}}$ for $\bar{\nu}_{0,\mathrm{a}}$ indicates, as expected, a significant excess scatter between the FCs, albeit at a lower level than seen for the 2S-6P$_{1/2}$ transition. Analogous to that transition, a scaled Doppler-averaged resonance frequency $\hat{\nu}_{0,\mathrm{a}} = \sqrt{\chi^2_{\mathrm{red}}}\bar{\nu}_{0,\mathrm{a}} \approx 1.18\,\bar{\nu}_{0,\mathrm{a}}$ is defined to test frequency differences within the data set.

As for the 2S-6P$_{1/2}$ transition, $\bar{\kappa}$ is in good agreement with zero with a similar uncertainty. $\bar{\kappa}$ includes a −1.8 Hz/(m/s) correction, consisting of a −4.8 Hz/(m/s), 1.3 Hz/(m/s), and 1.7 Hz/(m/s) correction for the LFS, QI line shifts, and SOD, respectively (see Fig. 6.4). Correspondingly, $\bar{\kappa}$ is also compatible with zero even if those corrections are not included.

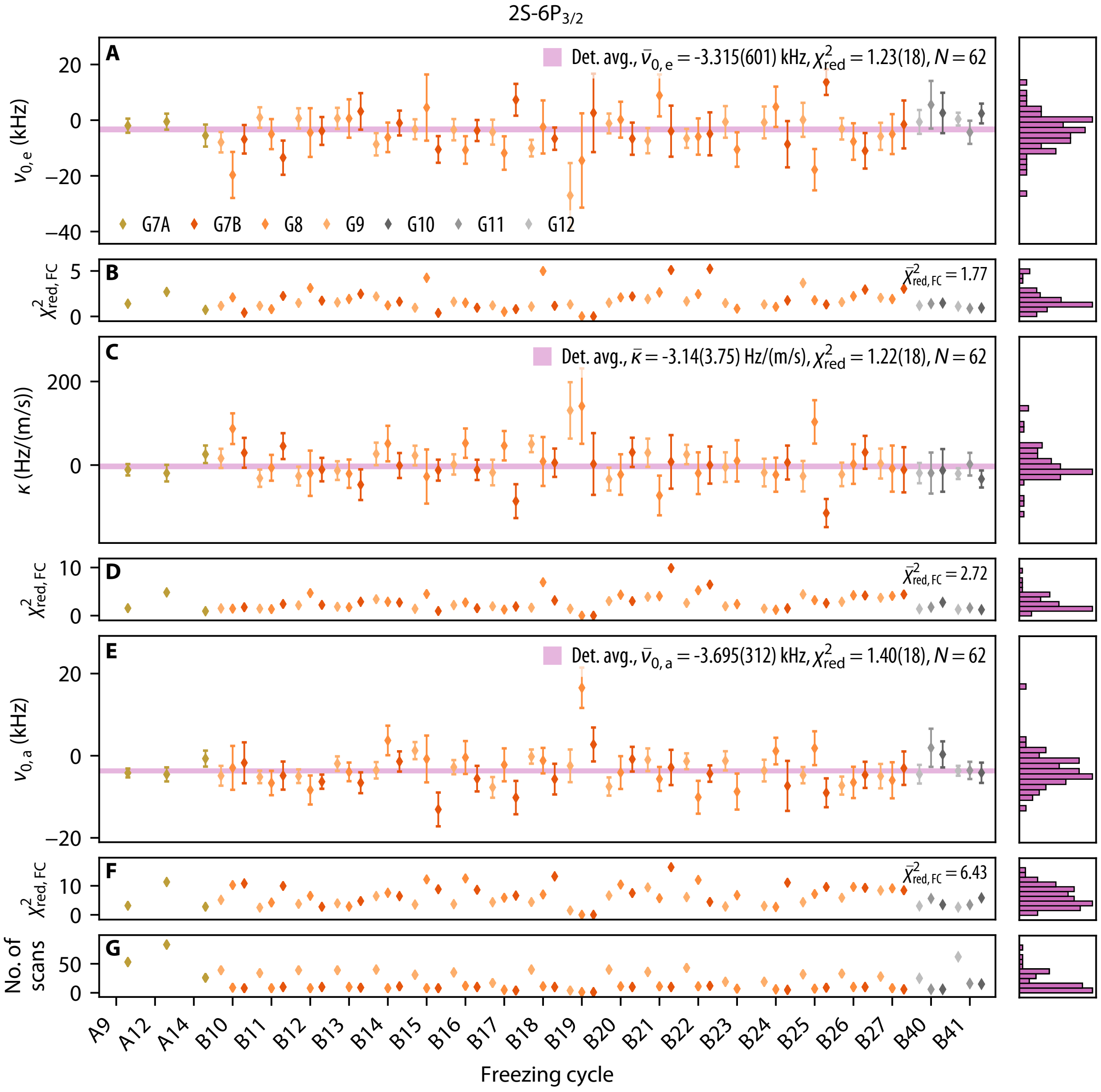

Figure 6.3: Similar to Fig. 6.2, but showing the blinded results of the measurement of the 2S-6P$_{3/2}$ transition for the different data groups (data groups G7–G12) and freezing cycles (FCs).

The measurement results for the 2S-6P$_{3/2}$ transition versus the spectroscopy laser power $P_{\text{2S-6P}}$ are shown in Fig. 6.5 (D–E). $\bar{\nu}_{0,\text{e}}$ and $\hat{\nu}_{0,\text{a}}$ for different $P_{\text{2S-6P}}$ are found to be compatible, whether the power-dependent corrections for the light force shift and the QI line shifts are taken into account or not.

#### 6.2.1.3 Combining the 2S-6P$_{1/2}$ and 2S-6P$_{3/2}$ transition

The blinded 2S-6P centroid is given by (see Section 6.2.4.6 and Eq. (6.21))

$$\tilde{\nu}_{\text{2S-6P}} = \frac{1}{3}\tilde{\nu}_{1/2} + \frac{2}{3}\tilde{\nu}_{3/2} = 5.922(432)\,\text{kHz}, \tag{6.7}$$

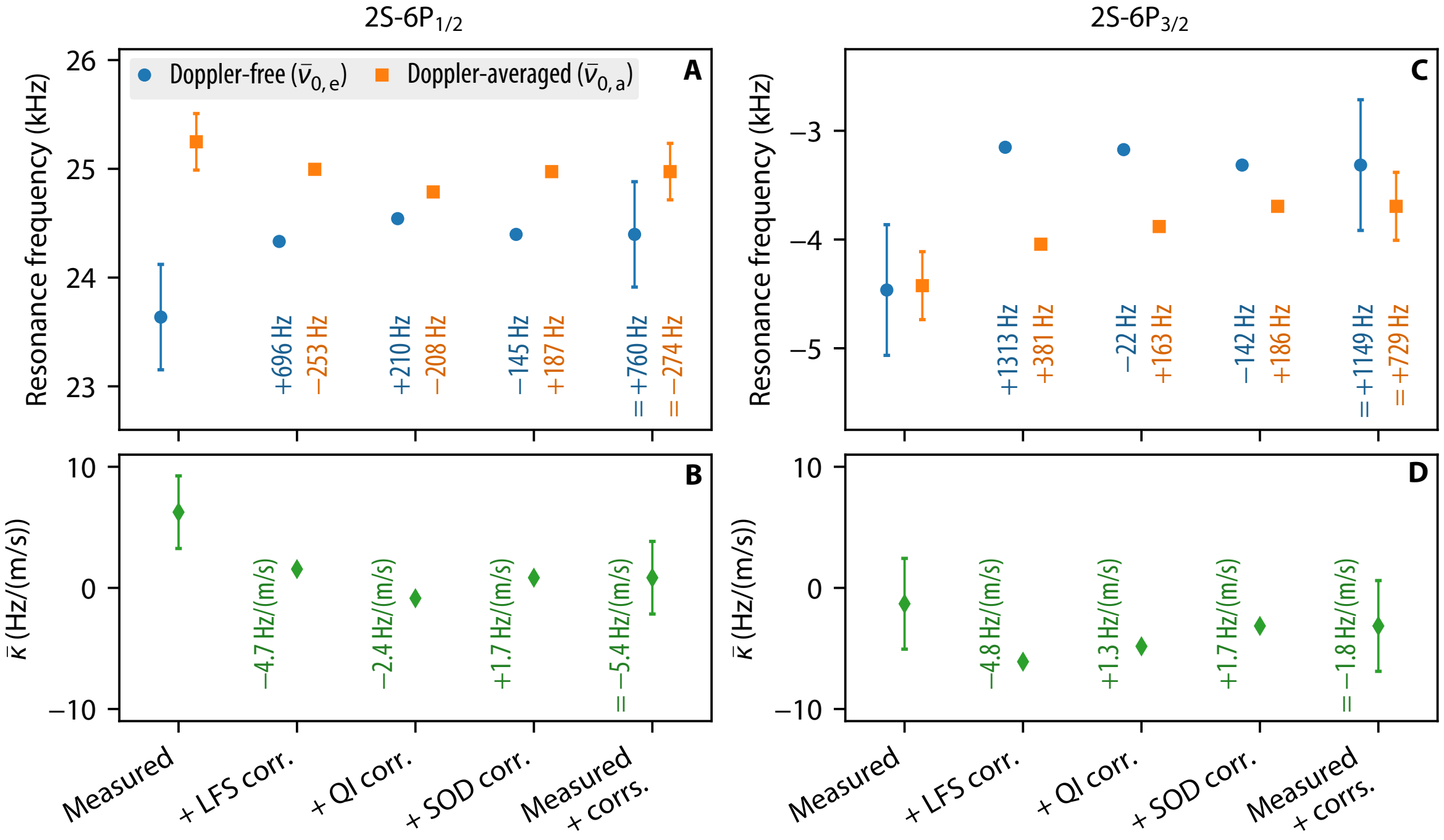

Figure 6.4: Simulation corrections included for the measurement of the (**A**, **B**) 2S-6P$_{1/2}$ and (**C**, **D**) 2S-6P$_{3/2}$ transition. All data, including for a nonzero $\alpha_0$, are taken into account (data groups G1–G14). Shown are the detector- and freezing-cycled-averaged (**A**, **C**) Doppler-free resonance frequency $\bar{\nu}_{0,\mathrm{e}}$ (blue circles), Doppler-averaged resonance frequency $\bar{\nu}_{0,\mathrm{a}}$ (orange squares), and (**B**, **D**) Doppler slope $\bar{\kappa}$ (green diamonds). Three corrections are applied in sequence: light force shift (LFS), quantum interference (QI) line shifts, and second-order Doppler shift (SOD) corrections. The error bars indicate the $1\sigma$ statistical uncertainty of the experimental data, but do not include the uncertainty of the corrections.

where the statistical uncertainties $\tilde{\nu}_{1/2}$ and $\tilde{\nu}_{3/2}$ have been assumed to be uncorrelated and the additional offset from Eq. (6.21) has been absorbed in the blind offset of $\tilde{\nu}_{\text{2S-6P}}$. The corrections applied to $\tilde{\nu}_{1/2}$ and $\tilde{\nu}_{3/2}$ are assumed to be fully correlated, leading to a correction of $\tilde{\nu}_{\text{2S-6P}}$ by 1019 Hz, consisting of a 1107 Hz, 55 Hz, and −144 Hz for the LFS, QI line shifts, and SOD, respectively. Thus, the corrections are dominated by the LFS, justifying the considerable effort that has been put into the derivation and verification of the LFS model. The statistical uncertainty of $\tilde{\nu}_{\text{2S-6P}}$ is a factor of 4.9 lower than the comparable uncertainty of the 4P centroid transition frequency of Appendix A.

One may average the values of the Doppler slope $\bar{\kappa}$ for the two transitions, assuming they are uncorrelated, to give a performance limit on the Doppler suppression of −0.71(2.34) Hz/(m/s). This corresponds to a Doppler shift of −142(468) Hz for an atom with $v_{\text{typ}} = 200\,\text{m/s}$.

### 6.2.2 Experimental test and uncertainty of light force shift model

The data discussed so far have been corrected for the light force shift (LFS), using the model developed in Chapter 3. This models predicts that the LFS changes from a negative shift to a positive shift for atoms flying along the laser beams with a (transverse) velocity $v_x$ exceeding the recoil velocity of $v_{\text{rec}} \approx 0.97\,\text{m/s}$. For the special case of $v_x \approx v_{\text{rec}}$, a resonance-like behavior occurs with a large LFS on the order of 200 kHz (see Fig. 3.3), which however only affects a small class of atoms. In the experiment, as shown in Fig. 5.4, most atoms have a transverse velocity below the recoil velocity, and the overall LFS of the Doppler-free and -averaged frequencies is negative. However, if the atomic beam and the spectroscopy laser beams do not cross at right angles, but instead at an offset angle $\alpha_0$ from the orthogonal, the transverse velocity increases and the negative LFS of some atoms can be balanced by the positive shift experienced by others, with eventually the total shift becoming positive.

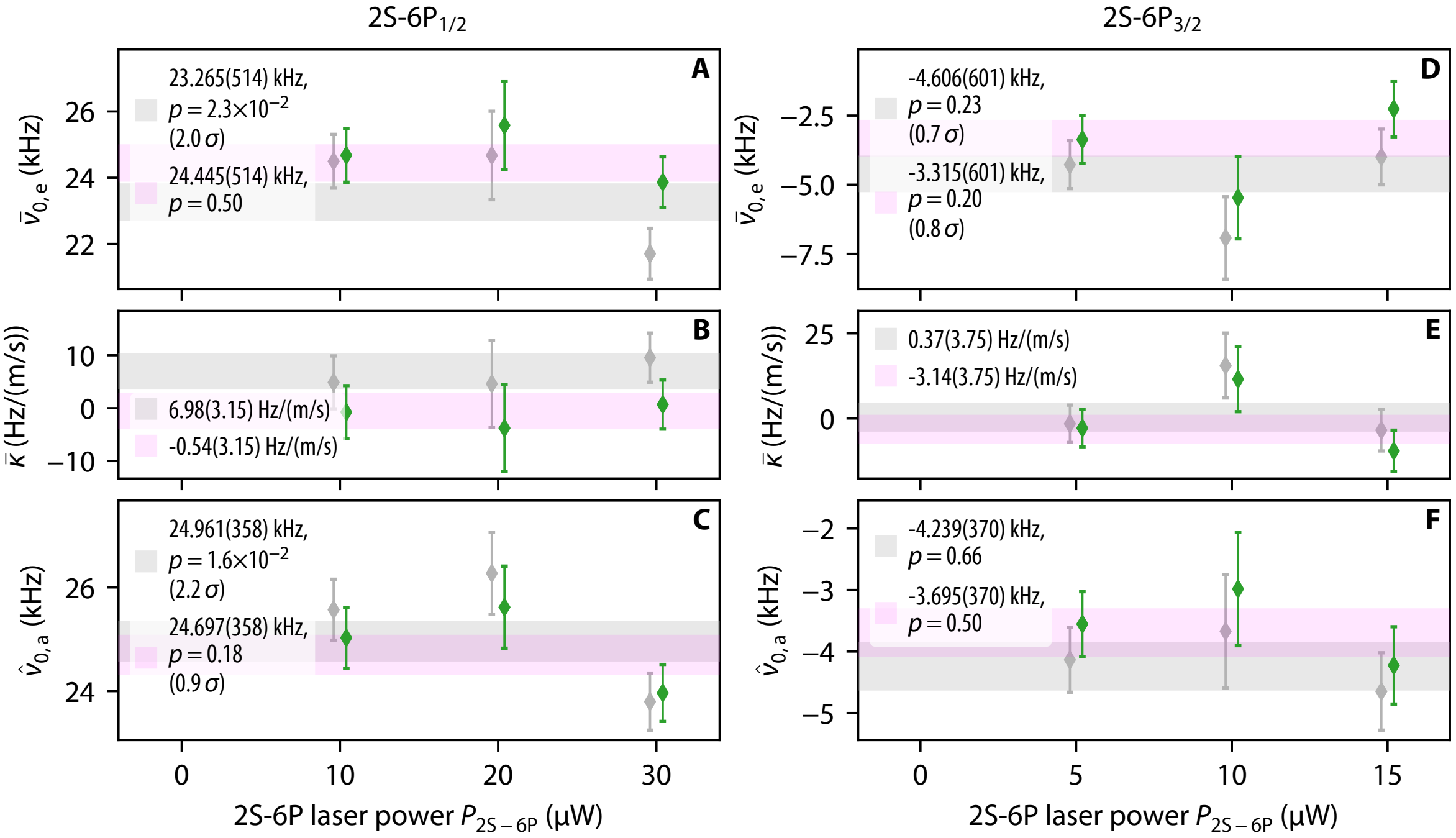

Figure 6.5: Blinded results of the measurement of the (**A**–**C**) 2S-6P$_{1/2}$ and (**D**–**F**) 2S-6P$_{3/2}$ transition versus the power of the 2S-6P spectroscopy laser, $P_{\text{2S-6P}}$. Only data for which $\alpha_0 = 0\,\text{mrad}$ are considered (data groups G1–G12). Shown are the detector- and freezing-cycle-averaged (**A, D**) Doppler-free resonance frequency $\bar{\nu}_{0,\text{e}}$, (**B, E**) Doppler slope $\bar{\kappa}$, and (**C, F**) scaled Doppler-averaged resonance frequency $\hat{\nu}_{0,\text{a}}$, without (gray diamonds) and with (green diamonds) corrections for quantum interference line shifts and light force shift. The gray and purple bands indicate the $1\sigma$ uncertainty region of weighted averages without and with corrections, respectively, with the $p$-value and significance $Z\sigma$ given in the legend. Only the statistical uncertainty, but not the uncertainty of the corrections, is taken into account. A small offset has been added along the $x$-axis for clarity.

To verify this prediction, data were taken for the 2S-6P$_{1/2}$ transition[1] at $\alpha_0 = 7.5\,\text{mrad}$ and 12.0 mrad (data groups G13 and G14, respectively), and using a spectroscopy laser power of $P_{\text{2S-6P}} = 30\,\mu\text{W}$ and a linear laser polarization angle of $\theta_{\text{L}} = 56.5°$. These data can then be compared to data taken for $\alpha_0 = 0\,\text{mrad}$ for the same values of $P_{\text{2S-6P}}$ and $\theta_{\text{L}}$, i.e., data groups G1A–G1C. The model predictions for the LFS as calculated for these data groups are shown versus the mean speed $\bar{v}$ of the delays in Fig. 6.6 (A). Only the fastest delays for zero $\alpha_0$ experience a positive LFS, since the beam divergence is large enough that these atoms exceed the recoil velocity in the transverse direction, while slower delays see a negative LFS of larger magnitude. Through the Doppler extrapolation to zero speed, this results in a negative Doppler-free resonance frequency $\nu_{0,\mathrm{e}}$ of the LFS prediction. For the data groups with $\alpha_0 > 0$, the LFS is always positive, as most atoms have a transverse velocity exceeding the recoil velocity, except for the last delay for $\alpha_0 = 7.5\,\mathrm{mrad}$, which however only has little statistical weight. Consequently, $\nu_{0,\mathrm{e}}$ of the LFS prediction is positive for $\alpha_0 > 0$.

[1] The reader might wonder why the 2S-6P$_{1/2}$ transition and not the 2S-6P$_{3/2}$ transition was used for this purpose, since the LFS is up to a factor of two higher for the latter (see Fig. 3.3). The reason is that at the time of the measurement, the influence of the back decay to the initial 2S level, which causes this difference, was not fully understood.

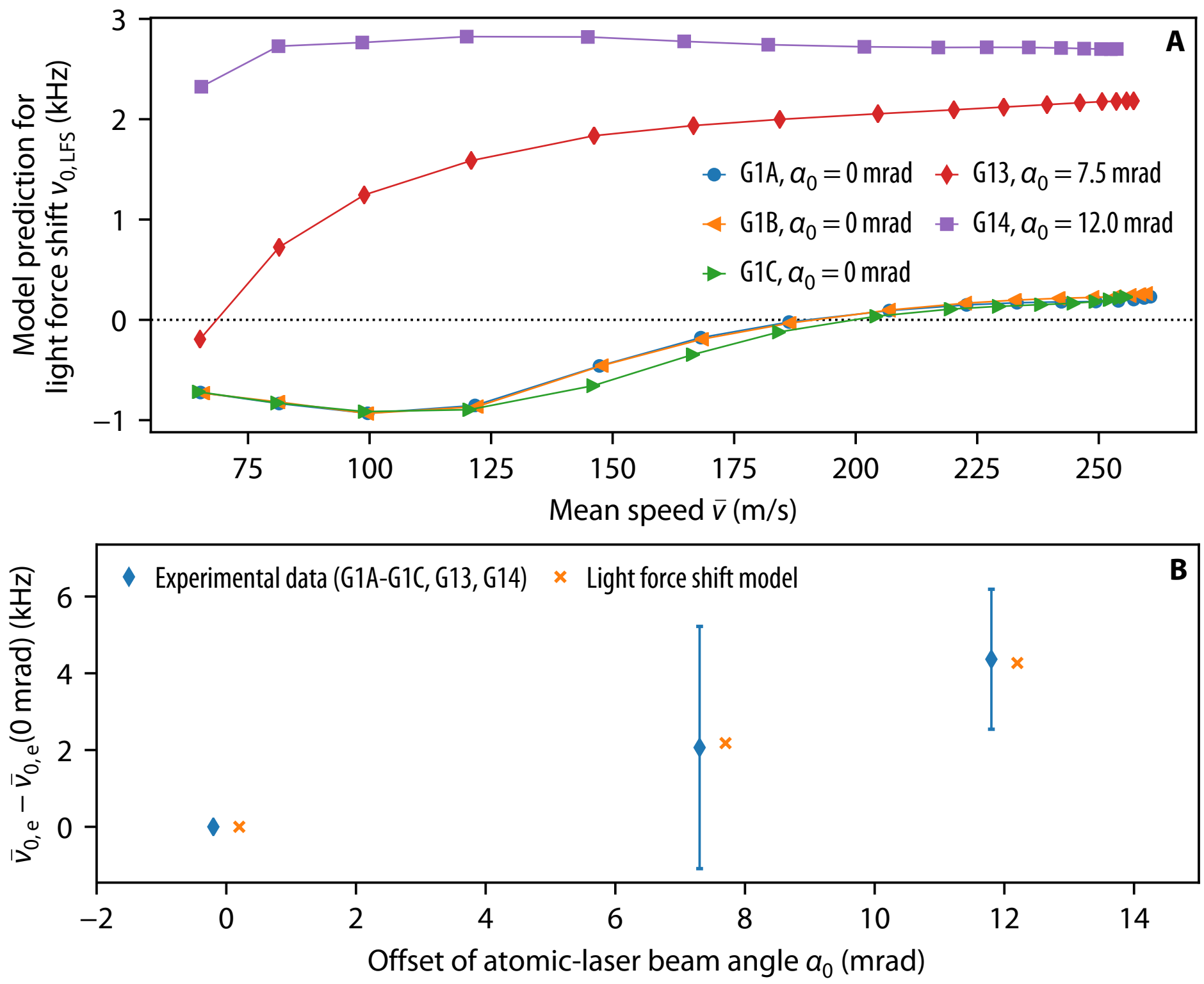

Figure 6.6: (**A**) Model prediction for the LFS $\nu_{0,\mathrm{LFS}}$ versus mean speed $\bar{v}$ of the delays, shown for data groups G1A–G1C, G13, and G14. The LFS is mainly negative for zero $\alpha_0$ (G1A–G1C, blue, orange, and green lines), but positive for $\alpha_0 = 7.5\,\mathrm{mrad}$ (G13, red line) and $\alpha_0 = 12.0\,\mathrm{mrad}$ (G14, purple line). This is because the LFS changes from a negative to a positive shift for transverse velocities above the recoil velocity $v_{\mathrm{rec}} \approx 0.97\,\mathrm{m/s}$, with the transverse velocity increasing as $\alpha_0$ is increased. The slight differences between G1A–G1C are mainly caused by the different powers of the 1S-2S preparation laser used for these data groups. (**B**) Measurement of the light force shift (LFS) of the 2S-6P$_{1/2}$ transition versus offset of atomic–laser beam angle, $\alpha_0$. The experimentally determined Doppler-free resonance frequency $\bar{\nu}_{0,\mathrm{e}}$ (blue diamonds) and the line shift predicted by the LFS model (orange crosses) are shown relative to their values for $\alpha_0 = 0\,\mathrm{mrad}$, where the predicted shift amounts to $-1.49\,\mathrm{kHz}$. The experimental data has been corrected for quantum interference line shifts and second-order Doppler shift, which are much smaller than the LFS observed here. The uncertainty of the experimental data for $\alpha_0 = 0\,\mathrm{mrad}$ has been absorbed into the uncertainty for $\alpha_0 \neq 0\,\mathrm{mrad}$ and is thus not shown. Only the statistical uncertainty, but not the uncertainty of the corrections, is taken into account. The model is in excellent agreement with the experimental data. The spectroscopy laser power $P_{\mathrm{2S\text{-}6P}} = 30\,\mu\mathrm{W}$ and linear laser polarization angle $\theta_{\mathrm{L}} = 56.5\,\mathrm{mrad}$ are the same for all data shown (data groups G1A–G1C, G13, G14). A small offset has been added along the $x$-axis for clarity.

The experimentally determined Doppler-free resonance frequency $\bar{\nu}_{0,\mathrm{e}}$ from these data groups is shown in Fig. 6.6 (B) (blue diamonds), along with the line shift predicted by the LFS model (orange crosses), versus $\alpha_0$. Both the experimental data and the prediction are shown relative to their values for $\alpha_0 = 0\,\mathrm{mrad}$, where the predicted shift amounts to $-1.49\,\mathrm{kHz}$. The experimental data has been corrected for quantum interference line shifts and second-order Doppler shift, which are much smaller than the LFS observed here, but not for the LFS as is done here elsewhere. The uncertainty of the experimental data for $\alpha_0 = 0\,\mathrm{mrad}$ has been absorbed into the uncertainty for $\alpha_0 \neq 0\,\mathrm{mrad}$. Because there is only one measurement day of data for $\alpha_0 = 7.5\,\mathrm{mrad}$, the uncertainty for this value is large and it does not contribute significantly to this test.

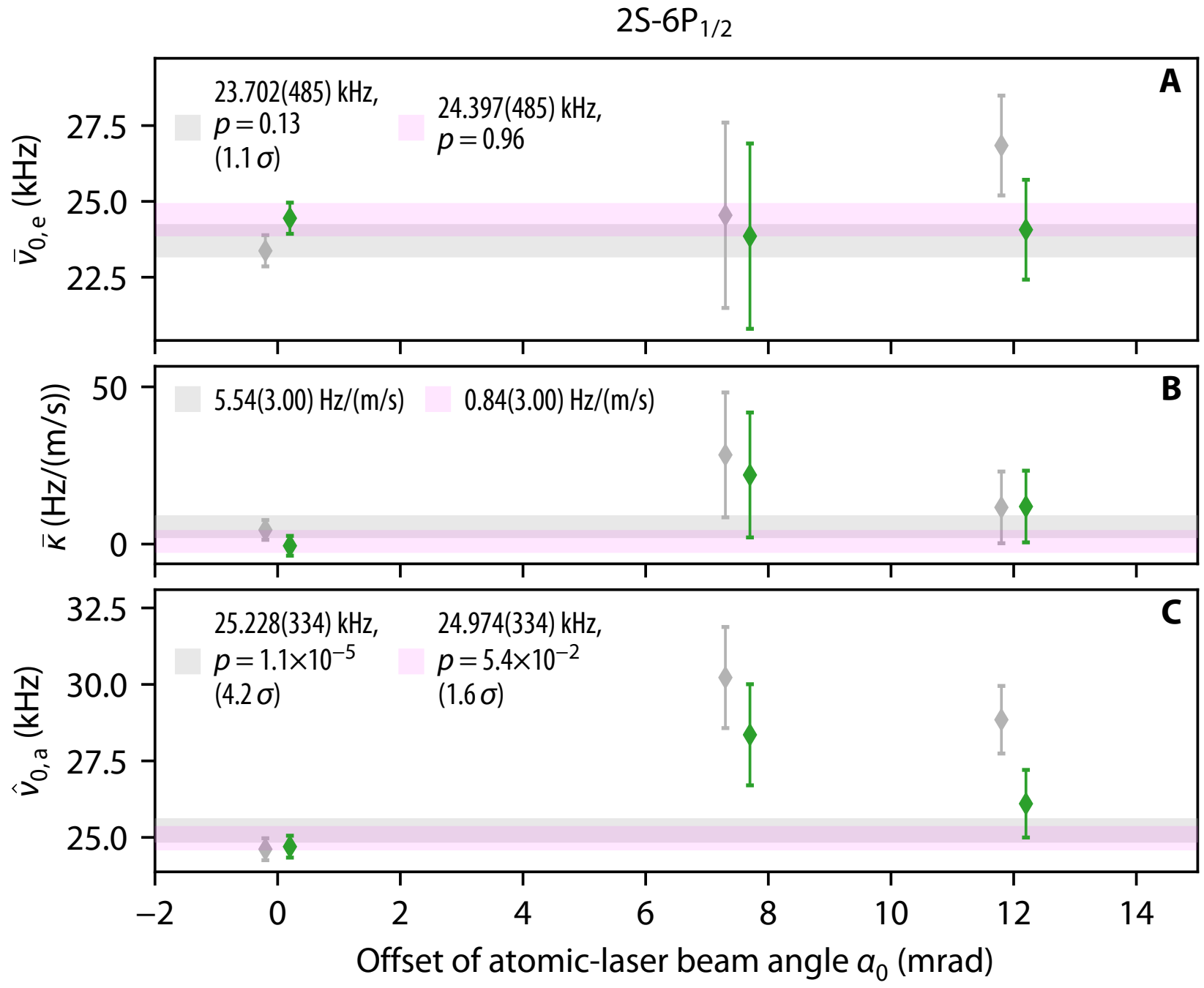

Figure 6.7: Blinded results of the measurement of the 2S-6P$_{1/2}$ transition versus offset of atomic–laser beam angle, $\alpha_0$ (data groups G1–G6, G13, G14). Shown are the detector- and freezing-cycle-averaged (**A**) Doppler-free resonance frequency $\bar{\nu}_{0,\mathrm{e}}$, (**B**) Doppler slope $\bar{\kappa}$, and (**C**) scaled Doppler-averaged resonance frequency $\hat{\nu}_{0,\mathrm{a}}$, without (gray diamonds) and with (green diamonds) corrections for the light force shift (LFS) taken into account. Both include corrections for quantum interference line shifts and second-order Doppler shift. The gray and purple bands indicate the $1\sigma$ uncertainty region of weighted averages without and with corrections, respectively, with the $p$-value and significance $Z\sigma$ given in the legend. Only the statistical uncertainty, but not the uncertainty of the corrections, is taken into account. A small offset has been added along the $x$-axis for clarity.

As can be seen from Fig. 6.6 (B), the LFS model is in excellent agreement with the experimental data. The frequency difference between the data for $\alpha_0 = 0\,\mathrm{mrad}$ and $\alpha_0 = 12\,\mathrm{mrad}$ is found to be

$$\bar{\nu}_{0,\mathrm{e}}(12.0\,\mathrm{mrad}) - \bar{\nu}_{0,\mathrm{e}}(0\,\mathrm{mrad}) = 4.37(1.82)\,\mathrm{kHz}, \tag{6.8}$$

which only contains negligible corrections of 173 Hz and 25 Hz for QI line shifts and the SOD, respectively. The corresponding prediction of the LFS model is

$$\nu_{0,\mathrm{e}}(12.0\,\mathrm{mrad}) - \nu_{0,\mathrm{e}}(0\,\mathrm{mrad}) = 4.27\,\mathrm{kHz}. \tag{6.9}$$

The uncertainty of the prediction, which should be found by varying the parameters entering the modeling of the atomic beam within their experimental constraints, has not yet been evaluated in detail, but is estimated to be below 20 % based on preliminary evaluations. Note that the parameters used for the prediction given here were not chosen to best reproduce the LFS, but are either given by geometry, measured laser parameters, or determined from the data using the big model, but not the LFS model, as described in Section 6.1.2.

Fig. 6.7 (A), (B), and (C) show the influence of the LFS corrections on the Doppler-free resonance frequency $\bar{\nu}_{0,\mathrm{e}}$, Doppler slope $\bar{\kappa}$, and scaled Doppler-averaged resonance frequency $\hat{\nu}_{0,\mathrm{a}}$, respectively, for all data for the 2S-6P$_{3/2}$ transition, versus $\alpha_0$. Including the LFS corrections (green diamonds and purple band) improves the consistency for $\bar{\nu}_{0,\mathrm{e}}$, $\bar{\kappa}$, $\hat{\nu}_{0,\mathrm{a}}$ over the case where only the corrections for quantum interference line shifts and second-order Doppler shift are included (gray diamonds and band).

### 6.2.3 Quantum interference line shift corrections

Quantum interference (QI) line shifts are strongly suppressed by the large solid angle of the fluorescence detection and the choice of the linear laser polarization angle $\theta_\mathrm{L}$. Fig. 6.8 shows a simulation of the QI line shifts of the Doppler-free resonance frequency $\nu_{0,\mathrm{e}}$ versus the linear laser polarization angle $\theta_\mathrm{L}$, for the 2S-6P$_{1/2}$ (solid lines) and 2S-6P$_{3/2}$ transition (dashed lines). The simulation uses the big model (see Section 2.3.2) in conjunction with simulations of the spatial detection efficiency (Table 4.2). The QI line shifts are shown for the four different detection efficiency simulations of Fig. 4.41, corresponding to different values for the transparency of the wire meshes inside the fluorescence detector assembly, for the top detector (colored lines) and the bottom detector (faint colored lines). The spectroscopy laser power is set to $P_{\text{2S-6P}} = 30\,\mu\mathrm{W}$ and $P_{\text{2S-6P}} = 15\,\mu\mathrm{W}$, with the remaining experimental parameters corresponding to those of data groups G1B and G10 for the 2S-6P$_{1/2}$ and 2S-6P$_{3/2}$ transition, respectively. For $P_{\text{2S-6P}} = 10\,\mu\mathrm{W}$ and $P_{\text{2S-6P}} = 5\,\mu\mathrm{W}$, the line shifts are approximately smaller by one-third.

As expected from the perturbative analysis of the 2S-$n$P transitions [29], the line shifts are approximately sinusoidal in $\theta_\mathrm{L}$, of opposite sign for the 2S-6P$_{1/2}$ and 2S-6P$_{3/2}$ transition, and approximately a factor of two larger for the 2S-6P$_{1/2}$ transition. Importantly, the sinusoidal behavior is not symmetric about zero line shift, i.e., a simple average over $\theta_\mathrm{L}$ leads to a systematic offset in $\nu_{0,\mathrm{e}}$. The line shifts are very similar for the different detection efficiency simulations and the two detectors, reaching an approximate amplitude of 5.1 kHz and 2.6 kHz for the 2S-6P$_{1/2}$ and 2S-6P$_{3/2}$ transition, respectively. The QI line shifts are approximately a factor of 8 smaller than for the 2S-4P measurement (see Fig. 4 (A, B) of Appendix A), which is a result of both the lower linewidth and larger detection solid angle of the 2S-6P measurement.

Most of the experimental data were taken for $\theta_\mathrm{L} = 56.5°$ (light blue bar), where the QI line shifts are approximately minimal, i.e., close to the magic angle. For the estimated alignment uncertainty of $\pm 3°$ of $\theta_\mathrm{L}$, the shifts are within $-250\,\mathrm{Hz}$ and $-1.4\,\mathrm{kHz}$ for the 2S-6P$_{1/2}$ transition and 125 Hz and 0.7 kHz for the 2S-6P$_{3/2}$ transition. Some data were also taken at $\theta_\mathrm{L} = 146.5°$ (light blue bar), where the line shifts are larger, reaching up to 3.8 kHz and 2.0 kHz for the two transitions. The shifts are also of opposite sign at $\theta_\mathrm{L} = 146.5°$ as compared to $\theta_\mathrm{L} = 56.5°$, leading to a partial cancellation of the QI line shifts if all data is averaged as done here to find the transition frequencies.

As was done for the 2S-4P measurement, one may take advantage of the fact that the line shifts are of opposite sign for the two transitions and combine their transition frequencies with appropriate weighting into the 2S-6P centroid as defined in Eq. (6.21). This reduces the observed shift to below 200 Hz (dotted lines) for all values of $\theta_{\mathrm{L}}$. For $\theta_{\mathrm{L}} = 56.5°$ (146.5°), the QI line shifts of the centroid are below 20 Hz (50 Hz).

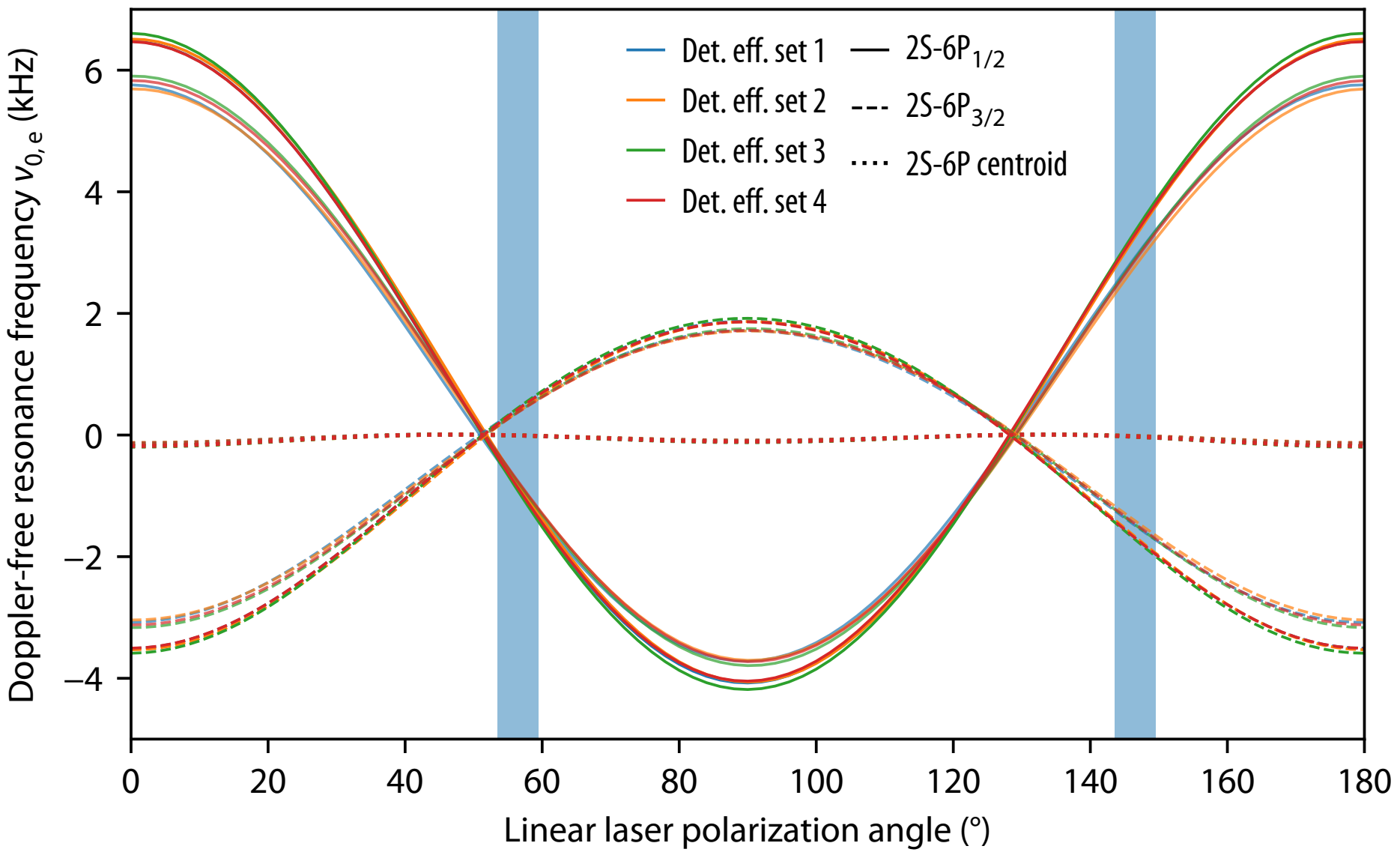

Figure 6.8: Simulation, using the big model, of the shift of the Doppler-free resonance frequency $\nu_{0,\mathrm{e}}$ from quantum interference (QI) versus the linear laser polarization angle $\theta_{\mathrm{L}}$, for the 2S-6P$_{1/2}$ (solid lines) and 2S-6P$_{3/2}$ transition (dashed lines). $\nu_{0,\mathrm{e}}$ is found by fitting Voigt line shapes to the individual delays and extrapolating to zero speed, as is done for the experimental data. The observed shift depends on the detection efficiency, which is determined from simulations, four of which for different parameters are shown here for the top detector (colored lines) and the bottom detector (faint colored lines) (see Fig. 4.41 and Table 4.2). Combining the 2S-6P$_{1/2}$ and 2S-6P$_{3/2}$ transitions into the 2S-6P centroid (see Eq. (6.21)) reduces the observed shift to below 200 Hz (dotted lines). The values of $\theta_{\mathrm{L}}$ used in the experiment, $\theta_{\mathrm{L}} = 56.5°$ (146.5°) (light blue bars), were chosen to minimize the QI line shifts, which are below 20 Hz (50 Hz) for the centroid. The spectroscopy laser power is set to $P_{\text{2S-6P}} = 30\,\mu\mathrm{W}$ and $P_{\text{2S-6P}} = 15\,\mu\mathrm{W}$, with the experimental parameters corresponding to those of data groups G1B and G10 for the 2S-6P$_{1/2}$ and 2S-6P$_{3/2}$ transition, respectively. The QI line shifts are approximately a factor of 8 smaller than for the 2S-4P measurement (see Fig. 4 (A, B) of Appendix A).

Fig. 6.9 (A–C) and (D–E) show the corrections for the QI line shifts as applied to the measurement of the 2S-6P$_{1/2}$ and 2S-6P$_{3/2}$ transition, respectively, versus $\theta_{\mathrm{L}}$. Data with nonzero $\alpha_0$ is excluded, as there is no such data for $\theta_{\mathrm{L}} = 146.5°$. With the QI corrections included, the consistency of the data generally increases, except for $\bar{\nu}_{0,\mathrm{e}}$ for the 2S-6P$_{3/2}$ transition, where the corrections lead to a slight strain in the data.

The uncertainties of the QI corrections are yet to be evaluated in more detail by varying more of the parameters entering the detection efficiency simulation. However, forming the 2S-6P centroid (see Section 6.2.1.3) reduces the QI corrections to below 100 Hz, which is well below the statistical uncertainty.

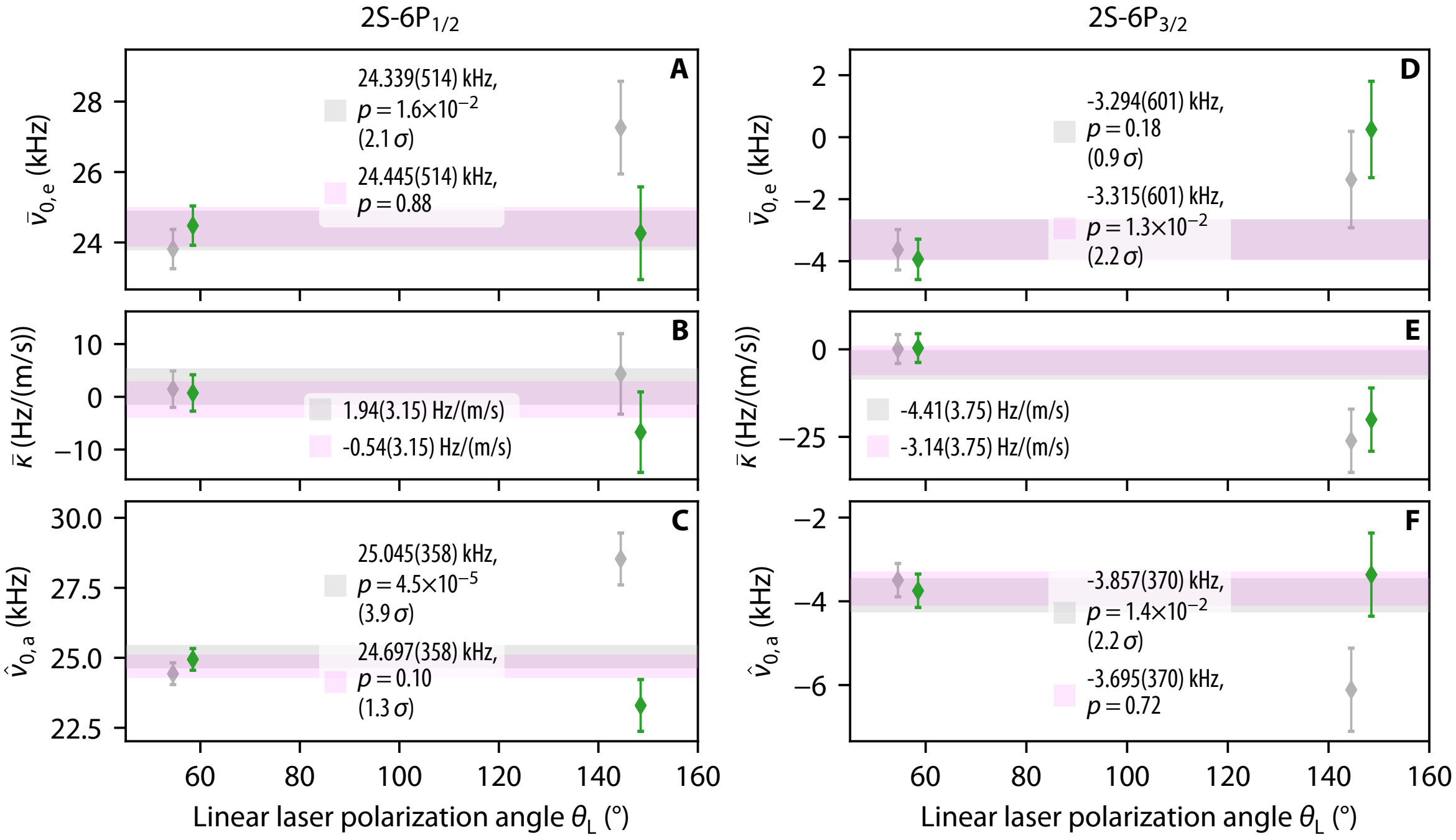

Figure 6.9: Blinded results of the measurement of the (**A**–**C**) 2S-6P$_{1/2}$ and (**D**–**F**) 2S-6P$_{3/2}$ transition versus the linear laser polarization angle of the 2S-6P spectroscopy laser, $\theta_\mathrm{L}$. Only data for which $\alpha_0 = 0\,\mathrm{mrad}$ are considered (data groups G1–G12). Shown are the detector- and freezing-cycle-averaged (**A, D**) Doppler-free resonance frequency $\bar{\nu}_{0,\mathrm{e}}$, (**B, E**) Doppler slope $\bar{\kappa}$, and (**C, F**) scaled Doppler-averaged resonance frequency $\hat{\nu}_{0,\mathrm{a}}$, without (gray diamonds) and with (green diamonds) corrections for quantum interference line shifts, while the light force shift corrections are always included. The gray and purple bands indicate the $1\sigma$ uncertainty region of weighted averages without and with corrections, respectively, with the $p$-value and significance $Z\sigma$ given in the legend. Only the statistical uncertainty, but not the uncertainty of the corrections, is taken into account. A small offset has been added along the $x$-axis for clarity.

### 6.2.4 Additional corrections and uncertainties

#### 6.2.4.1 Second-order Doppler shift

The second-order Doppler shift (SOD) is treated in Section 2.2.3. It leads to a shift of the transition frequency as seen in the laboratory frame of reference by $\Delta\nu_\mathrm{SOD}$, given in Eq. (2.8) for an atom flying through the spectroscopy laser beams with speed $v$. To describe the average SOD for a given delay, which contains a distribution of speeds, $v$ is replaced with the root mean square speed $\bar{v}_\mathrm{RMS}$. This results in

$$\Delta\nu_\mathrm{SOD} = -\frac{\nu_{\mathrm{A},0}}{2}\left(\frac{\bar{v}_\mathrm{RMS}}{c}\right)^2. \tag{6.10}$$

$\bar{v}_\mathrm{RMS}$ and $\Delta\nu_\mathrm{SOD}$ range, on average, between 274 m/s…67 m/s and −304 Hz…−18 Hz for delays 1…16.

The resonance frequency $\nu_0$ of each delay of each line scan is corrected for the SOD by subtracting $\Delta\nu_\mathrm{SOD}$ for the corresponding value of $\bar{v}_\mathrm{RMS}$, i.e., the correction leads to a higher resonance frequency. Consequently, the Doppler-averaged resonance frequency $\nu_{0,\mathrm{a}}$ is corrected upwards by 187 Hz (186 Hz) for the 2S-6P$_{1/2}$ (2S-6P$_{3/2}$) transition. However, the always negative SOD increases (quadratically) in magnitude with the speed, which in the linear Doppler extrapolation leads to an additional negative Doppler slope of $-1.7\,\mathrm{Hz/(m/s)}$ for both transitions. The effect of this Doppler slope outweighs the shift of the individual delays, and the Doppler-free resonance frequency $\nu_{0,\mathrm{e}}$ is corrected downwards, i.e., in the opposite direction as $\nu_{0,\mathrm{a}}$, by $-145\,\mathrm{Hz}$ ($-142\,\mathrm{Hz}$) for the 2S-6P$_{1/2}$ (2S-6P$_{3/2}$) transition. The corrections for the SOD are also shown in Fig. 6.4.

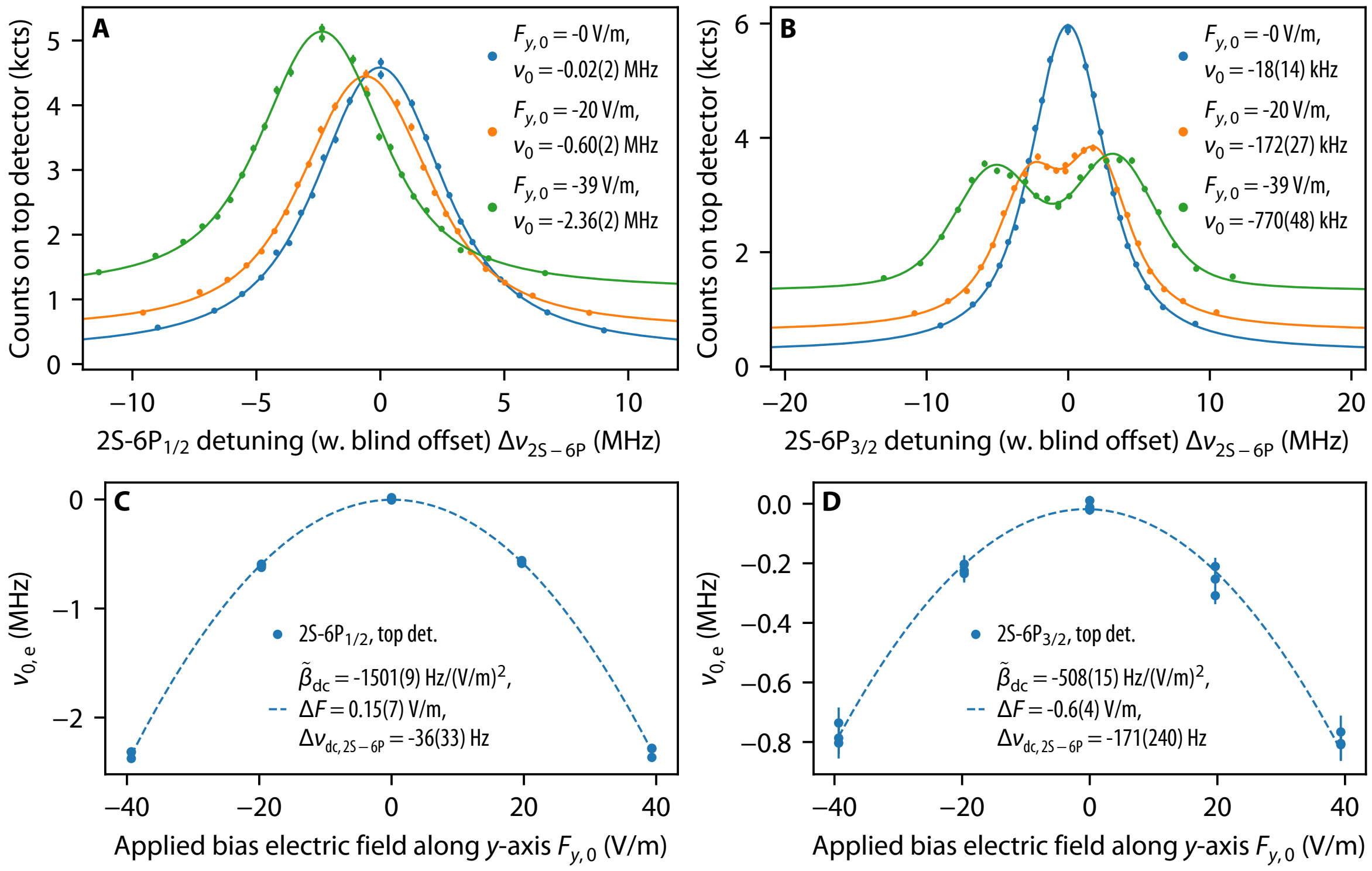

Figure 6.10: In situ determination of stray electric fields along the $y$-axis and the corresponding dc-Stark shift for the (**A**, **C**) 2S-6P$_{1/2}$ transition and the (**B**, **D**) 2S-6P$_{3/2}$ transition. Voltages $U_{+y}$ and $U_{-y} = -U_{+y}$ are applied to the top and bottom electrode, respectively, of the detector assembly's inner region, resulting in a bias electric field along the $y$-axis with field strength $F_{y,0} \propto U_{+y}$ at the center of the inner region (see Table 4.3). (**A, B**) Delay 13 of line scans for different values of $F_{y,0}$ (colored circles), along with fits (colored lines) of Voigt line shapes or Voigt doublet line shapes (for the 2S-6P$_{3/2}$ transition and nonzero $F_{y,0}$). The resonance of the 2S-6P$_{3/2}$ transition splits into two components if a large enough electric field is applied, resulting from the coupling to the 6D levels, which are separated in frequency by much less than the linewidth (see Section 2.4). The dc-Stark shift of the center of mass $\nu_0$, as given by the Voigt doublet fit, is much smaller than that of the individual resonances. The signal on the top detector is shown, which increases for negative $F_{y,0}$ as electrons are pulled from the inner region towards the top section of the detector assembly. (**C, D**) Doppler-free resonance frequency $\nu_{0,\mathrm{e}}$ versus field strength $F_{y,0}$. Parabolas are fit to these data (see Eq. (2.48)) to reveal the effective quadratic dc-Stark shift coefficients $\tilde{\beta}_{\mathrm{dc}}$, and the strengths of the stray electric field $\Delta F$, along the direction of the bias field. The dc-Stark shifts caused by the stray fields are then given by $\Delta\nu_{\mathrm{dc,2S\text{-}6P}} = \tilde{\beta}_{\mathrm{dc}}\Delta F^2$. The power of the 2S-6P spectroscopy laser was $P_{\mathrm{2S\text{-}6P}} = 30\,\mu\mathrm{W}$ ($15\,\mu\mathrm{W}$) for the 2S-6P$_{1/2}$ (2S-6P$_{3/2}$) transition.

#### 6.2.4.2 dc-Stark shift

Stray electric fields inside the detector assembly can lead to dc-Stark shifts of the observed transition frequency (see Section 2.4). The stray electric fields $\Delta F$ were determined in situ during the 2S-6P measurement by applying bias electric fields along the different directions (see Section 4.6.7) and determining the resulting dc-Stark shift of the 2S-6P transition. This procedure also gives the effective dc-Stark shift coefficients $\tilde{\beta}_{\text{dc}}$, i.e., the shift of the resonance observed in fluorescence as opposed to the shift of the atomic energy levels (see Section 2.4), from which the dc-Stark shifts caused by the stray fields can be determined as $\Delta\nu_{\text{dc,2S-6P}} = \tilde{\beta}_{\text{dc}}\Delta F^2$. An example of such a determination is shown in Fig. 6.10.

On most measurement days of the 2S-6P measurement, such determinations were performed at least once for each direction. In total, 269 determinations were performed, of which 223 and 46 used the 2S-$6P_{1/2}$ and 2S-$6P_{3/2}$ transition, respectively. For some freezing cycles dedicated to the measurement of the 2S-$6P_{3/2}$ transition, the stray electric field determination nevertheless used the 2S-$6P_{1/2}$ transition because the resulting line scans are simpler to analyze and the effective dc-Stark shift is larger, increasing the precision of the determination. These determinations have only been preliminary evaluated so far, with the stray fields found to be typically below 0.1 V/m (0.3 V/m) along the $x$- and $z$-axis ($y$-axis), corresponding to an observed total shift below 500 Hz. The uncertainty with which this shift can be corrected for is currently evaluated. The data shown here have not yet been corrected for the dc-Stark shift.

#### 6.2.4.3 Zeeman shift

The earth's magnetic field inside the beam apparatus is compensated by three orthogonal pairs of Helmholtz coils outside the vacuum chamber (see Fig. 4.2). In addition, the 2S-6P spectroscopy region is shielded from residual magnetic fields by a single-layer high-permeability metal (mu-metal) shield surrounding the high-vacuum enclosure (see Fig. 4.1). All components inside this enclosure, and the enclosure itself, are made from non-magnetic materials. Within a 5 mm-radius sphere centered in the 2S-6P spectroscopy region, the magnetic flux density $\boldsymbol{B}$ was measured to have a maximum magnitude $|\boldsymbol{B}|$ of 7 mG, with the field component $B_x$ along the $x$-axis below 1 mG.

For a given magnetic flux density $B$ along the quantization axis, the linear Zeeman effect shifts the energies of the magnetic sublevels with $m_F = \pm 1$ of the $6P_{1/2}^{F=1}$ manifold by

$$\Delta\nu = m_F g_F \mu_{\text{B}} B/h = m_F\, B \times (0.467\,\text{kHz/mG}). \tag{6.11}$$

Likewise, the sublevels of the $6P_{3/2}^{F=1}$ manifold are shifted by

$$\Delta\nu = m_F g_F \mu_{\text{B}} B/h = m_F\, B \times (2.33\,\text{kHz/mG}). \tag{6.12}$$

$g_F$ is the appropriate g-factor and $\mu_{\text{B}}$ is the Bohr magneton. Note that $g_F$ is independent of the principal quantum number $n$.

To analyze the Zeeman shifts for the experiment discussed here, we first assume that the magnetic field points along the propagation direction of the spectroscopy laser beams, i.e., along the $x$-axis of the apparatus, which is chosen as the atom's quantization axis. In this frame, the laser field is given by a combination of two spherical components that can be identified as the right- and left-handed circularly polarized components of the laser beams with amplitude $E_{\sigma^+}$ and $E_{\sigma^-}$, driving the transition to the $m_F = 1$ and $m_F = -1$ magnetic sublevels, respectively. The fluorescence signal from the $m_F = 1$ ($m_F = -1$) sublevel has an amplitude approximately proportional to $|E_{\sigma^+}|^2$ ($|E_{\sigma^-}|^2$), and is shifted in frequency by $g_F\mu_B B_x/h$ ($-g_F\mu_B B_x/h$), where $B_x$ is the magnetic flux density along the $x$-axis. The center of weight of the fluorescence signal, and thus the observed transition frequency, is then shifted by

$$\Delta\nu_{\text{Zeeman}} = \frac{|E_{\sigma^+}|^2 - |E_{\sigma^-}|^2}{|E_{\sigma^+}|^2 + |E_{\sigma^-}|^2} \frac{g_F\mu_B B_x}{h} = \frac{S_3}{S_0} \frac{g_F\mu_B B_x}{h}, \tag{6.13}$$

where $S_3/S_0$ is residual circularly polarized light fraction as defined in Section 4.4.6.

To analyze the situation where the propagation direction of the laser beams is not collinear with the magnetic field, the atom's quantization axis is kept along the magnetic field and the laser field is decomposed again in this frame. In general, this leads to three spherical components, $E_{\sigma^+}$, $E_{\sigma^-}$, and $E_\pi$, where $E_\pi$ drives the transition to the unshifted $m_F = 0$ sublevel. Note that $E_{\sigma^+}$ and $E_{\sigma^-}$ are now not identical to the right- and left-handed components of the laser beams. Taking, without loss of generality, that the magnetic field with flux density $B$ is oriented at an angle $\theta$ from the laser propagation direction, this decomposition results in $\Delta\nu_{\text{Zeeman}} = (S_3/S_0) g_F\mu_B B\cos\theta/h$. This is equivalent to Eq. (6.13), since $B\cos\theta \equiv B_x$, and thus only the magnetic field component along the laser beams needs to be considered.

The residual circularly polarized light fraction at the position of the atoms, $|(S_3/S_0)_{\text{atom}}|$, was measured in-situ during most line scans as detailed in Section 4.4.6. It is found to be below 10 % (see Fig. 4.24). Using this value, and the maximum value of $B_x$ as given above, results in an upper limit for the Zeeman shifts of $|\Delta\nu_{\text{Zeeman}}| = 47$ Hz and $|\Delta\nu_{\text{Zeeman}}| = 233$ Hz for the 2S-6P$_{1/2}$ and 2S-6P$_{3/2}$ transition, respectively. The Zeeman shift and the associated uncertainty is not included in the preliminary analysis shown here.

Towards the end of the 2S-6P measurement, some line scans of the 2S-6P$_{3/2}$ transition with an applied bias field of $B_x \approx 50$ mG were acquired to serve as an additional check on the estimation of the Zeeman shift. This data needs to be studied in detail, but in the preliminary analysis no shift of the observed transition frequency within the uncertainty of 3 kHz was found.

#### 6.2.4.4 Pressure shift

The interaction, or collisions, of the hydrogen atoms in the beam with other nearby particles can lead to pressure shifts of the observed transition frequency [161]. Two classes of collisions are distinguished here: intra-beam collisions with other atoms or hydrogen molecules in the beam, and collisions with particles from the background gas. The particles in the atomic beam are assumed to be at the temperature of the nozzle $T_N$, while the background gas is taken to be at room temperature. Furthermore, it is assumed that the background gas consists solely of hydrogen molecules.

For the 2S-6P measurement, a flow of hydrogen molecules into the dissociator of up to $Q_{H_2} = 0.35$ ml/min was used. The resulting number of atoms and molecules leaving the nozzle per second in the direction of the spectroscopy region is estimated in Section 4.5.2.3 to be $N'_{1S} = 1.6 \times 10^{16}$ atoms/s and $N'_{H_2} = 1.8 \times 10^{16}$ molecules/s, respectively. The background pressure in the spectroscopy region, and thus the pressure of room-temperature hydrogen molecules, is estimated to be $P_{HV1} = 1 \times 10^{-7}$ mbar (see Sections 4.2.4 and 4.2.10).

Arthur Matveev and colleagues have performed a Monte Carlo simulation of the pressure shift expected for the 2S-6P$_{1/2}$ and 2S-6P$_{3/2}$ transition from intra-beam collisions with other hydrogen atoms [162]. The simulation used the approximate parameters and geometry of the 2S-6P measurement, including the 1S-2S excitation and the time-resolved detection. From this, the pressure shift was found to be below 10 Hz for all delays. Scaling these results for the actually used parameters and geometry, especially the tenfold lower flux of hydrogen atoms, further reduces this upper limit to 1 Hz.

A. M. has also estimated the pressure shift from hydrogen molecules, both from intra-beam and background collisions. A corresponding publication is in preparation and the numbers reproduced here are preliminary. The estimates give an upper limit of the pressure shift from intra-beam collisions with hydrogen molecules of 0.2 Hz, and an upper limit of 5 Hz from collisions with the background gas of hydrogen molecules.

#### 6.2.4.5 Recoil shift

As detailed in Section 2.2.3, the measured laser frequency $\nu_{\mathrm{L},0}$ must be corrected for the recoil shift $\Delta\nu_{\mathrm{rec}}$ (see Eq. (2.6)) in order to determine the atom's resonance frequency $\nu_{\mathrm{A},0}$. Using the value of $\Delta\nu_{\mathrm{rec}}$ given in Table 2.1, the Doppler-free and Doppler-averaged resonance frequencies of the 2S-6P$_{1/2}$ and 2S-6P$_{3/2}$ transition need to be corrected for a recoil shift of

$$\Delta\nu_{\mathrm{rec}} = 1176.03\,\mathrm{kHz}. \tag{6.14}$$

As this correction is identical to all data groups, it is of no consequence to the blinded results given here.

#### 6.2.4.6 Fine- and hyperfine-structure corrections

In the measurement presented here, the two transition frequencies $\nu_{1/2}$ and $\nu_{3/2}$ from the hyperfine level $2\mathrm{S}_{1/2}^{F=0}$ to either of the hyperfine levels $6\mathrm{P}_{1/2}^{F=1}$ and $6\mathrm{P}_{3/2}^{F=1}$ are determined. However, some systematic effects, especially the quantum interference shifts, cancel to a large degree when combining these measured frequencies to what is here referred to as the 2S-6P centroid $\nu_{\text{2S-6P}}$. $\nu_{\text{2S-6P}}$ is the transition frequency from the 2S hyperfine centroid, i.e., the level energy if the nuclear spin was zero, to the 6P line-strength-averaged fine-structure centroid. The relevant levels and transition frequencies, and the corrections discussed below, are shown in Fig. 6.11.

First, the contribution from the hyperfine structure to the measured transition frequencies $\nu_{1/2}$ and $\nu_{3/2}$ is considered. The hyperfine interaction splits the fine-structure levels, of which $2\mathrm{S}_{1/2}$, $6\mathrm{P}_{1/2}$, and $6\mathrm{P}_{3/2}$ are here of interest, into doublets [39, 40]. Here, the resulting $2\mathrm{S}_{1/2}^{F=0}$, $6\mathrm{P}_{1/2}^{F=1}$, and $6\mathrm{P}_{3/2}^{F=1}$ hyperfine levels are probed, which are shifted from the fine-structure levels by the hyperfine energies $\Delta\nu_{\mathrm{HFS}}(2\mathrm{S}_{1/2}^{F=0})$, $\Delta\nu_{\mathrm{HFS}}(6\mathrm{P}_{1/2}^{F=1})$, and $\Delta\nu_{\mathrm{HFS}}(6\mathrm{P}_{3/2}^{F=1})$, respectively.

The value of $\Delta\nu_{\mathrm{HFS}}(2\mathrm{S}_{1/2}^{F=0})$ and its uncertainty can be obtained from a measurement of the 2S hyperfine splitting $\Delta\nu_{\mathrm{HFS}}(2\mathrm{S}_{1/2})$ [57] through

$$\Delta\nu_{\mathrm{HFS}}(2\mathrm{S}_{1/2}^{F=0}) = -(3/4)\Delta\nu_{\mathrm{HFS}}(2\mathrm{S}_{1/2}). \tag{6.15}$$

$\Delta\nu_{\mathrm{HFS}}(6\mathrm{P}_{1/2}^{F=1})$ and $\Delta\nu_{\mathrm{HFS}}(6\mathrm{P}_{3/2}^{F=1})$ can be obtained from hydrogen theory as detailed in [39, 40]. They include a small correction from off-diagonal elements in the hyperfine Hamiltonian, leading to an energy shift of hyperfine levels with the same value of $F$, but different values of $J$ [163, 164]. These corrections to the level energies amount to $\Delta\nu_{\mathrm{HFS}}^{\mathrm{o.d.}}(6\mathrm{P}_{1/2}^{F=1}) = -\Delta\nu_{\mathrm{HFS}}^{\mathrm{o.d.}}(6\mathrm{P}_{3/2}^{F=1}) = -92(1)\,\mathrm{Hz}$ [40].

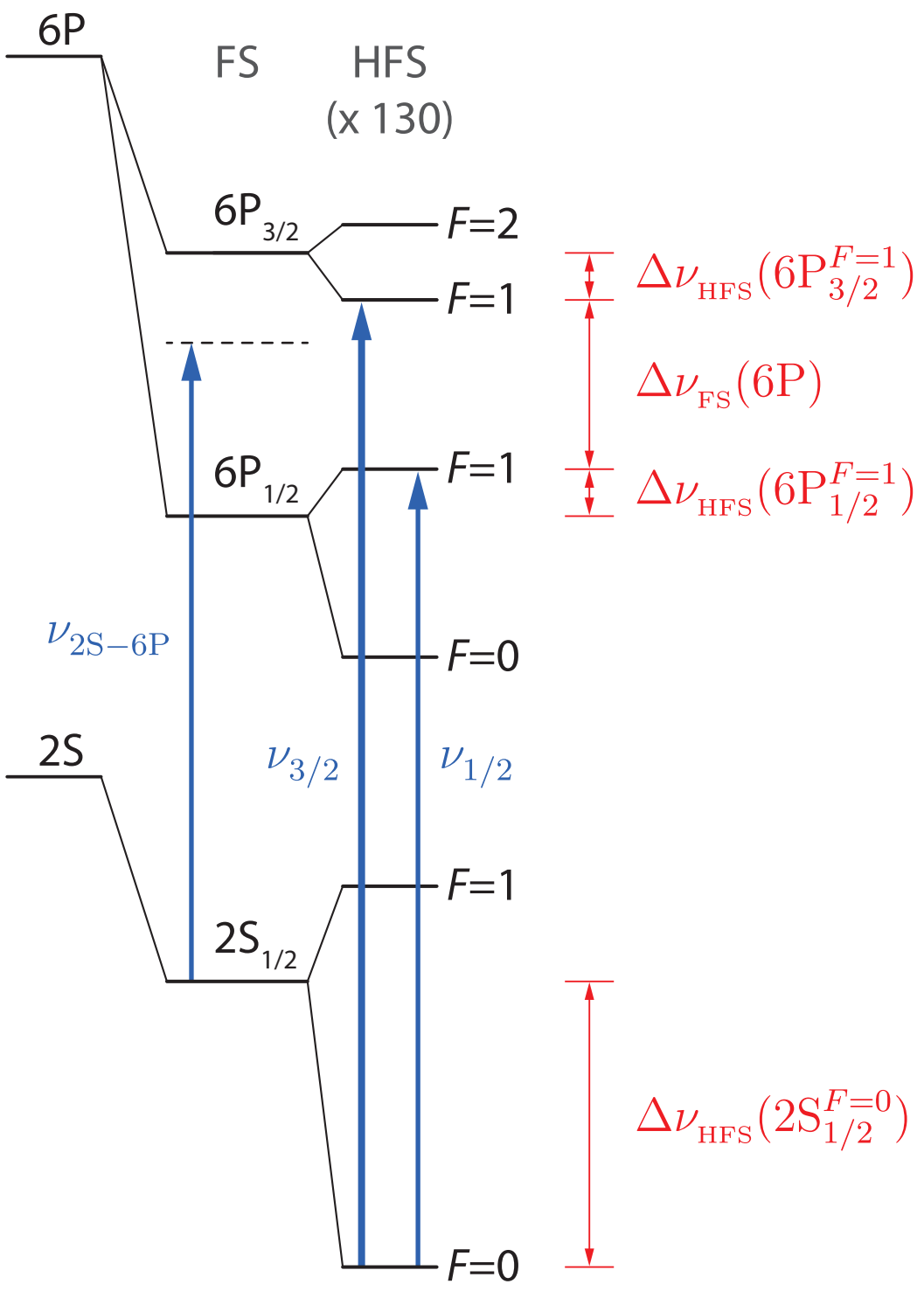

Figure 6.11: Level scheme (not to scale) of the 2S-6P transition including fine structure (FS) and hyperfine structure (HFS). The transition frequencies of the $2S^{F=0}_{1/2}-6P^{F=1}_{1/2}$ ($\nu_{1/2}$) and $2S^{F=0}_{1/2}-6P^{F=1}_{3/2}$ ($\nu_{3/2}$) transitions (line strength ratio 1:2) are experimentally determined. The transition frequency from the 2S HFS centroid to the line-strength-averaged 6P FS centroid, the 2S-6P centroid $\nu_{\text{2S-6P}}$, is determined by combining $\nu_{1/2}$ and $\nu_{3/2}$ and correcting for the hyperfine shifts $\Delta\nu_{\text{HFS}}(2S^{F=0}_{1/2})$, $\Delta\nu_{\text{HFS}}(6P^{F=1}_{1/2})$, and $\Delta\nu_{\text{HFS}}(6P^{F=1}_{3/2})$. The 6P fine-structure splitting $\Delta\nu_{\text{FS}}(6P)$ here corresponds to the energy difference of the $6P^{F=1}_{1/2}$ and $6P^{F=1}_{3/2}$ levels. In this drawing, the scale of the HFS is 130 times larger than for the FS, and the scale of the 6P levels is 40 times larger than for the 2S levels.

Since for each hyperfine doublet only one level is shifted by this effect, the center of gravity of the hyperfine doublets, as compared to the situation where the off-diagonal elements are neglected, is also shifted by $\Delta\nu^{\text{o.d.}}_{\text{HFS}}(6P_{1/2}) = (3/4)\Delta\nu^{\text{o.d.}}_{\text{HFS}}(6P^{F=1}_{1/2})$ and $\Delta\nu^{\text{o.d.}}_{\text{HFS}}(6P_{3/2}) = (3/8)\Delta\nu^{\text{o.d.}}_{\text{HFS}}(6P^{F=1}_{3/2})$. From Table 1 of [39], where the 6P hyperfine splittings $\Delta\nu_{\text{HFS}}(6P_{1/2})$ and $\Delta\nu_{\text{HFS}}(6P_{3/2})$ and the center-of-gravity shifts $\Delta\nu^{\text{o.d.}}_{\text{HFS}}(6P_{1/2})$ and $\Delta\nu^{\text{o.d.}}_{\text{HFS}}(6P_{3/2})$ are given, the values and uncertainties of the 6P hyperfine energies can be derived through the relations

$$\Delta\nu_{\text{HFS}}(6P^{F=1}_{1/2}) = \phantom{-}(1/4)\Delta\nu_{\text{HFS}}(6P_{1/2}) + \Delta\nu^{\text{o.d.}}_{\text{HFS}}(6P_{1/2}), \tag{6.16}$$

$$\Delta\nu_{\text{HFS}}(6P^{F=1}_{3/2}) = -(5/8)\Delta\nu_{\text{HFS}}(6P_{3/2}) + \Delta\nu^{\text{o.d.}}_{\text{HFS}}(6P_{3/2}). \tag{6.17}$$

Finally, the numerical values of the hyperfine energies are given by

$$\Delta\nu_{\text{HFS}}(2S^{F=0}_{1/2}) = -133\,167\,625.7(5.0)\text{ Hz}, \tag{6.18}$$

$$\Delta\nu_{\text{HFS}}(6P^{F=1}_{1/2}) = \phantom{-}547\,798(6)\text{ Hz}, \tag{6.19}$$

$$\Delta\nu_{\text{HFS}}(6P^{F=1}_{3/2}) = -547\,460(6)\text{ Hz}. \tag{6.20}$$

The hyperfine centroid is then defined as the level energy after removing, or correcting for, the contribution from the hyperfine structure, corresponding to a hypothetical situation where the nuclear spin is set to zero.

To determine the 2S-6P centroid $\nu_{\text{2S-6P}}$, the two hyperfine-corrected transition frequencies are averaged weighted by their line strengths, as given by the square of the dipole moment $\mu$. For $2\text{S}_{1/2}^{F=0}$–$6\text{P}_{1/2}^{F=1}$ and $2\text{S}_{1/2}^{F=0}$–$6\text{P}_{3/2}^{F=1}$, the line strength ratio is 1:2 (see Table 2.1). This results in

$$\begin{aligned}\nu_{\text{2S-6P}} &= \frac{1}{3}\left(\nu_{1/2} - \Delta\nu_{\text{HFS}}(6\text{P}_{1/2}^{F=1})\right) + \frac{2}{3}\left(\nu_{3/2} - \Delta\nu_{\text{HFS}}(6\text{P}_{3/2}^{F=1})\right) + \Delta\nu_{\text{HFS}}(2\text{S}_{1/2}^{F=0}) \\ &= \frac{1}{3}\nu_{1/2} + \frac{2}{3}\nu_{3/2} - 132\,985\,252(7)\,\text{Hz},\end{aligned} \tag{6.21}$$

with the given uncertainty assuming that the values given in Eqs. (6.18) to (6.20) are uncorrelated.

On the other hand, the difference of the measured transition frequencies gives the experimental value $\Delta\nu_{\text{FS}}^{\text{exp}}(6\text{P})$ of the 6P fine-structure splitting $\Delta\nu_{\text{FS}}(6\text{P})$ of the $6\text{P}_{1/2}^{F=1}$ and $6\text{P}_{3/2}^{F=1}$ levels as

$$\Delta\nu_{\text{FS}}^{\text{exp}}(6\text{P}) = \nu_{3/2} - \nu_{1/2}. \tag{6.22}$$

Note that here the 6P fine-structure splitting $\Delta\nu_{\text{FS}}(6\text{P})$ refers to the energy difference between the measured HFS levels, while other authors may use the same expression to refer to the energy difference of the hyperfine centroids. Of course, using the values given in Eqs. (6.19) and (6.20), the conversion between the two definitions is straightforward.

The predicted value $\Delta\nu_{\text{FS}}^{\text{pred}}(6\text{P})$ of $\Delta\nu_{\text{FS}}(6\text{P})$ may be obtained from the difference in the total binding energies of the $6\text{P}_{1/2}^{F=1}$ and $6\text{P}_{3/2}^{F=1}$ levels as given in Table IV of [40], resulting in

$$\Delta\nu_{\text{FS}}^{\text{pred}}(6\text{P}) = 405\,164.5(1)\,\text{kHz}. \tag{6.23}$$

This value serves as a reference to which the experimental value $\Delta\nu_{\text{FS}}^{\text{exp}}(6\text{P})$ can be compared after unblinding the data.

Due to the preliminary nature of the data analysis shown here, neither $\nu_{\text{2S-6P}}$ nor $\Delta\nu_{\text{FS}}^{\text{exp}}(6\text{P})$ are given at this point. Note that in order to assign uncertainties to these two combined frequencies, the correlations between the uncertainties of two measured frequencies need to be studied first.

# Chapter 7

# Conclusion and outlook

In the course of this thesis, the precision of laser spectroscopy of the 2S-$n$P transitions in atomic hydrogen has been improved significantly. This has been demonstrated by measurements of both the 2S-4P transition and the 2S-6P transition. The achieved precision advances the test of one of the fundamental theories of physics, quantum electrodynamics (QED), and sheds light on discrepancies such as the proton radius puzzle. It also allows the precise determination of physical constants, with the at the time best determination of the Rydberg constant $R_\infty$ and the proton radius $r_\mathrm{p}$ from atomic hydrogen achieved with the measurement of the 2S-4P transition. The recent measurement of the narrower 2S-6P transition is set to further significantly improve upon this result.

The 2S-4P measurement, which has been performed during the first half of the work underlying this thesis (published in 2017 [24], see Appendix A), surpassed the frequency uncertainty of other hydrogen measurements besides the much narrower 1S-2S transition by at least a factor of three. Furthermore, many of the systematic effects encountered for the dipole-allowed 2S-$n$P transitions are different from those of the more commonly probed two-photon transitions, making them an ideal check on the accuracy of and discrepancies within hydrogen spectroscopy itself.

The two prominent challenges of this experiment are the first-order Doppler shift and the large natural linewidth. The 2S-4P measurement and its analysis showed that both can be successfully addressed, through an intricate optical setup, and a very large experimental signal-to-noise ratio and a detailed understanding of the line shape, respectively. The importance of the latter was highlighted by the presence of subtle distortions of the line shape, caused by quantum interference of neighboring atomic resonances, which could lead to substantial line shifts if not properly accounted for. These line shifts, which were only significant because of the very large resolution relative to the linewidth, were directly observed, and could be removed by use of a line shape model based on perturbative calculations. With this, the 2S-4P transition frequency was found with a relative uncertainty of $3.7 \times 10^{-12}$, corresponding to 2.3 kHz in absolute terms or almost one part in 10 000 of the observed linewidth. Combining the 2S-4P and 1S-2S transition frequencies, $R_\infty$ could be determined with a relative uncertainty of $8.7 \times 10^{-12}$, making it one of the most precisely determined physical constants. The likewise determined value of $r_\mathrm{p}$ is compatible with the muonic result [22], a result also favored by most subsequent precision measurements [20, 25, 26].

To further improve the precision, the study of the 2S-6P transition was begun, with work starting in earnest towards the end of the 2S-4P measurement. The beam apparatus and its cryogenic beam of metastable atoms were improved substantially to harness the lower natural linewidth of the 2S-6P transition, resulting in an up to three times lower observed linewidth and a close to an order of magnitude larger flux of slow atoms as compared to the 2S-4P measurement. An improved precision also requires an improved suppression of the first-order Doppler shift, being the limiting effect in the previous measurement. To this end, the active fiber-based retroreflector was rebuilt for the new transition wavelength of 410 nm, including a newly designed fiber collimator offering excellent beam quality only limited by the fiber itself, polarization monitoring, and improved stabilization and alignment schemes. The detector assembly was redesigned to accommodate electrodes and to offer both a larger solid angle and efficiency for the fluorescence detection. Through these improvements, implemented in the course of this thesis, the fluorescence signal could be increased by up to an factor of 16 compared to the 2S-4P measurement (see Fig. 1.3).

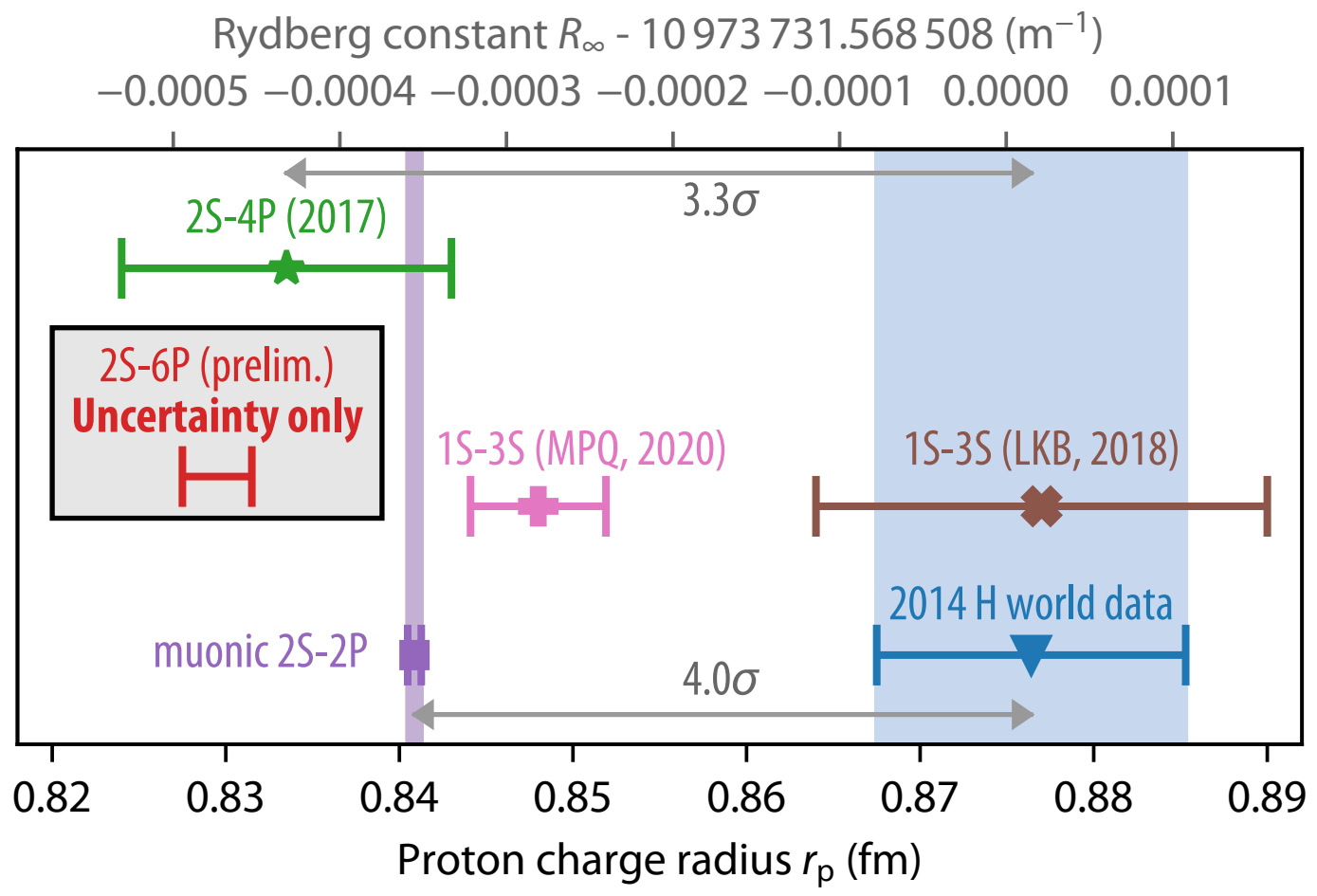

Figure 7.1: Determinations of the Rydberg constant $R_\infty$ (top axis) and proton charge radius $r_\mathrm{p}$ (bottom axis) from hydrogen spectroscopy, similar to Fig. 1.1, but including the estimated preliminary uncertainty (red error bar) of the 2S-6P measurement when combined with the 1S-2S measurement [23]. At the time of writing, the 2S-6P transition frequency is blinded and thus no values for $r_\mathrm{p}$ and $R_\infty$ derived from it can be given, and the shown error bar is placed at an arbitrary value of $r_\mathrm{p}$ and $R_\infty$.

Full use of the increased experimental resolution can however only be made if it is accompanied by a likewise improved understanding of the fluorescence line shape and modeling of the experiment. This is why a considerable effort was made to understand theoretically, to model numerically, and to measure experimentally the light force shift (LFS), which is the largest systematic correction for the 2S-6P measurement and had so far not been studied in detail or observed directly for the 2S-$n$P transitions. This lead to a model of the LFS which was subsequently experimentally tested in the 2S-6P measurement. Excellent agreement between the model and the experiment was found. The modeling of the cryogenic atomic beam, and especially its speed distribution, is also critical, as it serves as input to the LFS model, the extrapolation of the Doppler shift, and other simulation corrections. Consequently, an improved model of the beam was developed and tested in a series of measurements.

The 2S-6P transition was probed during three measurement runs in 2019. The statistical uncertainty of the resulting data set, including the uncertainty from the extrapolation of the Doppler shift, is 430 Hz, five times lower than for the 2S-4P measurement. This also implies that the Doppler shift suppression was successfully extended by a factor of five, now corresponding to a suppression of the full collinear shift by six orders of magnitude.

The data analysis is still in progress at the time of writing, and since the results are blinded to prevent experimenter bias, no preliminary results can be given here. However, factoring in the expected systematic uncertainties, an uncertainty of ≈600 Hz seems feasible, which is within a factor of two of the uncertainty of QED calculations as given in [4]. The corresponding relative uncertainty for a determination of $R_\infty$ and $r_\mathrm{p}$, when combined with the 1S-2S transition, is $1.6 \times 10^{-12}$ and $2.3 \times 10^{-3}$, respectively. As shown in Fig. 7.1, this would improve on the 2S-4P and recent 1S-3S values by a factor of five and two, respectively, and would be within a factor of five of the uncertainty of the determination of $r_\mathrm{p}$ from muonic hydrogen.

### Towards the 2S-6P transition frequency

Some work remains to be done for the results to be unblinded: first, the uncertainties of the simulation corrections, primarily for the LFS, need to be estimated by varying the various input parameters within their experimental constraints. Second, the experimental determinations of the dc-Stark shift performed at regular intervals during the 2S-6P measurement have to be evaluated and compared to simulations to derive a corresponding correction and uncertainty. Third, the excess scatter of the data should be studied more closely by correlating it with other experimental parameters and through simulations. Fourth, and closely related, one might be able to reduce the scatter by removing the drift of the fluorescence count rate during line scans, but care must be taken that no systematic shifts are inadvertently introduced by such a procedure, which should be tested with Monte Carlo simulations. These simulations are also necessary to constrain any systematic shifts from the sampling of the resonance. Fifth, the interplay between the two models used to find the simulation corrections, the big model and the LFS model, needs to be studied. To this end, it might be possible to combine both models.

The data set of the 2S-6P measurement also contains additional information that has so far not been looked into in detail. During measurement run B of the 2S-6P measurement, all line scans were acquired in pairs in an interlaced manner, resulting in dual scans that were effectively recorded within seconds of each other (see Section 4.7.2). Three types of pairs were recorded, using 10/30 µW, 10/20 µW, and 10/10 µW of spectroscopy laser power[1] for the first/second line scan. While the dual scans were treated like independent scans in the preliminary analysis, they may prove useful to disentangle residual Doppler shifts and power-dependent effects such as the LFS.

Moreover, 85 line scans of the 1S-2S transition were acquired during the 2S-6P measurement to set the frequency detuning of the preparation laser. As opposed to previous 1S-2S measurements in the same apparatus [23], where the 2S atoms where quenched with a static electric field and the resulting fluorescence was detected, here the fluorescence from the decay of the 6P level serves as signal. This results in an orders of magnitude larger signal, as shown in Fig. 7.2. However, the intracavity power used here is three times larger than the maximum power used in [23], and thus line shape distortions from inhomogeneous ac-Stark shifts are more pronounced here. Together with the second-order Doppler shift, this leads to shifts of approximately 1 kHz between the different delays, with the statistical uncertainty for each delay on the order of 10 Hz. These line scans could be useful in two ways: first, the frequency shift between the delays could be compared to the predictions from the simulations developed for the 2S-6P measurement. This constitutes a powerful check on the modeling of the experiment because the shifts are much larger than the statistical uncertainty. Second, to extract $R_\infty$ and $r_p$ from the 2S-6P measurement without losing any of its precision, the 1S-2S transition frequency only needs to be known within approximately 500 Hz, corresponding to a relative uncertainty of $2\times10^{-13}$. A determination of the 1S-2S transition frequency at this level of uncertainty from the data of the 2S-6P measurement seems feasible, and would allow the determination of $R_\infty$ and $r_p$ without any additional input. For much lower uncertainties, the determination of the laser frequency will most likely become the limiting factor.

[1]The powers given apply to the 2S-6P$_{1/2}$ transition, with the equivalent powers used for the 2S-6P$_{3/2}$ transition a factor of two lower.

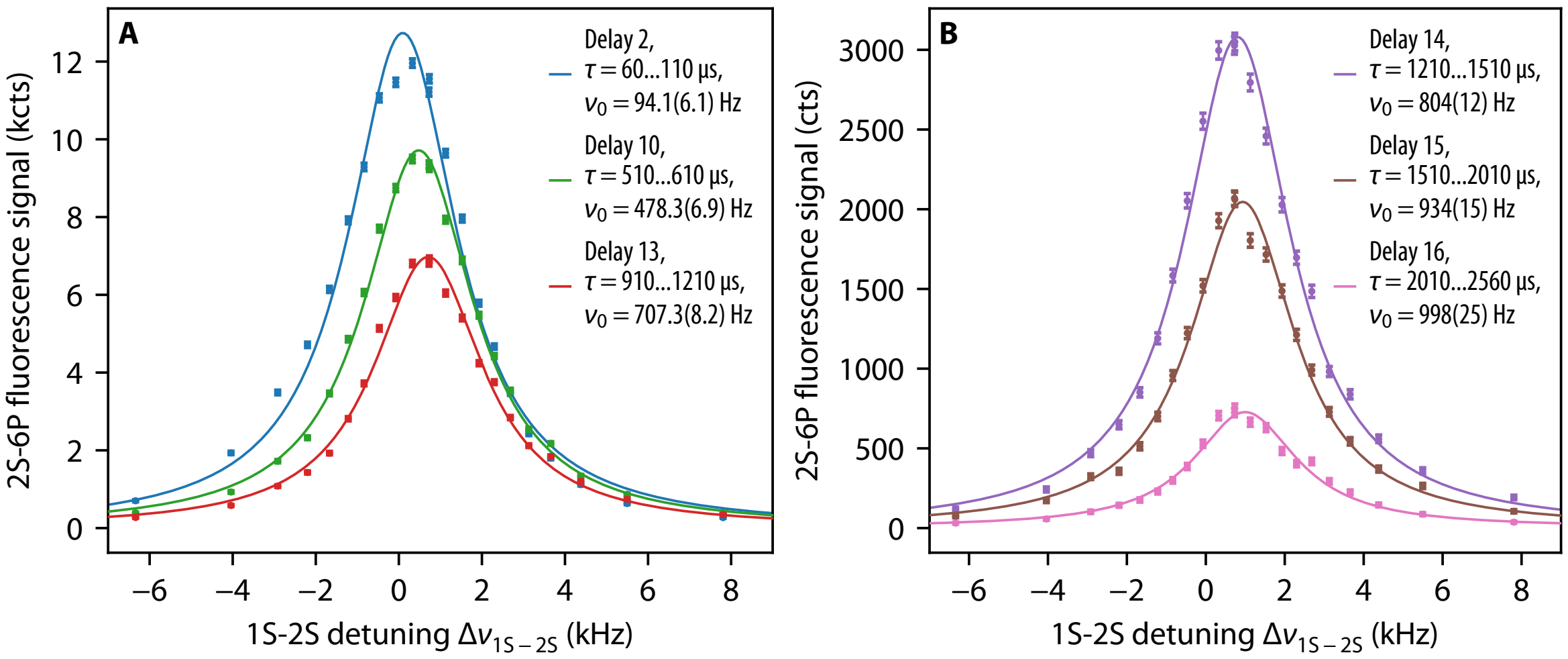

Figure 7.2: Typical line scan of the $1S_{1/2}^{F=0} - 2S_{1/2}^{F=0}$ transition, using an intracavity power of $P_{1S\text{-}2S} = 1.0\,\mathrm{W}$ for the 1S-2S preparation laser for (**A**) delays 2, 10 and 13, and (**B**) delays 14–16. The 2S-6P spectroscopy laser, with a power of $P_{2S\text{-}6P} = 30\,\mu\mathrm{W}$, is kept on resonance while the (atomic) detuning $\Delta\nu_{1S\text{-}2S}$ of the preparation laser is scanned. The counts (here shown for the top detector) from the fluorescence of the 6P level constitute the signal, and are binned into the same delays (colored circles) as used for the 2S-6P line scans (see Table 5.1). Lorentzian fits (colored lines) to the data reveal the resonance frequency $\nu_0$, with the statistical uncertainty $\sigma_{\nu_0}$ given in parentheses. Line distortions from inhomogeneous ac-Stark shifts and second-order Doppler shifts are especially pronounced for short delay times (see Section 2.2.6). The linewidth is $\Gamma_F \approx 3.3\,\mathrm{kHz}$. Compared to the signal observed at $P_{1S\text{-}2S} = 370\,\mathrm{mW}$ in the most recent 1S-2S measurement (see Fig. 2 of [23]), the signal here is 80 (45) times larger for atoms with mean speed $\bar{v} \approx 100\,\mathrm{m/s}$ ($\bar{v} \approx 160\,\mathrm{m/s}$). In total 85 of such 1S-2S line scans were recorded during the 2S-6P measurement.

Finally, for each freezing cycle of the 2S-6P measurement at least one alignment of the offset angle $\alpha_0$, which is the angle from the orthogonal between the atomic and laser beams, was performed, resulting in a total of 76 such alignments. For each alignment, line scans at various values of $\alpha_0$, with and without the shutter of the active fiber-based retroreflector blocking the returning spectroscopy laser beam, are acquired, which so far have only been used to set $\alpha_0$ close to zero during the measurement. 160 of these line scans were recorded for $\alpha_0 > 2.5\,\mathrm{mrad}$ and with the shutter open, i.e., the Doppler shift suppression active, with $\alpha_0 \approx 5\,\mathrm{mrad}$ used for most. These line scans could possibly also be used in the measurement of the LFS, albeit the difference in the LFS from data taken at $\alpha_0 = 0\,\mathrm{mrad}$ is rather small for these comparatively low values of $\alpha_0$.

The line scans for which the shutter was closed, on the other hand, contain information on the speed distribution through the now unsuppressed Doppler shift [28]. However, these line scans have so far not been studied in detail, because the approximately Maxwellian speed distribution of the atoms leads to line shape distortions for $\alpha_0 \neq 0\,\text{mrad}$, as the line shape of each atom is convoluted with the speed distribution of the atomic beam. These distortions need to be accounted for if the speed distribution is to be determined from these line scans with an accuracy below approximately 10 %. In a preliminary analysis, the speed distribution derived from these line scans was found to agree well with the speed distribution used in this work, which is based on an exponential suppression of low speeds and was verified using the observed delay-dependent line amplitudes.

### Future experiments using the 2S-$n$P apparatus

Having measured the 2S-4P and 2S-6P transition, an obvious choice for future experiments is to move to 2S-$n$P transition with an even higher $n$. Indeed, a laser system[1] capable of driving the transitions with $n = 8, 9, 10$ has already been set up in the laboratory by Florian Stehr as part of his master's thesis [165]. Using this laser system, the 2S-8P transition has been observed in an early test measurement. Yet, there are two principal problems with this strategy: first, the obvious advantage of a higher $n$ is the narrower natural linewidth of the transition. However, even for the 2S-6P transition the observed linewidth is limited by the divergence of the atomic beam, which cannot be easily reduced in the current setup. This is because decreasing the width of the apertures forming the beam will lead to large losses for the 243 nm enhancement cavity. To this end, the beam size of the cavity mode could be reduced, but this will also increase the ionization of the 2S atoms and lead to a lower effective excitation probability. Alternatively, one might implement apertures based on static electric or laser fields that selectively quench the metastable 2S atoms. Second, the dc-Stark shift increases as $n^7$ (see Section 2.4), which with the currently present electric bias fields would quickly limit the precision. Thus, the bias fields would need to be reduced, which might be most straightforwardly done by increasing the size of the spectroscopy region. Replacing the cryopump and the thermal gradients it causes, as discussed below, may also help. Finally, there is one problem particular to the 2S-8P$^{F=1}_{1/2}$ transition: since the fine-structure splitting of the 8P level closely matches that of the 2S level, the 2S$^{F=1}_{1/2}$ levels, which are populated by decays from the 8P level, are near-resonantly coupled to the 8P$^{F=1}_{3/2}$ levels, leading to overlapping resonances in the fluorescence signal. It thus might be a better strategy to improve the measurement of the 2S-6P transition with the strategies outline in the section below.

Another interesting experiment is the spectroscopy of the 2S-$n$P transitions in atomic deuterium. This is especially so since there is a measurement of the deuteron charge radius from laser spectroscopy of muonic deuterium [166]. As for the corresponding measurement of $r_\text{p}$, this measurement is much more precise than determinations from electronic deuterium, and, importantly, the muonic deuteron charge radius is likewise discrepant with these determinations. Thus, in addition to the proton radius puzzle there is also a deuteron radius puzzle [167]. However, as opposed to the former, no recent electronic deuterium spectroscopy results have become available to shed light on this puzzle. Therefore, a precise determination of the 2S-$n$P transition frequencies in atomic deuterium would be highly desirable.

[1] Toptica TA-SHG pro, consisting of an extended-cavity diode laser, a tampered amplifier, and a resonant second-harmonic generation stage. The output wavelength is tunable between 380 nm and 389 nm. Similar to the 1S-2S preparation laser and the 2S-6P spectroscopy laser, this system is frequency-stabilized to a high-finesse Fabry-Pérot cavity. A beat note with the frequency comb to determine the laser frequency is also available.

Such a measurement, compared to the hydrogen equivalent, comes with an additional challenge because the deuteron is a spin 1 particle. For hydrogen, angular momentum selection rules only allow a single hyperfine component of the 2S-6P$_{1/2}$ or 2S-6P$_{3/2}$ transition when using the $2S_{1/2}^{F=0}$ level as initial level, independent of the polarization of the laser. This is no longer the case in deuterium, and both hyperfine components are allowed and overlap within their linewidths. This has the following two consequences: first, the center of mass of the two components can depend on the experimental conditions. However, preliminary investigations reveal that, using again linear laser polarization and the fluorescence as signal, the center of mass is fixed to a certain value unless any two of the following three conditions is not fulfilled: there is no residual circular polarization of the laser beams, the two $2S_{1/2}^{F=1/2}$ initial levels are equally populated, and both $\sigma^+$ and $\sigma^-$ decays are detected with equal efficiency[1]. Within some uncertainty, these are quite reasonable assumption. Furthermore, they can be tested, e.g., by comparing the frequency found when using $\sigma^+$- and $\sigma^-$-polarized light for the excitation. To this end, the generation and monitoring of arbitrary polarization states inside the spectroscopy region has been preliminary investigated.

Second, quantum interference can now not only occur between different fine-structure components, as is the case for hydrogen, but additionally between the hyperfine components. Because of their small frequency difference, this can lead to much stronger distortions of the line shape, as demonstrated in [168, 169] for lithium. Fortunately, there is again some cancellation, with both an imbalance in the population of the initial levels and a different detection efficiency for the $\sigma^\pm$ decays needed for distortions of the line shape to occur. All in all, a measurement of deuterium seems feasible, and in fact such a measurement is currently in preparation by Vitaly Wirthl.

The line scans of the 1S-2S transition shown in Fig. 7.2 demonstrate the potential of measurements of this transition using the 2S-$n$P apparatus. However, currently, improving on the relative uncertainty of $4.2 \times 10^{-15}$ of the latest measurement of the 1S-2S transition [23] does not lead to an improved test of QED, as the much higher uncertainties of other transition frequencies are limiting. Furthermore, such an improvement implies knowing the speed distribution to better than 1 m/s, which is below the change in mean speed observed in the 2S-6P measurement during a single freezing cycle (see Fig. 6.1). One may however improve the measurement of the 2S hyperfine splitting [57], for which, when combined with the very precisely known 1S hyperfine splitting, QED predictions with a much smaller uncertainty than the current experimental uncertainty are available. This measurement relies on rapidly switching between the two hyperfine components of the 1S-2S transition, and the influence of the speed distribution is consequently reduced if only the frequency difference is of interest[2]. Looking further into the future, an improved measurement of the 1S-2S transition may become desirable as the equivalent measurement in antihydrogen continues to improve [170–172].

[1]Note that $\sigma^+$ and $\sigma^-$ decays have the same radiation pattern, but have a different polarization except along the direction orthogonal to the quantization axis.

[2]Interestingly, however, in the most recent measurement of the 2S hyperfine splitting of [57] a rather large difference in the speed distribution of the two hyperfine components was still observed. The reason for this is currently unclear, but could be caused by the different dc-Stark and Zeeman shifts of the two components.

Finally, laser cooling of hydrogen has been pursued for many decades and various schemes have been proposed [95, 173–179], but so far it could not be successfully implemented as part of a spectroscopic measurement. A strategy that seems promising and relatively straightforward to test is to laser cool the already slow atoms in the beam, with only a few recoils necessary to load them into an optical trap. This optical trap could be at a magic wavelength for the 1S-2S transition [180], i.e., a wavelength where the ac-Stark shifts of the 1S and 2S levels cancel. The understanding gained about the speed distribution of the atomic beam in this thesis could be helpful in that regard. A project investigating such a scheme is currently underway at the Laser Spectroscopy Division.

## Planned and suggested improvements of the hydrogen spectrometer

While the hydrogen spectrometer used to probe the 2S-$n$P transitions has been substantially improved in many aspects in the course of this work, there is still plenty of room for more improvement. It would be especially advantageous to increase the fraction of time spectroscopy data are taken during the operation of the apparatus by decreasing the time spent on preparing and aligning the spectrometer.

During the writing of this thesis, it was realized that the limiting factor for the degree of dissociation $\alpha_{\mathrm{dis}}$ of hydrogen is most likely not recombination inside the nozzle, as was assumed before, and which is difficult to improve within the given constraints on the nozzle geometry. Instead, the transport of hydrogen through PTFE (Teflon) tubing to the nozzle is thought to be limiting, with $\alpha_{\mathrm{dis}}$ at the input of the nozzle might possibly being as low as 1 % (see Sections 4.5.1, 4.5.2.3 and 5.3.3). That is, the particles arriving at the nozzle are mostly hydrogen molecules which freeze on the nozzle, eventually clogging it and leading to the experimental cycle based on 2 hour-long freezing cycles. Fortunately, this also means that it should be relatively straightforward to increase $\alpha_{\mathrm{dis}}$ by reducing the length of the tubing, by increasing the diameter of the tubing, or both. Increasing $\alpha_{\mathrm{dis}}$ at the nozzle input would greatly improve the experiment, since the freezing of the nozzle would take longer, while at the same time the number of hydrogen atoms in the beam would increase. The latter would then allow for a reduction in the flow of hydrogen into the system, further increasing the duration of the freezing cycle, while reducing collisions between particles and thus the loss of slow atoms. As each freezing cycle comes with a substantial overhead from the time needed to unfreeze the nozzle and realign the atomic beam, a longer freezing cycle would greatly improve the amount of data that can be acquired in a measurement day, not to mention reduce the workload of the experimenter. Additionally, as can seen in Fig. 6.1, the speed distribution of the atomic beam changes as the nozzle freezes. If this freezing can be slowed down, this effect might be substantially reduced as well.

Another limitation on the time of operation is the degradation of the mirrors of the 243 nm enhancement cavity. To this end, an upgrade of the apparatus is currently prepared that would place the mirrors inside an oxygen atmosphere to prevent this degradation, using multiple differential pumping stages to keep the partial pressure of oxygen at a sufficiently low level in the spectroscopy region. The reliability of the 1S-2S spectroscopy laser, especially its second-harmonic generation cavity producing the 243 nm light, was also an issue during the 2S-6P measurement. The laser has since been upgraded by the manufacturer and reliably outputs more than 100 mW of power at 243 nm.

The cryopump limits the experiment in several ways: first, it cools down the apparatus, leading to temperature drifts and gradients, which entail frequent realignments and, possibly, stray electric fields. Second, as it cannot be run while data are acquired, it consumes valuable measurement time during each freezing cycle. Third, the vibrations it produces while switched on disturb the active fiber-based retroreflector and the 243 nm enhancement cavity. Fourth, because of its large size, it cannot be mounted on an optical table and instead is placed on the laboratory floor, which couples environmental noise to the vacuum chamber. Its large pumping speed and capacity for hydrogen have made it so far challenging to find a suitable replacement for the cryopump. However, recently a non-evaporable getter material[1] has become available that offers a sufficiently large sorption capacity for hydrogen [181] to replace the cryopump. This upgrade should be relatively straightforward, as only the large bottom flange of the vacuum chamber needs to be replaced. In the long term, the vacuum chamber could then be redesigned without the design constraints currently required by the cryopump.

Finally, fluctuations of the nozzle temperature are thought to be mainly responsible for the noise present on the fluorescence signal. The cryostat used during the 2S-6P measurement has since been replaced with a different model[2] which offers a higher temperature stability. The alignment stage to which the cryostat is attached has also been upgraded to improve the ease and precision of the nozzle alignment, and can be fitted with motorized actuators if needed.

All in all, with a longer freezing cycle, an enhancement cavity not subject to mirror degradation, a reliable 243 nm laser source, and without the disturbances from the cryopump, it should be possible to automate the experiment to such a degree that many hours of data can be taken without the need for the experimenter to be present in the laboratory. This, in turn, could enable a higher precision for the spectroscopy of the 2S-$n$P transition, since a longer time of operation allows for more detailed experimental studies of systematic effects such as the light force shift. It also allows the power of the spectroscopy laser to be further reduced, thereby decreasing the size of the light force shift and other power-dependent systematic effects.

With this, a factor of two higher precision for the 2S-$n$P transitions seems feasible, which would correspond to an uncertainty comparable to that of the QED calculations. There are no decisive obstacles to even higher precision[3], although, as the history of the experiments performed here shows, a seemingly simple fluorescence line shape can be full of surprises.

[1] ZAO, which is a Zr-V-Ti-Al alloy, available from SAES Getters.

[2] Advanced Research Systems LT3, temperature stability specified as $\leq 2$ mK.

[3] Cesium fountain clocks achieve a precision corresponding to almost one part in $10^6$ of the observed linewidth [182], almost two orders of magnitude higher than what was achieved in the 2S-6P measurement.

# Appendix A

# 2S-4P transition frequency measurement

This appendix reprints the publication containing the description and results of the 2S-4P transition frequency measurement. The publication originally appeared as (referred to as [24] throughout this document):

A. Beyer, L. Maisenbacher, A. Matveev, R. Pohl, K. Khabarova, A. Grinin, T. Lamour, D. C. Yost, T. W. Hänsch, N. Kolachevsky, and Th. Udem, “The Rydberg constant and proton size from atomic hydrogen”, Science **358**, 79–85 (2017).

The supplementary materials of the publication, containing many details on the experimental method and the data analysis, are reprinted after the publication itself.

RESEARCH ARTICLE

ATOMIC PHYSICS

# The Rydberg constant and proton size from atomic hydrogen

**Axel Beyer,[1] Lothar Maisenbacher,[1]* Arthur Matveev,[1] Randolf Pohl,[1]† Ksenia Khabarova,[2,3] Alexey Grinin,[1] Tobias Lamour,[1] Dylan C. Yost,[1]‡ Theodor W. Hänsch,[1,4] Nikolai Kolachevsky,[2,3] Thomas Udem[1,4]**

**At the core of the "proton radius puzzle" is a four–standard deviation discrepancy between the proton root-mean-square charge radii ($r_p$) determined from the regular hydrogen (H) and the muonic hydrogen (µp) atoms. Using a cryogenic beam of H atoms, we measured the 2S-4P transition frequency in H, yielding the values of the Rydberg constant $R_\infty$ = 10973731.568076(96) per meter and $r_p$ = 0.8335(95) femtometer. Our $r_p$ value is 3.3 combined standard deviations smaller than the previous H world data, but in good agreement with the µp value. We motivate an asymmetric fit function, which eliminates line shifts from quantum interference of neighboring atomic resonances.**

The study of the hydrogen atom (H) has been at the heart of the development of modern physics. Precision laser spectroscopy of H is used today to determine fundamental physical constants such as the Rydberg constant $R_\infty$ and the proton charge radius $r_p$, defined as the root mean square (RMS) of its charge distribution. Owing to the simplicity of H, theoretical calculations can be carried out with astonishing accuracy, reaching precision up to the 12th decimal place. At the same time, high-resolution laser spectroscopy experiments deliver measurements with even higher accuracy, reaching up to the 15th decimal place in the case of the 1S-2S transition (*1*, *2*), the most precisely determined transition frequency in H.

The energy levels in H can be expressed as

$$E_{nlj} = R_\infty\left(-\frac{1}{n^2} + f_{nlj}\left(\alpha, \frac{m_e}{m_p}, \ldots\right) + \delta_{\ell 0}\frac{C_{NS}}{n^3}{r_p}^2\right) \quad (1)$$

where $n$, $l$, and $j$ are the principal, orbital, and total angular momentum quantum numbers, respectively. The first term describes the gross structure of H as a function of $n$ and was first observed in the visible H spectrum and explained empirically by Rydberg. Later, the Bohr model, in which the electron is orbiting a pointlike and, in simplest approximation, infinitely heavy proton, provided a deeper theoretical understanding.

The Rydberg constant $R_\infty = m_e\alpha^2c/2h$ links the natural energy scale of atomic systems and the SI unit system. It connects the mass of the electron $m_e$, the fine structure constant $\alpha$, Planck's constant $h$, and the speed of light in vacuum $c$. Precision spectroscopy of H has been used to determine $R_\infty$ by means of Eq. 1 with a relative uncertainty of 6 parts in $10^{12}$, making it one of the most precisely determined constants of nature to date and a cornerstone in the global adjustment of fundamental constants (*3*).

The second term in Eq. 1, $f_{nlj}(\alpha, \frac{m_e}{m_p}, \ldots) = X_{20}\alpha^2 + X_{30}\alpha^3 + X_{31}\alpha^3\ln(\alpha) + X_{40}\alpha^4 + \ldots$, accounts for relativistic corrections, contributions coming from the interactions of the bound-state system with the quantum electrodynamics (QED) vacuum fields, and other corrections calculated in the framework of QED (*3*). The electron-to-proton mass ratio $m_e/m_p$ enters the coefficients $X_{20}$, $X_{30}$, … through recoil corrections caused by the finite proton mass.

The last term in Eq. 1 with coefficient $C_{NS}$ is the leading-order correction originating from the finite charge radius of the proton, $r_p$ (*3*). It only affects atomic S states (with $l = 0$) for which the electron's wave function is nonzero at the origin. Higher-order nuclear charge distribution contributions are included in $f_{nlj}(\alpha, \frac{m_e}{m_p}, \ldots)$.

## The proton radius puzzle

The proton charge radius $r_p$ has been under debate for some time now because the very accurate value from laser spectroscopy of the exotic muonic hydrogen atom (µp) (*4*, *5*) yielded a value that is 4%, corresponding to 5.6σ, smaller than the CODATA 2014 value of $r_p$ (*3*) [see (*6*–*8*) for reviews on this issue]. The CODATA value is obtained from a combination of 24 transition frequency measurements in H and deuterium and several results from elastic electron scattering (*9*–*11*). The accuracy of the µp result is enabled by the fact that the muon's orbit is ~200 times smaller than the electron's orbit in H, resulting in a seven orders of magnitude larger influence of $r_p$ on the energy levels.

Here we study the spectroscopic part of the discrepancy, in particular the 4σ discrepancy between the µp value and the global average of all transitions measured in H (*12*) (H world data, Fig. 1). Recently, a similar discrepancy has arisen for the deuteron radius with a new result from laser spectroscopy of muonic deuterium (*13*).

Considering Eq. 1 and the fact that $f_{nlj}(\alpha, \frac{m_e}{m_p}, \ldots)$ is known with sufficiently high accuracy, one finds a very strong correlation between $R_\infty$ and $r_p$. CODATA quotes a correlation coefficient of 0.9891. Equation 1 involves two parameters, $R_\infty$ and $r_p$, which need to be determined simultaneously from a combination of at least two measurements in H. The 1S-2S transition frequency serves as a cornerstone in this procedure. Owing to its small natural line width of only 1.3 Hz, experimental determinations are one thousand times more accurate than for any other transition frequency in H, where typical line widths amount to 1 MHz or more.

Examining previous determinations of the value pairs [$R_\infty$, $r_p$] from H (Fig. 1, bottom), one notes that many of the individual measurements are in fact not in disagreement with the µp value. The discrepancy of 4σ appears when averaging all H values (µp versus H world data; Fig. 1, top).

## Principle of the measurement

Here we report on a measurement of the 2S-4P transition in H (Fig. 2A), yielding [$R_\infty$, $r_p$] with an uncertainty comparable to the aggregate H world data and significantly smaller than the proton radius discrepancy, which corresponds to 8.9 kHz in terms of the 2S-4P transition frequency. This uncertainty requires a determination of the resonance frequency to almost one part in 10,000 of the observed line width of 20 MHz (Fig. 2B).

The previous most accurate measurements [see, e.g., (*14*–*16*) and references therein] were limited by the electron-impact excitation used to produce atoms in the metastable 2S state. This excitation results in hot atoms with mean thermal velocities of 3000 m/s or more and an uncontrolled mixture of population in the four 2S Zeeman sublevels. In turn, this typically leads to corrections on the order of tens of kilohertz because of effects such as the second-order Doppler and ac-Stark shifts or the excitation of multiple unresolved hyperfine components.

Our measurement is essentially unaffected by these systematic effects (*17*) because we use the Garching 1S-2S apparatus (*1*, *2*) (Fig. 3) as a well-controlled cryogenic source of 5.8-K cold 2S atoms. Here, Doppler-free two-photon excitation is used to almost exclusively populate the $2S_{1/2}^{F=0}$ Zeeman sublevel without imparting additional momentum on the atoms.

The remaining main systematic effects in our experiment are the first-order Doppler shift and apparent line shifts caused by quantum interference of neighboring atomic resonances, both of which are suppressed by using methods specifically developed for this measurement and detailed below.

[1]Max-Planck-Institut für Quantenoptik, 85748 Garching, Germany. [2]P.N. Lebedev Physical Institute, 119991 Moscow, Russia. [3]Russian Quantum Center, 143025 Skolkovo, Russia. [4]Ludwig-Maximilians-Universität, 80539 München, Germany.
*Corresponding author. Email: lothar.maisenbacher@mpq.mpg.de
†Present address: Johannes Gutenberg-Universität Mainz, 55122 Mainz, Germany. ‡Present address: Colorado State University, Fort Collins, CO 80523, USA.

## Quantum interference

Line shape distortions caused by quantum interference from distant neighboring atomic resonances have recently come into the focus of the precision spectroscopy community (*18*). To the best of our knowledge, this effect has been considered in the analysis of only one of the previous H experiments and was found to be unimportant for that particular experimental scheme (*19*). The effect was found to be responsible for discrepancies in the value of the fine structure constant $\alpha$ extracted from various precision spectroscopy experiments in helium (*20*, *21*). The root of the matter is that natural line shapes of atomic resonances may experience deviations from a perfect Lorentzian when off-resonant transitions are present. One common way of dealing with these effects has been to perform sophisticated numerical simulations to correct the experimental results (*18*, *20*, *22*–*26*). These simulations require a highly accurate characterization of the experimental geometry if the line center needs to be determined with high accuracy relative to the line width, as is the case in this measurement. Here we remove this necessity and a source of potential inaccuracies by a suitable line shape model to compensate for the line shape distortions.

### *Two driven oscillators*

Within the framework of perturbation theory, the induced dipole moment $\vec{D}(\omega)$ of an atom driven by a laser field $\vec{E}$ at frequency $\omega$ is given by the Kramers-Heisenberg formula (*27*–*29*). For two resonances at $\omega_0$ and $\omega_0 + \Delta$ with identical damping constants $\Gamma$, the resulting dipole moment is given by

$$\vec{D}(\omega) \propto \frac{\vec{D}_0}{(\omega_0 - \omega) + i\Gamma/2} + \frac{\vec{D}_1}{(\omega_0 + \Delta - \omega) + i\Gamma/2} \qquad (2)$$

It is analogous to two coherently driven resonating classical dipoles $\vec{D}_0$ and $\vec{D}_1$. In the quantum description, each of these dipoles is constructed through an absorbing and an emitting dipole, connecting the initial state ($|i\rangle$) with the final state ($|f\rangle$) via the excited states ($|e\rangle$, $|e'\rangle$) (see Fig. 2A). With the atomic dipole matrix elements $d_{jk}$ with $j, k \in i, e, e', f$, the contributing dipole moments are given by $\vec{D}_0 \propto (\vec{E}\cdot\vec{d}_{ie})\vec{d}_{ef}$ and $\vec{D}_1 \propto (\vec{E}\cdot\vec{d}_{ie'})\vec{d}_{e'f}$. The induced dipole $\vec{D}(\omega)$ generates a field $\propto \left(\vec{r} \times \vec{D}(\omega)\right) \times \vec{r}/|\vec{r}^3|$ at position $\vec{r}$ whose power spectrum $P(\omega, \vec{r})$ is proportional to the square modulus of $\vec{D}(\omega)$. It consists of two real valued Lorentzians and a non-Lorentzian cross term. The latter depends not only on the relative orientation of $\vec{D}_0$ and $\vec{D}_1$ but also on the direction of the emitted radiation relative to the orientation of the dipoles. Because the orientation of the dipoles is itself a function of the laser polarization, i.e., the orientation of $\vec{E}$, the observed cross term will effectively depend on the orientation of the laser polarization relative to detection direction. If the detection is not pointlike, as is the case in our measurement, which is designed for an as-large-as-possible collection efficiency, the exact detection geometry will enter in the observed cross term. The relative strength of the cross term tends to decrease with increasing detection solid angle, with the cross term completely disappearing for detection of all radiation emitted, i.e., in a $4\pi$ solid angle.

For a sufficiently large separation of the two resonances ($\Gamma/\Delta << 1$), the second resonance at $\omega_0 + \Delta$ can be treated as a perturbation to the resonance at $\omega_0$ and the full line shape $P(\omega, \vec{r})$ can be expanded around the resonance at $\omega_0$ (*28*)

$$P(\omega, \vec{r}) \approx \frac{C}{(\omega - \omega_0)^2 + (\Gamma/2)^2} + a(\omega - \omega_0) + \frac{b(\omega - \omega_0)}{(\omega - \omega_0)^2 + (\Gamma/2)^2} \qquad (3)$$

The first term represents the Lorentzian line shape with amplitude $C$ of the isolated, unperturbed resonance at $\omega_0$, whereas the other two terms denote perturbations caused by the presence of the second resonance. The second term, linear in ω and of amplitude $a$, accounts for the resonance of interest sitting on the far-reaching Lorentzian tail of the perturbing resonance. The dispersive-shaped third term stems from the non-Lorentzian cross term and accounts for the quantum interference between the resonances, with the dependence of the cross term on the detection geometry now absorbed in the amplitude $b$. For a typical fluorescence-detection geometry, the line shifts caused by the coherent third term may be much larger than the ones caused by the incoherent second term.

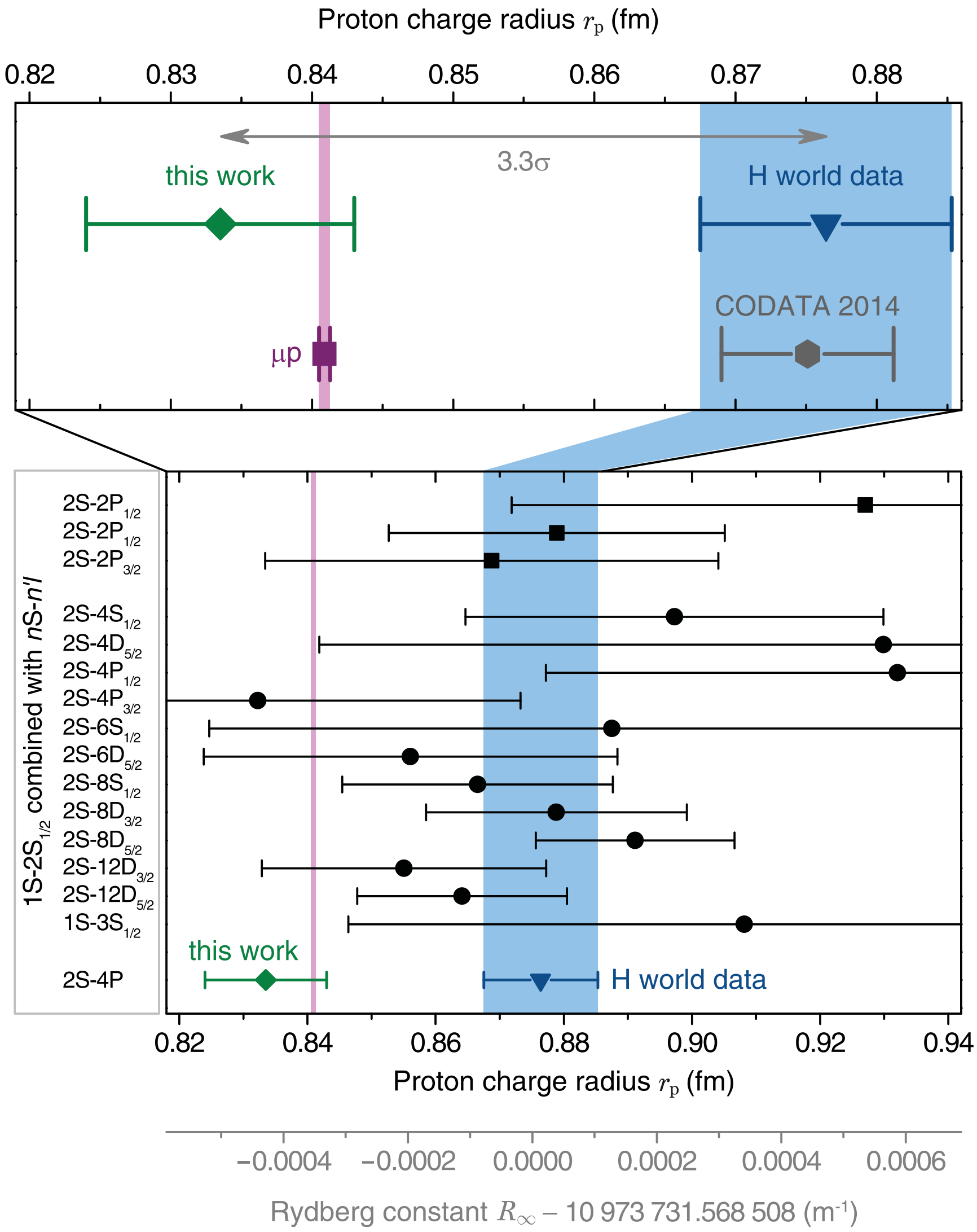

**Fig. 1. Rydberg constant $R_\infty$ and proton RMS charge radius $r_p$.** Values of $r_p$ derived from this work (green diamond) and spectroscopy of μp (μp; pink bar and violet square) agree. We find a discrepancy of 3.3 and 3.7 combined standard deviations with respect to the H spectroscopy world data (*12*) (blue bar and blue triangle) and the CODATA 2014 global adjustment of fundamental constants (*3*) (gray hexagon), respectively. The H world data consist of 15 individual measurements (black circles, optical measurements; black squares, microwave measurements). In addition to H data, the CODATA adjustment includes deuterium data (nine measurements) and elastic electron scattering data. An almost identical plot arises when showing $R_\infty$ instead of $r_p$ because of the strong correlation of these two parameters. This is indicated by the $R_\infty$ axis shown at the bottom.

The emergence of asymmetric line shapes because of interference between a resonant and a nonresonant process is perhaps best known from Fano resonances (*30*), where a background and a resonant scattering process interfere. It should not then be surprising that Eq. 3 is very similar to the line shape of Fano resonances.

Neglecting the influence of the perturbing resonance and thus the quantum interference between the resonances, e.g., by a fit of the spectrum $P(\omega, \vec{r})$ with a single Lorentzian, leads to apparent shifts of the determined line center of approximately (*28*)

$$\Delta\omega = \frac{b\Gamma^2}{4C} + \frac{a\Gamma^4}{8C} \approx -\frac{\vec{D}_0\cdot\vec{D}_1}{2D_0^2}\frac{\Gamma^2}{\Delta} + \mathcal{O}\left(\frac{\Gamma^4}{\Delta^3}\right) \quad (4)$$

Typical values of $\Gamma^2/\Delta$ are on the order of $10^{-2}$ Γ for the transitions listed for H in Fig. 1. This is one order of magnitude larger than the proton radius discrepancy, which amounts to about $10^{-3}$ Γ or less for all individual 2S-$nl$ measurements in Fig. 1. However, these measurements do not detect the emitted radiation (but rather the surviving 2S population), which diminishes the effect of quantum interference drastically at the cost of a reduced signal-to-noise ratio. The second term in Eq. 4, which stems from the term proportional to $a$ in Eq. 3, is much smaller (on the order of $10^{-6}$ Γ) and may be safely ignored at this point. Importantly, the shift changes sign when exchanging $\vec{D}_0$ and $\vec{D}_1$ and replacing Δ with –Δ, i.e., the two resonances always shift in opposite directions. Thus, by combining measurements of both resonances with appropriate weights, the shift may be drastically reduced or even canceled, a fact we will make use of below.

### *Atomic line shape model*

For the 2S-4P transition in H, the role of the mutually perturbing resonances is played by the two dipole-allowed transitions to the fine structure components of the excited state, 2S-$4P_{1/2}$ and 2S-$4P_{3/2}$ (Fig. 2). Somewhat analogous to Young's double-slit experiment, the atom can coherently evolve from the initial 2S state, through any of the two 4P fine structure components, before finally reaching the 1S ground state. Given the separation between the two components, Δ = 106 × Γ, and the natural line width, Γ = 2π × 12.9 MHz, Eq. 4 predicts apparent, geometry-dependent line shifts of up to ~120 kHz. With our large solid angle detectors, the maximum

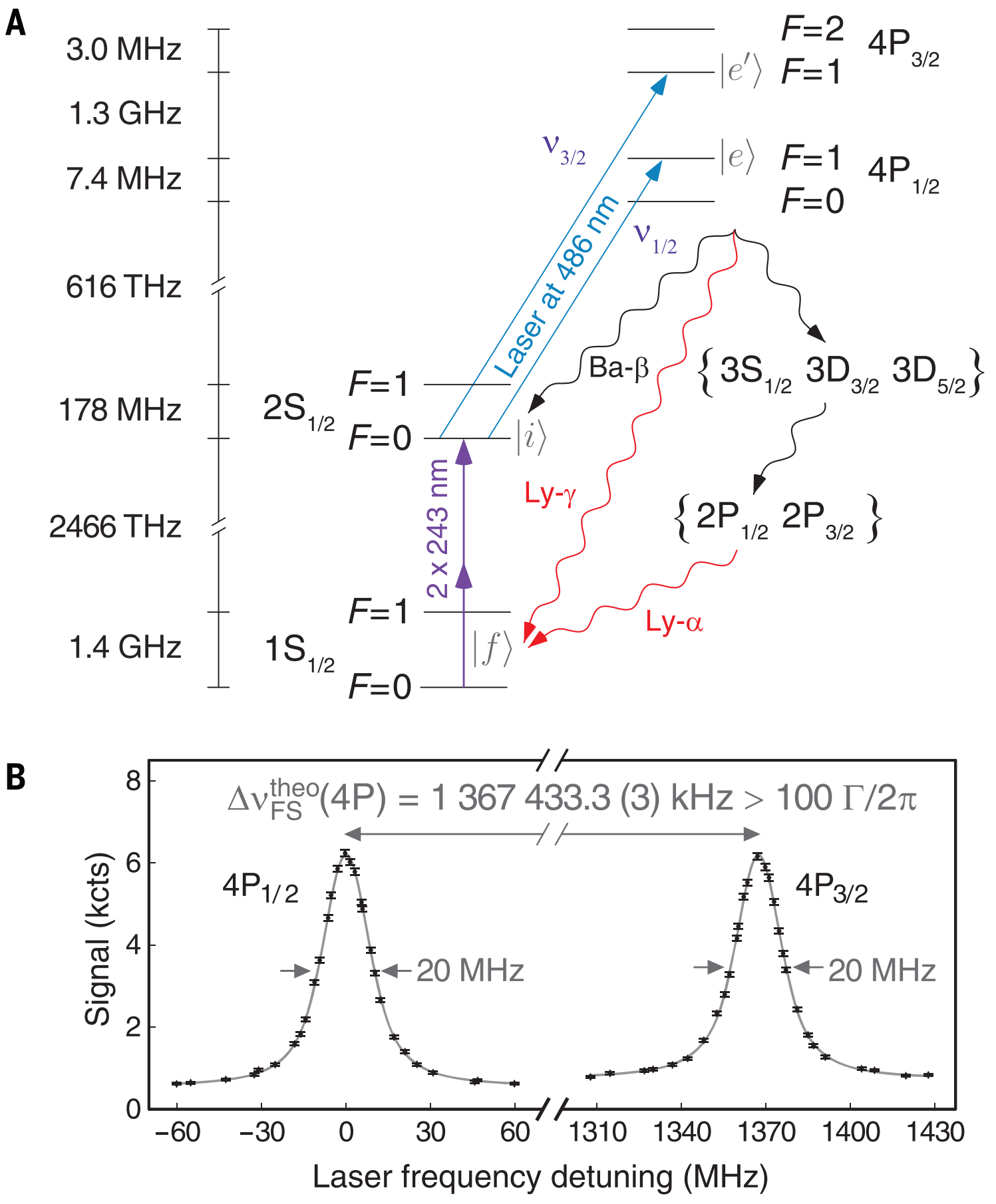

**Fig. 2. Hydrogen 2S-4P spectroscopy.** (**A**) Relevant energy levels for hydrogen 2S-4P spectroscopy are shown (not to scale). The atoms are prepared in the $2S_{1/2}^{F=0}$ metastable state (|*i*⟩) by two-photon excitation with a preparation laser at 243 nm. The spectroscopy laser at 486 nm drives the one-photon 2S-$4P_{1/2}$ and 2S-$4P_{3/2}$ transitions to the $4P_{1/2}^{F=1}$ (|*e*⟩) and $4P_{3/2}^{F=1}$ (|*e*′⟩) states to determine the transition frequencies $\nu_{1/2}$ and $\nu_{3/2}$, respectively. These states decay rapidly, predominantly to the 1S ground state (|*f*⟩) either directly through Lyman-γ fluorescence at 97 nm (Ly-γ, branching ratio 84%) or indirectly through the 3S, 3D, and 2P levels, yielding one Lyman-α photon at 121 nm (Ly-α, branching ratio 4%). The remaining 12% of the decays lead back to the 2S state through Balmer-β decay (Ba-β), with 4% decaying back to the initial $2S_{1/2}^{F=0}$ state. Excitations from the $2S_{1/2}^{F=0}$ to the $4P_{1/2}^{F=0}$ and $4P_{3/2}^{F=2}$ levels are forbidden by angular momentum conservation. (**B**) Typical experimental fluorescence signal from a single line scan over the 2S-$4P_{1/2}$ (left) and 2S-$4P_{3/2}$ (right) resonance (black diamonds). The observed line width (full width at half maximum) of ~2π × 20 MHz is larger than the natural line width Γ = 2π × 12.9 MHz because of Doppler and power broadening. The accuracy of our measurement corresponds to almost 1 part in 10,000 of the observed line width. The constant background counts are caused by the decay of 2S atoms inside the detector (*17*). kcts, kilocounts.

shift is reduced to 45 kHz, corresponding to five times the proton radius discrepancy.

One way to model this shift is to perform elaborate simulations of the entire experiment by numerical integration of the optical Bloch equations (OBE), including all relevant intermediate states and, importantly, the often-neglected cross-damping terms between them leading to quantum interference (*18*, *20*, *22*–*26*). The results of this simulation then have to be evaluated for the experimental geometry, a requirement that may be difficult to meet with sufficient accuracy. For the 2S→{$4P_{1/2}$, $4P_{3/2}$}→1S excitation spectrum considered here, this simulation consists of a total number of 2707 coupled, complex-valued ordinary differential equations. We have performed such an OBE simulation of the experiment using high-performance computation resources provided by the Max Planck Computing and Data Facility. By taking into account our experimental geometry with a sophisticated model, including particle tracing of the detected photoelectrons, the simulation is able to explain the measured data very well (see dashed line in Fig. 4, A and B). However, it is challenging to reliably estimate the uncertainty of the modeling of the detection geometry that dominates the simulation uncertainty.

Realizing that the natural line shape of the 2S→{$4P_{1/2}$, $4P_{3/2}$}→1S excitation spectrum can also be parametrized according to Eq. 3, a much simpler data analysis is possible. This only requires one additional free parameter, $b/C$, which encodes the experimental geometry (we have dropped the negligible term proportional to $a$). For sufficiently low excitation rates such as in this experiment, the influence of quantum interference will then lead to a nonzero $b/C$, but the extracted line center $\omega_0$ will not be shifted with respect to its unperturbed value. In contrast to the OBE simulation, the influence of the experimental geometry can be precisely extracted from the spectroscopy data, rather than required as an external input.

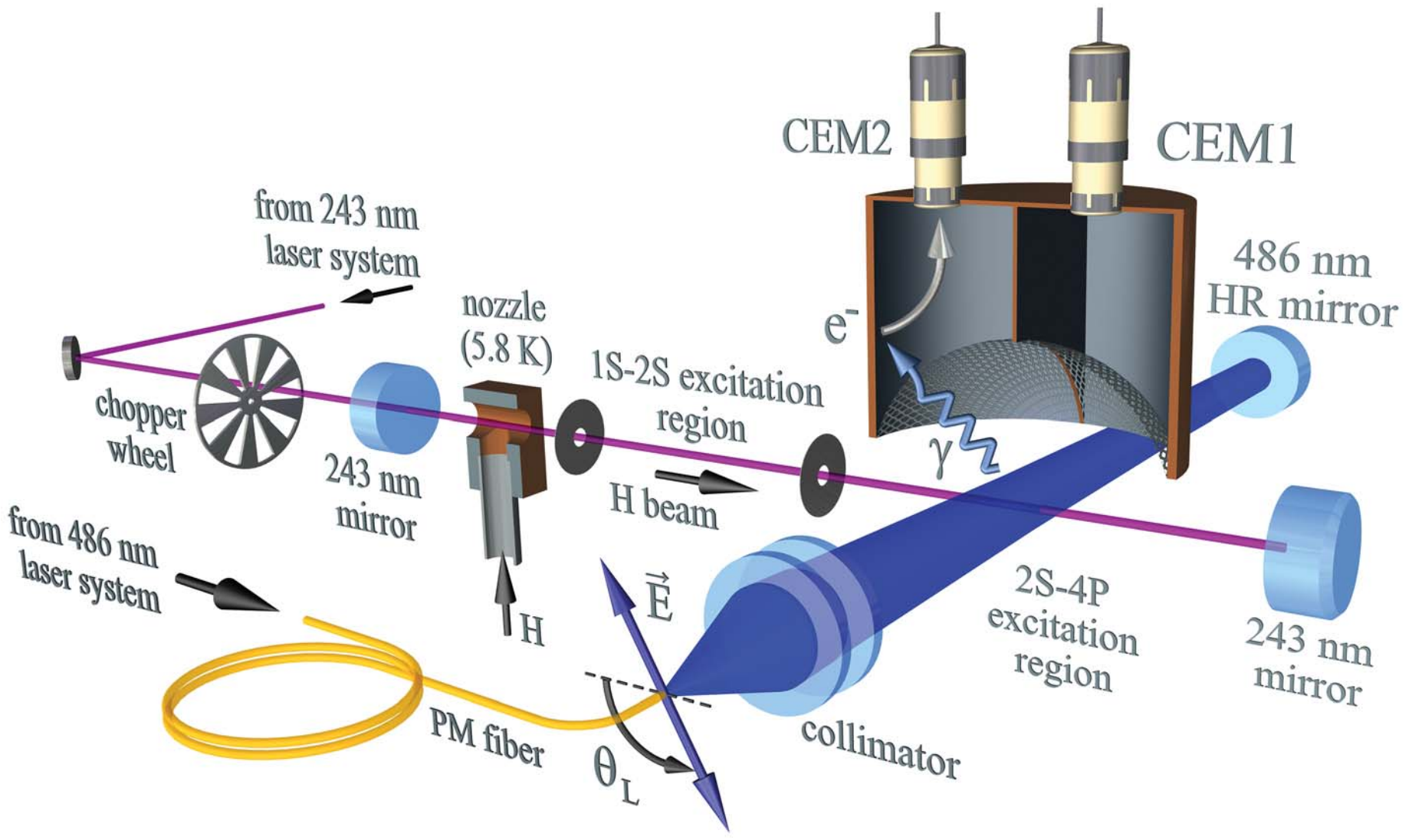

**Fig. 3. Experimental apparatus (not to scale).** A preparation laser at 243 nm is used to excite hydrogen atoms that emerge from the cold copper nozzle (5.8 K) from the ground state to the 2S state. The 2S-4P transition is driven with the spectroscopy laser at 486 nm. This laser is coupled to an active fiber-based retroreflector [consisting of polarization-maintaining (PM) fiber, collimator, and high-reflectivity (HR) mirror] oriented perpendicular to the atomic beam; this setup provides a large suppression of the first-order Doppler effect (*36*). In the dark phases of the chopper wheel, Lyman-γ fluorescence photons (γ) emitted upon the rapid 4P→1S decay are detected via photoelectrons ($e^-$) by channel electron multipliers CEM1 and CEM2. The two detectors are separated by a vertical wall along the direction of the 486-nm light propagation. The 2S-4P excitation region is shielded from stray electric fields (with dedicated meshes) and magnetic fields (with magnetic shielding, not shown), resulting in stray fields below 0.6 V/m and 1 mG, respectively (*17*). The blue double-sided arrow labeled $\vec{E}$ indicates the electric field of the 486-nm spectroscopy laser with orientation $\theta_L$ against the horizontal.

To also take into account Gaussian broadening mechanisms, such as the atomic beam divergence in our experiment detailed below, the expanded line shape (Eq. 3) is convolved with a Gaussian of width $\Gamma_G$ (full width at half maximum). Again omitting the small linear term proportional to $a$, this yields what we in the following will refer to as Fano-Voigt line shape (*17*, *31*, *32*)

$$\mathcal{F}(\omega) = A\{\mathrm{Re}[w(z)] + 2\eta\mathrm{Im}[w(z)]\} \quad (5)$$

where $w(z)$ denotes the Faddeeva function of the argument $z = 2\sqrt{\ln 2}[(\omega - \omega_0) + i\Gamma/2]/\Gamma_G$. Analogous to Eq. 3, the Fano-Voigt line shape consists of a Voigt profile, corresponding to the convolution of the Gaussian and the Lorentzian profile, and a dispersive-shaped perturbation. The asymmetry parameter $\eta = b\Gamma/4C$ measures the amplitude of this perturbation relative to the observed line strength $A$ and directly gives the line shift, in units of the observed line width, that is canceled by including the perturbation.

Additional line shifts caused by the interplay of quantum interference with both the back decay of the 4P state to the initial 2S state and the depletion of this initial state are not fully accounted for by the Fano-Voigt line shape but could in principle be removed by using an even more sophisticated line shape. However, those additional shifts are considerably smaller and less geometry-dependent than the shift removed by the Fano-Voigt line shape. Thus, we apply small model corrections to the data [1.3(2) kHz for the most affected 2S-4$P_{1/2}$ transition] determined by fitting the Fano-Voigt line shape to the OBE simulations (*17*). Note that these additional shifts also have opposite signs for the two mutually perturbing resonances. Although the bulk of the broadening caused by the atomic beam divergence and saturation effects is well described by the Fano-Voigt line shape, small deviations symmetric about the line center remain. In combination with an imperfectly symmetric experimental sampling of the resonance about its center, this can lead to a sampling bias in the determined line centers. We reduce this sampling bias by selectively removing a small amount of experimental points to enforce fair sampling (*17*). The remaining sampling bias is estimated with a Monte Carlo simulation using the experimental sampling and signal-to-noise ratio, leading to a maximum correction of 0.8(0.7) kHz.

## Experimental setup

To measure the 2S-4P transition frequency and study the effect of quantum interference, we use the dedicated setup depicted in Fig. 3 (*33*–*35*). A cryogenic beam of H in the metastable 2S state obtained from Doppler-free two-photon excitation of the 1S-2S transition is crossed at right angles with radiation from the spectroscopy laser at 486 nm, driving the 2S-4P transition. The hyperfine splitting in the 2S state is resolved in the 1S-2S excitation, so that the atoms are almost exclusively prepared in the $2S_{1/2}^{F=0}$ sublevel. From this state, only two dipole-allowed transitions may be driven as depicted in Fig. 2, either to the $4P_{1/2}^{F=1}$ state (2S-4$P_{1/2}$ transition) or to the $4P_{3/2}^{F=1}$ state (2S-4$P_{3/2}$ transition). The linear polarization of the spectroscopy laser is oriented at angle $\theta_L$ to the horizontal and defined by a polarization-maintaining (PM) fiber (intensity polarization extinction ratio 200:1). The polarization can be rotated about the laser beam axis by either making use of the two orthogonal PM axes of the fiber or rotating the fiber itself.

To observe the effects of quantum interference more clearly, we have split our large solid angle detector by a vertical wall along the spectroscopy laser beam, creating two detectors that observe the fluorescence of the 4P state from different directions, but with the same solid angle. The Lyman-γ extreme ultraviolet photons emitted upon this rapid decay of the short-lived 4P state to the 1S ground state release photoelectrons from the graphite-coated inner walls of the detectors, which are counted by two channel electron multipliers, CEM1 and CEM2; the output of these multipliers is our signal.

## Doppler shift

The mean thermal velocity of atoms in our cryogenic beam is about 300 m/s, 10 times smaller than in previous experiments. In addition, a high level of compensation of the first-order Doppler shift is achieved by using an active fiber-based retroreflector specifically developed for this experiment (*36*). The transition is driven by two phase-retracing antiparallel laser beams, leading to Doppler shifts of opposite sign and equal amplitude for atoms being excited by the respective beams. To verify this scheme, we probe atomic samples with mean velocities ranging from 295 down to 85 m/s. These low velocities are achieved by quickly switching off the 1S-2S excitation light at 243 nm and letting the fastest 2S atoms escape before acquiring data (time-of-flight resolved detection scheme). Any residual first-order Doppler shift can be constrained by extracting the rate of change of the observed transition frequency with the mean velocity of the atoms interrogated for each delay time. We extract this Doppler slope from the same data used to determine the transition frequencies presented here and find it to be compatible with zero for each transition after averaging all our data. The corresponding frequency uncertainty is found by multiplying the Doppler slope with the mean velocity of all atoms interrogated, 240 m/s, giving an uncertainty of 2.9 and 2.8 kHz for the 2S-4$P_{1/2}$ and the 2S-4$P_{3/2}$ transitions, respectively. The two antiparallel laser beams weakly couple different momentum eigenstates of the 2S atoms and can drive Raman transitions between them. Because the coupling is detuning dependent, it can lead to small line shifts, which we evaluate with an auxiliary OBE simulation that takes into account the recoil of the atoms. For our atomic beam geometry and excitation rates, this light force shift is found to be below 0.5 kHz for both transitions measured (*17*).

Although the laser beam configuration resembles the well-known saturated absorption configuration, the characteristic dip in the line shape expected for this configuration is not present here because the Doppler width of the atomic beam closely matches the natural line width of the 2S-4P transition and we work in the low-saturation regime.

## Observation of quantum interference line shifts

Figure 4 shows the effects of quantum interference line distortions for the 2S-4P$_{1/2}$ and 2S-4P$_{3/2}$ transitions. Data were acquired at different orientations $\theta_L$ of the linear laser polarization (see Fig. 3) and thus for different orientations of the induced atomic dipole relative to the field of view of the detectors. The data set consists of a total number of 22,928 and 25,064 individual resonances for the 2S-4P$_{1/2}$ and 2S-4P$_{3/2}$ transitions, respectively, with varying amounts of resonances recorded per $\theta_L$ setting.

By using the Voigt approximation ($\eta$ = 0 in Eq. 5) as line shape model, a $\theta_L$ dependence of the extracted resonance frequency is observed for both transitions and detectors (Fig. 4, A and B). The amplitudes of the shift of 40 and 20 kHz for the 2S-4P$_{1/2}$ and 2S-4P$_{3/2}$ transitions, respectively, are much larger than the proton radius discrepancy of 8.9 kHz. Although averaging over $\theta_L$ reduces the shift, it does not average to zero and a significant shift of 6.8 and –3.0 kHz remains for the two transitions, respectively, as determined from our simulation. As expected from Eq. 4, the shifts of the two mutually perturbing transitions are of opposite signs; as expected from the symmetry of the experimental geometry, exchanging the two detectors corresponds to mirroring the laser polarization about the vertical ($\theta_L$ = 90°). The behavior is well reproduced by our simulation (dashed lines in Fig. 4), confirming that the detection geometry has been correctly taken into account. This is a direct observation of a quantum interference line shift in the regime of large separation between the atomic resonances ($\Delta/\Gamma > 100$); for the unresolved D2 lines in lithium ($\Delta/\Gamma \approx 1$), similar effects have been observed before (*29*, *37*).

Fitting the resonances with the Fano-Voigt line shape, on the other hand, removes the $\theta_L$ dependence (Fig. 4, C and D), with the geometry dependence now absorbed in the asymmetry parameter $\eta$. The residual amplitudes ($A_{res}$, dotted lines in Fig. 4) of possible remaining quantum interference shifts are determined by fitting a parametrized version of our simulation to the data and are found to be well compatible with zero, except for a 3.2(1.2) kHz effect for the CEM1 data of the 2S-4P$_{1/2}$ transition. When averaging over $\theta_L$ and both detectors to determine the 2S-4P$_{1/2}$ and 2S-4P$_{3/2}$ transition frequencies, this results in insignificant residual shifts of 0.3(3) and 0.0(3) kHz, respectively (*17*). Because the Doppler shift uncertainty on the order of 10 kHz per point is correlated between the two line shape models, it is not included in the error bars shown in Fig. 4 to highlight the difference between the models. When including the Doppler shift and systematic uncertainties (red-shaded area in Fig. 4), we see only a small scatter of the data points, with the notable exception of the points for 0° and 90° for the 2S-4P$_{1/2}$ transition. These points were taken during the first two days of the measurement where the larger observed line width suggests a slight misalignment between the 2S-4P excitation laser and the atomic beam. However, discarding these data would only shift the final result (Eq. 9) by an insignificant 0.3 kHz.

## 2S-4P absolute transition frequency

Having removed the influence of quantum interference by using the Fano-Voigt line shape, we can now give the unperturbed transition frequencies by averaging over the different laser polarization settings and both detectors. The laser frequency has been determined with a frequency comb linked to a Global Positioning System (GPS)–referenced hydrogen maser. For the transition frequencies from the $2S_{1/2}^{F=0}$ state to the $4P_{1/2}^{F=1}$ state ($\nu_{1/2}$) and the $4P_{3/2}^{F=1}$ state ($\nu_{3/2}$), we find

$$\nu_{1/2} = 616520152555.1(3.0)\,\mathrm{kHz} \quad (6)$$

$$\nu_{3/2} = 616521519990.8(3.0)\,\mathrm{kHz} \quad (7)$$

where the given uncertainties include both statistical and systematic uncertainties and are dominated by the Doppler shift uncertainty. This result corresponds to an improvement of a factor of 4.9 and 3.3 in uncertainty, respectively, compared to the previous best measurements of the $2S\text{-}4P_{1/2}$ and $2S\text{-}4P_{3/2}$ transitions (16). The values in Eqs. 6 and 7 have been corrected for the recoil shift of 837.23 kHz. Details of the data analysis and a list of corrections and uncertainties are given in (17) (see table S2).

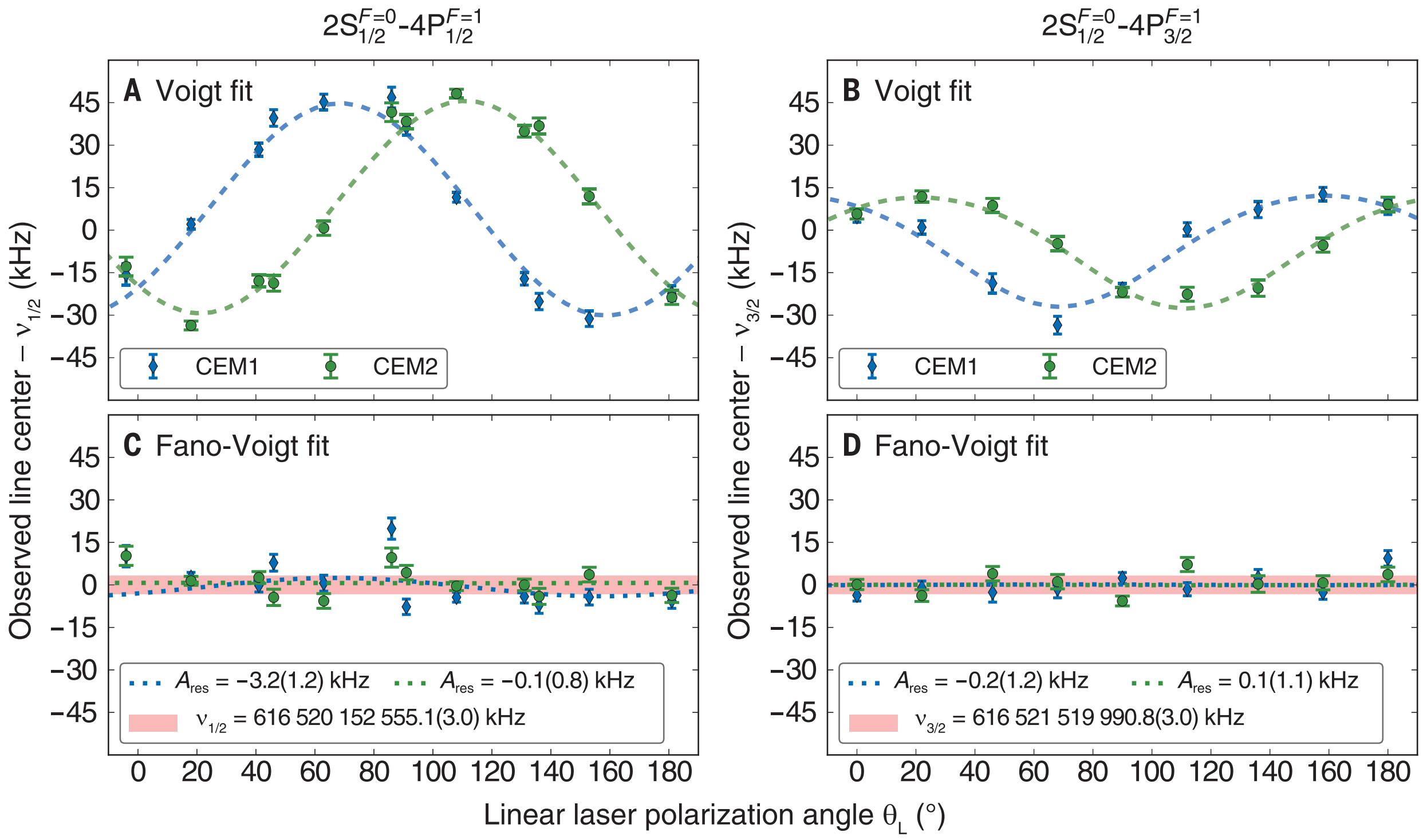

**Fig. 4. Observation of quantum interference.** Shown are apparent line shifts caused by quantum interference (**A** and **B**) and their suppression (**C** and **D**). Observed line centers of the 2S-4P$_{1/2}$ (A) and 2S-4P$_{3/2}$ (B) transitions determined with symmetric Voigt fits show a dependence on the direction of the linear laser polarization $\theta_L$ with an amplitude of up to 40 kHz in our geometry. Our numerical simulation (dashed lines) reproduces this behavior very well (*17*). Using the Fano-Voigt line shape (Eq. 5) removes the $\theta_L$ dependence (C) and (D). Blue and green symbols indicate data recorded with CEM1 and CEM2, respectively (see Fig. 3). Error bars indicate the statistical uncertainty only. The red-shaded areas (*17*) show the weighted mean of both detectors, including the final uncertainty dominated by the Doppler shift uncertainty. The dotted lines show an estimate of possible remaining line shifts with amplitude $A_{res}$.

**Table 1. List of corrections $\Delta\nu$ and uncertainties $\sigma$ for the determination of the 2S-4P fine structure centroid $\nu_{2S\text{-}4P}$. See (17) for details.**

| Contribution | $\Delta\nu$ (kHz) | $\sigma$ (kHz) |
|---|---|---|
| Statistics | 0.00 | 0.41 |
| First-order Doppler shift | 0.00 | 2.13 |
| Quantum interference shift | 0.00 | 0.21 |
| Light force shift | –0.32 | 0.30 |
| Model corrections | 0.11 | 0.06 |
| Sampling bias | 0.44 | 0.49 |
| Second-order Doppler shift | 0.22 | 0.05 |
| dc-Stark shift | 0.00 | 0.20 |
| Zeeman shift | 0.00 | 0.22 |
| Pressure shift | 0.00 | 0.02 |
| Laser spectrum | 0.00 | 0.10 |
| Frequency standard (hydrogen maser) | 0.00 | 0.06 |
| Recoil shift | –837.23 | 0.00 |
| Hyperfine structure corrections | –132,552.092 | 0.075 |
| Total | –133,388.9 | 2.3 |

We subtract $\nu_{1/2}$ from $\nu_{3/2}$ to obtain the 4P fine structure splitting $\Delta\nu_{FS}^{exp}(4P)$ between the $4P_{1/2}^{F=1}$ and the $4P_{3/2}^{F=1}$ states (17) (see table S3)

$$\Delta\nu_{FS}^{exp}(4P) = 1367435.7(4.3)\mathrm{kHz} \quad (8)$$

The fine structure splitting is essentially free from finite-size corrections and can therefore be calculated very precisely (38), yielding $\Delta\nu_{FS}^{theo}(4P)$ = 1367433.3(3) kHz. With a difference of 2.4 (4.3) kHz, our experimental result is in excellent agreement with the theoretical value. Furthermore, it represents the most accurately determined fine structure splitting in H inferred from an optical transition frequency measurement.

Even more importantly, because any shifts caused by quantum interference will be of opposite signs for the two resonances, the comparison of $\Delta\nu_{FS}^{exp}(4P)$ and $\Delta\nu_{FS}^{theo}(4P)$ provides a sensitive measure for residual quantum interference shifts and an independent test of the internal consistency of our data analysis. If not accounted for, the quantum interference line shifts would lead to a discrepancy of about 10 kHz between $\Delta\nu_{FS}^{exp}(4P)$ and $\Delta\nu_{FS}^{theo}(4P)$ in our measurement, when the signal is averaged over all polarization angles and both detectors. To increase the sensitivity of this test further, one can compare data for laser polarizations where the line shifts are largest, e.g., at $\theta_L \approx 110°$ for CEM2. Here, the difference of the splitting to the theory value is 10.0(16.9) kHz, after the Fano-Voigt line shape has compensated for an ~70-kHz shift.

To make use of the fact that the quantum interference effects, including those not compensated for by the Fano-Voigt line shape and accounted for by small model corrections, shift the two resonances in opposite directions, it is advantageous to determine the transition frequency from the 2S hyperfine centroid to the 4P fine structure centroid [i.e., the centroid of the hyperfine centroids; see eq. S16 in (17)] using Eqs. 6 and 7

$$\nu_{2S-4P} = 616520931626.8(2.3)\,\mathrm{kHz} \quad (9)$$

With this combination, the model correction and the upper limit on possible residual line shifts caused by quantum interference are reduced to a negligible 0.1(1) and 0.2 kHz, respectively. The final measurement uncertainty of 2.3 kHz is four times smaller than the proton radius discrepancy for the 2S-4P transition. The uncertainty is dominated by the first-order Doppler shift uncertainty, given by the weighted average of the corresponding uncorrelated uncertainties for the $2S\text{-}4P_{1/2}$ and $2S\text{-}4P_{3/2}$ transitions. A list of the corrections applied and the contributions to the total uncertainty is given in Table 1.

## Rydberg constant and proton charge radius

Following (3), we combine Eq. 9 with our previous measurement of the 1S-2S transition frequency (1, 2) to determine the value pair [$R_\infty$, $r_p$] using Eq. 1

$$R_\infty = 10973731.568076(96)\ \mathrm{m}^{-1} \quad (10)$$

$$r_p = 0.8335(95)\ \mathrm{fm} \quad (11)$$

providing the most accurate determination of these values from H spectroscopy with uncertainties equivalent to the aggregate H world data. We find good agreement with the μp value (4, 5), but a discrepancy of 3.3 combined standard deviations to the H world data (see Fig. 1) for both $R_\infty$ and $r_p$. Our new value for $R_\infty$ also agrees with the one obtained from the combination of the muonic deuterium measurement (13) and the 1S-2S transition frequency in electronic deuterium (39).

Previous H experiments almost exclusively used the depletion of the 2S initial state population to detect 2S-*nl* excitations in a number of different detection geometries. These schemes are generally much less prone to the effects of quantum interference. This advantage, however, comes at the price of a considerably reduced signal-to-noise ratio (compared to our fluorescence-detection scheme), which makes the identification and study of small systematic shifts much more difficult. Averaging results from various sources (i.e., geometries) may further be expected to cancel the potential residual shifts caused by quantum interference to some extent so that it seems rather unlikely that this effect can explain the observed discrepancy with the H world data.

The discrepancy of the results in this work with the H world data limits the precision of tests of bound-state QED. Provided that QED calculations are correct, new experiments with improved accuracy will help to understand the discrepancy. Several of such experiments using various approaches are currently under way (40–48). The tools developed in this work for 2S-4P spectroscopy can now be applied to other 2S-*n*P transitions to provide additional experimental data.

**ACKNOWLEDGMENTS**

The authors thank E. A. Hessels and U. D. Jentschura for insightful discussions and W. Simon, K. Linner, and H. Brückner for technical support. K.K. and N.K. acknowledge support from Russian Science Foundation 16-12-00096, R.P. from the European Research Council (ERC) Starting Grant #279765, and T.W.H. from the Max Planck Foundation. The data underlying this study are available from the corresponding author upon reasonable request.

**SUPPLEMENTARY MATERIALS**

www.sciencemag.org/content/358/6359/79/suppl/DC1
Materials and Methods
Fig. S1
Tables S1 to S3
References (*49*–*59*)

# Supplementary Materials for

## The Rydberg constant and proton size from atomic hydrogen

Axel Beyer, Lothar Maisenbacher,* Arthur Matveev, Randolf Pohl, Ksenia Khabarova, Alexey Grinin, Tobias Lamour, Dylan C. Yost, Theodor W. Hänsch, Nikolai Kolachevsky, Thomas Udem

*Corresponding author. Email: lothar.maisenbacher@mpq.mpg.de

**This PDF file includes:**

# Materials and methods

# 1 Data acquisition and analysis

## 1.1 Data acquisition

Here we briefly review the experimental setup as depicted in Fig. 3 in the main text. Excitation of the $2\mathrm{S}_{1/2}^{F=0}$ state takes place by Doppler-free two-photon excitation of the $1\mathrm{S}_{1/2}^{F=0}$-$2\mathrm{S}_{1/2}^{F=0}$ transition using a preparation laser at 243 nm. The line width of this excitation is $\sim$2 kHz (full width at half maximum (FWHM)) and is limited by time-of-flight broadening (*1*). Because this line width is much smaller than the 621 MHz laser detuning necessary to drive the $1\mathrm{S}_{1/2}^{F=1}$-$2\mathrm{S}_{1/2}^{F=1}$ transition by Doppler-free two-photon excitation, the $2\mathrm{S}_{1/2}^{F=1}$ states are only populated by Doppler-sensitive two-photon excitation. This leads to a negligible population of approx. $3 \times 10^{-7}$ in each of the three $2\mathrm{S}_{1/2}^{F=1}$ states relative to the population in the $2\mathrm{S}_{1/2}^{F=0}$ state, because only a small number of atoms are in the velocity class that is resonant with this transition. The atoms then travel from the 1S-2S excitation region to the 2S-4P excitation region, where they interact with the two counter-propagating beams of the spectroscopy laser at 486 nm. Detection of the 2S-4P fluorescence only takes place after the production of 2S atoms has been discontinued at time $\tau = 0$, i.e. the 1S-2S excitation light has been blocked by a chopper wheel running at 160 Hz. At an excitation light intensity of about 1.4 W/m$^2$ (0.6 W/m$^2$) per direction of the spectroscopy laser (beam waist $w_0 = 1.85$ mm), on average about 30% of the atoms in the $2\mathrm{S}_{1/2}^{F=0}$ state are excited to the $4\mathrm{P}_{1/2}^{F=1}$ ($4\mathrm{P}_{3/2}^{F=1}$) state. The Lyman-$\gamma$ (and, with lower efficiency, Lyman-$\alpha$) extreme ultraviolet photons emitted by the atoms upon decay release photoelectrons from the graphite-coated inner walls of the detectors, which are subsequently counted by the detectors CEM1 and CEM2. The detector assembly is differentially pumped with a cryopump to ensure a background gas pressure lower than $1 \times 10^{-7}$ mbar inside the 1S-2S and 2S-4P excitation regions.

The recorded counts are binned according to their arrival time $\tau$ into 10 delay time intervals $[\tau_i, \tau_{i+1}]$, ranging from delay times of 10 µs to 2621 µs. The width of the intervals was chosen in a way that provides comparable counting statistics to all 10 data sets and ranges from 50 µs for early delay times to 1711 µs for the longest delay time. The individual subsets of data obtained in this way sample different velocity groups of the excited atoms, which is used to constrain the first-order Doppler shift (see Sec. 2.1).

For each measurement setting, i.e. orientation of the linear laser polarization $\theta_\mathrm{L}$, the angle $\alpha$ between the spectroscopy laser beams and the atomic beam is adjusted to be close to $\alpha = 90^\circ$ before the actual data acquisition. This is achieved by blocking the spectroscopy laser beam before the HR mirror that otherwise retroreflects it and then minimizing the slope of the resulting first-order Doppler shift as a function of atomic velocity (*36*). Absolute frequency data for the 2S-4P transitions is subsequently acquired with active Doppler compensation, i.e. with counter-propagating, actively stabilized (HR mirror tip/tilt) laser beams, for a specific setting of the laser polarization angle $\theta_{\mathrm{L}}$.

A single resonance scan consists of 29 randomized frequency points and takes about 90 s (*35*). The resonance is scanned by alternating between the red- and blue-detuned sides to reduce a possible bias in the deduced line centers by slow drifts in the fluorescence count rate caused by a drift in the number of 2S atoms reaching the 2S-4P excitation region. At each frequency point, the signal is integrated over 170 cycles of the chopper wheel, then the polarization is rotated by $90\,^{\circ}$ (*36*) and the signal is integrated over another 170 cycles. This rotation of polarization is accomplished by coupling light from two paths into the two orthogonal polarization-maintaining axes of the fiber used for light delivery and successively switching between the paths. Typical resonances recorded with CEM2 for the first delay interval are shown in Fig. 2B in the main text.

For each measurement day, data are taken for a fixed value of $\theta_{\mathrm{L}}$ and $\theta_{\mathrm{L}} + 90\,^{\circ}$. In total, a typical measurement day consists of about 100 resonance scans per polarization direction, leading to a total number of about 4000 individual resonances per day (2 detectors, 2 polarization directions, 10 individual velocity subsets).

We observe background counts, i.e. counts when the spectroscopy laser is tuned off-resonance, caused by the decay of some of the metastable 2S atoms inside the detection region (the dark count rate of the detectors when no 2S atoms are present is negligible). The off-resonance background $y_0$, measured relative to the on-resonance amplitude $A$ above the background, is not identical for the two detectors. On average, $y_0/A \approx 0.21$ for CEM1 and $y_0/A \approx 0.10$ for CEM2, even though the amplitude $A$ is very similar for CEM1 and CEM2. This background is much larger than the minimum background expected from the decay of unperturbed 2S atoms flying through the detector, which we estimate contributes only $y_0/A \approx 0.002$ and is expected to be identical for both detectors. We attribute the increased background to secondary electron emission from 2S atoms striking the downstream detector wall near CEM1. This is because the small size of the opening in the detector wall, designed to allow efficient differential pumping and a large detection solid angle, does not allow all 2S atoms to leave the detection region, especially in the presence of unavoidable small misalignments. This process can also account for the different background levels of the two detectors, since the detection efficiency of the secondary electrons emitted from the downstream detector wall is not expected to be identical between the detectors. We have also considered the possibility of collisional quenching, by a background gas, or Stark quenching, by stray electric fields, of the 2S atoms. We estimate the pressure of $H_2$ molecules, the dominant background gas in the vacuum chamber, needed to account for the observed background to be $9 \times 10^{-6}$ mbar, by far exceeding the experimental upper limit on the pressure inside the detector, $1 \times 10^{-7}$ mbar. Similarly, an electric field of 50 V/m would be needed for Stark quenching to explain the background, much larger than the experimental limit of 0.6 V/m (see Sec. 2.7).

Even though the background seen by the two detectors is different by more than a factor of two, the transition frequencies determined using our full analysis, but only the data from either detector, are in excellent agreement. Using our numerical simulations, we have confirmed that the presence of a background due to the decay of 2S atoms does not influence the determined transition frequency. Additionally, we have also repeated the data analysis detailed below, but including a linear dependence of the background on laser detuning, and find the result to be in agreement with the result of the main analysis, which assumes only a constant background.

## 1.2 Line shape model

The line shape of an atom at rest which is subject to small distortions caused by quantum interference of one additional far-detuned resonance may be expressed by (*28*):

$$P(\omega,\vec{r}) \approx \frac{C}{(\omega-\omega_0)^2+(\Gamma/2)^2} + a(\omega-\omega_0) + \frac{b(\omega-\omega_0)}{(\omega-\omega_0)^2+(\Gamma/2)^2}, \quad \text{(S1)}$$

where $\omega_0$ is the frequency of the resonance of interest and $\Gamma$ the natural line width of this resonance. The geometry dependence expressed by $\vec{r}$ and the frequency separation $\Delta$ of the perturbing resonance from the resonance of interest is buried in the coefficients $C, a, b$. We drop the $a$ term, as discussed in the main text, because the corresponding shifts are smaller than the shifts due to the $b$ term by an additional factor of $\Gamma^2/\Delta^2 < 10^{-4}$ for the case of our 2S-4P spectroscopy, yielding

$$P(\omega,\vec{r}) \approx \frac{C}{(\omega-\omega_0)^2+(\Gamma/2)^2} + \frac{b(\omega-\omega_0)}{(\omega-\omega_0)^2+(\Gamma/2)^2}. \quad \text{(S2)}$$

To take into account the finite Doppler width caused by the transverse divergence of the atomic beam, we convolve this line shape with a Gaussian:

$$p(\Delta\omega) = \frac{2\sqrt{\ln 2}}{\sqrt{\pi}\Gamma_\mathrm{G}}\, e^{-4\ln 2 \frac{(\Delta\omega)^2}{\Gamma_\mathrm{G}^2}}, \quad \text{(S3)}$$

where $\Gamma_\mathrm{G}$ gives the FWHM of $p(\Delta\omega)$ that describes the probability to find an atom with its resonance frequency shifted by $\Delta\omega$ by the first-order Doppler shift. $\Gamma_\mathrm{G}$ can in principle be taken from the simulations of the atomic beam described in the discussion of the first-order Doppler shift, however there is a slight dependence on experimental parameters such as the angle between the atomic and laser beam. Since saturation effects and the resulting power-broadening of the resonance are not explicitly included in the line shape model, they will implicitly show up as an increase of both $\Gamma_\mathrm{G}$ and $\Gamma$ over the value expected just from atomic beam divergence. Note that for short delay times ($\tau < 400\,\mu\mathrm{s}$), Doppler-broadening dominates, while for longer delay times saturation effects are the dominant broadening mechanism. For these reasons, both $\Gamma_\mathrm{G}$ and $\Gamma$ are used as free parameters when fitting the experimental data. We find that $\Gamma_\mathrm{G}$ ($\Gamma$) ranges from 14 MHz (13 MHz) to 6 MHz (16 MHz) for the different delay times (see Sec. 2.1), leading to total FWHM line width ranging between 22 MHz and 17 MHz.

The convolution of Eq. S2 and Eq. S3 can be written as:

$$F(\omega) = C\frac{4\sqrt{\pi \ln 2}}{\Gamma_{\mathrm{G}}\Gamma}\left\{\mathrm{Re}[w(z)] + \frac{\Gamma}{2}\frac{b}{C}\,\mathrm{Im}[w(z)]\right\}, \tag{S4}$$

with $z = 2\sqrt{\ln 2}[(\omega - \omega_0) + i\Gamma/2]/\Gamma_{\mathrm{G}}$ and the Faddeeva function $w(z)$ given by (*49*)

$$w(z) \equiv e^{-z^2}\left(1 + \frac{2i}{\sqrt{\pi}}\int_0^z e^{t^2}\mathrm{d}t\right). \tag{S5}$$

Defining the asymmetry parameter as $\eta \equiv b\Gamma/4C$ and replacing the constant multiplicative prefactor with the free fit parameter $A = C\frac{4\sqrt{\pi \ln 2}}{\Gamma_{\mathrm{G}}\Gamma}$, we recover the Fano-Voigt line shape given in Eq. 5 of the main text. To account for the experimental background (see Sec. 1.1), a constant term $y_0$ is added to the line shape (to Eq. S2 and thus Eq. S4) as a free fit parameter.

We test the Fano-Voigt line shape by fitting it to the results of the OBE simulation described in the main text and find a high suppression of quantum interference line shifts. To test the robustness of the fit and to control possible biases, we conduct a Monte Carlo simulation with the experimental frequency sampling of the resonance and the observed signal-to-noise ratio, including slow drifts in the latter, applied to the line shape from the OBE simulation.

## 1.3 Data analysis

Each recorded resonance, consisting of $N$ pairs of laser frequency and number of fluorescence photons detected, is fit with the Fano-Voigt line shape with six free parameters (line center $\nu_0 = \omega_0/2\pi$, amplitude $A$, background $y_0$, Lorentzian line width $\Gamma$, Gaussian line width $\Gamma_{\mathrm{G}}$ and asymmetry parameter $\eta$). We assume that the uncertainty $\sigma_{y,i}$ on the number of fluorescence photons detected $y_i$ for each frequency point $i$ is dominated by shot noise, i.e. $\sigma_{y,i} \approx \sqrt{y_i}$ (since $y_i \gg 1$, we can approximate the Poisson distribution with a normal distribution). The optimal values of the free parameters are then found by minimizing $\chi^2$, with the uncertainty of the values corresponding to an increase of $\chi^2$ of 1 around the optimal values.

We use $\chi^2_{\mathrm{red}} = \chi^2/N_{\mathrm{DOF}}$, with $N_{\mathrm{DOF}} = N - 6$, as a measure of the goodness of fit. The resulting $\chi^2_{\mathrm{red}}$ distribution for the individual resonance fits deviates from the distribution expected for the assumed noise in two regards. First, the mode of the observed distribution is about 20% larger than expected, corresponding to a mean $\chi^2_{\mathrm{red}}$ of $\sim$1.20. Second, while the distribution follows the expected shape for $\chi^2_{\mathrm{red}} \lesssim 2$, there is an excess of resonance fits with $\chi^2_{\mathrm{red}} \gtrsim 2$.

The former is partly caused by the fact that while the Fano-Voigt line shape describes the observed line shape very well, there are small deviations between the two. The deviations can be decomposed in a contribution symmetric about the line center (up to 4% relative deviation) and a contribution asymmetric about the line center, more than an order of magnitude smaller than the symmetric contribution. The symmetric contribution is caused by (1) saturation effects related to the depletion of the initial $2\mathrm{S}_{1/2}^{F=0}$ state and (2) non-Gaussian Doppler-broadening, both not included in the Fano-Voigt line shape. This symmetric contribution can lead to a sampling bias, as discussed in Sec. 2.5. The asymmetric contribution is caused by quantum interference effects not described by the Fano-Voigt line shape and discussed in Sec. 2.4. These deviations increase the fit residuals over what would be expected from pure shot noise and shift the mode of the observed $\chi^2_{\text{red}}$ distribution to higher values. From our Monte Carlo simulation used to test the Fano-Voigt line shape (see Sec. 1.2), we expect this effect to increase the mode by 10%.

Additionally, there are other sources of noise present in the system. Besides shot noise, we expect the dominant noise contribution to be drifts in the number of 2S atoms reaching the 2S-4P excitation region caused by built-up of hydrogen ice on the cryogenic nozzle and drifts in the 243 nm laser power and in the condition of the RF discharge producing the hydrogen atoms. This additional noise also shifts the mode of the observed $\chi^2_{\text{red}}$ distribution to a higher value than expected for pure shot noise.

These drifts also cause an excess of resonance fits with $\chi^2_{\text{red}} \gtrsim 2$ as compared to the shot noise only situation, as can be deduced from our Monte Carlo simulation when properly modeling the drifts by interpolating the observed line amplitude and background as function of time. Another contribution are short (i.e. only affecting one or two frequency points), but large perturbations of the system, such as discharges in the detectors causing a short spike in count rate. To remove such events from the data analysis, a $\chi^2_{\text{red}}$ cut-off of 3 is introduced, i.e. individual resonance fits with a $\chi^2_{\text{red}} \geq 3$ are neglected in the data analysis. This results in a removal of less than 4% of the recorded resonances. The effect of this cut-off on the determined transition frequencies is within the final uncertainties.

By design, the free parameter $\eta$ of the Fano-Voigt line shape is correlated with the line center $\nu_0$. For the typical signal-to-noise ratios of the recorded resonances, this leads to a significantly larger uncertainty on the line center when fitting the Fano-Voigt line shape as compared to fitting the Voigt line shape (where $\eta = 0$). To decrease this uncertainty, $\eta$ is not treated as a free parameter for each resonance, but rather treated as one free parameter shared by a subset of resonances, effectively increasing the signal-to-noise ratio. Each subset only contains data taken for a specific laser polarization setting $\theta_{\text{L}}$ and for a single delay time interval and detector and thus subject to the same line distortions due to quantum interference corresponding to the same value of $\eta$. With this procedure, the uncertainty on the line center is reduced by about a factor of 2 and to the same level as when using the Voigt line shape. Since the Fano-Voigt line shape tends to be a numerically unstable fit for resonances where the Gaussian broadening is negligible ($\Gamma_{\text{G}} < 0.1$ MHz), we use a Fano-Lorentzian fit, i.e. the line shape given in Eq. S2 that does not include the Gaussian broadening of the Fano-Voigt. Both procedures, fitting using a shared $\eta$ and using the Fano-Lorentzian fit where appropriate, do not change the determined transition frequencies within the fit uncertainty.

Finally, to determine the transition frequencies given in the main text, small model, sampling bias and light force shift corrections (see Sec. 2.4, 2.5 and 2.3) are determined for each recorded resonance and applied to the extracted line center. The transition frequencies are deduced by a weighted average of the line centers for all laser polarization settings $\theta_L$, all delay times and both detectors, using the fit uncertainty on the line center as weight. The statistical uncertainty given corresponds to the uncertainty of the weighted average. The $\chi^2_{\text{red}}$ of this weighted average is 1.16 for both the $2S\text{-}4P_{1/2}$ and $2S\text{-}4P_{3/2}$ transitions.

# 2 Corrections and uncertainties

Lists of the corrections applied and the contributions (at the one standard deviation $\sigma$ level) to the total measurement uncertainty for two measured transitions $2S\text{-}4P_{1/2}$ and $2S\text{-}4P_{3/2}$ are given in table S2. We assume that the individual uncertainty contributions in each list are not correlated and add them in quadrature. The uncertainty contributions are, however, correlated between the two transition frequencies and we give the corresponding correlation coefficient $r$ ($r_{X,Y} = \frac{\mathrm{cov}(X,Y)}{\sigma_X \sigma_Y}$ for the contributions $X$, $Y$, with $\mathrm{cov}(X, Y)$ the covariance between $X$ and $Y$). The derived corrections and uncertainties for the 4P fine structure centroid $\nu_{2S-4P}$ and 4P fine structure splitting $\Delta\nu_{\mathrm{FS}}^{\mathrm{exp}}(4P)$ are given in Table 1 in the main text and table S3, respectively. In the following, each of the items in the lists is briefly discussed.

## 2.1 First-order Doppler shift

The suppression of the first-order Doppler shift using an active fiber-based retroreflector (AFR) and the experimental constraint on this suppression have been discussed in the main text and at length in (*36*). In contrast to (*36*), where a symmetric line shape was used to determine the line center, we here analyze our data as detailed in Sec. 1.3 using the Fano-Voigt line shape and small corrections from simulations. This is important because there is a dependence of the line distortions due to quantum interference on the interaction time with the spectroscopy beam and thus the mean velocity of the atoms (see Sec. 2.4). If those line distortions are not properly accounted for, they can result in an apparent shift of the transition frequency as a function of delay time, thus mimicking a Doppler shift.

The velocity distribution of the atoms for the different detection delay times is not only given by the initial velocity distribution of the atoms leaving the nozzle and the subsequent collimation by the apertures (see Fig. 3 of the main text), but is also influenced by the beam radius, power and detuning of the 243 nm laser that excites the ground state atoms to the $2S_{1/2}^{F=0}$ level. To model this, we perform a Monte Carlo simulation of atomic trajectories, taking into account the 1S-2S and 2S-4P excitation. To estimate the uncertainty of this approach, we vary the input parameters, including the initial velocity distribution, and compare the simulation results with experimentally accessible parameters such as the signal amplitude and line width for different delay times, which are both highly sensitive to the velocity distribution. The mean velocity $v_i$ of atoms excited to the 4P level is found to range between 295(40) m/s and 85(10) m/s for the ten different delay time intervals, leading to an overall mean velocity of $\bar{v}$ = 240(30) m/s for all delay times considered. The transverse velocity distribution, i.e. along the direction of the 486 nm beams and relevant for Doppler broadening as opposed to a shift, is approximately Gaussian with a FWHM that ranges from 6(1) m/s down to 1.9(2) m/s.

Using these mean velocities $v_i$ and the measured transition frequencies $\nu_i$ for the ten different delay time intervals, the rate of change $m$ of the transition frequency as a function of mean velocity, or Doppler slope, can be extracted for different subsets of the data. This is done by fitting a linear model $\nu = mv + \delta\nu$ to the data, with the uncertainty on the Doppler slope derived from the uncertainty on the measured transition frequencies. If $\chi^2_{\text{red}} = \chi^2/8$ of the determination is found to be above 1, the Doppler slope uncertainty is scaled with $\sqrt{\chi^2_{\text{red}}}$ to arrive at a conservative estimate of the Doppler slope uncertainty $\sigma_m$. Finally, the corresponding Doppler uncertainty $\sigma_{\nu,\text{D}}$ on the transition frequency averaged over the delay times is found by multiplying the Doppler slope uncertainty with overall mean velocity of all delay times, $\sigma_{\nu,\text{D}} = \sigma_m \bar{v}$.

When comparing the Doppler slopes extracted in this way for different measurement days, we find some excessive day-to-day scatter. We attribute this to slight misalignments in the AFR, such as in the angle $\alpha$ between the spectroscopy laser beams and the atomic beam or in the position of the beam waist of the laser beams, which should ideally coincide with HR mirror surface (*36*). Indeed, for a few measurement days the observed line width was slightly larger than on average, hinting at a possible slight misalignment of $\alpha$. These misalignments can cause a residual Doppler shift with a sign depending on the direction of misalignment. Since the AFR is re-adjusted for most measurement days and since we expect the misalignments to be in a random direction, we expect the Doppler shift to compatible with zero when averaging over a sufficient number of days or, equivalently, re-alignments.

For the complete data set for each transition measured, we find the Doppler slopes to be $m = 0.7(12.1)$ Hz/(m/s) and $m = 9.5(11.8)$ Hz/(m/s) for the $\text{2S-4P}_{1/2}$ and $\text{2S-4P}_{3/2}$ transitions, respectively, and thus compatible with zero. The excessive scatter manifests itself in an increased $\chi^2_{\text{red}}$ of 1.83 and 1.47 for the two transition and has already been taken into account for the uncertainties by scaling with $\sqrt{\chi^2_{\text{red}}}$. Since the Doppler slopes are found to be compatible with zero, we do not apply a correction to the transition frequencies, but only include the uncertainty on the slopes. Finally, the Doppler uncertainty on the transition frequencies $\nu_{1/2}$ and $\nu_{3/2}$ is found to be 2.92 kHz and 2.84 kHz, respectively. We assume the uncertainties to be uncorrelated for the two transitions ($r = 0$) and thus find a Doppler uncertainty of 2.13 kHz for $\nu_{\text{2S}-\text{4P}}$.

## 2.2 Quantum interference

While we have thoroughly tested the compensation of line shifts due to quantum interference with the Fano-Voigt line shape using our simulations (see Sec. 1.2, 1.3 and 2.4), we here estimate the level of compensation directly from the experimental data. The basic idea is that any residual line shifts should follow the same functional behavior $\Delta\nu \equiv \Delta\nu(\theta_{\text{L}})$, where $\theta_{\text{L}}$ is again the linear laser polarization angle, as the uncompensated line shifts seen when using the Voigt line shape for data analysis (see Fig. 4 (A and B) in the main text). For this purpose, we parametrize $\Delta\nu(\theta_{\text{L}})$, starting from the analytical expression derived in the perturbative limit. For the $\text{2S-4P}_{1/2}$ transition, $\Delta\nu(\theta_{\text{L}})$ is a simple sinusoidal function, while for the $\text{2S-4P}_{3/2}$ transition $\Delta\nu(\theta_{\text{L}})$ is a more complicated function that can be approximated by a power series in sinusoidal functions (as will be detailed elsewhere). The parametrization $\Delta\nu(\theta_{\mathrm{L}}, A_{\mathrm{res}}, \Delta\nu_0)$ has two free parameters, amplitude $A_{\mathrm{res}}$ and offset $\Delta\nu_0$, while the phase and period are fixed to the value of the uncompensated line shift and $180\,^\circ$, respectively. $\Delta\nu(\theta_{\mathrm{L}}, A_{\mathrm{res}}, \Delta\nu_0)$ is then fit to the data analyzed with the Fano-Voigt line shape, taking into account only the statistical uncertainty.

We first analyze the difference between the detectors, i.e. the difference in the observed line centers, which is somewhat more sensitive to quantum interference effects than the signal from the individual detectors. Here, we find a residual amplitude of $A_{\mathrm{res}} = 4.46(1.36)$ kHz (compared to 54.8(1.3) kHz for the Voigt line shape) for the $2\mathrm{S}\text{-}4\mathrm{P}_{1/2}$ transition and a residual amplitude of $A_{\mathrm{res}} = -1.68(2.00)$ kHz (compared to 28.6(2.0) kHz for the Voigt line shape) for the $2\mathrm{S}\text{-}4\mathrm{P}_{3/2}$ transition. Thus, the residual amplitude of the $2\mathrm{S}\text{-}4\mathrm{P}_{3/2}$ transition is compatible with zero, while there is small residual effect for $2\mathrm{S}\text{-}4\mathrm{P}_{1/2}$ transition.

To identify the origin of the residual amplitudes and to estimate the related uncertainty, we analyze the data of the individual detectors (see Fig. 4 (C and D) in the main text). The amplitudes of residual line shifts $A_{\mathrm{res}}$ are found to be well compatible with zero for detector CEM2 for the $2\mathrm{S}\text{-}4\mathrm{P}_{1/2}$ transition ($A_{\mathrm{res}} = -0.09(0.84)$ kHz) and for both detectors CEM1 and CEM2 for the $2\mathrm{S}\text{-}4\mathrm{P}_{3/2}$ transition ($A_{\mathrm{res}} = -0.16(1.23)$ kHz and $A_{\mathrm{res}} = 0.11(1.06)$ kHz, respectively). For detector CEM1 and the $2\mathrm{S}\text{-}4\mathrm{P}_{1/2}$ transition, we find a small residual amplitude of $A_{\mathrm{res}} = 3.23(1.16)$ kHz with a goodness of fit of $\chi^2_{\mathrm{red}} = 6.6$. Thus, the residual amplitude seen in the difference is solely caused by the data from one of the detectors, CEM1. We note that otherwise the data from the two detectors, including the line shifts due to quantum interference (see Fig. 4 (A and B) in the main text), are very similar. To cross-check the compensation of asymmetries with the Fano-Voigt line shape, we have also analyzed the residual amplitudes using the Voigt line shape combined with our simulations (resulting in large corrections of tens of kilohertz) and find very similar results. The significance of this nonzero residual amplitude should be contrasted with the fact that a simple weighted average describes the data equally well ($\chi^2_{\mathrm{red}} = 6.7$). Furthermore, as evident in the large $\chi^2_{\mathrm{red}}$, we have so far neglected residual Doppler shifts, which contribute an uncertainty as large as the residual amplitude and are expected to cause the day-to-day scatter seen here (with data for different $\theta_{\mathrm{L}}$ values recorded on different days) (see Sec. 2.1). Note that discarding the data for $\theta_{\mathrm{L}} = 0\,^\circ$ and $\theta_{\mathrm{L}} = 90\,^\circ$ for the $2\mathrm{S}\text{-}4\mathrm{P}_{1/2}$ transition, which shows increased scatter as discussed in the main text, reduces the significance of the nonzero residual amplitude only slightly.

The transition frequencies given in the main text are determined by averaging over $\theta_{\mathrm{L}}$. To estimate residual quantum interference line shifts of these transition frequencies, the relation of the amplitude $A_{\mathrm{res}}$ and offset $\Delta\nu_0$ is investigated using our OBE simulations, using the experimental sampling of $\theta_{\mathrm{L}}$. We find $\Delta\nu_0/A_{\mathrm{res}} \approx 0.2$ and $\Delta\nu_0/A_{\mathrm{res}} \approx -0.2$ for the $2\mathrm{S}\text{-}4\mathrm{P}_{1/2}$ and $2\mathrm{S}\text{-}4\mathrm{P}_{3/2}$ transitions. With this, and after averaging over the two detectors, the residual line shifts are determined to be 0.29(33) kHz and 0.00(26) kHz for the $2\mathrm{S}\text{-}4\mathrm{P}_{1/2}$ and $2\mathrm{S}\text{-}4\mathrm{P}_{3/2}$ transitions, respectively, where $\chi^2_{\mathrm{red}} > 1$ has been taken into account by scaling the uncertainties with $\sqrt{\chi^2_{\mathrm{red}}}$.

We have tried to reproduce, using our simulations, the occurence of a residual amplitude in only one of the two detectors, but were not able to come up with a satisfactory explanation. One potential reason for a broken symmetry between the detectors is that the 2S-4P excitation region is not pointlike, but corresponds to the laser beam size (beam waist $w_0 = 1.85\,\text{mm}$, detector radius 28 mm). This in turn leads to the two detectors observing slightly different solid angles and atomic velocities, and we indeed observe a slight differential Doppler shift between the two detectors.

However, when averaging over the two detectors and $\theta_\text{L}$, both the line shifts due to this residual amplitude, as shown above, and the residual Doppler shift (see Sec. 2.1) are found to be compatible with zero, which is why we are confident that this residual effect does not influence our final results obtained by this average.

Thus, we assign an uncertainty of 0.33 kHz and 0.26 kHz due to residual quantum interference line shifts to the measured transition frequencies $\nu_{1/2}$ and $\nu_{3/2}$ respectively. Since the uncertainties are assumed to be limited by statistics, they are uncorrelated ($r = 0$) for the two transitions, resulting in an uncertainty of 0.20 kHz for $\nu_{2\text{S}-4\text{P}}$.

## 2.3 Light force shift

Atoms that are on a classical trajectory through a near-resonant standing wave may be subject to forces that are attractive to the nodes or anti-nodes for red or blue detuning, respectively (*50*). In this simple classical view, it seems obvious that the observed resonance can be distorted because this effect can enhance the red wing of the resonance while reducing the signal on the blue side, or, depending on the trajectory, vice versa. In fact one may model these line distortions by solving the OBEs with a position-dependent Rabi frequency simultaneously with Newton's equation of motion (*51, 52*). However, this simple classical description of atomic motion is attached to several conditions (*52*). One of them is that the atoms need to be sufficiently localized in order to assign a position-dependent force to them. In our case this condition is violated for two reasons: the transverse temperature of the 2S atoms is low enough to yield a coherence length of the atomic matter wave comparable to the periodicity of the optical potential ($\lambda/2 = 243\,\text{nm}$). In addition, a single photon recoil is enough to separate the ground state part of the wave function by several half wavelengths from the excited state part while the atom crosses the 2S-4P spectroscopy beams.

The system then has to be described by including the atom's transverse momentum $p$ along the 2S-4P spectroscopy laser beams in the quantum mechanical model together with the atom's internal dynamics. Interaction with the laser beams changes the atom's momentum by $\pm n\hbar k$, corresponding to the exchange of $n$ photons with momentum $\hbar k$, and thus couples the corresponding momentum states, while spontaneous decay leads to a random recoil which averages to zero. In this picture, the coupling of the momentum states modifies the line shape of the transition probed and leads to a coherent superposition of the momentum states, corresponding to a partial localization similar to the classical picture. An analytic solution to this problem can be obtained with the effective Hamiltonian approach (see e.g. (*53*)), using the Wigner function to describe the initial momentum state of the atoms emerging from the nozzle and flying through the apertures in the apparatus (subject of an upcoming publication). Because of the rapid decay of the 4P excited state to 1S ground state, the otherwise infinite momentum space can be reduced and $n = 4$ is found to be sufficient to describe the system. This analytic solution ignores the back decay of the excited 4P state to the initial $2S_{1/2}^{F=0}$ state, but since the branching ratio is only 4% this approximation is adequate and has been confirmed with more sophisticated models including this back decay. While it should in principle be possible to include effects that influence the initial momentum state such as the 1S-2S excitation in this analytical solution, we could also show that it is sufficient to describe the initial momentum state as a (fully delocalized) momentum eigenstate $|p\rangle$. We can then employ our OBE simulation to describe the system in a realistic experimental setting by including the momentum eigenstates $|p + n\hbar k\rangle$ along with the internal states of the atom. Using the atomic trajectories discussed in Sec. 2.1 as input for the OBE simulation, a Monte Carlo simulation can then be used to estimate shifts of the observed line center caused by the coupling of momentum states.

In this way, we find a light force shift of -0.43(40) kHz and -0.26(25) kHz for the $2S\text{-}4P_{1/2}$ and $2S\text{-}4P_{3/2}$ transitions, respectively. The $\nu_{1/2}$ and $\nu_{3/2}$ transition frequencies have been corrected for this shift. The uncertainty is limited by our knowledge of the atomic velocity distribution, which is correlated for the two transitions. Thus the uncertainties are fully correlated ($r = 1$) for the two transitions, leading to a correction of -0.32(30) kHz for $\nu_{2S-4P}$.

## 2.4 Model corrections

The Fano-Voigt line shape is derived from the perturbative description of quantum interference, i.e. it does not account for effects such as back decay from the excited 4P state to the initial $2S_{1/2}^{F=0}$ state and the depletion of this initial state, leading to a saturation of the observed transition. In our system, 4% of the decays of the 4P state lead back to the initial $2S_{1/2}^{F=0}$ state and we excite about 30% of the 2S atoms to the 4P state. Under these conditions, the line asymmetries due to quantum interference can increase more than two-fold over the perturbative regime for the slowest probed atoms, i.e. those with longest interaction times with the spectroscopy laser. However, we find that the Fano-Voigt line shape is still a good description for our system, since the bulk of the increased line asymmetry is matched by it and thus can be accounted for by fitting the Fano-Voigt line shape to the observed resonances. Only a small residual asymmetry that does not match the asymmetry of the Fano-Voigt line shape remains. Importantly, this residual asymmetry is significantly less detection geometry-dependent than the asymmetry removed by the Fano-Voigt line shape, since it mainly stems from the small modification of the $2S_{1/2}^{F=0}$ population caused by the back decay of the 4P state to this state, which is independent of the detection geometry. This allows us to model this residual asymmetry without the need to accurately describe the detection geometry. The line shifts associated with this residual asymmetry depend on the intensity of and interaction time with the spectroscopy laser and are on the order of 1 kHz. As all effects stemming from quantum interference, the shifts are of opposite sign for the two perturbing resonances. Since the intensities in the experiment were chosen such that the Rabi frequencies are approximately identical for the two transitions probed, the shifts are of ratio 2:1 for the $2\text{S-}4\text{P}_{1/2}$ and the $2\text{S-}4\text{P}_{3/2}$ transitions. Thus, for the 4P fine structure centroid $\nu_{2\text{S}-4\text{P}}$ they largely cancel.

Apart from these effects related to quantum interference, there is also an AC Stark shift of the atomic levels involved and a line shift associated with off-resonant excitations caused by optical pumping into the $2\text{S}_{1/2}^{F=1}$ states. Both contributions lead to small shifts of the observed line center on the order of $0.10\,\text{kHz}$. The initial population in the $2\text{S}_{1/2}^{F=1}$ states from Doppler-sensitive two-photon excitation is negligible compared to the population accumulated in these states through optical pumping.

All these effects are included in the OBE simulations of our system. For each recorded resonance, a corresponding line shape is generated from the OBE simulation. Experimental noise, consisting of shot noise and slow drifts in the number of 2S atoms contributing to the signal, is added and the resulting resonance is fit with the Fano-Voigt line shape, a process that is repeated multiple times in the fashion of a Monte Carlo simulation. The line center determined from these fits is then used a model correction for the recorded resonances. With this, we find a model correction of 1.34(23) kHz and -0.50(10) kHz for the $\nu_{1/2}$ and $\nu_{3/2}$ transition frequencies, respectively. The uncertainty is estimated by varying the input parameters to the OBE simulation within the experimental constraints. Some of the contributions to the uncertainty, such as the spectroscopy laser power, are uncorrelated, while others, such as the atomic velocity distribution, are correlated for the two transition measurements. The total uncertainty is found to be partially anticorrelated ($r = -0.65$) for the two measurements, resulting in a model correction of 0.11(6) kHz for $\nu_{2\text{S}-4\text{P}}$.

## 2.5 Sampling bias

As has been detailed in Sec. 1.3, there are small deviations symmetric about the line center between the observed line shape and the Fano-Voigt line shape used for data analysis (the asymmetric deviations are much smaller and included in the model corrections discussed in Sec. 2.4). In combination with an experimental frequency sampling of the resonances that is not quite symmetric about the line center, this can lead to a bias in the line center determined from the fit to the resonance. We use our Monte Carlo simulations discussed in Sec. 2.4 to estimate this bias and find it to be up to 2.1 kHz for the $2\text{S-}4\text{P}_{3/2}$ transition.

To reduce this bias, we enforce fair sampling of the resonance by selectively removing up to two experimental frequency points for each resonance. The procedure is the following: First, the simulated resonance is fit with the Fano-Voigt line shape for all available simulation points and with no experimental noise added. Then, it is fit for the experimental frequency points, with the difference in extracted line centers giving the bias to be reduced. Next, experimental frequency points are removed and the simulated resonance is fit again. This is repeated for all combinations of removing one or two points. Finally, we choose to remove the one or two experimental points without which the bias is lowest (no point is removed if the bias is lowest for full sampling), resulting on the removal of 1.94 frequency points on average. This procedure is applied to every recorded resonance. The experimental data is then fit again with these points removed, leading to a slight increase in the statistical uncertainty of approximately 4%.

After applying this procedure, the sampling bias is, again using the Monte Carlo simulations, found to be 0.34 kHz and 0.83 kHz for the 2S-4P$_{1/2}$ and 2S-4P$_{3/2}$ transitions, respectively. The transition frequencies $\nu_{1/2}$ and $\nu_{3/2}$ have been corrected for this bias and an uncertainty of 0.40 kHz and 0.70 kHz, respectively, is assigned to these corrections. This uncertainty accounts for the fact that a separate OBE simulation is used to estimate the light force shift (see Sec. 2.3) and that this simulation shows very similar symmetric deviations (since it also includes the same saturation effects and same Doppler-broadening as the quantum interference OBE simulation) and thus leads to a very similar sampling bias. In order to avoid a double counting of this bias, we only correct for the bias found for the quantum interference OBE simulation, but add the bias of the light force shift OBE simulation as uncertainty. The uncertainty for the two transitions is uncorrelated ($r = 0$), and thus $\nu_{2S-4P}$ includes a correction of 0.44(49) kHz.

## 2.6 Second-order Doppler shift

The second-order Doppler effect is not canceled by the excitation of the 2S-4P transition utilizing phase-retracing beams. However, the signal weighted, mean squared velocity $\overline{v^2}$ = 255(30) m/s provided by the cryogenic source of 2S atoms and estimated using the Monte Carlo simulations described above, is sufficiently small so that the second-order Doppler shift only amounts to

$$\Delta\nu_{\text{SOD}} = -\frac{1}{2}\frac{\overline{v^2}}{c^2}\nu_{2S-4P} = -0.22(5)\,\text{kHz}. \tag{S6}$$

The measured transitions frequencies $\nu_{1/2}$ and $\nu_{3/2}$ are corrected for the second-order Doppler shift by subtracting $\Delta\nu_{\text{SOD}}$. The uncertainty is fully correlated ($r = 1$) for the two transitions, and thus the correction and uncertainty for $\nu_{2S-4P}$ is the same as for the individual frequencies.

## 2.7 dc-Stark shift

Special care was taken to suppress stray electric fields in the 2S-4P excitation region to avoid the associated line shifts due to the dc-Stark effect. A grounded Faraday cage made from stainless steel mesh with two wires (diameter 30 μm) per millimeter shields the excitation region from the static electric fields created by the channel electron multiplier input surfaces (+270 V) (see Fig. 3 in the main text). The Faraday cage and all surfaces in the excitation region are spray-coated with colloidal graphite to suppress the built-up of patch charges and to avoid fields due to contact potentials. An upper limit of 0.6 V/m has been obtained for the field strength of stray electric fields using 1S-2S spectroscopy in a dedicated setup with similar dimensions and identical coating (*1, 2, 54*). We use this upper limit on the field strength to estimate line shifts for the 2S-4P transitions.

The dc-Stark effect shifts the energies of the $4P_{1/2}^{F=1}$ and $4P_{3/2}^{F=1}$ levels (the shift of the $2S_{1/2}^{F=0}$ level is negligible on the current level of accuracy) in the presence of a static electric field $\vec{F}$, with the energy shift $\Delta\omega = 2\pi \times a_{j,m_F}F^2$ proportional to the square of the field strength $F = |\vec{F}|$. The coefficients $a_{j,m_F}$ depends on the orientation of the static electric field $\vec{F}$ with respect to the quantization axis of the atom, given by the linear laser polarization $\vec{E}$. The coefficients $a_{j,m_F}$, derived by diagonalizing the atomic Hamiltonian in the presence of a static electric field and fitting the resulting energy shifts with a quadratic function, are shown in table S1.

The upper limit for the dc-Stark shift of the transition frequencies for the worst case orientation of the stray electric fields is $-0.03$ kHz and $-0.49$ kHz for the $2\text{S-}4\text{P}_{1/2}$ and $2\text{S-}4\text{P}_{3/2}$ transitions, respectively. However, we assume that the orientation of possible stray electric fields $\vec{F}$ is not correlated with the orientation of the linear polarization $\vec{E}$ of the spectroscopy laser. Since the transition frequencies $\nu_{1/2}$ and $\nu_{3/2}$ are determined from data taken for different orientations of the laser polarization (see Fig. 4 in the main text), the upper limit on the dc-Stark shift for these transitions is further reduced by averaging over the different orientations of stray electric fields. With this, we estimate an uncertainty due to the dc-Stark shift of 0.03 kHz and 0.30 kHz for the $\nu_{1/2}$ and $\nu_{3/2}$ transition frequencies, respectively. The uncertainty for the two transitions is uncorrelated ($r = 0$) and the combined uncertainty is 0.20 kHz for $\nu_{2\text{S}-4\text{P}}$.

Furthermore, we note that the shift of the line center extracted by fitting the observed resonance may be smaller than the shift of the energy levels, since the electric field does not only shift the energy of the involved levels, but mixes different atomic levels. In this way, excitations of the 4S and 4D levels (which now have some admixture of the 4P level) become dipole-allowed. These transitions tend to cancel out the dc-Stark shift of the 4P levels when the transitions are within the recorded laser frequency range. We have however not fully investigated this cancellation, which depends on the excitation dynamics of the system, and thus here use the shift of the energy levels as upper limit for the shift of the resonances.

## 2.8 Zeeman shift

The earth's magnetic field is pre-compensated by three orthogonal pairs of Helmholtz coils outside the vacuum chamber. In addition, the 2S-4P excitation region is shielded from residual magnetic fields by a double-layer high-permeability metal (mu-metal) shield. Thereby, magnetic fields are suppressed to less than 1 mG in a volume of about 15 $\text{cm}^3$ around the 2S-4P excitation region.

For a given magnetic flux density $B$, the linear Zeeman effect shifts the energies of the magnetic sublevels $m_F = \pm 1$ of the $4\text{P}_{1/2}^{F=1}$ ($4\text{P}_{3/2}^{F=1}$) state by $\Delta E/\hbar = g_F \mu_B B m_F/\hbar = 2\pi \times 0.467$ kHz/mG ($\Delta E/\hbar = 2\pi \times 2.33$ kHz/mG), using the appropriate g-factor $g_F$ and the Bohr magneton $\mu_B$. The observed transition frequency, i.e. the center of weight of the signals from the different magnetic sublevels, will only be shifted if the $m_F = \pm 1$ sublevels contribute with different amplitudes. This situation requires some circularly polarized light about the direction of the magnetic field, i.e. a mismatch in the intensities $|E_L|^2$ and $|E_R|^2$ of the left- and right-polarized components as given by the Stokes parameter $v = (|E_L|^2 - |E_R|^2)/(|E_L|^2 + |E_R|^2)$. An upper limit for the shift of the observed transition frequency of $\Delta\omega = \Delta E/\hbar \times v$ is obtained by assuming that the magnetic field is aligned along the spectroscopy laser beam.

We determine $v$ from the measured intensity polarization extinction ratio $P_{\text{er}} = 1/200$, limited by the polarization-maintaining fiber, and find $v = 0.14$. Thus, we estimate the uncertainty due to the Zeeman shift to be below 0.07 kHz and 0.33 kHz for the $\nu_{1/2}$ and $\nu_{3/2}$ transition frequencies, respectively. The shift is uncorrelated ($r = 0$) for the two measured transitions and a combined uncertainty of 0.22 kHz is assigned for $\nu_{\text{2S-4P}}$. The quadratic Zeeman effect that would also affect the initial $2\text{S}_{1/2}^{F=0}$, $m_F = 0$ and the $4\text{P}_{1/2}^{F=1}$, $m_F = 0$ and $4\text{P}_{3/2}^{F=1}$, $m_F = 0$ states is negligible at our current level of accuracy.

## 2.9 Pressure shift

To estimate the pressure shift we use the impact approximation for binary collisions (*55*). The interaction energy between the perturbed and the perturbing atom is required as input for this theory and is given by the near-field dipole-dipole interaction:

$$\hat{V} = \frac{1}{4\pi\varepsilon_0 R^3}\left(\hat{d}_{1x}\hat{d}_{2x} + \hat{d}_{1y}\hat{d}_{2y} - 2\hat{d}_{1z}\hat{d}_{2z}\right), \tag{S7}$$

where $R$ is the distance between the atoms and $\hat{d}_{ij}$ with $j = x, y, z$ are the components of the electric dipole moment for the perturbing ($i = 1$) and perturbed ($i = 2$) atoms. The energy shift of the product state $|n\rangle = |n_1\rangle \otimes |n_2\rangle = |n_1, n_2\rangle$ of the perturbing and perturbed atoms due to the Van-der-Waals interaction can be calculated using second-order perturbation theory:

$$E_{\text{VdW}}(n) = \sum_{m, E_n \neq E_m} \frac{|\langle n|\hat{V}|m\rangle|^2}{E_n - E_m} = \frac{C_6}{R^6}, \tag{S8}$$

where $|m\rangle$ are all possible product states of the two atoms and $E_m$ is the energy of state $|m\rangle$. The interaction energy is thus $\propto 1/R^6$ and the strength of the interaction is expressed by the coefficient $C_6$. For resonant interactions, $E_n - E_m = 0$, and Eq. S8 is not valid anymore. Furthermore, in our case the perturbing and perturbed atoms can be connected by dipole-allowed transition (e.g. colliding 4P and 1S atoms) and thus a resonant interaction of $|n_1, n_2\rangle$ with $|n_2, n_1\rangle$ and non-vanishing $\langle n_1, n_2|\hat{V}|n_2, n_1\rangle$ is possible, corresponding to an excitation exchange between the atoms. Nevertheless, it is well known that these interactions only cause a line broadening, but do not cause a line shift (*56*). Hence we can estimate the largest contribution with the smallest possible $E_n - E_m$ and use the analytic expressions for the dipole matrix elements for colliding hydrogen atoms in any states. Furthermore, we assume that all perturbing particles are 1S atoms, since the density of 2S atoms is only $\sim 10^{-3}$ of the 1S atom density and the contribution due to collisions with background molecules can be neglected as there is no close degeneracy. The perturbation of atoms in the 2S initial state causes a negligible line shifts on the current level of accuracy, leaving only the perturbation of the excited $4\text{P}_J$ atoms caused by collisions with 1S atoms to consider. We approximate the sum in Eq. S8 with the minimum combined internal energy difference given by the hyperfine splitting of $4P_j$ states (7.39623(7) MHz for $j = 1/2$ and 2.95647(3) MHz for $j = 3/2$, see Fig. 2A in the main text) and obtain

$$C_6(4\text{P}_{1/2} - 1\text{S}) = 1.9 \times 10^5 hcR_\infty a_0^6, \quad C_6(4\text{P}_{3/2} - 1\text{S}) = 1.9 \times 10^6 hcR_\infty a_0^6, \tag{S9}$$

with the Bohr radius $a_0$. Within the impact approximation the $C_6$ Van-der-Waals interaction leads to a frequency shift of (*55*)

$$\Delta\omega \approx 2.9 \left(\frac{C_6}{\hbar}\right)^{2/5} v^{3/5} N, \tag{S10}$$

where $v$ is velocity of colliding atoms and $N$ is the density of the perturbing atoms. From our experimental data, we estimate the concentration of 1S atoms in the beam to be $N_{\text{beam}} \approx 2.2 \times 10^{15}$ atoms/m$^3$ at temperature 5 K and the concentration of background atoms to be not more than $N_{\text{bkg}} \approx 2.4 \times 10^{15}$ atoms/m$^3$ at temperature $300\,\text{K}$. Estimating the collisional velocity of the atoms by their mean velocity, we find the contributions to the pressure shifts of the two measured transitions to be $\Delta\omega(4\text{P}_{1/2} - 1\text{S}, \text{beam-beam}) \approx 2\pi \times 3\,\text{Hz}$, $\Delta\omega(4\text{P}_{1/2} - 1\text{S}, \text{beam-background}) \approx 2\pi \times 9\,\text{Hz}$, $\Delta\omega(4\text{P}_{3/2} - 1\text{S}, \text{beam-beam}) \approx 2\pi \times 8\,\text{Hz}$ and $\Delta\omega(4\text{P}_{3/2} - 1\text{S}, \text{beam-background}) \approx 2\pi \times 23\,\text{Hz}$. Thus, the pressure shift is estimated to be below 0.01 kHz for the $2\text{S-}4\text{P}_{1/2}$ transition and below 0.03 kHz for the $2\text{S-}4\text{P}_{3/2}$ transition. The uncertainty is assumed to be fully correlated ($r = 1$) for the two transitions, resulting in an uncertainty of 0.02 kHz for $\nu_{2\text{S}-4\text{P}}$.

## 2.10 Laser spectrum and frequency calibration

The two laser systems used in the measurement, the spectroscopy laser for 2S-4P excitation at 486 nm and the preparation laser for 1S-2S excitation at 243 nm, share a similar design. Both laser systems consist of an external cavity diode laser as master oscillator running at 972 nm (*57*). The frequency of the lasers is stabilized to high-finesse Fabry-Pérot cavities (*58*), which reduces the laser line width to a few Hz. However, the delta-shaped laser line sits on a weak but broad noise pedestal (*58*). After power amplification with a tapered amplifier, the light is frequency doubled (frequency doubled twice) to 486 nm (243 nm) for the spectroscopy (excitation) laser system. The spectral purity of the lasers is routinely monitored by a beat note between the two systems at 486 nm.

Asymmetries of the noise pedestal that might fold into the observed 2S-4P line shapes are small because of the use of long external cavity diode lasers (see (*57, 58*)). We obtain an upper limit of 0.10 kHz by artificially removing the noise pedestal on one side of the measured laser spectrum, numerically folding it into the 2S-4P line shape, and fitting the result. The same laser was used for both transition measurements and thus the upper limit is fully correlated ($r = 1$) for the two measurements.

Both laser systems are phase-coherently linked to an Er-doped fiber frequency comb which is referenced to an active hydrogen maser. The maser serves as the frequency reference for the experiment and is calibrated (steered) via the global positioning system (GPS), resulting in a fractional frequency uncertainty of 1 part in $10^{13}$. The maser calibration uncertainty translates to an uncertainty of 0.06 kHz for the 2S-4P transition frequencies and is fully correlated ($r = 1$) for the two measured transitions.

The absolute frequencies of the laser systems are determined with a beat note with the frequency comb at 972 nm. The frequency of the 2S-4P spectroscopy light $\nu_{\rm Laser}^{486}$ at 486 nm at a given time $t$ can be deduced from the beat note data with

$$\nu_{\rm Laser}^{486} = 2 \times (N \times \nu_{\rm rep} + 2\nu_{\rm CEO} - \nu_{\rm LO}) + 2 \times (\nu_{\rm beat}(t) - \nu_{\rm AOM}(t)), \quad \text{(S11)}$$

where $\nu_{\rm rep} = 250$ MHz denotes the repetition rate of the the frequency comb, $\nu_{\rm CEO} = 30$ MHz the carrier-envelope offset frequency, $\nu_{\rm LO}$ the frequency of an additional local oscillator used to mix down the frequency of the beat note to $\nu_{\rm beat}(t) \approx 20$ MHz and $\nu_{\rm AOM}(t) \approx 350\pm30$ MHz the frequency of the acousto-optic modulator (AOM) used for scanning over the atomic resonance. The comb mode numbers are $N = 1\,233\,042$ and $N = 1\,233\,044$ for the measurement of the $2\text{S-}4\text{P}_{1/2}$ and $2\text{S-}4\text{P}_{3/2}$ transitions, respectively.

The laser frequencies are determined with a linear fit of the comb beat note data $\nu_{\rm beat}(t)$ and using Eq. S11, leading to an uncertainty in the laser frequency determination of less than 0.10 kHz for each recorded resonance. This leads to a negligible uncertainty for the determined transition frequencies.

## 2.11 Recoil shift

Energy and momentum conservation require the absorbed photon energy to be larger than the atomic resonance frequency $\nu$ by the recoil shift of the atom upon absorption. The corresponding recoil shift can be written as

$$\Delta\nu_{\rm recoil} = \frac{h}{2M_{\rm H}}\left(\frac{\nu}{c}\right)^2 \approx 837.23\,\text{kHz}, \quad \text{(S12)}$$

with the mass of the hydrogen atom $M_{\rm H}$. $\Delta\nu_{\rm recoil}$ is known with much smaller uncertainty than required here. $h/M_{\rm H}$ can be calculated using the experimental values of the ratio $h/m_e$ of the Planck constant $h$ and the electron mass $m_e$, the binding energy of the H atom and the mass of the proton and the electron in atomic mass units (*3*). The transition frequencies $\nu_{1/2}$, $\nu_{3/2}$ and $\nu_{\rm 2S-4P}$ given in the main text have been corrected for the recoil shift.

## 2.12 Hyperfine corrections

In order to obtain the transition frequency from the 2S hyperfine centroid to the 4P fine structure centroid (see fig. S1), the measured transition frequencies $\nu_{1/2}$ and $\nu_{3/2}$ have to be corrected for the hyperfine shift of the $2\text{S}_{1/2}^{F=0}$, $4\text{P}_{1/2}^{F=1}$ and $4\text{P}_{3/2}^{F=1}$ states (table II in (*38*))

$$\Delta\nu_{\rm HFS}(2\text{S}_{1/2}^{F=0}) = -133\,167.6257(51)\ \text{kHz}, \quad \text{(S13)}$$

$$\Delta\nu_{\rm HFS}(4\text{P}_{1/2}^{F=1}) = +1848.8(1)\ \text{kHz}, \quad \text{(S14)}$$

$$\Delta\nu_{\rm HFS}(4\text{P}_{3/2}^{F=1}) = -1847.7(1)\ \text{kHz}. \quad \text{(S15)}$$

These shifts have been obtained experimentally (for 2S, see (*59*)) and by extrapolation to higher $n$ and include a small off-diagonal term of $\Delta\nu_{\text{HFS}}^{\text{o.d.}} = \pm 0.313$ kHz for the 4P states. The transition frequency from the 2S hyperfine centroid to the 4P fine structure centroid, $\nu_{2\text{S}-4\text{P}}$ (Eq. 9 in the main text), is obtained by a weighted average of the hyperfine centroids

$$\begin{aligned}\nu_{2\text{S}-4\text{P}} &= \frac{1}{3}\left(\nu_{1/2} - \Delta\nu_{\text{HFS}}(4\text{P}_{1/2}^{F=1})\right) + \frac{2}{3}\left(\nu_{3/2} - \Delta\nu_{\text{HFS}}(4\text{P}_{3/2}^{F=1})\right) + \Delta\nu_{\text{HFS}}(2\text{S}_{1/2}^{F=0}) \\ &= \frac{1}{3}\nu_{1/2} + \frac{2}{3}\nu_{3/2} - 132\,552.092(75)\ \text{kHz}. \qquad \text{(S16)}\end{aligned}$$

The fine structure splitting $\Delta\nu_{\text{FS}}^{\text{theo}}(4\text{P})$ of the $4\text{P}_{1/2}^{F=1}$ and $4\text{P}_{3/2}^{F=1}$ states may readily be obtained from the difference in the total binding energies of the $4\text{P}_{1/2}^{F=1}$ and $4\text{P}_{3/2}^{F=1}$ states given in table IV in (*38*)

$$\Delta\nu_{\text{FS}}^{\text{theo}}(4\text{P}) = 1\,367\,433.3\,(3)\,\text{kHz}. \qquad \text{(S17)}$$

## Supplementary figures

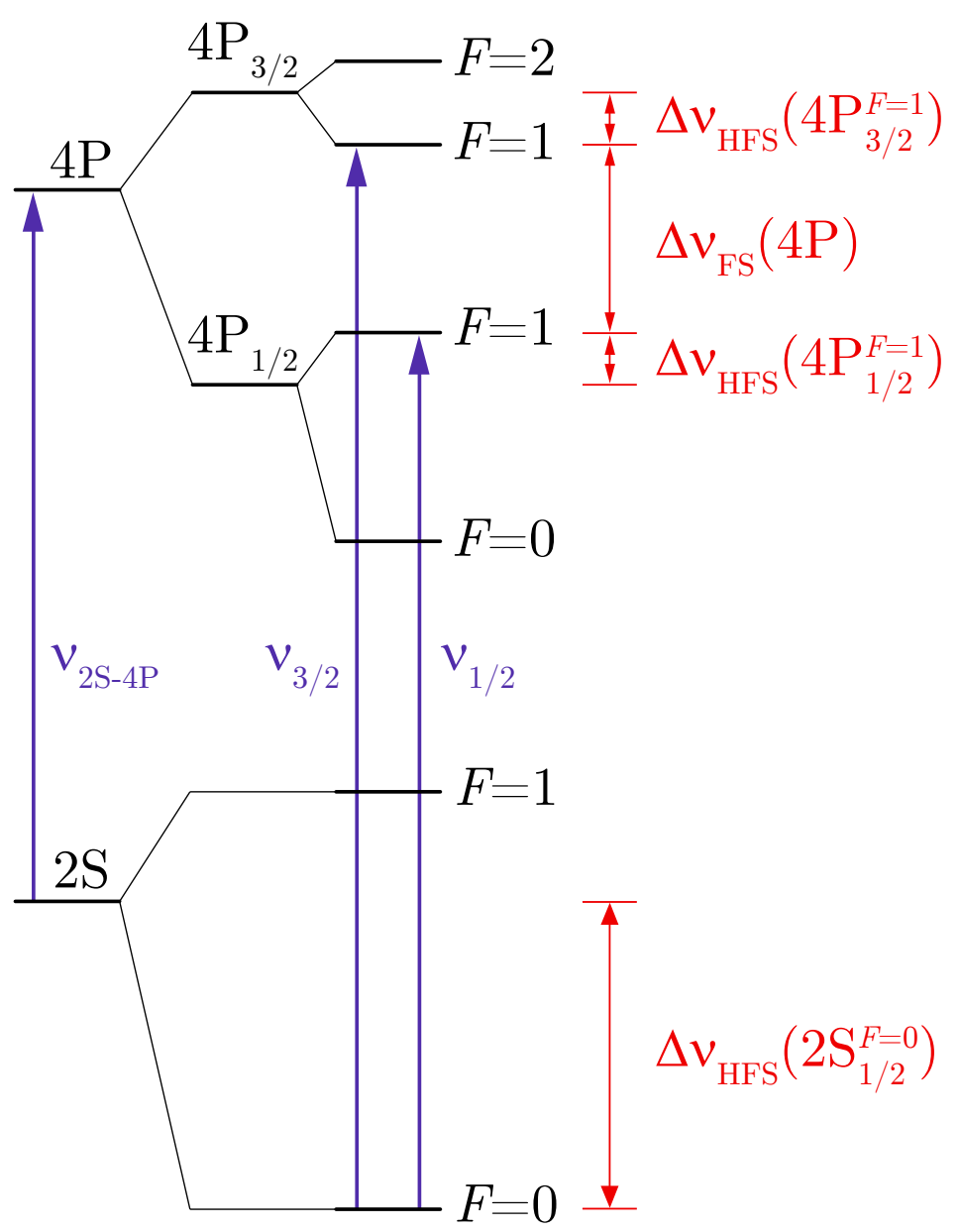

Figure S1: **Hydrogen 2S-4P level scheme (not to scale).** The transition frequencies of the $2S_{1/2}^{F=0}$-$4P_{1/2}^{F=1}$ ($\nu_{1/2}$) and $2S_{1/2}^{F=0}$-$4P_{3/2}^{F=1}$ ($\nu_{3/2}$) transition are experimentally determined. The transition frequency from the 2S hyperfine structure centroid to the 4P fine structure centroid, $\nu_{2S-4P}$, is determined by combining $\nu_{1/2}$ and $\nu_{3/2}$ and correcting for the hyperfine shifts $\Delta\nu_{\text{HFS}}(2S_{1/2}^{F=0})$, $\Delta\nu_{\text{HFS}}(4P_{1/2}^{F=1})$ and $\Delta\nu_{\text{HFS}}(4P_{3/2}^{F=1})$. The fine structure splitting $\Delta\nu_{\text{FS}}(4P)$ corresponds to the energy difference of the $4P_{1/2}^{F=1}$ and $4P_{3/2}^{F=1}$ states.

# Supplementary tables

Table S1: List of quadratic coefficients for the dc-Stark effect for the relevant atomic levels and different orientations of the static electric field $\vec{F}$ with respect to the laser polarization $\vec{E}$.

| Level | Orientation of $\vec{F}$ | $a_{j,m_F}$ (Hz/(V/m)$^2$) |
|---|---|---|
| $4P_{1/2}^{F=1}$ | $\vec{F} \parallel \vec{E}$ | -97 |
| $4P_{1/2}^{F=1}$ | $\vec{F} \perp \vec{E}$ | -83 |
| $4P_{3/2}^{F=1}$ | $\vec{F} \parallel \vec{E}$ | 186 |
| $4P_{3/2}^{F=1}$ | $\vec{F} \perp \vec{E}$ | -1354 |

Table S2: List of corrections $\Delta\nu$ and uncertainties $\sigma$ for the determination of the $2S_{1/2}^{F=0}$-$4P_{1/2}^{F=1}$ ($\nu_{1/2}$) and $2S_{1/2}^{F=0}$-$4P_{3/2}^{F=1}$ ($\nu_{3/2}$) transition frequencies and the correlation coefficient $r$ of the uncertainties for the two transitions.

| Contribution | $2S_{1/2}^{F=0}$-$4P_{1/2}^{F=1}$ ($\nu_{1/2}$) | | $2S_{1/2}^{F=0}$-$4P_{3/2}^{F=1}$ ($\nu_{3/2}$) | | Correlation coefficient |
|---|---|---|---|---|---|
| | $\Delta\nu$ (kHz) | $\sigma$ (kHz) | $\Delta\nu$ (kHz) | $\sigma$ (kHz) | $r$ |
| Statistics | 0.00 | 0.48 | 0.00 | 0.56 | 0 |
| First-order Doppler shift | 0.00 | 2.92 | 0.00 | 2.84 | 0 |
| Quantum interference shift | 0.00 | 0.33 | 0.00 | 0.26 | 0 |
| Light force shift | -0.43 | 0.40 | -0.26 | 0.25 | 1 |
| Model corrections | 1.34 | 0.23 | -0.50 | 0.10 | -0.65 |
| Sampling bias | -0.34 | 0.40 | 0.83 | 0.70 | 0 |
| Second-order Doppler shift | 0.22 | 0.05 | 0.22 | 0.05 | 1 |
| dc-Stark shift | 0.00 | 0.03 | 0.00 | 0.30 | 0 |
| Zeeman shift | 0.00 | 0.07 | 0.00 | 0.33 | 0 |
| Pressure shift | 0.00 | 0.01 | 0.00 | 0.03 | 1 |
| Laser spectrum | 0.00 | 0.10 | 0.00 | 0.10 | 1 |
| Frequency standard (hydrogen maser) | 0.00 | 0.06 | 0.00 | 0.06 | 1 |
| Recoil shift | -837.23 | 0.00 | -837.23 | 0.00 | n/a |
| Total | -836.4 | 3.0 | -836.9 | 3.0 | 0.011 |

Table S3: List of corrections $\Delta\nu$ and uncertainties $\sigma$ for the determination of the 4P fine structure splitting $\Delta\nu_{\mathrm{FS}}^{\mathrm{exp}}(4\mathrm{P})$.

| Contribution | $\Delta\nu$ (kHz) | $\sigma$ (kHz) |
|---|---|---|
| Statistics | 0.00 | 0.74 |
| First-order Doppler shift | 0.00 | 4.07 |
| Quantum interference shift | 0.00 | 0.42 |
| Light force shift | 0.17 | 0.15 |
| Model corrections | -1.84 | 0.30 |
| Sampling bias | 1.17 | 0.81 |
| Second-order Doppler shift | 0.00 | 0.00 |
| dc-Stark shift | 0.00 | 0.30 |
| Zeeman shift | 0.00 | 0.34 |
| Pressure shift | 0.00 | 0.02 |
| Laser spectrum | 0.00 | 0.10 |
| Frequency standard (hydrogen maser) | 0.00 | 0.06 |
| Recoil shift | 0.00 | 0.00 |
| Total | -0.5 | 4.3 |

# Appendix B

# Additional Figures and Tables

## B.1 Properties of the 2S-4P transition

Table B.1: Atomic properties of the 2S-4P transition. See Table 2.1, the corresponding Table for the 2S-6P transition, for details.

| $J$ (excited level) | $1/2$ ($\|e_1\rangle$) | $3/2$ ($\|e_2\rangle$) |
|---|---|---|
| Transition frequency $\nu_{A,0}$ (kHz) | 616 520 152 558.5 | 616 521 519 991.8 |
| Transition wavenumber $K_L$ (1/m) | 12 921 307 | 12 921 336 |
| Dipole moment $\mu$ ($e\,a_0$) | $-\frac{m_e+m_p}{m_p}\frac{512}{2187}\sqrt{\frac{10}{3}}$ | $\frac{m_e+m_p}{m_p}\frac{512}{2187}\sqrt{\frac{2\cdot 10}{3}}$ |
| Dipole moment $\mu$ ($10^{-30}$ C m) | $-3.626$ | 5.128 |
| Rabi frequency $\Omega_0$ (krad/s (W/m$^2$)$^{-1/2}$) | $2\pi \times 150.2$ | $2\pi \times 212.4$ |
| Natural linewidth $\Gamma$ (MHz) | 12.93 | |
| Decay rates $A$ (Mdcy/s) | | |
| $\gamma_{\text{e-2S}}$: $\|e_{1/2}\rangle \rightarrow$ 2S manifold | $2\pi \times 1.539$ | |
| $\gamma_{\text{ei}}$: $\|e_{1/2}\rangle \rightarrow \|i\rangle$ | $2\pi \times 0.5129$ | $2\pi \times 1.026$ |
| $\Gamma_{\text{e-1S}}$: $\|e_{1/2}\rangle \rightarrow$ 1S manifold | $2\pi \times 11.40$ | |
| $\Gamma_{\text{det}}$: Detected signal (Ly-$\gamma$)[a] | $2\pi \times 10.85$ | |
| Non-resonant ac-Stark shift coefficient[b] $\beta_{\text{ac},0}$ (Hz/(W/m$^2$)) | $3.184 \times 10^{-4}$ | $3.212 \times 10^{-4}$ |
| Photoionization coefficient[c] $\beta_{\text{ioni}}$ (Hz/(W/m$^2$)) | $1.053 \times 10^{-4}$ | $1.224 \times 10^{-4}$ |
| Mass of hydrogen atom $m_H$ (kg) | $1.673\,533 \times 10^{-27}$ | |
| Recoil shift $\Delta\nu_{\text{rec}}$ (Hz) | 837 230 | 837 234 |
| Recoil velocity $v_{\text{rec}}$ (m/s) | 0.814 232 | 0.814 234 |

[a] In the experiment, only Lyman decays are detected, with Ly-$\gamma$ photons accounting for ∼99 % of the signal (see Tables 4.1 and B.2).

[b] This coefficient is derived from a perturbative calculation (see Appendix C.1) and does not include near-resonant contributions.

[c] See Appendix C.2. Only the excited level can be photoionized by light of frequency $\nu_{A,0}$.

Table B.2: Probability $p$ and number $N$ of Lyman (Ly) and Balmer (Ba) decay paths, i.e., decay cascades with the final decay leading from the $n'$P level to the $n$S level for $n = 1$ and $n = 2$, respectively, for an atom initially in the $4P_{1/2}^{F=1}$, $m_F = 0$ level. The decay paths are grouped by the spherical component $q = -\Delta m_F = 0, \pm 1$ of the final decay, corresponding to $\pi, \sigma^{\pm}$ decays, respectively. $p$ is the ratio of the strengths of the considered decay paths to the total strength of all dipole-allowed decay paths leading to the 1S and 2S level. $N$ includes the number of possible paths leading to the $n'$P level from which the final decay starts. There are no decay paths to the 3P level and thus no Ly-$\beta$ or Ba-$\alpha$ decays.

| | $n$ | $n'$ | Energy | $\pi$ ($q = 0$) | | $\sigma^{\pm}$ ($q = \pm 1$) | | Sum ($\sigma^- + \pi + \sigma^+$) | |
|---|---|---|---|---|---|---|---|---|---|
| | | | (eV) | $p$ (%) | $N$ | $p$ (%) | $N$ | $p$ (%) | $N$ |
| Ly-$\alpha$ | 1 | 2 | 10.20 | 1.268 | 54 | 1.466 | 53 | 4.199 | 160 |
| Ly-$\gamma$ | 1 | 4 | 12.75 | 27.968 | 1 | 27.968 | 1 | 83.904 | 3 |
| Ly | 1 | | | 29.236 | 55 | 29.434 | 54 | 88.103 | 163 |
| Ba-$\beta$ | 2 | 4 | 2.55 | 3.966 | 1 | 3.966 | 1 | 11.897 | 3 |
| Ba | 2 | | | 3.966 | 1 | 3.966 | 1 | 11.897 | 3 |
| Sum | | | | 33.201 | 56 | 33.399 | 55 | 100.000 | 166 |

Table B.3: Same as Table B.2, but for an atom initially in the $4P_{3/2}^{F=1}$, $m_F = 0$ level.

| | $n$ | $n'$ | Energy | $\pi$ ($q = 0$) | | $\sigma^{\pm}$ ($q = \pm 1$) | | Sum ($\sigma^- + \pi + \sigma^+$) | |
|---|---|---|---|---|---|---|---|---|---|
| | | | (eV) | $p$ (%) | $N$ | $p$ (%) | $N$ | $p$ (%) | $N$ |
| Ly-$\alpha$ | 1 | 2 | 10.20 | 1.375 | 67 | 1.412 | 65 | 4.199 | 197 |
| Ly-$\gamma$ | 1 | 4 | 12.75 | 55.936 | 1 | 13.984 | 1 | 83.904 | 3 |
| Ly | 1 | | | 57.311 | 68 | 15.396 | 66 | 88.103 | 200 |
| Ba-$\beta$ | 2 | 4 | 2.55 | 7.931 | 1 | 1.983 | 1 | 11.897 | 3 |
| Ba | 2 | | | 7.931 | 1 | 1.983 | 1 | 11.897 | 3 |
| Sum | | | | 65.242 | 69 | 17.379 | 67 | 100.000 | 203 |

Table B.4: Probability $p$ and number $N$ of Lyman (Ly) and Balmer (Ba) decay paths, i.e., decay cascades with the final decay leading from the $n'$P level to the $n$S level for $n = 1$ and $n = 2$, respectively, for an atom initially in the $4\mathrm{P}_{1/2}^{F=1}$, $m_F = 0$ level. Similar to Table B.2, but with the decay paths grouped by the final hyperfine level reached.

| | $n$ | $n'$ | $n\mathrm{S}_{1/2}^{F=0}$, $m_F = 0$ | | $n\mathrm{S}_{1/2}^{F=1}$, $m_F = 0$ | | $n\mathrm{S}_{1/2}^{F=1}$, $m_F = \pm 1$ | |
|---|---|---|---|---|---|---|---|---|
| | | | $p$ (%) | $N$ | $p$ (%) | $N$ | $p$ (%) | $N$ |
| Ly-$\alpha$ | 1 | 2 | 1.187 | 36 | 0.784 | 38 | 1.114 | 43 |
| Ly-$\gamma$ | 1 | 4 | 27.968 | 1 | 0.000 | 0 | 27.968 | 1 |
| Ly | 1 | | 29.155 | 37 | 0.784 | 38 | 29.082 | 44 |
| Ba-$\beta$ | 2 | 4 | 3.966 | 1 | 0.000 | 0 | 3.966 | 1 |
| Ba | 2 | | 3.966 | 1 | 0.000 | 0 | 3.966 | 1 |

Table B.5: Same as Table B.4, but for an atom initially in the $4\mathrm{P}_{3/2}^{F=1}$, $m_F = 0$ level.

| | $n$ | $n'$ | $n\mathrm{S}_{1/2}^{F=0}$, $m_F = 0$ | | $n\mathrm{S}_{1/2}^{F=1}$, $m_F = 0$ | | $n\mathrm{S}_{1/2}^{F=1}$, $m_F = \pm 1$ | |
|---|---|---|---|---|---|---|---|---|
| | | | $p$ (%) | $N$ | $p$ (%) | $N$ | $p$ (%) | $N$ |
| Ly-$\alpha$ | 1 | 2 | 1.823 | 43 | 0.681 | 48 | 0.848 | 53 |
| Ly-$\gamma$ | 1 | 4 | 55.936 | 1 | 0.000 | 0 | 13.984 | 1 |
| Ly | 1 | | 57.759 | 44 | 0.681 | 48 | 14.832 | 54 |
| Ba-$\beta$ | 2 | 4 | 7.931 | 1 | 0.000 | 0 | 1.983 | 1 |
| Ba | 2 | | 7.931 | 1 | 0.000 | 0 | 1.983 | 1 |

## B.2 Detection efficiency

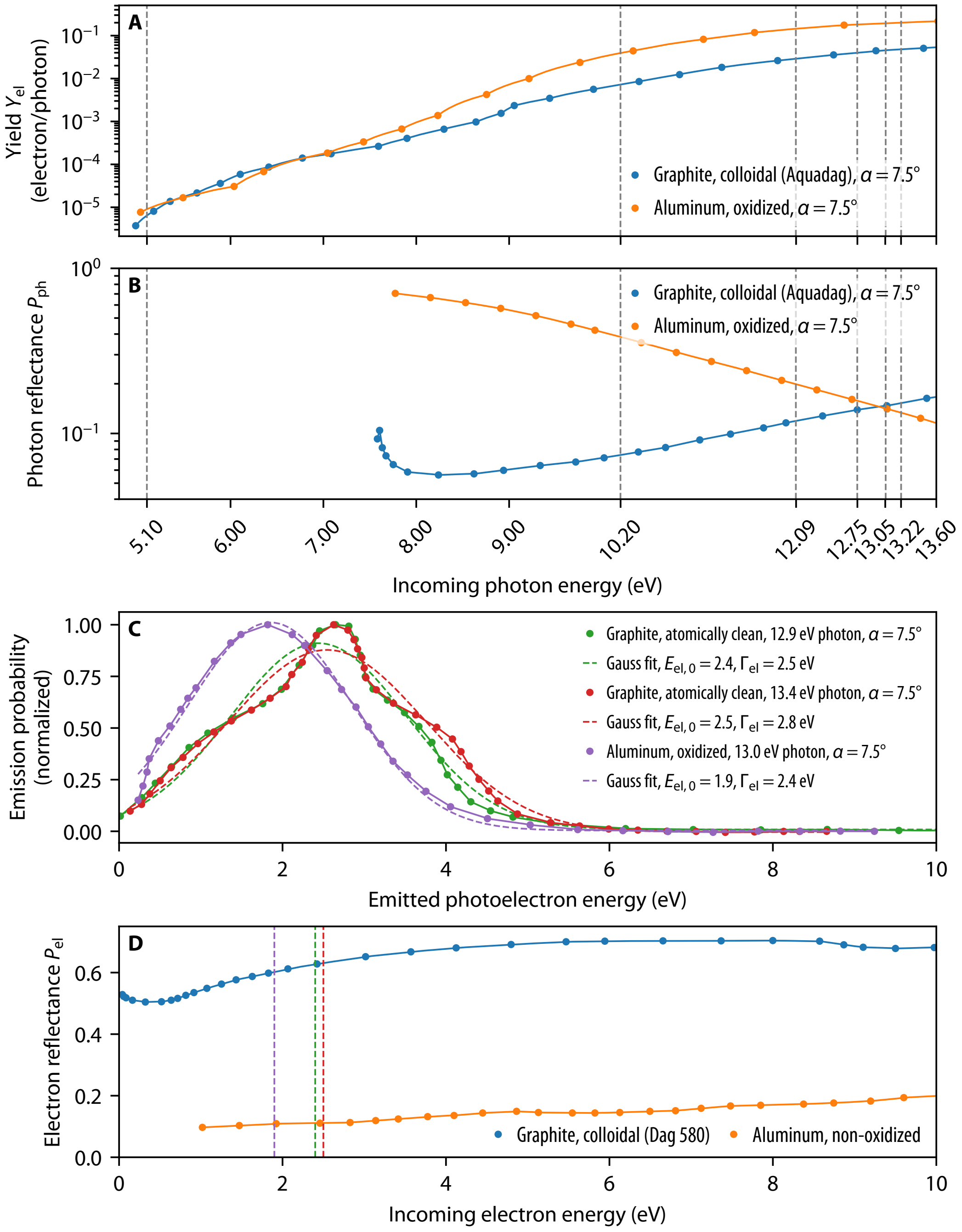

Figure B.1: Measurements of the properties of the materials used in the detector relevant to modeling the detection efficiency, as summarized in Table 4.1. The measurements shown in (**A**–**C**) are taken from [133], while the measurements of (**D**) are taken from [140] for graphite and from the results of [151] as reproduced in [152] for aluminum.

# Appendix C

# Non-resonant ac-Stark shift and photoionization coefficients

The ac-Stark shift is the shift in energy of atomic levels caused by the interaction with electromagnetic radiation. Experimentally, it manifests itself as a shift in the observed transition frequency as a function of laser intensity, with typically only effects first order in intensity relevant. The shift of the transition frequency is the combined effect of the ground and excited level of the transition shifting in energy. However, this shift can and generally does contain various contributions, typically requiring different theoretical descriptions. One contribution is the non-resonant coupling of the two levels of the transition to other levels, which is discussed here and can be treated in a perturbative fashion. This is typically the dominant contribution in the spectroscopy of two-photon transitions, such as for the 1S-2S transition used in this work. For one-photon transitions, such as the 2S-$n$P transitions, resonant contributions become dominant. Then, a non-perturbative treatment such as solving the optical Bloch equations discussed in Section 2.3.1 is necessary. Note that in this case, optical pumping effects that lead to an intensity-dependent frequency shift will also be included in the ac-Stark shift coefficient determined in this way.

In the next section, we discuss the contribution to the ac-Stark shift through non-resonant coupling, since it is not included in our optical Bloch equations. The closely related effect of photoionization is discussed in Appendix C.2.

## C.1 Non-resonant ac-Stark shift

### C.1.1 Light shift operator

We define the light shift operator $Q_q^{\mathrm{LS}}$ of an atom interacting with a single-frequency laser field, following [58] but extending the description to arbitrary light polarizations, as

$$Q_q^{\mathrm{LS}} = \sum_{\pm} (-1)^q \, r_{\pm q} \frac{1}{H_0 - E_\phi \pm \hbar\omega_{\mathrm{L}}} r_{\mp q}, \tag{C.1}$$

where $r_q$, $q = -1, 0, 1$, are the spherical components of the electron position operator $\boldsymbol{r}$, $H_0$ is the unperturbed atomic Hamiltonian, and $\omega_{\mathrm{L}}$ is the angular frequency of the laser light. $E_\phi$ is the energy of the atomic level under consideration, $|\phi\rangle = |n, l, m\rangle$, completely described by the principal quantum number $n$, the orbital angular momentum quantum number $l$, and the magnetic quantum number $m$, i.e., for now we neglect fine and hyperfine structure.

Within the electric dipole approximation [38, 64], the electric dipole operator is given by $\boldsymbol{\mu} = e\boldsymbol{r}$. The interaction between the atomic dipole and the electric field $\tilde{\boldsymbol{E}}$ of the laser light is described by the operator $V = -\boldsymbol{\mu}\tilde{\boldsymbol{E}} = -e\boldsymbol{r}\tilde{\boldsymbol{E}} = -e\sum_q(-1)^q r_{-q}\tilde{E}_q$. Thus, the components $q = -1, 0, 1$ of the light shift operator form a complete spherical basis of the light polarization. Throughout this section, without loss of generality, we take the $z$-axis as the quantization axis of the atom, i.e., $m$ measures the angular momentum projection onto the $z$-axis. Then, the component $q = 0$ corresponds to linear polarization along the $z$-axis, while the components $q = \pm 1$ correspond to circular ($\sigma^\pm$) polarizations about the $z$-axis, i.e., with the electric field vector rotating in the $x$-$y$-plane. Any polarization state can then be described by decomposing it into these spherical components.

The ordering of the components of the operator $r_{\pm q}$ in Eq. (C.1) can be understood as follows: the "$-$" term corresponds to an absorption of a photon with polarization $q$, described with the operator $r_{+q}$, changing the $z$-component of the atom's angular momentum by $\Delta m = q$ and leading to an excitation of a higher-lying virtual level. This is followed by the emission of a photon ($\Delta m = -q$), described with the operator $r_{-q}$, leading back to the initial level. The "+" term, on the other hand, corresponds to the emission of a photon ($\Delta m = -q$) and the excitation of a lower-lying virtual level, followed by the absorption of a photon ($\Delta m = q$) leading back to the initial level.

The choice of the relative $\pm$ signs of the denominator and the operators $r_{\pm q}$ in Eq. (C.1) can be motivated by decomposing into spherical components the real-valued electric field $\tilde{\boldsymbol{E}}_\pm(t, z)$ of circularly polarized laser light. As we define it here, $\tilde{\boldsymbol{E}}_+$ rotates counterclockwise and $\tilde{\boldsymbol{E}}_-$ rotates clockwise about the $z$-axis with frequency $\omega_\mathrm{L}$, corresponding to $\sigma^+$ and $\sigma^-$ polarizations[1], respectively. The decomposition of $\tilde{\boldsymbol{E}}_\pm(t, z)$ results in

$$\begin{aligned}\tilde{\boldsymbol{E}}_\pm(t, z = 0) &= \frac{\tilde{E}_0}{\sqrt{2}}\left(\cos(\omega_\mathrm{L}t)\hat{\boldsymbol{x}} \pm \sin(\omega_\mathrm{L}t)\hat{\boldsymbol{y}}\right) \\ &= \frac{\tilde{E}_0}{2\sqrt{2}}e^{-i\omega_\mathrm{L}t}(\hat{\boldsymbol{x}} \pm i\hat{\boldsymbol{y}}) + \frac{\tilde{E}_0}{2\sqrt{2}}e^{i\omega_\mathrm{L}t}(\hat{\boldsymbol{x}} \mp i\hat{\boldsymbol{y}}) \\ &= \mp\frac{\tilde{E}_0}{2}e^{-i\omega_\mathrm{L}t}\hat{\boldsymbol{r}}_\pm \pm \frac{\tilde{E}_0}{2}e^{i\omega_\mathrm{L}t}\hat{\boldsymbol{r}}_\mp, \end{aligned} \tag{C.2}$$

where $\hat{\boldsymbol{x}}, \hat{\boldsymbol{y}}$ are the Cartesian basis vectors in the plane of the electric field, $\hat{\boldsymbol{r}}_\pm = \mp\frac{1}{\sqrt{2}}(\hat{\boldsymbol{x}} \pm i\hat{\boldsymbol{y}})$ are the spherical basis vectors, and $\tilde{E}_0$ is the electric field amplitude. Thus, the real-valued electric field actually contains two counter-rotating components in the spherical basis[2]. Realizing that $e^{-i\omega_\mathrm{L}t}$ and $e^{i\omega_\mathrm{L}t}$ correspond to the virtual absorption and emission of photons

[1] Note that some authors adopt the convention in which the sense of rotation is measured with respect to the propagation direction the laser beam, and not with respect to the laboratory frame of reference as done here. E.g., in the first convention, $\sigma^\pm$ polarization turns into $\sigma^\mp$ polarization upon retroreflection, while in our convention, it does not.

[2] When an atomic transition is resonantly driven, only the $e^{-i\omega_\mathrm{L}t}$ part of Eq. (C.2) contributes to the transition probability in an appreciable way, and the $e^{i\omega_\mathrm{L}t}$ part is usually neglected through the rotating wave approximation (see Section 2.3.1). Within this approximation, $\tilde{\boldsymbol{E}}_\pm$ is then proportional to $\hat{\boldsymbol{r}}_\pm$, and in this context $\hat{\boldsymbol{r}}_\pm$ is often used synonymously with circular polarization. In the case of the light shift, however, both oscillating parts of Eq. (C.2) generally correspond to off-resonant virtual excitations and we thus cannot neglect either part.

[65], respectively, in combination with the conclusions from the previous paragraph, then dictates the structure of Eq. (C.1).

The dynamic polarizability $P_{\omega_{\mathrm{L}},q}(\phi)$ of the atomic level $|\phi\rangle$ is given by the corresponding matrix element of $Q_q^{\mathrm{LS}}$,

$$P_{\omega_{\mathrm{L}},q}(\phi) = \langle\phi|Q_q^{\mathrm{LS}}|\phi\rangle. \tag{C.3}$$

$P_{\omega_{\mathrm{L}},q}(\phi)$ is in general complex, with the real part describing an energy shift and the imaginary part corresponding to a population loss through photoionization [58, 64]. Consequently, we define the ac-Stark shift $\Delta\nu_{\mathrm{ac}}$ as the energy shift, measured in frequency units (Hz), of the level $|\phi\rangle$ when the atom is placed in a laser beam with intensity $I_{\mathrm{loc}}$ as

$$\Delta\nu_{\mathrm{ac}}(\phi) = -\frac{e^2}{2\epsilon_0 hc} I_{\mathrm{loc}} \operatorname{Re}\Big[P_{\omega_{\mathrm{L}},q}(\phi)\Big] = \beta_{\mathrm{ac},q}(\phi) I_{\mathrm{loc}}. \tag{C.4}$$

Here, $\beta_{\mathrm{ac},q}(\phi)$ is the (real-valued) ac-Stark shift coefficient, given in units of $\mathrm{Hz}/(\mathrm{W}/\mathrm{m}^2)$.

To evaluate $P_{\omega_{\mathrm{L}},q}(\phi)$, we insert a complete basis of atomic levels, $\sum\!\!\!\!\!\!\int_u |u\rangle\langle u| = \mathbf{1}$, $|u\rangle = |n_u, l_u, m_u\rangle$, where the sum here symbolizes both a sum over the discrete bound levels of the atom and an integral over the continuum levels, i.e., for a free electron and ionized core,

$$P_{\omega_{\mathrm{L}},q}(\phi) = \sum\!\!\!\!\!\!\!\int_u \sum_{\pm} \frac{\langle\phi|(-1)^q\, r_{\pm q}|u\rangle\langle u|r_{\mp q}|\phi\rangle}{E_u - E_\phi \pm \hbar\omega_{\mathrm{L}}} = \sum\!\!\!\!\!\!\!\int_u \sum_{\pm} \frac{|\langle u|r_{\mp q}|\phi\rangle|^2}{E_u - E_\phi \pm \hbar\omega_{\mathrm{L}}}, \tag{C.5}$$

where we have used $H_0|u\rangle = E_u|u\rangle$, with $E_u \equiv E_u(n_u, l_u)$ the energy of level $|u\rangle$. The matrix elements $\langle u|r_q|\phi\rangle$ are to good approximation constant over the range over which $\omega_{\mathrm{L}}$ is varied when probing the atomic resonance.

### C.1.2 Perturbation theory and on-resonance contributions

The expression given in Eq. (C.5) can be derived using second-order time-dependent perturbation theory [58, 64, 183] and taking the operator $V$ defined above as perturbation. During this derivation, terms oscillating with twice the optical frequency ($\propto e^{\pm i2\omega_{\mathrm{L}}t}$) occur. However, the contribution to the ac-Stark shift from these terms will average out for time scales much longer than the inverse optical frequency, which is almost always fulfilled in spectroscopy experiments. The terms are thus dropped to arrive at Eq. (C.5).

From this point of view, it is clear that Eq. (C.5) is only valid for the case of a small perturbation that does not change the state of the atom substantially. Clearly, this is not the case when driving a dipole-allowed transition with resonant laser light, leading to a substantial depopulation and repopulation through decay of the initial state, as is the case in this work for 2S-$n$P transitions. In Eq. (C.5), this results in a divergence, since for a laser resonant with the transition from $|\phi\rangle$ to $|\phi'\rangle$, i.e., $\hbar\omega_{\mathrm{L}} = E_{\phi'} - E_\phi$, the "$-$" term in $P_{\omega_{\mathrm{L}},q}(\phi)$ diverges as $E_u - E_\phi - \hbar\omega_{\mathrm{L}} = 0$ for $u = \phi'$, and likewise the "$+$" term in $P_{\omega_{\mathrm{L}},q}(\phi')$ diverges as $E_u - E_{\phi'} + \hbar\omega_{\mathrm{L}} = 0$ for $u = \phi$. This divergence may be removed by including the linewidth $\Gamma_u$ of the perturbing level $|u\rangle$ by the replacement[1] $E_u - E_\phi \pm \hbar\omega_{\mathrm{L}} \to E_u - E_\phi \pm \hbar\omega_{\mathrm{L}} + i\Gamma_u/2$ in Eq. (C.5). Additionally, when $|E_u - E_\phi| \approx \hbar\omega_{\mathrm{L}}$, we can no longer neglect the fine and hyperfine structure as done here, since, e.g., for the 2S-$n$P transitions there is a nonzero shift from the $n$P $J = 1/2$ level when resonantly driving the transition to the $n$P $J = 3/2$ level and vice versa. However, even when including the linewidth and the fine and hyperfine structure, the perturbative treatment cannot account for optical pumping effects, which are dominant contributions to the effective ac-Stark shift for the 2S-$n$P transitions. Therefore, we have to use non-perturbative methods such the optical Bloch equations discussed in Section 2.3.1 to determine the near-resonant contributions to the ac-Stark shift. In the following, these contributions are excluded from $P_{\omega_\mathrm{L},q}$, thus giving only the non-resonant ac-Stark shift. For the 2S-$n$P transitions, this means that the level shift of the 2S ($n$P) level calculated in this way includes the coupling to all levels except the resonant coupling to the $n$P (2S) level. On the other hand, for two-photon transitions starting from the 1S or $2l$ levels, there are no diverging contributions and hence all terms are included in $P_{\omega_\mathrm{L},q}$.

[1] See [65], where the linewidth of the intermediate level in the related Kramers-Heisenberg equation is taken into account in the same way.

### C.1.3 Relation to the Bloch-Siegert shift

We note that, returning again to the case of a laser with polarization $q$ and resonant with the transition from $|\phi\rangle = |n_\phi, l_\phi, m_\phi\rangle$ to $|\phi'\rangle = |n_{\phi'}, l_{\phi'}, m_{\phi'} = m_\phi + q\rangle$, the non-diverging "+" term in $P_{\omega_\mathrm{L},q}(\phi)$ for $|u\rangle = |n_{\phi'}, l_{\phi'}\rangle$ and the "−" term in $P_{\omega_\mathrm{L},q}(\phi')$ for $|u\rangle = |n_\phi, l_\phi\rangle$ can be evaluated and are included in the results given here. In fact, these terms together constitute the Bloch-Siegert shift, which is a model error built into the master equation of atom–light interaction when using the rotating wave approximation (RWA). For optical transitions, the RWA is well motivated, and is used in this thesis, as detailed in Section 2.3.1. Evaluating the terms of the Bloch-Siegert shift explicitly, we find

$$\Delta\nu_\mathrm{ac}^{\phi'}(\phi) = -\frac{e^2}{2\epsilon_0 hc} I_\mathrm{loc} \frac{|\langle n_{\phi'}, l_{\phi'}, m_\phi - q | r_{-q} | \phi\rangle|^2}{E_{\phi'} - E_\phi + \hbar\omega_\mathrm{L}}, \tag{C.6}$$

$$\Delta\nu_\mathrm{ac}^{\phi}(\phi') = -\frac{e^2}{2\epsilon_0 hc} I_\mathrm{loc} \frac{|\langle n_{\phi}, l_{\phi}, m_{\phi'} + q | r_{q} | \phi'\rangle|^2}{E_{\phi} - E_\phi' - \hbar\omega_\mathrm{L}}. \tag{C.7}$$

It is instructive to look at the case for $q = 0$. Noting $|\langle n_{\phi'}, l_{\phi'}, m_\phi | r_0 | \phi\rangle|^2 = |\langle n_{\phi}, l_{\phi}, m_{\phi'} | r_0 | \phi'\rangle|^2$ and using the definition of the Rabi frequency $\Omega_{\phi\to\phi'}$ of the transition from $|\phi\rangle$ to $|\phi'\rangle$ as given in Eqs. (2.30) and (2.31), we find

$$\Delta\nu_\mathrm{ac}^{\phi'}(\phi) = -\Delta\nu_\mathrm{ac}^{\phi}(\phi') = -\frac{1}{2\pi}\frac{\Omega_{\phi\to\phi'}^2}{8\omega_\mathrm{L}}. \tag{C.8}$$

The Bloch-Siegert shift of the transition is then $\Delta\nu_\mathrm{ac}^{\phi}(\phi') - \Delta\nu_\mathrm{ac}^{\phi'}(\phi) = (1/2\pi)\,\Omega_{\phi\to\phi'}^2/4\omega_\mathrm{L}$, which matches[1] the expression given in Eq. (2.45) of [184]. For the 2S-4P and 2S-6P transitions and linear laser polarization, the contribution of the Bloch-Siegert shift to the non-resonant ac-Stark shift is about 6 % and 4 %, respectively.

### C.1.4 Spherical tensor decomposition

It is instructive to decompose $Q_q^\mathrm{LS}$ into two sets of spherical tensors of rank $k \leq 2$ with $2k+1$ components each, $Q_{q'}^{(k,\pm)}$, $q' = -k, -k+1, \ldots, k$ (see [185] for a similar decomposition of the two-photon transition operator). $Q_{q'}^{(k,\pm)}$ are given by the tensor product of two spherical vectors (spherical tensors of rank 1, here the two position operators $r_q$) and a scalar (a spherical tensor of rank 0, here $\frac{1}{H_0 - E_\phi \pm \hbar\omega_L}$) as (see Eq. (14.45) of [161])

$$Q_{q'}^{(k,\pm)} = \sum_{q_1,q_2} \langle 1\, q_1\, 1\, q_2 | k\, q' \rangle\, r_{q_1} \frac{1}{H_0 - E_\phi \pm \hbar\omega_{\mathrm{L}}} r_{q_2}. \tag{C.9}$$

[1] A different expression for the Bloch-Siegert shift is given in [29]. However, the contribution calculated there is from driving to resonance separated in energy with a single field, which is different from the situation here, where a single resonance is driven with an on-resonance and off-resonance field.

For a given laser polarization $q$, the light shift operator $Q_q^{\mathrm{LS}}$ is then given by

$$Q_q^{\mathrm{LS}} = \sum_{k,q'} \sum_{\pm} a_{q'}^{(k,\pm)}(q)\, Q_{q'}^{(k,\pm)}, \tag{C.10}$$

$$\text{with} \quad a_{q'}^{(k,\pm)}(q) = (-1)^q \langle 1\ \pm q\, 1\ \mp q | k\, q' \rangle$$
$$= \begin{cases} (-1)^q \langle 1\ \pm q\, 1\ \mp q | k\, 0 \rangle & q' = 0 \\ 0 & q' \neq 0. \end{cases} \tag{C.11}$$

The numerical values of the coefficients $a_0^{(k,\pm)}(q)$ are given by:

| $k$ | $a_0^{(k,\pm)}(-1)$ | $a_0^{(k,\pm)}(0)$ | $a_0^{(k,\pm)}(+1)$ |
|---|---|---|---|
| 0 | $-\sqrt{1/3}$ | $-\sqrt{1/3}$ | $-\sqrt{1/3}$ |
| 1 | $\pm\sqrt{1/2}$ | 0 | $\mp\sqrt{1/2}$ |
| 2 | $-\sqrt{1/6}$ | $\sqrt{2/3}$ | $-\sqrt{1/6}$ |

Inspecting $a_0^{(k,\pm)}(q)$, we can further simplify Eq. (C.10) to

$$Q_q^{\mathrm{LS}} = \sum_k a^{(k)}(q) \sum_{\pm} (\mp 1)^k Q_0^{(k,\pm)} = \sum_k a^{(k)}(q)\, Q^{(k)}, \tag{C.12}$$

$$\text{with} \quad a^{(k)}(q) = a_0^{(k,-)}(q), \tag{C.13}$$

$$Q^{(k)} = \begin{cases} Q_0^{(k,-)} + Q_0^{(k,+)} & k = 0, 2 \\ Q_0^{(1,-)} - Q_0^{(1,+)} & k = 1. \end{cases} \tag{C.14}$$

Therefore, for $q = 0$, corresponding to linear polarization along the $z$-axis, $Q_q^{\mathrm{LS}}$ only contains spherical tensors of rank 0 and 2, while for $q = \pm 1$, corresponding to circular polarization about the $z$-axis, $Q_q^{\mathrm{LS}}$ additionally contains rank-1 spherical tensors.

The calculation of the dynamic polarizability is then shifted to the evaluation of the matrix elements of $Q_0^k$, since

$$P_{\omega_{\mathrm{L}},q}(\phi) = \sum_k a^{(k)}(q) \sum_{\pm} (\mp 1)^k \langle \phi | Q_0^{(k,\pm)} | \phi \rangle. \tag{C.15}$$

The problem can be further simplified by the use of the Wigner-Eckart theorem (see Eq. (14.14) of [161]), giving

$$P_{\omega_{\mathrm{L}},q}(\phi) = \sum_k a^{(k)}(q) \sum_{\pm} (\mp 1)^k (-1)^{l-m} \langle n, l || Q^{(k,\pm)} || n, l \rangle \begin{pmatrix} l & k & l \\ -m & 0 & m \end{pmatrix}, \tag{C.16}$$

Table C.1: Numerical values of ac-Stark shift coefficients $\tilde{\beta}_{\rm ac}^{(k)}$ for the 2S-$n$P transitions, with the laser frequency chosen such that one-photon excitation is resonant. The coefficients include the coupling of the 2S ($n$P) level to all other levels except the $n$P (2S) level, see text for details. The coefficient $\beta_{{\rm ac},q}(\tilde{\phi})$ for a given HFS level can be found by applying Eq. (C.20) and the finite nucleus mass can by included by multiplying with the factor given in Eq. (C.22). As for the numbers given in Table 2.1, relativistic effects have been neglected and the results are only expected to be accurate on the order of $\alpha^2$, but a higher precision is given to facilitate comparison with results given elsewhere.

| 2S-$n$P | $\tilde{\beta}_{\rm ac}^{(0)}(2{\rm S})$ (Hz/(W/m$^2$)) | $\tilde{\beta}_{\rm ac}^{(0)}(n{\rm P})$ (Hz/(W/m$^2$)) | $\tilde{\beta}_{\rm ac}^{(1)}(n{\rm P})$ (Hz/(W/m$^2$)) | $\tilde{\beta}_{\rm ac}^{(2)}(n{\rm P})$ (Hz/(W/m$^2$)) |
|---|---|---|---|---|
| 2S-3P | $6.064\,87 \times 10^{-4}$ | $-3.988\,32 \times 10^{-3}$ | $1.784\,63 \times 10^{-3}$ | $-1.499\,42 \times 10^{-3}$ |
| 2S-4P | $-4.710\,19 \times 10^{-4}$ | $-1.769\,53 \times 10^{-3}$ | $2.625\,75 \times 10^{-4}$ | $-1.858\,76 \times 10^{-4}$ |
| 2S-5P | $-5.230\,83 \times 10^{-4}$ | $-1.334\,00 \times 10^{-3}$ | $8.888\,53 \times 10^{-5}$ | $-5.865\,28 \times 10^{-5}$ |
| 2S-6P | $-5.200\,67 \times 10^{-4}$ | $-1.166\,71 \times 10^{-3}$ | $4.168\,79 \times 10^{-5}$ | $-2.651\,68 \times 10^{-5}$ |
| 2S-7P | $-5.128\,16 \times 10^{-4}$ | $-1.082\,56 \times 10^{-3}$ | $2.319\,98 \times 10^{-5}$ | $-1.445\,54 \times 10^{-5}$ |
| 2S-8P | $-5.067\,49 \times 10^{-4}$ | $-1.033\,58 \times 10^{-3}$ | $1.432\,61 \times 10^{-5}$ | $-8.832\,02 \times 10^{-6}$ |
| 2S-9P | $-5.021\,97 \times 10^{-4}$ | $-1.002\,30 \times 10^{-3}$ | $9.471\,67 \times 10^{-6}$ | $-5.825\,90 \times 10^{-6}$ |
| 2S-10P | $-4.988\,37 \times 10^{-4}$ | $-9.810\,11 \times 10^{-4}$ | $6.556\,14 \times 10^{-6}$ | $-4.060\,22 \times 10^{-6}$ |

where the last term is the Wigner 3-$j$ symbol and $\langle n,l||Q^{(k,\pm)}||n,l\rangle$ is the reduced matrix element[1] of $Q_{q'}^{(k,\pm)}$. This reduced matrix element can be expressed in terms of the reduced matrix elements of the position operator $\boldsymbol{r}$, from which the tensor $Q_{q'}^{(k,\pm)}$ was constructed from, as (see Eq. (7.1.1) of [186])

$$\langle n,l||Q^{(k,\pm)}||n,l\rangle = \sqrt{2k+1}\,(-1)^{k+2l} \sum_{n_u,l_u} \begin{Bmatrix} 1 & 1 & k \\ l & l & l_u \end{Bmatrix} \frac{|\langle n_u,l_u||\boldsymbol{r}||n,l\rangle|^2}{E_u(n_u,l_u)-E_\phi(n,l)\pm\hbar\omega_{\rm L}}, \quad \text{(C.17)}$$

where the curly brackets is the Wigner 6-$j$ symbol, and again the sum over $n_u$ includes both a sum over the bound and the continuum levels.

### C.1.5 ac-Stark shift coefficient for fine and hyperfine levels

The coupling of the orbital angular momentum $\boldsymbol{L}$ with the electron spin $\boldsymbol{S}$ results in fine structure (FS) sublevels with angular momentum $\boldsymbol{J}=\boldsymbol{L}+\boldsymbol{S}$. Since $\boldsymbol{L}$ and $\boldsymbol{S}$ commute, the reduced matrix elements of $Q_{q'}^{(k,\pm)}$ with respect to the FS sublevel $|n,l,J\rangle$ are given by (see Eq. (14.69) of [161])

$$\langle n,l,J||Q^{(k,\pm)}||n,l,J\rangle = (-1)^{l+S+J+k}\langle n,l||Q^{(k,\pm)}||n,l\rangle(2J+1)\begin{Bmatrix} l & J & S \\ J & l & k \end{Bmatrix}. \quad \text{(C.18)}$$

[1]The reduced matrix element is defined by and determined through the Wigner-Eckart theorem, which is given with different phase and normalization factors by different authors, thus resulting in different values of the reduced matrix element. The definition of [161] matches that of [186] and [58], but not of [64].

Table C.2: Numerical values of ac-Stark shift coefficients $\tilde{\beta}_{\mathrm{ac}}^{(k)}$ as in Table C.1, but for a selection of 1S-$n$S, 1S-$n$D, and 2S-$n$D transitions relevant for precision spectroscopy, with the laser frequency chosen such that two-photon excitation is resonant. These results extend the values given in Tables IV, VI, and VII in [58] to arbitrary laser polarizations.

| 1S/2S-$nl$ | $\tilde{\beta}_{\mathrm{ac}}^{(0)}(1\mathrm{S}/2\mathrm{S})$ (Hz/(W/m$^2$)) | $\tilde{\beta}_{\mathrm{ac}}^{(0)}(nl)$ (Hz/(W/m$^2$)) | $\tilde{\beta}_{\mathrm{ac}}^{(1)}(nl)$ (Hz/(W/m$^2$)) | $\tilde{\beta}_{\mathrm{ac}}^{(2)}(nl)$ (Hz/(W/m$^2$)) |
|---|---|---|---|---|
| 1S-2S | $4.638\,90 \times 10^{-5}$ | $-2.423\,61 \times 10^{-4}$ | — | — |
| 1S-3S | $5.232\,59 \times 10^{-5}$ | $-1.698\,88 \times 10^{-4}$ | — | — |
| 1S-4S | $5.513\,13 \times 10^{-5}$ | $-1.500\,80 \times 10^{-4}$ | — | — |
| 1S-3D | $5.232\,59 \times 10^{-5}$ | $-3.661\,18 \times 10^{-4}$ | $-4.152\,06 \times 10^{-6}$ | $1.600\,28 \times 10^{-5}$ |
| 1S-4D | $5.513\,13 \times 10^{-5}$ | $-3.296\,35 \times 10^{-4}$ | $-1.895\,42 \times 10^{-6}$ | $5.766\,42 \times 10^{-6}$ |
| 2S-3D | $1.244\,99 \times 10^{-3}$ | $2.038\,59 \times 10^{-2}$ | $-2.641\,00 \times 10^{-2}$ | $6.122\,09 \times 10^{-3}$ |
| 2S-4D | $1.641\,64 \times 10^{-3}$ | $-9.483\,44 \times 10^{-3}$ | $1.561\,98 \times 10^{-3}$ | $-2.009\,10 \times 10^{-4}$ |
| 2S-5D | $2.024\,50 \times 10^{-3}$ | $-6.778\,24 \times 10^{-3}$ | $1.960\,43 \times 10^{-4}$ | $2.326\,76 \times 10^{-4}$ |
| 2S-6D | $2.362\,14 \times 10^{-3}$ | $-5.903\,79 \times 10^{-3}$ | $-1.012\,21 \times 10^{-5}$ | $2.009\,71 \times 10^{-4}$ |
| 2S-7D | $2.647\,76 \times 10^{-3}$ | $-5.494\,76 \times 10^{-3}$ | $-4.724\,59 \times 10^{-5}$ | $1.492\,97 \times 10^{-4}$ |
| 2S-8D | $2.884\,51 \times 10^{-3}$ | $-5.264\,86 \times 10^{-3}$ | $-4.901\,52 \times 10^{-5}$ | $1.101\,76 \times 10^{-4}$ |
| 2S-9D | $3.079\,12 \times 10^{-3}$ | $-5.120\,65 \times 10^{-3}$ | $-4.296\,99 \times 10^{-5}$ | $8.259\,88 \times 10^{-5}$ |
| 2S-10D | $3.238\,84 \times 10^{-3}$ | $-5.023\,32 \times 10^{-3}$ | $-3.595\,84 \times 10^{-5}$ | $6.313\,19 \times 10^{-5}$ |
| 2S-11D | $3.370\,27 \times 10^{-3}$ | $-4.954\,10 \times 10^{-3}$ | $-2.971\,58 \times 10^{-5}$ | $4.916\,07 \times 10^{-5}$ |
| 2S-12D | $3.478\,96 \times 10^{-3}$ | $-4.902\,87 \times 10^{-3}$ | $-2.455\,86 \times 10^{-5}$ | $3.893\,73 \times 10^{-5}$ |

Likewise, the coupling of $\boldsymbol{J}$ with the nuclear spin $\boldsymbol{I}$ to $\boldsymbol{F} = \boldsymbol{J} + \boldsymbol{I}$ results in hyperfine structure (HFS) levels $|n, l, J, F\rangle$. $\boldsymbol{J}$ and $\boldsymbol{I}$ also commute, giving

$$\langle n, l, J, F||Q^{(k,\pm)}||n, l, J, F\rangle = (-1)^{J+I+F+k}\langle n, l, J||Q^{(k,\pm)}||n, l, J\rangle(2F+1)\begin{Bmatrix} J & F & I \\ F & J & k \end{Bmatrix}. \tag{C.19}$$

Finally, we can give an expression of the ac-Stark shift coefficient $\beta_{\mathrm{ac},q}(\tilde{\phi})$ of the HFS level $|\tilde{\phi}\rangle = |n, l, J, F, m_F\rangle$ and for laser polarization $q$ as

$$\begin{aligned} \beta_{\mathrm{ac},q}(\tilde{\phi}) = &\sum_k (-1)^{l+S+2J+I+2F+2k-m_F}(2J+1)(2F+1) \\ &\times \begin{pmatrix} F & k & F \\ -m_F & 0 & m_F \end{pmatrix} \begin{Bmatrix} l & J & S \\ J & l & k \end{Bmatrix} \begin{Bmatrix} J & F & I \\ F & J & k \end{Bmatrix} \\ &\times a^{(k)}(q)\, \tilde{\beta}_{\mathrm{ac}}^{(k)}(nl), \end{aligned} \tag{C.20}$$

where we have defined in analogy to Eq. (C.14)

$$\tilde{\beta}_{\mathrm{ac}}^{(k)}(nl) = \begin{cases} \langle n, l||Q^{(k,-)}||n, l\rangle + \langle n, l||Q^{(k,+)}||n, l\rangle & k = 0, 2 \\ \langle n, l||Q^{(k,-)}||n, l\rangle - \langle n, l||Q^{(k,+)}||n, l\rangle & k = 1. \end{cases} \tag{C.21}$$

Note that for atomic S levels ($l = 0$), $\tilde{\beta}_{\mathrm{ac}}^{(k)}(nl)$ is nonzero only for $k = 0$.

To facilitate the comparison of our results, e.g., with [58], the values given for $\tilde{\beta}_{\mathrm{ac}}^{(k)}(n,l)$ here correspond to the case of an infinitely heavy nucleus. To take the finite nucleus mass $m_{\mathrm{N}}$ into account, the coefficients $\tilde{\beta}_{\mathrm{ac}}^{(k)}(n,l)$ have to be multiplied by a factor

$$\left(\frac{m_{\mathrm{N}}+m_{\mathrm{e}}}{m_{\mathrm{N}}}\right)^3, \tag{C.22}$$

where $m_{\mathrm{e}}$ is the electron mass. For atomic hydrogen [46], this factor is equal to 1.001 63.

In Table C.1 we give numerical values of $\tilde{\beta}_{\mathrm{ac}}^{(k)}(nl)$ for the one-photon 2S-$n$P transitions considered in this work. Results for the two-photon 1S-$n$S, 1S-$n$D, and 2S-$n$D transitions is also given in Table C.2, extending the results of [58] for arbitrary laser polarizations. Note that the ac-Stark shift coefficients $\beta_{\mathrm{ac}}^{(k)}(nl)$ given in [58] already include the laser polarization factor $a^{(k)}(q)$ for linear polarization ($q=0$), i.e., $\beta_{\mathrm{ac}}^{(k)}(nl) = a^{(k)}(0)\tilde{\beta}_{\mathrm{ac}}^{(k)}(nl)$.

### C.1.6 ac-Stark shift for counter-propagating laser beams

So far, we have only considered the interaction with a light field with constant intensity $I_{\mathrm{loc}}$. An important special case is the interaction with the superposition of two counter-propagating light beams with identical frequencies.

#### C.1.6.1 Counter-propagating beams with parallel linear polarizations

We first consider the case where the two light beams are both linearly polarized along the same axis, here taken to be the $z$-axis and thus corresponding to $q=0$. Assuming both beams have the same electric field amplitude $\tilde{E}_0$, the combined electric field is given by

$$\begin{aligned}\tilde{\boldsymbol{E}}_{\mathbf{lin}\|\mathbf{lin}}(t,x) &= \tilde{E}_0\hat{\boldsymbol{z}}\left(\cos(\omega_{\mathrm{L}}t - K_{\mathrm{L}}x) + \cos(\omega_{\mathrm{L}}t + K_{\mathrm{L}}x)\right)\\ &= 2\tilde{E}_0\hat{\boldsymbol{z}}\cos(\omega_{\mathrm{L}}t)\cos(K_{\mathrm{L}}x),\end{aligned} \tag{C.23}$$

where $K_{\mathrm{L}}$ is the wavenumber of the laser beams, and the beams are taken to propagate along the $x$-axis. The resulting total intensity is $I_{\mathrm{lin}\|\mathrm{lin}}(x) = 4I\cos^2(K_{\mathrm{L}}x)$, with the intensity $I = c\epsilon_0\tilde{E}_0^2/2$ of each of the beams. Thus, the total intensity $I_{\mathrm{lin}\|\mathrm{lin}}(x)$ is spatially modulated, with the two beams forming an intensity standing wave. Both the excitations of the 1S-2S transition and the 2S-$n$P transition take place in such a configuration in the experiments described in this work.

Since the position and time dependence of the electric field $\tilde{\boldsymbol{E}}_{\mathbf{lin}\|\mathbf{lin}}(t,x)$ can be separated and $Q_q^{\mathrm{LS}}$ does not act on the atom's position, the ac-Stark shift at position $x$ is then described by using $I_{\mathrm{loc}} = I_{\mathrm{lin}\|\mathrm{lin}}(x)$ and $q=0$ in Eq. (C.4).

While Eq. (C.4) describes the ac-Stark shift of the atomic levels at any position and time, in the experiment we obtain an integrated signal of many atoms, traveling with a range of velocities $v_x$ along the intensity standing wave. We are interested in the ac-Stark shift of the observed transition frequency, derived from this signal. In many cases, the modulation of the ac-Stark shift due to intensity modulations will then average out[1], e.g., by the atoms moving through many modulations as the signal is collected or probing a range of intensities simultaneously. To describe the ac-Stark shift of the signal, it is advantageous to work in the atom's reference frame, where the two beams have their frequencies Doppler-shifted by the equal and opposite amount, $\Delta\nu_{\mathrm{D}} = \pm K_{\mathrm{L}} v_x/2\pi$, from the frequency in the laboratory frame.

[1]Note that intensity modulations on a larger spatial case, such as from the intensity profile of the Gaussian laser beams, typically cannot be averaged over and are taken into account explicitly in this work.

For the excitation of the 1S-2S transition, this situation has been discussed in [58] for very similar experimental parameters as used in this work. It is found that in the case where the Doppler shift is much larger than any other characteristic frequencies such as the Rabi frequency and the linewidth (see Eq. (19) of [58]), applicable here, the problem can be treated as if there is a constant ac-Stark shift from each of the beams. The total ac-Stark shift of the signal is then just given by setting $I_{\mathrm{loc}} = 2I$. Put another way, for averaging to occur, the time scale of the intensity modulation, given by $\pi/v_x K_{\mathrm{L}} = \pi/\Delta\nu_{\mathrm{D}}$, must be much shorter than the excitation time scale, given by the Rabi frequency and the linewidth.

When probing the 2S-$n$P transitions in this work, on the other hand, the Doppler shift can be much smaller than these characteristic frequencies, since the standing wave is oriented perpendicular to the atomic beam. For a complete picture, Eq. (C.4) then needs to be included in the optical Bloch equations derived in Chapters 2 and 3. However, the intensity modulations are still expected to average out to a great extend, since we observe the signal from many atoms, all sampling different trajectories through the standing light wave[1]. Furthermore, as discussed in Chapter 3, the atoms are partially delocalized over the standing wave. Finally, the non-resonant ac-Stark shift is orders of magnitude below the measurement uncertainty, and thus we here can set $I_{\mathrm{loc}} = 2I$ to estimate its contribution.

#### C.1.6.2 Counter-propagating beams with orthogonal circular polarizations

Another configuration of interest is the superposition of two counter-propagating light beams with orthogonal circular polarizations, i.e., with their fields rotating in the opposite sense in the laboratory frame. This configuration is, e.g., used in the spectroscopy of the 1S-3S transition in atomic hydrogen in [137]. Taking the beam with $\sigma^+$ ($\sigma^-$) polarization, corresponding to $q = +1$ ($q = -1$), and electric field amplitude $\tilde{E}_{\sigma^+}$ ($\tilde{E}_{\sigma^-}$) to be propagating in the positive (negative) direction along the $z$-axis, the combined electric field reads (see Eq. (C.2))

$$\begin{aligned}\tilde{\boldsymbol{E}}_{\boldsymbol{\sigma}^+,\boldsymbol{\sigma}^-}(t,z) = {} & \frac{\tilde{E}_{\sigma^+}}{2}\left(-e^{-i\omega_{\mathrm{L}}t+iK_{\mathrm{L}}z}\hat{\boldsymbol{r}}_+ + e^{i\omega_{\mathrm{L}}t-iK_{\mathrm{L}}z}\hat{\boldsymbol{r}}_-\right) \\ & + \frac{\tilde{E}_{\sigma^-}}{2}\left(e^{-i\omega_{\mathrm{L}}t-iK_{\mathrm{L}}z}\hat{\boldsymbol{r}}_- - e^{i\omega_{\mathrm{L}}t+iK_{\mathrm{L}}z}\hat{\boldsymbol{r}}_+\right),\end{aligned} \tag{C.24}$$

where $K_{\mathrm{L}}$ is the wavenumber of the laser beams. The total intensity $I_{\sigma^+,\sigma^-} = I_{\sigma^+} + I_{\sigma^-}$ of this configuration is simply the sum of the individual intensities of the two beams, $I_{\sigma^\pm} = c\epsilon_0\tilde{E}^2_{\sigma^\pm}/2$, and thus there is no spatial modulation as in the case of parallel linear polarizations.

To calculate the resulting ac-Stark shift from such a field, we evaluate the effect of the perturbation $V = -e\boldsymbol{r}\tilde{\boldsymbol{E}}_{\boldsymbol{\sigma}^+,\boldsymbol{\sigma}^-}(t,x)$ within second-order time-dependent perturbation theory. Again neglecting terms oscillating at twice the optical frequency, the only remaining terms proportional to $\tilde{E}_{\sigma^+}\tilde{E}_{\sigma^-}$, corresponding to a virtual absorption of a photon from one beam and a virtual emission of a photon into the other beam (and vice versa), contain the combination of dipole matrix elements

$$\langle\phi|r_{\pm q}|u\rangle\langle u|r_{\pm q}|\phi\rangle. \tag{C.25}$$

[1] One caveat is that atoms experiencing a greater ac-Stark shift because of larger local intensity generally tend to contribute more to the detected signal, and thus the ac-Stark shift of the averaged signal does not necessarily equal the intensity-averaged ac-Stark shift.

Table C.3: Numerical values of photoionization coefficients $\tilde{\beta}_{\text{ac}}^{(k)}$ for the 2S-$n$P transitions, with the laser frequency chosen such that one-photon excitation is resonant, analogous to the the ac-Stark shift coefficients given in Table C.1.

| 2S-$n$P | $\tilde{\beta}_{\text{ioni}}^{(0)}(n\text{P})$ (Hz/(W/m$^2$)) | $\tilde{\beta}_{\text{ioni}}^{(1)}(n\text{P})$ (Hz/(W/m$^2$)) | $\tilde{\beta}_{\text{ioni}}^{(2)}(n\text{P})$ (Hz/(W/m$^2$)) |
|---|---|---|---|
| 2S-3P | $-2.356\,55 \times 10^{-3}$ | $1.710\,15 \times 10^{-3}$ | $-7.830\,85 \times 10^{-4}$ |
| 2S-4P | $-3.153\,85 \times 10^{-4}$ | $2.159\,28 \times 10^{-4}$ | $-1.148\,32 \times 10^{-4}$ |
| 2S-5P | $-1.039\,75 \times 10^{-4}$ | $6.947\,13 \times 10^{-5}$ | $-3.918\,58 \times 10^{-5}$ |
| 2S-6P | $-4.824\,30 \times 10^{-5}$ | $3.182\,99 \times 10^{-5}$ | $-1.849\,45 \times 10^{-5}$ |
| 2S-7P | $-2.673\,68 \times 10^{-5}$ | $1.751\,05 \times 10^{-5}$ | $-1.035\,06 \times 10^{-5}$ |
| 2S-8P | $-1.652\,02 \times 10^{-5}$ | $1.076\,85 \times 10^{-5}$ | $-6.434\,87 \times 10^{-6}$ |
| 2S-9P | $-1.098\,67 \times 10^{-5}$ | $7.138\,67 \times 10^{-6}$ | $-4.297\,22 \times 10^{-6}$ |
| 2S-10P | $-7.706\,17 \times 10^{-6}$ | $4.995\,77 \times 10^{-6}$ | $-3.022\,91 \times 10^{-6}$ |

These terms however give no contribution, since $\langle u|r_{\pm q}|\phi\rangle$ implies $m_u = m_\phi \pm 1$, while $\langle \phi|r_{\pm q}|u\rangle$ implies $m_u = m_\phi \mp 1$, corresponding to an unphysical situation. Thus, the expression for the ac-Stark shift only contains terms proportional to $\tilde{E}_{\sigma^+}\tilde{E}_{\sigma^+}$ or $\tilde{E}_{\sigma^-}\tilde{E}_{\sigma^-}$, but not to $\tilde{E}_{\sigma^+}\tilde{E}_{\sigma^-}$.

In other words, when absorbing a photon with polarization $q = \pm 1$, leading to an excitation of a virtual level $|u\rangle$ with $m_u = m_\phi \pm 1$, the stimulated emission leading back to the initial level must correspond to the emission of a photon with polarization $q = \pm 1$ such that $m_\phi = m_u \mp 1$. This is only possible if both photons are from the same beam.

Thus, in this case the total ac-Stark shift $\Delta\nu_{\text{ac}}(\tilde{\phi})$ of the atomic levels $|\tilde{\phi}\rangle$ is the sum of the ac-Stark shifts from each of the beams as given in Eq. (C.4), i.e.,

$$\Delta\nu_{\text{ac}}(\tilde{\phi}) = \beta_{\text{ac},+1}(\tilde{\phi})I_{\sigma^+} + \beta_{\text{ac},-1}(\tilde{\phi})I_{\sigma^-} \tag{C.26}$$

with the coefficients $\beta_{\text{ac},\pm}(\tilde{\phi})$ as given in Eq. (C.20). In the special case where $|\tilde{\phi}\rangle$ is an atomic S level, $\beta_{\text{ac},-1}(\tilde{\phi}) = \beta_{\text{ac},1}(\tilde{\phi}) = \beta_{\text{ac},0}(\tilde{\phi})$, since $\tilde{\beta}_{\text{ac}}^{(k)}(nl) = 0$ for $k \neq 0$.

## C.2 Photoionization

Photoionization is the ionization of the atom through absorption of a photon of sufficient energy, resulting in a free electron and an ionized atom. This effect is contained in the light shift operator defined in the previous section (Eq. (C.1)) and manifests itself through an imaginary part in the dynamic polarizability $P_{\omega_{\text{L}},q}$ (Eq. (C.3)) [58, 64]. The rate $\gamma_{\text{ioni}}(\phi)$ at which the level $|\phi\rangle$ is ionized is then given by

$$\gamma_{\text{ioni}}(\phi) = -\frac{e^2}{\hbar\epsilon_0 c} I_{\text{loc}} \,\text{Im}\Big[P_{\omega_{\text{L}},q}(\phi)\Big] = 2\pi\beta_{\text{ioni}}(\phi)I_{\text{loc}}. \tag{C.27}$$

In analogy to the ac-Stark shift coefficient $\beta_{\text{ac},q}(\phi)$, $\beta_{\text{ioni}}(\phi)$ is the (real-valued) ionization coefficient, given in units of Hz/(W/m$^2$).

Table C.4: Numerical values of photoionization coefficients $\tilde{\beta}_{\mathrm{ac}}^{(k)}$ for 1S-$n$S, 1S-$n$D, and 2S-$n$D transitions, with the laser frequency chosen such that two-photon excitation is resonant, analogous to the the ac-Stark shift coefficients given in Table C.2.

| 1S/2S-$nl$ | $\tilde{\beta}_{\mathrm{ioni}}^{(0)}(nl)$ (Hz/(W/m$^2$)) | $\tilde{\beta}_{\mathrm{ioni}}^{(1)}(nl)$ (Hz/(W/m$^2$)) | $\tilde{\beta}_{\mathrm{ioni}}^{(2)}(nl)$ (Hz/(W/m$^2$)) |
|---|---|---|---|
| 1S-2S | $-2.082\,06 \times 10^{-4}$ | — | — |
| 1S-3S | $-3.502\,91 \times 10^{-5}$ | — | — |
| 1S-4S | $-1.231\,12 \times 10^{-5}$ | — | — |
| 1S-3D | $-6.364\,11 \times 10^{-6}$ | $5.874\,56 \times 10^{-6}$ | $-2.590\,44 \times 10^{-6}$ |
| 1S-4D | $-2.411\,19 \times 10^{-6}$ | $2.193\,84 \times 10^{-6}$ | $-9.934\,92 \times 10^{-7}$ |
| 2S-4D | $-3.318\,77 \times 10^{-3}$ | $3.056\,57 \times 10^{-3}$ | $-1.353\,48 \times 10^{-3}$ |
| 2S-5D | $-1.012\,77 \times 10^{-3}$ | $9.129\,92 \times 10^{-4}$ | $-4.205\,07 \times 10^{-4}$ |
| 2S-6D | $-4.550\,79 \times 10^{-4}$ | $4.057\,29 \times 10^{-4}$ | $-1.906\,56 \times 10^{-4}$ |
| 2S-7D | $-2.479\,88 \times 10^{-4}$ | $2.196\,75 \times 10^{-4}$ | $-1.044\,32 \times 10^{-4}$ |
| 2S-8D | $-1.516\,84 \times 10^{-4}$ | $1.338\,18 \times 10^{-4}$ | $-6.408\,42 \times 10^{-5}$ |
| 2S-9D | $-1.002\,15 \times 10^{-4}$ | $8.816\,71 \times 10^{-5}$ | $-4.243\,13 \times 10^{-5}$ |
| 2S-10D | $-6.997\,25 \times 10^{-5}$ | $6.144\,02 \times 10^{-5}$ | $-2.967\,20 \times 10^{-5}$ |
| 2S-11D | $-5.093\,60 \times 10^{-5}$ | $4.466\,09 \times 10^{-5}$ | $-2.162\,38 \times 10^{-5}$ |
| 2S-12D | $-3.830\,86 \times 10^{-5}$ | $3.355\,27 \times 10^{-5}$ | $-1.627\,69 \times 10^{-5}$ |

The evaluation of the integrals giving the imaginary part of $P_{\omega_{\mathrm{L}},q}$ is not always straightforward. Instead, we here directly calculate the photoionization coefficient using first-order time-dependent perturbation theory[1], which can be shown to be formally equivalent to the former approach in this case [58, 64]. This approach gives

$$\beta_{\mathrm{ioni}}(\phi) = \frac{\pi e^2}{h\epsilon_0 c} \sum\!\!\!\!\!\!\int_u |\langle u|r_q|\phi\rangle|^2. \tag{C.28}$$

The sum is over all continuum levels $|u\rangle = |n_u, l_u, m_u\rangle$ with energy $E_u = E_\phi + \hbar\omega_{\mathrm{L}}$ and a nonzero dipole matrix element $\langle\phi|r_q|u\rangle$, as determined by the dipole selection rules (e.g., a P level is connected to S and D continuum levels).

Since the structure of the underlying operator is identical for both the ac-Stark shift and photoionization, the photoionization coefficient $\beta_{\mathrm{ioni}}(\tilde{\phi})$ of any HFS level $|\tilde{\phi}\rangle$ can also be given as a combination of the coefficients $\tilde{\beta}_{\mathrm{ioni}}^{(k)}(nl)$. $\beta_{\mathrm{ioni}}(\tilde{\phi})$ can then be calculated using Eq. (C.20) by substituting $\tilde{\beta}_{\mathrm{ac}}^{(k)}(nl)$ with $\tilde{\beta}_{\mathrm{ioni}}^{(k)}(nl)$. Note that since only the "$-$" in Eq. (C.1) contributes to photoionization, $\tilde{\beta}_{\mathrm{ioni}}^{(k)}(nl) = \langle n,l||Q^{(k,-)}||n,l\rangle$, as opposed to the corresponding definition of $\tilde{\beta}_{\mathrm{ac}}^{(k)}(nl)$ in Eq. (C.21). The correction factor given in Eq. (C.22) and the discussion regarding intensity modulations also apply. Numerical values of $\tilde{\beta}_{\mathrm{ioni}}^{(k)}(nl)$ are given in Table C.3 for the one-photon 2S-$n$P transitions, and in Table C.4 for the two-photon 1S-$n$S, 1S-$n$D, and 2S-$n$D transitions, again extending the results of [58] for arbitrary laser polarizations.

[1] Note that in first-order perturbation theory, the amplitude of the photoionized level $|u\rangle$ perturbing the atomic level $|\phi\rangle$ is proportional to $\langle u|r_q|\phi\rangle$, but the probability to find the system in $|u\rangle$ is proportional to $|\langle u|r_q|\phi\rangle|^2$.

# List of changes in this revised version

- Table 2.1 (and related Table B.1) and multiple sections: added reduced-mass correction in dipole moments and updated values of Rabi frequencies, linewidths, decay rates, ac-Stark shift and photoionization coefficients accordingly; updated caption, adding reference [47]; updated 2S-6P linewidth throughout text.
- Table 2.1, footnote a: corrected Ly-$\epsilon$ signal contribution from $\sim$99 % to $\sim$97 %.
- Section 2.4: added details on simulations and note on spectral detection efficiency dependence of dc-Stark shift coefficients.
- Table 2.6: scaled perturbative dc-Stark shift coefficients to account for reduced-mass correction in dipole moments; added additional simulation parameters in caption.
- Section 3.3.4: clarified effect of propagation distance on transverse coherence length.
- Sections 3.3.4 and 3.4.2: corrected some numerical values from Wigner function with intermediate aperture.
- Section 3.3.5: clarified wording on sum over signals from fully delocalized atoms.
- Section 3.4.1, Eq. (3.44): corrected $(\Delta p_\mathrm{D}/\hbar K_\mathrm{K}) \rightarrow (\Delta p_\mathrm{D}/\hbar K_\mathrm{K})^2$.
- Fig. 4.1: added missing label “AB” and corrected label “RC”.
- Section 4.3: added definition of 1S-2S preparation laser atomic detuning $\Delta\nu_\mathrm{1S\text{-}2S}$ (and using wording “atomic detuning” throughout).
- Sections 4.4 and 4.4.1: added definition of 2S-6P spectroscopy laser detuning $\Delta\nu_\mathrm{2S\text{-}6P}$.
- Section 4.4.6: added note that even in the absence of circular birefringence ($\phi = 0$), the linear laser polarization angle cannot be reconstructed.
- Section 4.6.2: corrected CEM bias current equation and direction.
- Section 4.6.3: corrected figure reference (Fig. 4.37 → Fig. 4.36).
- Section 4.8: corrected laser drift rates; removed parts of incomplete discussion of laser spectra.
- Section 5.1.1.5: corrected definition of standard deviation of reduced chi-squared distribution.
- Table 5.3: corrected distance of 243 nm waist to nozzle orifice.
- Section 6.2.3: corrected quantum interference shift amplitudes to 5.1 kHz and 2.6 kHz for the $2S\text{-}6P_{1/2}$ and $2S\text{-}6P_{3/2}$ transitions, respectively; clarified that data group G1B belongs to $2S\text{-}6P_{1/2}$ transition.
- Section 6.2.4.6: corrected equation references.
- Tables B.3 and B.5: corrected Ba-$\gamma$ → Ba-$\beta$.
- Bibliography: added erratum [36] to reference [35]; corrected author lists of [27, 35]; there are two additional references in total, shifting reference numbers throughout document.
- Corrected various typos and grammatical issues throughout document.